\documentclass[a4paper,two side,openright,11pt]{report}

\usepackage{etex}
\makeatletter
\def\@font@warning#1{} 
\makeatother

\usepackage[utf8]{inputenc}
\usepackage[T1]{fontenc}
\usepackage[english]{babel} 
\makeatletter
\let\asme@citex\@citex 
\makeatother
\makeatletter
\let\@citex\asme@citex 
\makeatother

\usepackage{mathpazo}	
	
\usepackage{eurosym} 	
\usepackage{textcomp}	
\usepackage{fontawesome}
\usepackage{textalpha}

\usepackage{soul} 				
\usepackage[outline]{contour} 	
\contourlength{5pt}				
\usepackage{fancybox} 			
\usepackage{enumitem} 			
\usepackage{setspace} 			
\usepackage{parskip}
\usepackage{color} 				
\usepackage{colortbl} 			
\usepackage[dvipsnames,svgnames,rgb]{xcolor} 
\newcommand{\changed}[1]{{\color{black}#1}} 

\usepackage{amsthm}		
\usepackage{amsmath}	
\allowdisplaybreaks
\usepackage{amssymb}	
\numberwithin{equation}{section}
\usepackage{sansmath}	
\usepackage{bm} 		
\usepackage{thmtools} 	
\usepackage{cases} 		
\usepackage{empheq}		
\usepackage{cancel} 	
\usepackage{mathrsfs}   

\usepackage{graphicx} 	
\usepackage{pict2e}		
\usepackage{wrapfig} 	
\usepackage{picins} 	
\usepackage{float}		
\usepackage{subfig}		
\usepackage{rotating} 	
\usepackage{epstopdf}	
\usepackage{graphpap}	
\usepackage[justification=centering,textfont=it]{caption}	
\usepackage{pgfplots}
\pgfplotsset{compat=newest}
\usepackage{mdframed}

\usepackage{tikz} 								
\usetikzlibrary{decorations.pathmorphing} 		
\usetikzlibrary{arrows,decorations.markings}	
\usetikzlibrary{calc}							
\usetikzlibrary{backgrounds}					
\usetikzlibrary{shapes.geometric}				
\usepackage{pgfplots}							
\usepgfplotslibrary{polar}						
\usetikzlibrary{decorations.text}				
\usetikzlibrary{arrows.meta}					
\usetikzlibrary{patterns}						

\usepackage{multirow} 	
\usepackage{colortbl} 	
\usepackage{makecell} 	
\usepackage{booktabs}	

\usepackage[top=2.5cm, bottom=3.5cm, left=2.5cm, right=2.5cm, footskip=2cm,headheight=1437pt]{geometry}
\usepackage{fancyhdr} 	
\usepackage{titlesec} 	
\usepackage{xspace} 	
\usepackage{layout} 	
\usepackage{multicol}	
\usepackage[final]{pdfpages} 	
\usepackage{tocbibind}

\usepackage[bookmarksnumbered]{hyperref} 	
\hypersetup{
	colorlinks=true, 	
	breaklinks=true, 	
	urlcolor= blue, 	
	linkcolor= blue, 	
	citecolor= BrickRed, 	
	linktocpage=true	
}

\colorlet{headbgcolor}{black!40}
\def\headrulefill{{\color{black}\leaders\hrule width 0pt height 3pt depth -2.8pt \hfill}}

\fancypagestyle{plain}{
	\fancyhf{}
}

\titleformat{\chapter}[frame]
{\normalfont \color{black}}{\sffamily \large \enspace 
\chaptertitlename \enspace \large \thechapter \enspace}
{20pt}{\fontfamily{pzc}\selectfont \Huge \filcenter}
\titlespacing{\chapter}{0cm}{-1cm}{1.5cm}
\titleformat{\section}  
{\sffamily \fontsize{14}{16} \bfseries}{\thesection}{1em}{}[{\titlerule[0.5pt]}\vspace{5pt}]
\titleformat{\subsection}
{\sffamily \fontsize{12}{15}\bfseries}{\thesubsection}{1em}{}
\titleformat{\subsubsection}
{\sffamily \fontsize{11}{11}\bfseries\itshape}{\thesubsubsection}{1em}{}
\let\originalparagraph\paragraph
\renewcommand{\paragraph}[2][.]{ \originalparagraph{\sffamily #2#1}}

\usepackage{tcolorbox} 
\usepackage[absolute,overlay]{textpos}
\usepackage{blindtext}

\newcommand{\D}{\text d}
\newcommand{\p}{\partial}
\newcommand{\ndelta}{\delta\hspace{-0.50em}\slash\hspace{-0.05em} }
\newcommand{\cI}{\mathscr I}
\newcommand{\Lie}{\mathcal L}
\newcommand{\rmg}{\sqrt{-g}}
\newcommand{\pref}{\frac{\sqrt{q}}{16\pi G}}
\newcommand{\Order}[1]{\mathcal{O}(r^{-#1})}
\newcommand{\prefwg}{\frac{1}{16\pi G}}
\newcommand{\mn}{{\mu\nu}}
\newcommand{\eps}{\epsilon}

\global\mdfdefinestyle{resultat}{
	linecolor=black!40,%
	linewidth=3pt,%
	topline=false,%
	bottomline=false,%
	rightline=false,%
	align=right,%
	backgroundcolor=black!3,%
	userdefinedwidth=0.97\textwidth,%
	innerleftmargin=10,%
	innertopmargin=10,%
	innerbottommargin=10,%
	innerrightmargin=10,
	skipabove=15,%
}
\newcommand{\resu}[2]{
\begin{mdframed}[style=resultat]
\textbf{#1.}  #2
\end{mdframed}
\vspace{8pt}
}
\newcommand{\resuwt}[1]{
\begin{mdframed}[style=resultat]
#1
\end{mdframed}
\vspace{8pt}
}

\makeatletter
\DeclareRobustCommand{\loplus}{\mathbin{\mathpalette\dog@lsemi{+}}}
\newcommand{\dog@rsemi}[2]{\dog@semi{#1}{#2}{-90,90}}
\newcommand{\dog@lsemi}[2]{\dog@semi{#1}{#2}{270,90}}
\newcommand{\dog@semi}[3]{%
  \begingroup
  \sbox\z@{$\m@th#1#2$}%
  \setlength{\unitlength}{\dimexpr\ht\z@+\dp\z@\relax}%
  \makebox[\wd\z@]{\raisebox{-\dp\z@}{%
    \begin{picture}(1,1)
    \linethickness{\variable@rule{#1}}
    \roundcap
    \put(0.5,0.5){\makebox(0,0){\raisebox{\dp\z@}{$\m@th#1#2$}}}
    \put(0.5,0.5){\arc[#3]{0.5}}
    \end{picture}%
  }}%
  \endgroup
}
\newcommand{\variable@rule}[1]{%
  \fontdimen8  
  \ifx#1\displaystyle\textfont3\else
    \ifx#1\textstyle\textfont3\else
      \ifx#1\scriptstyle\scriptfont3\else
        \scriptscriptfont3\relax
  \fi\fi\fi
}
\makeatother

\usepackage{cite}
\addto{\captionsenglish}{}

\usepackage{adjustbox}

\begin{document}

%
%

\thispagestyle{empty}

\begin{tikzpicture}[overlay,remember picture]
	\draw [xshift=4mm,line width=1.5pt,rounded corners=15pt]
        ($ (current page.north west) + (.5cm,-.5cm) $)
        rectangle
        ($ (current page.south east) + (-.5cm,.5cm) $);
    \draw [line width=1pt,rounded corners=15pt,black!75]
        ($ (current page.north west) + (.75cm,-.75cm) $)
        rectangle
        ($ (current page.south east) + (-.75cm,.75cm) $);
    \draw [line width=0.5pt,rounded corners=15pt,black!50]
        ($ (current page.north west) + (1cm,-1cm) $)
        rectangle
        ($ (current page.south east) + (-1cm,1cm) $);
\end{tikzpicture}


\begin{figure}[h!]
\begin{center}
\includegraphics[width=0.3\textwidth]{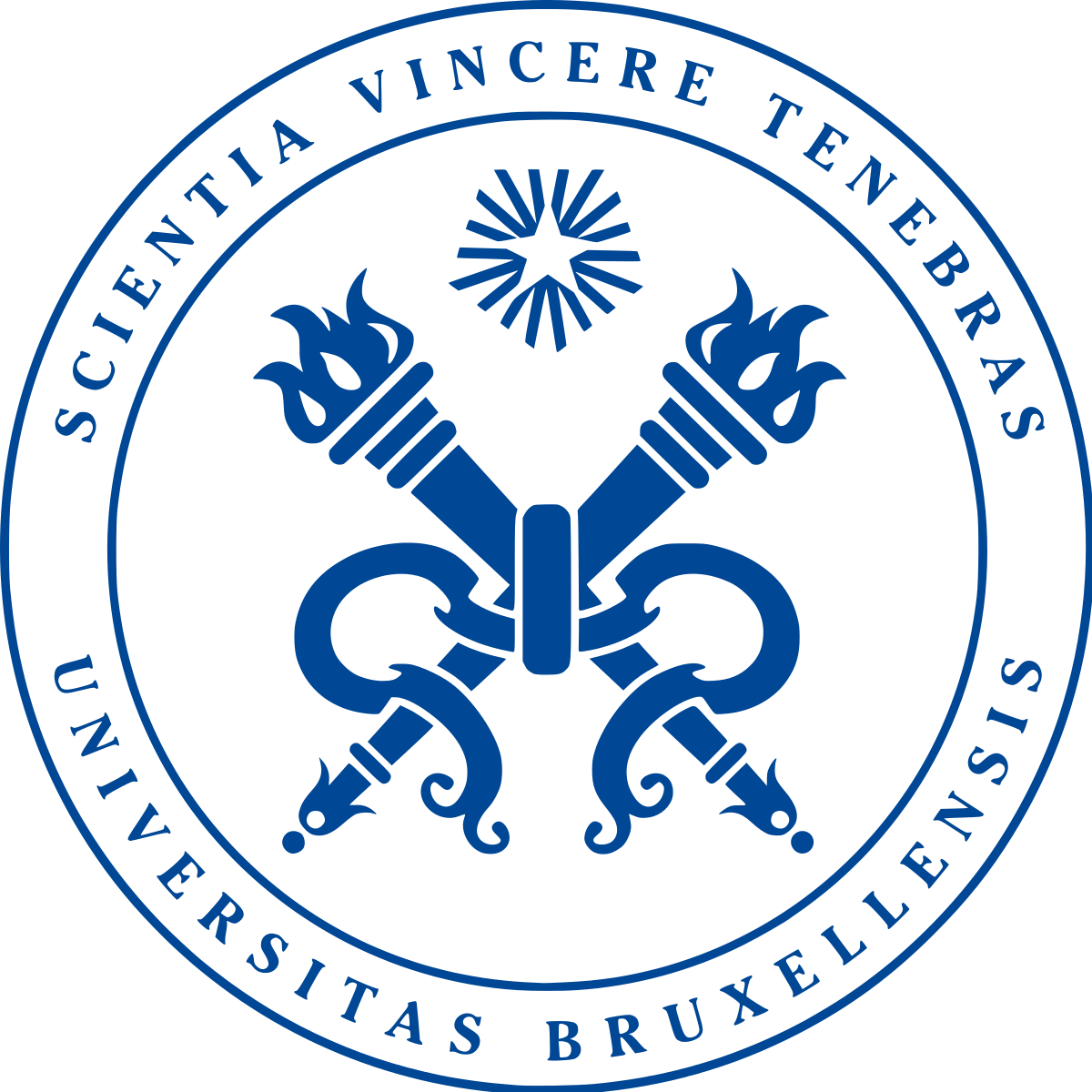}
\end{center}
\end{figure}

\vspace{25pt}

\begin{center}

\begin{minipage}[c]{0.65\textwidth}
\begin{center}
\textit{Universit\'e Libre de Bruxelles and International Solvay Institutes
CP 231, B-1050 Brussels, Belgium}
\end{center}
\end{minipage}

\vspace{25pt}

\begin{minipage}[c]{0.5\textwidth}
\begin{center}
Thesis submitted in fulfilment of the requirements of the \emph{PhD Degree in Sciences}. Academic year 2020-2021.
\end{center}
\end{minipage}

\vspace{35pt}

\begin{adjustbox}{minipage=[c][0.12\textheight]{0.75\textwidth},fbox}
	\begin{center}
	{\bfseries\huge Leaky covariant phase spaces}\\[5pt]
	{\bfseries\Large\itshape Theory and application to $\bm\Lambda$-BMS symmetry}
	\end{center}
\end{adjustbox}

\vspace{35pt}
{\LARGE \textbf{\centering Adrien Fiorucci}}

\vspace{20pt}

\textbf{Supervisors:} \\[7pt] Prof. Geoffrey Comp\`ere \&  Prof. Glenn Barnich.  

\vspace{10pt}

\textbf{Thesis jury:}\\[7pt]
Prof. Frank Ferrari (\textit{Universit\'e Libre de Bruxelles}),\\
Prof. Andr\`es Collinucci (\textit{Universit\'e Libre de Bruxelles}),\\
Prof. Daniel Grumiller (\textit{Technische Universit\"at Wien}),\\
Prof. Simone Speziale (\textit{Centre de Physique Th\'eorique Marseille}).

\end{center}

\newpage\thispagestyle{empty}
$ $
\newpage
\renewcommand{\thepage}{\Roman{page}}

%
%

\thispagestyle{empty}

\begin{center}
\textbf{--- Abstract ---}
\end{center}

The present thesis aims at providing a unified description of radiative phase spaces in General Relativity for any value of the cosmological constant using covariant phase space methods. We start by considering generic asymptotically locally (A)dS spacetimes with leaky boundary conditions in the Starobinsky/Fefferman-Graham gauge and in arbitrary dimensions. The boundary structure is allowed to fluctuate and plays the role of source yielding some flux of gravitational radiation at the boundary. The holographic renormalization procedure is employed to remove divergences from the presymplectic structure, which leads to finite surface charges for the whole class of boundary diffeomorphisms and Weyl rescalings. The charge algebra represents this asymptotic symmetry algebra under the Barnich-Troessaert bracket, up to a field-dependent 2-cocycle in odd spacetime dimensions. We then propose a boundary gauge fixing isolating the radiative components among the boundary degrees of freedom without constraining the Cauchy problem in asymptotically de Sitter spacetimes. This additional gauge fixing reduces the set of allowed boundary diffeomorphisms to the infinite-dimensional $\Lambda$-BMS algebroid, which is the counterpart to the Generalized BMS algebra of smooth supertranslations and super-Lorentz transformations in the flat limit. 

In a second round, the analysis is repeated in the Bondi gauge, which is better suited to discuss radiative phenomena as well as construct a flat limit process at the level of the solution space. Thanks to a diffeomorphism we translate the results previously obtained in the Starobinsky/Fefferman-Graham coordinates and identify the analogues of the Bondi news, mass and angular momentum aspects in the presence of a non-vanishing cosmological constant. We give a prescription to perform the flat limit at the level of the phase space and demonstrate how to use this connection to renormalize the corresponding phase space of asymptotically locally flat spacetimes at null infinity. The latter is made necessary as soon as the boundary structure of the gravitational field is allowed to vary under arbitrary super-Lorentz transformations. 

The last part of the manuscript is devoted to discussing the various implications of these super-Lorentz transformations as genuine asymptotic symmetries of asymptotically flat Einstein's gravity. In particular, we derive a closed-form expression of the orbit of gravitational vacua under the Generalized BMS symmetries. Transitions among these vacua are related to the refraction/velocity kick memory effect and the displacement memory effect. Finally, we give a physical prescription to define finite Hamiltonian generators canonically conjugated to Generalized BMS transformations on the subclass of physical solutions that are stationary at early and late times and comment on the enhancement of the infrared structure of gravity in the presence of super-Lorentz transformations.

\newpage\thispagestyle{empty}
\ \vfill 
\begin{center}
\begin{minipage}[b]{0.75\textwidth}
\begin{center}
The research compiled in this doctoral thesis was supported by the \textit{FRS-F.N.R.S. Belgium}. It was carried out at the \textit{Université de Bruxelles} (Belgium) from the 1st of October 2017 to the 1st of May 2021.  \\[10pt]
The present document was typeset in \LaTeX\ \texttt{report} class and written using \emph{Texmaker} (v.5.0.4) running on \emph{Microsoft Windows 10 Pro}. The figures are vector graphics drawn with the creation tools PGF/Ti\textit{k}Z. Some symbolic computations were performed by means of the \textit{Wolfram Mathematica} software (v.12.2). \\[10pt]
[ arXiv version ] \\[20pt]
\copyright\ \textbf{Adrien Fiorucci, 2021}.
\end{center}
\end{minipage}
\end{center}

%
%

\chapter*{Acknowledgements}
\markboth{Acknowledgements}{Acknowledgements}
\addcontentsline{toc}{chapter}{Acknowledgements}

Completing a doctoral thesis is as much a personal scientific challenge as a shared human adventure. Therefore, it seems fair to me to begin this manuscript by paying tribute to the various people who, from near or far, have contributed to making this first research experience truly magical.

First, I wish to show all my gratitude to \textit{Geoffrey Compère} for his exemplary supervision. Since early in my master studies, Geoffrey has been a passionate guide in the discovery of my future profession. Gifted with an impressive intuition, an extensive knowledge on both theoretical and experimental aspects of modern gravitational physics and a formidable mathematical ease, he represents the ideal of a brilliant researcher towards which I will strive. His unquenchable thirst for new scientific challenges, his constant enthusiasm and his attentive support always acted as boosters that made me progress and surpass myself. For his constant availability to discuss deep conceptual issues and provide concrete help in concluding a computation, for his patience, for his pedagogical qualities and finally for providing me with many project opportunities and stimulating ideas during five years of collaboration, I give him my heartfelt thanks. 

Furthermore, I am grateful to my co-supervisor, \textit{Glenn Barnich}, for the numerous discussions we had, during which he shared to me his prodigious knowledge on asymptotic symmetries in gauge theories among many other areas of physics. Working with Glenn is a true immersion into the finest subtleties in the field from which I always came back enriched. His ability to directly detect the core of the problem, formalize it in a mathematically robust and unequivocal way and never relax concentration before fully understanding the outcome are all rare qualities to be treasured in science. His powerful, rigorous and independent vision of scientific research will definitively inspire me in my future career. 

The third main actor who participated in the success of my early scientific enterprise was my colleague and friend \textit{Romain Ruzziconi}, in whom I discovered the most wonderful collaborator. I could spend several pages detailing the importance of his contribution to the numerous ideas developed in this manuscript, or even remembering the passionate and stimulating discussions we had. I thank him for the fantastic human he is, for the scientific journey we made together and for the future of our fruitful collaboration, which already promises to be rich in exciting projects.

In parallel to this research, I was able to free up a few hours to devote myself to another of my greatest passions: teaching. Therefore, I would also like to thank \textit{Glenn Barnich}, \textit{Nicolas Boulanger}, \textit{Frank Ferrari} and \textit{Serge Massar} for giving me the opportunity to teach alongside them during the past three years. I am delighted to have benefited from their advice and teaching experience to improve my pedagogical methods. 

I gratefully acknowledge the support of \textit{Riccardo Argurio} and \textit{Andrès Collinucci}, who formed my accompaniment committee with Geoffrey and Glenn. Moreover, I would like to express my sincere thanks to \textit{Andrès Collinucci, Daniel Grumiller, Frank Ferrari} and \textit{Simone Speziale} for agreeing to be my examiners for the final step of this Ph.D. adventure, and for the precious time they will dedicate to reading this document.

On the scientific level, I thank \textit{Abhay Ashtekar, Luca Ciambelli, Laura Donnay, Laurent Freidel, Marc Geiller, Marc Henneaux, Yannick Herfray, Stefan Hollands, Yegor Korovin, Charles Marteau, Prahar Mitra, Kévin Nguyen, Blagoje Oblak, Roberto Oliveri, Sabrina Pasterski, Daniele Pranzetti, Simone Speziale, Cédric Troessaert, Amitabh Virmani} and \textit{Céline Zwikel} for useful discussions and correspondence during my Ph.D. (apologies if I unintentionally forgot someone). I would also like to thank all the members of my research group for the stimulating dynamic they created around me. 

Looking towards the past, I would also like to pay tribute to \textit{Didier Salmon}, my first physics teacher (Athénée Royal de Gembloux, Belgium), who shared with me a passion for mathematical rigor as well as the wonder of scientific discovery. I also thank my teachers at the University of Namur (Belgium) for providing me the necessary push towards my final choice when I was undecided between an artistic and a scientific career. I feel especially indebted to \textit{André Füzfa} for having ignited in me the spark of love for General Relativity. 

On the private level, I would really like to thank my friends for their unwavering support and for making my life more exciting. My thanks goes particularly to \textit{Thomas Evrard} and \textit{Marvyn Gulina} for our unbreakable friendship which has lasted for 10 years now and to \textit{Antoine Pasternak} for countless fantastic evenings spent between science and entertainment, in Brussels or abroad, that will remain among the best memories of my youth.

Finally, I would like to express my infinite gratitude to \textit{my parents}, who are undoubtedly the determining agents of my success. From the beginning, with love, attention and benevolence, they made every effort to provide a peaceful and comfortable work space for me to evolve and progress. They also demonstrated a boundless patience facing my mood swings, my passionate outbursts and sometimes difficult moments. They served as solid examples of upright, courageous and respectable people, instilling in my mind determination and refusal of failure that characterizes quite well my personality. That marks a path that I am committed to follow during my whole life. 

My work was supported by the \textit{FRS-F.N.R.S. Belgium} (2017-2021), to which I address my earnest thanks for giving me the possibility to perform this exciting research.\hfill{\color{black!40}$\blacksquare$}

%
%

\chapter*{Conventions and notations}
\markboth{Conventions and notations}{Conventions and notations}
\addcontentsline{toc}{chapter}{Conventions and notations}

\paragraph{Geometry and units} We choose units such that the speed of light $c=1$ but we keep the Newton-Cavendish constant $G$ explicit. Time and lengths are therefore measured in SI units of meters. The spacetime manifold is denoted by the couple $(\mathscr M,g_{\mu\nu})$. The signature of the Lorentzian metric $g_{\mu\nu}$ obeys the mostly plus convention $(-,+,+,\dots)$. Bulk hypersurfaces and boundaries in $\mathscr M$ are denoted by curly letters $\mathscr B,\mathscr I,\mathscr U$ \textit{etc.} The spacetime dimension is $n\in\mathbb N_0$ ($n>1$). When a natural foliation is defined in the bulk, we write explicitly $n=d+1$ where $d$ represents the number of transverse dimensions with respect to the foliation.  

\paragraph{Indices} The notation of spacetime coordinates is as follows. Greek indices $\mu,\nu,\dots$ span the full dimension of spacetime $n=d+1$, so $\mu \in \lbrace 0,\dots, d \rbrace$. The index $0$ represents the coordinate along the direction of foliation if present. Latin indices label the transverse coordinates $x^a$ with $a \in \lbrace 1,\dots, d \rbrace$. Capital Latin letters are used to denote compact (angular) coordinates $x^A$ among the transverse coordinates. We use the Einstein convention for summation: repeated indices are supposed to be summed over their full range.

\paragraph{Affine connection and curvature} Concerning the definition of geometrical objects used in General Relativity, we follow the conventions adopted in the book \textit{Gravitation} \cite{Misner_1973} by Wheeler, Thorne and Misner. In particular, the Riemann-Christoffel tensor is determined in terms of the Christoffel symbols ${\Gamma^\mu}_{\nu\rho}$ as 
\[
R^\mu_{\phantom{\mu}\nu\alpha\beta} = \partial_\alpha \Gamma^\mu_{\phantom{\mu}\nu\beta} - \partial_\beta \Gamma^\mu_{\phantom{\mu}\nu\alpha} + \Gamma^\mu_{\phantom{\mu}\kappa\alpha} \Gamma^\kappa_{\phantom{\kappa}\nu\beta} - \Gamma^\mu_{\phantom{\mu}\kappa\beta} \Gamma^\kappa_{\phantom{\kappa}\nu\alpha}.
\]
In this convention, the $n$-sphere has positive Ricci scalar curvature, $R=R^\alpha_{\phantom{\alpha}\alpha}$, where the Ricci tensor is $R^{\alpha}_{\phantom{\alpha}\mu\alpha\nu}$. Accordingly, the Ricci curvature for vacuum configurations has the same sign as the cosmological constant, \textit{i.e.} $R>0$ for the de Sitter spacetime and $R<0$ for the anti-de Sitter spacetime. The Levi-Civita connection with respect to the bulk geometry on $\mathscr M$ is denoted by $\nabla_\mu$, while the induced connection on codimension 1 hypersurfaces is written $\mathcal D_a$. Additionally, the Levi-Civita connection with respect to the boundary geometry and its restriction to (codimension 2) angular coordinates will be denoted by $D_a$ and $D_A$ respectively.

\paragraph{Differential forms} In some coordinate patch $\{ x^\mu \}$, we define the basis of $(n-k)$ differential forms (also codimension $k$ forms) as
\[
(\D^{n-k}x)_{\mu_1\dots\mu_k} = \frac{1}{k!(n-k)!}\varepsilon_{\mu_1\dots\mu_k\nu_1\dots\nu_{n-k}}\D x^{\nu_1}\wedge\cdots\wedge\D x^{\nu_{n-k}},
\]
where $\varepsilon_{\mu_1\dots\mu_n}$ denotes the (numerically invariant) Levi-Civita symbol in $n$ dimensions. The $(n-k)$ differential forms on $\mathscr M$ are denoted by boldface symbols
\[
\bm A = A^{\mu_1\dots\mu_k}(\D^{n-k}x)_{\mu_1\dots\mu_k} \in \Omega^{n-k}(\mathscr M),
\]
where $\Omega^p(\mathscr M)$ represents the space of $p$-forms on $\mathscr M$. The convention for exterior derivative is $\D = \D x^\sigma\partial_\sigma$, which gives
\[
\D\bm A = \partial_\sigma A^{\mu_1\dots\mu_{k-1}\sigma}(\D^{n-k+1}x)_{\mu_1\dots\mu_k-1}.
\]
The volume form is $\sqrt{|g|} (\D^{n}x)$. If $r,t$ are two coordinates on $\mathscr M$, the integration measure on constant $r$ and constant $r,t$ hypersurfaces are given respectively by $(\D^{n-1}x)_r \equiv (\D^{n-1}x)$ and $(\D^{n-2}x)_{rt} \equiv (\D^{n-2}x)$ where the subscripts are dropped for the sake of conciseness if the submanifold on which we integrate is clearly specified.

\paragraph{Covariant phase space formalism} We use the conventions of \cite{Barnich:2007bf} regarding the formal tools of the covariant phase space formalism. Unless otherwise stated, $\phi = (\phi^i)$ represents a collection of fields and background structures needed to define the theory under consideration (with arbitrary index structure summarized by the compact index $i$ including tensorial index structure as well as labels of the various fields). $S[\phi]$ denotes the action integral of the theory, built from the Lagrangian top-form $\bm L[\phi]$. The variational operator on the phase space is denoted as $\delta$. Assuming general covariance, a field $\phi^i$ is modified under an infinitesimal diffeomorphism generated by the vector $\xi = \xi^\mu\partial_\mu$ by the Lie derivative $\delta_{\xi} \phi^i =+\Lie_\xi \phi^i$. Finally, the Poisson bracket is defined as $\{ H_{\xi_1},H_{\xi_2}\} = \delta_{\xi_2}H_{\xi_1}$ in order to represent the algebra of symmetry generators with correct signs. \hfill{\color{black!40}$\blacksquare$}

%
%

\tableofcontents

%
%


%
%

\listoffigures

%
%

\chapter{Introduction}
\renewcommand{\thepage}{\arabic{page}}
\setcounter{page}{1}

\section{A brief history of gravitational waves}
Einstein's theory of General Relativity \cite{1915SPAW.......778E} is often considered as one of the most exceptional achievements of science. Based on a small number of natural and robust postulates, among which one finds the equivalence principles or the principle of general covariance, it has initiated the modern way of geometrizing physics from first principles with the broad success we all celebrate, for instance in the building of the standard model of particle physics. Although refractory to quantification and technically very intricate, Einstein's paradigm of gravity is still capable of describing with astonishing precision all observable gravitational phenomena, whether at astrophysical or cosmological scales. This is true more than 100 years after the theory's development from the pure genius of a man guided by powerful intuition and the observationally-driven \textit{Gedankenexperimenten} he liked so much. There is no surprise that General Relativity still inspires extensive research today and that during the past century, its fantastic predictions have fed the imagination of many authors and filmmakers, which has also contributed to the notoriety of the theory among the public. The new century has opened new routes of active research on General Relativity, as new technology, experimental equipment and computer performance levels now allow for probing the deepest abysses of the cosmos and ultimately testing the efficiency and the limits of the theory. For this reason, the interest of cutting-edge researchers in General Relativity is more justified than ever.

Among the stunning predictions of General Relativity, one finds the existence of \textit{gravitational waves}, which consist of small curvature oscillations around a given steady solution propagating at the speed of light and are conceivable as soon as the spacetime itself is endowed with a dynamical structure. Einstein considered very early linearized perturbations around the Minkowski flat vacuum spacetime \cite{1918SPAW.......154E}. He proved that gravitational radiation can emerge from periodic motions of massive bodies, such as binaries orbiting around each other and presenting at least a quadrupolar component. This is a major difference with electromagnetic waves that can be excited by a dipole. The second difference is that gravitational waves are ripples of the spacetime itself and are thus not stopped by ambient matter, with an obvious crucial interest in multi-source astronomy. The modes deform circular arrangements of test masses by quadrupolar oscillations, with two independent polarizations separated by an angle of 45$^\circ$. Einstein also derived the famous quadrupole formula linking the sources of radiation with the energy flux they are losing by gravitational emission. As interesting as it was, Einstein doubted his discovery and the potential for its physical realization, changing several times his point of view. His hesitation was the origin of a controversy concerning the status of gravitational waves (see \textit{e.g.} \cite{Kennefick:1997kb} for a review) : were they an artifact of linear theory that could be eliminated by a change of coordinates, or could we observe gravitational waves carrying energy over long distances in the full nonlinear theory as well? The debate raged for 40 long years mainly because of the innate difficulty of addressing this question in a suitable framework to offer a meaningful answer. In particular, the isolation of the propagating degrees of freedom among the 10 components of the metric field was a long-standing problem in the absence of a canonical prescription that General Relativity cannot provide.
 
It was not until the 60's that Bondi, Metzner, van der Burg \cite{Bondi:1962px} and Sachs \cite{Sachs:1962wk,Sachs:1962zza} finally proposed a sturdy framework to address the question in an unambiguous and elegant manner. They introduced a set of coordinates exploiting the fact that gravitational radiation travels along light cones together with a set of boundary conditions assuming that the sources of emission are localized in the bulk of spacetime. The formalism provided a parametrization of the metric field and a well-controlled expansion in the asymptotic zone, far from the emitting sources, in which the separation between kinematical and radiative degrees of freedom could be explicitly achieved. In particular, they were capable to derive the \textit{Bondi mass loss formula}, proving the existence of gravitational waves in the fully non-linear theory. They showed that some fields in their parametrization are responsible for a net flux of energy propagating at very large distances and identified them as encoded in what is broadly known as the \textit{Bondi news tensor}. Fall-offs of the components of the Weyl tensor in the presence of radiation (\textit{peeling}) were worked out by Sachs \cite{Sachs:1962wk,Sachs:1962zza}, providing constraints to firmly and completely describe the gravitational field as observed remotely from the source. Shortly after, the construction was confirmed by Penrose, who elaborated conformal techniques to model relativistic infinities in a fully covariant manner \cite{Penrose:1964ge}, leading to the characterization of fundamental properties of null asymptotes, the conformal compactness of the spacetime or the asymptotic flatness at null infinity heuristically theorized by Bondi and coworkers. 

Although these arguments finally ended the controversy among theoretical physicists, there was not yet any experimental evidence available. After Weber's unsuccessful attempts in 1969 \cite{Weber:1969bz}, the first indirect experimental clue in favor of the existence of gravitational waves came from the observation of the binary pulsar PSR B1913+16 by Hulse and Taylor in 1974 \cite{Hulse:1974eb}. After monitoring for several years radio pulses emitted by the system and following the evolution of its orbital parameters, they could deduce the loss of energy causing the inspiral motion of the system and established a perfect match with the prediction of General Relativity. Their discovery was awarded the Nobel Price of Physics in 1993. During the subsequent decades, gravitational radiation physics in the asymptotically flat context has benefited from strong interest. In the wake of Penrose, purely geometrical tools were developed -- namely by Geroch, Ashtekar and collaborators \cite{1977asst.conf....1G,Ashtekar:1978zza,Ashtekar:1981bq} -- in order to discuss the asymptotic structure at null infinity and the associated radiative data through coordinate independent objects. New frameworks have been successfully proposed to complement the Bondi-Sachs formalism and have proven to be more convenient for studying gravitational wave generation as well as for applications to the data analysis of possible experimental events. For instance, in order to find a connection between the material sources generating the waves and the asymptotic structure of the radiative gravitational field, Blanchet and Damour developed the multipolar post-Minkowskian formalism \cite{Blanchet:1987wq,Blanchet:1992br,Blanchet:1985spa,Blanchet:1986dk,Blanchet:1998in} 
combining the multipole expansion for the field in the near zone around the source, naturally defined for the linearized theory, with an expansion of the gravitational field in powers of the gravitational constant, perturbatively encapsulating the intrinsic non-linearity of gravitational dynamics. At linear order, it reduces to the solution presented by Thorne in 1980 \cite{Thorne:1980ru} as the general solution of the linearized Einstein equations in terms of gravitational multipoles, for which Einstein's quadrupolar result is the leading order approximation. Despite constant efforts in the field and outstanding theoretical improvements, the experimental evidence is long overdue. 

The main issue encountered by experimenters is the notable weakness of the gravitational interaction compared to the other three fundamental forces (about 40 orders of magnitude weaker). As a result, the supposed characteristic wave length of the gravitational radiation makes detection by light interferometry impossible without the use of gigantic experimental devices spanning over kilometers. In September 2015, nearly one century after Einstein's prediction, gravitational waves were finally declared to be directly detected by LIGO (\textit{Laser Interferometer Gravitational-Wave Observatory}, U.S.A.) \cite{2016PhRvL.116f1102A} while observing the inspiral and merging of two black holes orbiting at 1,3 billion light-years from Earth. Officially confirmed in February 2016, this spectacular observation earned Thorne, Weiss and Barish the 2017 Nobel Prize for their previous works on gravitational waves as well as their involvement in the LIGO project and also paved the way for gravitational astronomy. The technological endeavor is now focused on improving the accuracy of measurements. For instance, space probe interferometers currently in project such as eLISA (\textit{Laser Interferometer Space Antenna}, Europa) \cite{elisa}, which is scheduled for 2032 and designed to span over 2,5 million kilometers in the interstellar vacuum, should give us access to more precise acquisitions free of earthquake noise.

\section{Asymptotic symmetries in General Relativity}

\subsection{Gauge theories with boundaries}

The study of dynamics at null infinity initiated by Bondi and coworkers has raised many questions and mobilized constant efforts over decades, at the junction of geometrical formalization and physical intuitions. One central question regards the symmetries of General Relativity in the presence of boundaries and the conserved charges associated with them, in the spirit of Noether's theorem \cite{Noether:1918zz}. From the beginning, Einstein's theory was meant to be invariant under arbitrary changes of coordinates (diffeomorphisms), in accordance with the covariance principle establishing the equivalence of all observers. Believed to hold for any theory of gravitational interaction, this statement can be rephrased as gravity being a gauge theory under the gauge group of diffeomorphic automorphisms over the spacetime manifold. Two solutions for Einstein's field equations that are related by diffeomorphism are physically indistinguishable, as they are just two equivalent interpretations of the same physical system described by two different observers and the diffeomorphism shall carry no bulk charge. However, when analyzing the asymptotic behavior of the gravitational field around a boundary, be it repelled to infinity, the situation drastically changes. Indeed, when a boundary is present, the general covariance is explicitly broken by the choice of a particular class of observers that all agree with the position of the boundary and the assorted set of boundary conditions for the dynamical fields (gravity or matter) under consideration or even live on it. For instance, when the arms of LIGO are measuring gravitational radiation, they lay at future null infinity of a modelized asymptotically flat spacetime where the sources of radiation emit from deep bulk regions. By the fixation of a boundary and boundary conditions describing the behavior of the physical fields when approaching the boundary, a class of gauge transformations play a physical role as soon as they acquire a non-vanishing charge. While a majority of transformations still describe a pure redundancy of the theory with zero charge, the charged symmetries among the residual gauge transformations preserving the structure around the boundary become physical symmetries of the theory, and go under the name of \textit{asymptotic symmetries}. 

\subsection{The BMS group and its extensions}
In the context of radiative Einstein's gravity in four dimensions, the second important outcome of the seminal work of Bondi, Metzner, Sachs and van der Burg \cite{Bondi:1962px,Sachs:1962wk,Sachs:1962zza} was the derivation of the asymptotic symmetry group of asymptotically flat spacetimes at null infinity. Far from the sources of gravitation, as the spacetime geometry decays back to flat space, one could naively expect to find only the exact isometries of Minkowski space -- the Poincaré group. Bondi and coworkers demonstrated that the structure of the asymptotic group was richer than Poincaré, enhancing the four translations of the latter into an infinite-dimensional normal subgroup of supertranslations (\textit{i.e.} angular-dependent translations) and earning the name of ``BMS group.'' In fact, the inclusion of supertranslations was unavoidable to encompass the interesting sector of radiative solutions. Indeed, any reduction to Poincaré by freezing the pure supertranslations amounts to canceling the radiative boundary degrees of freedom \cite{Ashtekar:1981bq}. This result contrasts with previous analyses by Arnowitt, Deser and Misner at spatial infinity \cite{Arnowitt:1959ah}, where Poincaré invariances were necessary and sufficient to define total (ADM) mass and angular momentum of spacetime. The BMS group enjoyed a constant interest in the subsequent decades, unveiling its algebraic properties in a fully covariant language \cite{Penrose:1962ij,Newman:1966ub,Newman:1968uj,Ashtekar:1980nik,1977asst.conf....1G} and discussing associated radiative phase space and fluxes of charges (see \textit{e.g.} \cite{Ashtekar:1981bq,Dray:1984rfa,Barnich:2011mi,Wald:1999wa,Barnich:2019vzx}). Thereafter, BMS symmetries proved their relevance in many other physical contexts, distinct from future null infinity, such as at spatial infinity of asymptotically flat spacetimes \cite{Henneaux:2018cst,Henneaux:2019yax,Troessaert:2017jcm,Compere:2017knf,Compere:2020lrt}, in three-dimensional gravity \cite{Barnich:2014kra,Barnich:2015uva,Barnich:2012aw,Barnich:2006av} and at the black hole horizon \cite{Donnay:2015abr,Hawking:2016msc,Hawking:2016sgy}.

As the content of the asymptotic symmetry group is obviously constrained by the precise boundary conditions under consideration, it can be further enlarged by weakening the definition of asymptotic flatness. The natural extension of the BMS group lies in enhancing the Lorentz factor into an infinite-dimensional bunch of \textit{super-Lorentz transformations} (in the terminology of \cite{Compere:2018ylh}). Barnich and Troessaert recently introduced such an extension into the infinite-dimensional group of conformal transformations on the two dimensional sphere at infinity \cite{Barnich:2010eb,Barnich:2009se,Barnich:2011mi,Barnich:2011ct}, which they called ``\textit{superrotations}''. Their primary motivation was an interest in two dimensional conformal quantum field theories \cite{Belavin:1984vu} for holographic perspectives like in asymptotically AdS$_3$ gravity \cite{Brown:1986ed} (see section \ref{sec:holo}) and resulted a bit later in the celestial sphere holography program \cite{Pate:2019lpp,Ball:2019atb,Pasterski:2016qvg,He:2017fsb,Kapec:2016jld,Kapec:2014opa,Strominger:2017zoo} in asymptotically flat spacetimes. As generic holomorphic functions of the complex stereographic coordinates on the sphere, Barnich-Troessaert's superrotations are singular beyond the Lorentz transformations which are the six globally-defined Laurent modes. They introduce poles on the sphere that destroy the common sense of asymptotic flatness on a countable set of points; by consistency of the BMS algebra, the supertranslations become singular as well. Discarding these singularities requires focusing only on local properties of conformal isometries or trading the sphere for the cylinder as topology of the null generators of null infinity.

Campiglia and Laddha later proposed another definition of super-Lorentz extension by allowing the whole class of smooth diffeomorphisms on the celestial sphere as part of the asymptotic symmetry group \cite{Campiglia:2014yka,Campiglia:2015yka,Compere:2018ylh,Flanagan:2019vbl}, leading to the Generalized BMS group. The motivation for this wark came from infrared properties of quantum scattering amplitudes involving low-energy gravitons (see section \ref{sec:quantum}). Here the smooth super-Lorentz modify the leading-order structure at null infinity, which requires a firm relaxing of the definition of asymptotic flatness. The more general boundary conditions have to include solutions that are no longer strictly asymptotically Minkowski but with a fluctuating boundary metric, allowing for the choice of any arbitrary frame on the celestial sphere. We comprehensively discuss this extension later in the manuscript.

\subsection{Formal tools and covariant phase spaces}
At the theoretical level, the determination of produced and flowing amounts of energy in the presence of gravitational radiation poses an intricate problem. As a diffeomorphism covariant theory, and in accordance with the equivalence principle, there is no notion of local stress-energy for the gravitational field, because the latter can always be locally suppressed by a change of coordinates. For instance, the Hamiltonian associated with the gravitational field is a pure boundary term and vanishes on-shell. The definition of conserved charges by the usual Noetherian procedure is thus doomed to failure in General Relativity, even in the presence of exact isometries, as we review in the main text. Due to the gauged nature of the invariance under diffeomorphisms, non-trivial gravitational charges are codimension two (``surface'') objects living on the boundary of the spacetime volume whose energy, linear and angular momenta, for instance, have to be evaluated. For observers far away from the sources of gravitation, the useful boundaries are repelled towards conformal infinity and the desired dynamical quantities are canonically conjugated with asymptotic symmetries.

At spatial infinity, the Arnowitt-Deser-Misner (ADM) Hamiltonian analysis allows for computing the total Poincaré charges of any asymptotically flat spacetime \cite{Arnowitt:1959ah}. With suitable asymptotically flat boundary conditions and assuming a parity condition on the fields, Regge and Teitelboim prescribed the boundary terms to supply to the gravitational Hamiltonian that reproduce the Poincaré charges at spatial infinity \cite{Regge:1974zd}. These Hamiltonian charges are integrable on the solution space and conserved in time because no radiation can escape through spatial infinity. They form an algebra of phase space functionals representing the diffeomorphism algebra of asymptotic symmetries up to an eventual central extension, as shown by Brown and Henneaux in \cite{Brown:1986ed}. The result was immediately verified in three-dimensional \cite{Brown:1986nw} and higher-dimensional \cite{Henneaux:1985ey,Henneaux:1985tv} Einstein gravity with negative cosmological constant. Boundary structure at the spatial infinity of asymptotically flat spacetimes was further discussed in \cite{Mann:2005yr,Compere:2011ve,Compere:2011db,Ashtekar:1990gc,Ashtekar:1991vb,Ashtekar:1978zza,Hawking:1995fd,Beig:1987zz,1982CMaPh..87...65B}, culminating with the recent realization of enhancement of Poincaré to BMS symmetry also at spatial infinity \cite{Henneaux:2018cst,Henneaux:2019yax,Troessaert:2017jcm,Compere:2017knf}.

Things change substantially when discussing an equivalent notion of ``conserved charges'' from the perspective of null infinity, mainly because energy fluxes cannot be discarded and the evolution parameter along the boundary is a null coordinate. The system genuinely exchanges energy with its environment when radiative modes are turned on, causing potential non-integrability, but surely non-conservation, of the surface charges. As previously noted, the first promising attempts to quantify the radiated flux of energy carried by gravitational radiation crossing null infinity date back to Trautman \cite{Trautman:2016xic} and Bondi \cite{Bondi:1962px} analyses. The definitions of the so-called Bondi mass and momenta ensure that they decay to the ADM quantities near spatial infinity and that the associated fluxes provide a negative balance from the bulk point of view, asserting that energy is escaping through future null infinity while the Bondi mass decreases in (retarded) time, as it should according to elementary physical intuition. Ashtekar and Steubel \cite{Ashtekar:1981bq} later observed that the BMS transformations act as canonical transformations on the exact radiative modes in the full non-linear theory characterized covariantly and intrinsically at null infinity \cite{Ashtekar:1978zza,Ashtekar:1981hw}. The generating Hamiltonians are the integrated fluxes of energy on the whole boundary, which offered no information about local charges defined on constant-time sections of null infinity until Dray and Streubel \cite{Dray:1984rfa} introduced a definition of BMS momenta and supermomenta (for proper supertranslations) whose fluxes agree with \cite{Ashtekar:1981bq}.

With the aim of deducing the expression of local surface charges at null infinity, covariant phase space methods -- initiated by \cite{Crnkovic:1986be,Crnkovic:1986ex,Ashtekar:1990gc} -- were invoked and refined over the years. The formalism consists of a smart fusion between Hamiltonian and Lagrangian methods, incorporating the powerful features of phase space analyses into a flexible language that does not abandon covariance. A considerable research effort has been pushed in this direction by Wald and coworkers in \cite{Lee:1990nz,Iyer:1994ys,Wald:1993nt}. The construction of surface charges at the boundary of any Lagrangian covariant theory of fields was then routed through an algorithm. The latter is inspired by the symplectic methods for Hamiltonian classical mechanics, and provides as a first step the symplectic structure to be equipped on the solution space from the variational principle. The charges are then computed as a contraction of this symplectic structure with a gauge transformation. Since the integrability of the charges is no longer guaranteed in such a non-equilibrium physics experienced at null infinity, the definition of a Hamiltonian is ambiguous until a clear separation has been made between physical charges and physical fluxes. Wald and Zoupas introduced a prescription to fix this ambiguity by demanding that the flux of suitable charges is zero on stationary configurations in the absence of radiation \cite{Wald:1999wa}. 

Another definition for the surface charges was developed in parallel by Barnich and Brandt in \cite{Barnich:2001jy,Barnich:2003xg}, but relying on the equations of motion rather than the variational principle. BRST techniques are at the core of this second definition, mainly motivated by cohomogical approaches of asymptotic symmetries \cite{Anderson:1996sc} discussed with Henneaux in \cite{Barnich:1994db,Barnich:1995ap}. Free of the various ambiguities in boundary terms arising from Wald's definition, Barnich and Brandt's prescription fixes the local charges as soon as the dynamical equations for the fields are given, and is applicable to a more general class of gauge theories than the only generally covariant theories. The non-linear theory for exact and asymptotic symmetries in this framework was developed in \cite{Barnich:2007bf,Compere:2008us}. 

Echoing the representation theorem obtained in Hamiltonian formalism, a natural question to ask was, are the BMS symmetries represented at the level of the charges? Barnich and Troessaert \cite{Barnich:2011mi} observed that the presence of non-vanishing physical flux through null infinity represents an obstruction for the BMS surface charges to close an algebra under the usual covariant Poisson bracket \cite{Koga:2001vq}. They introduced a new definition of the charge bracket that includes a supplementary term vanishing in the absence of Bondi news and proved that this extension was sufficient to obtain an avatar of the representation theorem for BMS charges, up to a field-dependent ``central extension.'' Despite many checks in various contexts (see \textit{e.g.} \cite{Barnich:2019vzx,Compere:2020lrt,Fiorucci:2020xto}) and theoretical progress (see \textit{e.g.} \cite{Troessaert:2015nia,Wieland:2020gno,Wieland:2021eth,Freidel:2021cbc,Freidel:2021yqe,Chandrasekaran:2020wwn}), the right framework into which a general proof of this generalized representation theorem for open gauged systems could be formulated is still missing and mobilizes some current research efforts \cite{ToAppear}. 

As the title of the thesis suggests, we review and discuss the advances and difficulties related to this problem later in the manuscript. The following sub-sections return to physics.

\subsection{Memory effects}
An important aspect of asymptotic symmetries is that they govern the kinematics of physically measurable effects recast as transitions among solutions in the phase space. In General Relativity for instance, the passage of gravitational radiation causes a permanent shift in the relative angular positions of a detector comprised of a pair of free-falling observers in the asymptotic region. Since the geometry keeps track of the radiative phenomenon after its extinction, this constitutes a \textit{gravitational memory effect}. Before and after the non-stationary phase, the gravitational field is supposed to be in a steady vacuum configuration, but the radiation has driven a vacuum transition from the initial vacuum to another inequivalent one. The closure of the solution space under the action of the asymptotic symmetry group implies that both vacua are diffeomorphic to each other, and the permanent shift experienced by the inertial observers defined above is precisely controlled by the amplitude of the involved asymptotic symmetry. In the context of asymptotically flat spacetimes in four dimensions, the displacement memory effect \cite{Blanchet:1987wq,1987Natur.327..123B,Thorne:1992sdb,Christodoulou:1991cr,Blanchet:1992br} induced by null matter or gravitational radiation reaching null infinity is understood as a vacuum transition driven by a proper supertranslation \cite{Strominger:2014pwa,Compere:2016jwb}. At the dynamical level, the flux-balance law for the BMS supertranslation charge offers the key to relate the amplitude of the total displacement memory effect and the strain of radiation among other observable parameters (such as the orbital parameters of the black hole merger producing the waves). 

Many other memory effects have been identified in four-dimensional General Relativity, such as the spin memory effect \cite{Pasterski:2015tva}, concerning the phase shift experienced by photonic loops under the action of gravitational waves; the center-of-mass memory effect \cite{Nichols:2018qac}, modifying the angular momentum of spacetime in the center-of-mass frame; the velocity kick memory effect \cite{Podolsky:2002saa,Podolsky:2010xha}, predicting a cumulative geodesic deviation and finally the black hole memory effect \cite{Donnay:2018ckb}, which is the dynamical response of a black hole horizon when subjected to some incoming gravitational burst. In comparison to the ripples recorded by LIGO for typical black hole mergers, these memory effects appear as subleading deviations in the inspiral-merging curve profiles and are currently beyond the experimental tolerance of detectors. The hope is that these limitations will soon be lifted by the use of space probe interferometers. The rich panel of gravitational memories, even for higher dimensional configurations \cite{Pate:2017fgt}, has inspired a reinforced interest to discuss similar effects in other gauge theories, namely in Maxwell \cite{Susskind:2015hpa,Pasterski:2015zua,Bieri:2013hqa,Mao:2019sph} and Yang-Mills theories \cite{Pate:2017vwa,Ball:2018prg}.

\subsection{Quantum aspects}
\label{sec:quantum}
The exploration of the structure of asymptotic symmetry groups in General Relativity and the algebraic properties of the associated conserved charges is also crucial to address the problem of building a quantum theory of gravity. Without even evoking very precise motivations like in holography that we later discuss (see sub-section \ref{sec:holo}), the existence of rich symmetry structures in a classical theory and the control of their representations tells us a lot about how the Hilbert space of quantum states should be organized in the quantized theory. In particular, the charges become observables whose canonical commutation with the fields generate the symmetry at the quantum level. This is particularly significant for gravity, for which the quantum theory is still beyond our reach. The key point is that asymptotic symmetries provide a clear and robust anchorage for progress towards the quantification while lacking the full non-linear quantum theory.

An important research program started by Strominger in 2013 set out to study the quantum aspects of the low-energy regime of gravity. The program sought to discuss the scattering problem of gravitational waves in asymptotically flat spacetimes at both past and future null infinity, where two different BMS groups seem to act independently in the naive picture. The scattering process is well-defined if one selects the diagonal subgroup of BMS transformations antipodally identified near spatial infinity \cite{Strominger:2013jfa}. This yields an infinite tower of conservation laws for the supermomenta whose quantum versions state nothing but the invariance of the gravitational S-matrix under the BMS transformations. The Ward identity associated with supertranslations is then equivalent to the Weinberg soft graviton theorem \cite{He:2014laa}, which is an old result of pure quantum field theory \cite{Weinberg:1965nx}, according to which the contributions of soft particles (\textit{i.e.} massless particles with nearly zero momentum) are universal and factorize in quantum scattering amplitudes. Interestingly, performing a Fourier transform on the universal soft factor appearing in this result reproduces the metric deviation involved in the kinematics of the displacement memory effect \cite{Strominger:2014pwa}. 

The presence of the supertranslation symmetry at the quantum level in the linearized theory sheds new light on the infrared (low-energy) structure of gravity, organized as a triangle whose vertices bear the symmetry itself, the associated Ward identity in the form of Weinberg's theorem and the conjugated dynamical process that is the displacement memory effect \cite{Strominger:2017zoo}. It is remarkable that this tripartite structure has also been subsequently uncovered for other gauge theories, such as electrodynamics \cite{Lysov:2014csa,Mao:2017tey} and chromodynamics \cite{Pate:2017vwa}. In gravity, the enhancement of the BMS group has led to subleading connections involving super-Lorentz transformations \cite{Barnich:2010eb,Barnich:2009se,Barnich:2011ct,Campiglia:2014yka,Campiglia:2015yka,Compere:2018ylh} whose Ward identity was shown to be equivalent to a subleading version of the soft gravitation theorem \cite{Cachazo:2014fwa,Kapec:2014opa,Himwich:2019qmj}, and assorted with new memory effects such as the spin memory effect \cite{Pasterski:2015tva} or the center-of-mass memory effect \cite{Nichols:2018qac}. Similar discussions for Maxwell's theory were again quick to follow \cite{Conde:2016csj,Campiglia:2016hvg,Lysov:2014csa}.

In any physical process, one expects to observe a certain amount of emitted soft gravitons. In particular this emission could occur during the quantum evaporation of a black hole, as predicted by Hawking's seminal computation \cite{Hawking:1974sw} leading to the famous black hole information paradox \cite{Hawking:1976ra}, which results from the deterioration of the pure states inside the black hole into thermal mixed outgoing Hawking radiation. The counting and matching of degrees of freedom in the various attempts to solve this paradox must take into account the information encoded in the emitted soft modes, and the infrared triangular structure is very helpful in that task. Indeed, thanks to the correspondence between soft gravitons and BMS invariances, black holes are classically dressed with BMS charges (``soft hairs''), whose associated flux-balance laws at the horizon may be of crucial importance in the flow of quantum information \cite{Hawking:2016msc,Hawking:2016sgy,Haco:2018ske} (see also \cite{Haco:2019ggi,Mirbabayi:2016xvc}).

\section{Exploring other types of asymptotics}

\subsection{Positive cosmological constant and universal expansion}
\label{sec:dSintro}
The program started by the pioneer works of Bondi and coworkers focuses its attention on asymptotically flat spacetimes. However, during the course of the past century, we discovered strong evidence that our Universe experiences a phase of accelerated expansion, understood in cosmology as the contribution of a small, but non-vanishing positive cosmological constant $\Lambda$ in Einstein's equations \cite{Hubble:1929ig,Riess:1998cb,Perlmutter:1998np}. $\Lambda$ is evaluated around $10^{-52} m^{-2}$ according to the final results of the Planck probe mission \cite{Ade:2015xua}. Again assuming the localization of the sources of gravity (stars, planets, black holes, interstellar dust \textit{etc.}) deep in the bulk, the geometry of the Universe is shown to asymptote to the \textit{de Sitter spacetime} \cite{deSitter:1916zza,deSitter:1916zz,deSitter:1917zz}, which is the equivalent of the Minkowski vacuum when $\Lambda>0$. The class of radiative asymptotically de Sitter spacetimes is surely worth studying as soon as it would be possible to perform experimental detections of gravitational waves emitted by dynamical sources separated from Earth by cosmological-scale distances.

Physics in de Sitter presents notable peculiarities in comparison of the well-known flat case. Observers in the asymptotic region of de Sitter have no access to the entire spacetime because of the constantly accelerated cosmological expansion: for each observer, a (past) cosmological horizon exists, limiting the description of the physics to a half-causal wedge. In the flat space, in contrast, any asymptotic Bondi observer has access to the whole history of the spacetime up to its current (retarded) time. Moreover, the (future) conformal boundary where null rays come to an end is now a timelike surface, drawn as a horizontal line segment in the conformal diagram. There is no chronology provided on the boundary; therefore, the initial value problem for boundary data as well as the discussion of the loss of energy carried out by gravitational radiation crossing the boundary are more intricate. For that reason, one cannot distinguish between translations and rotations among the diffeomorphisms acting on the boundary as is possible for the BMS symmetries in the flat case. This explains why the de Sitter case is less covered in the literature and why any step forward is tricky and time-consuming. 

A renewal of interest around de Sitter asymptotics has been observed during the two last decades, mainly due to the success of holography in the presence of negative cosmological constant (see sub-section \ref{sec:holo}). Several attempts have been made to define boundary conditions at the future conformal boundary, including \cite{Strominger:2001pn,Anninos:2010zf,Anninos:2011jp}. However, as recently emphasized in \cite{Ashtekar:2014zfa,Ashtekar:2015lla,Anninos:2011jp}, that enterprise comes with a strong drawback: imposing future boundary conditions amounts to restricting the initial data backwards in time and therefore the bulk dynamics. Allowing generic initial data and, in particular, generic bulk gravitational waves prevents the imposition of any boundary conditions at the future boundary of de Sitter. An interesting question is then how to extend the notions of the radiative data present in Bondi formalism to asymptotically locally de Sitter spacetimes (see \textit{e.g.} \cite{Smalley:1978qd,Abbott:1981ff,Balasubramanian:2001nb,Kastor:2002fu,Bishop:2015kay,Ashtekar:2015ooa,Szabados:2015wqa,Chrusciel:2016oux,Saw:2016isu,Saw:2017amv,Szabados:2018erf,He:2018ikd,Poole:2018koa,Balakrishnan:2019zxm,Mao:2019ahc}). A considerable step forward in the understanding of gravitational radiation in the presence of a cosmological constant has been made very recently by Ashtekar and coworkers in a series of papers discussing the extension of the fully-geometrical asymptotic treatment to de Sitter \cite{Ashtekar:2014zfa}, inventorying the properties of linearized gravitational waves near the future conformal boundary while precisely quantifying the information lost by imposing further boundary conditions \cite{Ashtekar:2015lla} and deriving a generalization of Einstein's quadrupole formula in this new context \cite{Ashtekar:2015lxa}. The cosmological version of the displacement memory was also unveiled afterwards \cite{Tolish:2016ggo}. The Hamiltonian phase space for gravity in de Sitter has also been worked out \cite{Kelly:2012zc} in the spirit of Regge-Teitelboim's analysis in flat space \cite{Regge:1974zd} (see also \cite{Henneaux:2018cst,Henneaux:2019yax}). One of the main goals of this thesis is to take advantage of this \textit{impetus} to repeat the Bondi analysis for asymptotically de Sitter spacetimes and to discuss the relation with the well-known flat case through a flat limit process. The construction of radiative phase spaces in the presence of a cosmological constant is appealing and mainly motivated by uncovering the realization of a BMS-like symmetry and defining the right notion of Bondi news in de Sitter.

\subsection{Negative cosmological constant and holography}
\label{sec:holo}
In contrast to the de Sitter case, Einstein's gravity with negative cosmological constant has enjoyed stimulating research, mainly because of its crucial role in the holographic principle \cite{tHooft:1993dmi,Susskind:1994vu}.

The paradigm of holography is a promising road to understand quantum gravity. It is based on the hope that a theory of quantum gravity can be described in terms of a lower-dimensional dual quantum field theory. Even if this statement has not been proven yet in general, evidence in favor of holography has been accumulated during the last 20 years in the context of the AdS/CFT correspondence \cite{Maldacena:1997re}, according to which any asymptotically anti-de Sitter solution of gravity is dual to a conformal field theory living on the conformal codimension one boundary of the spacetime. The holographic dictionary provides a translation guide between gravitation and fields living in the bulk of spacetime and the dynamics of dual boundary quantum fields. At the level of the symmetries, the asymptotic symmetries of the bulk theory are dual to the global symmetries acting on the ultraviolet modes of the boundary theory, and the study of asymptotic symmetries on the gravity side of the duality has been shown to be extremely efficient in extracting some patterns of the holographic correspondence. 

One of the crucial properties of asymptotically anti-de Sitter solutions that make them the natural breeding ground for holographic investigations is that their conformal boundaries look like timelike infinite cylinders, on which a notion of chronology can be defined. This feature contrasts with the de Sitter case, where no such boundary time direction even exists, and the flat case for which the asymptotic boundary is a null manifold. However, from the point of view of the bulk theory, this peculiar feature allows the null rays to reach the conformal boundary after a finite amount of coordinate time: a well-posed initial value problem requires to supplement the data on any Cauchy slice from which the null motion originates by a set of boundary conditions predicting the fate of the null rays after they encountered the conformal boundary. Satisfactory boundary conditions for holographic purposes are Dirichlet boundary conditions that aim at reflecting the null rays back towards the bulk. This causally links the interior of the spacetime with its boundary and ensures the unitarity of the quantum dual theory. In bulk dimensions greater than three, Dirichlet boundary conditions reduce the asymptotic symmetry group to the lower-dimensional conformal group, which is the (finite-dimensional) isometry group of vacuum anti-de Sitter solution \cite{Hawking:1983mx , Ashtekar:1984zz , Henneaux:1985tv , Henneaux:1985ey , Ashtekar:1999jx}. In three dimensions, the asymptotic symmetry algebra under Dirichlet boundary conditions is infinite-dimensional: this is the conformal group in two dimensions consisting of a double copy of the Virasoro algebra. The associated charge algebra was shown to represent this double Virasoro symmetry with the famous Brown-Henneaux central extension \cite{Brown:1986nw}. The latter provided a first encouraging hint leading to the holographic conjecture. Indeed, this central extension was later related to the conformal anomaly of the dual conformal field theory \cite{Henningson:1998gx}. Furthermore, it was used to reproduce the BTZ black hole entropy using the Cardy formula \cite{Strominger:1997eq}. Many other boundary conditions in asymptotically anti-de Sitter spacetimes have been proposed in the literature \cite{Troessaert:2013fma , Compere:2013bya , Papadimitriou:2005ii , Grumiller:2016pqb , Perez:2016vqo , Aros:1999kt , Alessio:2020ioh}, leading to different asymptotic symmetries and holographic dualities. As in the flat case, the analysis of conserved charges associated with the asymptotic symmetries in AdS has attracted considerable attention (see \textit{e.g.} \cite{Ashtekar:1984zz,Ashtekar:1999jx,Hawking:1995fd,Chrusciel:2000dd,Chrusciel:2001qr,Cai:2006az,Hollands:2005wt,Abbott:1981ff,Henneaux:1985tv,Brown:1986nw}).

Nevertheless, the motivation to fix these various classes of boundary conditions was the existence of a well-defined variational principle \cite{Ishibashi:2004wx} in the usual sense that the action is stationary on any solution of the equations of motion. This condition was guaranteed by the absence of flux through the conformal boundary, in complete contrast with what is observed for radiative asymptotically flat or de Sitter spacetimes. This feature presents several drawbacks if one pays attention to recent developments in the field. A first example occurs when trying to address the black hole information paradox in the language of the AdS/CFT correspondence. Indeed, reflexive boundary conditions preventing leaks at the conformal boundary do not allow for dynamical black hole evaporation \cite{Lowe:1999pk} as the energy flow is permanently re-injected into the bulk, invariably leading to eternal black holes \cite{Maldacena:2001kr}. In particular, it was recently shown in the ``islands'' program \cite{Almheiri:2019yqk , Almheiri:2019qdq , Almheiri:2020cfm} that the usual boundary conditions prevent from deriving the Page curve \cite{Page:1993wv,Page:2013dx} under the assumption of unitarity from quantum gravity path integral arguments. However, the Page curve is a central ingredient in the discussion, as it is assumed to govern the time evolution of the total quantum entropy  during the evaporation (see \textit{e.g.} section 5 of \cite{Almheiri:2020cfm} for a review). In this context, it has been useful to allow some radiation to escape the spacetime boundary so that the black hole can evaporate in anti-de Sitter as it does in flat space. This method was implemented in practice by gluing an asymptotically flat region to the conformal boundary, providing a reservoir for the outgoing radiation, and coupling the dual theory to a thermal bath. Another example appears when considering brane worlds interacting with ambient higher-dimensional spacetimes with negative cosmological constant in the spirit of Randall and Sundrum \cite{Randall:1999ee, Randall:1999vf}. This picture naturally yields holographic dualities with permeable boundaries, mathematically implemented by a dynamical boundary metric, or equivalently said, quantum gravity on the boundary \cite{Compere:2008us}. These examples are certainly appealing to investigate a more general class of boundary conditions that allow for some flux at infinity and fluctuating boundary structure.

\section{Original contributions}
The purpose of this thesis is to provide a unified description of radiative covariant phase spaces with leaky boundary conditions whatever the value of the cosmological constant. The present manuscript -- conceived to introduce the original results achieved during these four last years of research, contextualize them in the state of the art and describe our current understanding about them -- is based on the following papers: 
\begin{enumerate}[leftmargin=1.5cm,label={[P\arabic*]}, ref={[P\arabic*]}]
\item G.~Comp\`ere, A.~Fiorucci, and R.~Ruzziconi, \textit{Superboost transitions,
  refraction memory and super-Lorentz charge algebra}, {\em JHEP} {\bf 11}
  (2018) 200,
\href{http://www.arXiv.org/abs/1810.00377}{{\tt 1810.00377}}.  \label{paper:P1}
\item G.~Comp\`ere, A.~Fiorucci, and R.~Ruzziconi, \textit{The $\Lambda$-BMS$_4$ group of dS$_4$ and new boundary conditions for AdS$_4$}, {\em Class. Quant. Grav.} {\bf 36} (2019), no.~19, 195017, \href{http://www.arXiv.org/abs/1905.00971}{{\tt 1905.00971}}.  \label{paper:P2}
\item G.~Comp\`ere, A.~Fiorucci, and R.~Ruzziconi, \textit{The $\Lambda$-BMS$_4$ charge   algebra}, {\em JHEP} {\bf 10} (2020) 205, \href{http://www.arXiv.org/abs/2004.10769}{{\tt 2004.10769}}.  \label{paper:P3}
\item A.~Fiorucci and R.~Ruzziconi, \textit{Charge Algebra in Al(A)dS$_n$ Spacetimes}, {\em JHEP} {\bf 05} (2021) 210, \href{http://www.arXiv.org/abs/2011.02002}{{\tt 2011.02002}}.  \label{paper:P4}
\end{enumerate}
Some review parts of the manuscript also reproduce the following lecture notes:
\begin{enumerate}[leftmargin=1.5cm,label={[P\arabic*]}, ref={[P\arabic*]}]
\setcounter{enumi}{4}
\item G.~Comp\`ere and A.~Fiorucci, {\textit{Advanced Lectures on General Relativity}},   \href{http://www.arXiv.org/abs/1801.07064}{{\tt 1801.07064}}. \label{paper:lect}
\end{enumerate} 
The following sub-sections summarize the content of these articles, focusing on the most striking results, in order to familiarize the reader with the original content of the thesis and provide a brief overview of the origin of each argument and development detailed later in the main text. 

\subsection{Super-Lorentz extension of the BMS group}

\paragraph{Super-Lorentz symmetry and memory effects} Staged in the asymptotic flat land in four dimensions, the first paper \ref{paper:P1} addresses the several issues encountered when the BMS$_4$ group is augmented with an infinite super-Lorentz extension as well as the new memory effects related to these overleading symmetries. We started with the derivation of the orbit of Minkowski spacetime under arbitrary Diff$(S^2)$ super-Lorentz transformations and supertranslations in the Bondi gauge, in the spirit of \cite{Compere:2016hzt,Compere:2016jwb,Compere:2016gwf}. We determined that such vacua are labelled by a set of three fields on the celestial sphere, which we identified as the superboost, superrotation and supertranslation fields. The terminology ``superboost'' and ``superrotation'' is reserved for transformations that are driven by curl-free and divergence-free vectors on the sphere, respectively. The knowledge gained from this careful study of the vacuum structure of the theory was used to identify the memory effects, which are understood as transitions between vacua generated by non-stationary phases occurring when gravitational radiation or more exotic cosmic events perturb the metric irreversibly. We showed how impulsive transitions \cite{Strominger:2016wns,Penrose:1972aa,Nutku:1992aa} among vacua are related to the velocity kick/refraction memory effect \cite{Podolsky:2002sa,Podolsky:2010xh,Podolsky:2016mqg} and the displacement memory effect \cite{Christodoulou:1991cr,Thorne:1992sdb,Zeldovich:1974gvh,Blanchet:1987wq,Blanchet:1992br,Strominger:2014pwa}. 

\paragraph{Renormalized phase space} The second main goal of the paper was to construct, in a rigorous manner and from first principles, a covariant phase space whose asymptotic symmetry group consists of arbitrary Diff$(S^2)$ super-Lorentz transformations and smooth supertranslations. The motivations for doing so are numerous, but it is important to note that besides its relevance for memory effects, the smooth super-Lorentz extension is required to prove \cite{Campiglia:2014yka,Campiglia:2015yka,Campiglia:2020qvc} that the Ward identities for the transverse sector of the BMS$_4$ transformations are equivalent to the Cachazo-Strominger subleading soft graviton theorem \cite{Cachazo:2014fwa}. Since these symmetries modify the leading structure of the gravitational field at null infinity, their inclusion in the phase space introduces divergences in the on-shell action as well as in the symplectic structure and finally the surface charges. Therefore, a renormalization of the symplectic structure is necessary. The renormalization procedure was performed heuristically using non-covariant objects due to the background structure induced by the Bondi gauge fixing as well as the boundary conditions. Nevertheless, the renormalization scheme was consolidated by taking the flat limit of the radiative asymptotically (A)dS phase spaces in \ref{paper:P3}. 

After obtaining radially finite charges, we defined a subset of solutions that contain temporary non-stationary effects, coming from a steady gravitational field in the past to relax back to another steady field in the future. This is the natural frame to formalize transitions leading to memory effects as well as realistic localized sources. For this class of solutions, we defined finite Hamiltonian conjugated to the Generalized BMS$_4$ diffeomorphisms by controlling the time divergence at the corners of null infinity and requiring that their canonical fluxes were zero in the absence of gravitational radiation, extending the prescription of \cite{Wald:1999wa} to general super-Lorentz frames. Finally, we showed that our flux expressions are consistent with the leading \cite{Weinberg:1965nx,Strominger:2014pwa,He:2014laa} and subleading \cite{Cachazo:2014fwa} soft graviton theorems and computed the charge algebra under the Barnich-Troessaert bracket \cite{Barnich:2010eb}. We concluded by contrasting the leading BMS triangle structure \cite{Strominger:2017zoo} with the mixed overleading/subleading BMS square structure. 

\subsection{Radiative phase spaces with cosmological constant}

As previously mentioned, the radiative phase spaces with a non-vanishing cosmological constant $\Lambda$ are significantly less studied in the literature compared to the $\Lambda=0$ case. In the AdS ($\Lambda<0$) case, some radiative configurations, such as Robinson-Trautman solutions \cite{Robinson:1960zzb}, have been considered \cite{Bakas:2008zg,Bicak:1999ha,Bicak:1999hb} as concrete models for non-equilibrium holography, but overall the conformal boundary of AdS has remained resolutely impermeable, which is suitable for the sake of unitarity of the dual quantum theory. The dS case ($\Lambda>0$), although relevant from cosmological perspectives, has not received much interest and only recent analyses \cite{Ashtekar:2014zfa,Ashtekar:2015lla,Ashtekar:2015lxa} address the questions of consistent boundary conditions at the future conformal boundary of dS compatible with generic radiative sources in the bulk (see sub-section \ref{sec:dSintro}). 

The second line of research pursued in this thesis proposes to repeat the Bondi analysis, which was very successful in the flat case, to identify the fields encoding the radiation flux at conformal infinity as well as to have an infinite-dimensional symmetry group mimicking the features of the (Generalized) BMS group of flat space. Such a group shall encompass more general diffeomorphisms than the conformal isometries of the boundary, and thus act non-trivially on the boundary metric. However, as soon as the symplectic flux in (A)dS is known to be sourced by fluctuations of the boundary metric \cite{Compere:2008us,Papadimitriou:2005ii}, the search for a non-trivial infinite-dimensional asymptotic symmetry group becomes equivalent to the search for leaky boundary conditions in (A)dS. From this program, we can expect advances in two ways: first, we can import techniques and results derived in the well-mastered asymptotically flat context to (A)dS asymptotics, with some necessary precautions about the physical differences between flat null infinity and (A)dS conformal boundaries; second, if we build a mathematically robust flat limit process allowing for linking the results obtained for $\Lambda\neq 0$, where one has better control on holographic features, to the flat case, we can expect to gain a better understanding of radiative spacetimes with fluctuating boundary structures for $\Lambda=0$.

\paragraph{BMS-like symmetries in presence of cosmological constant} In a first paper \ref{paper:P2} inspired by the recent work \cite{Poole:2018koa}, we derived the general four-dimensional solution space with asymptotically locally (A)dS boundary conditions in the Bondi gauge, and gave a holographic interpretation to the Bondi fields thanks to a diffeomorphism to the Starobinsky/Fefferman-Graham \cite{Starobinsky:1982mr,Fefferman:1985aa} gauge, where all the holographic features of (A)dS spacetimes are more transparent. It is worth emphasizing that diffeomorphisms breaking the SFG gauge are not considered hereafter, although they might be associated with further non-trivial charges as in three-dimensional Einstein gravity (see \textit{e.g.} \cite{Grumiller:2016pqb,Ciambelli:2020ftk,Ciambelli:2020eba}). Using this dictionary between the gauges, we proved that the dynamical equations in the Bondi gauge admit a well-defined flat limit $\Lambda\to 0$. More crucially, we also identified the analogues of the Bondi news, Bondi mass and Bondi angular momentum aspects at the conformal boundary. 

This finding led us to separate the boundary degrees of freedom between dynamical (or radiative) and kinematical pieces by the choice of a particular universal boundary structure, consisting of a foliation and a measure on the transverse spaces. We argued that this fixation is always reachable by diffeomorphism and is thus merely a boundary gauge fixing that does not rule out any solution for the phase space and does not constrain the initial value problem, more particularly in dS. We introduced the $\Lambda$-BMS$_4$ group as the set of residual symmetries of the Bondi gauge after this additional boundary gauge fixing. It consists of infinite-dimensional non-abelian ``supertranslations'' and ``superrotations'' but it reduces in the asymptotically flat limit to the Generalized BMS$_4$ group \cite{Campiglia:2014yka,Campiglia:2015yka,Campiglia:2020qvc,Compere:2018ylh} . Given that the constraints among the diffeomorphism parameters are field-dependent, one should adopt a precise terminology stating that $\Lambda$-BMS$_4$ is a Lie groupoid, associated with a Lie algebroid of infinitesimal diffeomorphisms closing with solution-dependent ``structure constants'' \cite{Crainic,Barnich:2010xq,Barnich:2017ubf}. 

Finally, we presented new boundary conditions for asymptotically locally AdS$_4$ spacetimes insuring a well-posed Cauchy problem, that admit $\mathbb R$ times the group of area-preserving diffeomorphisms as the asymptotic symmetry group. This conservative subset of the $\Lambda$-BMS$_4$ boundary conditions amounts to fixing two components of the holographic stress-tensor while allowing two components of the boundary metric to fluctuate. They correspond to a deformation of a holographic conformal field in three dimensions, which is coupled to a fluctuating spatial metric of fixed area.

\paragraph{Radiative phase spaces with a cosmological constant and flat limit} We devoted a second paper \ref{paper:P3} to deriving the surface charge algebra associated with these novel $\Lambda$-BMS$_4$ symmetries. We started with a derivation of the surface charge (and algebra) of generic asymptotically locally (A)dS$_4$ spacetimes without matter. The computations were performed without assuming any further boundary conditions than the existence of a conformal completion \cite{Penrose:1964ge,Ashtekar:1984zz}. Surface charges associated with boundary Weyl rescalings \cite{Imbimbo:1999bj,deHaro:2000vlm,Schwimmer:2008yh} were found to be vanishing, while the boundary diffeomorphism charge algebra was non-trivially represented without central extension under the Barnich-Troessaert bracket \cite{Barnich:2010eb}. We particularized the result by specifying a boundary foliation and a boundary measure and obtained the $\Lambda$-BMS$_4$ charge algebra. 

Pursuing the investigations of \ref{paper:P2}, we considered the feasibility of a well-defined flat limit at the level of the phase space. The procedure requires the incorporation of corner terms in the action principle and symplectic structure that are defined from the boundary foliation and measure to exempt them from poles in $1/\Lambda$. The flat limit then reproduces the Generalized BMS$_4$ phase space and charge algebra of supertranslations and super-Lorentz transformations acting on the class of asymptotically locally flat spacetimes studied in \ref{paper:P1}. In particular, the renormalization procedure needed in the flat case when super-Lorentz diffeomorphisms act at null infinity is morally equivalent to the completely covariant corner subtraction of $1/\Lambda$ poles in the (A)dS case, confirming the result of \ref{paper:P1} in a somewhat indirect and non-trivial way. Finally, we again added a tiny layer of refinement to the results of \ref{paper:P1} and proposed a second prescription for the finite Hamiltonians associated with Generalized BMS$_4$ transformations (in agreement with \cite{Hawking:2016sgy,Campiglia:2020qvc}), which enjoy the desirable property of representing the Generalized BMS$_4$ algebra without central extension at the corners of null infinity under the standard Poisson bracket. This feature implies that the BMS$_4$ flux algebra admits no non-trivial central extension.

\paragraph{Generalization to arbitrary dimensions} The series is completed by a third paper \ref{paper:P4} in which the analysis of \ref{paper:P3} was extended to generic asymptotically locally (A)dS spacetimes in $n\geq 3$ dimensions, with the only exception that the computations were performed within the Starobinsky/Fefferman-Graham gauge. Again, we obtained the gravitational charge algebra for boundary diffeomorphisms and Weyl rescalings without assuming any further boundary condition than the minimal falloffs allowing for conformal compactification. In particular, the whole boundary structure was free to fluctuate and play the role of source yielding some symplectic flux at the boundary. 

Using the holographic renormalization procedure \cite{deHaro:2000vlm,Bianchi:2001kw,Skenderis:2000in}, the divergences in the holographic coordinate were removed from the symplectic structure in any dimension, leading to finite expressions \cite{Compere:2008us,Papadimitriou:2005ii}. The charges associated with boundary diffeomorphisms were found to be generically non-vanishing, non-integrable and not conserved, which was interpreted as the manifestation of the leaks of gravitational radiation through the conformal boundary, eventually collected by an outer reservoir \cite{Almheiri:2019qdq,Almheiri:2019yqk,Almheiri:2020cfm}. However, the charges associated with boundary Weyl rescalings were non-vanishing only in odd dimensions, which we interpreted as the presence of Weyl anomalies in the dual theory \cite{Henningson:1998gx,deHaro:2000vlm,Papadimitriou:2005ii}. 

Due to the presence of leaks, the charge algebra was computed using the Barnich-Troessaert bracket \cite{Barnich:2010eb} and exhibited a field-dependent $2$-cocycle in odd dimensions. When the general framework was restricted to three-dimensional asymptotically AdS spacetimes with Dirichlet boundary conditions, the $2$-cocycle reduced to the Brown-Henneaux central extension \cite{Brown:1986nw}. The analysis was finally specified to leaky boundary conditions in asymptotically locally (A)dS spacetimes that lead to the $\Lambda$-BMS asymptotic symmetry group, generalizing the boundary conditions introduced in \ref{paper:P2}. In the flat limit, the latter contracts into the Generalized BMS group in $n$ dimensions, including smooth supertranslations and arbitrary smooth codimension 2 vectors on the $(n-2)$-dimensional celestial sphere.

\section{Reading guide and plan of the thesis}
We structure our presentation as follows. 

Chapter \ref{chapter:Charges} consists of a pedagogical introduction to the notion of asymptotic symmetries and their associated charges in gauge theories, with a main focus on General Relativity, leaning on the first and third chapters of the lecture notes \ref{paper:lect}. We provide the fundamental definitions and results to address problems of gravitational physics in the presence of boundaries, including the well-definiteness of the variational principle, the fixation of consistent boundary conditions and the associated residual gauge symmetries among which some gauge generators turn out to be canonically charged. We then discuss the mathematical structure of the solution spaces promoted as covariant phase spaces using a notion of presymplectic form that directly derives from the variational principle. We show how to use the functionalities of this particular kind of field space to compute canonical surface charges and address the crucial questions of their conservation and integrability. Finally, we review how the set of conserved integrable charges form an algebra representing the asymptotic symmetry algebra, and explain how to generalize this result when the charges are neither conserved nor integrable, features in which we are mainly interested in this thesis. 

Throughout the chapter, the concepts defined and articulated in various theorems are exemplified on the class of radiative asymptotically flat spacetimes with fluctuating boundary structure. Besides the obvious pedagogical utility, this basis provides a rich ground on which to build the ideas and methods developed in the subsequent chapters for other classes of radiative phase spaces. The asymptotic symmetry group under consideration is the Generalized BMS$_4$ symmetry group, consisting of a double infinite-dimensional extension of the Poincaré group into smooth supertranslations and super-Lorentz transformations. We describe the various features in the Bondi gauge, which are particularly well-adapted to treat radiative problems in gravity. The central notions of the Bondi analysis are reviewed in our notations and framework before presenting the renormalized phase space discussed in \ref{paper:P1}, including the overleading super-Lorentz symmetries as genuine asymptotic symmetries.

In Chapter \ref{chapter:AdSd}, after reviewing interesting facts about geometry and physics in dS and AdS spacetimes, we study the most general solution space with asymptotically locally (A)dS boundary conditions and the residual gauge transformations in the Starobinsky/Fefferman-Graham gauge. We use the holographic renormalization procedure to control and remove the divergences from the presymplectic structure, leading to the definition of general radiative phase spaces with a non-vanishing cosmological constant, without assuming any further boundary condition than the minimal fall-offs ensuring conformal compactness. Within this framework, we compute the surface charges associated with the residual gauge symmetries as well as their algebra. We finally apply the formalism to well-known sub-cases of conservative boundary conditions at conformal infinity. 

We extend the discussion in Chapter \ref{chapter:LambdaBMS}, where we identify a consistent boundary gauge fixing that singles out a set of residual gauge diffeomorphisms whose structure is very similar to the BMS group of flat space and reduces to it at flat limit. It therefore receives the name of $\Lambda$-BMS algebroid, because of the unavoidable field-dependence in the gauge parameters. We explain why the boundary conditions imposing this further gauge fixing are suitable to describe genuine radiative modes at conformal infinity, especially in asymptotically dS spacetimes where the Cauchy problem is left unconstrained. We then detail some aspects of the $\Lambda$-BMS phase space in four dimensions, principally in the Bondi gauge. We take care to derive the most general solution of Einstein's equations in this coordinate system. Thanks to a diffeomorphism between both gauges, we translate the dynamical quantities and the symmetry parameters from Starobinsky/Fefferman-Graham to Bondi coordinates, and identify the analog of the Bondi news tensor for radiative phase spaces with a non-vanishing cosmological constant. After describing an interesting sub-sector of conservative boundary conditions, we discuss the flat limit of the $\Lambda$-BMS$_4$ phase space and explain our method to regulate poles in $\Lambda$ in the presymplectic structure by addition of covariant corner terms at the boundaries of conformal infinity.

The flat limit process impressively gives the phase space of radiative asymptotically locally flat spacetimes and confirms the renormalization procedure previously discussed in Chapter \ref{chapter:Charges}. To close the loop opened at the beginning of the thesis, Chapter \ref{chapter:Flat} recovers the Einstein gravity with a vanishing cosmological constant and aims at answering several questions raised by the super-Lorentz extension of the BMS$_4$ group. First we derive the class of vacua obtained by exponentiating the Generalized BMS$_4$ group around Minkowski space, and identify the set of fields needed to parametrize this gauge orbit. We make good use of this knowledge to study the memory effects observed for vacuum transitions driven by non-stationary events localized in time and make contact with well-known effects such as the displacement memory effect. Finally, we give a prescription to define finite Generalized BMS$_4$ Hamiltonians for the important sub-class of asymptotically flat solutions without radiation at early and late times, which is the natural arena to model standard astrophysical configurations as well as the above-mentioned memory effects. We show that the integrated fluxes of these Hamiltonians are compatible with the soft theorems obtained at quantum level for the linearized theory and comment on the enhancing of the infrared structure of gravity via super-Lorentz symmetries. 

The manuscript is concluded in Chapter \ref{chapter:Conclusion}, which gives an overview of the results derived together with an outlook on the many interesting questions raised by our research program. The thesis also contains Appendices \ref{app:ExactVectors}--\ref{New massive gravity}, which gather additional technical material referenced in the main text.

A considerable effort was made to ensure that this thesis is self-contained, whether at the conceptual level or at the level of notations. Only some lengthy mathematical developments were not reproduced to provide a smoother reading experience. In order to be accessible to a broader readership beyond specialists, numerous well-known facts and results have been reviewed before discussing the original contributions of this research. Although resulting in the lengthening of the manuscript, we hope that this choice improves the experience of the reader who does not necessarily have extensive base knowledge, and thus would have the unpleasant task of interrupting reading for incessant dives into the abundant literature in the field. Having outlined the thesis plan, we now progress to the formulation of covariant phase spaces in General Relativity.\hfill{\color{black!40}$\blacksquare$}

%
%

\chapter{Covariant phase spaces in General Relativity}
\label{chapter:Charges}

This second chapter is a self-contained and duly contextualized introduction to the notion of covariant phase spaces in generally covariant theories. In section \ref{sec:Notion of asymptotic symmetries}, we rigorously define the notion of asymptotic symmetries in general gauge theories within the gauge fixing approach. The problem of fixing consistent boundary conditions and the formulation of the variational principle are discussed. We then give the extensive example of asymptotically flat Einstein's gravity in four dimensions, which allows us to highlight several results and features that are widely used in the following chapters and to pedagogically illustrate the theoretical concepts introduced thus far. We primarily consider the BMS$_4$ asymptotic symmetry group and its various super-Lorentz extensions. Section \ref{sec:Surface charges} subsequently discusses how to build canonical surface charges associated with asymptotic symmetries in gravity within the framework of the covariant phase space formalism, which is directly inspired by the Hamiltonian formulation of classical mechanics that we also briefly review. We address the crucial questions of the conservation and the integrability of the charges and particularize again to the radiative asymptotically flat spacetimes. Section \ref{sec:Surface charge algebra} concludes with an analysis of the algebraic properties of the surface charges and describes in which sense they represent the asymptotic symmetry algebra in full generality, without assuming conservation or integrability. Finally, we show that the surface charge algebra in asymptotically flat spacetimes contains deep physical information on the flux of gravitational radiation at null infinity and make several comments on the physics of thermodynamically open gauge systems.

This chapter consists of an updated and augmented version of the lecture notes \cite{Compere:2018aar} with additional elements and precisions gleaned in \cite{Barnich:2010xq,Barnich:2010eb,Barnich:2018gdh,Henneaux:1992ig,Barnich:2000zw,Barnich:2007bf,Lee:1990nz,Iyer:1994ys,Barnich:2001jy,Barnich:1994db,Ruzziconi:2020cjt,Ruzziconi:2019pzd,Compere:2007az} in order to improve pedagogical efficiency and to cover all of the features that are observed in the concrete realizations explained further in the manuscript. Although it looks like an oriented review of covariant phase space methods, this hybrid introductory part also contains original results. Sections \ref{sec:Application to asymptotically locally flat spacetimes at null infinity}, \ref{sec:Asymptotically locally flat radiative phase spaces} and \ref{sec:The Generalized BMS$_4$ charge algebra} are essentially reproduced from \cite{Compere:2018ylh}, while section \ref{sec:Non-integrable case: the Barnich-Troessaert bracket} is based on future work to appear \cite{ToAppear}. In summary, the aim of this formal prolegomenon -- which comprises the longest chapter of this thesis -- is to present the current understanding about the central notion of asymptotic symmetries and their surface charges in a fluid and pedagogical way and to offer our contribution to the field providing the basis for the more evolved considerations that are explored in the following chapters.

\section{Notion of asymptotic symmetries}
\label{sec:Notion of asymptotic symmetries}

This first section is aimed at defining the notion of asymptotic symmetries in a precise way, within the gauge fixing approach. The structure of the presentation exactly reflects the rigorous algorithmic procedure used to derive and study the asymptotic structure of any gauge theory, exemplified throughout with the 4$d$ Einstein theory of gravity with (not so standard) asymptotically flat boundary conditions. We successively discuss the definition of the action integral for gauge theories, the fixation of a suitable gauge in which boundary conditions can be formulated in a consistent and meaningful manner, the derivation of the solution space in this gauge for the considered boundary conditions and finally the uncovering of the residual gauge symmetries which are tangent to this solution space, which we eventually call \textit{asymptotic symmetries}. 

We stress that we restrain ourselves to the gauge fixing approach which was the route chosen in the various works compiled in this thesis, but alternative ways to study the asymptotic behavior of gauge theories also exist. The fully geometric approach, mostly relevant for gravity \cite{Penrose:1962ij,Penrose:1964ge,Ashtekar:1978zz,Ashtekar:1981bq,Ashtekar:2014zsa,Herfray:2020rvq}, has the privilege to remain manifestly gauge invariant and does not rely on any coordinate system in the discussion of asymptotic symmetries. The price to pay for this elegance is some lack of flexibility in imposing boundary conditions, implemented by assuming the existence of universal boundary structures that must be preserved by the symmetries. The modification of these boundary structures is often involved and consequently first worked out in the gauge fixing approach and re-understood later in terms of covariant structures. This holds also for the original content of this thesis, in which we try to give a purely geometrical equivalent to the various boundary conditions we are considering. Another successful approach that is worth mentioning is the Hamiltonian formalism, for which a slicing of the spacetime by constant time hypersurfaces is assumed to exist. This framework is logically well-adapted to treat spacelike infinities in gravity (see \textit{e.g.} \cite{Regge:1974zd,Brown:1986nw,Henneaux:1985tv,Henneaux:2019yax}) without requiring any further gauge fixing than the existence of a time coordinate, but suffers from some weaknesses when treating other kinds of boundaries, namely null boundaries. As a consequence, in the following, we always work in particular systems of coordinates in order to gain in pedagogical, interpretative and computational efficiency.

\subsection{Preliminary definitions}
We begin by providing basic definitions and notations that are necessary to formulate the notion of asymptotic structure in a mathematically robust way. Although our main concern is invariably Einstein's gravity, we keep the concepts and the notations sufficiently general to include a large class of Lagrangian theories with gauge invariances.

\subsubsection{Gauge symmetries}
Consider a classical field theory defined on some $n$-dimensional spacetime manifold $(\mathscr M,g)$, with a certain amount of fields $\phi^i$ collectively denoted by $\phi = (\phi^i)$. $\mathscr M$ is generically covered by local coordinates $\{ x^\mu \}$ (at least in a sufficiently large open neighborhood of the events under consideration). The dynamics of the theory is encoded into the action integral
\begin{equation}
S[\phi] = \int_{\mathscr M} \bm L[\phi,\partial_\mu\phi,\partial_\mu\partial_\nu\phi,\dots] \label{action general}
\end{equation}
where $\bm L = L\,\D^n x$ is the Lagrangian $n$-form. A general transformation of the fields $\phi = (\phi^i)$ is defined as
\begin{equation}
\delta_Q \phi^i = Q^i. \label{transfo}
\end{equation}
The characteristics $Q$ is generically a collection of \textit{local functions}, which means that each $Q^i$ depends on the spacetime coordinates as well as on the fields and their derivatives. Such a transformation is a \textit{symmetry} of the theory if and only if it preserves the action integral \eqref{action general}, or in other words
\begin{equation}
\delta_Q \bm L = \D \bm B_Q \label{symmetry condition}
\end{equation}
under the transformation \eqref{transfo}, for some codimension 1 form $\bm B_Q$. Among all the transformations acting on the system, we can find some of them that depend on parameters $\lambda = (\lambda^\alpha)$ that are arbitrary spacetime functions: they are coined as \textit{gauge transformations}. The number of independent gauge parameters in $\lambda$ is denoted as $n_g\in\mathbb N_0$. At infinitesimal level, we represent the action of a gauge transformation on the fields by \cite{Henneaux:1992ig}
\begin{equation}
\delta_\lambda \phi^i = R^i[\lambda] = \sum_{(\mu)} R^{i(\mu)}_\alpha \partial_{(\mu)}\lambda^\alpha \label{infinitesimal transformation}
\end{equation}
where the characteristics $R^{i(\mu)}_\alpha$, $\forall i,\alpha, k$, are also local functions of the spacetime coordinates, the fields and their derivatives. $(\mu)$ is a multi-index notation representing sets of implicitly symmetrized indices $(\mu) = (\mu_1,\mu_2,\dots,\mu_k)$. The cardinal of $(\mu)$ is denoted as $|\mu|=k$ and $\partial_{(\mu)} = \partial_{\mu_1}\partial_{\mu_2}\dots\partial_{\mu_k}$ in that case. The writing \eqref{infinitesimal transformation} assumes that the characteristics of the transformation $R[\lambda]$ depend linearly and homogeneously on the gauge parameters $\lambda$. The gauge transformation defined by the set of paramters $(\lambda^\alpha)$ is a gauge symmetry of the theory if and only if they obey the condition \eqref{symmetry condition}, \textit{i.e.} there exists a boundary term $\bm B_\lambda \in \Omega^{n-1}(\mathscr M)$ such that $\delta_\lambda \bm L = \D \bm B_\lambda$ under the infinitesimal transformation \eqref{infinitesimal transformation}. 

As a pedagogical example, we can consider free Maxwell theory, whose the only-field is the 4-potential 1-form $\bm A = A_\mu \D x^\mu$. The Lagrangian is built up from the Faraday's tensor $F_{\mu\nu} = 2\partial_{[\mu}A_{\nu]}$ and reads as $\bm L[\bm A] = -\frac{1}{4}F_{\mu\nu}F^{\mu\nu}\D^n x$. The transformation $\delta_\lambda \bm A = \D\lambda$ for any smooth function $\lambda(x)$ is a gauge symmetry of the theory: it simply leaves the Lagrangian invariant. Now let us consider Einstein's gravity, for which the dynamical field is $g$. The action integral is given by the Einstein-Hilbert functional
\begin{equation}
S_{EH}[g] = \frac{1}{16\pi G} \int_{\mathscr M} (R-2\Lambda)\sqrt{-g}\,\D^n x \label{general Einstein Hilbert action}
\end{equation}
where $R$ is the Ricci scalar curvature, $\sqrt{-g}\D^n x$ the volume form on $\mathscr M$ and $\Lambda$ the cosmological constant. A diffeomorphism $\xi$ is a gauge transformation which acts infinitesimally on the metric by a Lie derivative, $\delta_\xi g_{\mu\nu} = \mathcal L_\xi g_{\mu\nu} = \xi^\rho\partial_\rho g_{\mu\nu} + 2 g_{\rho(\mu}\partial_{\nu)}\xi^\rho$. It is straightforward to show that the gauge transformation generated by $\xi$ is a symmetry of the theory, in accordance with the general covariance principle. It is remarkable that for gravity, as well as for electromagnetism, the functions $R^{i(\mu)}_\alpha$ are vanishing for any $|\mu|>1$.

\subsubsection{Gauge fixing conditions}
\label{sec:Gauge fixing conditions}
In the presence of gauge symmetries, a physical theory has natural redundancies along the gauge orbits. Since we have $n_g$ parameters at our disposal, we can use them to perform a collective gauge transformation on the fields in order to satisfy some set of \textit{gauge fixing conditions}
\begin{equation}
\mathcal G[\phi] = \left\{ c_1[\phi] = 0,c_2[\phi]=0,\dots, c_{n_g}[\phi] = 0 \right\} \label{generic gauge fixing}
\end{equation}
where the constraints $c_i[\phi] = 0$ are assumed to be mutually independent. They can be either algebraic or differentiable but must be reachable by a gauge transformation included in the set of symmetries of the theory. For example, the Weyl gauge fixing in electromagnetism is algebraic and amounts to setting $A_0 = 0$ which is always reachable since $A_0 + \partial_0 \lambda = 0$ has always a solution for $\lambda$. The most famous gauge fixing, the Lorenz gauge, is differential and requires that $\partial_\mu A^\mu = 0$, reachable by a transformation satisfying $\partial_\mu A^\mu + \partial_\mu \partial^\mu\lambda=0$ which admits a solution for any field $\bm A$. 

In gravity, any gauge fixing consists in providing a set of coordinates to parametrize the metric field. When interested in gravitational wave physics, since the latter travel along light cones, one can use a coordinate system well-adapted to treat null radiation, \textit{i.e.} in which, concretely, null congruences live on constant-coordinate slices. Let us consider a family of null hypersurfaces labeled by constant $u$ coordinate in a $4$-dimensional spacetime ($n=4$). The normal vector of these hypersurfaces $n^\mu = g^{\mu\nu} \partial_\nu u$ is null by construction, so we fix $g^{uu} = 0$. We define angular coordinates $x^A = (\theta,\phi)$ such that the directional derivative along the normal $n^\mu$ is zero, $n^\mu \partial_\mu x^A = 0 \Rightarrow g^{uA} = 0$. We finally select the radial coordinate $r$ to be the luminosity distance, \textit{i.e.} we fix $\partial_r (\det g_{AB}/r^4) = 0$. The so defined coordinates $x^\mu = (u,r,x^A)$ are known in the literature as the \textit{Bondi-Sachs} coordinate system or \textit{Bondi gauge} \cite{Bondi:1962px,Sachs:1962wk,Sachs:1962zza}. $u$ has the meaning of a retarded time, hence the Bondi gauge we have just defined should be qualified as \textit{retarded Bondi gauge} since null rays at constant $u$ are outgoing in a sense that will be precised after giving a notion of asymptotics and imposing boundary conditions. After lowering the indices, we obtain the Bondi gauge fixing conditions on $g$ reading as $\{c_i[\phi] = 0\}_{i=1}^4$ with
\begin{equation}
g_{rr} = 0, \qquad g_{rA} = 0,\qquad \partial_r \left(\frac{\det g_{AB}}{r^4}\right) = 0. \label{Bondi gauge conditions}
\end{equation}
Let us notice that the conditions $c_1=0$, $c_2=0$ and $c_3=0$ are algebraic while $c_4=0$, coined as the \textit{Bondi determinant condition}, is differential. It is worth emphasizing that the condition $c_4$ is weaker than the historical determinant condition demanding that $\det g_{AB} = r^2 \det \mathring q_{AB}$ \cite{Bondi:1962px,Sachs:1962wk,Sachs:1962zza}. As we will see later, this slight extension allows to modify the boundary volume element by performing boundary Weyl rescalings \cite{Barnich:2010eb,Freidel:2021cbc}. In the Bondi parametrization, the line element takes the form
\begin{equation}
\D s^2 = e^{2\beta}\frac{V}{r}\D u^2 - 2 e^{2\beta}\D u \D r+g_{AB}(\D x^A - U^A \D u)(\D x^B - U^B \D u) \label{Bondi line element}
\end{equation}
where $\beta, U^A$ and $V$ are arbitrary functions of the coordinates. As a gauge fixing always reachable by diffeomorphism \cite{Sachs:1962wk,Sachs:1962zza}, any metric field can be expressed in these coordinates. For example, the Minkowski line element can be written as \eqref{Bondi line element} with $\beta = 0$, $U^A = 0$, $\frac{V}{r}=-1$ and $g_{AB} = \mathring q_{AB}$ the unit-round sphere metric on the 2-sphere,
\begin{equation}
\mathring q_{AB} \D x^A \D x^B = \D\theta^2 + \sin^2\theta \D\phi^2.
\end{equation}
Note crucially that fixing the gauge in gravity explicitly breaks general covariance by selecting a preferred class of observers who will express the physical quantities and perform calculations. In other words, the gauge fixing can be seen as the adjunction of a background structure while defining the dynamical fields of the theory. For Bondi gauge, this background structure is represented by a set of fixed null hypersurfaces and a particular tangent vector for their null generators. 

Another choice for the radial coordinate is to take it as an affine parameter $\rho$ along the null generators of the constant $u$ hypersurfaces. This defines the \textit{Newmann-Unti gauge} \cite{Newman:1962cia} with gauge conditions $g_{\rho\rho}=0$, $g_{\rho A} = 0$ and $g_{u\rho} = -1$, coming with the advantage to be completely algebraic. This gauge fixing exists as an alternative to the Bondi gauge and is particularly well-suited to make the link between the second order formulation of gravity that we present here and the first order tetrad formalism \cite{Newman:1968uj}.

\subsubsection{Residual gauge transformations}
\label{sec:Residual gauge transformations}
When the gauge fixing conditions \eqref{generic gauge fixing} are imposed, the set of allowed gauge transformations is reduced, sometimes drastically, but not completely destroyed. For example, fixing the Lorenz gauge in electromagnetism still allows for gauge transformations generated by harmonic functions ($\partial_\mu\partial^\mu\lambda=0$). This is a very simple example of \textit{residual gauge transformations} defined as gauge transformations that preserve the gauge fixing conditions \eqref{generic gauge fixing}, namely $\delta_\lambda c_i[\phi] = 0$, $i=1,\dots,n_g$. The generators $\lambda$ that solve this set of conditions can be re-parametrized in terms of local functions of $(n-1)$ coordinates, \textit{i.e.} $\lambda = \lambda(s)$ for $s=\{s_1,s_2,...\}$. 

Turning back to our favorite example of gravity in the Bondi gauge, infinitesimal diffeomorphisms $\xi$ that preserve the gauge fixing conditions \eqref{Bondi gauge conditions} have to satisfy \cite{Bondi:1962px,Barnich:2010eb,Compere:2018ylh}
\begin{equation}
\mathcal{L}_\xi g_{rr} = 0, \quad \mathcal{L}_\xi g_{rA} = 0, \quad g^{AB} \mathcal{L}_\xi g_{AB} = 4 \omega(u, x^C).
\label{eq:GaugeConstraints}
\end{equation}
While the two first constraints are readily understandable, the third one needs some work. The Bondi determinant condition can be integrated in $\det( g_{AB}) = r^4\chi(u,x^C)$ where $\chi$ is an arbitrary function of $(u,x^C)$. So
\begin{equation}
g^{AB}\mathcal L_\xi g_{AB} = \delta_\xi \ln [\det(g_{AB})] = \delta_\xi \ln r^4 + \delta_\xi \ln\chi = 4\omega(u,x^C)
\end{equation}
where $\omega(u,x^C)$ is another arbitrary function of $(u,x^C)$ and the prefactor $4$ is introduced for convenience. Developing the Lie derivatives in \eqref{eq:GaugeConstraints} and integrating the resulting partial differential equations yields
\begin{equation}
\begin{split}
\xi^u &= f, \\
\xi^A &= Y^A + I^A, \quad I^A = -\partial_B f \int_r^\infty \D r'  (e^{2 \beta} g^{AB}),\\
\xi^r &= - \frac{r}{2} (\mathcal{D}_A Y^A - 2 \omega + \mathcal{D}_A I^A - \partial_B f U^B + \frac{1}{2} f g^{-1} \partial_u g) ,\\
\end{split}
\label{eq:xir}
\end{equation} 
where $\partial_r f = 0 = \partial_r Y^A$ and $g= \det (g_{AB})$ \cite{Barnich:2010eb,Barnich:2013sxa}. The covariant derivative $\mathcal{D}_A$ is associated with the $2$-dimensional metric $g_{AB}$. The residual gauge transformations are thus parametrized by 4 functions $\omega$, $f$ and $Y^A$ of $(u,x^A)$. Let us remark that some components of the metric field explicitly appear in \eqref{eq:xir}! This is a generic result: since the gauge fixing conditions $c_i[\phi]=0$ involve the fields, the preservation conditions $\delta_\lambda c_i[\phi] = 0$ may have field-dependent solutions. As we have seen, the most striking example of that is gravity, because the spacetime itself has a dynamical structure and the gauge field is the metric: simply raising some indices in one of these conditions is sufficient to bring field-dependence into the game.

\subsection{Solution space and boundary conditions}
\label{sec:Theory solution space}
For any local theory such as \eqref{action general}, the dynamical equations for the fields $\phi$ are partial differential equations which need \textit{boundary conditions} in order to be solved. This is the unavoidable step to precise the kinematics of the system. Like the gauge fixing, the choice of boundary conditions is completely free but should be motivated by the class of physical phenomena one wants to describe. This is the most ``artistic'' part in the path leading to the formulation of a phase space, since there is neither mathematical nor physical generic prescription to define the right boundary conditions for the right purpose. Their deduction, or better, discovery, is often the achievement of a quest summarized by the optimization problem that we describe now.

\subsubsection{The art of fixing boundary conditions}
Who says ``boundary conditions'' says in particular ``boundary,'' a concept that can appear under multiple avatars: a black hole horizon, an hypersurface located somewhere in the spacetime, or also some notion of infinity at large distances. Let us define a boundary $\mathscr B$ in $\mathscr M$ and $\mathscr U$ an open neighborhood of $\mathscr B$. While discussing asymptotics, $\mathscr B$ can be repelled at infinity in a sense that we will precise in a few lines. Giving boundary conditions in the vicinity of $\mathscr B$ consists in prescribing the behavior of the fields in the neighborhood $\mathscr U$ (\textit{fall-off conditions}) as well as their value on $\mathscr B$ -- for those that do not decay sufficiently fast while traveling $\mathscr U$ towards $\mathscr B$. We can choose a coordinate system covering $\mathscr U$ such that one of these coordinates, say $r$, tends to infinity when approaching $\mathscr B$. Fall-off conditions dictate the behavior in $r$ of the various fields when $r$ is large. The latter may be singular on $\mathscr B$ but this issue can often be cured by a change of coordinates on $\mathscr U$, or by taking another picture of the situation if $\mathscr B$ stands for some notion of infinity (for example, through a conformal compactification process \cite{Penrose:1964ge}). A set of boundary conditions, encompassing fall-off conditions in $r$ as well as codimension 1 conditions on the fields pulled back to $\mathscr B$, is considered to be physically relevant if the following requirements are met:
\begin{itemize}[label=$\rhd$]
	\item The conditions are sufficiently weak to have a non-trivial associated set of solutions while keeping a sufficient amount of symmetries acting on it. The conserved quantities in the evolution of the system should be finite while $r$ runs to infinity and generically non-vanishing for a non-trivial subset of solutions.
	\item The conditions are sufficiently strong to avoid dealing with unphysical solutions (in a sense that obviously depends on the situation) and infinite conserved quantities. 
\end{itemize}
We can help ourselves by a careful study of the asymptotic structure of some well-known solutions that we want to include in the solution space compatible with the boundary conditions. Anyways, the best boundary conditions must minimize the tension between the two requirements enunciated here above, although the task to derive them is highly non-trivial. Let us finally emphasize that any set of boundary conditions underlies a fixation of a \textit{boundary structure} including the boundary itself. The latter is said \textit{universal} because it is shared by all of the solutions and can take different forms: a boundary foliation, a metric tensor fixed on $\mathscr B$, a codimension 1 source for the fields, \textit{etc.} We will particularize this quite philosophical discussion to concrete examples in sections \ref{sec:Application to asymptotically locally flat spacetimes at null infinity} and \ref{Boundary conditions and asymptotic symmetry algebra LBMS}. 

\subsubsection{Solution space and equations of motion}
\label{sec:Solution space and equations of motion}
Let us consider again our general theory \eqref{action general} for which we have imposed the gauge fixing \eqref{generic gauge fixing} and given some boundary conditions. The elementary pieces of kinematics are right in place to study the solution of the equations of motion that prescribe the evolution of the fields $\phi$. These are the Euler-Lagrange equations, written formally as
\begin{equation}
\frac{\delta \bm L}{\delta \phi^i} = \sum_{(\mu)}(-1)^{|\mu|} \partial_{(\mu)} \left( \frac{\partial \bm L}{\partial \partial_{(\mu)}\phi^i} \right) = 0. \label{EulerLagrange}
\end{equation}
The \textit{solution space} $\mathcal S$ of the theory is defined as the set of field configurations that solve the Euler-Lagrange equations while satisfying the gauge fixing conditions as well as the boundary conditions. After formal resolution, any solution $\phi\in\mathcal S$ can be parametrized by arbitrary functions of $(n-1)$ coordinates $p = \{p_1,p_2,\dots\}$, \textit{i.e.} $\phi = \phi(p)$. We will observe this feature on the solution space of asymptotically locally flat spacetimes.

\subsection{Variational principle}
\label{sec:Variational principle theory}

Since the pioneer works of d'Alembert and Lagrange, we know that the dynamical equations \eqref{EulerLagrange} are in fact a consequence of a variational principle which is traditionally formulated as a optimization problem involving the action integral \eqref{action general} as a functional on the fields $\phi$. The \textit{variations} $\delta\phi$ are defined as infinitesimal perturbations of $\phi\in\mathcal S$ such that $\phi+\delta\phi$ also belongs to $\mathcal S$. This means in particular that $\delta\phi$ must obey the boundary conditions imposed in the vicinity of $\mathscr B$. Considering an arbitrary variation $\delta\phi$, we have \cite{Lee:1990nz,Iyer:1994ys}
\begin{equation}
\begin{split}
\delta \bm L = \delta\phi^i \frac{\partial\bm L}{\partial\phi^i} + \partial_\mu \delta\phi^i \frac{\delta \bm L}{\delta \partial_\mu\phi^i}+\cdots = \delta\phi^i \frac{\delta\bm L}{\delta\phi^i} + \D \bm\Theta[\phi;\delta\phi]. \label{delta L theo}
\end{split}
\end{equation}
The second equality is obtained by performing inverse Leibniz rules on the spacetime derivatives $\partial_\mu,\partial_\mu\partial_\nu \dots$ in order to recognize \eqref{EulerLagrange} in front of the arbitrary variation $\delta\phi$. The residue of these integrations by part is a boundary term appearing as the codimension 1 form $\bm \Theta[\phi;\delta\phi]$ depending linearly on $\delta\phi$, $\partial_\mu\delta\phi$, $\partial_\mu\partial_\nu\delta\phi\dots$ It goes under the name of \textit{presymplectic potential} \cite{Lee:1990nz} for reasons that will be clarified in section \ref{sec:Covariant phase space formalism}. Integrating over $\mathscr M$, we get
\begin{equation}
\delta S[\phi] = \int_{\mathscr M} \delta\phi^i \frac{\delta\bm L}{\delta\phi^i} + \int_{\mathscr B} \bm \Theta[\phi;\delta\phi] 
\end{equation}
thanks to the Stokes theorem. The cancellation of the bulk term for any set of arbitrary variations leads to the equations of motion \eqref{EulerLagrange}, so much so that the \textit{on-shell variational principle}, \textit{i.e.} considered for variations $\delta\phi$ around a point of $\mathcal S$, is given by \cite{Lee:1990nz}
\begin{equation}
\delta S[\phi] = \int_{\mathscr B} \bm \Theta[\phi;\delta\phi]. \label{symplectic flux general}
\end{equation}
Contributions coming from the boundaries of $\mathscr B$ (``\textit{corners}'') will be ignored in the current discussion for the sake of simplicity. Although they are crucial for the definition of the complete variational principle, we are mainly interested in dynamical (radiative) parts of $\mathscr B$ and not the corners that are often rigid boundaries without dynamics. The problem to incorporate these corner contributions will be addressed later in this thesis (see section \ref{sec:Some residual ambiguities} for details concerning codimension 2 ambiguity terms in the presymplectic potential and sections \ref{sec:Flat variational principle} and \ref{sec:Flat limit of the action and corner terms} for concrete examples).

The result \eqref{symplectic flux general} is an important observation: the pull-back of the presymplectic potential on the boundary $\mathscr B$ controls the on-shell variational principle. The right-hand side of this equation is called the \textit{presymplectic flux} through $\mathscr B$. The precise value of it depends on the fixation of the boundary conditions around $\mathscr B$. That is where physics comes into the setup! In many situations (as in standard classical mechanics) we are interested in systems without presymplectic flux at the boundary, but we will argue that sometimes allowing for some non-vanishing presymplectic flux through the boundary is mandatory in order to preserve the whole dynamics of the system. We are thus facing a dichotomous choice in the boundary conditions, that can either forbid or allow for a non-vanishing presymplectic flux. 

\subsubsection{Conservative boundary conditions} 
\label{sec:Conservative boundary conditions theo} 
We qualify as \textit{conservative boundary conditions} a set of conditions imposed on $\mathscr B$ such that the presymplectic flux vanishes for any variation. In other words, the action $S[\phi]$ is \textit{stationary} on-shell,
\begin{equation}
\delta S[\phi] = 0. \label{action stationary}
\end{equation}
This is the case, for instance, if $\bm\Theta[\phi;\delta\phi]$ decays sufficiently fast when approaching $\mathscr B$, or if the variations $\delta\phi$ are designed to have sufficiently strong fall-offs to cancel out on the boundary. The requirement of vanishing symplectic flux is obviously defined up to a redefinition of the action integral \eqref{action general} by pure boundary terms that do not affect the Euler-Lagrange equations \eqref{EulerLagrange}. Therefore in general, a set of boundary conditions is declared to be conservative if and only if there exists a codimension 1 form $\bm B[\phi|\bar\chi]$ depending on the fields $\phi$ as well as the background structures $\bar\chi$ needed to fix the boundary conditions at $\mathscr B$, such that
\begin{equation}
S\to S'=S + \int_{\mathscr B} \bm B[\phi|\bar\chi], \label{S+int B}
\end{equation}
is stationary on-shell, $\delta S'[\phi] = 0$. Again using Stokes theorem, we see that $\bm L'[\phi] = \bm L[\phi] + \D\bm  B[\phi|\bar\chi]$. Performing a variation as in \eqref{delta L theo}, we see that $\delta S'[\phi] = 0$ is ensured if and only if
\begin{equation}
\bm\Theta[\phi;\delta\phi]\Big|_{\mathscr B} = -\delta\bm B[\phi|\bar\chi], \label{theta is exact}
\end{equation}
(see \textit{e.g.} \cite{Julia:2002df,Compere:2007az}). Conservative boundary conditions describe theories without any dynamics through the boundary $\mathscr B$, since traveling $\mathcal S$ from $\phi$ to $\phi+\delta\phi$ conserves the total action integral of the system which contains all the information about it. When a covariant phase space will be rigorously defined, we will see that this kind of boundary conditions lead to conserved charges, which justifies \textit{a posteriori} the terminology. 

\subsubsection{Leaky boundary conditions and open action principles} 
\label{sec:Leaky boundary conditions theo}
In contrast, boundary conditions allowing for some presymplectic flux through the $\mathscr B$ will be called \textit{leaky boundary conditions}. In these configurations, the condition \eqref{theta is exact} must be negated: it implies that there is no boundary term that can supplement the action integral in order to have a refined action which is stationary on solutions. In that context, what we consider to be a ``well-defined variational principle'' must be revised, since the stationarity is no longer a criterion that can be imposed. We say that the variational principle \eqref{symplectic flux general} is \textit{well-defined} if the two following assumptions are met\cite{Fiorucci:2020xto,Ruzziconi:2020wrb}:
\begin{enumerate}[leftmargin=1.5cm,label={(A\arabic*)}]
\item The variation of the action $\delta S[\phi]$ on $\mathcal S$ is \textit{finite} on-shell, for the sake of definiteness and regularity of the presymplectic flux. \label{well defined dS 1}
\item The stationarity of the action is restored for any sub-class of non-radiative solutions, up to the incorporation of a finite counter-term like \eqref{S+int B}. 
 \label{well defined dS 2}
\end{enumerate}
In other words, \ref{well defined dS 2} stipulates that a subset of conservative boundary conditions must exist as a limiting case of leaky boundary conditions when turning off the ``leaking fields'' that source the presymplectic flux and demanding that $\bm\Theta[\phi;\delta\phi]$ is selected in such a way that $\delta S[\phi]=0$ when these sources are absent. In general, the definition of a renormalization scheme will be necessary in order to obey the assumption \ref{well defined dS 1}, while an adjustment of the finite part of the renormalized presymplectic potential will easily ensure the second assumption \ref{well defined dS 2}. In our sense, a well-defined variational principle is therefore more general than \eqref{action stationary}: the on-shell action is not supposed to be preserved when moving from $\phi$ to $\phi+\delta\phi$: the discrepancy between the two values $S[\phi]$ and $S[\phi]+\delta S[\phi]$ is precisely controlled by the presymplectic flux through $\mathscr B$. There is an exchange between the system under consideration, enclosed in $\mathscr B$, and the exterior \textit{environment} outside $\mathscr B$. Leaky boundary conditions are therefore unavoidable when treating dissipative phenomena as well as open systems. In particular, the various charges that are defined on the system (such as the mass, the energy, the electric charge \textit{etc.}) will be associated with some flux passing through $\mathscr B$ and, therefore, will not be expected to be conserved. In this thesis, we will give several examples of such open systems for Einstein's gravity in various dimensionalities and for different values of the cosmological constant. We will motivate the importance of considering open variational systems for which $S[\phi]$ is not stationary on-shell for a fundamental study of the radiative sector of the theory. 

At this point, we insist on the fact that the statement ``$\delta S\neq 0$ on-shell'' is \textit{not} at all a rejection of the principle of least action, the latter being without any doubt one of the most robust pillars of any theoretical approach of Physics. The non-stationarity is due to the fact that we formulate the variational principle for the system with boundary $\mathscr B$ without telling more about what is beyond $\mathscr B$ and what is collecting the symplectic flux going through this boundary. In this point of view, \eqref{symplectic flux general} is nothing but an avatar of some continuity equation through the interface $\mathscr B$ and characterizes how far we are from an equilibrium state without leaks. When the action principle is well-defined in the sense of the hypotheses \ref{well defined dS 1} and \ref{well defined dS 2}, we can also say that it is \textit{compatible} with a global stationary variational principle encompassing the dynamics of the system as well as its environment. Indeed, when these two assumptions are obeyed, it is possible to define a complete variational principle for the system and one particular environment whose dynamics has to be precised. In particular, if the system and the environment form together a closed system, the stationarity of the complete on-shell action can be restored.

This temporarily closes our discussion on the variational principle. We will address more precisely the questions on what is physical or not in the presymplectic flux and what are the laws controlling the non-conservation of the dynamical quantities of the system at the boundary $\mathscr B$ when canonical charges associated with gauge symmetries have been defined in section \ref{sec:Surface charges}. This will need a bunch of formal tools that we do not want to introduce now in order to get directly into the notion of asymptotic symmetries.

\subsection{Asymptotic symmetries}
We have enough material to define rigorously the notion of asymptotic symmetries in our approach. Let us quickly recall the main steps and the necessary ingredients. We started from a Lagrangian theory of fields \eqref{action general} with some gauge symmetries \eqref{infinitesimal transformation} parametrized by $\lambda$. In order to eliminate a considerable part of trivial gauge transformations mapping equivalent physical situations to each other, we imposed the consistent set of gauge fixing conditions $\mathcal G[\phi]$ (see \eqref{generic gauge fixing}) which are always reachable by a gauge transformation and bring generically constraints among the gauge parameters because $\delta_\lambda \mathcal G[\phi]=0$ by definition. The transformations compatible with these constraints have been coined as residual gauge transformations and are parametrized as $\lambda = \lambda(s)$ where $s$ denote a collection of codimension 1 functions. Next, we can impose boundary conditions and derive the associated class $\mathcal S$ of solutions to Euler-Lagrange equations \eqref{EulerLagrange}. The elements of $\mathcal S$ are parametrized as $\phi = \phi(p)$ where $p$ denotes another set of codimension 1 functions. Finally, the last question we want to address is: what are the gauge symmetries that survive on the solution space after the imposition of the boundary conditions? In other words, are \textit{all} the parameters in $s$ still pertinent to describe the residual gauge transformations, or do some of them have been ruled out by the boundary conditions?

\subsubsection{General definition}
A gauge transformation $\delta_\lambda \phi = R[\lambda]$ is \textit{tangent} to the solution space $\mathcal S$ if it preserves the gauge fixing conditions \eqref{generic gauge fixing} and the boundary conditions defining $\mathcal S$. The first requirement implies that $\lambda$ has to be found among the residual gauge transformations. The fact that $\delta_\lambda \phi$ is tangent to the solution space is obviously an on-shell notion and gives some constraints among the parameters $s$ which are generically dependent on the field configuration $\phi$. The intuitive point of view is that such a gauge transformation preserves the location of the limiting hypersurface $\mathscr B$ defined in section \eqref{sec:Theory solution space} and the boundary structure defined on $\mathscr B$ to implement the boundary conditions. Because this kind of residual gauge transformation is designed to preserve the universal boundary structure leading to the boundary conditions under consideration, it receives the name of \textit{asymptotic symmetry}. In gravity, when the boundary structure is a fixed induced metric, a particular class of asymptotic symmetries, that generalize the concept of exact isometries, are diffeomorphisms $\xi$ that solve the Killing equation $\nabla_{(\mu}\xi_{\nu)} = 0$ asymptotically. For this reason they go under the name of \textit{asymptotic Killing vectors} although this nomenclature is not appropriate for more general sets of boundary conditions. Even though our interest is regarding only to asymptotic boundaries, we mention that numerous classes of asymptotic symmetries around boundaries located at finite distance in the bulk of spacetime have been studied in the literature, namely at the black hole horizon \cite{Hawking:2016sgy,Hawking:2016msc,Haco:2018ske,Donnay:2016ejv,Donnay:2019jiz,Donnay:2019zif,Grumiller:2019fmp,Koga:2001vq,Carlip:2002be,Silva:2002jq,Koga:2006ez}. The study of asymptotic symmetries benefits also of a popular research movement in other gauge theories (see \textit{e.g.} \cite{Barnich:2013sxa,Strominger:2013lka} for Yang-Mills theories or \cite{He:2014cra,Lu:2019jus,Afshar:2018apx} for electrodynamics and $k$-forms theories).

\subsubsection{Asymptotic symmetry algebra}
\label{sec:Asymptotic symmetry algebra}
By definition, the asymptotic symmetries act internally on the solution space. As usual, two symmetries can be performed successively on the physical system described by the action integral \eqref{action general} and the operation is equivalent to another transformation performed in a single stroke. Each symmetry is meant to be inverted and there exists a neutral transformation representing the absence of transformation. In short, there is a notion of group of transformations underlying the class of asymptotic symmetries. The composition of two gauge transformations is translated at infinitesimal level by a commutation law for the generators $\lambda$. Given a couple of generators $(\lambda_1,\lambda_2)$, we define the \textit{modified bracket} as \cite{Barnich:2010xq,Barnich:2010eb,Barnich:2018gdh}
\begin{equation}
[\lambda_1,\lambda_2]_\star \equiv [\lambda_1,\lambda_2] - \delta_{\lambda_1}\lambda_2 + \delta_{\lambda_2}\lambda_1. \label{modified bracket general}
\end{equation}
The first $[\lambda_1,\lambda_2]$ represents the standard Lie bracket between gauge parameters generically defined as \cite{Barnich:2001jy}
\begin{equation}
[\lambda_1,\lambda_2]\equiv \sum_{(\mu),(\nu)} C_{\alpha\beta}^{(\mu)(\nu)} \partial_{(\mu)}\lambda_1^\alpha \partial_{(\nu)}\lambda_2^\beta
\end{equation}
where the structure functions $C_{\alpha\beta}^{(\mu)(\nu)}$ are anti-symmetric in $(\alpha,\beta)$ and possibly field-dependent. For example, the vector Lie bracket on the tangent bundle of $\mathscr M$, $[\xi_1,\xi_2] = (\xi_1^\mu \partial_\mu \xi_2^\nu - \xi_2^\mu \partial_\mu \xi_1^\nu)\partial_\nu$, falls into this definition. The additional terms $- \delta_{\lambda_1}\lambda_2 + \delta_{\lambda_2}\lambda_1$ in \eqref{modified bracket general} must be included as soon as the generators are field-dependent. They bring a correction to extract the true action of one generator, say $\lambda_1$, on the other, $\lambda_2$, regardless of the peculiar action of $\lambda_1$ on the field-dependence of $\lambda_2$. It can be shown that the so-defined modified bracket is a Lie bracket (\textit{i.e.} it acts as a antisymmetric bilinear operator on the vector space of generators and satisfies the Jacobi identity) since the standard bracket is itself a Lie bracket. This implies that the asymptotic symmetry generators form an algebra, coined as the \textit{asymptotic symmetry algebra.}

\subsubsection{Representation on the solution space}
\label{sec:Representation on the solution space}
The asymptotic symmetries act infinitesimally on the solution space $\mathcal S$ as $\delta_\lambda \phi = \phi[p+\delta_\lambda p] - \phi[p]$ where the difference is evaluated by retaining only the linear order in $\delta_\lambda p$. This is the statement that the formal structure of the solution $\phi(p)$ is preserved in order to keep solving the equations of motion for the prescribed the boundary conditions after that the transformation has been performed. Only the parameters $p$ vary non-trivially under the action of the generators $\lambda$ and this transformation is fixed by the preservation requirement we have just mentioned. The algebra of generators under the modified bracket \eqref{modified bracket general} is therefore represented on $\mathcal S$ because
\begin{equation}
\delta_{\lambda_1} R[\lambda_2] - \delta_{\lambda_2} R[\lambda_1] = -R\left[ [\lambda_1,\lambda_2]_\star\right] \label{bracket of variations}
\end{equation}
holds on-shell, \textit{i.e.} for any $\phi\in\mathcal S$ (see \textit{e.g.} \cite{Henneaux:1992ig,Barnich:2000zw,Barnich:2007bf} for a proof). The left-hand side defines a bracket on the variations as
\begin{equation}
[\delta_{\lambda_1},\delta_{\lambda_2}]\phi \equiv \left[ R[\lambda_1],R[\lambda_2] \right]
\end{equation}
while the right-hand side is nothing but the variation under the transformation generated by the modified bracket of generators, \textit{i.e.} $\delta_{[\lambda_1,\lambda_2]_\star}\phi = R[[\lambda_1,\lambda_2]_\star]$. Hence the representation result can be written as
\begin{equation}
\boxed{
[\delta_{\lambda_1},\delta_{\lambda_2}]\phi = -\delta_{[\lambda_1,\lambda_2]_\star}\phi.
}
\label{representation on solution space general}
\end{equation}
Because of the generic field-dependence of the asymptotic symmetries, there is a tight link between them and the solution space from which they emerge. For that reason, the natural structure appearing here is not properly a Lie algebra, but rather a \textit{Lie algebroid} \cite{Crainic,Barnich:2010xq,Barnich:2017ubf} of asymptotic symmetries. 

The axiomatic definition is as follows. Let $\mathscr N$ be a differentiable manifold. A Lie algebroid over $\mathscr N$ consists of a vector bundle $\mathscr A$ together with an \textit{anchor map} $\rho_{\mathscr A} : \mathscr A \to T\mathscr N$ from $\mathscr A$ onto the tangent bundle of $\mathscr N$ and a Lie bracket $[\cdot,\cdot]_{\mathscr A}$ on the space of sections $\Gamma(\mathscr A)$, satisfying the Leibniz rule $[\alpha,f\beta]_{\mathscr A} = f[\alpha,\beta]_{\mathscr A} + \beta \mathcal L_{\rho_{\mathscr A}(\alpha)}f$ for all $\alpha,\beta$ in the vector space $\Gamma(\mathscr A)$ and all smooth $f$ on $\mathscr N$. This formalizes the idea that a Lie algebroid is a Lie algebra which depends on the point of $\mathscr N$ we want to consider in the section $\Gamma(\mathscr A)$. The most intuitive example of a Lie algebroid is when $\mathscr A = T\mathscr N$ with the identity map as anchor and $[\cdot,\cdot]_{\mathscr A}$ is the vector commutator. The tangent space in a point $P\in\mathscr N$ is obviously depending on $P$ for general geometrical configurations since it is a local notion defined from the tangent directions to curves intersecting at $P$. A Lie algebra is a limiting case of Lie algebroid for which $\mathscr N$ is reduced to a singleton. In the context that concerns us, the base manifold $\mathscr N$ is the solution space $\mathcal S$ and the vector bundle $\mathscr A$ is populated by the gauge parameters $\lambda$ whose each section is equipped with the modified Lie bracket \eqref{modified bracket general}. The characteristics $R[\lambda]$ of the asymptotic symmetries driven by $\lambda$ belong to the tangent space of $\mathcal S$, as we explained earlier. The anchor map is given by $\lambda\to R[\lambda]$. Considering a particular section of $\mathscr A$ localizes the field-dependent gauge parameters around a point $\phi\in\mathcal S$ and each section is a Lie algebra closing under the modified Lie bracket \eqref{modified bracket general}. The last terms $- \delta_{\lambda_1}\lambda_2 + \delta_{\lambda_2}\lambda_1$ gain a new interpretation: their presence is mandatory to ensure that the bracket is internal to the section $\Gamma(\mathscr A)$ taken for the solution $\phi\in\mathcal S$, or in a more pedestrian way, that we remain stuck at $\phi$ when one generator acts involuntarily on the field-dependence of its partner. Finally, in this picture, one can say that the anchors map guarantees that the solution space forms a representation of the Lie algebroid of asymptotic symmetries, which is precisely the statement \eqref{representation on solution space general}. This quite robust result can be verified explicitly on the example of asymptotically locally flat solution space of Einstein's gravity, although the check is computationally intricate and will not presented here.

\subsection{Application to asymptotically locally flat spacetimes at null infinity}
\label{sec:Application to asymptotically locally flat spacetimes at null infinity}

For the purpose of illustrating this technical chapter, we are interested in solutions of Einstein's gravity written in the Bondi gauge $\{u,r,x^A\}$ described by observers analyzing gravitational waves and other null wave phenomena emitted by remote sources situated at astrophysical distances from them. These observers are living in the radiation zone at large $r$ where these null waves leave their imprint on spacetime far from the sources of emission. Below cosmological scales, a very good approximation willing to model this kind of situation is the notion of asymptotically flat spacetime at future null infinity \cite{Penrose:1962ij,Ashtekar:1978zz,1977asst.conf....1G}. The latter is defined as the far future of light, described by the limit $r\to\infty$ while keeping the retarded time $u$ fixed.

\subsubsection{Boundary conditions}
\label{sec:Flat BC}
The requirement of asymptotic flatness (at future null infinity), giving some notion of localized sources of emission, implies the following fall-off conditions on the metric field in the Bondi gauge \cite{Bondi:1962px,Sachs:1962wk,Sachs:1962zza,Penrose:1965am,Barnich:2010eb,Compere:2018ylh}:
\begin{equation}
\beta = o(r^0), \qquad \frac{V}{r} = o(r^2), \qquad U^A = o(r^{0}), \qquad g_{AB} = r^2\, q_{AB}+r\, C_{AB}+\mathcal O(r^0). \label{preliminary BC Bondi}
\end{equation}
In these notations, $q_{AB}$, $C_{AB}$ and the subleading coefficients in $g_{AB}$ are codimension 2 symmetrical tensors whose components are functions of $(u,x^A)$. The fourth requirement in \eqref{preliminary BC Bondi} has a separate status because it implies that the angular metric $g_{AB}$ has a second order pole at null infinity. This problem can be cured by performing a conformal compactification \cite{Penrose:1964ge} with the conformal factor $\Omega = 1/r$ in order to have a well-behaved though unphysical metric field $\Omega^2 g$ when $r\to\infty$ (see section \ref{sec:Conformally compact manifolds} for more details on the conformal compactification process). The limiting surface at $\Omega = 0$ is the conformal boundary of the unphysical spacetime and is denoted as $\mathscr I^+$. By identification with the region $r\to\infty$ in the physical spacetime, one denotes $\mathscr I^+ = \{r=+\infty\}$ and abusively calls it \textit{future null infinity}. The pull-back of the Bondi line element to $\mathscr I^+$ gives the degenerate line element $0\times \D u^2 + q_{AB}\D x^A \D x^B$. The boundary conditions are imposed geometrically by fixing a null normal vector $\bm T$ to $\mathscr I^+$ (which is also tangential because $\mathscr I^+$ is null) and a codimension 2 metric $q_{AB}$ on the space of null generators of $\mathscr I^+$ \cite{Ashtekar:1978zza,1977asst.conf....1G}. This space is topologically $S^2$ and goes under the name of \textit{celestial sphere}, denoted by $S_\infty$. Incorporating the null direction spanned by the coordinate $u$, $\mathscr I^+$ has the topology of a null cylinder $\mathbb R\times S^2$. Every outgoing null ray in the spacetime ends at $\mathscr I^+$ for some retarded time $-\infty < u < +\infty$. The 2-sphere obtained in the limit $u\to -\infty$ and denoted as $\mathscr I^+_-$ represents spatial infinity. Indeed, recalling that $u = t_M-r_M$ in Minkowski spacetime where $t_M$ is the global time and $r_M$ the radius of constant time 2-spheres, $\mathscr I^+_-$ describes the asymptotic sphere such that $r_M\to\infty$ for any time $t_M$. This is the end-point of any spacelike curve in $\mathscr M$. Conversely, the asymptotic sphere $\mathscr I^+_+$, reached in the limit $u\to\infty$ on $\mathscr I^+$, or $t_M\to\infty$ for any $r_M$, is recognized as the future timelike infinity, where any timelike curve in $\mathscr M$ terminates.

\begin{figure}[ht!]
\centering
\begin{tikzpicture}
\draw[opacity=0] (-1,-1) -- (-1,5) -- (5,5) -- (5,-1) -- cycle;
\draw[thick,black] (0.5,4.5)node[circle,fill=black,inner sep=2pt]{} -- (4.5,0.5)node[circle,fill=black,inner sep=2pt]{};
\draw[] (0.5,4.5)node[anchor=south east]{$\mathscr I^+_+$};
\draw[] (4.5,0.5)node[anchor=north west]{$\mathscr I^+_-$};
\draw[thick,red,-Latex] (1.5,1.5)node[anchor=north,rotate=-45]{$u=\text{cst}$} -- (2.35,2.35);
\draw[thick,red,-Latex] (1.8,1.2) -- (2.65,2.05);
\draw[thick,red,-Latex] (1.2,1.8) -- (2.05,2.65);
\draw[-Latex,orange,very thick] (2.2,2.8) -- (1.3,3.7)node[anchor=south west]{$\bm T$};
\draw[] (2.5,2.5)node[circle,fill=blue,inner sep=2pt]{};
\draw[] (2.8,2.2)node[circle,fill=blue,inner sep=2pt]{};
\draw[] (2.2,2.8)node[circle,fill=blue,inner sep=2pt]{};
\draw[blue] (2.6,2.6)node[anchor=south west]{$S^2$};
\draw[] (0.5,0.5)node[draw=black,fill=black!20,text width=2cm,align=center,rotate=-45,rounded corners=.2cm]{Localized sources};
\draw[] (2.22,2.28)node[align=center,rotate=-45,black!50]{\textit{Wave zone}{}\qquad\qquad{}\textit{Flat space}};
\end{tikzpicture}
\caption{Asymptotic flatness at $\mathscr I^+$ in the Bondi gauge.}
\label{fig:Asymptotic flatness}
\end{figure}
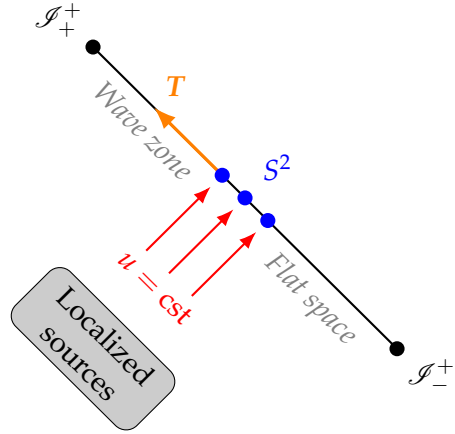
The strongest notion of asymptotic flatness consists in demanding that the Bondi metric \eqref{Bondi line element} with the boundary conditions \eqref{preliminary BC Bondi} approaches the Minkowski metric when $r$ runs to infinity. This yields the complete fixing of the boundary metric and 
\begin{equation}
q_{AB} = \mathring q_{AB}. \label{qAB fixed in Bondi}
\end{equation}
This is the historical set of boundary conditions \cite{Bondi:1962px,Sachs:1962wk,Sachs:1962zza,Ashtekar:1978zza,Ashtekar:1981bq}. We can be less restrictive and demand that the metric field does not approach rigorously the Minkowski metric at large $r$ but that the local volume is asymptotically Minkowskian. This gives another constraint on the transverse boundary metric \cite{Compere:2018ylh,Flanagan:2019vbl,Campiglia:2014yka,Campiglia:2015yka}
\begin{equation}
\sqrt{q} = \sqrt{\bar q}, \label{sqrt q fixed in Bondi}
\end{equation}
where $\bar q$ is an arbitrarly chosen volume element, generically depending on $(u,x^A)$. The requirements \eqref{preliminary BC Bondi} and \eqref{sqrt q fixed in Bondi} lead to the concept of \textit{asymptotically locally flat spacetime}. The boundary structure induced by these boundary conditions encompasses the null vector $\bm T$ defined earlier and a fixed volume form $\sqrt{q}\D^2 x$ on the transverse spaces \cite{Campiglia:2015yka,Flanagan:2019vbl}. In this thesis, we will abundantly discuss this weaker notion of asymptotic flatness and see what are its physical implications.

\subsubsection{Solution space in the Bondi gauge}
\label{sec:Solution space in Bondi gauge}
Given the gauge fixing \eqref{Bondi line element} and the boundary conditions \eqref{preliminary BC Bondi}-\eqref{sqrt q fixed in Bondi}, we can solve Einstein's field equations for the metric tensor $g_{\mu\nu}$ in the vacuum. In the absence of a cosmological constant, they reduce to $R_{\mu\nu} = 0$ since $R=0$. The procedure to obtain the solution space in the Bondi gauge is inspired by \cite{Barnich:2010eb,Tamburino:1966zz} and will be discussed in full details with more generality in section \ref{sec:Bondi Organization of Einstein's equations}. Hence we just sketch the development here. We write the angular components of the metric as
\begin{equation}
g_{AB} = r^2\,q_{AB}+r\,C_{AB}+\,D_{AB}+\frac{1}{r}E_{AB}+\mathcal O(r^{-2}) \label{polynomial gAB}
\end{equation}
in accordance with \eqref{preliminary BC Bondi}. This expansion being polynomial in $1/r$, we do not include logarithmic branches of solutions in our analysis (see \textit{e.g.} \cite{Winicour1985,Chrusciel:1993hx} for a detailed discussion including polyhomogeneous expansions). Apart from that, the hypothesis \eqref{polynomial gAB} does not reduce the set of residual gauge diffeomorphisms \eqref{eq:xir}. The determinant condition constraining the Bondi coordinates implies $\det g_{AB} = r^4 \det q_{AB}$ which is frozen to $r^4 \bar q$ because of \eqref{sqrt q fixed in Bondi}. An equivalent statement of the determinant condition is $g^{AB}\partial_r g_{AB} = 4/r$ which can be solved order by order in $r$ to fix the trace of the various coefficients in \eqref{polynomial gAB}. The leading order gives that $C_{AB}$ is traceless, \textit{i.e.} $q^{AB}C_{AB} = 0$. The subleading pieces give successively
\begin{equation}
\begin{split}
&D_{AB} = \frac{1}{4} q_{AB} C^{CD} C_{CD} + \mathcal{D}_{AB} (u,x^C),  \\
&E_{AB} = \frac{1}{2} q_{AB} \mathcal{D}_{CD}C^{CD} + \mathcal{E}_{AB} (u,x^C), \\
&F_{AB} = \frac{1}{2} q_{AB} \Big[ C^{CD}\mathcal{E}_{CD} + \frac{1}{2} \mathcal{D}^{CD}\mathcal{D}_{CD} - \frac{1}{32} (C^{CD}C_{CD})^2 \Big] + \mathcal{F}_{AB}(u,x^C),
\end{split} \label{traces of D E F}
\end{equation}
with $q^{AB} \mathcal{D}_{AB} = q^{AB} \mathcal{E}_{AB} = q^{AB} \mathcal{F}_{AB} = 0$. The datum of the expansion \eqref{polynomial gAB} and the various traces fixed by the Bondi determinant condition is necessary and sufficient to solve the vacuum Einstein's equations assuming only polynomial expansions in $1/r$ of the Bondi metric coefficients. We take the following convention: when the $r$-dependencies have been fully explicited, the angular indices $A,B,...$ are respectively lowered and raised by the boundary metric $q_{AB}$ and its inverse $q^{AB}$. The Levi-Civita connection associated with $q_{AB}$ is denoted by $D_A$. The radial constraint $R_{rr} = 0$ uniquely determines the radial expansion of $\beta$ in terms of $g_{AB}$ as
\begin{align}
\beta(u,r,x^A) &= \frac{1}{r^2} \Big[ -\frac{1}{32} C^{AB} C_{AB} \Big] + \frac{1}{r^3} \Big[ -\frac{1}{12} C^{AB} \mathcal{D}_{AB} \Big] \label{eq:EOM_beta flat} \\
&\qquad + \frac{1}{r^4}\Big[ - \frac{3}{32} C^{AB}\mathcal{E}_{AB} - \frac{1}{16} \mathcal{D}^{AB}\mathcal{D}_{AB} + \frac{1}{128} (C^{AB}C_{AB})^2 \Big] + \mathcal{O}(r^{-5}). \nonumber
\end{align}
Notice that the integration ``constant'' with respect to $r$ has been set to zero to obey the boundary condition $\beta = o(r^0)$. The cross-term constraint $R_{rA} = 0$ completely determines the radial expansion of $U^A$ in terms of $g_{AB}$ as
\begin{equation}
\begin{split}
U^A = &-\frac{1}{2} D_B C^{AB} \frac{1}{r^2} - \frac{2}{3} \Big[ N^A - \frac{1}{2} C^{AB} D^C C_{BC} - \frac{1}{3} D_B \mathcal{D}^{AB}  \Big] \frac{1}{r^3} - \frac{2}{3} D_B \mathcal{D}^{AB} \frac{\ln r}{r^3} + o(r^{-3})
\end{split} \label{eq:EOM_UA flat}
\end{equation}
and satisfies the boundary condition $U^A = o(r^0)$. $N^A(u,x^B)$ is an integration ``constant'' which is left arbitrary at this point by Einstein's equations. It goes under the name of \textit{Bondi angular momentum aspect}, nomenclature that will become transparent when we will be computing the charges. Note that when $\beta$ and $U^A$ are decaying at null infinity, the $\mathcal O(r^{-1})$ terms are set to zero by the equations of motion. Furthermore, in order to be consistent with the starting hypothesis discarding logarithmic branches, one has to require that \cite{Barnich:2010eb}
\begin{equation}
D^A \mathcal D_{AB} = 0 \label{eq:DivDAB=0}
\end{equation}
which is a weaker assumption than in \cite{Sachs:1962wk} where $\mathcal D_{AB}$ was fixed to zero. As a side remark, the precise definition of $N^A$ as an integration constant is conventional. In this thesis, we follow the convention of \cite{Barnich:2010eb,Barnich:2011mi}. It differs from \cite{Flanagan:2015pxa} where $N_A^{\text{\cite{Flanagan:2015pxa}}} = N_A + \frac{1}{4}C_{AB}D_C C^{BC} + \frac{3}{32}\partial_A (C_{CD}C^{CD})$ and \cite{Hawking:2016sgy} where $N_A^{\text{\cite{Hawking:2016sgy}}} = N_A^{\text{\cite{Flanagan:2015pxa}}} - u\partial_A M$.

The asymptotic behavior of $V/r$ is now uncovered by solving $R_{ur}=0$. Since the equations $R_{rr} = 0$ and $R_{rA}$ have already been solved, one can show that an equivalent constraint is $g^{AB}R_{AB} = 0$. It gives
\begin{equation}
\frac{V}{r} = -rl - \frac{1}{2}R[q] + \frac{2M}{r} + o(r^{-1}) \label{eq:EOM_Vr flat}
\end{equation}
where $l = \partial_u \ln\sqrt{q}$, $R[q]$ is the Ricci scalar curvature associated with $q_{AB}$ and $M(u,x^A)$ is another arbitrary function of angles and retarded time, called the \textit{Bondi mass aspect}. The last algebraic relations to be derived on the Bondi parameters are hidden in the purely angular constraint $R_{AB} = 0$. Since the cancellation of the trace $g^{AB}R_{AB}=0$ is already ensured by \eqref{eq:EOM_Vr flat}, it just remains to solve $R_{AB} - \frac{1}{2}(g^{CD}R_{CD})g_{AB} = 0$. Writing \eqref{polynomial gAB} as $g_{AB} = r^2\sum_{n\geq 0}g_{AB}^{(n)}r^{-n}$, this equation provides an algebraic expression of the coefficient $g_{AB}^{(k)}$ in terms of the others Bondi parameters at each order $\mathcal O(r^{-k})$. For some $k\in\mathbb N$, the contribution at order $\mathcal O(r^{-k})$ does not involve the coefficient $g_{AB}^{(k)}$ and this tensor is declared partially constrained: it will potentially contain some undetermined functions. For $k=0$, the equation is tautologic, which is not a surprise since otherwise it would have fixed the boundary metric itself. At the next order $k=1$, instead of collecting an algebraic constraint on $C_{AB}$, the time evolution of the boundary metric gets controlled as
\begin{equation}
\partial_u q_{AB} = l \, q_{AB}, \label{EOM qAB time evolution}
\end{equation}
meaning that the time dependence of $q_{AB}$ is entirely encoded in a conformal factor. In other words, there is only one dynamical quantity among the three components of $q_{AB}$. More crucially, $C_{AB}$ is left completely free by Einstein's equations. This symmetric traceless tensor on the celestial sphere represents the \textit{asymptotic shear} of null geodesic congruences traveling to $\mathscr I^+$. The two degrees of freedom encoded in $C_{AB}$ are precisely the two polarization modes of the strain measured by a gravitational wave detector at large distance from the emitting source \cite{Thorne:1982cv}. Its time derivative, $N_{AB} = \partial_u C_{AB}$, is the \textit{Bondi news tensor} \cite{Bondi:1962px} which is aimed at encoding the flux of gravitational radiation as we will review below. For any $k>1$, the remaining tower of subleading pieces of the angular constraint equations imposes an evolution equation of the form $\partial_u g_{AB}^{(k-1)} + (...) = 0$ where the dots indicate some function of the previous coefficients $g_{AB}^{(k-2)}, g_{AB}^{(k-3)},...$ In particular, it comes that $\partial_u \mathcal D_{AB}=0$. But going further the tensors $\mathcal E_{AB},\mathcal F_{AB},...$ are not totally constrained by Einstein's equations: only their time evolution is prescribed on $\mathscr I^+$ \cite{Barnich:2010eb,Tamburino:1966zz,10.2307/2415610}.

As long as we are willing to talk about time evolution, recall that we are not done in solving Einstein's equations, because $R_{uu} = 0$ and $R_{uA} = 0$ remain unsolved at this point. Instead of performing a long and fastidious development in $1/r$ as needed for the previous components, one can show thanks to a beautiful argument that, apart of collecting some time evolution constraints of the aforementioned integration ``constants'' $M$ and $N_A$, there is no more dynamical parameter to be uncovered in these equations \cite{Barnich:2010eb,Tamburino:1966zz}. This argument is based on the contracted Bianchi identities $\nabla_\mu G^{\mu\nu} = 0$ for the Einstein tensor. Assuming that the equations $R_{rr}=0$, $R_{rA}=0$, $R_{ur} =0$ and $R_{AB}=0$ have been solved, the Bianchi identities simplify to $\partial_r (r^2 R_{uu}) =0$ and $\partial(r^2 R_{uA}) = 0$, which implies that if $r^2 R_{uu} = 0$ and $r^2 R_{uA} = 0$ for a certain value of $r$, the same holds for any $r$. Hence only the $r$-independent part of $r^2 R_{uu}=0$ and $r^2 R_{uA}=0$ will bring some new piece of information. As a result, it is sufficient to demand that the $\mathcal O(r^{-2})$ terms in $R_{uu}$ and $R_{uA}$ vanish on-shell. These pieces yield respectively
\begin{equation}
\begin{split}
(\partial_u + \frac{3}{2}l )  M +\frac{1}{8} N_{AB} N^{AB} - \frac{1}{8} l N_{AB} C^{AB} + \frac{1}{32} l^2 C_{AB} C^{AB} - \frac{1}{8} D_A D^A R[q] \\
- \frac{1}{4} D_A D_B N^{AB} + \frac{1}{4} C^{AB} D_A D_B l + \frac{1}{4} \partial_{(A} l D_{B)} C^{AB} + \frac{1}{8} l D_A D_B C^{AB} = 0,
\end{split}\label{duM flat}
\end{equation}
and 
\begin{align}
(\partial_u + l)  N_A &- \partial_A  M - \frac{1}{4} C_{AB} \partial^B R[q] - \frac{1}{16} \partial_A (N_{BC}C^{BC}) \nonumber \\
&- \frac{1}{32} l \partial_A (C_{BC}C^{BC}) +\frac{1}{4} N_{BC} D_A C^{BC} + \frac{1}{4} D_B (C^{BC} N_{AC} - N^{BC} C_{AC}) \nonumber \\
&+\frac{1}{4} D_B (D^B D^C C_{AC} - D_A D_C C^{BC}) = 0. \label{duNA flat}
\end{align}
These boundary evolution equations come together with the tower of subleading equations of motion constraining the time derivative of the traceless transverse tensors $\mathcal E_{AB}$, $\mathcal F_{AB}$,\dots as mentioned earlier.

In summary, the solution space $\bar{\mathcal S}_0$ for Einstein's gravity in the Bondi gauge \eqref{Bondi line element} with asymptotically locally flat boundary conditions \eqref{preliminary BC Bondi}, \eqref{sqrt q fixed in Bondi} and \eqref{polynomial gAB} is parametrized by a countable set of codimension 1 functions (\textit{i.e.} depending on the boundary coordinates $u$ and $x^A$ but no longer in the radial coordinate $r$). 
\begin{equation}
\bar{\mathcal S}_0 = \left\{ g_{\mu\nu}\left[q_{AB},C_{AB},M,N_A,\mathcal D_{AB},\mathcal E_{AB},\mathcal F_{AB},\dots\right] \ \Big| \ R_{\mu\nu}[g] = 0, \, \sqrt{q}=\sqrt{\bar q} \right\}. \label{S0}
\end{equation}
First-order time evolution equations constrain all parameters except $C_{AB}$. The Cauchy problem requires thus the following set of initial data
\begin{equation}
q_{AB}(u_0,x^C) ,  M(u_0,x^C) ,  N_A(u_0,x^C) ,  \mathcal E_{AB}(u_0,x^C) ,  \mathcal F_{AB}(u_0,x^C),\dots
\end{equation} at some retarded time $u_0$, complemented with a time-independent and divergence-free tensor $\mathcal D_{AB}(x^C)$ and the two codimension 1 functions encoded in the asymptotic shear $C_{AB}(u,x^C)$ on the whole boundary $\mathscr I^+$. 

Before closing this discussion, let us see how the equations simplify when the boundary structure is strengthened. Note that the global Minkowski vacuum is included in the solution space if and only if the fixed boundary volume $\sqrt{\bar q}$ defined in \eqref{sqrt q fixed in Bondi} is the volume of the unit-round sphere metric \cite{Campiglia:2015yka,Compere:2018ylh}
\begin{equation}
\sqrt{q} = \sqrt{\mathring q} \label{sqrt q round sphere}
\end{equation}
This is an hypothesis we will assume from now on. In that case, the asymptotically locally flat solution space
\begin{equation}
\mathring{\mathcal S}_0 = \left\{ g_{\mu\nu}\left[q_{AB},C_{AB},M,N_A,\mathcal D_{AB},\mathcal E_{AB},\mathcal F_{AB},\dots\right] \ \Big| \ R_{\mu\nu}[g] = 0, \, \sqrt{q}=\sqrt{\mathring q} \right\}. \label{S0ring}
\end{equation}
contains all asymptotically Minkowskian manifolds satisfying the more restrictive boundary condition \eqref{qAB fixed in Bondi}
\begin{equation}
\mathring{\mathcal S}_0^{\text{Mink}} = \left\{ g_{\mu\nu}\left[C_{AB},M,N_A,\mathcal D_{AB},\mathcal E_{AB},\mathcal F_{AB},\dots\right] \ \Big| \ R_{\mu\nu}[g] = 0, \, q_{AB}=\mathring q_{AB} \right\} \subset \mathring{\mathcal S}_0. \label{S0Mink}
\end{equation}
The condition \eqref{sqrt q round sphere} implies in particular that $\partial_u \sqrt{q} = 0$, so $l = 0$ and 
\begin{equation}
\partial_u q_{AB} \Big|_{\mathring{\mathcal S}_0} = 0 \label{pu qAB = 0}
\end{equation}
by virtue of \eqref{EOM qAB time evolution}. The equations \eqref{eq:EOM_beta flat} and \eqref{eq:EOM_UA flat} for $\beta$ and $U^A$ are unchanged and \eqref{eq:EOM_Vr flat} implies that $\frac{V}{r}$ is now finite in $r$
\begin{equation}
\frac{V}{r}\Big|_{\mathring{\mathcal S}_0} = -\frac{1}{2}R[q] - \frac{2M}{r}+o(r^{-1}). \label{Vr flat with sphere det}
\end{equation}
The evolution equations for Bondi mass and angular momentum aspects $M$, $N_A$ simplify drastically in the absence of $l$, namely
\begin{align}
\partial_u M\Big|_{\mathring{\mathcal S}_0} &= - \frac{1}{8} N_{AB} N^{AB} + \frac{1}{4} D_A D_B N^{AB} + {\frac{1}{8} D_A D^A R[q]},\label{duM} \\ 
\partial_u N_A\Big|_{\mathring{\mathcal S}_0} &= D_A M + \frac{1}{16} D_A (N_{BC} C^{BC}) - \frac{1}{4} N^{BC} D_A C_{BC} \nonumber \\
&\quad -\frac{1}{4} D_B (C^{BC} N_{AC} - N^{BC} C_{AC}) - \frac{1}{4} D_B D^B D^C C_{AC} \label{EOM1} \\
&\quad + \frac{1}{4} D_B D_A D_C C^{BC} + { \frac{1}{4} C_{AB} D^B R[q]}. \nonumber
\end{align}
Geometrically, these equations are written for an arbitrary time-independent metric on the sphere. The sub-space \eqref{S0Mink} of solutions constrained by the boundary condition \eqref{qAB fixed in Bondi} is simply obtained by freezing the arbitrary frame on the boundary geometry to $q_{AB} = \mathring q_{AB}$ for which $R[\mathring q] = 2$. Any term involving the boundary Ricci curvature is thus meant to disappear and, in particular, $\frac{V}{r}$ tends to the Minkowskian value $-1$ when $r\to\infty$. This is the historical solution space studied by Bondi, Metzner, Sachs and van der Burg (BMS) in \cite{Bondi:1962px,Sachs:1962wk,Sachs:1962zza}.

\subsubsection{Asymptotic symmetries and the BMS\texorpdfstring{$_4$}{4} group}
\label{sec:Asymptotic symmetries and the BMS4 group}
Assuming the Bondi gauge fixing \eqref{Bondi line element} and the preliminary boundary conditions \eqref{preliminary BC Bondi}, Einstein's equations are solved by any metric field belonging to $\bar{\mathcal S}_0$. The goal of this section is to discuss the asymptotic symmetries preserving these boundary conditions and particularize the general discussion to asymptotically flat spacetimes, first in the strict sense of \eqref{qAB fixed in Bondi} but later in the weaker sense of \eqref{sqrt q round sphere}. 

Considering \eqref{eq:xir}, the most general diffeomorphism $\xi$ preserving the Bondi gauge defined by \eqref{Bondi gauge conditions} is parametrized by two boundary scalar fields $f$ and $\omega$ and a time-dependent vector field $Y^A$ on the celestial sphere. The preservation of the asymptotic behavior in $1/r$ of $\beta$ and $U^A$, as prescribed in \eqref{preliminary BC Bondi}, yields additional constraints on this set of codimension 1 parameters. Developing the finite order of $\mathcal L_{\xi} g_{ur} = \mathcal O(r^{-2})$ and $\mathcal L_{\xi} g_{uA} = \mathcal O(r^{-2})$ gives the constraints 
\begin{equation}
\partial_u f = \frac{1}{2}D_A Y^A-\omega + \frac{1}{2}lf, \quad \partial_u Y^A = 0
\label{most general BMS}
\end{equation}
whose general solution is
\begin{equation}
f = \sqrt[4]{q} \left[ \tilde T(x^A) + \frac{1}{2} \int^u_{-\infty} \D u' \frac{1}{\sqrt[4]{q}}(D_A Y^A-2\omega)\right], \quad Y^A = Y^A (x^B)
\label{most general BMS solution}
\end{equation}
where $q = \det q_{AB}$ as usual. The most general asymptotic symmetries respecting the fall-offs \eqref{preliminary BC Bondi} are parametrized by two fields $\tilde T$ and $Y^A$ living on the spherical sections of $\mathscr I^+$, in addition to the parameter $\omega$ which remains unconstrained. We write the asymptotic vector as $\xi = \xi(f,Y^A,\omega)$ keeping the parameter $f$ instead of expliciting things in terms of $\tilde T$ for convenience. For two sets of gauge parameters $\xi(f_1,Y^A_1,\omega_1)$ and $\xi(f_2,Y^A_2,\omega_2)$, we compute the modified Lie bracket \eqref{modified bracket general} to get some information about the asymptotic symmetry algebra. We have \cite{Barnich:2010eb}
\begin{equation}
[\xi(f_1,Y^A_1,\omega_1),\xi(f_2,Y_2^A,\omega_2)]_\star = \xi(\hat f,\hat Y^A,\hat \omega)
\end{equation}
for the commutation relations
\begin{align}
\hat f &= Y_1^A\partial_A f_2 + \frac{1}{2} f_1 D_A Y_2^A - (1\leftrightarrow 2), \label{commu f flat general} \\
\hat Y^A &= Y_1^B\partial_B Y^A_2- (1\leftrightarrow 2), \label{commu Y flat general} \\
\hat \omega &= 0. \label{commu omega flat general}
\end{align}
It is worth noticing that only the local information on the parameters $f,Y^A,\omega$, coming from \eqref{most general BMS}, is necessary to establish the algebra. The parameters $f$ and $Y^A$ describe diffeomorphisms on the null hypersurface $\mathscr I^+$, the first one acting as reparametrizations $u \to u + f$ along the integral lines of the null normal $\bm T$ and the second one acting as codimension 2 conformal diffeomorphisms on the celestial sphere. Indeed, examining the leading order of $\mathcal L_\xi g_{AB} = r^2 \delta_\xi q_{AB} + \mathcal O(r)$, we can prove that
\begin{equation}
\delta_{\xi}q_{AB} = \mathcal L_Y q_{AB} - (D_C Y^C - 2\omega)q_{AB}.
\label{delta qAB general flat}
\end{equation}
Hence we observe that the transformation of the boundary metric $q_{AB}$ under the action of the boundary vector field $Y^A$ is not reduced to the covariant term $\mathcal L_Y q_{AB}$ but also involves a conformal transformation driven by the divergence of $Y^A$ as well as a \textit{Weyl rescaling} generated by $\omega$ (\textit{i.e.} $q_{AB}\to e^{2\omega}q_{AB}$ for the finite transformation). Interestingly, we find the contribution of the second and third terms in \eqref{delta qAB general flat} in the radial component of the boundary diffeomorphism: indeed, if we develop $\xi^r$, given by \eqref{eq:xir}, in $r$ near infinity, we get
\begin{equation}
\xi^r = -\frac{r}{2}(D_C Y^C -2\omega) + \mathcal O(r^0) \Rightarrow \delta_{\xi}q_{AB} - \mathcal L_Y q_{AB} = \lim_{r\to\infty} \left(\frac{2}{r}\xi^r\right) q_{AB}.  \label{xir interpretation}
\end{equation}
Therefore, the supplementary terms in \eqref{delta qAB general flat} aim at ensuring that the boundary diffeomorphism actually preserves the location of constant $r$ hypersurfaces such as $\mathscr I^+$ to be compatible with the boundary conditions imposed around infinity in terms of power series in $1/r$. The modification due to the $r^2$ factor in front of $q_{AB}$ in the physical metric must thus be exactly compensated by a conformal transformation on $q_{AB}$ and this is precisely what \eqref{xir interpretation} does. We do not plan to give such a meaning for all terms appearing in the variations of the dynamical fields in the Bondi gauge, and discussed in section \ref{sec:Solution space in Bondi gauge}. Nevertheless, it seemed important to us to exemplify how the preserving of the boundary structure at infinity constrains the transformations of the fields.

Apart of this precision, the algebra \eqref{commu Y flat general} of the $Y^A$ parameters is the $\text{Diff}(S^2)$ algebra under the standard vector bracket \cite{Campiglia:2014yka,Flanagan:2019vbl,Compere:2018ylh}. The class of Weyl rescaling symmetries \cite{Barnich:2010eb,Freidel:2021yqe} driven by codimension 1 functions $\omega$, is obviously abelian, see \eqref{commu omega flat general}. Considering \eqref{commu f flat general}, we can show that the codimension 1 generators $f$ form an abelian ideal of the asymptotic symmetry algebra, but transforms non-trivially under the action of the $\text{Diff}(S^2)$ generators $Y^A$. The first term represents the natural diffeomorphic action of $Y^A$ on the scalar field $f$ while the second is again needed to keep the leading order of $g_{ur}$ at its asymptotic value $-1$ even under the action of the $r$ transformation pulled back on $\mathscr I^+$ like in \eqref{xir interpretation}. 

\paragraph{Asymptotically Minkowskian spacetimes and the BMS$_4$ group} Let us now incorporate the remaining boundary conditions that constrain the boundary metric $q_{AB}$. Fixing the boundary volume as in \eqref{sqrt q fixed in Bondi} forbids to perform any Weyl rescaling on the celestial sphere metric, \textit{i.e.}
\begin{equation}
\omega = 0. \label{no weyl}
\end{equation}
This is a consequence of \eqref{delta qAB general flat}, since the trace of this variation gives $\delta_\xi\sqrt{q} = 4\omega \sqrt{q}$. Requiring further that $q_{AB}$ is fixed as \eqref{qAB fixed in Bondi} singles out the solutions in $\bar{\mathcal S}_0$ that are asymptotically Minkowski when $r$ approaches infinity, \textit{i.e.} belong to $\mathring{\mathcal S}_0^{\text{Mink}}$. The diffeomorphisms we are searching for are tangent to $\mathring{\mathcal S}_0^{\text{Mink}}$. They thus have to preserve the universal structure formed by the boundary null foliation $\bm T$ and the fixed transverse metric $q_{AB}$, which translates the usual BMS universal structure \cite{1977asst.conf....1G,Ashtekar:1978zza,Geroch:1978ub} relying on conformal classes in the gauge fixing language. These diffeomorphisms are easy to find: now that $\partial_u \sqrt{q} = 0$, the integration in \eqref{most general BMS solution} can be performed explicitly. We get
\begin{equation}
f = T(x^A) + \frac{u}{2}D_A Y^A, \qquad Y^A = Y^A(x^B) \label{BMS fY}
\end{equation}
for arbitrary functions $T(x^A)$ on the sphere. The set of allowed boundary diffeomorphisms $Y^A$ is consequently reduced because \eqref{qAB fixed in Bondi} implies that
\begin{equation}
\delta_\xi q_{AB} = 0 \Rightarrow D_A Y_B + D_B Y_A = D_C Y^C q_{AB}. \label{Killing eq for rotations}
\end{equation}
The solutions of \eqref{Killing eq for rotations} for $Y^A$ are the six conformal Killing vectors of the unit-round sphere $S^2$ generating the (proper orthochron) \textit{Lorentz algebra} $SO(3,1)$. In terms of the codimension 2 parameters $T$ and $Y^A$, the commutation relations \eqref{commu f flat general}-\eqref{commu Y flat general} become
\begin{equation}
\begin{split}
[\xi(T_1,Y^A_1),\xi(T_2,Y^A_2)] &= \xi(\hat T,\hat Y^A),\\
\hat T &= Y_1^A\partial_A T_2 + \frac{1}{2} T_1 D_A Y_2^A - (1\leftrightarrow 2), \\
\hat Y^A &= Y_1^B\partial_B Y^A_2- (1\leftrightarrow 2), 
\end{split}
\label{eq:BMSCommu}
\end{equation}
These relations define the global BMS$_4$ algebra, named after Bondi, Metzner, Sachs and van der Burg \cite{Bondi:1962px,Sachs:1962wk,Sachs:1962zza}, which stands for the asymptotic algebra of asymptotically flat spacetimes compatible with the asymptotically Minkowskian boundary conditions. The elements of this algebra, parametrized as $\xi  =\xi(T,Y^A)$, are genuine asymptotic Killing vectors as $\nabla_{(\mu}\xi_{\nu)}\to 0$ when $r\to+\infty$. By exponentiating them to get finite amplitude transformations internal to $\mathring{\mathcal S}_0^{\text{Mink}}$, one would obtain the BMS$_4$ group. In this text, we will encounter asymptotic symmetry algebras whose exponentiation is either problematic or poorly known, hence we will limit ourselves to discuss about algebras instead of their group counterpart.

\begin{figure}[ht!]
    \begin{center}
        \begin{tikzpicture}[scale=0.9]
        		\coordinate (centre) at (0,0.4);
        		\draw[OrangeRed,densely dashed,thick,fill=OrangeRed!10] (centre) ellipse (2.57 and 0.3);
        		\draw[densely dashed] (0, -3.5) -- (0,4);
        		\path[->, bend right = 25] (-0.5,3.6) edge (0.5,3.6);       
				\draw[densely dotted,line width=0.1mm,gray!75] (-0.5,0.11) -- (0,3);
				\draw[densely dotted,line width=0.1mm,gray!75] (-1.0,0.13) -- (0,3);
				\draw[densely dotted,line width=0.1mm,gray!75] (-1.5,0.16) -- (0,3);
				\draw[densely dotted,line width=0.1mm,gray!75] (-2.0,0.22) -- (0,3);
				\draw[densely dotted,line width=0.1mm,gray!75] (-2.5,0.33) -- (0,3);
				\draw[densely dotted,line width=0.1mm,gray!75] (0.5,0.11) -- (0,3);
				\draw[densely dotted,line width=0.1mm,gray!75] (1.0,0.13) -- (0,3);
				\draw[densely dotted,line width=0.1mm,gray!75] (1.5,0.16) -- (0,3);
				\draw[densely dotted,line width=0.1mm,gray!75] (2.0,0.22) -- (0,3);
				\draw[densely dotted,line width=0.1mm,gray!75] (2.5,0.33) -- (0,3);
				\draw[->,>=stealth,blue,thick] (-0.5,0.11) -- (-0.39,0.75);
				\draw[->,>=stealth,blue,thick] (-1.0,0.13) -- (-0.695,1.0);
				\draw[->,>=stealth,blue,thick] (-1.5,0.16) -- (-0.74,1.60);
				\draw[->,>=stealth,blue,thick] (-2.0,0.22) -- (-1.30,1.2);
				\draw[->,>=stealth,blue,thick] (-2.5,0.33) -- (-2.105,0.75);
				\draw[->,>=stealth,blue,thick] (0,0.1) -- (0,0.5);
				\draw[->,>=stealth,blue,thick] (0.5,0.11) -- (0.26,1.5);
				\draw[->,>=stealth,blue,thick] (1.0,0.13) -- (0.70,1.0);
				\draw[->,>=stealth,blue,thick] (1.5,0.16) -- (0.82,1.45);
				\draw[->,>=stealth,blue,thick] (2.0,0.22) -- (1.44,1.0);
				\draw[->,>=stealth,blue,thick] (2.5,0.33) -- (1.96,0.91);
        		\draw[OrangeRed,thick] (2.57,0.4) arc (0:-180:2.57 and 0.3);
        		\draw[thick] (0, 3) -- (3, 0) node[right] {\footnotesize $i^0$} -- (0,-3) -- (-3,0) -- cycle;
	            \node[] (1) at (1.5, 2.5) {\footnotesize $\cI^+$};
    	        \node[] (2) at (1.5, -2.5) {\footnotesize $\cI^-$};
				\fill[black] ( 3, 0) circle [radius=2pt];
				\fill[black] ( 0, 3) circle [radius=2pt];
				\fill[black] ( 0,-3) circle [radius=2pt];
				\fill[black] (-3, 0) circle [radius=2pt]; 
				\node[] at (-1.5,-0.25) {{\color{OrangeRed} {\footnotesize $S^2_\infty(u_0)$}}};
				\node[above] at (2.57,0.43) {{\color{ForestGreen} {\footnotesize $\qquad \qquad u=u_0$}}};
				\fill[ForestGreen] (2.57,0.43) circle [radius=2pt];
				\node[above] at (-2.3,1.6) {{\color{blue} {\footnotesize $T(x^A) \partial_u$}}};
				\draw[->,black] (-2.3,1.6) -- (-1.7,0.8);
        \end{tikzpicture}
    \end{center}
    \caption{BMS supertranslations.}
    \label{fig:Supertranslations}
\end{figure}
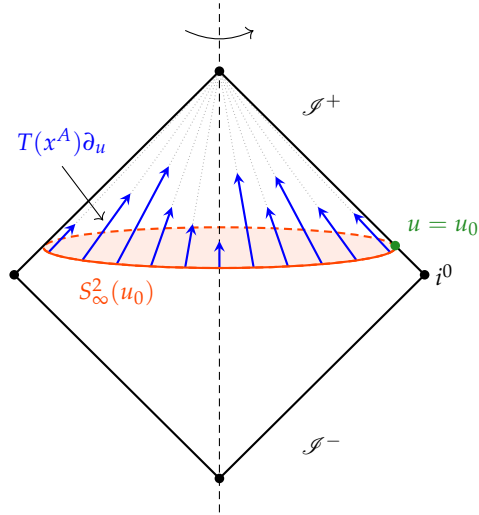

\paragraph{Supertranslations and degeneracy of the gravitational vacua} The vectors generated by $T(x^A)$ are asymptotically given by
\begin{equation}
\xi(T,0) = T(x^C) \partial_u - \frac{1}{r} D^A T(x^C) \partial_A + \frac{1}{2} D_A D^A T(x^C) \partial_r + \cdots
\end{equation}
They form an abelian ideal $s$ of the global BMS$_4$ algebra and represent angular-dependent translations $u\to u+T(x^A)$ in the time direction on $\mathscr I^+$ (see figure \ref{fig:Supertranslations}), called \textit{supertranslations}. 

Their existence came as a surprise when they have been discovered in the sixties, since it goes against the following naive intuition. Since the spacetime looks like the Minkowski vacuum at large distances, one could expect to recover the exact symmetry algebra of Minkowski as asymptotic symmetry algebra, namely the \textit{Poincaré algebra}
\begin{equation}
ISO(3,1) = SO(3,1) \loplus t
\end{equation}
where $t$ is the finite-dimensional abelian ideal of translations in semi-direct sum with the Lorentz algebra $SO(3,1)$. The semi-direct structure comes from the fact that translation parameters can be seen as components of some vector in the Minkowski spacetime, on which the rotations and the boosts of $SO(3,1)$ have non-trivial action. But since $T(x^A)$ is a completely arbitrary scalar field on the sphere, the BMS$_4$ algebra encompasses the infinite-dimensional abelian ideal $s$ of supertranslations:
\begin{equation}
\text{BMS}_4 = SO(3,1) \loplus s
\end{equation}
This infinite-dimensional extension of translation is crucial from the perspective of gravitational radiation as we will recall very soon. However it can be shown \cite{Sachs:1962wk} that $s$ admits one unique finite sub-ideal that reproduces exactly the Poincaré translations $t$. The associated generators are built from the first 4 spherical harmonics in the decomposition of $T$, which verify $D_A D_B T - \frac{1}{2}\mathring q_{AB} D_CD^C T = 0$, namely $T(x^A) = a_{0,0} Y^0_0 (x^A) + a_{1,m} Y^m_1 (x^A), \: m \in \lbrace -1,0,+1 \rbrace$. As an example, the vertical translation $\partial_z$ is reproduced by $a_{1,0} \neq 0, a_{0,0}=a_{1,-1}=a_{1,1}=0$. Indeed, we have $\partial_z =  \cos \theta \partial_r - \frac{1}{r} \sin \theta \partial_\theta$ in spherical static coordinates $(t,r,\theta,\phi)$, so in retarded  coordinates $\partial_z = - \cos \theta \partial_u +\cos \theta \partial_r - \frac{1}{r} \sin \theta \partial_\theta$ and $Y^0_1(\theta,\phi) \propto \cos\theta$.

As asymptotic symmetries, the supertranslations act internally on $\mathring{\mathcal S}_0^{\text{Mink}}$, transforming an asymptotically Minkowskian geometry into another one, physically inequivalent. The variations of the various fields under a supertranslation $\xi(T,0)$ can be extracted from $\mathcal L_\xi g_{\mu\nu}[p_i] = g_{\mu\nu}[p_i+\delta_\xi p_i]-g_{\mu\nu}[p_i]$ where the $p_i$ are the (codimension 1) parameters of $\mathring{\mathcal S}_0^{\text{Mink}}$ described in \eqref{S0Mink}. In particular, we find
\begin{align}
\delta_T C_{AB} &= T \partial_u C_{AB} - 2 D_A D_B T + \mathring q_{AB} D_C D^C T ,\quad \delta_T N_{AB} = T\partial_u N_{AB} \label{eq:STField} \\
\delta_T M &= T \partial_u M + \frac{1}{4} \left[ N^{AB} D_A D_B T + 2 D_A N^{AB} \partial_B T \right]. 
\end{align}
by inspecting respectively the $\mathcal O(r^1)$ contribution of $\mathcal L_\xi g_{AB}$ and the $\mathcal O(r^{-1})$ contribution of $\mathcal L_\xi g_{uu}$. From these relations, it is obvious that a supertranslation cannot create inertial mass or gravitational radiation. Indeed, if we apply a supertranslation on the Minkowski global vacuum $M = 0$, $C_{AB}=N_{AB} = 0$ we get $\delta_T M = 0$ and $\delta_T N_{AB} = 0$. The only field that can be shifted is $C_{AB}$, but since the Bondi news $N_{AB}$ remain zero, we are left with a stationary configuration for which $\delta_T C_{AB} = - 2 D_A D_B T + \mathring q_{AB} D^2 T$. As a symmetric traceless tensor on the sphere, $C_{AB}$ falls into a representation of the $SO(3)$ algebra and takes the general form
\begin{equation}
C_{AB} = -2D_AD_B C+\mathring q_{AB}D^2 C + \mathring \varepsilon_{C(A}D_{B)}D^C\Psi, \label{CAB parametrized}
\end{equation}
which is the sum of a curl-free (``electric'') part parametrized by a scalar field $C(x^A)$ and a divergence-free (``magnetic'') part parametrized by another scalar field $\Psi(x^A)$, involving $\mathring \varepsilon_{AB}$, the Levi-Civita tensor on the sphere. $C$ and $\Psi$ are the two natural degrees of freedom of $C_{AB}$ that transform under supertranslations as
\begin{align}
\delta_T C(x^A) = T(x^A), \quad \delta_T\Psi(x^A)   = 0. \label{C field variation}
\end{align}
Since $C_{AB}=0$ in the Minkowski spacetime, we have simply
\begin{equation}
C_{AB} = -2D_AD_B C+\mathring q_{AB}D^2 C  \label{electric part BMS}
\end{equation}
after the transformation \eqref{C field variation}. The orbit of the Minkowski global vacuum under supertranslations is thus an infinite-dimensional class of flat metrics parametrized by $C$, the latter being called the \textit{supertranslation field} for obvious reasons. The fixation of $C$ is equivalent to a spontaneous breaking of the supertranslation invariance among the gravitational vacua and, as a consequence, $C$ is the Goldstone boson which accompanies the selection of a vacuum \cite{Strominger:2013jfa,Compere:2016jwb}. It is noteworthy that the four Poincaré translations are not concerned by the breaking, because the four lowest spherical harmonics of $T(x^A)$ are annihilated by the differential operator $-2D_A D_B + \mathring q_{AB} D^2$. Therefore, up to the translation ambiguity, $C$ labels the various degeneracies of the gravitational field. Moreover, since supertranslations commute with the (retarded) time translation, their associated charges are expected to commute with the Hamiltonian, which means in turn that all of these degenerate states have the same energy \cite{Strominger:2013jfa}. In the last years, several authors attempted to define a more precise notion of distinct gravitational vacua including \cite{Compere:2016jwb}. We will come back on the determination of the vacua orbit later in the text.

\paragraph{Extension by super-Lorentz transformations} The second set of generators of the BMS$_4$ algebra is constituted by the generators with $T=0$, given asymptotically by
\begin{equation}
\xi(0,Y) = Y^A (x^C)\partial_A - \frac{r+u}{2} D_A Y^A (x^C) \partial_r + \frac{u}{2} D_A Y^A (x^C) \partial_u + \cdots
\end{equation}
As announced, these vectors constitute the six globally well-defined conformal Killing vectors on the celestial sphere, \textit{i.e.} $Y^A$ is solution of \eqref{Killing eq for rotations}. This reduction to a finite-dimensional sub-algebra is the relic of the Poincaré algebra for the sector concerning the diffeomorphisms on the celestial sphere, while the other part, living in the null direction spanned by $\bm T$, has been extended into a infinite-dimensional ideal. We can cure this unpleasant asymmetry by trading the boundary condition \eqref{qAB fixed in Bondi} for \eqref{sqrt q fixed in Bondi} or even \eqref{sqrt q round sphere} for definiteness. The larger solution space under consideration is thus $\mathring{\mathcal S}_0$ parametrized as \eqref{S0ring}. In that case, the boundary structure $\bm T$ is no longer supplemented by a fixed transverse metric but only a fixed area on the transverse spheres \cite{Campiglia:2015yka,Flanagan:2019vbl}. This maintains the triviality of the Weyl rescalings \eqref{no weyl} on the boundary but now allows any (smooth) diffeomorphism on the celestial sphere, which we call \textit{super-Lorentz transformations} \cite{Compere:2018ylh}. This leads us to the \textit{Generalized} BMS$_4$ \textit{algebra} \cite{Campiglia:2014yka}
\begin{equation}
\text{Generalized BMS}_4 = \text{Diff}(S^2)\loplus s. \label{generalized BMS 4 def}
\end{equation}
Any Generalized BMS$_4$ diffeomorphism $\xi$ acts on the boundary as $\xi = f\partial_u + Y^A\partial_A$ where the only differential restrictions on $f$ and $Y^A$ are
\begin{equation}
\partial_u f = \frac{1}{2}D_A Y^A , \quad \partial_u Y^A = 0, \label{generalized BMS 4 differential eq}
\end{equation} 
readily solved as \eqref{BMS fY} for an arbitrary scalar field $T(x^A)$ and a arbitrary smooth vector field $Y^A(x^B)$ on the celestial sphere, parametrizing $\xi = \xi(T,Y)$. This natural extension of the BMS$_4$ algebra has been proposed first in \cite{Campiglia:2014yka,Campiglia:2015yka} motivated by semi-classical arguments \cite{Cachazo:2014fwa} on which we will come back in the final chapter. Vectors generating the Generalized BMS$_4$ transformations are not asymptotically Killing but still obey $\nabla_\mu \xi^\mu \to 0$ when approaching $\mathscr I^+$ \cite{Campiglia:2015yka} because of \eqref{sqrt q fixed in Bondi}. In contrast to the supertranslations, the Diff($S^2$) sector modifies the metric field at leading order, since $\delta_\xi q_{AB} = 0$ \eqref{Killing eq for rotations} is not obeyed anymore for all $Y^A$. So their inclusion into the solution space does not go without consequences. For instance, one expects that the fluctuations of the boundary metric $q_{AB}$ would be responsible for radial divergences in the action principle as well as in the definition of the phase space \cite{Compere:2018ylh} which will be developed in section \ref{sec:Asymptotically locally flat radiative phase spaces}. We will show there how to treat these divergences and confirm the status of the celestial Diff($S^2$) as rightful asymptotic symmetries of asymptotically flat gravity. In chapter \ref{chapter:Flat}, we will study some of the physical properties of these overleading symmetries and give particular processes for which their inclusion in the phase space are mandatory. 

Let us now mention that \eqref{generalized BMS 4 def} was not the first proposal for an extension by super-Lorentz transformations. The first attempt is found in \cite{Barnich:2009se,Barnich:2011ct,Barnich:2010eb} where it is proposed to maintain the conformal Killing equation \eqref{Killing eq for rotations} nearly everywhere on the celestial sphere, except at some points. The solution of this equation is found easily by introducing complex stereographic coordinates on the sphere $z = e^{i\phi} \cot (\theta/2),\: \bar{z} = z^*$, in which the unit round metric on $S^2$ is simply the off-diagonal line element $\D s^2_{S^2} = 4(1+z\bar{z})^{-2} \D z\D \bar{z}$. \eqref{Killing eq for rotations} requires thus that $\partial_z Y^z = 0$ and $\partial_{\bar z}Y^{\bar z}=0$, solved by holomorphic functions $Y^z = Y^z(z)$ and $Y^{\bar z}(\bar z)$ their antiholomorphic counterpart. These can be expanded in Laurent series and appear thus as a sum of monomial terms $Y^z = z^k, \: k\in \mathbb{Z}$. Taking $k=0,1,2$ yields 6 globally well-defined vectors on the sphere spanning the Lorentz algebra $SO(3,1)$. Any transformation for $k>2$ still solve the conformal Killing equations except that these define singular functions which have poles on the sphere ($z^k$ for $k>2$ is singular at the south pole $z=0$, $\theta=\pi$, while $z^k$ for $k<0$ is singular at the north pole $z=\infty$, $\theta=0$). The proposal of \cite{Barnich:2009se} is to allow the full range of $k$ in the Laurent spectrum of $Y^z(z)$. Although very natural from the conformal point of view \cite{Belavin:1984vu,Barnich:2009se}, it leads in fact to an extension of the BMS$_4$ algebra by \emph{meromorphic superrotations}
\begin{equation}
\text{Extended BMS}_4 = \left[\text{Diff}(S^1)\oplus\text{Diff}(S^1)\right]\loplus s^*, \label{extended bms 4}
\end{equation} 
for which, by consistency of the commutation relations \eqref{eq:BMSCommu}, supertranslations now also contain poles on the sphere and are generated with meromorphic functions $T$, spanning the abelian ideal $s^*$ instead of $s$. Then, one can show that all of these meromorphic supertranslations admit infinite conserved charges for the Kerr black hole \cite{Barnich:2011mi}. The issue of singularity can be fixed by working on the punctured complex plane instead of the celestial sphere, but it is not what we did in this thesis. We desire considering asymptotically flat phase spaces with a null infinity equipped with the standard topology $\mathbb R\times S^2$ and admitting smooth asymptotic symmetries. We will thus favorize \eqref{generalized BMS 4 def} to the detriment of \eqref{extended bms 4} for extending the BMS$_4$ symmetries, and many points raised by the corresponding study would also be relevant in the presence of meromorphic super-Lorentz transformation.

\paragraph{Variations of the solution space} The Generalized BMS$_4$ vectors $\xi(T,Y)$ preserve the solution space $\mathring S_0$ in the sense that, infinitesimally,
\begin{equation}
\mathcal{L}_{\xi(T,Y)} g_{\mu\nu} [p_i] = g_{\mu\nu} [ p_i + \delta_\xi p_i ] - g_{\mu\nu} [ p_i ],
\end{equation}
where $p_i = \{ q_{AB}, C_{AB}, M, N_A \}$ denotes the collection of relevant fields that describe the metric in the Bondi gauge, discarding the tower of subleading dynamical fields $\mathcal E_{AB},\mathcal F_{AB},\dots$ whose role in the boundary dynamics is neglectable. As asymptotic symmetries, tangent to $\mathring{\mathcal S}_0$, the action of the Generalized BMS$_4$ vectors preserve the form of the metric but modify the fields $p_i$, in such a way that the above equation is verified. We can show that \cite{Barnich:2010eb,Compere:2018ylh}
\begin{align}
\delta_{\xi} q_{AB} &= [\mathcal L_Y - D_C Y^C]q_{AB},\label{dqAB} \\
\delta_{\xi} C_{AB} &= [f \partial_u + \mathcal{L}_Y - \frac{1}{2} D_C Y^C ] C_{AB} - 2 D_A D_B f + q_{AB} D_C D^C f,\label{dCAB}\\
\delta_{\xi} N_{AB} &= [f\partial_u + \mathcal{L}_Y] N_{AB} - (D_A D_B D_C Y^C - \frac{1}{2} q_{AB} D_C D^C D_D Y^D),\label{dNAB}\\
\delta_{\xi} M &= [f \partial_u + \mathcal{L}_Y + \frac{3}{2} D_C Y^C] M + \frac{1}{4} D_A f D^A \mathring{R} + \frac{1}{4} N^{AB} D_A D_B f \nonumber \\
&\quad + \frac{1}{2} D_A f D_B N^{AB}   + \frac{1}{8}D_A D_B D_C Y^C C^{AB}  , \label{dM}\\
\delta_{\xi} N_A &= [f\partial_u + \mathcal{L}_Y + D_C Y^C] N_A + 3 M D_A f - \frac{3}{16} D_A f N_{BC} C^{BC} + \frac{1}{2} D_B f N^{BC} C_{AC} \nonumber \\
&\quad - \frac{1}{32} D_A D_B Y^B C_{CD}C^{CD} + \frac{1}{4} (D^B f R[q] + D^B D_C D^C f) C_{AB} \nonumber \\
&\quad - \frac{3}{4} D_B f (D^B D^C C_{AC} - D_A D_C C^{BC}) + \frac{3}{8} D_A (D_C D_B f C^{BC}) \nonumber \\
&\quad + \frac{1}{2} (D_A D_B f - \frac{1}{2} D_C D^C f q_{AB}) D_C C^{BC}.   \label{dNA}
\end{align} 
Each variation of the fields comes with a regular pattern. Except for $q_{AB}$, which does not depend on time in $\mathring{\mathcal S}_0$, any field transforms \textit{homogeneously} as a scalar under supertranslations, $f\partial_u(\dots)$, a codimension 2 tensor under super-Lorentz transformations, $\mathcal L_Y(\dots)$, sometimes with a non-zero conformal weight, $D_C Y^C(\dots)$ and, possibly, with some \textit{inhomogeneous} transformation by means of which the symmetries can source the field. For example, \eqref{dCAB} indicates that performing a supertranslation generated by $T$ on a spacetime where $C_{AB} = 0$ will source the $C$-field defined in \eqref{CAB parametrized} as \eqref{C field variation}. Note also that the news tensor $N_{AB}$ transforms homogeneously under supertranslations but not under super-Lorentz transformations. Indeed, the two last terms in \eqref{dNAB} can produce some non-vanishing $N_{AB}$ when a super-Lorentz generator acts on a stationary configuration $N_{AB} = 0$. Remark that this inhomogeneous part is trivially zero if $Y^A$ is a conformal Killing vector on the celestial sphere, \textit{i.e.} $N_{AB}$ transforms homogeneously in the subspace $\mathring{\mathcal S}_0^{\text{Mink}}$. These considerations will be important in chapter \ref{chapter:Flat} when we will be discussing the physical implications of super-Lorentz symmetries.

\section{Surface charges}
\label{sec:Surface charges}
Leaving behind us our discussion about asymptotic symmetries, we are now about to examine how to define associated canonical charges. The first step consists in constructing a notion of covariant phase space from the solution spaces of fields previously described and analyzing its underlying algebraic structure. This formalism provides a set of powerful tools allowing to define rigorously infinitesimal charges evaluated between two solutions differing by an infinitesimal variation and address the questions of conservation in time and integrability on the phase space. We will see that in the presence of gauge invariances, neither the first question nor the second one, even the determination of the charges, have easy answers. We will present the covariant phase space formalism mainly in the language of Iyer, Wald and collaborators \cite{Lee:1990nz,Iyer:1994ys,Wald:1999wa}, which is the immediate generalization of the Hamiltonian formalism for classical mechanics and has the benefit to be very flexible (notably when some renormalization of the action principle is needed) and pedagogically illuminating. The canonical charges naturally defined in this formalism, which we call \textit{Iyer-Wald charges} for that reason, are analogous to the Hamiltonian generators in classical mechanics except that they are surface charges instead of volume charges and they are defined infinitesimally around each point of the solution space. As we did for the asymptotic symmetries, we will apply the formalism to the particularly instructive case of asymptotically flat gravity and review how long-time celebrated results in the field could be derived in that language. Again, there are different ways to proceed and we do not plan to cover more than we actually need to describe and explain the results obtained during this thesis. Nevertheless, we will mention that another prescription for the charges, robustly defined from the equations of motion rather than from the (ambiguous) presymplectic structure of the covariant phase space, has been formulated by Barnich, Brandt and Henneaux in \cite{Barnich:1994db,Barnich:2000zw,Barnich:2001jy,Barnich:2003xg}, and many points raised in this section hold for both formulations (see \textit{e.g.} \cite{Compere:2018aar,Ruzziconi:2019pzd,Ruzziconi:2020cjt,Compere:2007az} for reviews).

\subsection{The puzzle to define canonical charges for gauge theories}
In physics, the quest for conserved quantities associated with the classical motion of a system needs as premise the catalog of the symmetries acting on this system. The tight relation existing between both concepts is the core of one of the most famous statements ever established in modern science: the Noether first theorem \cite{Noether:1918zz}. Considered as a ``monument of mathematical thought'' by Einstein himself, it is not abusive to say that most of progresses in modern physics rely on this result. We present it here without proof, but in a quite updated form that needs some preliminary definitions. Next, we discuss the consequences of that theorem while dealing with gauge symmetries and show that the Noether charge is not the right object to define canonical conserved quantities for this kind of symmetries. The puzzle can be solved by considering lower-degree conservation laws as we explain afterwards.

\subsubsection{Noether's first theorem and trivial currents}

We consider again the physical system whose action integral is \eqref{action general} and for which we have discussed the gauge invariances at a theoretical level. The concern of Noether's first theorem is more properly the \textit{global symmetries} which have to be defined with a bit of care in the presence of gauge symmetries. Let us pick two continuous symmetries with characteristics $Q_1$ and $Q_2$, in the sense that they depend upon continuous parameters. If $Q_2$ differs from $Q_1$ only by a gauge transformation $R[\lambda]$ and a local function $F$ that vanishes on-shell, one cannot distinguish them when the equations of motion are satisfied and a gauge invariance is operating on the system. This defines the following equivalence relation among symmetries:
\begin{equation}
Q_1 \sim Q_2 \Leftrightarrow Q_2 = Q_1 + R[\lambda] + F\left[\frac{\delta\bm L}{\delta\phi}\right] \label{global symm}
\end{equation}
whose equivalence classes $[Q]$ are naturally defined as \textit{global symmetries} of the theory. From that perspective, a gauge symmetry $Q = R[\lambda]$ belongs to the equivalence class of the trivial transformation, and can therefore be assimilated with a trivial global symmetry. Owing to this definition of (continuous) global symmetries, Noether's first theorem can be enunciated as follows \cite{Barnich:2000zw} (see \textit{e.g.} \cite{Barnich:2001jy,Barnich:2018gdh} for proofs in this updated language).

\resu{Noether's first theorem}{
Take any physical theory described by a Lagrangian $\bm L$ defined on a spacetime manifold $(\mathscr M,g)$ which admits global symmetries, some of which might be gauge invariances.
There exists a bijection between :
\begin{itemize}[label=$\rhd$]
\item The equivalence classes of global continuous symmetries of $\bm L$ and
\item The equivalence classes of conserved codimension 1 forms $\bm J$ or \textit{Noether currents}.
\end{itemize}
}
The equivalence relations on Noether currents is the mirror of \eqref{global symm}: two currents $\bm J_1$ and $\bm J_2$ are declared to be equivalent if and only if they differ by a trivial current:
\begin{equation}
\bm J_1 \sim \bm J_2 \Leftrightarrow \bm J_2 = \bm J_1 + \D\bm k + \bm t \label{current equivalence class}
\end{equation}
where $\bm k = k^{\mu\nu}(\D^{n-2}x)_{\mu\nu}$ is a codimension 2 form and $\bm t$ is a codimension 1 form that vanishes on-shell. We have $\D\bm J_1=0=\D\bm J_2$ when the equations of motion are solved. The standard representative $\bm J_Q = \bm B_Q - \bm\Theta[\phi;\delta_Q\phi]$ of the Noether current associated with a continuous global symmetry $[Q]$ can be obtained by developping \eqref{symmetry condition}, see \eqref{Noether for Q}. The associated Noether charge is defined as
\begin{equation}
H_Q[\phi] = \int_\Sigma \bm J_Q[\phi] = \int_\Sigma (\D^{n-1}x)_\mu J^\mu_Q[\phi] \label{noether charge general}
\end{equation}
for a codimension 1 spacelike hypersurface $\Sigma$ with boundary $\partial\Sigma$. Using Stokes theorem, we observe that the definition \eqref{noether charge general} does not depend on the particular representative in $[\bm J_Q]$ if and only if the ambiguity $\bm k$ decays sufficiently rapidly in the vicinity of $\partial\Sigma$. 

Using this formulation of Noether's theorem, a thorny problem immediately arises. Imagine that you have a pure gauge theory, \textit{i.e.} a gauge theory with no non-trivial global symmetry at disposal. In that case, there exists only one equivalence class of conserved currents as a result of Noether's first theorem: the trivial ones. In particular, for generally covariant theories, any transformation like $x^\mu \rightarrow x^\mu + \xi^\mu$ is pure gauge, so the natural symmetries, also called \textit{isometries}, are associated with trivial currents in a similar way. Since, for a gauge symmetry, $\bm J_Q = \D\bm k$ on-shell, the Noether charge reads simply as $H_Q = \int_\Sigma \bm J_Q = \oint_{\partial\Sigma} \bm k$ when the equations of motion hold. $H_Q$ is manifestly completely arbitrary, because $\bm k$ is totally unconstrained!

\subsubsection{Lower degree conservation laws}
One is able to sketch a solution to this puzzle simply by considering more carefully the expression of the arbitrary Noether charge $H_Q = \oint_{\partial\Sigma} \bm{k}$. We see that it reduces to the flux of $\bm{k}$ through the boundary $\partial\Sigma$ and depends only on the properties of this codimension 2 form in the vicinity of $\partial\Sigma$. This suggests to invoke lower degree conservation laws \cite{Anderson:1996sc,Barnich:2001jy,Barnich:1994db}, involving conserved $(n-2)$-forms instead of conserved $(n-1)$-forms. Indeed, let us imagine that we have some $(n-2)$-form $\bm{k} = k^{[\mu\nu]} (\D^{n-2}x)_{\mu\nu}$ such that $\D\bm{k} = 0 \Leftrightarrow \partial_\nu k^{[\mu\nu]} = 0$. Thanks to such an object, we can define an integral charge $H_Q = \int_{\partial\Sigma} \bm{k}$ which will be conserved in the following sense:
\begin{equation}
H_Q\Big|_{\partial\Sigma_2} - H_Q\Big|_{\partial\Sigma_1} = \oint_{\partial\Sigma_2} \bm{k} - \oint_{\partial\Sigma_1} \bm{k} = \int_{\Sigma_{12}} \D \bm k = 0
\end{equation}
where $\Sigma_{12}$ is a codimension 1 hypersurfaces enclosed by the boundaries $\partial\Sigma_1$ and $\partial\Sigma_2$. Seeking for conserved $(n-2)$-forms is thus the right path to obtain a canonical notion of integral constants of motion in gauge theories. It remains to fix two indeterminations: the first one is the link between the symmetries and the conserved quantities built up from closed codimension 2 forms $\bm k$, while the second one concerns the actual expression of $\bm k$ when the first puzzle has been solved.

While Noether's first theorem maps each symmetry to a class a conserved currents (or equivalently closed $(n-1)$-forms $\bm{J}$), there exists a generalized version which focuses on lower degree conserved forms and precisely involves $(n-2)$-forms. This result was established by Barnich, Brandt and Henneaux \cite{Barnich:1994db,Barnich:1995ap} using cohomological methods and we present it here also without proof. Before stating the result, we just need one more conceptual definition. 

Among the gauge symmetries of the theory \eqref{action general}, we assume that there is a subclass of non-trivial gauge parameters $\bar\lambda$ that verify $\delta_{\bar\lambda} \phi = R[\bar\lambda]=0$ on-shell. They are called \textit{reducibility parameters} because they generate gauge directions that are trivial when the equations of motion hold. A natural equivalence relation $\bar\lambda_1\sim \bar\lambda_2$ is defined by the requirement that $\bar\lambda_1=\bar\lambda_2$ on-shell: in this way, they generate the same trivial gauge direction. One important thing to notice is that the condition $R[\bar\lambda]=0$ can be solved in some contexts for field-independent parameters as an off-shell condition. This is the case in electromagnetism (also in Yang-Mills theories), where the reducibility parameters have to satisfy $\partial_\mu \bar\lambda=0$, constraint solved by the constants $\bar\lambda\in\mathbb R_0$ with no intervention of the equations of motion. The same peculiarity occurs for gravity, where the reducibility parameters are simply the isometries of the metric field, \textit{i.e.} diffeomorphisms $\bar\xi$ such that $\mathcal L_{\bar\xi} g_{\mu\nu} = 0$ (\textit{Killing vectors}). These are particular cases of \textit{exact reducibility parameters} solving the off-shell constraints $R[\bar\lambda]=0$. 

Given this definition, we can enunciate the \textit{Generalized Noether theorem} as follows:
\resu{Generalized Noether theorem}{
Take any physical theory described by a Lagrangian density $L$ defined on a spacetime manifold $(\mathscr M,g)$ which admits global symmetries, some of which might be gauge invariances. There exists a bijection between :
\begin{itemize}[label=$\rhd$]
\item The set of equivalence classes of reducibility parameters $\bar\lambda$ (such as the variations of fields $\delta_{\bar\lambda}\phi$ vanish on-shell) and
\item The set of equivalence classes of $(n-2)$-forms $\bm{k}$ that are closed on-shell ($\D\bm{k} = 0$) but not exact : it is impossible to find a $(n-3)$-form $\bm{l}$ such that $\bm{k} = \D\bm{l}$ on-shell.
\end{itemize}
}
The existence of equivalence classes of closed codimension 2 forms $\bm k$ describes the fact that the expression of the conserved $(n-2)$-forms actually remains ambiguous. Indeed, we can always add to $\bm{k}$ the divergence of a $(n-3)$-form $\bm b$ and a $(n-2)$-form $\bm a$ which vanishes on-shell
\begin{equation}
\bm k_1 \sim \bm k_2 \Leftrightarrow \bm k_2 = \bm k_1 + \D\bm b + \bm a.
\end{equation}
The total derivative ambiguity $\bm b$ is irrelevant for integrations on codimension 2 closed surfaces. Now $\bm{k}$ is no longer totally arbitrary and we can give a physical sense to the integral charge $H_Q$ which is by right a pure surface integral, \textit{i.e.} a \textit{surface charge}, because the integrand $\bm{k}$ is a $(n-2)$-form. It is nothing but than the outline of what we need to do: we must understand how to construct these surface charges out of the theory, see the assumptions under which they are actually conserved or not and finally discuss their properties and algebra. 

As an example, let us show how we can reconstruct the electric charge in classical electrodynamics with this new magnificent tool. In that context, we recall that the exact reducibility parameters are the non-vanishing constants $\bar\lambda = c\in\mathbb R_0$. By virtue of the vacuum Maxwell equations, the Faraday tensor $F^{\mu\nu}$ is divergence-free. It turns out, as we will see below, that the generalized Noether theorem precisely gives the conserved $(n-2)$-form: $\bm{k}_c [A] = c F^{\mu\nu} (\D^{n-2} x)_{\mu\nu}$, $\D\bm{k}_c = 0$ on-shell. Now we can integrate $\bm{k}_{c=1}$ on a sphere $S$ of constant time $t$ and radius $r$ to get the electric charge $Q_E = \oint_S \bm{k}_{c=1} = \oint_S \vec{E}\cdot \vec{e}_r \: \D S$. We immediately verify that it is conserved in time:
\begin{equation}
\frac{\D}{\D t} Q_E = \oint_S \partial_t k^{tr} \, 2(\D^{n-2} x)_{tr} = -\oint_S \partial_A k^{Ar} \D S = 0.
\end{equation} 
The last equality follows from the fact that the integration of a closed form on a sphere is vanishing (assuming of course that the field strength is regular on $S$, \textit{i.e.} for instance, the trajectories of charged particles do not cross $S$). Another point to notice is that 
\begin{equation}
\frac{\D}{\D r} Q_E = \oint_S \partial_r k^{tr} \, 2(d^{n-2} x)_{tr} = -\oint_S \partial_A k^{tA} \D S = 0
\end{equation} 
after using the time component of $\D\bm{k}=0$, namely $\partial_r k^{tr} + \partial_A k^{tA}=0$ and after assuming again that the field strength is regular on $S$. More generally, we obtain the Gauss law, stipulating that only the homology class of the integration surface matters (\textit{i.e.} the sources enclosed by $S$). 

\subsubsection{Surface charges in generally covariant theories}
\label{sec:Surface charges in generally covariant theories}
In electrodynamics, we have just seen that there exists exactly one class of reducibility parameters: the global gauge transformation by $1$ everywhere in spacetime, which is associated with the conserved electric charge and always available for any solution space. This particularity is helped by the fact that Maxwell theory is an easy linear theory. In General Relativity, which is a non-linear theory, life is not so easy. The reducibility parameters, or exact Killing vectors, are conversely pretty rare, because for a general spacetime, the metric $g_{\mu\nu}$ has few or more often no isometries at all. Hence the generalized Noether theorem cannot be applied to any generally defined diffeomorphism $\xi$. Correspondingly, it seems hopeless to write a formula describing a conserved $(n-2)$-form for any diffeomorphism in a generally covariant theory.

One way out is to make good use of the linearized theory around a suitably chosen solution. Let us consider a solution $\bar{g}_{\mu\nu}$ of General Relativity as \textit{background field} which we perturb by adding an infinitesimal contribution $g_{\mu\nu} = \bar{g}_{\mu\nu} + h_{\mu\nu}$. It is not difficult to show that the Einstein-Hilbert Lagrangian linearized around $\bar{g}_{\mu\nu}$ and expressed in terms of $h_{\mu\nu}$ is gauge-invariant under the linearized diffeomorphisms $\xi$ acting as $\delta_\xi h_{\mu\nu} = \Lie_\xi \bar{g}_{\mu\nu}$. So, if the background admits some Killing symmetries, their generators also define a set of reducibility parameters for the linearized theory, \textit{i.e.} if $\bar\xi$ satisfies $\mathcal L_{\bar\xi} \bar g_{\mu\nu}=0$, we have $\delta_{\bar\xi} h_{\mu\nu}=0$. Fortunately, this promotion of background isometries as exact reducibility parameters of the linearized theory allows to exploit the generalized Noether theorem to claim the existence of a set of conserved $(n-2)$-forms $\bm{k}_\xi [\bar{g};h]$ if $h_{\mu\nu}$ satisfies the linearized equations of motion around $\bar{g}_{\mu\nu}$ \cite{Barnich:1994db,Anderson:1996sc,Barnich:2004ts}. For instance, in asymptotically flat gravity at spatial infinity, the integration of $\bm{k}_\xi [\bar{g};h]$ for the Poincaré generators $\xi$ acting on the Minkowski background spacetime $\bar g_{\mu\nu} = \eta_{\mu\nu}$ gives the (local) ADM charges \cite{Arnowitt:1959ah} of linearized gravity (see \cite{Barnich:2001jy} for a proof).

The integration of these $(n-2)$-forms on a codimension 2 hypersurface yields thus a set of dynamical invariants for the linearized theory. The construction of the surface charges from the knowledge of the exact reducibility parameters at linear level gives a procedure to define infinitesimal surface charges associated with any diffeomorphism, including the asymptotic symmetries, around a target solution. We consider $\mathcal S$, a solution space assorted with some boundary conditions. Let us pick an arbitrary metric $g_{\mu\nu}\in\mathcal S$ and a reference or background solution $\bar{g}_{\mu\nu} \in \mathcal S$. Since General Relativity is a non-linear theory, we do not expect in general that all asymptotic charges only depend upon $\bar{g}_{\mu\nu}$ and the linearized perturbation $h_{\mu\nu} =g_{\mu\nu}-\bar g_{\mu\nu}$, \textit{i.e.} $\oint_S \bm k_\xi [\bar{g};g-\bar{g}]$ and the charges might depend non-linearly on $g_{\mu\nu}$. The better way to define $\bm k_\xi$ is in a local sense on $\mathcal S$ \cite{Barnich:2001jy}. We can linearize the theory around each $g_{\mu\nu}\in\mathcal S$, by considering an abstract field variation $\delta g_{\mu\nu}$ and evaluating the \textit{infinitesimal surface charge} between the solution $g_{\mu\nu}$ and $g_{\mu\nu}+\delta g_{\mu\nu}$. The expression for $\bm k_\xi$ is suspected to be the same as before and just waits to be integrated on a codimension 2 surface $S$ as
\begin{equation}
\ndelta H_\xi[g] = \oint_S \bm k_\xi [g;\delta g]. \label{theory infinitesimal charge}
\end{equation}
The \textit{finite surface charge} evaluated for the solution $g_{\mu\nu}$ with respect to the corresponding charge for the reference $\bar g_{\mu\nu}$ (which fixes the ``zero'' in some sense) might be obtained by integration on a path in $\mathcal S$ joining $\bar g_{\mu\nu}$ and $g_{\mu\nu}$ and we would like to write the following expression
\begin{equation}
H_\xi[g] = \int_{\bar g}^{g} \oint_S \bm k_\xi [g;\delta g]. \label{theory finite infinitesimal charge}
\end{equation}
The notation $\ndelta H_\xi$ in \eqref{theory infinitesimal charge} reflects our ignorance about the path-dependence or not of the integration on the solution space leading from \eqref{theory infinitesimal charge} to \eqref{theory finite infinitesimal charge}, or in other terms if the infinitesimal charge \eqref{theory infinitesimal charge} is \textit{integrable} or not. The charge $H_\xi$ will be conserved as long as the closure condition $\D\bm{k}_\xi [g;\delta g] = 0$ is obeyed on-shell for any solution $g_{\mu\nu} \in \mathcal S$ and for any variations that are ``tangent'' to $\mathcal S$, \textit{i.e.} the $\delta g_{\mu\nu}$ solving the linearized equations of motion around $g_{\mu\nu}$. 

In the following, we review how to give more mathematical aplomb to this intuitive construction. In particular, we need to discover in which sense $\delta$ could be seen as an exterior derivative on field spaces like $\mathcal S$ (on which the integration would make sense), how ``tangent to $\mathcal S$'' can be defined rigorously, how to discuss mathematically the (non-)integrability as well as the (non-)conservation of the charges,\dots but more importantly, how to compute from first principles the codimension 2 forms $\bm k_\xi[g;\delta g]$ whose existence is suggested (and promised in linearized theory) by the above reasoning. But before entering deeper into the formalism, let us mention a last but important conceptual point. The fact that the energy, in particular, is a surface charge in General Relativity can be interpreted as gravity being holographic! Indeed, in quantum gravity, the energy levels of all states of the theory can be found by quantizing the gravitational Hamiltonian. In the classical limit, this Hamiltonian is a surface charge. If this remains true at quantum level (as it does for example in the AdS/CFT correspondence) knowing the field on the surface bounding the bulk of spacetime will allow to know all possible states in the bulk of spacetime. 

\subsection{Covariant phase space formalism}
\label{sec:Covariant phase space formalism}

This section aims at introducing the key ingredients of the covariant phase space formalism \cite{Crnkovic:1986be,Crnkovic:1986ex,Lee:1990nz,Wald:1990idc,Iyer:1994ys,Wald:1999wa} which gives a procedure and a mathematical framework to define rigorously the infinitesimal surface charges \eqref{theory finite infinitesimal charge} from first principles and characterize their properties. The presentation is constructive and elaborated from preconceptions coming from the well-known Hamiltonian mechanics.

\subsubsection{Hamiltonian mechanics}
\label{sec:Hamilton}

We present here a brief review of classical mechanics from the point of view of symplectic geometry. The latter is the canonical abstraction of Hamiltonian mechanics that allows to consider it in an intrinsic way, \textit{i.e.} independently of a chosen set of symplectic conjugated coordinates $(q^i,p_j)$. This will help us in understanding the construction of the covariant phase space formalism, which is simply the generalization of the mechanical concept of phase space to classical covariant field theories. Here we restrain ourselves to fields on a manifold reduced to the time direction. More details and explanations can be gleaned \textit{e.g.} in \cite{1978mmcm.book.....A}.

Let $\mathcal P$ be the phase space of a mechanical autonomic system whose configuration space has dimension $n\in\mathbb N_0$. The $2n$ coordinates on $\mathcal P$ are denoted by $z=\{ z^A \}_{A=1}^{2n}$. To match with the conventional notations of variations, we denote as $\delta$ the exterior derivative on $\mathcal P$. The system is supposed to be described by the first order Hamiltonian action 
\begin{equation}
S[z] = \int_{t_1}^{t_2} \D t \left( \theta_A(z)\dot z^A - \mathcal H(z) \right)  \label{action hamilton}
\end{equation}
where $t$ is the time parameter of the trajectories $z(t)$ in $\mathcal P$, $\dot z \equiv \frac{\D z}{\D t}$, $\theta_A(z)$ are the components of the Liouville one-form $\bm\theta = \theta_A(z)\delta z^A \in\Omega^1(\mathcal P)$ and $\mathcal H(z)$ is the Hamiltonian function. Taking one more exterior derivative of $\bm\theta$ yields the \textit{presymplectic form}
\begin{equation}
\bm\omega = \delta\bm\theta = (\partial_A \theta_B - \partial_B \theta_A ) \delta z^A\wedge\delta z^B \in\Omega^2(\mathcal P).
\end{equation}
This definition motivates the appellation \textit{presymplectic potential} for $\bm \theta$. The presymplectic form is closed by construction, $\delta\bm\omega = 0$, but non-necessarily non-degenerate. We admit the latter assumption in prevision of the treatment of gauge symmetries that define invariance directions in the kernel of $\bm\omega$. The epithet ``presymplectic'' instead of ``symplectic'' used to qualify $\bm\theta$ and $\bm\omega$ indicates that we leave room for gauge transformations, which are simply considered here as transformations parametrized by arbitrary functions of time. Given $\bm\omega$, one can promote $\mathcal P$ into a real symplectic manifold $(\mathcal P,\bm\omega)$ of dimension $2n$. Assuming the non-degeneracy of $\bm\omega$, the Darboux theorem states that there exists a basis of coordinates $\{ (q^i,p_j) \}$ in which one can ``diagonalize'' the symplectic form as $\bm \omega = \delta p_i \wedge \delta q^i$. Hence $\bm\theta = p_i\delta q^i$, $\theta_A = (p_1,\dots,p_n,0,\dots,0)$ and the Hamiltonian action takes the usual form 
\begin{equation}
S[q^i,p_j] = \int_{t_1}^{t_2} \D t \left( p_i\dot q^i - \mathcal H(q^i,p_j) \right).
\end{equation}
Under the transformation $z^A \to z^A + \delta z^A$ the action \eqref{action hamilton} is varied as
\begin{equation}
\delta \mathcal S = \int_{t_1}^{t_2} \D t \left[ \left(\omega_{AB}\dot z^B -  \partial_A \mathcal H\right) \delta z^A + \frac{\D}{\D t}\left(\theta_A\delta z^A\right) \right]
\label{action principle}
\end{equation}
with $\dot z^A$ given by the equations of motion $\partial_A \mathcal H = \omega_{AB}\dot z^B$ as the symplectic gradient of $\mathcal H(z)$. The on-shell variational principle reads as $\delta S = [\bm\theta]^{t_2}_{t_1}$. The presymplectic potential appears thus as a boundary term in the action principle. For Dirichlet boundary conditions $\theta_A\delta z^A|_{t_1} = 0 = \theta_A\delta z^A|_{t_2}$ (or $\delta q^i|_{t_1} = 0 = \delta q^i|_{t_2}$ in Darboux coordinates), the action is stationary on solutions of the equations of motion, \textit{i.e.} $\delta  S= 0$. 

Let us consider some vector field $X = X^A\partial_A$ on $\mathcal P$. It generates the infinitesimal transformation $\delta_X z^A = X^A(z)$ on the phase space coordinates. $X$ is a \textit{symmetry} if and only if it preserves the Hamilonian action of the system up to a total derivative
\begin{equation}
\delta_X \left[ \theta_A(z)\dot z^A - \mathcal H(z) \right] = \frac{\D }{\D t}B_X(z). \label{symmetry classical}
\end{equation}
By virtue of Noether's first theorem, there must exist a \textit{conserved charge} associated with the continuous symmetry spanned by $X$. Developing \eqref{symmetry classical} yields a representative of this charge:
\begin{equation}
\frac{\D}{\D t} \left(B_X -i_X \bm\theta\right) = X^A (\omega_{AB}\dot z^B-\partial_A \mathcal H), \label{conservation HX}
\end{equation}
where $i_X\bm\theta$ denotes the interior product of the presymplectic potential one-form $\bm \theta$ by the vector field $X$. As a result
\begin{equation}
H_X(z) \equiv B_X(z) -i_X \bm\theta(z) + H_0 \label{Hamilton Noether charge}
\end{equation}
is conserved on-shell, \textit{i.e.} $\frac{\D}{\D t}H_X=0$ when the equations of motion hold. It is defined up to a real constant $H_0$ that fixes the ``zero'' of the charge. The flow of the symmetry $X$ preserves the Hamiltonian, \textit{i.e.} $\mathcal L_X \mathcal H = X^A\partial_A\mathcal H = 0$. Hence the conservation equation \eqref{conservation HX} for $H_X$ can be rewritten as $\frac{\D}{\D t}H_X = \partial_A H_X \dot z^A = X^A \omega_{AB}\dot z^B$ or in a covariant way
\begin{equation}
\boxed{ i_X\bm\omega = \delta H_X.} \label{def hamilton}
\end{equation}
Among all vector fields defined on $\mathcal P$, the symmetries are the vector fields $X$ for which there exists a smooth function $H_{X}$ on $\mathscr P$ such that \eqref{def hamilton} holds, meaning that $X$ is a \textit{Hamiltonian vector field}. The charge $H_X$ appearing in \eqref{def hamilton} is said to be \textit{integrable} because the 1-form $\delta H_X$ is exact, so $H_X$ is known up to a real constant (assuming that the topology of $\mathcal P$ is trivial). The simplest example of a Hamiltonian vector field is $X_t = \dot z^A \partial_A$ for an autonomous system such as \eqref{action hamilton}. Indeed, the equations of motion imply $i_{X_t}\bm \omega = \delta \mathcal H$ and the associated charge is precisely the total mechanical energy of the system.

The idea behind the symplectic formalism is to understand the motion of the mechanical system as the flow of the Hamiltonian vector fields $X$ on $\mathcal P$. In order to make this idea more precise, one defines the \textit{Poisson bracket} $\{\cdot,\cdot\}$ for functions $H_{X}$ canonically conjugated to Hamiltonian vector fields $X$ by contracting two times the presymplectic form. Let us take $X_1$, $X_2$ two Hamiltonian vector fields on $\mathcal P$, such that $i_{X_{1}}\bm\omega = \delta H_{X_{1}}$, $i_{X_{2}}\bm\omega = \delta H_{X_{2}}$. Then
\begin{equation}
\{ H_{X_1},H_{X_2} \} \equiv i_{X_2} i_{X_1}\bm\omega. \label{Poisson bracket}
\end{equation}
By definition, we have
\begin{equation}
\{ H_{X_1},H_{X_2} \} = \delta_{X_2}H_{X_1} = X_2[H_{X_1}] \label{Poisson 2}
\end{equation}
where $X[F]$ represents the application of the vector $X$ on the function $F$ on $\mathcal P$. The Poisson bracket is directly bilinear and antisymmetric and it can be quickly checked that it satisfies the Jacobi identity. As a consequence of \eqref{Poisson 2}, we can rewrite the evolution of any function $F$ on the flow of $X$ as $X[F] = \{ F,H_{X} \}$ which encodes the equations of motion in a covariant way with respect to the phase space geometry when $X = X_t$. In that case we have $\dot F = X_t[F] = \{ F,\mathcal H\}$.

An important feature of the Poisson bracket is that it promotes the space of local functions on $\mathcal P$ to a Lie algebra for which the Lie bracket is \eqref{Poisson bracket}. Indeed, it is known that the commutator of two Hamiltonian vector fields is also a Hamiltonian vector field. The proof relies only on structural properties of the differential manifold $\mathcal P$, namely the identity $i_{[X_1,X_2]}(\cdot) = [\mathcal L_{X_1},i_{X_2}](\cdot)$ and the Cartan's magic formula $\mathcal L_{X}(\cdot) = i_X \delta (\cdot) + \delta i_X (\cdot)$. The outcome is precisely $i_{[X_1,X_2]}\bm\omega = \delta(\{H_{X_1},H_{X_2}\})$. Hence, there exists a function $H_{[X_1,X_2]}$ on $\mathcal P$ such that $i_{[ X_1, X_2]}\bm\omega \equiv \delta H_{[ X_1, X_2]}$. Coupling both pieces of information, one obtains
\begin{equation}
\{ H_{X_1},H_{X_2}\} = H_{[X_1,X_2]} + K_{X_1,X_2} \label{algebra integrable case}
\end{equation}
where $K_{X_1,X_2}$ is a real constant. Note crucially that $K_{X_1,X_2}$ is not arbitrary and beside of depending directly upon $X_1$ and $X_2$, it is subjected to two properties: it is antisymmetric under the exchange of vectors (\textit{i.e.} $K_{X_2,X_1} = -K_{X_1,X_2}$) and satisfies the condition $K_{X_1,[X_2,X_3]} + \text{cyclic(1,2,3)} = 0$ because \eqref{Poisson bracket} is a Lie bracket. A real constant $K_{X_1,X_2}$ constrained in that way forms a \textit{Lie algebra 2-cocycle} over the vector algebra. Since it Poisson-commutes with any function on $\mathcal P$, it represents a \textit{central extension} of the algebra of these functions. Therefore the result \eqref{algebra integrable case} can be phrased as follows: \textit{the algebra of conserved charges represents the algebra of Hamiltonian vector fields by means of the Lie bracket defined in \eqref{Poisson bracket}}. Note crucially that the charge algebra is fundamental because it encodes the whole algebraic structure as well as the dynamics of the physical system. For instance, \eqref{algebra integrable case} reproduces the conservation laws $\frac{\D}{\D t}H_X=0$ when we evaluate the Poisson bracket of $H_X$ with the Hamiltonian $\mathcal H$. This closes our brief review on symplectic methods for classical Hamiltonian systems.

\subsubsection{Field fibration and jet bundle}
After the introduction of fundamentals about phase spaces in classical mechanics, we review the construction of the covariant phase space formalism \cite{Lee:1990nz,Iyer:1994ys,Crnkovic:1986ex,Wald:1993nt} as a generalization the basic concepts we introduced in the previous section. The coordinates on the covariant phase space are not only dependent of time but on the whole set of coordinates defined on a base differentiable manifold. We will see how this enlargement brings new features and opens new horizons in the formalism.

We work again on a base spacetime $\mathscr M$ which is a Lorentzian manifold provided with a set of coordinates $\{ x^\mu\}$. In contrast to the simple timeline for classical mechanics, here the base space is a differentiable manifold with much richer structure. Let us set aside the metric tensor $g_{\mu\nu}$ for the moment. Vector fields tangent to $\mathscr M$ are decomposed in a natural coordinate basis $\lbrace \partial_\mu \rbrace$ while differential forms are decomposed in the dual natural basis $\lbrace \D x^\mu \rbrace$. The tangent structure of $\mathscr M$ also contains a countable tower of vector spaces populated by multilinear $k$-forms, which we write $\Omega^k(\mathscr M)$, for $k\in\mathbb N$. By reflexivity, vector fields $\xi \in T\mathscr M$ can be seen as functions on $\Omega^1(\mathscr M)$ thanks to the interior product $\iota_\xi : \Omega^1(\mathscr M)\to\mathbb R : \bm w \mapsto \xi^\mu \partial_\mu \bm w$. We can extend this definition to promote the interior product to an operator $\iota_\xi : \Omega^k  (\mathscr M)\rightarrow \Omega^{k-1} (\mathscr M)$ by requiring that $\iota_\xi \bm w \equiv \xi^\mu \frac{\partial}{\partial \D x^\mu} \bm w, \: \forall \bm w \in \Omega^k (\mathscr M)$. We have also at our disposal a differential operator $\D = \D x^\mu \partial_\mu$, the \textit{exterior derivative} that induces the \textit{De Rham complex}. Starting from scalars ($0$-forms), successive applications of $\D$ lead to higher order forms :
\begin{equation}
\Omega^0 (\mathscr M) \rightarrow \Omega^1 (\mathscr M) \rightarrow \Omega^2 (\mathscr M) \rightarrow \cdots \rightarrow \Omega^{n-1} (\mathscr M) \rightarrow \Omega^n (\mathscr M) \rightarrow 0.
\end{equation}
In summary, we have a first space which is the manifold $\mathscr M$ with local coordinates $\{ x^\mu\} $ and equipped with a natural differential operator $\D$. Using it, we get forms of higher degree, since $\D : \Omega^k(\mathscr M) \rightarrow \Omega^{k+1}(\mathscr M)$. In the other hand, we can use the interior product $\iota_\xi$ to ascend the chain of $\Omega$'s and, consequently, travel the full set of spaces $\lbrace \Omega^k(\mathscr M) \: |\: k = 0,1,...,n \rbrace$.

Let us now introduce fields $\phi = (\phi^i)$ on $\mathscr M$, including the metric tensor $g_{\mu\nu}$ as well. At the beginning, it is highly convenient to consider them as abstract entities without dependence in the coordinates. We define the \textit{jet space} $\mathscr J$ as the collection of fields $\phi^i$ and their symmetrized derivatives $\phi^i_\mu$, $\phi^i_{(\mu\nu)}\dots$ We denote a ``point'' of $\mathscr J$ by $(\phi_{(\mu)})$. Around such a point, we can define the cotangent space as the collection of abstract variations $(\delta\phi^i_{(\mu)}) = (\delta\phi^i,\delta\phi_\mu^i, \delta\phi_{\mu\nu}^i,\dots )$ of the fields. The natural definition of an exterior derivative on $\mathscr J$ is
\begin{equation}
\delta = \sum_{(\mu)}\delta \phi^i_{(\mu)} \frac{\partial}{\partial \phi^i_{(\mu)} } = \delta \phi^i \frac{\partial}{\partial \phi^i} +\delta \phi_\mu^i \frac{\partial}{\partial \phi_\mu^i} +\delta \phi_{\mu\nu}^i \frac{\partial}{\partial \phi_{\mu\nu}^i} + \cdots \label{vertical derivative}
\end{equation}
which is perfectly reminiscent of $\D = \D x^\mu \partial_\mu$ on $\mathscr M$. In this definition, the symmetrized derivatives with respect to the fields are meant to satisfy 
\begin{equation}
\frac{\partial \phi^i_{(\mu)}}{\partial \phi^j_{(\nu)}} = \delta^{(\mu)}_{(\nu)} \delta^i_j, \quad \delta^{(\mu)}_{(\nu)} = \delta^{\mu_1}_{(\nu_1}\dots \delta^{\mu_k}_{\nu_k)}, \, k = |\mu|=|\nu|.
\end{equation}
We take all the $\delta\phi^i_{(\mu)}$ to be Grassmann odd, which implies that $\delta^2 = 0$, an expected property for a exterior derivative operator. 

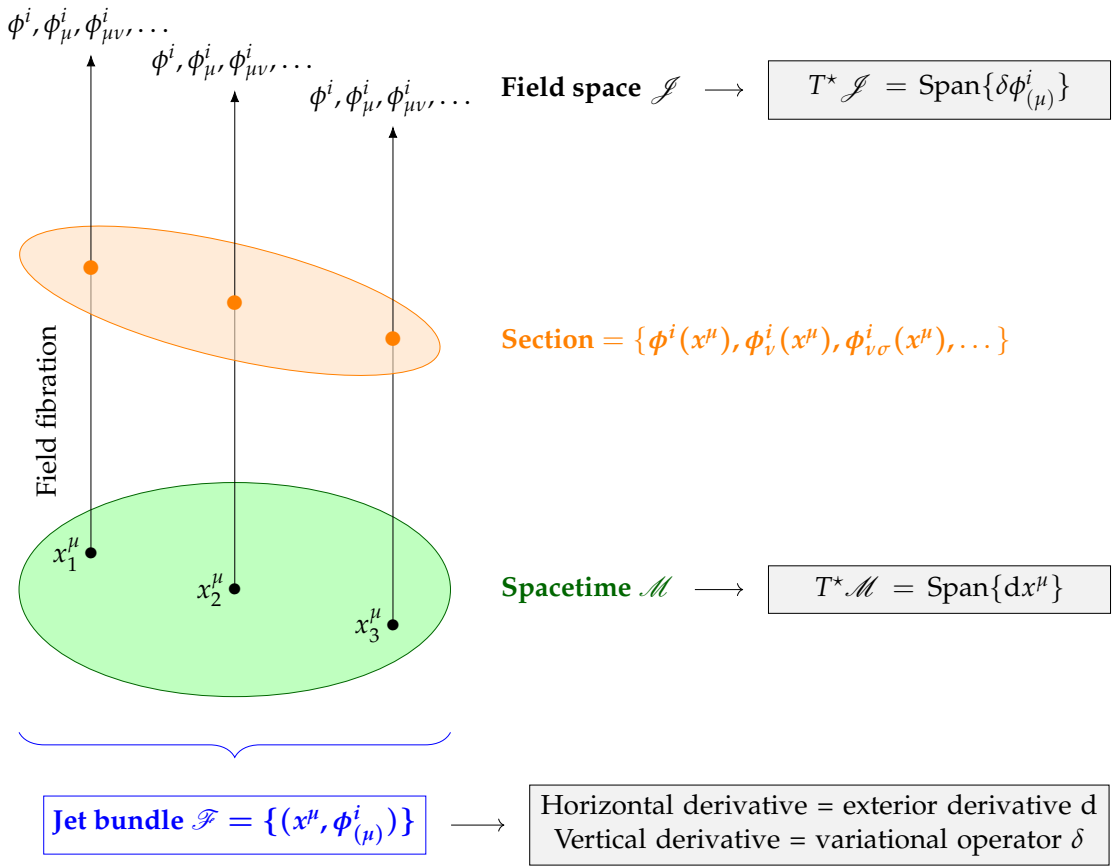
\begin{figure}[ht!]
\centering
\resizebox{\textwidth}{!}{%
\begin{tikzpicture}
	\draw[green!40!black,fill=green!25] (0,-5) circle (3 and 1.5);
	\coordinate (Xone) at (-2,-4.5);
	\coordinate (Xtwo) at (0,-5);
	\coordinate (Xthree) at (2.2,-5.5);
	\draw (Xone) node {$\bullet$};
	\draw (Xtwo) node {$\bullet$};
	\draw (Xthree) node {$\bullet$};
	\draw (Xone) node[left] {$x_1^\mu$};
	\draw (Xtwo) node[left] {$x_2^\mu$};
	\draw (Xthree) node[left] {$x_3^\mu$};
	\def\offset{3.4};
	\def\decal{6.4};
	\coordinate (labelM) at ($(Xtwo)+(\offset,0)$);
	\node[green!40!black,outer sep=5pt,right] (nlabelM) at (labelM) {\textbf{Spacetime $\mathscr{M}$}};
	\coordinate (TM) at ($(labelM)+(\decal,0)$);
	\node[draw=black,fill=gray!10,text width=4.5cm,align=center,outer sep=10pt] (nTM) at (TM) {$T^\star \mathscr M = \text{Span} \lbrace \D x^\mu \rbrace$};
	\draw[->] (nlabelM) -- (nTM);
	\coordinate (Sone) at ($(Xone)+(0,4)$);
	\coordinate (Stwo) at ($(Xtwo)+(0,4)$);
	\coordinate (Sthree) at ($(Xthree)+(0,4)$);
	\draw (Xone) -- (Sone);
	\draw (Xtwo) -- (Stwo);
	\draw (Xthree) -- (Sthree);
	\node[rotate=90,above,outer sep=10pt] at ($(Xone)!0.5!(Sone)$) {Field fibration};
	\draw[orange,fill=orange!25,rotate=-13.5,fill opacity=0.6] ($(Xone)+(1.05,3.9)$) circle (3 and 0.8);
	\coordinate (labelS) at ($(Stwo)+(\offset,-0.5)$);
	\node[orange,outer sep=5pt,right] (nlabelS) at (labelS) {\textbf{Section }$ =  \lbrace \bm{\phi^i (x^\mu), \phi^i_{\nu} (x^\mu), \phi^i_{\nu\sigma}(x^\mu),}\dots \rbrace$};
	\coordinate (Eone) at ($(Sone)+(0,3)$);
	\coordinate (Etwo) at ($(Stwo)+(0,3)$);
	\coordinate (Ethree) at ($(Sthree)+(0,3)$);
	\draw[-Latex] (Sone) -- (Eone);
	\draw[-Latex] (Stwo) -- (Etwo);
	\draw[-Latex] (Sthree) -- (Ethree);
	\draw (Sone) node[orange]{\Large $\bullet$};
	\draw (Stwo) node[orange]{\Large $\bullet$};
	\draw (Sthree) node[orange]{\Large $\bullet$};
	\node[above,align=center,text width=2.5cm] at (Eone) {$\phi^i , \phi_\mu^i ,\phi^i_{\mu\nu},\dots$};
	\node[above,align=center,text width=2.5cm] at (Etwo) {$\phi^i , \phi_\mu^i ,\phi^i_{\mu\nu},\dots$};
	\node[above,align=center,text width=2.5cm] at (Ethree) {$\phi^i , \phi_\mu^i ,\phi^i_{\mu\nu},\dots$};
	\coordinate (labelF) at ($(Etwo)+(\offset,0)$);
	\node[outer sep=5pt,right] (nlabelF) at (labelF) {\textbf{Field space $\mathscr{J}$}};
	\coordinate (TJ) at ($(labelF)+(\decal,0)$);
	\node[draw=black,fill=gray!10,text width=4.5cm,align=center,outer sep=10pt] (nTJ) at (TJ) {$T^\star \mathscr J = \text{Span} \lbrace \delta \phi^i_{(\mu)} \rbrace$};
	\draw[->] (nlabelF) -- (nTJ);
	\draw[blue,decorate,decoration={brace,amplitude=10pt}] (3.0,-7.0) -- (-3.0,-7.0);
	\def\z{-8.3}
	\node[blue,draw=blue,outer sep=10pt] (m) at (0.0,\z) {$\bm{ \text{\textbf{Jet bundle }} \mathscr F = \lbrace (x^\mu,\phi^i_{(\mu)} )\rbrace}$};
	\node[right,fill=gray!10,text width=7.8cm,draw=black,align=center,outer sep=10pt] (n) at (3.75,\z)  {Horizontal derivative = exterior derivative $\D$ Vertical derivative = variational operator $\delta$};
	\draw[->] (m) -- (n);
	
\draw[opacity=0] (-3.8,-9.6) -- (13,-9.6) -- (13,3.7) -- (-3.8,3.7) -- cycle;
\end{tikzpicture}
}
\caption{Structure of the jet bundle.}
\label{fig:Bicomplex}
\end{figure}

The last step consists in putting the manifold and the field space together. Doing so, one gets the \textit{jet bundle} $\mathscr F$ \cite{Kolar_naturaloperations,saunders1989geometry,olver_1995,Anderson0}. This is practically achieved by anchoring the fields of $\mathscr J$ at points of the manifold $\mathscr M$, as depicted schematically on the figure \ref{fig:Bicomplex}. Mathematically, one says that $\mathscr F$ is the fiber bundle whose local trivialization is the pairing $(x^\mu,\phi,\phi_\mu,\phi_{\mu\nu},$ $\dots)$. This gives local coordinates on $\mathscr F$ which, in turn, looks like the product $\mathscr M \times \mathscr J$. Taking a section of this fiber bundle amounts to giving a map $x\in\mathscr M \to (\phi(x),\phi_\mu(x),\phi_{\mu\nu}(x),\dots)$ providing the coordinate-dependent fields of the theory. The usual differential operator $\D$ is still defined on $\mathscr M$ but must be continued to the full jet bundle $\mathscr F$ as $\D = \D x^\mu \partial_\mu$ with
\begin{equation}
 \partial_\mu \equiv \frac{\partial}{\partial x^\mu} + \sum_{(\nu)}\phi^i_{\mu(\nu)} \frac{\partial}{\partial\phi^i_{(\nu)}} =  \frac{\partial}{\partial x^\mu} + \phi^i_\mu \frac{\partial}{\partial \phi^i} +  \phi^i_{\mu\nu} \frac{\partial}{\partial \phi_\nu^i}+\cdots \label{horizontal derivative}
\end{equation}
Thenceforth we have two Grassman-odd differential operators at our disposal: $\D$ and $\delta$ respectively coined as the \textit{horizontal} and the \textit{vertical derivative}. They anti-commute, $\{ \delta,\D \} = 0$, as it can be checked from \eqref{vertical derivative} and \eqref{horizontal derivative}. Vector spaces of multilinear differential forms tangent to $\mathscr F$ can be defined in the standard way and are denoted by $\Omega^{p,q}(\mathscr F)$. They contain $(p,q)$-forms which are $p$-forms with respect to the base manifold $\mathscr M$ and $q$-forms with respect to the jet space $\mathscr J$. Roughly speaking, they contain $p$ $\D x^\mu$ and $q$ $\delta \phi^i_{(\mu)}$.

A crucial point that has to be mentioned here is that the horizontal derivative $\D$ takes the presence of the field fibration along $\mathscr J$ into account and has for that reason a reacher cohomology structure than the exterior derivative on $\mathscr M$. For the latter, the \textit{Poincaré lemma} states that in a simply connected open subset of $\mathscr M$, the de Rham cohomology class $\texttt{H}^p_{\mathscr M}$, \textit{i.e.} the set of equivalence classes of closed $p$-forms modulo the exact forms, is empty for $0<p\leq n$ and is $\mathbb R$ for $p=n$:
\begin{equation}
\texttt{H}^p_{\mathscr M} = \left\lbrace
\begin{array}{lcl}
\mathbb R &\text{if}& p=0,\\
0 &\text{if}& 0<p\leq n.
\end{array}
\right.
\end{equation}
That means that, locally, every closed $p$-form is exact, except for the trivial case $\D c = 0$ for any real constant $0$-form $c$. This property cannot be verified by the operator \eqref{horizontal derivative}, simply because this would mean that every $n$-form is exact. Hence any Lagrangian $\bm L$ would be equivalent to a boundary term for which the Euler-Lagrange equations are trivially satisfied. Fortunately, the following result can be proven (see \textit{e.g.} \cite{Barnich:2018gdh}):
\resu{Algebraic Poincaré lemma}{
The cohomology class $\texttt{H}^p_{\mathscr F}$ for the horizontal derivative operator $\D$ is given by
\[
\texttt{H}^p_{\mathscr F} = \left\lbrace
\begin{array}{lcl}
\mathbb R &\text{if}& p=0,\\
0 &\text{if}& 0<p<n,\\
\left[\bm \omega^n\right] &\text{if}& p=n,
\end{array}
\right.
\]
where $[\bm\omega^n]$ denotes equivalence classes of $n$-forms such that
\[
[\bm\omega^n] = \left\lbrace \bm\omega \sim \bm\omega' \in\Omega^n(\mathscr M) \ \Big| \ \bm\omega' = \bm\omega+\D \bm b,\, \bm b\in\Omega^{n-1}(\mathscr M) \Leftrightarrow \frac{\delta}{\delta\phi}(\bm\omega'-\bm\omega) = 0\right\rbrace.
\]
}
For $p<n$, these cohomology classes are the same as for the exterior derivative, but the crucial difference shows up for $p=n$. $\texttt{H}^n_{\mathscr F}$ is not trivial but defines equivalence classes of $n$-forms that are non-exact even though closed and differ by a boundary term that does not change the equations of motion. 

To complete the picture, we need a notion of interior product in the jet bundle $\mathscr F$. This will allow to compute particular variations of the fields from the abstract ones tangent to $\mathscr J$. A \textit{variation} $\delta_Q\phi^i = Q^i$ under a transformation of characteristic $Q$ is defined as
\begin{equation}
\delta_Q F = \sum_{(\mu)} \left[ \partial_{(\mu)} Q^i\frac{\partial F}{\partial \phi^i_{(\mu)}} + \partial_{(\mu)}\delta Q^i \frac{\partial F}{\partial \delta \phi^i_{(\mu)}} \right] \label{variation definition}
\end{equation}
for any function $F$ on $\mathscr F$. By definition, $Q$ is a vector tangent to the jet space and $\delta_Q$ represents the Lie derivative along $Q$ -- we could technically write it as $\mathcal L_Q$ though we prefer to keep the usual notation $\delta_Q$ for the sake of clarity. Like the interior product on spacetime differential forms $\iota_\xi = \xi^\mu \frac{\partial}{\partial \D x^\mu}$, aimed at replacing $\D x^\mu$ by $\xi^\mu$ to obtain the contraction with the vector $\xi$, we define the interior product on the jet space as
\begin{equation}
i_Q =\sum_{(\mu)}\partial_{(\mu)} Q^i\frac{\partial}{\partial \delta\phi^i_{(\mu)}} \label{interior product jet space}
\end{equation}
which replaces in some differential form on $\mathscr J$ the arbitrary variations $\delta\phi^i_{(\mu)}$ by the transformation of characteristic $Q$. The operators \eqref{variation definition} and \eqref{interior product jet space} obey the following properties
\begin{align}
[\delta_Q,\D] &= [\delta_Q,\delta] = 0, \label{commu on jet space}\\
\delta_Q &= i_Q\delta + \delta i_Q. \label{Cartan for jet bundle}
\end{align}
The crucial identity \eqref{Cartan for jet bundle} is nothing but Cartan's magic formula for the jet bundle. The characteristics of transformations form a Lie algebra under the Lie bracket $[Q_1,Q_2]=\delta_{Q_1}Q_2 - \delta_{Q_2}Q_1$, for which we can prove the additional identities
\begin{align}
i_{[Q_1,Q_2]} &= [i_{Q_1},\delta_{Q_2}], \label{Lie with inner product} \\
[ \delta_{Q_1},\delta_{Q_2}] &= -\delta_{[Q_1,Q_2]}. \label{Lie commutation}
\end{align}
This closes our exposition of the mathematical properties of $\mathscr F$, which is the natural and robust cadre in which we are going to stage the little theater of classical fields!

\subsubsection{Presymplectic structure}
Let us go back to our theory \eqref{action general}, whose Lagrangian $\bm L$ encodes the classical physics of the fields $\phi$. Assorted with a suitable set of boundary conditions, the set of allowed fields has been denoted as $\mathcal S$, the solution space. The fields in $\mathcal S$ collectively define the jet space $\mathscr J$ whereas the spacetime $\mathscr M$ provides the base manifold on which the fields live. $\bm L$ is the most basic field of the jet bundle bi-differential structure and turns out to be a $n$-form with respect to $\mathscr M$ and a $0$-form with respect to $\mathscr J$. Taking some variation $\delta\phi$ on the fields, one gets
\begin{equation}
\boxed{
\delta \bm{L} = \delta \phi^i \frac{\delta \bm{L}}{\delta \phi^i} - \D\bm{\Theta} [\phi;\delta \phi], \label{variation lagrangian theory}
}
\end{equation}
where the first term involves the Euler-Lagrange derivatives defined as \eqref{EulerLagrange}, isolated after after iterative applications of the inverse Leibniz rule. The relics of these ``integrations by parts'' are gathered in the boundary term $\bm{\Theta}[\phi;\delta\phi]$, called \textit{presymplectic potential} \cite{Lee:1990nz}, which is a $(n-1)$-form with respect to $\mathscr M$ and a $1$-form with respect to the fields. The unusual minus sign in \eqref{variation lagrangian theory} is due to our convention about the Grassmann parity of $\delta$ and $\D$. Contracting with any particular (Grassmann-even) variation $\delta_a\phi$ tangent to $\mathcal S$, we recover \eqref{action principle}. Indeed,
\begin{align}
i_{\delta_a}\delta \bm L &= \delta_a \phi^i \frac{\delta \bm{L}}{\delta \phi^i} - i_{\delta_a}\D\bm{\Theta} [\phi;\delta \phi] \nonumber\\
\Rightarrow \delta_a \bm L &= \delta_a \phi^i \frac{\delta \bm{L}}{\delta \phi^i} + \D\bm{\Theta} [\phi;\delta_a \phi]. \label{variation lagrangian grassmann even}
\end{align}
This observation justifies the employment of $\bm\Theta[\phi,\delta\phi]$ as presymplectic potential. This is also in accordance with the formula \eqref{delta L theo} in which $\delta$ is the standard Grassmann even variational operator. As in Hamiltonian mechanics, the \textit{(Lee-Wald) presymplectic current} is defined as a variation of the presymplectic potential \cite{Lee:1990nz}
\begin{equation}
{
\boxed{ \bm\omega [\phi;\delta \phi,\delta \phi ] = \delta \bm\Theta[\phi;\delta \phi]. }
} \label{omega grassmann odd}
\end{equation}
It is a $(n-1,2)$-form by construction. It gives the \textit{presymplectic form} $\mathcal W$ of the covariant phase space after integration on a codimension 1 spacelike hypersurface $\Sigma$,
\begin{equation}
\boxed{
\mathcal W[\phi;\delta\phi,\delta\phi] = \int_{\Sigma} \bm\omega [\phi;\delta \phi,\delta \phi ],
} \label{presymplectic form}
\end{equation}
which is a $(0,2)$-form on the jet bundle. Since the presymplectic current \eqref{omega grassmann odd} is closed on-shell, the definition \eqref{presymplectic form} is invariant upon smooth deformations of $\Sigma$ once the boundaries $\partial\Sigma$ are fixed. Again, starting from \eqref{omega grassmann odd}, we can go back to a notation where variations are the more familiar Grassmann even quantities. To do this, we contract both sides of the equation with the interior products $i_{\delta_2}i_{\delta_1}$. The operator $i_{\delta_1}$ hits either the first or second $\delta$. There are two terms: in each case the remaining $\delta$ is replaced by $\delta_2$. Taking into account the sign obtained by anticommuting the $\delta$'s, we obtain
\begin{equation}
i_{\delta_2}i_{\delta_1}\bm\omega \triangleq  \bm\omega[\phi;\delta_1 \phi,\delta_2 \phi] = \delta_1 \bm\Theta [\phi;\delta_2 \phi] - \delta_2 \bm\Theta [\phi;\delta_1 \phi]. \label{omega grassmann even}
\end{equation}
Our little review of covariantized Hamiltonian mechanics in section \ref{sec:Hamilton} has recalled us the tight link between conserved charges and (pre)symplectic form in classical mechanics, see \eqref{def hamilton}. The main goal of the covariant phase space formalism consists in relating the presymplectic form we have just defined on the jet bundle with the set of conserved $(n-2)$-forms we announced before, in section \ref{sec:Surface charges in generally covariant theories}. But before that, we need to make some \textit{intermezzo} about Noether's second theorem, which will be used afterwards as a lemma! 

\subsubsection{Noether's second theorem: an important lemma}
For any symmetry of characteristic $Q$, we saw that $\delta_Q \bm L = \D \bm B_Q$, by definition. Contracting \eqref{variation lagrangian grassmann even} with the variation $\delta_Q\phi = Q$ yields
\begin{equation}
\D \bm B_Q = Q^i \frac{\delta\bm L}{\delta \phi^i} + \D (i_Q\bm\Theta) \Rightarrow \boxed{ Q^i \frac{\delta\bm L}{\delta \phi^i} = \D\bm J_Q, } \label{Noether for Q}
\end{equation}
which provides a particular representative of the \textit{Noether current} $\bm J_Q = \bm B_Q - i_Q\bm\Theta$ for this transformation and whose expression is completely reminiscent to the Noether charge in classical mechanics, see \eqref{Hamilton Noether charge}. It is conserved on-shell, \textit{i.e.} $\D \bm J_Q = 0$ when the equations of motion $\frac{\delta\bm L}{\delta \phi^i} = 0$ hold.  For a gauge transformation $\delta_\lambda\phi = R[\lambda]$, the equation \eqref{Noether for Q} can be reworked thanks to some integrations by parts on the first member:
\begin{align}
R^i[\lambda]\frac{\delta\bm L}{\delta\phi^i} &= (R^i_\alpha\lambda^\alpha + R^{i\mu}_\alpha\partial_\mu\lambda^\alpha + R^{i(\mu\nu)}_{\alpha}\partial_\mu\partial_\nu \lambda^\alpha+\cdots)\frac{\delta\bm L}{\delta\phi^i} = \lambda^\alpha N_\alpha \left[\frac{\delta\bm L}{\delta\phi^i}\right] + \D \bm S_\lambda\left[\frac{\delta\bm L}{\delta\phi^i}\right], \label{Ri} \\
N_\alpha \left[\frac{\delta\bm L}{\delta\phi^i}\right] &\equiv R^i_\alpha \frac{\delta\bm L}{\delta\phi^i} - \partial_\mu \left( R^{i\mu}_\alpha \frac{\delta\bm L}{\delta\phi^i}\right) + \partial_\mu\partial_\nu \left( R^{i(\mu\nu)}_\alpha \frac{\delta\bm L}{\delta\phi^i}\right)+\cdots,\\
\bm S_\lambda \left[\frac{\delta\bm L}{\delta\phi^i}\right] &\equiv \lambda^\alpha \left[ R^{i\mu}_\alpha \frac{\delta\bm L}{\delta\phi^i} - \partial_\nu \left(R^{i(\mu\nu)}_\alpha \frac{\delta\bm L}{\delta\phi^i}\right)+\cdots\right](\D^{n-1}x)_\mu. \label{S weakly}
\end{align}
Details of the derivations can be found \textit{e.g.} in \cite{Barnich:2001jy,Barnich:2018gdh}. We are left with
\begin{equation}
 \lambda^\alpha N_\alpha \left[\frac{\delta\bm L}{\delta\phi^i}\right] = \D(\bm J_\lambda - \bm S_\lambda). \label{eq temp 2}
\end{equation}
This equation can tell us even more. Indeed, let us take a Euler-Lagrange derivative with respect to the gauge parameters $\lambda^\alpha$, which are arbitrary functions on $\mathscr M$. The right-hand side is a total derivative on which any Euler-Lagrange derivative vanishes identically. We obtain as conclusion that
\begin{equation}
\boxed{ N_\alpha \left[\frac{\delta\bm L}{\delta\phi^i}\right] = 0, } \label{Noether identities}
\end{equation}
which constitute the sets of \textit{Noether identities}. Notice that they hold without imposing the equations of motion: these are off-shell differential relations between the Lagrangian equations of motion and the characteristics of the gauge transformation. There is one Noether identity for each gauge parameter and we will give a concrete example of it soon. But before doing that, let us observe that a second current has appeared in the $(n-1)$-form $\bm S_\lambda$. Employing the Noether identities on \eqref{Ri} we see that, just like $\bm J_\lambda$, $\bm S_\lambda$ is conserved on-shell, but it also vanishes on-shell considering \eqref{S weakly}. For that reason, $\bm S_\lambda$ goes under the name of \textit{Noether weakly vanishing current} and is the concern of Noether's second theorem \cite{Noether:1918zz} for gauge theories:

\resu{Noether's second theorem}{
Given a Lagrangian $n$-form $\bm{L}=L\, \D^n x$ describing a gauge theory with parameters $\lambda = (\lambda^\alpha)$ one has 
\[ \delta_\lambda \phi^i\frac{\delta \bm{L}}{\delta \phi^i} = \D \bm{S}_\lambda \left[ \frac{\delta L}{\delta \phi} \right] \]
where $\bm{S}_\xi$ is the Noether weakly vanishing current given by \eqref{S weakly} and such that $\bm S_\lambda = 0$ on-shell.
}
The proof is immediate if one insert the Noether identities \eqref{Noether identities} into \eqref{Ri} \cite{Barnich:2018gdh}. It formalizes the idea that each gauge symmetry of a Lagrangian theory gives rise to an identity among its equations of motion. In other words, it relies directly on the existence of a Noether identity for each gauge transformation which is due to the degeneracy induced by gauge invariance in the Lagrangian system.

Let us show how this Noether's second theorem works for Einstein's gravity. Let $\delta_\xi$ be generating some diffeomorphism. The left-hand side of Noether's second theorem is
\begin{align}
\frac{\delta {\bm L}}{\delta g_{\mu\nu}} \delta_\xi g_{\mu\nu} &= \frac{1}{16\pi G} \: \D^n x \: \sqrt{-g} \: \Big[ \frac{1}{\sqrt{-g}} \frac{\delta}{\delta g_{\mu\nu}}\left(\sqrt{-g} (R-2\Lambda)\right) \Big] \delta_\xi g_{\mu\nu} \\
 &= -\frac{1}{16\pi G}\:  \D^n x \: \sqrt{-g} \: (G^{\mu\nu}+\Lambda g^{\mu\nu}) \mathcal{L}_\xi g_{\mu\nu} \\
 &= -\frac{1}{8\pi G} \: \D^n x \: \sqrt{-g} \: (G^{\mu\nu}+\Lambda g^{\mu\nu}) \nabla_\mu \xi_\nu \\
 &= -\frac{1}{8\pi G} \: \D^n x \: \sqrt{-g} \: \nabla_\mu \Big[ (G^{\mu\nu}+\Lambda g^{\mu\nu}) \xi_\nu \Big] + \frac{1}{8\pi G} \: \D^n x \: \sqrt{-g} \: \nabla_\mu G^{\mu\nu} \xi_\nu  \\
 &= \D^n x \: \partial_\mu \Big[ -\frac{1}{8\pi G} \: \sqrt{-g} \: (G^{\mu\nu}+\Lambda g^{\mu\nu}) \xi_\nu \Big] + \frac{1}{8\pi G} \: \D^n x \: \sqrt{-g} \: \nabla_\mu G^{\mu\nu} \xi_\nu
\end{align}
The Noether identities associated with the general covariance are the \textit{contracted Bianchi identities} $\nabla_\mu G^{\mu\nu} =0$ for the Einstein tensor, while the weakly vanishing current is given by
\begin{equation}
\boxed{ \bm{S}_\xi = -\frac{1}{8\pi G} \: (\D^{n-1} x)_\mu \: \sqrt{-g} \: (G^{\mu\nu}+\Lambda g^{\mu\nu})\: \xi_\nu . } 
\label{eq:Noether2ndThmEinstein}
\end{equation}
A similar computation for the Einstein-Maxwell theory can be found \textit{e.g.} in \cite{Compere:2018aar}.

\subsubsection{Fundamental theorem of covariant phase space formalism}
We have enough material and knowledge about the covariant phase space formalism to talk about surface charges. Let us again consider the Lagrangian $\bm L$ describing a field theory which is required to be generally covariant. The diffeomorphisms $\xi$ are thus among the gauge symmetries of the theory and general covariance requires \cite{Wald:1993nt}
\begin{equation}
\delta_\xi \bm L = \mathcal L_\xi\bm L = \D (\iota_\xi\bm L) \Rightarrow \bm B_\xi = \iota_\xi \bm L .
\end{equation}
Furthermore, the variation of the Lagrangian is given by the contraction of \eqref{variation lagrangian theory} on a transformation by diffeomorphism, \textit{i.e.}
\begin{equation}
\delta_\xi\bm L = \frac{\delta\bm L}{\delta \phi^i} \delta_\xi\phi^i + \D(i_{\delta_\xi} \bm \Theta[\phi;\delta\phi]) = \frac{\delta\bm L}{\delta \phi^i} \delta_\xi\phi^i + \D\bm\Theta[\phi;\delta_\xi\phi].
\end{equation}
By virtue of Noether's second theorem, we have
\begin{equation}
\D(\iota_\xi \bm L) = \D\bm S_\xi \left[\frac{\delta\bm L}{\delta\phi}\right] + \D\bm\Theta[\phi;\delta_\xi\phi] \Rightarrow \D(\bm J_\xi - \bm S_\xi) = 0 \label{dJ dS}
\end{equation}
where
\begin{equation}
\bm J_\xi = \iota_\xi \bm L - \bm\Theta[\phi;\delta_\xi\phi] = \bm B_\xi[\phi] - i_{\delta_\xi} \bm \Theta[\phi;\delta\phi] \label{JNoetherWald}
\end{equation}
is the Noether current \cite{Lee:1990nz} (this is a particular case of \eqref{Noether for Q}). According to the algebraic Poincaré lemma telling us that, in particular, $\texttt{H}^1_{\mathscr F} = 0$, we can integrate \eqref{dJ dS} as 
\begin{equation}
\bm J_\xi = \bm S_\xi + \D\bm Q_\xi \label{JSdQ}
\end{equation}
with some $(n-2,0)$-form $\bm Q_\xi$ on the jet bundle. To proceed further, let us assume that there exists an \textit{homotopy operator} $I_\xi^p : \Omega^{p,q}\to\Omega^{p-1,q}$ that formally integrates $\bm\alpha_\xi = \D\bm\beta_\xi$ to $\bm\beta_\xi = I_\xi^p\bm\alpha_\xi$, for $\bm\alpha_\xi\in\Omega^{p,q}$ and $\bm\beta_\xi\in\Omega^{p-1,q}$ depending differentially on $\xi$. This operator has to satisfy $I^{p+1}_\xi \D +\D I^p_\xi = \text{id}$ in order to have the desired property
\begin{equation}
\D\bm\beta_\xi = \D I^p_\xi\bm\alpha_\xi = \bm\alpha_\xi - I^{p+1}_\xi \D\bm\alpha_\xi = \bm\alpha_\xi.
\end{equation}
One can show that \cite{Barnich:2007bf}
\begin{equation}
\forall\bm\omega_\xi\in\Omega^p(\mathscr M) : I^p_\xi \bm\omega_\xi = \sum_{(\mu)} \frac{|\mu|+1}{n-p+|\mu|+1}\partial_{(\mu)} \left( \xi^\alpha \frac{\partial}{\partial \partial_\nu \partial_{(\mu)}\xi^\alpha} \frac{\partial}{\partial\D x^\nu}\bm\omega_\xi\right). \label{homotopy operator wald}
\end{equation}
Using this new tool, we can write $\bm Q_\xi = I_\xi^{n-1}(\bm J_\xi - \bm S_\xi)$. Since only terms depending on at least one derivative of $\xi^\alpha$ matter, and neither $\bm{S}_\xi $ nor $\iota_\xi \bm{L}$ contain derivatives of $\xi^\mu$, we have $I_\xi^{n-1}\bm{S}_\xi =I_\xi^{n-1} \iota_\xi \bm{L}=0$ so 
\begin{equation}
\boxed{
\bm{Q}_\xi [\phi] = -I_\xi^{n-1} \bm\Theta [\phi;\delta_\xi\phi].
}
\label{eq:NoetherWaldCharge}
\end{equation}
We call this $(n-2)$-form the \textit{Noether-Wald surface charge} \cite{Wald:1993nt,Iyer:1994ys}. We are now ready to state and prove the following fundamental theorem \cite{Lee:1990nz} that finally defines the $(n-2)$-forms leading to surface charges!

\resu{Fundamental theorem of the covariant phase space formalism}{
Contracting the presymplectic form \eqref{omega grassmann odd} with the transformation $\delta_{\xi} \phi^i$, there exists a $(n-2,1)$-form $\bm{k}_\xi [\delta\phi,\phi]$ that satisfies the identity
\begin{equation}
\boxed{ \bm\omega[\phi; \delta_\xi\phi,\delta\phi] = \D\bm{k}_\xi [\phi;\delta \phi] }\label{FUNDAMENTAL THEOREM}
\end{equation}
when $\phi^i$ solves the equations of motion and $\delta\phi^i$ solves the linearized equations of motion around the solution $\phi^i$. The $(n-2)$-form $\bm{k}_\xi [\delta\phi,\phi]$ is unique, up to total derivatives that do not affect the equality above, and is given in terms of the Noether-Wald surface charge \eqref{eq:NoetherWaldCharge} and the presymplectic potential by the following relation:
\begin{equation}
\boxed{
\bm{k}_\xi [\phi;\delta\phi] = \delta \bm{Q}_\xi [\phi] - \bm{Q}_{\delta \xi} [\phi] - \iota_\xi \bm\Theta [\phi;\delta\phi] + \D(\cdot). 
} \label{IYERWALD}
\end{equation}
}
The quantity $\bm k_\xi [\phi;\delta\phi]$ defines the \textit{Iyer-Wald codimension 2 form} \cite{Iyer:1994ys} associated with the diffeomorphism $\xi$. By hypothesis, the fields $\phi^i$ belong to the solution space $\mathcal S$ and the variations $\delta\phi^i$ are tangent to $\mathcal S$. We have argued in sections \ref{sec:Gauge fixing conditions} and \ref{sec:Theory solution space} that any gauge fixing as well as the imposition of boundary conditions may bring field-dependence in the gauge parameters ($\xi$ here). $\delta\xi$ represents precisely the variation $\delta\xi[\phi^i] \equiv \xi[\delta\phi^i]$ where the variational operator $\delta$ hits the fields implictly present in $\xi$ preserving the gauge fixing conditions \eqref{generic gauge fixing} and the boundary conditions defining $\mathcal S$. Hence $\bm Q_{\delta\xi}$ \cite{Barnich:2004uw} represents the Noether-Wald charge computed from $\delta\xi$ instead of $\xi$. Note also that the codimension 2 form $\bm k_\xi[\phi;\delta\phi]$ is fundamentally defined by the relation \eqref{FUNDAMENTAL THEOREM} which is blind to the addition of any exact $(n-3)$-form. This ambiguity does not play any role, as we will see, since the actual surface charge will be obtained by integration on a closed codimension 2 surface to which any total derivative term does not contribute.

Now let us explain how to prove this fantastic theorem \cite{Wald:1993nt,Barnich:2001jy,Wald:1999wa} in the spirit of \cite{Compere:2018aar}. The demonstration only invokes the intrinsic properties of the jet bundle $\mathscr F$ and relies on Noether's second theorem. The latter provides the weakly vanishing current $\bm S_\xi = \bm J_\xi - \D \bm Q_\xi$ \eqref{JSdQ} on which we can compute an arbitrary variation.
\begin{align}
\delta\bm S_\xi \left[\frac{\delta\bm L}{\delta\phi}\right] &= \delta \bm J_\xi [\phi;\delta\phi] - \delta \D \bm Q_\xi [\phi;\delta\phi] \\
&= \delta \iota_\xi \bm L - \delta \bm \Theta[\phi;\delta_\xi\phi] +\D \delta  \bm Q_\xi [\phi] \\
&= -\iota_\xi \delta\bm L +\iota_{\delta\xi}\bm L - \delta \bm \Theta[\phi;\delta_\xi\phi] +\D \delta  \bm Q_\xi [\phi] \\
&= -\iota_\xi \left(\frac{\delta\bm L}{\delta\phi}\delta\phi - \D\bm\Theta[\phi;\delta\phi]\right) +\iota_{\delta\xi}\bm L - \delta \bm \Theta[\phi;\delta_\xi\phi] +\D \delta  \bm Q_\xi [\phi] \\
&= \mathcal L_\xi \bm\Theta[\phi;\delta\phi] - \delta\bm\Theta[\phi;\delta_\xi\phi] + \D\left(\delta\bm Q_\xi[\phi]-\iota_\xi\bm\Theta[\phi;\delta\phi]\right) +\iota_{\delta\xi}\bm L. \label{eqdem}
\end{align}
The second and third equalities use $\{\D,\delta\}=0$, the fourth equality holds because of \eqref{variation lagrangian theory} and the last equality imposes the equations of motion $\frac{\delta\bm L}{\delta\phi}=0$ and the Cartan's magic formula on $\bm\Theta[\phi;\delta \phi]$. Note that as soon as the theory is assumed to be generally covariant, we have $\delta_\xi \phi^i = \mathcal L_\xi \phi^i$ by definition, hence $\delta_\xi \bm F[\phi] = \mathcal L_\xi \bm F[\phi]$ for any $(k,0)$-form $\bm F[\phi]$ on $\mathscr J$. Using \eqref{commu on jet space} or $[\delta,\delta_\xi] = 0$, we get $\delta_\xi\delta\phi^i = \delta(\mathcal L_\xi \phi^i) = \mathcal L_\xi \delta\phi^i + \mathcal L_{\delta\xi}\phi^i$, where the second term is only necessary for field-dependent diffeomorphism parameters. We thus have $\delta_\xi \bm\Theta[\phi;\delta\phi] = \mathcal L_\xi \bm\Theta[\phi;\delta\phi] + \bm\Theta[\phi;\mathcal L_{\delta\xi}\phi]$ for any $(k,1)$-form $\bm \Theta[\phi;\delta\phi]$ on $\mathscr J$. The generalization to arbitrary $(k,p)$-forms is straightforward. Taking benefit from this new piece of information, we massage \eqref{eqdem} to get
\begin{align}
\delta\bm S_\xi &= \left(\delta_\xi \bm\Theta[\phi;\delta\phi] - \delta \bm \Theta[\phi;\delta_\xi\phi]\right) + \D \left( \delta  \bm Q_\xi [\phi] - \iota_\xi\bm\Theta[\phi;\delta\phi]\right) + \left( \iota_{\delta\xi}\bm L - \bm\Theta[\phi;\mathcal L_{\delta\xi}\phi] \right) \\
&= \bm\omega[\phi;\delta\phi,\delta_\xi\phi] + \D \left( \delta  \bm Q_\xi [\phi] - \iota_\xi\bm\Theta[\phi;\delta\phi]\right) + \bm J_{\delta\xi}[\phi] \\
&= \D \left( \delta  \bm Q_\xi [\phi] - \bm Q_{\delta\xi}[\phi] - \iota_\xi\bm\Theta[\phi;\delta\phi]\right) - \bm\omega[\phi;\delta_\xi\phi,\delta\phi] + \bm S_{\delta\xi},
\end{align}
successively because of \eqref{omega grassmann even} together with \eqref{JNoetherWald} and \eqref{JSdQ} written for $\delta\xi$ instead of $\xi$. In conclusion, the on-shell variation of the weakly vanishing current gives the condition
\begin{equation}
\boxed{\delta \bm S_\xi - \bm S_{\delta\xi} = \D \bm k_\xi [\phi;\delta\phi] - \bm\omega[\phi;\delta_\xi\phi,\delta\phi],}
\end{equation}
where $\bm k_\xi [\phi;\delta\phi]$ is defined as \eqref{IYERWALD}. Assuming that the fields $\phi$ are on-shell and their variations $\delta\phi$ obey the linearized equations of motion, we have that the weakly vanishing current and its on-shell variation vanish, which concludes the proof of the theorem \eqref{FUNDAMENTAL THEOREM}.

\subsubsection{Some residual ambiguities}
\label{sec:Some residual ambiguities}
Translating the fundamental theorem in the usual Hamiltonian language, \eqref{FUNDAMENTAL THEOREM} allows to provide (up to irrelevant exact forms) the infinitesimal surface charge $\bm{k}_\xi$ from the presymplectic form. But is the definition of the presymplectic current unambiguous? 
\begin{itemize}[label=$\rhd$]
\item First, we notice that the presymplectic potential $\bm\Theta$ is defined ambiguously from the action principle \cite{Lee:1990nz}. Indeed, if we add a boundary term $\D\bm{A}$ to the Lagrangian $\bm{L} \rightarrow \bm{L} + \D\bm{A}$ we do not modify the action principle but we will get $\bm\Theta \rightarrow \bm\Theta -\delta \bm{A}$ because $\{\delta,\D\} = 0$ in our conventions. However, since $\bm\omega = \delta\bm\Theta$, this transformation has no effect on the presymplectic current because $\delta^2 = 0$ by design. As a result, the presence of $\bm A$ does not affect the codimension 2 form provided by the fundamental theorem \eqref{FUNDAMENTAL THEOREM}.
\item Second, $\bm\Theta$ is defined from a prescription involving some inverse Leibniz rules and which gives the canonical definition of $\bm\Theta$. The derivation goes through by modifying $\bm\Theta \rightarrow \bm\Theta + \D\bm{Y}$ and therefore $\bm\omega \rightarrow \bm\omega - \D\bm\omega_b$ where $\bm\omega_b\equiv \delta\bm Y$. $\bm Y$ is a $(n-2,1)$-form on $\mathscr F$ coined as the \textit{Iyer-Wald ambiguity} \cite{Iyer:1994ys}. It reflects our ignorance on how to select the boundary terms in the presymplectic current. Conversely to $\bm A$, the Iyer-Wald ambiguity has a non-trivial impact on the charges. Indeed,
\begin{equation}
\bm\omega\to\bm\omega - \D\delta\bm Y \Rightarrow \bm k_\xi \to \bm k_\xi + \delta\bm Y[\phi;\delta_\xi\phi] - \bm Y[\phi,\delta_{\delta\xi}\phi] - \iota_\xi \D \bm Y[\phi,\delta\phi].
\end{equation}
It is fortunately irrelevant for charges associated with exact symmetries of the fields (Killing symmetries in the case of Einstein's theory) because $i_{\delta_\xi} \bm\omega_b$ $=$ $\bm\omega_b[\phi;\delta_\xi\phi,\delta\phi]$ $=$ $0$ when $\delta_\xi\phi = 0$ by linearity, hence $\bm k_\xi \to \bm k_\xi$.
\end{itemize}
The presence of the Iyer-Wald ambiguity could sounds as a bad feature of the covariant phase space formalism, but it allows to have a better control on the asymptotic structure of the theory as well as the formulation of the variation principle. Since the pull-back of $\bm\Theta$ to the boundary controls the on-shell variational principle (see \textit{e.g.} \eqref{symplectic flux general}), one can use the freedom on $\bm Y$ to adjust the value of the symplectic flux through the boundary for particular cases of boundary conditions \cite{Harlow:2019yfa}, or renormalize the on-shell action \cite{Papadimitriou:2005ii,Compere:2008us}. Two concrete examples of such procedures will be given later in this manuscript. We can also mention that in the edge mode program \cite{Freidel:2015gpa,Hopfmuller:2016scf,Donnelly:2016auv,Geiller:2017xad,Horowitz:2019dym}, the formulation of the variational principle on entangled wedges in the bulk of spacetime as well as the computation of corner charges at the apex of the wedge are obviously sensitive to $\bm Y$ and its determination is central in these discussions.

\subsection{Iyer-Wald surface charges}

This section finally provides the definition of the surface charges in the sense of Iyer and Wald \cite{Iyer:1994ys}. Let us recap the key steps. We start from the action integral \eqref{action general} defined from the Lagrangian $\bm L$ describing a generally-covariant theory. The variational principle gives a set of equations of motion driven by the Euler-Lagrange derivatives \eqref{EulerLagrange} and a boundary term $\bm\Theta[\phi;\delta\phi]$. The latter is chosen to be the potential for the presymplectic current $\bm\omega$ \eqref{omega grassmann odd} in the covariant phase space, leading itself to the definition \eqref{presymplectic form} of the presymplectic form, evaluated on a codimension 1 hypersurface $\Sigma$. In Hamiltonian mechanics, a geometrical way to build canonical conserved charges is to extract them from a contraction of the presymplectic form, see \eqref{def hamilton}: this is precisely the definition we are about to consider here in the context of covariant field theories. For diffeomorphisms, the local value of the contracted presymplectic current, by virtue of the fundamental theorem \eqref{FUNDAMENTAL THEOREM}, defines the codimension 2 form \eqref{IYERWALD} which will lead to local surface charges.

\subsubsection{Definition of the charges}
Inspired by the definition \eqref{def hamilton} of the Hamiltonian charges in classical mechanics, one is tempted to propose the following definition of the gravitational charges \cite{Wald:1993nt}
\begin{equation}
\boxed{
\ndelta H_\xi [\phi;\delta\phi] \equiv \mathcal W[\phi;\delta_\xi\phi,\delta\phi] = \int_\Sigma \bm\omega [\phi;\delta_\xi\phi,\delta\phi]
} \label{charges a la Hamilton}
\end{equation}
resulting from the contraction of the presymplectic form with a gauge transformation $\delta_\xi\phi$. We select $\Sigma$ such that the codimension 2 boundary of $\partial\Sigma \equiv S$ belongs to the boundary $\mathscr B$. Using the fundamental relation \eqref{FUNDAMENTAL THEOREM} holding on-shell and the Stokes theorem, the right-hand side of \eqref{charges a la Hamilton} can be expressed as the surface term \cite{Barnich:2001jy,Barnich:2003xg}
\begin{equation}
\boxed{
\ndelta H_\xi [\phi;\delta\phi] = \oint_S \bm{k}_\xi [\phi;\delta\phi].
} \label{ndelta H xi}
\end{equation}
The existence of $\bm k_\xi$ thus provides a tentative definition for the $(n-2)$-forms canonically conjugated with gauge transformation, in analogy with the Generalized Noether theorem. Therefore, the quest may end here: we have reviewed how to construct the surface charge \eqref{ndelta H xi} as the result of the integration of the codimension 2 form $\bm {k}_\xi$ given by \eqref{IYERWALD} on a closed surface $S$ of codimension 2 (\textit{e.g.} a sphere at time and radius fixed).

Rigorously speaking, the quantity $\ndelta H_\xi$ is a $(0,1)$-form on the jet bundle $\mathscr F$, \eqref{ndelta H xi} and gives the \textit{local variation of the surface charge} between the two solutions $\phi^i$ and $\phi^i+\delta\phi^i$, where $\phi^i\in\mathcal S$ (\textit{i.e.} is a solution of the equations of motion and satisfies the boundary conditions) and $\delta\phi^i$ is tangent to $\mathcal S$ (\textit{i.e.} is a solution of the linearized equations of motion around $\phi^i$ and compatible with the boundary conditions). The surface charge associated with the field configuration $\phi\in\mathcal S$ is obtained by integrating \eqref{ndelta H xi} on a path $\gamma$ in the phase space that links a reference solution $\bar\phi$ to the target solution $\phi$,
\begin{equation}
\mathcal Q_\xi [\phi] = \int_\gamma \oint_S \bm k_\xi [\phi;\delta\phi] + N_\xi [\bar\phi]. \label{integrated surface charge}
\end{equation}
$N_\xi [\bar\phi]$ is the surface charge for the reference solution which can be adjusted as an off-set for measuring the numerical values of the charges. The notation $\ndelta H_\xi$ in \eqref{ndelta H xi} emphazises that the result of the integration on $S$ of the $(n-2,1)$-form $\bm k_\xi$ is not necessarily an exact $(0,1)$-form, \textit{i.e.} in general, $\ndelta H_\xi$ is not the ``variation of something,'' that would have been written $\delta H_\xi$. If it is the case, the infinitesimal charge is said to be \textit{integrable} and the integration on $\gamma$ written in \eqref{integrated surface charge} is path-independent. Otherwise the charge is declared \textit{non-integrable} and the integral on $\gamma$ may yield different results if one trade a path from $\bar\phi$ to $\phi$ for another. Given these definitions, let us now spend some time to discuss the properties of the surface charge \eqref{integrated surface charge}.

\subsubsection{Conservation of the charges}
\label{sec:Conservation of the charges theo}
A natural question that arises at this point is: are the charges \eqref{integrated surface charge} conserved? Answering to this question with a bit of care is crucial to understand the radiative phase spaces in General Relativity (which are the main concern of this thesis) and allows also to address the question of integrability in the covariant phase space formalism.

Let $S_1,S_2$ be two codimension 2 sections of $\mathscr B$ delimiting a non-trivial codimension 1 portion $\mathscr C\subset \mathscr B$. For example, if $\mathscr B = \mathscr I^+$ in 4$d$ asymptotically flat gravity, $S_1$ and $S_2$ can be taken as spheres for two values $u_1<u_2$ of the retarded time defining the interval $\mathscr C=[u_1,u_2]\times S^2\subset\mathscr B$. From the fundamental theorem, we learn that the presymplectic current controls the non-conservation of $\ndelta H_\xi$ on-shell under continuous deformations of $S$, because of the \textit{local flux-balance law} \eqref{FUNDAMENTAL THEOREM}. Indeed, given two codimension 2 surfaces $S_1$ and $S_2$ enclosing the codimension 1 hypersurface $\mathscr C\subset\mathscr B$, we have \cite{Wald:1999wa,Barnich:2001jy}
\begin{equation}
\left. \ndelta H_\xi \right|_{S_2} - \left. \ndelta H_\xi \right|_{S_1} = \oint_{S_2} \bm{k}_\xi - \oint_{S_1} \bm{k}_\xi =  \int_\mathscr{C} \D\bm{k}_\xi = \int_\mathscr{C} i_{\delta_\xi\phi} \bm\omega. \label{conservation criterion}
\end{equation}
Therefore the conservation criterion for the charge associated with the diffeomorphism $\xi$ is that $i_{\delta_\xi\phi}\bm\omega$ vanishes on $\mathscr C$. In that case, the numerical value of $\ndelta H_\xi$ will not be affected by jumping from a surface $S_1$ to another surface $S_2$. This weak requirement can be obeyed by some $\xi$ but not the others. For instant, considering some exact (Killing) symmetry $\xi$, we have $\delta_\xi g_{\mu\nu} = \Lie_\xi g_{\mu\nu} = 0$ so $\bm\omega[g;\delta_\xi g, \delta g] = 0$. Therefore, any Killing symmetry is associated with a conserved surface charge everywhere in the bulk of spacetime \cite{1963PhRv..129.1873K}, because the presymplectic form vanishes identically on $\mathscr M$ for such diffeomorphisms, hence the surfaces $S_1$ and $S_2$ can be walked around without affecting the conservation criterion. For a particular class of asymptotic symmetries such that the Killing equation $\Lie_\xi g_{\mu\nu} = 0$ is only verified asymptotically, we have also $\bm\omega[\delta g, \delta_\xi g , g] \rightarrow 0$ in asymptotic regions. As a consequence, the charges associated with $\xi$ will be conserved at infinity. However in many cases, asking for conservation on the whole boundary of spacetime is too stringent and the asymptotic symmetries are not generalically in the kernel of the presymplectic current. Consistent boundary conditions at least requires conservation at spatial infinity, far from sources and radiation (see \textit{e.g.} \cite{Arnowitt:1959ah,Regge:1974zd}).

\subsubsection{Integrability of the charges}
\label{sec:Integrability of the charges theo}

Another important aspect in the definition \eqref{integrated surface charge} is of course the integrability of the infinitesimal surface charge on the phase space. The latter is guaranteed as soon as $\bm k_\xi$ is an exact $1$-form on the jet space. In that case $\ndelta H_\xi = \delta H_\xi$ and $\mathcal Q_\xi[\phi] = H_\xi[\phi]$ up to the contribution of the reference solution. We are brought back to the situation \eqref{def hamilton} in classical mechanics where the Hamiltonian function associated with a Hamiltonian vector field is integrable by definition. In the present context, the necessary condition allowing for the existence of such an Hamiltonian generator $H_\xi$ associated with $\xi$ is the following \textit{integrability condition} \cite{Barnich:2007bf}
\begin{equation}
\delta\oint_S \bm k_\xi [\phi;\delta\phi] \stackrel{!}{=} 0.
\label{eq:Integrability}
\end{equation}
It is also a sufficient condition if the space of fields does not have any topological obstruction, which is one of our hypotheses from the beginning. The form of the Iyer-Wald codimension 2 form $\bm k_\xi[\phi,\delta\phi] = \delta \bm Q_\xi [\phi] - \bm Q_{\delta\xi}[\phi] - i_\xi \bm\Theta[\phi;\delta\phi]$ provides a good basis to discuss, at least qualitatively, the sources of non-integrability. Let us proceed term by term.

\paragraph{Field-dependence} The Noether-Wald term $\delta \bm Q_\xi [\phi]$ being natively integrable, the first potential source of non-integrability is hidden in the second term $\bm Q_{\delta\xi}[\phi]$ which is the correction to the Noether-Wald charge brought by the field-dependence of the diffeomorphism parameters. A rightful question to be addressed is the possibility of performing a field-dependent redefinition of the parameters in order to render the charges integrable. Indeed, there are several examples in the literature (see \textit{e.g.} \cite{Adami:2020ugu,Compere:2017knf,Ruzziconi:2020wrb,Alessio:2020ioh,Grumiller:2019fmp}) where such a redistribution between the parameters determining the diffeomorphism provides integrable charges. This way to cure non-integrability was beautifully presented in \cite{Barnich:2007bf} as the integration of a Pfaff system on the solution space $\mathcal S$. We recall that the solutions in $\mathcal S$ are parametrized as $\phi = \phi(x,p)$, in the notations of section \ref{sec:Solution space and equations of motion}, for any set of coordinates $\{x^\mu\}$ on the base manifold $\mathscr M$. The field-dependent residual diffeomorphisms are parametrized as $\xi = \xi(x,p,s)$ for $s = (s^i)_{i=1}^{n_s}$ defined in section \ref{sec:Residual gauge transformations}. A basis of the Lie algebroid of diffeomorphisms tangent to $\mathcal S$ at some $\phi\in\mathcal S$ is given by the $n_s$ generators $e_i(x, a) = \frac{\partial}{\partial s^i} \xi(x,p,s)$, $i=1,\dots,n_s$. They are associated with $1$-forms on the fields $\ndelta H_{e_i}[\phi(x,p);\delta\phi(x,p)]$. From the perspective of the Frobenius theorem, this set of $n_s$ $1$-forms is integrable if and only if it exist a (field-dependent) invertible $n_s\times n_s$ matrix ${M^i}_j(p)$ (\textit{integrating factors}) redistributing the generators as $f_j = e_i{M^i}_j(p)$ such that
\begin{equation}
\ndelta Q_{e_i}[\phi;\delta\phi] {M^i}_j(p) = \delta H_{f_j}[\phi]. 
\end{equation}
When a set of integrating factors exists, we can trade the generators $(e_i)$ for the refined ones $(f_j)$. It just means that the previous choice of parametrization was not the best to compute the charges, therefore the apparent non-integrability was an artefact of, say, a ``bad'' choice of basis. When there is no such integrating factors, the non-integrability cannot be cured by a mere change of basis and has thus deeper origins. 

\paragraph{Dynamics at the boundary} A second source of non-integrability comes from the third term in the Iyer-Wald charge $- i_\xi \bm\Theta[\phi;\delta\phi]$, involving the presymplectic potential $\bm\Theta[\phi;\delta\phi]$. The contribution of this term only can be extracted directly from \eqref{eq:Integrability} supposing for the moment that the diffeomorphism parameters are field-independent \textit{i.e.} $\delta\xi = 0$. Indeed, a simple calculation shows that \cite{Wald:1999wa}
\begin{equation}
\delta\oint_S \bm k_\xi [\phi;\delta\phi] = \oint_S \delta\iota_\xi\bm\Theta[\phi;\delta\phi] = -\oint_S \iota_\xi\delta\bm\Theta[\phi;\delta\phi] = -\oint_S\iota_\xi\bm\omega[\phi;\delta\phi,\delta\phi]. \label{eq:integrability for IyerWald}
\end{equation}
The value of the presymplectic current on the boundary now appears as an obstruction for the charges being integrable, in addition to generate their non-conservation, see \eqref{conservation criterion}. A sufficient condition for integrability is thus $\bm\omega[\phi;\delta \phi,\delta \phi]|_{\mathscr B} = 0$ when $\delta\xi = 0$. This observation feeds the sensation that non-integrability and non-conservation are intimately related, which is transparent in the covariant phase space formalism. Even though the analysis is more convoluted when $\delta\xi\neq 0$, in many examples (see \cite{Henneaux:1985ey,Compere:2017knf,Adami:2020amw,Henneaux:2018cst,Donnay:2016ejv} to cite only a few references), it was observed that the condition $\bm\omega[\phi;\delta \phi,\delta \phi]|_{\mathscr B} = 0$ (whose physical meaning will be debated in more details in the next section) is sufficient to ensure integrability, up to some modification in the diffeomorphism parameters as discussed earlier. 

\paragraph{Definition of the Hamiltonian generator} When the infinitesimal charges \eqref{ndelta H xi} are non-integrable, we can give an explicit \textit{split} between integrable and non-integrable parts,
\begin{equation}
\ndelta H_\xi[\phi] = \delta H_\xi [\phi] + \Xi_\xi [\phi;\delta\phi] \label{split theory}
\end{equation}
and consider the integrable term $H_\xi[\phi]$ as the Hamiltonian canonically associated with $\xi$. In contrast to the integrable case for which the Hamiltonian function is unique (up to the normalization of the reference solution), the split \eqref{split theory} is ambiguous \cite{Barnich:2011mi}, because we can refine 
\begin{equation}
H_\xi \to H_\xi - \Delta H_\xi, \qquad \Xi_\xi \to \Xi_\xi + \delta \Delta H_\xi \label{split ambiguity N}
\end{equation}
for some $\Delta H_\xi = \Delta H_\xi[\phi]$ without changing the infinitesimal charge. Hence the canonical Hamiltonian $H_\xi$ conjugated to $\xi$ cannot be extracted directly. The fixation of the integrable part of the infinitesimal surface charge \eqref{split theory} requires an additional input that comes from the particular physics of the problem. The prescription, althrough physically motivated, must be compatible with the set of residual gauge transformations as well as the boundary conditions. One can imagine that a suitable choice of integrable part may call for finiteness of the integrated charge on the whole $n-1$ coordinates of $\mathscr B$ (in order to have well-defined integrated flux through $\mathscr B$), or also for conservation for stationary field configurations \cite{Wald:1999wa}. We will explore this kind of choice more concretely for two cases of asymptotics below.

\subsubsection{Physical content of the presymplectic flux}
\label{sec:Physical content of the presymplectic flux}
In order to make ideas more precise and see the physical consequences of a non-vanishing presymplectic current at the boundary $\mathscr B$, let us go back to fundamentals, namely the variational principle \eqref{symplectic flux general} itself that we assume to be well-defined in the sense that it obeys both requirements \ref{well defined dS 1} and \ref{well defined dS 2}. 

\paragraph{Conservative boundary conditions} For a set of conservative boundary conditions as defined in section \ref{sec:Conservative boundary conditions theo}, the presymplectic flux is trivial and can be canceled by a suitable choice of boundary terms ensuring that the action is stationary on-shell. The boundary terms must obey \eqref{theta is exact} which translates into the vanishing of the presymplectic current pulled back to $\mathscr B$,
\begin{equation}
\bm \omega [\phi ; \delta_1 \phi, \delta_2 \phi] \Big|_{\mathscr B} = 0, \label{conservative omega}
\end{equation}
because of the definition \eqref{omega grassmann odd}. This is the local statement for the boundary conditions to be conservative. Such boundary conditions lead to a solution space on which any infinitesimal charge \eqref{ndelta H xi} is conserved. Indeed, because any contraction of \eqref{conservative omega} with a diffeomorphism parameter gives zero, the codimension 2 form defining the charge is closed by virtue of \eqref{FUNDAMENTAL THEOREM} and \eqref{conservation criterion} indicates that $\ndelta H_\xi$ is conserved. This sheds a new light on our terminology for the boundary conditions. Up to some fancy redefinition of the diffeomorphism parameters, the charges can also be made integrable because \eqref{conservative omega}, through \eqref{eq:integrability for IyerWald}, implies the integrability condition \eqref{eq:Integrability}. In conclusion, if one wants strict conservation of the whole set of surface charges at $\mathscr B$ and a stationary on-shell action, one must build the phase space from conservative boundary conditions. As a corollary, dealing with open systems with leaks through $\mathscr B$ which are responsible for some non-conservation of the charges goes hand in hand with considering non-stationary variational principles and non-conservative boundary conditions.

\paragraph{Leaky boundary conditions} In contrast to \eqref{conservative omega}, leaky boundary conditions in the terminology of section \ref{sec:Leaky boundary conditions theo}, \textit{i.e.} allowing for some presymplectic flux through the conformal boundary $\mathscr I$, are defined as
\begin{equation}
\bm \omega [\phi ; \delta_1 \phi, \delta_2 \phi] \Big|_{\mathscr B} \neq 0. \label{leaky omega}
\end{equation}
In these configurations, the charges are no longer expected to be conserved and there is some flux of charges through the boundary $\mathscr B$, as it can be seen from \eqref{conservation criterion}: the difference between the infinitesimal charge evaluated in $S_1$ and the same charge evaluated in $S_2$ is precisely equal to the flux crossing the portion $\mathscr C$ of $\mathscr B$ delimited by $S_1$ and $S_2$. This flux of charge is encoded in the presymplectic current pulled-back to $\mathscr B$ and the non-conservation is controlled by the fundamental relation \eqref{FUNDAMENTAL THEOREM}. In addition, the charges are not expected to be integrable, because \eqref{leaky omega} gives an obstruction to integrability -- see \eqref{eq:integrability for IyerWald} -- and since the statement \eqref{leaky omega} does not involve the diffeomorphism parameters, it does not exist any redefinition of these which would be able to make the charges integrable. This shows that the existence of a non-vanishing symplectic flux through $\mathscr B$ yields some non-equilibrium physics and, therefore, the integral on the covariant phase space depends explicitly on the chosen path: it is thus hardly surprising to be faced with non-integrability in that context. 

For leaky boundary conditions, $\mathscr B$ is no longer a rigid frontier but rather a permissive interface opening onto the environment encircling the physical system. The latter acts as a reservoir collecting the flux of charges going through this boundary, according to \eqref{FUNDAMENTAL THEOREM}. In this point of view, the statement of the fundamental theorem is seen as a set of local \textit{flux-balance laws} on any section of $\mathscr B$. Note crucially that we did not give a sign for the flux, hence the leaks can be for the benefit of the physical system or the environment: we can imagine either \textit{outgoing} or \textit{ingoing leaky boundary conditions}. In the first case, the flux of charges is positive in \eqref{conservation criterion} while in the second it is negative. For example, we will see below that the presence of gravitational waves in the bulk of any asymptotically flat spacetime is responsible for some outgoing flux at future null infinity. If the gravitational system is sourced by another and can be approximated by an asymptotically flat spacetime, one can also imagine that some gravitational waves are free to enter through past null infinity, describing a toy-model of ingoing radiative conditions.

\begin{figure}[ht!]
\vspace{5pt}
\centering
\begin{tikzpicture}
\tikzset{snake it/.style={decorate,decoration={snake,segment length=5,amplitude=1.5pt}}};
\def\xsize{8};
\def\ysize{3};
\def\dsize{3};
\def\eps{0.15};
\coordinate (A) at (0,0);
\coordinate (B) at (\xsize,0);
\coordinate (C) at (\xsize,\ysize);
\coordinate (D) at (0,\ysize);
\coordinate (Ap) at ($(A)+(-\dsize,1.25)$);
\coordinate (Bp) at ($(B)+( \dsize,1.25)$);
\coordinate (Cp) at ($(C)+( \dsize,\dsize)$);
\coordinate (Dp) at ($(D)+(-\dsize,\dsize)$);
\coordinate (TC) at ($(C)!0.5!(D)$);
\fill[pattern=dots, pattern color=black!20] (Ap) -- (Bp) -- (Cp) -- (Dp) -- cycle;
\draw[green!50!black,thick,fill=white] (B) -- (C) -- (D) -- (A);
\foreach \k in {-4,-2,...,4}{
\draw[-Latex,red] ($(TC)+(\k*\eps,0)-(0,1.2)$) -- ($(TC)+(\k*\eps,0)+(0,1.2)$);
\draw[orange] ($(TC)+(\k*\eps,0)$) node[fill,circle,inner sep=2pt] {};
\draw[ultra thick,-Latex,orange] ($(TC)+(\k*\eps,0)-(0,0.5)$) -- ($(TC)+(\k*\eps,0)+(0,0.7)$);
}
\foreach \l in {0.1,0.3,0.7,0.9}{
\coordinate (temp) at ($(D)!\l!(C)$);
\draw[blue] (temp) node[fill,circle,inner sep=2pt] {};
\draw[ultra thick,Latex-Latex,blue] ($(temp)-(0.7,0)$) -- ($(temp)+(0.7,0)$);
}
\draw[] ($(TC)+(4*\eps,0)+(0,0.5)$)node[right]{$\leftarrow$ {\color{orange}Dynamical BDoF  (\textit{sources})}};
\draw[] ($(TC)-(0,1.2)$)node[green!50!black,draw,below,align=center,outer sep=5pt]{\textit{\textbf{Radiative open system}}} ($(TC)+(0,1.2)$)node[red,above,text width=3cm,align=center]{Outgoing radiation};
\draw[] ($(D)!0.1!(C)$)node[above,align=center,text width=3cm,outer sep=3pt]{{\color{blue}Kinematical BDoF}\\$\downarrow$};
\draw[green!50!black] (B)node[anchor=south west]{$\mathscr B$};
\draw[green!50!black] ($(A)!0.5!(B)$)node[above]{$\delta S_{sys} \neq 0$};
\draw[] (Dp)node[black!70,anchor=north west,text width=3cm,outer sep=3pt]{\textit{\textbf{Environment}}\\$\delta S_{env} \neq 0$};
\draw [decorate,decoration={brace,amplitude=10pt,mirror}]
($(A)+(-2,-0.3)$) -- ($(B)+(2,-0.3)$) node[black,midway,below,yshift=-10pt] {$\delta S_{tot}=0$};
\end{tikzpicture}
\caption{Kinematical \emph{vs.} dynamical boundary degrees of freedom (BDoF).}
\label{fig:DOF}
\end{figure}
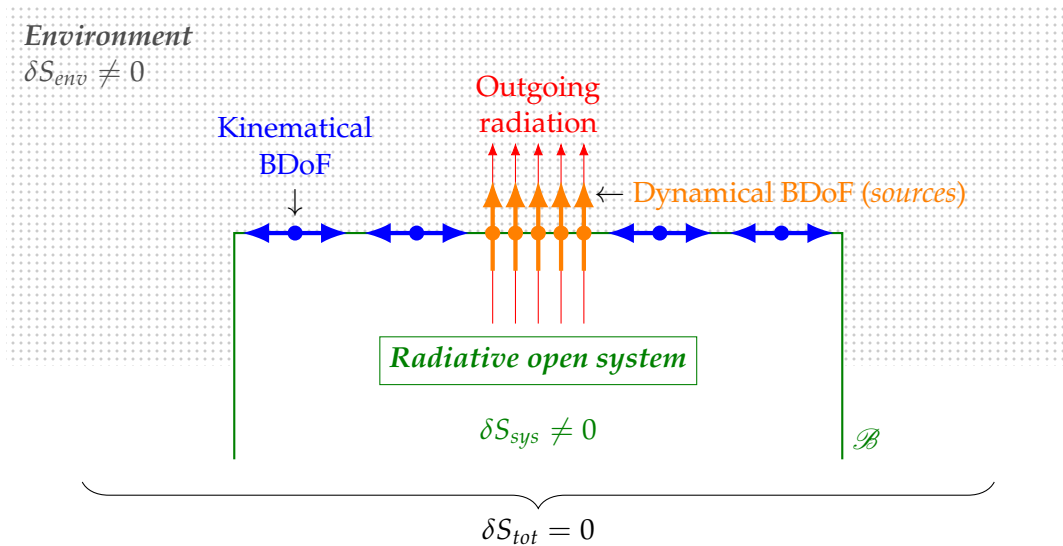

\paragraph{Boundary degrees of freedom and physical flux} We will show below, by means of concrete examples, that it is convenient to interpret the primary causes of the leaks as external sources or ``boundary degrees of freedom.'' From an asymptotic analysis considering only the physical system with action principle defined by $S$ and a set of boundary conditions at $\mathscr B$, the dynamics of these boundary degrees of freedom is unconstrained. However, if one would include the environment into the analysis, then they would acquire some constrained dynamics. The latter would be dictated by the precise nature of the environment under consideration. As a result, the concatenation between both the leaky physical system and its environment would restore a well-defined variational principle. In the statement \eqref{symplectic flux general}, there is merely an apparent violation of this because we are tracing on the states of the environment in the analysis of asymptotics. This is a local statement that reflects our ignorance about the outer region which collects the outcoming flux of charges or produces some ingoing sources for the charges. The most beautiful thing is that the framework we are discussing in this chapter has enough strength to give the possibility to perform the complete analysis of the asymptotics without knowing the precise nature of the environment and the dynamics of the external sources. For instance, in gravitational astronomy, the experimenters and their experimental devices act as observers sitting at the future null infinity of a spacetime where the phenomenon under interest has occurred. In order to provide them with predictions, we just need to get some information about the dynamics at $\mathscr B$, but what lies beyond is out of our scope and our interest. This is the point of view we have adopted in the present thesis.

\paragraph{Radiative and kinematical degrees of freedom} There is however a subtlety to be mentioned here regarding the physical interpretation of the symplectic flux. A non-vanishing symplectic current on $\mathscr B$ does not necessarily mean that physical dynamical degrees of freedom contribute to the genuine flux through the boundary. For example, one can have a non-radiative gravitational system enclosed by $\mathscr B$ with a fluctuating boundary structure producing non-trivial contributions in $\bm\omega[\phi;\delta\phi,\delta\phi]|_\mathscr{B}$. This feature was observed \textit{e.g.} in \cite{Ruzziconi:2020wrb} for 3$d$ gravity where there is no gravitational propagating degree of freedom (the Weyl tensor is zero), but a non-vanishing presymplectic flux at infinity is observed due to fluctuations of the metric field at leading order. Furthermore, the charges can be made integrable by changing the slicing of the phase space. This leads us to a natural distinction between the \textit{dynamical part} of $\bm\omega[\phi;\delta\phi,\delta\phi]|_\mathscr{B}$ which encodes genuine radiation flowing through $\mathscr B$ and the \textit{kinematical part} of $\bm\omega[\phi;\delta\phi,\delta\phi]|_\mathscr{B}$ which is related to intrinsic boundary degrees of freedom that are not sourced by the radiative modes (see figure \ref{fig:DOF}). To express this with a tensorial object, it suffices to compute the Weyl curvature and check that its asymptotic value only involves the dynamical boundary degrees of freedom.

The natural conjecture that can be derived from this discussion is the following: ``\textit{when the dynamical part of the presymplectic flux at $\mathscr B$ vanishes, the surface charges are integrable''} \cite{Adami:2020ugu,Ruzziconi:2020wrb}. Up to our knowledge, there is no counterexample that invalidates this statement and the various phase spaces studied in this thesis are also consistent with this. Since only the dynamical part of the symplectic flux is assumed to vanish, the conservation of the charges is not guaranteed in general. The example of 3$d$ gravity is again very instructive, because the dynamical symplectic flux is genuinely zero. In \cite{Alessio:2020ioh}, the boundary structure contains kinematical degrees of freedom, namely the Weyl rescalings on the circle. These are responsible for a non-vanishing presymplectic flux leading to some breaking of the conservation laws, yet without any gravitational radiation reaching the boundary. This is nicely interpreted in our framework: the variation of the bulk action $S$ under a Weyl rescaling $\sigma$ reads as $\delta_\sigma S = \int_{\mathscr B} \mathcal A \sigma$ on-shell, which gives rise to the Weyl anomaly $\mathcal A$ in three dimensions. The latter generates some kinematical symplectic flux at infinity although gravitational radiation is absent. Consequently, the presymplectic current is sourced by $\mathcal A$ and the charges cannot be conserved when $\mathcal A$ is non-vanishing. 

The kinematical boundary degrees of freedom can be recognized among others because we can get rid of them by using the residual gauge freedom. For instance, in asymptotically flat gravity, the boundary area $\sqrt{q}$ can be fixed for free by using the diffeomorphism freedom among the residual gauge diffeomorphism in the Bondi gauge without ruling out any solution \cite{Compere:2018ylh,Compere:2020lrt}. It is not the same for $C_{AB}$ which enters in the symplectic flux as a dynamical degree of freedom, as we will show in section \ref{sec:Asymptotically locally flat radiative phase spaces}. Apart from a pure-gauge part \eqref{electric part BMS} which can be brought to zero by a suitable BMS$_4$ supertranslation, the two functions forming the rank 2 symmetric traceless tensor $C_{AB}$ cannot be gauged away by a choice of coordinates in the presence of gravitational radiation \cite{Ashtekar:1981bq,Ashtekar:2014zfa,Thorne:1982cv}. The latter, escaping from $\mathscr M$ through future null infinity, sets the open system in a non-equilibrium state, for which the integrability of the charges can only be restored as soon as $N_{AB}=0$, or $\partial_u C_{AB}=0$ \cite{Wald:1999wa,Barnich:2011mi}.

Let us conclude this quite philosophical section by a crucial comment. The fact that a kinematical boundary degree of freedom can be fixed does not mean that it is not related to interesting asymptotic symmetries. This is the case, for instance, for the Weyl rescalings that will be fixed on the boundary throughout this thesis. The Weyl symmetry is always present as soon as the manifold under consideration is conformally compact \cite{Penrose:1964ge} (see \ref{sec:Conformally compact manifolds} for a rigorous definition) because it emerges from the ambiguity on the choice of the conformal factor in the treatment of asymptotics through the conformal compactification process. Generically conjugated with non-trivial charges (see \cite{Alessio:2020ioh,Fiorucci:2020xto,Freidel:2021yqe,Barnich:2019vzx}), it is definitely a non-trivial asymptotic symmetry. Freezing it thanks to a boundary gauge fixing (namely demanding that boundary transformations preserve the area of the celestial sphere) will make these charges disappear but no solution will be rejected from the phase space. Hence the choice to maintain arbitrary Weyl rescalings on the celestial sphere or ruling them out is a matter of how general we want to be. In this perspective, which is the status of the remaining degrees of freedom in $q_{AB}$ after the fixation of the area and why not also fix them? For vacuum solutions, it is true that the boundary condition \eqref{sqrt q round sphere} can always be reached by a gauge transformation starting from any time-dependent area element $\sqrt{q}$, because the Weyl parameter $\omega$ has also an arbitrary time dependence, see \eqref{eq:GaugeConstraints}. In this case, \eqref{EOM qAB time evolution} implies that $\partial_u q_{AB} = 0$ and the remaining allowed transformations on the celestial sphere are smooth diffeomorphisms $Y^A(x^B)$ by virtue of \eqref{most general BMS solution}. We can thus select some $Y^A$ to send the arbitrary $q_{AB}(x^C)$ to $\mathring q_{AB}(x^C)$ and get back to asymptotically Minkowskian conditions. However, we will see in chapter \ref{chapter:Flat} that the coupling with matter can induce transitions in $q_{AB}$ and subsequent interesting gravitational memory effects. While it is always possible to fix the area element to $\sqrt{\mathring q}$ thanks a time-dependent Weyl rescaling on the whole boundary, the super-Lorentz transformations $Y^A(x^B)$ are no longer sufficient to trivialize the boundary metric everywhere since $\partial_u Y^A=0$. It shows that the two boundary degrees of freedom in $q_{AB}$ which survive to the Weyl fixing are dynamical in general and have to be included explicitly in the flux. We close here our theoretical discussion on leaky boundary conditions: more precise considerations will be given during the study of concrete examples of radiative phase spaces.

\subsubsection{Expressions in General Relativity} 
Let us devote this little subsection to particularize the fundamental quantities developed in the covariant phase space formalism for Einstein's theory without matter, whose action integral is \eqref{general Einstein Hilbert action}. Under a linearized variation $\delta g_{\mu\nu}$ around $g_{\mu\nu}$, this action transforms as
\begin{equation}
\delta S_{EH}[g] = \int_{\mathscr M} \D^n x \, \left[ \frac{\sqrt{-g}}{16\pi G} (G_{\mn}+\Lambda g_\mn) \delta g^\mn + \partial_\mu \Theta^\mu_{EH} [g;\delta g] \right] \label{EH variational principle}
\end{equation}
for the presymplectic potential \cite{Iyer:1994ys}
\begin{equation}
\Theta^\mu_{EH} [g;\delta g] = \frac{\sqrt{-g}}{16\pi G}\left(\nabla_\nu (\delta g)^{\mu\nu}-\nabla^\mu {(\delta g)^\nu}_\nu \right). \label{EH potential}
\end{equation}
Note that $(\delta g)^{\mu\nu}$ is obtained by varying the metric $g_{\mu\nu}$ \textit{before} raising the indices with $g^{\mu\nu}$, hence $(\delta g)^{\mu\nu} = -\delta g^{\mu\nu} = -g^{\mu\alpha}g^{\nu\beta}\delta g_{\alpha\beta}$. If the variation is contracted with the action of a diffeomorphism $\delta_\xi g_{\mu\nu} = \mathcal L_{\xi} g_{\mu\nu} = 2\nabla_{(\mu}\xi_{\nu)}$, we get
\begin{equation}
\Theta^\mu_{EH} [g;\delta_\xi g] = \frac{\sqrt{-g}}{8\pi G}\left( \nabla_\nu \nabla^{(\mu}\xi^{\nu)} - \nabla^\mu \nabla_\nu\xi^\nu\right).
\end{equation}
Using Einstein's field equations, we get $\nabla^\mu\nabla_\nu \xi^\nu = \nabla_\nu\nabla^\mu\xi^\nu + {{{R^\nu}_\alpha}^\mu}_\nu \xi^\alpha = \nabla_\nu\nabla^\mu\xi^\nu + \frac{2\Lambda}{n-2}\xi^\mu$, so
\begin{equation}
\Theta^\mu_{EH} [g;\delta_\xi g] = \frac{\sqrt{-g}}{16\pi G} \left[ \nabla_\nu \left(  \nabla^\nu\xi^\mu-\nabla^\mu\xi^\nu\right) - \frac{4\Lambda}{n-2}\xi^\mu \right].
\end{equation}
This presymplectic potential gives us access to the Noether-Wald charge \eqref{eq:NoetherWaldCharge} after the application of the homotopy operator \eqref{homotopy operator wald}. The latter only hits terms with at least one derivative on $\xi^\mu$, so the last term involving the cosmological constant disappears. A little calculation gives \cite{Iyer:1994ys}
\begin{equation}
\bm Q^{EH}_\xi[g] = \frac{\sqrt{-g}}{16\pi G} (\nabla^\mu\xi^\nu - \nabla^\nu\xi^\mu) \, (\D^{n-2}x)_{\mu\nu}. \label{NoetherWald Explicit}
\end{equation}
The contribution \eqref{NoetherWald Explicit} in the Iyer-Wald codimension 2 form \eqref{IYERWALD} is often called the \textit{Komar term}, in reference to the Komar integrals \cite{1963PhRv..129.1873K} that give the mass and angular momentum of simple spacetimes when $\frac{\sqrt{-g}}{8\pi G}\nabla^{[\mu}\xi^{\nu]}$ for an isometry $\xi$ is evaluated on an asymptotic 2-sphere. The last ingredient we need to build the local charge is
\begin{equation}
\iota_\xi \bm\Theta_{EH} = \xi^\mu \frac{\partial}{\partial\D x^\mu} \Theta_{EH}^\nu (\D^{n-1}x)_\nu = (\xi^\mu\Theta_{EH}^\nu-\xi^\nu\Theta_{EH}^\mu)\,(\D^{n-2}x)_{\mu\nu}.
\end{equation}
Denoting $h_{\mu\nu} = \delta g_{\mu\nu}$ and $h\equiv {h^\mu}_\mu$, we obtain \cite{Iyer:1994ys} (see also \cite{Compere:2007az,Compere:2018aar})
\begin{equation}
\bm k_{\xi}^{EH}[g;h] = \frac{\sqrt{-g}}{8\pi G}\left[ \xi^\mu \nabla_\sigma h^{\nu\sigma} - \xi^\mu \nabla^\nu h + \xi_\sigma\nabla^\nu h^{\mu\sigma} + \frac{h}{2}\nabla^\nu\xi^\mu - h^{\rho\nu}\nabla_\rho\xi^\nu\right] \,(\D^{n-2}x)_{\mu\nu} \label{IyerWald explicit in GR}
\end{equation}
after some tensorial algebra. This gives the explicit expression of the Iyer-Wald codimension 2 form in General Relativity. Notice that the terms in $\delta\xi$ appearing in $\delta \bm Q_\xi[g]$ have been canceled by $\bm Q_{\delta\xi}[g]$ as expected. At the pedagogical level, it is very illuminating to take a concrete although pretty simple example. For instance, let us evaluate \eqref{IyerWald explicit in GR} for the time translation symmetry of the Schwarzschild black hole. In spherical coordinates $(t,r,\theta,\phi)$, the region outside the horizon is described by the line element
\begin{equation}
g_{\mu\nu} [M] = -\left(1 - \frac{2M}{r}\right) \D t^2 + \left(1 - \frac{2M}{r}\right)^{-1} \D r^2 + r^2 (\D\theta^2+ \sin^2\theta \D\phi^2).
\end{equation}
The only parameter that labels the metric is the radius $M$, so $h_{\mu\nu} [M;\delta M] = \frac{\partial g_{\mu\nu}}{\partial M} \delta M$ at first order. Picking $\xi = \partial_t$, a little massage of \eqref{IyerWald explicit in GR} shows that $\bm k_\xi[M;\delta M] = \frac{\delta M}{4\pi G} \delta_t^\mu \delta_r^\nu (\D^{n-2}x)_{\mu\nu}$. Integrating on a sphere $S$ of constant $t,r$ gives
\begin{equation}
\delta H_\xi = \oint_S (\D^{n-2}x)_{tr} \frac{\delta M}{4 \pi G} = \int_0^{2\pi} \D\phi \: \int_0^\pi \D\theta \: \sin \theta \frac{\delta M}{4 \pi G} = \frac{\delta M}{G} = \delta \mathcal M \label{M sch}
\end{equation}
where $\mathcal M = M/G$ is the total mass of spacetime. The charge is trivially integrable and, after a simple path integration between the Minkowski metric ($M=0$) and a target metric with given $M> 0$, we get the right result according to which $\mathcal M$ is the total energy of the Schwarzschild black hole. Note that in this simple case, \eqref{M sch} contains only the local expression of the Komar charge \cite{1963PhRv..129.1873K}.

\subsubsection{Link with the Barnich-Brandt formalism}
We conclude the presentation of the formalism for the surface charge by quickly reviewing the alternative formulation provided by Barnich and Brandt in \cite{Barnich:2001jy,Barnich:2003xg} and making the link with the Iyer-Wald definition. 

The great advantage of the Barnich-Brandt definition for the local codimension 2 forms $\bm k_\xi^{BB}[\phi;\delta\phi]$ consists in \textit{not} requiring that the theory under consideration is generally covariant. In contrast with the Iyer-Wald definition, it is applicable for any Lagrangian gauge theory and leads to a particular prescription to fix the boundary ambiguity in the presymplectic form. Moreover, the definition of the presymplectic current relies only on the datum of some equations of motion, regardless of whether the gauge theory admits a Lagrangian formulation or not! Up to this major difference, the Barnich-Brandt method functions in a way similar to Iyer-Wald's procedure and also relies on the link between the symplectic structure on $\mathscr J$ and lower-degree conserved currents. 

We first introduce a more formal procedure for performing integration by parts on expressions depending on the fields $\phi^i$ but not necessarily on the gauge parameters. It involves \textit{Anderson's homotopy operator} $I^p_{\delta\phi}$ \cite{Tulczyjew1980TheER,Anderson0}, which bears some resemblance with the operator $I^p_\xi$ constructed above. Using the Grassmann odd convention for $\delta$, it can be defined as follows :
\begin{align}
- \D I^n_{\delta\phi} + \delta \phi^i \frac{\delta}{\delta\phi^i} &= \delta \quad \text{ when acting on }n\text{-forms ;} \label{eq:Anderson} \\
- \D I^p_{\delta\phi} + I^{p+1}_{\delta\phi} \D &= \delta \quad \text{ when acting on }p\text{-forms }(p<n). \label{eq:Anderson2}
\end{align}
The cases $p=n$ and $p=n-1$ can be worked out explicitly as
\begin{align}
I^n_{\delta\phi} &= \left[ \delta \phi^i \frac{\partial}{\partial \partial_\mu \phi^i} - \delta \phi^i \partial_\nu \frac{\partial}{\partial \partial_\mu \partial_\nu \phi^i} + \partial_\nu  \delta \phi^i \frac{\partial}{\partial \partial_\mu \partial_\nu \phi^i} + \cdots \right] \frac{\partial}{\partial \D x^\mu}, \\
I^{n-1}_{\delta\phi} &= \left[ \frac{1}{2} \delta \phi^i \frac{\partial}{\partial \partial_\mu \phi^i} - \frac{1}{3} \delta \phi^i \partial_\nu \frac{\partial}{\partial \partial_\mu \partial_\nu \phi^i} + \frac{2}{3} \partial_\nu  \delta \phi^i \frac{\partial}{\partial \partial_\mu \partial_\nu \phi^i} + \cdots \right] \frac{\partial}{\partial \D x^\mu}.
\end{align}
Applying \eqref{eq:Anderson} on the Lagrangian $\bm L$, we get 
\begin{equation}
\delta \bm{L} = \delta \phi^i \frac{\delta \bm{L}}{\delta \phi^i} - \D I^n_{\delta\phi} \bm{L}
\label{eq:deltaL2}
\end{equation}
implying that $\bm\Theta = I^n_{\delta \phi} \bm{L}$ thanks to \eqref{variation lagrangian theory}. Using (\ref{eq:Anderson2}), we can apply $I_{\delta\phi}^n$ on both sides of (\ref{eq:deltaL2}) to get :
\begin{align}
&I^n_{\delta\phi} \delta \bm{L} = I^n_{\delta\phi} \left( \delta \phi^i \frac{\delta \bm{L}}{\delta \phi^i} \right) - I_{\delta\phi}^n \D I_{\delta\phi}^n \bm{L} = I^n_{\delta\phi} \left( \delta \phi^i \frac{\delta \bm{L}}{\delta \phi^i} \right) - \delta I_{\delta\phi}^n \bm{L} - \D I^{n-1}_{\delta\phi}I^n_{\delta\phi} \bm{L} \\
&\Rightarrow I^n_{\delta\phi} \delta \bm{L} + \delta I_{\delta\phi}^n \bm{L} = I^n_{\delta\phi} \left( \delta \phi^i \frac{\delta \bm{L}}{\delta \phi^i} \right) - \D I^{n-1}_{\delta\phi}I^n_{\delta\phi} \bm{L}.
\end{align}
Since $[\delta,I^n_{\delta\phi}] = 0$ because $\delta^2=0$, the left-hand side is nothing but $2\delta I^n_{\delta\phi} \bm{L} = 2\bm\omega[\delta\phi,\delta\phi,\phi]$ by definition, \eqref{omega grassmann odd}. We conclude that \cite{Barnich:2007bf}
\begin{equation}
\bm\omega[\phi;\delta\phi,\delta\phi] = \bm{W}[\phi;\delta\phi,\delta\phi] + \D\bm{E}[\phi;\delta\phi,\delta\phi], \label{w and W}
\end{equation}
where we have isolated the \textit{invariant presymplectic current}
\begin{equation}
\bm{W}[\phi;\delta\phi,\delta\phi] = \frac{1}{2} I^n_{\delta\phi} \left( \delta \phi^i \frac{\delta \bm{L}}{\delta \phi^i} \right).
\end{equation}
It differs from Iyer-Wald's presymplectic current by a boundary term that reads as
\begin{equation}
\bm{E}[\phi;\delta\phi,\delta\phi] = \frac{1}{2} I^{n-1}_{\delta\phi}I^n_{\delta\phi} \bm{L}.
\label{eq:AmbiguityE_BarnichBrandt}
\end{equation}
Barnich and Brandt proposed to choose $\bm{W}$ instead of $\bm\omega$ as symplectic form to build conserved surface charges \cite{Barnich:2001jy}. $\bm{W}$ is called ``invariant'' because it is defined in terms of the equations of motion and does not depend upon the boundary terms added to the action. No ambiguity appears in this formulation, which can be thought as a force as well as a weakness. Indeed, we will show later in this thesis, on (at least) two concrete examples, that the Iyer-Wald ambiguity remains nonetheless very useful when renormalization of the presymplectic structure is mandatory. The offer of this freedom allows to bring easily the results of the renormalization scheme defined for the on-shell variational principle at the level of the presymplectic structure and consequently to the local surface charges. 

Let us consider a diffeomorphic transformation $\xi$ existing for a covariant field theory $\bm L[\phi^i]$. $\delta_\xi\phi^i = \Lie_\xi\phi^i$ by hypothesis. Restarting again from Noether's second theorem
\begin{equation}
\D\bm{S}_\xi = \frac{\delta \bm{L}}{\delta \phi^i} \Lie_\xi \phi^i , 
\end{equation}
we can apply the homotopy operator $I^n_{\delta\phi}$ to obtain
\begin{align}
I^n_{\delta\phi} \D\bm{S}_\xi &= I^n_{\delta\phi} \left( \frac{\delta \bm{L}}{\delta \phi^i} \Lie_\xi \phi^i \right) = \delta\bm{S}_\xi + d I^{n-1}_{\delta\phi} \bm{S}_\xi.
\end{align}
After some non-trivial steps, it can be proved that the contraction of $\bm{W}$ on-shell with the transformation $\delta_\xi\phi$ reads as
\begin{equation}
\bm{W}[\phi;\Lie_\xi \phi,\delta\phi] = i_{\Lie_\xi \phi} \bm{W}[\phi;\delta\phi,\delta\phi] = I^n_{\Lie_\xi \phi} \left( \frac{\delta \bm{L}}{\delta \phi^i} \delta \phi^i \right) = - I^n_{\delta\phi} \left( \frac{\delta \bm{L}}{\delta \phi} \Lie_\xi \phi \right), 
\end{equation}
see property 13 of \cite{Barnich:2007bf}. Compiling both pieces of information and recalling that $\delta \bm{S}_\xi$ vanishes as soon as $\phi^i$ and $\delta \phi^i$ are on-shell, we get Barnich-Brandt's version of \eqref{FUNDAMENTAL THEOREM}:
\begin{equation}
\boxed{
\bm{W}[\phi;\delta_\xi \phi, \delta\phi] = \D\bm{k}^{BB}_\xi [\phi;\delta\phi], 
}
\label{Fundamental BB}
\end{equation}
which holds on-shell and involves the \emph{invariant} or \textit{Barnich-Brandt codimension 2 form} \cite{Barnich:2001jy} 
\begin{equation}
\boxed{
\bm{k}^{BB}_\xi [\phi;\delta\phi] = -I^{n-1}_{\delta\phi} \bm{S}_\xi \left[ \frac{\delta \bm L}{\delta \phi},\phi \right].
}
\end{equation}
The surface charges are obtained by integration on a codimension 2 surface and on the phase space, as before. The computation of the Barnich-Brandt charge for General Relativity can be performed thanks to this formula and with the mere knowledge of $\bm{S}_\xi$ reading as (\ref{eq:Noether2ndThmEinstein}). Here we can take a shortcut by gathering the relations \eqref{FUNDAMENTAL THEOREM}, \eqref{Fundamental BB} and \eqref{w and W} to get 
\begin{equation}
\bm{k}^{IW}_\xi [\phi;\delta\phi] = \bm{k}^{BB}_\xi [\phi;\delta\phi] + \bm{E}[\phi;\delta_\xi\phi,\delta\phi] + \D (\cdot),
\end{equation}
again dropping the irrelevant total derivatives. Using the definition of the ambiguity $\bm E$ in terms of $\bm L$ and assuming that the latter is the Einstein-Hilbert Lagrangian, we have
\begin{equation}
\bm{E}[\delta g,\delta g ; g]= \frac{1}{32\pi G} (\delta g)^\mu_{\phantom{\mu}\alpha} \wedge (\delta g)^{\alpha \nu} (\D^{n-2} x)_{\mu\nu}.
\label{eq:AmbiguityForRG}
\end{equation}
Denoting again $\phi = g_{\mu\nu}$ and $\delta\phi = h_{\mu\nu}$, we get
\begin{equation}
\begin{split}
\bm{k}^{\mu\nu}_\xi [g;h] = \frac{\rmg}{8\pi G} \Big[& \xi^\mu \nabla^\sigma h^\nu_\sigma - \xi^\mu \nabla^\nu h + \xi^\sigma \nabla^\nu h^{\mu}_{\sigma} \\
&+ \frac{1}{2} h \nabla^\nu \xi^\mu + \frac{1}{2}h^{\nu}_{\sigma}(\nabla^\mu\xi^\sigma - \nabla^\sigma\xi^\mu) \Big](\D^{n-2}x)_{\mu\nu}.
\end{split}
\label{def:ch2}
\end{equation}
This formula was also obtained by Abbott and Deser by a similar procedure involving integrations by parts \cite{Abbott:1981ff,Deser:2002rt,Deser:2002jk}, without using the formal operators \eqref{eq:Anderson}-\eqref{eq:Anderson2}. To conclude on very good news at practical level, one can check that the boundary term \eqref{eq:AmbiguityForRG} is often zero when contracted with the variation $\delta_\xi g_{\mu\nu}$. This is true for exact (respectively asymptotic) Killing vectors, for which $\delta_\xi g_{\mu\nu} = 0$ everywhere (respectively in the asymptotic region). Moreover, in many coordinate systems (and particularly in the Bondi, Newman-Unti and Starobinsky/Fefferman-Graham gauges used in this thesis), $\bm E[g;\delta_\xi g,\delta g] = 0$ for any residual diffeomorphism $\xi$. This last observation allows us to trade one formalism for the other without worrying so much about the slight difference, irrelevant in the coordinate systems we will work in practice. We close this short digression to return back to concrete considerations regarding asymptotically locally flat radiative phase spaces.

\subsection{Asymptotically locally flat radiative phase spaces}
\label{sec:Asymptotically locally flat radiative phase spaces}

In section \ref{sec:Application to asymptotically locally flat spacetimes at null infinity}, we have discussed the boundary conditions and the solution space $\mathring{\mathcal S}_0$ for Einstein's gravity with asymptotically locally flat behavior at future null infinity, as well as the group of asymptotic symmetries coined as the Generalized BMS$_4$ group. We connected this extended solution space to the historical set of asymptotically Minkowskian solutions $\mathring{\mathcal S}^{\text{Mink}}_0$, whose asymptotic group is the global BMS$_4$ group. Here, we promote these solution spaces into phase spaces thanks to the covariant phase space formalism, compute asymptotic surface charges and discuss their properties. Note that, sometimes, we will reserve the full discussion including super-Lorentz transformations for the more involved investigations summarized in chapter \ref{chapter:Flat}. At this point, besides providing new sources for leaks at infinity and increasing the complexity of the analysis, they are not needed to tackle the core of the discussion, since all of the interesting features of the surface charges (namely non-conservation and non-integrability) are yet observable for solutions in $\mathring{\mathcal S}^{\text{Mink}}_0$.

\subsubsection{Variational principle and renormalization}
\label{sec:Flat variational principle}
\begin{figure}[ht!]
\begin{center}
\begin{tikzpicture}[scale=0.8]
\coordinate (A) at (0,4);
\coordinate (B) at (4,0);
\draw[black] (A) node[anchor=south east]{$\mathscr{I}^+_+$} -- (4,0) node[anchor=north west]{$\mathscr{I}^+_-$};
\draw[black] ($(A)!0.5!(B)$) node[above right] {$\mathscr{I}^+$};
\coordinate (C) at ($(A)!0.25!(B)-(0.5,0.5)$);
\coordinate (D) at ($(A)!0.75!(B)-(0.5,0.5)$);
\coordinate (E) at ($(C) - (2,2)$);
\coordinate (F) at ($(D) - (2,2)$);
\draw[blue] (E) -- (C) -- (D) -- (F);
\node[blue,rotate=-45,above] at ($(C)!0.5!(D)$) {$r=r^+$};
\node[blue,rotate=45,above] at ($(C)!0.5!(E)$) {$u=u^+$};
\node[blue,rotate=45,below] at ($(D)!0.5!(F)$) {$u=u^-$};
\coordinate (A1) at ($(C)!0.5!(D)-(0.5,0.5)$);
\coordinate (A2) at ($(A1)+(-0.3,0.3)$);
\coordinate (A3) at ($(A1)-(-0.3,0.3)$);
\draw[red,-latex,thick] ($(A1)-(1.3,1.3)$)node[below,rotate=-45] {Radiation} -- (A1);
\draw[red,-latex,thick] ($(A2)-(1.3,1.3)$) -- (A2);
\draw[red,-latex,thick] ($(A3)-(1.3,1.3)$) -- (A3);
\end{tikzpicture}
\end{center}\caption{Schematic contour for the variational principle.}\label{FigV}
\end{figure}
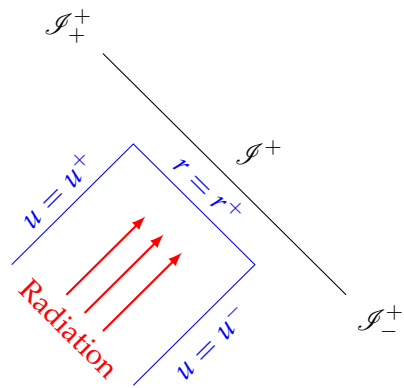

Let us consider $g_{\mu\nu}\in\mathring{\mathcal S}_0$. The Einstein-Hilbert action $S_{EH}[g]$ is given as \eqref{general Einstein Hilbert action} with $\Lambda=0$. In order to prevent some eventual divergences in the variational principle, we evaluate the action integral on a volume $\mathscr V$ of $\mathscr M$ bounded by the three hypersurfaces $\mathscr U^- = \{u = u^-,r\leq r^+\}$, $\mathscr I^+_r = \{u^-\leq u\leq u^+, r = r^+\}$, $\mathscr U^+ = \{u = u^+,r\leq r^+\}$ for some radial cut-off $r^+ < +\infty$ and two time cut-offs $-\infty < u^- < u^+ < +\infty$, as represented in Figure \ref{FigV}. We will consider the limit $r^+ \rightarrow +\infty$, $u^\pm \rightarrow \pm\infty$ afterwards. 

Using \eqref{EH variational principle} and \eqref{EH potential}, the on-shell variational principle is evaluated to be
\begin{equation}
\delta S_{EH} = \int_{\partial\mathscr V} \bm\Theta_{EH}[g,\delta g] = \int_{\mathscr U^-} \text{d}r \, \text d^2 x \, \Theta_{EH}^u + \int_{\mathscr I^+_r} \text{d}u \, \text d^2 x \, \Theta_{EH}^r + \int_{\mathscr U^+} \text{d}r \, \text d^2 x \, \Theta_{EH}^u
\end{equation}
with
\begin{align}
\Theta^u_{EH} &= r \,\Theta^u_{(div)} + \Theta^u_{(fin)} + \Order{2}\label{th1}, \\
\Theta^r_{EH} &= r \,\Theta^r_{(div)} + \Theta^r_{(fin)} + \Order{1}\label{th2}.
\end{align}
in the large $r$ limit. We have $\Theta^u_{(div)} \propto \delta \sqrt{q}$, therefore vanishing because of the boundary condition \eqref{sqrt q fixed in Bondi}. The other components are
\begin{align}
\Theta^u_{(fin)} &= \pref \frac{1}{2} C_{AB} \delta q^{AB}  ,\\
\Theta^r_{(div)} &=-\pref \delta R[q] - \frac{1}{2} \pref  N_{AB} \delta q^{AB} , \\
\Theta^r_{(fin)} &= \pref \, \delta \left[ -\frac{1}{8}C^{AB}N_{AB} + 2 M + D_A \mathcal{U}^A \right] \nonumber \\
& + \bar \Theta_{flux} - \pref D_A (\mathcal{U}_B \delta q^{AB} ), \label{theta fin flat} 
\end{align}
where $\mathcal U^A \equiv -\frac{1}{2}D_B C^{AB}$ and we define with hindsight the important quantity
\begin{equation}
\bar \Theta_{flux}[q_{AB},C_{AB};\delta q^{AB},\delta C^{AB}] \equiv  \pref \Big[ \frac{1}{2}  N_{AB}\delta C^{AB} - \frac{1}{4}  R[q] C_{AB} \delta q^{AB} +  \mathcal{U}_B D_A \delta q^{AB} \Big].\label{thetaflux}
\end{equation}
Some useful identities for the computation of $\Theta^u_{EH}$ and $\Theta^r_{EH}$, involving variations of symmetric traceless tensors as well as a fluctuating boundary metric, can be found in the appendix \ref{Useful relations}. The bare action principle is divergent on-shell because of the pole in $\mathcal O(r)$ in the presymplectic flux when $r\to+\infty$. The latter is due to the fluctuations of the metric on celestial sphere and disappear when we restrict the analysis to $\mathring{\mathcal S}_0^{\text{Mink}}$. Allowing variations of the boundary structure calls for a renormalization procedure of the on-shell variational principle, at least for the radial divergences. At this point, it is useful to remark that a total derivative and a total variation can be isolated in the divergent pieces:
\begin{eqnarray}
\Theta^u_{(fin)} &=&- \p_r Y^{ur}, \label{d1}\\
r\, \Theta^r_{(div)}&=& -\p_u Y^{ru} + \delta (-\sqrt{q} \mathring{R}\; r) = -\p_u Y^{ru} -\p_A Y^{rA}\label{d2}
\end{eqnarray}
where
\begin{equation}
Y^{ur} =-Y^{ru}= -r \frac{1}{2} \pref C_{AB} \delta q^{AB}, \label{Yur}
\end{equation}
and
\begin{equation}
Y^{rA} = r \, \prefwg \, \Theta^A_{2d}(\delta q ; q) \label{YrA}
\end{equation}
is $r$ times the presymplectic potential of the 2-dimensional Einstein-Hilbert action, $\p_A \Theta^A_{2d} =  \delta (\sqrt{q} \mathring{R})$. Since the boundary of a boundary is zero, the corner terms $\propto\, Y^{ur}$ in the variational principle drop. After integration over the sphere, the total derivatives $\propto\, \p_A v^A$ also drop. The radially divergent contribution to the action is therefore the integrable term
\begin{equation}
\frac{1}{4} \, \delta \left[ r \int_{u^-}^{u^+} \text du\, \chi[q] \right], \qquad \chi \equiv \frac{1}{4\pi} \oint_{S_\infty} \sqrt{q}  R[q] 
\end{equation}
where $\chi[q]$ is the topological Euler number of $q_{AB}$. We have $\chi[\mathring q] = 2$ for the unit-round sphere metric, which is assumed to be part of the phase space since $\sqrt{q}=\sqrt{\mathring q}$ on $\mathring{\mathcal S}_0$. Since this condition is diffeomorphic and Weyl invariant, it is consistent with the action of super-Lorentz transformations and supertranslations. Under an infinitesimal smooth variation, a topological number cannot change. The action of boundary diffeomorphisms as well as Weyl transformations preserves $\chi[q]=2$, hence $\delta_Y \chi[q] = 0$ under any smooth super-Lorentz transformation. If we allow singular infinitesimal changes, such as the ones generated by singular super-Lorentz transformations that arise in the snapping of cosmic strings, the boundary topology changes and one would require to add a boundary term in the action to cancel this divergence. Here we can discard this divergent term by imposing the supplementary boundary condition
\begin{equation}
\delta \left[ \frac{1}{4\pi} \oint_{S_\infty} \sqrt{q}  R[q] \right] = 0 \label{BC2}
\end{equation}
which is perfectly compatible with the action of super-Lorentz transformations. 

Let us now evaluate the symplectic flux. In \eqref{theta fin flat}, the total derivative terms will drop after integration on the sphere, hence the radial symplectic flux contains a integrable term which has no influence on the presymplectic current and the charges. We can remove it if we supplement the Einstein-Hilbert action by the following boundary term:
\begin{equation}
S_{ren} = S_{EH} - \prefwg \int_{\mathscr I^+} \text{d}u \, \text{d}^2\Omega \, \Big( 2M- \frac{1}{8}C^{AB}N_{AB}\Big)
\end{equation}
where $\text d^2\Omega \equiv \sqrt{q}\, \text d^2 x = \frac{1}{2}\sqrt{q}\, \epsilon_{AB}\text dx^A \text dx^B$. Notice that the last term, $D_A\mathcal U^A$, has not been included since it gives no contribution after integration on the sphere. We do not provide a covariant formulation of this boundary term, or the boundary terms $Y^{\mu\nu}$, which would require geometrical tools on boundary null surfaces \cite{Parattu:2015gga,Lehner:2016vdi,Wieland:2017zkf,Duval:2014uva} or a prescription from holographic renormalization \cite{Kraus:1999di,Mann:2005yr}. The renormalized variational principle in $\mathscr V$ reads thus as
\begin{equation}
\delta S_{ren} = \int_{\mathscr U^-} \text{d}r \, \text d^2 x \, \Theta_{in}^u + \int_{\mathscr I^+_r} \text{d}u \, \text d^2 x \, \bar\Theta_{flux} + \int_{\mathscr U^+} \text{d}r \, \text d^2 x \, \Theta_{in}^u
\end{equation}
where $\Theta_{in}^u = \mathcal O(r^{-2})$. In the limit $|u|\to +\infty$, the first term drops since we assume that there are no incoming gravitational radiation at $\mathscr I^-$. Ignoring for the moment the corner divergences in $u$ whose discussion is postponed to chapter \ref{chapter:Flat}, the variational principle gives
\begin{equation}
\delta S_{ren} = \int_{\mathscr I^+} \text{d}u \, \text d^2 x \, \bar\Theta_{flux} \label{deltaS}
\end{equation}
in the limit $r\to+\infty$. An asymptotically locally flat spacetime is thus considered in general as an open system with physical flux leaking through the boundary $\mathscr I^+$. This leak is indispensable for allowing gravitational radiation in the asymptotic region. Observing the form of \eqref{thetaflux}, there are two sources for the flux: the asymptotic shear $C_{AB}$ which is related to standard gravitational radiation \cite{Bondi:1962px,Sachs:1962wk,Sachs:1962zza,Ashtekar:2014zsa,Barnich:2010eb} and $q_{AB}$ which can be varied when one includes \textit{cosmic events}, whose action modify the metric tensor at leading order at $\mathscr I^+$ \cite{Strominger:2016wns,Compere:2018ylh}, in the phase space. Notice that even in the simplest case of the asymptotically Minkowskian phase space where $\delta q_{AB}=0$, the symplectic flux is non-vanishing, hence the action $S_{ren}$ is not stationary on solutions! This suggests that suitable boundary conditions allowing for modifications of the asymptotic shear $C_{AB}$ are naturally leaky. We will verify that this is the case and make the link with the presence of gravitational waves in the bulk of $\mathscr M$.

\subsubsection{Symplectic structure and charges}
\label{sec:Symplectic structure and charges FLAT}
In the previous paragraph, we have motivated that the divergences in the Einstein-Hilbert presymplectic potential could be compensated by the incorporation of corner terms $Y^{ur}$ and $Y^{rA}$, see \eqref{d1}-\eqref{d2}. We can make good use of the Iyer-Wald ambiguity $\bm\Theta_{EH}\to\bm\Theta_{ren} = \bm\Theta_{EH}+\D\bm Y$ to bring the renormalization at the level of the presymplectic potential. The components $Y^{ur}$ and $Y^{rA}$ of the codimension 2 form $\bm Y$ have already been identified in \eqref{Yur}-\eqref{YrA}. Since $Y^{rA}$ is a $\delta$-exact term, it does not contribute to the presymplectic current. After renormalization, the presymplectic potential reads as
\begin{equation}
\Theta^u_{ren} = \mathcal O(r^{-2}), \quad \Theta^r_{ren} = \bar \Theta_{flux} + \mathcal O(r^{-1}),
\end{equation}
leading to the following presymplectic current evaluated on $\mathscr I^+$
\begin{equation}
\begin{split}
&\bm\omega_{ren}[\phi;\delta_1\phi,\delta_2\phi]\Big|_{\mathscr I^+} = \\
&\qquad\boxed{
\frac{\D u\,\D^2\Omega}{16\pi G} \left[ \frac{1}{2}\delta_1\left(N^{AB}+\frac{1}{2}R[q]q^{AB}\right)\wedge \delta_2 C_{AB} - \delta_1 (D_{(A}\mathcal U_{B)})\wedge \delta_2 q^{AB}\right].
}
\end{split}\label{omega flat complete}
\end{equation}
Here, the notation $\phi = \{g_{\mu\nu},q_{AB},C_{AB},\bm T,\sqrt{\mathring q}\}$ encompasses the sources for the presymplectic flux and the background structures. The latter, like the renormalization procedure, are defined in the specific radial foliation given in the Bondi gauge: the definitions of $q_{AB}$, $C_{AB}$ and the boundary term $\bm Y$ are not meant to be bulk covariant because they depend on the additional background structure close to $\mathscr I^+$ (such as the null foliation $\bm T$ or the fixed codimension 2 volume form). Although a gauge-invariant formulation of the renormalization process would be beneficial, we do not plan to discuss that further here, since we are just using the asymptotically flat case as an illustrative example for the covariant phase space methods and a kind of prequel story for our discussion in (A)dS in the following chapters \ref{chapter:AdSd} and \ref{chapter:LambdaBMS}. Moreover, we will not need to detail these boundary structures in the following discussion and will just start from now by assuming that the presymplectic current is given by \eqref{omega flat complete}, which defines the following symplectic structure at $\mathscr I^+$
\begin{align}
\mathcal W_{flux}[\delta_1\phi,\delta_2\phi] &\equiv \frac{1}{16\pi G} \int_{\mathscr I^+} \D u\D^2\Omega \left[ \frac{1}{2}\delta_1\left(N^{AB}+\frac{1}{2}R[q]q^{AB}\right)\wedge \delta_2 C_{AB} - \delta_1 (D_{(A}\mathcal U_{B)})\wedge \delta_2 q^{AB}\right] \nonumber \\
&= \int_{\mathscr I^+} \D u\D^2\Omega \left( \delta_1\bar\Theta_{flux}[\phi;\delta_2\phi]-\delta_2\bar\Theta_{flux}[\phi;\delta_1\phi]\right).
\end{align}
In the second equality, $\bar\Theta_{flux}$ is still defined as \eqref{thetaflux} but a total derivative has been discarded. 

By means of the fundamental theorem of the covariant phase space formalism, the surface charges associated with the Generalized BMS$_4$ symmetries $\xi(T,Y)$ are defined from the codimension 2 forms $\bm k^{\text{GBMS}_4}_\xi[\phi;\delta\phi]$ obeying $\D \bm k_\xi[\phi;\delta\phi] = \bm\omega_{ren} [\phi;\delta_\xi\phi,\delta\phi]$ with $\bm\omega_{ren}$ given by \eqref{omega flat complete}. Since we need to compute charges evaluated on spheres $S$ of $u$ at infinity, we are interested in the radial component of the current \textit{i.e.} $\partial_u k^{ur}_\xi[\phi;\delta\phi] + \partial_A k^{Ar}_\xi[\phi;\delta\phi] = \omega^r_{ren}[\phi;\delta_\xi\phi,\delta\phi]$. The contribution of the angular components $k^{Ar}_\xi$ disappears after the integration on $S$, so we just need $k^{ur}_\xi[\phi;\delta\phi]$. The computation of this quantity is outlined in \cite{Compere:2018ylh} and we just present the final result here. The surface charges are
\begin{equation}
\ndelta H_\xi^{\text{GBMS}_4} [\phi] = \oint_{S_\infty} \bm k_\xi^{\text{GBMS}_4}[\phi;\delta\phi] \equiv \delta H^{\text{GBMS}_4}_\xi[\phi] + \Xi^{\text{GBMS}_4}_\xi [\phi;\delta\phi] \label{decomposition charge}
\end{equation}
for the natural split between the integrable part \cite{Compere:2018ylh,Barnich:2011mi} (see also \cite{Flanagan:2015pxa,Hawking:2016sgy})
\begin{equation}
H^{\text{GBMS}_4}_\xi [\phi] = \prefwg \oint_{S_\infty} \text{d}^2 \Omega  \left[ 4 f M + 2 Y^A N_A + \frac{1}{16} Y^A D_A (C_{BC}C^{BC}) \right] \label{FNCharges}
\end{equation}
and the non-integrable part \cite{Compere:2018ylh}
\begin{equation}
\begin{split}
\Xi_\xi^{\text{GBMS}_4}[\phi,\delta \phi] &= \frac{1}{16\pi G} \oint_{S_\infty} \text{d}^2\Omega \left[ \frac{1}{2}f \left(N^{AB}+\frac{1}{2}q^{AB} R[q] \right) \delta C_{AB}- 2\partial_{(A} f \mathcal U_{B)} \delta q^{AB}\right. \\
&\qquad \qquad \qquad \qquad \left. - f D_{(A} \mathcal U_{B)}\delta q^{AB} -\frac{1}{4}D_C D^C f C_{AB}\delta q^{AB}\right],
\end{split} \label{FNCharges non int}
\end{equation}
keeping $f = T+\frac{1}{2}uD_AY^A$ instead of $T$ for the sake of readability. $M$ and $N_A$ appear as conjugated to $T$ (or equivalently $f$) and $Y^A$ in the expression \eqref{FNCharges}, which justify their denominations as mass and angular momentum aspects. For the Kerr black hole for instance, the surface charges associated with the Killing vectors $\partial_u$ and $\partial_\phi$ are just given by these integrable pieces, providing the total mass and angular momentum of the black hole as the integral on sphere of $M$ and $N_A$ respectively. When $\delta q_{AB}=0$ and $Y^A$ is a conformal Killing vector on $S_\infty$ (except at poles), we recover the result of \cite{Barnich:2011mi}. 

Clearly, the charges are \textit{non-integrable} and we observe that the sources $(\delta q^{AB},\delta C_{AB})$ for the symplectic flux are responsible for this non-integrability. Therefore, the split \eqref{decomposition charge} is ambiguous and the integrable piece \eqref{FNCharges} is just provided here as the natural choice coming out from the computation without giving more justification. The fixation of the finite Hamiltonian canonically conjugated with $\xi(T,Y)$ needs some additional physical inputs, such as regularity conditions at the corners $|u|\to+\infty$ \cite{Christodoulou:1993uv,Compere:2018ylh} and conditions on the flux of charges \cite{Wald:1999wa,Compere:2018ylh}. An important choice of Hamiltonian enjoying numerous suitable properties will be presented in section \ref{sec:Generalized BMS4 finite charges}.

Moreover, since the contraction of \eqref{omega flat complete} on a diffeomorphism parameter $\xi(T,Y)$ is non-zero at null infinity, the charges are also \textit{not conserved} in (retarded) time:
\begin{equation}
\frac{\D}{\D u} \ndelta H^{\text{GBMS}_4}_\xi[\phi;\delta\phi] = \oint_{S_\infty} \omega^r_{ren}[\phi;\delta_\xi\phi,\delta\phi] \neq 0.
\end{equation} 
The integrated flux of charges associated with the diffeomorphism $\xi$ is defined as
\begin{equation}
\begin{split}
F^{\text{GBMS}_4}_\xi [\phi] &= \int_\gamma \int_{-\infty}^{+\infty} \D u \,\partial_u \ndelta H_\xi^{\text{GBMS}_4} [\phi;\delta\phi] \\
&= \int_\gamma \int_{\mathscr I^+} \D \bm k_\xi^{\text{GBMS}_4} [\phi;\delta\phi] = \int_\gamma \mathcal W_{flux}[\phi;\delta_\xi\phi,\delta \phi].
\end{split}
\end{equation}
Hence the presence of the presymplectic flux $\bar\Theta_{flux}$ in \eqref{deltaS} is responsible for the non-conservation of the surface charges, \textit{i.e.} $F_\xi^{\text{GBMS}_4}[\phi]\neq 0$. 

Let us briefly discuss one important particular case. The (retarded) time translation $\xi(1,0)$ $=\partial_u$ being clearly part of the asymptotic group, we consider the flux-balance law for the associated surface charge $\ndelta H_{\xi(1,0)}^{\text{GBMS}_4}$ that would play the role of the Hamiltonian if $u$ was a timelike coordinate. We have explicitly
\begin{equation}
\frac{\D}{\D u} \ndelta H_{\xi(1,0)}^{\text{GBMS}_4}[\phi;\delta\phi] = \frac{1}{32\pi G} \oint_{S_\infty}\D^2\Omega \left[ \partial_u N_{AB}\delta C^{AB} - \frac{1}{2}\delta (N_{AB}N^{AB}) \right]
\end{equation}
using the variations \eqref{dqAB}--\eqref{dNA} with $\xi(1,0)$ and $2N_{AB}\delta N^{AB} = \delta(N_{AB}N^{AB})$ when $\delta\sqrt{q}=0$. The left-hand side reads as $\frac{1}{32\pi G}[8 \delta \partial_u M + \partial_u N_{AB}\delta C^{AB} + \frac{1}{2}\delta(N_{AB} N^{AB})]$ and the non-integrable pieces involving $\partial_u N_{AB}$ cancel out. We get rid of $\delta$'s quite easily to obtain
\begin{equation}
\frac{\D}{\D u} H^{\text{GBMS}_4}_{\xi(1,0)} = \frac{\D}{\D u} \oint_{S_\infty} \D^2\Omega \frac{1}{4\pi G} M(u,x^C) = -\frac{1}{32\pi G} \oint_{S_\infty} \D^2\Omega \, N_{AB}N^{AB}.
\end{equation}
Recalling the definition of the \textit{Bondi mass} as $\mathcal M = \oint_{S_\infty} \D^2\Omega \, \frac{M}{4\pi G}$, we have in fact re-derived the long-time celebrated \textit{Bondi mass loss formula} \cite{Bondi:1962px}
\begin{equation}
\boxed{
\frac{\D}{\D u}\mathcal M = -\frac{1}{32\pi G}\oint_{S_{\infty}} \D^2\Omega \, N_{AB}N^{AB}.
}\label{Bondi mass loss formula}
\end{equation}
As a check, we can observe that \eqref{Bondi mass loss formula} is consistent with the time evolution for the Bondi mass aspect \eqref{duM}. Historically, this flux-balance formula for the Hamiltonian $H^{\text{GBMS}_4}_{\xi(1,0)}$ canonically conjugated to the (retarded) time translation at $\mathscr I^+$ was used as a pretty convincing argument that gravitational waves exist beyond the linear approximation of the theory \cite{Bondi:1962px} and, therefore, are not an artifact of linearization. The latter would precisely kill the quadratic term appearing in \eqref{Bondi mass loss formula}. This formula allows also to give the physical meaning of the Bondi news $N_{AB}$, as the analog of the Maxwell field strengh for gravitational radiation, since its square is proportional to the energy flux carried out by the gravitational waves across $\mathscr I^+$. This discussion clarifies a bit the role of the source $C_{AB}$ for non-integrability and non-conservation and establishes the boundary conditions leading to asymptotically flat phase spaces as leaky in our terminology. In the asymptotically Minkowskian case, $C_{AB}$ is the unique source to be turned on. Imposing a subset of conservative boundary conditions in that context requires to restrict the phase space to stationary spacetimes where $\partial_u C_{AB} = N_{AB} = 0$, which completely kills the dynamics of gravitational waves. When we generalize to asymptotically locally flat spacetimes where a second set of sources is turned on in the boundary metric, the notion of stationarity is a bit less evident and we need more work to interpret the physical flux at $\mathscr I^+$. We postpone these more evolved considerations to chapter \ref{chapter:Flat}.

\section{Surface charge algebra}
\label{sec:Surface charge algebra}
As we reviewed symplectic methods for Hamiltonian mechanics, we have seen that Hamiltonian charges form an algebra of smooth functions on the phase space under the Poisson bracket. Moreover, they represent the algebra of symmetries, defined as Hamiltonian vectors on the phase space. Here we show how to import these results in the framework of the covariant phase space formalism, first for integrable charges. The task is not so difficult if we focus on the charges \eqref{charges a la Hamilton} rather than the codimension 2 form $\bm k_\xi[\phi;\delta\phi]$. In particular, the proof of the so-called representation theorem \cite{Brown:1986nw,Brown:1986ed} (see also \cite{Koga:2001vq}) proceeds in the same way that in classical mechanics and relies only on the structural properties \eqref{commu on jet space}--\eqref{Lie commutation} of the jet bundle. Local considerations involving $\bm k_\xi[\phi;\delta\phi]$ can be found among the profound discussions of \cite{Barnich:2007bf} but will not be part of the developments presented here below. 

\changed{
When the integrability of the charges \eqref{charges a la Hamilton} cannot be assumed, the representation theorem looses its central hypothesis and must be partially extended. The failure has to be ascribed to the Poisson bracket that is no longer well-suited to compute the charge algebra in the presence of non-integrability. In \cite{Barnich:2011mi}, Barnich and Troessaert heuristically introduced a new definition for the charge bracket while attempting to compute the algebra formed the BMS$_4$ non-integrable charges in asymptotically flat spacetimes. They showed that this algebra is closed under their modified bracket up to a field-dependent ``central extension.'' Since then, similar computations were reproduced in different contexts (see \textit{e.g.} \cite{Compere:2018ylh,Compere:2020lrt,Troessaert:2015nia,Distler:2018rwu,Chandrasekaran:2020wwn,Freidel:2021cbc}), accumulating evidences in favor this generalized charge algebra. The good news is that it is possible to give a completely general proof of the representation theorem for non-integrable charges involving the Barnich-Troessaert bracket. As before, it relies only on the existence of a differential phase space with trivial topological features and the datum of an infinitesimal charge split between integrable and non-integrable pieces. The result will only be mentioned here and its proof will be discussed elsewhere \cite{ToAppear}.
}

\subsection{Integrable case: the Poisson bracket}
\label{sec:Integrable case: the Poisson bracket}
Recalling the definition \eqref{charges a la Hamilton} of the infinitesimal surface charges, the hypothesis of integrability yields
\begin{equation}
i_{\delta_\xi}\mathcal W[\phi;\delta\phi,\delta\phi]=\delta H_\xi[\phi]
\label{Hamiltonian in covariant phase space}
\end{equation}
This states that $\delta_{\xi}\phi$ is a Hamiltonian vector field on the jet bundle associated with the Hamiltonian function $H_\xi[\phi]$, in analogy with \eqref{def hamilton} in classical mechanics. The dynamics of the system is completely encoded into the transformations of the charges under gauge transformations: it is sufficient to know $\delta_{\xi_2}H_{\xi_1}$ for any couple of diffeomorphisms $\xi_1,\xi_2$ to have complete control on the evolution of the physical system under consideration. For instance, a rotation in the azimuthal angle $\varphi$ translates as $\partial_\varphi H_\xi + \delta_{\partial_{\varphi}}H_\xi$ at the level of the charges $H_\xi$ for any $\xi$; the evolution of the system itself is modelized by the collection of time derivatives $\partial_t H_\xi +\delta_{\partial_t} H_\xi$ for a timelike coordinate $t$. As in classical mechanics, the first term represents the explicit coordinate dependence while the second term is the net variation under the action of the symmetry. Using \eqref{Cartan for jet bundle}, \eqref{Hamiltonian in covariant phase space} can be contracted a second time to give
\begin{equation}
\begin{split}
\delta_{\xi_2}H_{\xi_1}[\phi] &= i_{\delta_{\xi_2}}\delta H_{\xi_1}[\phi] = i_{\delta_{\xi_2}}i_{\delta_{\xi_1}}\mathcal W[\phi;\delta\phi,\delta\phi].
\end{split}
\end{equation}
It is not hard to show that this expression defines a Lie bracket between Hamiltonian functions associated with diffeomorphisms. This is the exact equivalent to the classical \textit{Poisson bracket} for field theories 
\begin{equation}
\{ H_{\xi_1} [\phi] , H_{\xi_2} [\phi] \} \equiv i_{\delta_{\xi_2}}i_{\delta_{\xi_1}}\mathcal W[\phi;\delta\phi,\delta\phi] = \delta_{\xi_2} H_{\xi_1}[\phi], \label{Poisson bracket jet space}
\end{equation}
see \cite{Brown:1986nw,Brown:1986ed,Henneaux:1992ig} for the pure Hamiltonian version and \cite{Barnich:2007bf,Koga:2001vq} for the covariant phase space version. As announced, a strong link exists between the Lie algebra of diffeomorphisms $\xi$ and the corresponding Lie algebra of canonically conjugated Hamiltonian functions $H_\xi[\phi]$: this is the concern of the following theorem \cite{Brown:1986ed,Barnich:2007bf}

\resu{Charge representation theorem (integrable case)}{
Assuming integrability, the conserved charges associated with a Lie algebra of diffeomorphisms also form an algebra under the Poisson braket $\{ H_{\xi_1},H_{\xi_2} \} = \delta_{\xi_2}H_{\xi_1}$ which is isomorphic to the Lie algebra of diffeomorphisms up to a central extension depending only on a reference solution $\bar\phi$:
\begin{equation}
\{ H_{\xi_1}[\phi],H_{\xi_2}[\phi]\} = H_{[\xi_1,\xi_2]_\star}[\phi] + K_{\xi_1,\xi_2}[\bar\phi]. \label{Rep theo integrable}
\end{equation}
}
The proof is similar to the representation theorem for Hamiltonian charges in classical mechanics and is developed as follows. We contract the presymplectic form with the variation along the vector $[\xi_1,\xi_2]_\star$: 
\begin{equation}
\begin{split}
&i_{\delta_{[\xi_1,\xi_2]_\star}}\mathcal W [\phi;\delta\phi,\delta\phi] \\
&= -i_{[\delta_{\xi_1},\delta_{\xi_2}]}\mathcal W [\phi;\delta\phi,\delta\phi] \\
&= -i_{\delta_{\xi_1}} \delta_{\xi_2} \mathcal W [\phi;\delta\phi,\delta\phi] + \delta_{\xi_2} i_{\delta_{\xi_1}} \mathcal W [\phi;\delta\phi,\delta\phi] \\
&= -i_{\delta_{\xi_1}} i_{\delta_{\xi_2}}\delta \mathcal W [\phi;\delta\phi,\delta\phi] -i_{\delta_{\xi_1}} \delta i_{\delta_{\xi_2}} \mathcal W [\phi;\delta\phi,\delta\phi] + \delta_{\xi_2} i_{\delta_{\xi_1}} \mathcal W [\phi;\delta\phi,\delta\phi] \\
&= \delta_{\xi_2}\mathcal W [\phi;\delta_{\xi_1}\phi,\delta\phi] -i_{\delta_{\xi_1}} \delta \mathcal W [\phi;\delta_{\xi_2}\phi,\delta\phi]
\end{split} \label{proof algebra general relation}
\end{equation}
The first equality uses \eqref{Lie commutation}, the second one is due to \eqref{Lie with inner product}, the third one uses Cartan's magic formula \eqref{Cartan for jet bundle} while the fourth one needs $\delta\mathcal W=0$ because $\mathcal W$ is $\delta$-exact. We use now the assumption of integrability \eqref{Hamiltonian in covariant phase space}. Using $\delta^2=0$ a couple of times, \eqref{proof algebra general relation} becomes
\begin{equation}
i_{\delta_{[\xi_1,\xi_2]_\star}}\mathcal W [\phi;\delta\phi,\delta\phi] 
= \delta_{\xi_2} \delta H_{\xi_1}[\phi] = \delta i_{\delta_{\xi_2}} \delta H_{\xi_1}[\phi] = 
 \delta ( \delta_{\xi_2} H_{\xi_1}[\phi] )
 \label{proof algebra integrable case 1}
\end{equation}
again thanks to \eqref{Cartan for jet bundle}. Now injecting the definition of the Poisson bracket \eqref{Poisson bracket jet space}, we have proven that $i_{\delta_{[\xi_1,\xi_2]_\star}}\mathcal W$ is integrable and
\begin{equation}
\delta H_{[\xi_1,\xi_2]_\star}[\phi] \equiv i_{\delta_{[\xi_1,\xi_2]_\star}}\mathcal W [\phi;\delta\phi,\delta\phi] = \delta\left( \{ H_{\xi_1} [\phi] , H_{\xi_2} [\phi] \} \right),
\end{equation}
which can be integrated on the phase space between a reference solution $\bar\phi$ to a target solution $\phi$ as \eqref{Rep theo integrable} up to some functional $K_{\xi_1,\xi_2}[\bar\phi]$ depending only on $\bar\phi$. Because $\delta K_{\xi_1,\xi_2}=0$, it commutes with any surface charge $H_\xi$ under the Poisson bracket \eqref{Poisson bracket jet space}, so it belongs to the center of the charge algebra: we obtain thus a \textit{central extension} when we consider the charge algebra instead of the associated vector algebra. Evaluating \eqref{Rep theo integrable} for the reference solution $\bar\phi$, one gets
\begin{equation}
K_{\xi_1,\xi_2}[\bar\phi] = \delta_{\xi_2}N_{\xi_1}[\bar\phi] - N_{[\xi_1,\xi_2]_\star}[\bar\phi]. \label{eq:central charge explicit in N}
\end{equation}
By consistency of the charge algebra, the central extension must be antisymmetric under the exchange $\xi_1\leftrightarrow\xi_2$ and satisfy
\begin{equation}
K_{[\xi_1,\xi_2]_\star,\xi_3}[\bar\phi] + \text{cyclic}(1,2,3) = 0 \label{cocycle condition field independent}
\end{equation}
as a consequence of the Jacobi identity for the bracket \eqref{Poisson bracket jet space}. In other words, $K_{\xi_1,\xi_2}[\bar\phi]$ forms a 2-\textit{cocycle} on the Lie algebra of diffeomorphisms, which is said to be \textit{non-trivial} if and only if it cannot be absorbed by a normalization of the charges for the reference $\bar\phi$, see \eqref{eq:central charge explicit in N}. 

\subsection{Non-integrable case: the Barnich-Troessaert bracket}
\label{sec:Non-integrable case: the Barnich-Troessaert bracket}
Since the result \eqref{Rep theo integrable} relies on the integrability of the charge, it cannot be directly extended to general charges that do not satisfy the integrability condition \eqref{eq:Integrability}. Defining a split between integrable and non-integrable parts as \eqref{split theory}, we analyze here how to extend the result of the representation theorem in this more general context.

\subsubsection{Generalized representation theorem}
The generalization first requires a new definition for the bracket between charges $\{ H_{\xi_1},H_{\xi_2}\}_\star$, which will involve the integrable parts of the charges only to conserve some similarities with \eqref{Rep theo integrable}. Keeping in mind the definition of the Poisson bracket \eqref{Poisson bracket jet space}, let us compute
\begin{equation}
\begin{split}
i_{\delta_{\xi_2}} i_{\delta_{\xi_1}} \mathcal W[\phi;\delta\phi,\delta\phi] 
&= i_{\delta_{\xi_2}} (\delta H_{\xi_1}[\phi] + \Xi_{\xi_1}[\phi;\delta\phi]) = \delta_{\xi_2} H_{\xi_1}[\phi] + \Xi_{\xi_1}[\phi;\delta_{\xi_2}\phi]. \label{fail poisson}
\end{split}
\end{equation}
The double contraction of the presymplectic form $\mathcal W$ fails to give a good prescription for the charge bracket because of the non-integrable piece. But the aforementioned obstruction delivers some inspiration to define the following modified bracket
\begin{equation}
\{ H_{\xi_1}[\phi],H_{\xi_2}[\phi]\}_\star \equiv \mathcal W[\phi;\delta_{\xi_1}\phi,\delta_{\xi_2}\phi] - \Xi_{\xi_1}[\phi;\delta_{\xi_2}\phi] + \Xi_{\xi_2}[\phi;\delta_{\xi_1}\phi]
\end{equation}
which is manifestly antisymmetric under the exchange $\xi_1\leftrightarrow\xi_2$ and for which the Jacobi identity can be checked. Using \eqref{fail poisson}, we get
\begin{equation}
\boxed{
\{ H_{\xi_1}[\phi],H_{\xi_2}[\phi]\}_\star = \delta_{\xi_2}H_{\xi_1}[\phi] +  \Xi_{\xi_2}[\phi;\delta_{\xi_1}\phi].
} \label{BT definition}
\end{equation}
\changed{
The modification $\Xi_{\xi_2}[\phi;\delta_{\xi_1}\phi]$ to the ``standard Poisson bracket'' $\delta_{\xi_2}H_{\xi_1}[\phi]$ was first introduced in \cite{Barnich:2011mi} in the context of asymptotically flat spacetimes and the bracket \eqref{BT definition} goes nowadays under the name of the \textit{Barnich-Troessaert bracket}. Using the techniques involved in the proof of \eqref{Rep theo integrable}, we can demonstrate the following result \cite{ToAppear}:
}
\resu{Charge representation theorem (non-integrable case)}{
Let $\ndelta H_\xi = \mathcal W[\phi;\delta_\xi\phi,\delta\phi]$ be the (non-necessarly integrable) infinitesimal surface charges associated with diffeomorphisms $\xi$ evaluated for the solution $\phi$. Let also $\delta H_\xi[\phi] + \Xi_\xi [\phi;\delta\phi]$ be an arbitrary split of $\ndelta H_\xi$ into integrable and non-integrable parts $H_\xi[\phi]$ and  $\Xi_\xi [\phi;\delta\phi]$. The phase space functions $H_\xi[\phi]$ associated with the Lie algebra of diffeomorphisms $\xi$ also form an algebra under the Barnich-Troessaert bracket $\{ H_{\xi_1}[\phi],H_{\xi_2}[\phi]\}_\star = \delta_{\xi_2}H_{\xi_1}[\phi] +  \Xi_{\xi_2}[\phi;\delta_{\xi_1}\phi]$ which is isomorphic to the Lie algebra of diffeomorphisms up to a field-dependent 2-cocycle:
\begin{equation}
\{ H_{\xi_1}[\phi],H_{\xi_2}[\phi]\}_\star = H_{[\xi_1,\xi_2]_\star}[\phi] + K_{\xi_1,\xi_2}[\phi]. \label{Rep theo general}
\end{equation}
}
\changed{
We do not present the detailed proof in this manuscript. The main difference with \eqref{Rep theo integrable} is that the ``central extension'' $K_{\xi_1,\xi_2}[\phi]$ reads now as a \textit{field-dependent} object whose main properties are discussed in the next paragraph.
}

\subsubsection{Field-dependent 2-cocycle}
\label{sec:Field-dependent 2-cocycle}
\changed{
The charge algebra under the Barnich-Troessaert bracket reveals a field-dependent extension that is not central because $\delta K_{\xi_1,\xi_2}[\phi] \neq 0$ in general. This function on the phase space is not completely arbitrary, because it must obey the following conditions, in order to maintain the consistency of the algebra \eqref{Rep theo general}:
}
\begin{align}
&K_{\xi_1,\xi_2}[\phi] = -K_{\xi_2,\xi_1}[\phi], \\
&K_{[\xi_1,\xi_2]_\star,\xi_3}[\phi] - \delta_{\xi_3}K_{\xi_1,\xi_2}[\phi] + \text{cyclic}(1,2,3) = 0. \label{cocycle condition theory}
\end{align}
\changed{
This couple of conditions define a Lie algebroid 2-cocycle on the Lie algebroid formed by the diffeomorphisms. The second condition \eqref{cocycle condition theory}, and necessary to ensure the Jacobi identity for the Barnich-Troessaert bracket \eqref{BT definition} \cite{Barnich:2011mi}, is the generalization of \eqref{cocycle condition field independent} taking the field-dependence of $K_{\xi_1,\xi_2}[\phi]$ into account. The algebraic properties are completed by the following identities \cite{Barnich:2011mi}
\begin{align}
&\{ K_{\xi_1,\xi_2}[\phi],H_{\xi_3}[\phi] \}_\star = -\{ H_{\xi_3}[\phi], K_{\xi_1,\xi_2}[\phi] \}_\star = \delta_{\xi_3} K_{\xi_1,\xi_2}[\phi], \\
&\{ K_{\xi_1,\xi_2}[\phi], K_{\xi_3,\xi_4}[\phi]\}_\star = 0. 
\end{align}
The first one is quite logical since it amounts to say that the action of the charge on any phase space function is meant to generate the symmetry. The second one states that the extension is ``central'' in the sense that it does not generate any symmetry.
}

This is perhaps the good moment to discuss the impact on the algebra of the ambiguity \eqref{split ambiguity N} of the split \eqref{split theory} between integrable and non-integrable parts. Let us set
\begin{equation}
H'_\xi[\phi] = H_\xi[\phi] - \Delta H_\xi[\phi],\qquad \Xi'_\xi[\phi;\delta\phi] = \Xi_\xi[\phi;\delta\phi] + \delta \Delta H_\xi[\phi] \label{shift theo}
\end{equation}
such that $\mathcal W[\phi,\delta_\xi\phi,\delta\phi] = \delta H_\xi[\phi] + \Xi_\xi[\phi;\delta\phi] = \delta H'_\xi[\phi] + \Xi'_\xi[\phi;\delta\phi]$. Defining the Barnich-Troessaert bracket \eqref{BT definition} with primed objects
\begin{equation}
\{ H'_{\xi_1}[\phi],H'_{\xi_2}[\phi]\}_\star = \delta_{\xi_2}H'_{\xi_1}[\phi] +  \Xi'_{\xi_2}[\phi;\delta_{\xi_1}\phi],
\end{equation}
we can show that \eqref{Rep theo general} is preserved up to some modification in the 2-cocycle. Indeed, 
\begin{align}
\{ H'_{\xi_1}[\phi],H'_{\xi_2}[\phi]\}_\star &= \{ H_{\xi_1}[\phi],H_{\xi_2}[\phi]\}_\star - \delta_{\xi_2}\Delta H_{\xi_1}[\phi] + \delta_{\xi_1}\Delta H_{\xi_2}[\phi] \nonumber\\
&= H_{[\xi_1,\xi_2]_\star}[\phi] + K_{\xi_1,\xi_2}[\phi] - \delta_{\xi_2}\Delta H_{\xi_1}[\phi] + \delta_{\xi_1}\Delta H_{\xi_2}[\phi] \nonumber\\
&= H'_{[\xi_1,\xi_2]_\star}[\phi] + \left( K_{\xi_1,\xi_2}[\phi] - \delta_{\xi_2}\Delta H_{\xi_1}[\phi] + \delta_{\xi_1}\Delta H_{\xi_2}[\phi] + \Delta H_{[\xi_1,\xi_2]_\star}[\phi] \right)\nonumber\\
&\equiv H'_{[\xi_1,\xi_2]_\star}[\phi] +  K'_{\xi_1,\xi_2}[\phi]
\end{align}
where the new 2-cocycle
\begin{equation}
K'_{\xi_1,\xi_2}[\phi] \equiv K_{\xi_1,\xi_2}[\phi] - \delta_{\xi_2}\Delta H_{\xi_1}[\phi] + \delta_{\xi_1}\Delta H_{\xi_2}[\phi] + \Delta H_{[\xi_1,\xi_2]_\star}[\phi] \label{cocycle shift}
\end{equation}
directly satisfies the cocycle condition \eqref{cocycle condition theory} if and only if $K_{\xi_1,\xi_2}[\phi]$ is also a 2-cocycle. As a result, the structure of the algebra is left intact by the change of prescription for the split \eqref{split ambiguity N} while only the 2-cocycle is affected \cite{Barnich:2011mi}. We say that $K_{\xi_1,\xi_2}[\phi]$ is \textit{trivial} if there exists some function $\Delta H_\xi[\phi]$ such that $\delta K'_{\xi_1,\xi_2}[\phi]=0$, \textit{i.e.} it is possible to absorb the field-dependence of the 2-cocycle by a smart choice of the integrable part. Otherwise, it is non-trivial and consequently non-absorbable.

\subsection{The Generalized BMS\texorpdfstring{$_4$}{4} charge algebra}
\label{sec:The Generalized BMS$_4$ charge algebra}
As we mentioned, the first occurrence of the generalized result \eqref{Rep theo general} was found in \cite{Barnich:2011mi} in the context of asymptotically flat gravity with the restricted boundary conditions $\delta q_{AB}$ but allowing for meromorphic super-Lorentz transformations. It can be checked explicitly that the result holds also with the inclusion of smooth Diff($S^2$) super-Lorentz symmetries on the phase space and the proof, similar to the original derivation extensively detailed in \cite{Barnich:2011mi}, is lead as follows. Starting from the integrable part \eqref{FNCharges}, we can compute $\delta_{\xi_2}H_{\xi_1}^{\text{GBMS}_4}$ for any generators $\xi_1$, $\xi_2$ of the Generalized BMS$_4$ owing to the transformations \eqref{dqAB}--\eqref{dNA} on the asymptotically locally flat solution space $\mathring{\mathcal S}_0$. During the computations, any total derivative on the sphere can be ignored since we required that all fields, as well as the transformations, are smooth functions of the angles $x^A$. These variations can also be plugged into the non-integrable part \eqref{FNCharges non int} to reconstitute the Barnich-Troessaert bracket 
\begin{equation}
\{H_{\xi_1}^{\text{GBMS}_4}[\phi],H_{\xi_2}^{\text{GBMS}_4}[\phi]\}_\star = \delta_{\xi_2}H^{\text{GBMS}_4}_{\xi_1}[\phi] + \Xi^{\text{GBMS}_4}_{\xi_2}[\phi;\delta_{\xi_1}\phi].
\end{equation}
Again, $\phi$ designates here the metric field $g$ with any background structure needed to fix the gauge as well as the boundary conditions, or equivalently the solution space parameters appearing in \eqref{S0ring}. Secondly, we have to evaluate the charge \eqref{FNCharges} for $\xi = [\xi_1,\xi_2]_\star$ with parameters given as \eqref{commu f flat general}--\eqref{commu Y flat general}. After an involved computation, we get 
\begin{equation}
\{H_{\xi_1}^{\text{GBMS}_4}[\phi],H_{\xi_2}^{\text{GBMS}_4}[\phi]\}_\star = H^{\text{GBMS}_4}_{[\xi_1,\xi_2]_\star}[\phi]+ K^{\text{GBMS}_4}_{\xi_1,\xi_2}[\phi]
\label{eq:Algebra}
\end{equation}
with the particular 2-cocycle 
\begin{equation}
\begin{split}
{K}^{\text{GBMS}_4}_{\xi_1,\xi_2} [\phi] = \prefwg \oint_{S_\infty} \text{d}^2\Omega \, \left[ \frac{1}{2} f_2 D_A f_1 D^A R[q] + \frac{1}{2} C^{BC} f_2 D_B D_C D_D Y^D_1 \right]  - (1\leftrightarrow 2), \label{cocycle flat}
\end{split}
\end{equation} 
\changed{
generalizing the expression of \cite{Barnich:2011mi} to arbitrary metrics on the celestial sphere. This is a particular realization of the result \eqref{Rep theo general} and this is not the last discussed in this manuscript. The reference solution $\bar\phi$ is taken to be the global Minkowski vacuum with identically vanishing charges. The 2-cocycle \eqref{cocycle flat} is believed to be non-trivial thanks to the following arguments.
}
\begin{enumerate}
\item Two supertranslations commute (see \eqref{commu f flat general}) and therefore the 2-cocycle $K^{\text{GBMS}_4}_{\xi_1(T_1);\xi_2(T_2)} \neq 0$, $\xi(T)\equiv \xi(T,0)$, does not transform upon shifting the integrable charge according to \eqref{split ambiguity N} with a supertranslation charge that depends on $q_{AB}$ and not $C_{AB}$, \textit{i.e.} $K^{'\text{GBMS}_4}_{\xi_1(T_1);\xi_2(T_2)}$ $=$ $K^{\text{GBMS}_4}_{\xi_1(T_1);\xi_2(T_2)}$ using \eqref{cocycle shift} and \eqref{dqAB} implying $\delta_T q_{AB} = 0$. Since $K^{\text{GBMS}_4}_{\xi_1(T_1);\xi_2(T_2)} \neq 0$ and only depends upon $q_{AB}$ the first term in the 2-cocycle is non-trivial.
\item A cohomological formulation of the second term of \eqref{cocycle flat} can be found in \cite{Barnich:2017ubf}. After semi-classical quantization, the second term of the 2-cocyle can also be related to the non-commutativity of the double soft limit of gravitons \cite{Distler:2018rwu}.
\end{enumerate}
When restricting to the subspace \eqref{S0Mink} of asymptotically Minkowskian solutions, ${K}^{\text{GBMS}_4}_{\xi_1,\xi_2} [\phi]$ vanishes identically for any parameters since $R[q]=2$ and $Y^A$ is solution of the conformal Killing equation \eqref{Killing eq for rotations}. Hence, the Barnich-Troessaert bracket allows the BMS$_4$ charge algebra to represent the BMS$_4$ algebra without central extension. Because of the split ambiguity \eqref{split ambiguity N}, this statement is obviously dependent on the particular representative for the integrable part $H_\xi^{\text{GBMS}_4}[\phi]$, but this is also a motivation to choose \eqref{FNCharges} as Hamiltonians among the infinite bunch of possibilities. Nevertheless, we will argue later that it is not the best option from physical perspectives in section \ref{sec:Generalized BMS4 finite charges} and will give additional prescriptions to fix the expression of the Hamiltonians for the Generalized BMS$_4$ group. 

Now let us illustrate and confirm the intuitions given in the opening of section \ref{sec:Integrable case: the Poisson bracket} to motivate the study of the charge algebra, namely that the algebra \eqref{eq:Algebra} completely controls the dynamics of the gravitational system and contains useful pieces of information, even for practical purposes. For instance, as observed in the seminal work \cite{Barnich:2011mi}, it controls the non-conservation of the charges through the presence of the modification term involving the non-integrable part. The reasoning is as follows. As in Hamiltonian mechanics, the total evolution of the quantity $H_\xi^{\text{GBMS}_4}[\phi]$ is given by
\begin{equation}
\frac{\D }{\D u} H_{\xi}^{\text{GBMS}_4}[\phi] = \frac{\partial}{\partial u}H_{\xi}^{\text{GBMS}_4}[\phi] + \delta_{\partial_u} H_{\xi}^{\text{GBMS}_4}[\phi], \label{flux from algebra 1}
\end{equation}
where the first term represents the explicit dependence on $u$ while the second term gives the net evolution along the flow of the vector $\partial_u$. The first contribution is easy to derive since only the parameter $\xi$ has some explicit dependence in $u$. The fields involved in the charge have a constrained time evolution on-shell which is considered as an implicit dependence already included in the second term. For example, we see from \eqref{dM} that $\delta_{\partial_u}M = \delta_{\xi(1,0)}M = \partial_u M$, given on-shell by \eqref{duM flat}. So we have
\begin{equation}
\frac{\partial}{\partial u}H_{\xi}^{\text{GBMS}_4}[\phi] = H^{\text{GBMS}_4}_{\partial_u \xi}[\phi] = -H^{\text{GBMS}_4}_{[\xi,\partial_u]_\star}[\phi]. \label{flux from algebra 2}
\end{equation}
We insert \eqref{flux from algebra 2} and \eqref{eq:Algebra} in \eqref{flux from algebra 1} to obtain
\begin{equation}
\frac{\D}{\D u}H_\xi^{\text{GBMS}_4}[\phi] = -\Xi^{\text{GBMS}_4}_{\partial_u}[\phi;\delta_\xi\phi] - K^{\text{GBMS}_4}_{\partial_u,\xi}[\phi] = -\Xi^{\text{GBMS}_4}_{\partial_u}[\phi;\delta_\xi\phi],
\end{equation}
since \eqref{cocycle flat} is not present when $\xi_1$ is $\partial_u$. A quick comparison between \eqref{thetaflux} and \eqref{FNCharges non int}, at the price to perform some integrations by parts on the celestial sphere $S$, finally gives \cite{Compere:2018ylh,Compere:2020lrt}
\begin{equation}
\boxed{
\frac{\D}{\D u}H_\xi^{\text{GBMS}_4}[\phi] = -\oint_{S_\infty} \iota_{\partial_u}{\bm\Theta}_{ren}[\phi;\delta_\xi\phi] = -\oint_{S_\infty} \D^2 \Omega\, \bar\Theta_{flux}[q_{AB},C_{AB};\delta_\xi q^{AB},\delta_\xi C^{AB}].
} \label{flux from algebra final}
\end{equation}
This generalizes the flux formula previously obtained in \cite{Wald:1999wa} assuming $\delta q_{AB}=0$. The leading contribution of the presymplectic flux, sourced by the traditional radiative degrees of freedom in $C_{AB}$ and the additional kinematics encoded in $\delta q_{AB}$, is responsible for the non-conservation of the surface charges associated with the diffeomorphisms. The key point is that the flux-balance laws \eqref{flux from algebra final} are a direct consequence of the charge algebra \eqref{eq:Algebra}, hence hold independently of the representative of the integrable part of the charge. Taking in particular $\xi = \partial_u$, $H^{\text{GBMS}_4}_\xi$ is nothing but the Bondi mass and \eqref{flux from algebra final} implies \cite{Barnich:2011mi}
\begin{equation}
\frac{\D }{\D u}\mathcal M = -\frac{1}{32\pi G} \oint_{S_\infty} \D^2\Omega\, (N_{AB} \delta_{\partial_u} C^{AB}) = -\frac{1}{32\pi G} \oint_{S_\infty} \D^2\Omega\, (N_{AB} N^{AB})
\end{equation}
which is exactly the Bondi mass loss formula \eqref{Bondi mass loss formula} stating that the time derivative of the Bondi mass is negative definite and vanishes if and only if there is no Bondi news.

This aesthetic discussion about the algebraic properties of the gravitational surface charges marks the end point of this quite long chapter devoted to the covariant phase space formalism. We leave for the moment the flat land to explore more spectacular and fascinating asymptotics in the presence of a cosmological constant\dots\hfill{\color{black!40}$\blacksquare$}

%
%

\chapter{Asymptotically locally (A)dS spacetimes}
\label{chapter:AdSd}

In this chapter, we progress our exploration towards the asymptotic regions of solutions of Einstein's gravity in the presence of a non-vanishing cosmological constant. We export there the various features, techniques and nomenclatures discussed in the previous chapter and define genuine radiative phase spaces for these crucial kinds of asymptotics, broadly highlighted in holography as well as in modern cosmology. 

First, we review the causal and asymptotic properties of the (A)dS spacetimes, in order to get inspired for the definition of phase spaces with solutions that share the same asymptotic properties. In the spirit of Penrose's analysis of relativistic infinities \cite{Penrose:1964ge}, we leverage the conformal compactness of these solutions to write the covariant phase space in the very convenient Starobinsky/Fefferman-Graham coordinate system \cite{Starobinsky:1982mr,Fefferman:1985aa}. We perform the computations and interpret the outcomes in any bulk dimension and without assuming further hypotheses. In particular, the boundary structure is allowed to fluctuate and plays the role of source yielding some symplectic flux at the conformal boundary \cite{Papadimitriou:2005ii,Compere:2008us}. The class of spacetimes under consideration is thus generically radiative and the necessity to consider leaky boundary conditions even when the cosmological constant is present is debated at the same time. The most famous conservative sub-cases \cite{Brown:1986nw,Ashtekar:1984zz,Henneaux:1985tv,Henneaux:1985ey,Ashtekar:1999jx} are reviewed within our formalism in order to anchor our results in the extensive literature in the field and show the improvements that we have introduced thus far. 

Using the holographic renormalization procedure \cite{deHaro:2000vlm,Bianchi:2001kw,Skenderis:2000in}, we show how to remove the divergences from the symplectic structure \cite{Papadimitriou:2005ii,Compere:2008us}. The charges associated with the whole class of residual gauge diffeomorphisms in the Starobinsky/Fefferman-Graham gauge are computed. All non-trivial boundary diffeomorphisms are shown to be canonically conjugated with non-vanishing, non-integrable and not conserved surface charges, while those associated with boundary Weyl rescalings are non-vanishing only in odd dimensions, due to the presence of Weyl anomalies in the dual theory \cite{Henningson:1998gx,deHaro:2000vlm,Papadimitriou:2005ii}. Finally, the charge algebra under the Barnich-Troessaert bracket \cite{Barnich:2011mi} is derived and its main features are scrutinized. In particular, a field-dependent $2$-cocycle manifests in odd dimensions. When the general framework is restricted to three-dimensional asymptotically AdS spacetimes with Dirichlet boundary conditions, this cocycle reduces to the well-known Brown-Henneaux central extension \cite{Brown:1986nw}.

Apart from the review sections \ref{sec:Asymptotic properties of (A)dS spacetimes} and \ref{sec:Conformally compact manifolds}, this chapter essentially reproduces \cite{Fiorucci:2020xto} assorted with some elements taken from \cite{Compere:2019bua,Compere:2020lrt}.

\section{Gravity in Starobinsky/Fefferman-Graham gauge}

In this introductive section, we review how to define the notion of (A)dS asymptotics through the conformal compactification process. We start by reviewing some key features of global de Sitter and anti-de Sitter spacetimes and developing about their asymptotic structure. We continue by recalling some general statements about conformally compact spacetimes, for which one can perform a conformal transformation to add meaningfully the closure of the unphysical spacetime at finite distance and consider this as a robust idealization of a manifold at infinity. Finally, building on these considerations, we review the notion of asymptotically locally (anti-) de Sitter spacetime (Al(A)dS for short) within the common and powerful coordinate system often used to explore the physical features of these kinds of asymptotics. For the purpose of our dissertations, we restrict ourselves to coordinate-dependent discussions and directly provide explicit formulas. As for the flat case, completely covariant formalisms were developed in parallel of the various progressions in the gauge-fixing approach (see \textit{e.g.} \cite{Ashtekar:1984zz,Ashtekar:1999jx,Ashtekar:2014zfa}). They manage to confirm the results obtained by explicit methods thanks to elegant reformulations using fundamental geometrical objects, but we will not review them here. The mathematical developments of this section summarize the seminal works \cite{deHaro:2000vlm,Henningson:1998gx,Balasubramanian:1999re,Bianchi:2001kw,Skenderis:2000in} and allow us to install our conventions while defining rigorously the general framework in which we will construct the phase space.

\subsection{Asymptotic properties of (A)dS spacetimes}
\label{sec:Asymptotic properties of (A)dS spacetimes}

This opening section consists of a selective review of the interesting properties of the de Sitter and anti-de Sitter spacetimes. These are the fundamental solutions whose we want to study the asymptotic behavior in order to define a phase space of solutions mimicking (at least locally) their asymptotic structure at infinity. For the interested reader, more details can be found in the reviews \cite{Spradlin:2001pw,Compere:2018aar,Fischetti:2012rd,Graham:1999jg,Anderson:2004yi} and references therein.

\subsubsection{The de Sitter solution}
The \textit{de Sitter spacetime} dS$_{d+1}$ (in $n=d+1$ dimensions, $d>1$) is the only maximally symmetrical Lorentzian manifold of positive Ricci curvature \cite{deSitter:1916zza,deSitter:1916zz,deSitter:1917zz}. In the global coordinate system $\{ \tau,\chi,x^A \}$ covering the whole spacetime ($\tau\in\mathbb R,\chi\in [0,\pi]$ and $x^A$, $A=2,\dots,d$, are coordinates on the $S^{d-1}$ sphere), its line element reads as
\begin{equation}
\D s^2_{\text{dS}} = -\D \tau^2 + \ell^2 \cosh^2(\tau/\ell) \left(\D\chi^2 +\sin^2\chi \,\mathring q_{AB}\D x^A \D x^B\right). \label{dS global patch}
\end{equation}
The real parameter $\ell>0$ controls the Ricci curvature as $R = d(d+1)/\ell^2$ and $\mathring q_{AB}$ represents the unit-round metric on the $S^{d-1}$ sphere. At the physical level, \eqref{dS global patch} describes an Einsteinian manifold solving the vacuum Einstein field equations for a positive value of the cosmological constant given by
\begin{equation}
\Lambda = \frac{d(d-1)}{2\ell^2}.\label{Lambda dS}
\end{equation}
In that sense, this solution is the exact analog of the Minkowski gravitational vacuum in the presence of a strictly positive cosmological constant. By inspection of \eqref{dS global patch}, we learn that the de Sitter spacetime has the topology of $\mathbb R\times S^{d}$ and can be seen as a $d$-sphere first contracting then expanding when the ``evolution parameter'' $\tau$ travels the real line $\mathbb R$. Any fixation of the couple $(\tau,\chi)$ describes a $(d-1)$-sphere of radius $\sin\chi$ at fixed $\tau$ except for the extremal values $\chi=0$ and $\chi=\pi$ which represent actual points located at the north and south poles of the $d$-sphere. Geometrically, the de Sitter space can be thought of as an hyperboloid $\mathscr H_{\text{dS}}$ embedded in the $(d+2)$-dimensional Minkowski spacetime, defined as the geometrical locus of points in $\mathbb R^{d+1,1}$ at a fixed proper distance $\ell$ from the origin. For that reason, $\ell$ goes under the name of de Sitter radius. The isometry group of $\mathscr H_{\text{dS}}$ is then the subgroup of Minkowski isometries in $d+2$ dimensions that stabilizes the origin, \textit{i.e.} the $(d+2)$-dimensional-Lorentz group $SO(d+1,1)$. This includes rotations on the sections of $\mathscr H_{\text{dS}}$ (which are $S^d$ spheres at constant $\tau$) as well as boosts along the revolution axis of the hyperboloid. 

\begin{figure}[!t]
\centering
\begin{tikzpicture}[scale=0.6]
\draw[white] (-7,-7) -- (-7,7) -- (7,7) -- (7,-7) -- cycle;
	\coordinate (tl) at (-5, 5);
	\coordinate (tr) at ( 5, 5);
	\coordinate (bl) at (-5,-5);
	\coordinate (br) at ( 5,-5);
	\fill[opacity=0.1,blue] (tr) -- (bl) -- (br) -- cycle;
	\fill[opacity=0.1,red] (tl) -- (tr) -- (br) -- cycle;
	\draw[thick] (tl) -- (tr) -- (br) -- (bl) -- cycle;
	\draw[thick,red ] (br) -- (tl);
	\draw[thick,blue] (bl) -- (tr);
	\draw[magenta!50!black] ($(tr)!0.5!(br)-(0.3,0)$) node[above,rotate= 90]{\textit{Southern causal diamond}};
	\draw[blue!50!black] ($(bl)!0.5!(br)+(0,0.3)$) node[above]{\textit{Causal past} $\mathcal O^-$};
	\draw[red!50!black] ($(tl)!0.5!(tr)-(0,0.3)$) node[below]{\textit{Causal future} $\mathcal O^+$};
	\draw[] ($(tr)!0.5!(br)$) node[above,rotate=-90]{South pole};
	\draw[] ($(tl)!0.5!(bl)$) node[above,rotate= 90]{North pole};
	\draw[] ($(tl)!0.5!(tr)$) node[above]{$\mathscr{I}^+_{\text{dS}}$};
	\draw[] ($(bl)!0.5!(br)$) node[below]{$\mathscr{I}^-_{\text{dS}}$};
	\draw[thick] (tl) -- (tr) -- (br) -- (bl) -- cycle;
	\draw[] (tl)node[anchor=south east]{\scriptsize $(T,\chi)=(\frac{\pi}{2},0)$};
	\draw[] (tr)node[anchor=south west]{\scriptsize $(T,\chi)=(\frac{\pi}{2},\pi)$};
	\draw[] (bl)node[anchor=north east]{\scriptsize $(T,\chi)=(-\frac{\pi}{2},0)$};
	\draw[] (br)node[anchor=north west]{\scriptsize $(T,\chi)=(-\frac{\pi}{2},\pi)$};
    \draw[ultra thick,green!70!black] (br)node[circle,fill,inner sep=2.5pt]{} -- (tr)node[circle,fill,inner sep=2.5pt]{};
    \draw[green!70!black] ($(br)!0.2!(tr)$)node[right]{$\bm{\mathcal O}$};
\end{tikzpicture}
\caption{Causal structure of dS$_{d+1}$ spacetime.}
\label{fig:dS PenroseCarter}
\end{figure}
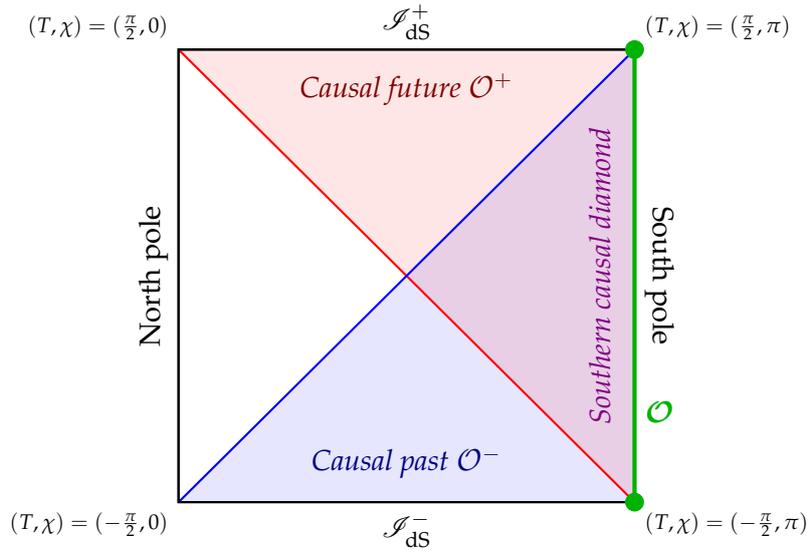

To investigate the causal structure of de Sitter and figure out what is its asymptotic behavior, it is convenient to trade the global time coordinate $\tau$ for the conformal time coordinate $T$ such that $\D s^2_{\text{dS}} = f(T)^2(-\D T^2 + \D \chi^2 + \sin^2\chi \D\Omega_{d-1}^2)$. The defining function of the conformal rescaling reads as $f(T) = \ell \cosh(\tau/\ell)$ where $\tau = \tau(T)$ and $\D \tau = f\D T$. Integrating this condition yields $T = \arctan\sinh(\tau/\ell)$ if $T(\tau=0)=0$. The real line $\tau\in\mathbb R$ is mapped onto $T\in ]-\pi/2,\pi/2[$. In the conformal coordinates $\{T,\chi,x^A\}$ the metric is manifestly conformal to a portion of the flat spacetime, which is smooth everywhere including the boundary represented here by extremal values of the conformal coordinate $T$. This portion is a square in the $(T,\chi)$ plane because both coordinates have an equal finite range of $\pi$: the Penrose-Carter diagram is represented by the Figure \ref{fig:dS PenroseCarter}. Each point of the diagram is a $S^{d-1}$ except the lateral edges which represent the poles of the $S^d$ slices of constant $T$ (or constant $\tau$). These lines are thus collections of true points. Strictly speaking, they are not part of the conformal boundary $\mathscr I_{\text{dS}} \equiv \mathscr I_{\text{dS}}^- \dot\cup \mathscr I_{\text{dS}}^+$ which reunites the past conformal boundary at $T = -\pi/2$ and the future conformal boundary at $T=+\pi/2$. These are spacelike hypersurfaces where any null geodesic has its end-points, starting from a point of $\mathscr I_{\text{dS}}^-$ and terminating at a point of $\mathscr I_{\text{dS}}^+$. Therefore, $\mathscr I_{\text{dS}}^+$ is the future null infinity of de Sitter, which will be of major interest in this thesis. The radial light rays, which have no motion along the $S^{d-1}$ spheres, travel the square at 45 degrees: such a ray originating from the north pole will exactly cross the diagram to reach the south pole. Other light rays with non-trivial motion in the angular directions travel the diagram at less than 45 degrees. 

One important although peculiar feature of the de Sitter spacetime is the presence of two frame-dependent horizons for any observer. It is a fact: an isolated observer in de Sitter cannot have access to the whole story of the spacetime events and cannot send a (light-driven) message to any future region. This contrasts drastically with the asymptotically flat case where a timelike observer can have access to the entire history of the spacetime if he waits for a sufficiently long time, enough to allow its past light cone to intercept all of the observable past events. In the Penrose-Carter diagram \ref{fig:dS PenroseCarter}, we represented the accessible regions for an observer $\mathcal O$ sitting at south pole (for example). His causal past $\mathcal O^-$ is determined by the union of all his past light cones, which are below the null direction connecting $\mathscr I_{\text{dS}}^-$ at north pole to $\mathscr I_{\text{dS}}^+$ at south pole. The latter thus defines the past horizon associated with $\mathcal O$ that prevents him from getting some information about the regions beyond. Similarly, the other diagonal null direction, connecting $\mathscr I_{\text{dS}}^-$ at south pole to $\mathscr I_{\text{dS}}^+$ at north pole, describes a future horizon for $\mathcal O$ because it delimits the region $\mathcal O^+$ to which the observer can send a message, identified as the union of all his future light cones. The intersection $\mathcal O^- \cap \mathcal O^+$ is coined as the southern causal diamond and represents the portion of de Sitter spacetime fully accessible to $\mathcal O$ living at the south pole. These causal properties are transparent in the coordinate chart $\{ t,r,x^A \}$ adapted to the southern observer $\mathcal O$, in which the de Sitter line element is written as
\begin{equation}
\D s_{\text{dS}}^2 = -\left( 1-\frac{r^2}{\ell^2}\right) \D t^2 + \frac{\D r^2}{\left( 1-\frac{r^2}{\ell^2}\right)} + r^2 \, \mathring q_{AB}\D x^A \D x^B. \label{dS static}
\end{equation}
One calls this set of coordinates the \textit{static coordinates} of de Sitter because $\partial_t$ is here a manifest Killing vector and \eqref{dS static} does not contain any cross-term involving $\D t$. It also allows to give another interpretation to the parameter $\ell$. Indeed, choosing $r$ negligible to $\ell$ transforms \eqref{dS static} into the Minkowski line element in spherical coordinates. This means that at small distances the de Sitter spacetime looks locally like the flat space. The cosmological effects become important when the phenomena we want to consider occur for large radii, commensurable to $\ell$. The static coordinates cover only one of the four causal diamonds delimited by the horizons $r^2=\ell^2$ since \eqref{dS static} becomes singular for these values, but it can be defined in each of them. For instance, the superior diamond adapted for observers at infinity is defined for $r>\ell$. It will particularly interest us later on in this thesis. To make contact with our previous discussion, let us cover the southern causal diamond. The past horizon is determined as $r = \ell$ while the future horizon is located at $r=-\ell$. Between these values, the norm of the Killing vector $\partial_t$ is negative, hence its flow correctly describes the time evolution. $\partial_t$ becomes null on the horizons and spacelike in the upper and lower causal diamonds. The absence of a global timelike Killing vector in the de Sitter vacuum has deep and unavoidable consequences for the evolution problem in this spacetime. Without entering into details about the complexity of defining the Hamiltonian in field theories living on de Sitter -- which would be desirable from the point of view of quantum theory, we will just mention some aspects of this difficulty below in the text, even at classical level while dealing with radiative asymptotically de Sitter spacetimes. 

\subsubsection{The anti-de Sitter solution}
Conversely, one defines the \textit{anti-de Sitter} spacetime in $n$ dimensions as the only maximally symmetrical Lorentzian manifold of negative Ricci curvature. In the set of hyperbolic coordinates $\{ \tau,\chi,x^A \}$ ($\tau\in\mathbb R$, $\chi\in\mathbb R^+$ and $x^A$ again parametrize the $S^{d-1}$ sphere), its line element reads as
\begin{equation}
\D s^2_{\text{AdS}} = -\cosh^2\chi \D\tau^2 + \ell^2\left( \D\chi^2 + \sinh^2\chi\,\mathring q_{AB}\D x^A \D x^B \right) \label{AdS global patch}
\end{equation}
for the real parameter $\ell>0$ defined this time as the \textit{AdS radius}. \eqref{AdS global patch} is manifestly static in this patch and solves the Einstein field equations for a negative value of the cosmological constant given by
\begin{equation}
\Lambda = -\frac{d(d-1)}{2\ell^2},\label{Lambda AdS}
\end{equation}
or, equivalently, $R = -d(d+1)/\ell^2$. In contrast to the de Sitter solution, the topology of the AdS spacetime is $\mathbb R\times H^d$ where $H^d$ denotes the upper layer of the $d$ dimensional hyperboloid whose line element is $\D\chi^2 + \sinh^2\chi\,\mathring q_{AB}\D x^A \D x^B$. The latter has the topology of $\mathbb R^d$, hence AdS has in fact the topology of $\mathbb R^{d+1}$, which is non-compact in all directions. This constitutes the major difference between dS and AdS global vacua and will impact the conformal compactification which will only be partial in the second case. 

Geometrically, the AdS spacetime can also be seen as an hyperboloid $\mathscr H_{\text{AdS}}$ of radius $\ell$, but with more peculiar features, since we have to embed it in the Riemannian flat space $\mathbb R^{d,2}$ provided with two time directions $t_1$ and $t_2$. The isometries of AdS$_{d+1}$ form the homogeneous subgroup of isometries of $\mathbb R^{d,2}$, that is, $SO(d,2)$. It encompasses the full rotation group in the hyperspace for constant $t_1,t_2$ directions as well as two copies of the proper orthochronous Lorentz group $SO(d,1)$ mixing $t_1$ or $t_2$ and transverse spacelike coordinates.

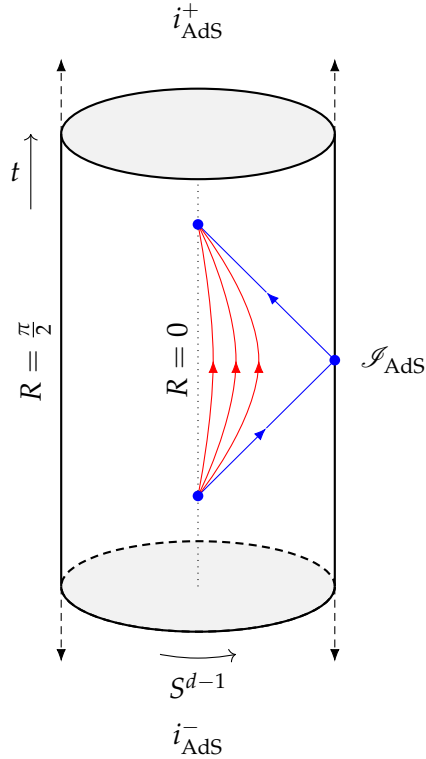
\begin{figure}[!ht]
\centering
\begin{tikzpicture}[scale=1]
\draw[opacity=0] (-3.5,-5.5) -- (-3.5,5.0) -- (3.5,5.0) -- (3.5,-5.5) -- cycle;
\coordinate (top) at (0,3);
\coordinate (bottom) at (0,-3);
\coordinate (centre) at (0,0);
\draw[densely dashed,thick,fill=gray!10] (bottom) ellipse (1.8 and 0.6);
\draw[dotted] (bottom) -- (top);
\draw[thick,fill=gray!10] (top) ellipse (1.8 and 0.6);
\draw[thick] ( $(bottom)+(-1.8,0)$ ) -- ( $(top)+(-1.8,0)$ );
\draw[thick] ( $(bottom)+( 1.8,0)$ ) -- ( $(top)+( 1.8,0)$ );
\draw[thick] (bottom)+(1.8,0) arc (0:-180:1.8 and 0.6);
\path (bottom) + (0,1.2) coordinate (A);
\draw[blue,-Latex] (A) -- ( $(A)+(0.9,0.9)$ );
\draw[blue] ( $(A)+(0.9,0.9)$ ) -- ( $(A)+(1.8,1.8)$ );
\node[right] at ( $(A)+(2.0,1.8)$ ) {$\cI_{\text{AdS}}$};
\draw[blue,-Latex] ( $(A)+(1.8,1.8)$ ) -- ( $(A)+(0.9,2.7)$ );
\draw[blue] ( $(A)+(0.9,2.7)$ ) -- ( $(A)+(0,3.6)$ );
\draw[red] plot [smooth,tension=0.8] coordinates { (A) ( $(A)+(0.8,1.8)$ ) ( $(A)+(0,3.6)$ )};
\draw[red,-Latex] ( $(A)+(0.8,1.8)$ );
\draw[red] plot [smooth,tension=0.8] coordinates { (A) ( $(A)+(0.5,1.8)$ ) ( $(A)+(0,3.6)$ )};
\draw[red,-Latex] ( $(A)+(0.5,1.8)$ );
\draw[red] plot [smooth,tension=0.8] coordinates { (A) ( $(A)+(0.2,1.8)$ ) ( $(A)+(0,3.6)$ )};
\draw[red,-Latex] ( $(A)+(0.2,1.8)$ );
\fill[blue] (A) circle [radius=2pt];
\fill[blue] (A)+(1.8,1.8) circle [radius=2pt];
\fill[blue] ( $(A)+(0,3.6)$ ) circle [radius=2pt];
\draw[densely dashed,-Latex] ( $(bottom)+(-1.8,0)$ ) -- ( $(bottom)+(-1.8,-1)$ );
\draw[densely dashed,-Latex] ( $(bottom)+( 1.8,0)$ ) -- ( $(bottom)+( 1.8,-1)$ );
\draw[densely dashed,-Latex] ( $(top)+(-1.8,0)$ ) -- ( $(top)+(-1.8, 1)$ );
\draw[densely dashed,-Latex] ( $(top)+( 1.8,0)$ ) -- ( $(top)+( 1.8, 1)$ );
\coordinate (legend) at ( $(bottom)+(0,-1)$ );
\node[below] at (legend) {$S^{d-1}$};
\draw[->] plot [smooth,tension=1] coordinates {($(legend)+(-0.5,0.1)$) ($(legend)+(0,0.05)$) ($(legend)+(0.5,0.1)$)};
\draw[<-] ($(top)+(-2.2,0)$) -- ($(top)+(-2.2,-1)$);
\node[left] at ($(top)+(-2.2,-0.5)$) {$t$};
\node[] at ($(top)+(0,1.5)$){$i^+_\text{AdS}$};
\node[] at ($(bottom)-(0,2.0)$){$i^-_\text{AdS}$};	
\node[rotate=90,above] at (centre){$R=0$};
\node[rotate=90,above] at ($(centre)+(-1.8,0)$){$R=\frac{\pi}{2}$};
\end{tikzpicture}
\caption{Causal structure of AdS$_{d+1}$ spacetime.}
\label{fig:AdS PenroseCarter}
\end{figure}

Another practical coordinate system, adapted to static observers in AdS spacetime, is the global static patch $\{t,r,x^A\}$, $t\in\mathbb R$ being the global time and $r\in\mathbb R^+$ a radial coordinate, for which the line element \eqref{AdS global patch} becomes
\begin{equation}
\D s_{\text{AdS}}^2 = -\left( 1+\frac{r^2}{\ell^2}\right) \D t^2 + \frac{\D r^2}{\left( 1+\frac{r^2}{\ell^2}\right)} + r^2 \, \mathring q_{AB}\D x^A \D x^B. \label{AdS static}
\end{equation}
This is the analytical continuation of \eqref{dS static} for $\ell\to i\ell$. It shows explicitly that AdS is also locally flat for distances $r\ll \ell$. In this second coordinate system, it is easy to compactify the radial direction by proposing the change of coordinate $R = \arctan(r/\ell)$. While $r$ runs on the semi-line $[0,+\infty[$, $R$ is limited to $[0,\pi/2[$ and the line element is $\D s_{\text{AdS}}^2 = \ell^2\cos^{-2} R(-\D t^2/\ell^2 + \D R^2+\sin^2 R\, \mathring q_{AB}\D x^A\D x^B)$. The AdS metric is thus conformal to the $\mathbb R\times S^d$ cylinder. The conformal boundary is the spatial infinity $r\to\infty$ located at the double pole $R=\pi/2$ of the metric. We denote it as $\mathscr I_{\text{AdS}}$, whose induced geometry is a $d$ dimensional cylinder aligned on the time direction with $S^{d-1}$ spheres as constant $t$ sections. It is noteworthy that, since the range of the timelike coordinate $t$ remains infinite while the range of the conformal radius $R$ is finite, there is no way to compress further the AdS$_{d+1}$ spactime into a finite range of coordinates for both time and radius if one wants to preserve the condition that null rays evolve along straight lines at 45 degrees. If one attempts to perform another conformal transformation to reduce $t$ in a finite range, the spheres generating the conformal boundary will be mapped to points, since the interval of the conformal radius $R$ will be squeezed to a single point. The Penrose-Carter diagram of AdS$_{d+1}$, depicted in Figure \ref{fig:AdS PenroseCarter}, is thus an infinite cylinder, whose upper and lower boundaries $i^+_\text{AdS}$ and $i^-_\text{AdS}$, rejected to infinity, are respectively future and past timelike infinities. These are not part of the conformal boundary: the AdS$_{d+1}$ spacetime is only partially conformally compact because the relativistic infinities cannot be brought back to finite distance by a conformal transformation.

A static observer whose proper time is $t$, sitting at the coordinates $(r,x^A)$, has neither past nor future horizon and has thus access to the whole spacetime history. This pleasant feature contrasting with the dS case is however counter-balanced by the major oddness of the AdS$_{d+1}$ spacetime, which is that null rays (in blue on Figure \ref{fig:AdS PenroseCarter}) reach the conformal boundary after a finite amount of coordinate time. The data given on a spacelike slice (or \textit{Cauchy slice}) in the past are not sufficient to describe causally what is happening after the rays have encountered the conformal boundary. Hence the Cauchy problem is ill-defined until boundary conditions on $\mathscr I_{\text{AdS}}$ are imposed to decide of the fate of the light ray. For the standard reflexive boundary conditions, the null geodesic bounces on the boundary to dig back to the center of the spacetime, as represented in the Figure \ref{fig:AdS PenroseCarter}. This property connecting the conformal boundary and the bulk of the spacetime manifold is peculiar to AdS and spearheads the AdS/CFT correspondence \cite{Maldacena:1997re,Witten:1998qj} debated in string theory for decades now. AdS is thus a prototypic ``spacetime in a box'' where any test body or radiation is naturally aimed to be repelled by the boundary towards the bulk. This is the case for massive particles whose worldlines are schematized in red on the Penrose-Carter diagram \ref{fig:AdS PenroseCarter}: the Christoffel symbols contain a counter-potential in $\mathcal O(r^2)$ whose importance grows as one approaches $\mathscr I_{\text{AdS}}$ and repulse timelike geodesics back to the center of the spacetime. This observation concludes our review of the asymptotic behavior of AdS et dS spacetimes. We thought it would be valuable to briefly summarize their properties and contrast them with the asymptotic behavior of flat space before discussing with more details the desirable boundary conditions that we are lead to impose in practice.

\subsection{Conformally compact manifolds}
\label{sec:Conformally compact manifolds}
Now that we have reviewed the asymptotic properties we want to focus on, let us describe generically the class of spacetimes for which the conformal compactification can be performed (at least partially, as for the AdS$_{d+1}$ spacetime). This is Penrose's powerful idea to treat relativistic infinities by bringing them back to finite distances and describe asymptotics as geometric properties of the boundary of the conformally compactified unphysical spacetime \cite{Penrose:1964ge}. In that context, a gauge-invariant and robust definition of ``asymptotically (A)dS'' spacetime -- which is merely a conformally compact manifold whose conformal boundary is spacelike (for dS) or timelike (for AdS) -- can be obtained. This constitutes the last step before defining a coordinate system appropriate to discuss the solution space of Al(A)dS$_{d+1}$.

Let us consider a Lorentzian manifold $(\mathscr M,g)$ with boundary $\mathscr I$ on which the metric tensor $g$ is possibly singular (due to the intrinsic nature of infinity, located ``at large distances'') but with the crucial hypothesis that this singularity on $\mathscr I$ is a double pole. This latter holds for the Minkowski vacuum as well as for the (A)dS global solutions. Hence, there exists a smooth function $F$ on $\bar{\mathscr M} = \mathscr M \dot\cup \mathscr I$ such that (1) $F >0$ on $\mathscr M$, $F|_{\mathscr I} = 0$, $\D F |_{\mathscr I} \neq 0$ and (2) $\bar{ g} = F^2 g$ can be extended smoothly on the whole manifold $\bar{\mathscr M}$. The evaluation $|_{\mathscr I}$ denotes the pull-back to the boundary $\mathscr I$ and the unphysical metric $\bar{ g}$ is supposed to be continuously differentiable up to a sufficiently large order allowing us to take as many derivatives as we need. If $(\mathscr M, g)$ satisfies all of these hypotheses, it is called a \textit{conformally compact manifold} \cite{Penrose:1964ge,Penrose:1986ca}.  The first hypothesis (1) allows to choose $F$ as a coordinate on the unphysical spacetime $(\bar{\mathscr M},\bar{ g})$ while the second (2) locates the boundary $\mathscr I$ at the poles of the physical metric $g$. We can thus identify the spacetime infinity with the boundary manifold $\mathscr I$. Obviously, the conformal factor $F$ is not uniquely fixed by the above definition and can be rescaled by a non-vanishing smooth function on $\mathscr M$, $F\to e^{f}F$ without withdrawing any of the previous hypotheses. As a consequence, the unphysical metric is only defined up to the choice of a conformal frame and the conformal compactification process precisely associates a physical metric $ g$ to a conformal class of unphysical metrics $\bar{ g} \sim e^{2f}\bar{ g}$. We will see later how to treat this ambiguity in practice: for instance, in a particular set of coordinates, one can fix the boundary metric $\bar{ g}|_{\mathscr I}$ without loosing this conformal invariance, which will be translated into a residual gauge transformation.

The property of being conformally compact is kinematical and does not bring more constraints on the factor $F$. Now assuming that $(\mathscr M, g)$ is a solution of Einstein gravity, the physical metric tensor $ g = F^{-2} \bar{ g}$ has to solve the dynamical Einstein's field equations. This requirement fixes the norm of the vector $N = (\bar g^{\mu\nu}\partial_\nu F)\partial_\mu$ normal to $\mathscr I$, \textit{i.e.} the quantity $N^2 = \bar g^{\mu\nu}\partial_\mu F\partial_\nu F$ which extends smoothly to $\mathscr I$. Indeed, a straightforward evaluation of the curvature tensor associated with $ g$ yields (see \textit{e.g.} \cite{Skenderis:2002wp})
\begin{equation}
R_{\mu\nu\alpha\beta} = -N^2(g_{\mu\alpha}g_{\nu\beta} - g_{\mu\beta}g_{\nu\alpha}) + \mathcal O(F^{-3}), \label{riemann at infinity}
\end{equation}
which implies that a conformally compact manifold asymptotes (when $F\to 0$) to a constant curvature manifold. The first term is $\mathcal O(F^{-4})$ since the divergence of $g$ at infinity is a double pole. Focusing on this leading term and imposing the Einstein field equations $G_{\mu\nu} + \Lambda g_{\mu\nu} = 0$, it is not hard to show that
\begin{equation}
N^2\Big|_{\mathscr I} = \frac{\eta}{\ell^2}, \label{normal at infinity}
\end{equation}
where $\eta = -\text{sgn}(\Lambda)$ and the length parameter $\ell$ is related to $\Lambda$ as \eqref{Lambda dS} and \eqref{Lambda AdS}, or 
\begin{equation}
\Lambda = -\eta \frac{d(d-1)}{2\ell^2}.
\end{equation}
The Lorentzian manifold $(\mathscr M, g)$ is \textit{asymptotically locally (A)dS$_{d+1}$} if it is conformally compact and satisfies \eqref{normal at infinity} for $\eta = -1$ (dS case) or $\eta=1$ (AdS case) \cite{Ashtekar:1984zz,Ashtekar:1999jx,deHaro:2000vlm,Papadimitriou:2005ii}. As an immediate consequences of the definitions, we see that $\mathscr I$ is timelike when $\eta=1$ and spacelike when $\eta=-1$. This matches perfectly with the asymptotic analysis for global AdS and dS spacetimes. In particular, the global AdS et dS metrics (in $d+1$ dimensions) are maximally symmetric solutions of Einstein's gravity, hence constant curvature manifolds. The equation \eqref{riemann at infinity} is thus the statement, in a covariant way, that the metric $ g$ approaches the (A)dS metric near infinity. But this asymptote is just local because \eqref{riemann at infinity} is a pointwise equation. It does not impose such rigid notion of strict asymptotically (A)dS spacetimes where the metric at infinity \textit{is} exactly the (A)dS line element. It brings also no restriction on the conformal structure or the topology of the boundary. For instance, the boundary metric on $\mathscr I$ is completely free since \eqref{riemann at infinity} is just the property of being conformally compact phrased by means of the curvature tensor.

\subsection{Starobinsky/Fefferman-Graham gauge}
The task now is to make good use of the features discussed in the previous section to define a set of coordinates appropriate to the treatment of Al(A)dS$_{d+1}$ spacetimes with the aim of defining an associated phase space later on. Historically, it was first discovered heuristically by Starobinsky \cite{Starobinsky:1982mr} while studying the asymptotic properties of non-isotropic cosmological expansions tending towards isotropization, then established mathematically by Fefferman and Graham \cite{Fefferman:1985aa} as the patch exploiting at best the conformal compactness in the presence of a non-vanishing cosmological constant. We call it the \textit{Starobinsky/Fefferman-Graham coordinate system} (SFG for short) and this is how it is built.

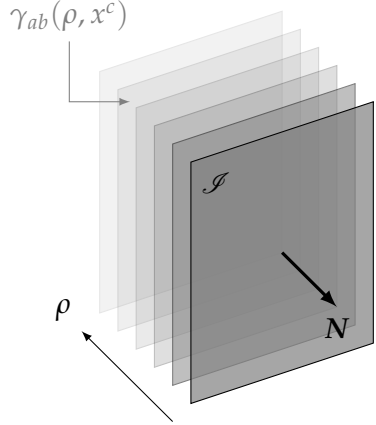
\begin{figure}[!ht]
\centering
\begin{tikzpicture}[scale=0.8]
\draw[opacity=0] (-2.5,-4.5) -- (3.5,-4.5) -- (3.5,3) -- (-2.5,3) -- cycle;
\coordinate (A) at (0,-4);
\coordinate (B) at (3,-3);
\coordinate (C) at (3,1);
\coordinate (D) at (0,0);
\coordinate (M) at ($(C)!0.5!(A)$);
\def\dx{-0.3};
\def\dy{0.3};
\foreach \k in {5,4,...,1}
{
\def\fillopa{0.5/\k};
\def\opa{0.5/\k}
\draw[black,opacity=\opa,fill=black!50,fill opacity=\fillopa] ($(A)+(\k*\dx,\k*\dy)$) -- ($(B)+(\k*\dx,\k*\dy)$) -- ($(C)+(\k*\dx,\k*\dy)$) -- ($(D)+(\k*\dx,\k*\dy)$) -- cycle;
}
\draw[black,fill=black!50,fill opacity=0.7] (A) -- (B) -- (C) -- (D) -- cycle;
\draw[very thick,black,-latex] (M) -- ($(M)-3*(\dx,\dy)$) node[below]{$\bm N$};
\draw[] (D)node[anchor=north west]{$\mathscr I$};
\draw[black!50,latex-] (-1,1) -| (-2,2) node[black!50,anchor=south]{$\gamma_{ab}(\rho,x^c)$};
\coordinate (R) at ($(A)-(-\dx,\dy)$);
\draw[-latex] (R) -- ($(R)+5*(\dx,\dy)$) node[anchor=south east]{$\rho$}; 
\end{tikzpicture}
\caption{Schematic picture of the \\Starobinsky/Fefferman-Graham foliation.}
\label{fig:FG}
\end{figure}

Let $\mathscr U$ be a finite neighborhood of $\mathscr I$ in the unphysical manifold $\bar{\mathscr M}$ set in the conformal frame defined by $F$. The latter being not fixed, one can choose another frame in the conformal class in order to get a modified defining function $\rho \equiv e^f F$ satisfying $|\D \rho|^2 = \eta/\ell^{2}$ as in \eqref{normal at infinity}. The most simple way to achieve this construction is to freeze the original conformal frame and set $f|_{\mathscr I} = 0$. Recalling the first point of the definition of conformal compactness, $\rho$ can be seen as a coordinate on $\mathscr U$. The patch is completed by $d$ coordinates $\{ x^a \}$ defined as orthogonal to $\rho$ in $\mathscr U$ according to $\bar{g}$. The orthogonality can always be demanded by gauge-fixing $d+1$ functions in $g$ (\textit{i.e.} fixing the lapse and the shift with respect to displacements along the $\rho$ coordinate), which amounts to build asymptotically a set of Gaussian normal coordinates around infinity $\mathscr I$, located at $\rho\to 0$ by definition. This is precisely what the SFG coordinate system is and the line element of any Al(A)dS$_{d+1}$ spacetime can be written as
\begin{equation}
\D s^2 = \eta\frac{\ell^2}{\rho^2}\D\rho^2 + \gamma_{ab}(\rho,x^c)\D x^a \D x^b.
\label{FG gauge}
\end{equation}
In \eqref{FG gauge}, $\gamma_{ab}(\rho,x^c)$ is the induced metric on leaves of constant $\rho$ and verifies $\gamma_{ab} = \mathcal O(\rho^{-2})$ by assuming conformal compactness .The local geometry around $\mathscr I$ is schematized in Figure \ref{fig:FG}. We call $\rho\geq 0$ the \textit{holographic coordinate} because it lifts the information intrinsically defined on the boundary $\mathscr I$ located at $\rho = 0$ into the bulk $\rho >0$. We choose it to have units of inverse of length and the transverse coordinates $x^a$ are taken to be dimensionless. $\rho$ is timelike for dS asymptotics ($\eta < 0$) and spacelike for AdS asymptotics ($\eta >0$). As a consequence, the boundary $\mathscr I$ is spacelike when $\Lambda>0$ and timelike when $\Lambda<0$.

The induced metric $\gamma_{ab}(\rho,x^c)$ represents the only dynamical field that will be constrained by the Einstein field equations. Apart of its asymptotic behavior, the construction above does not give more information on $\gamma_{ab}$. In order to solve the equations of motion, one has to start by assigning a power-series in $\rho$ to this field, see what occurs practically for the various coefficients and check the consistency of the ansatz. One sets \cite{deHaro:2000vlm,Henningson:1998gx}
\begin{equation}
\gamma_{ab} = \frac{1}{\rho^2} g_{ab}^{(0)} + \frac{1}{\rho} g_{ab}^{(1)} + g_{ab}^{(2)}+ \dots + \rho^{d-2} g_{ab}^{(d)}  + \rho^{d-2} \ln \rho^2 \tilde{g}_{ab}^{[d]}   +  \mathcal{O}(\rho^{d-1}) ,
\label{preliminary FG}
\end{equation}
where the logarithmic term appears only for even $d$. We will review how to derive the solution space in great details in the next section, but at this point, we can already get some flavor of what we have to expect. First of all, since \eqref{riemann at infinity} is imposed by design, we do not expect to get more constraints on the leading order coefficient $g^{(0)}_{ab}(x^c) \equiv \lim_{\rho\to 0}\rho^2\gamma_{ab}(\rho,x^c)$. We call this field the \textit{boundary metric}: it constitutes the first set of free data that we have at disposal. To be precise, from the conformal compactification process, $g^{(0)}_{ab}$ is only defined up to a conformal rescaling and we should call it a representative of the conformal class $[g^{(0)}_{ab}] = \{g^{(0)}_{ab} \sim e^{2\tilde f} g^{(0)}_{ab} \}$ where $\tilde f$ is a smooth boundary field. The first term in \eqref{FG gauge} being invariant under $\rho \to e^{\tilde f}\rho$, the conformal frame can be set on the boundary and the possibility to perform conformal transformations will be translated into a residual gauge diffeomorphism acting as a boundary Weyl rescaling, see section \ref{sec:Residual gauge transformations}. So there is nothing to worry about the fixation of the representative of the boundary metric, which is thus a well-defined object.

Since the Einstein equations are second-order partial differential equations, starting from $g^{(0)}_{ab}$, one gets, among other constraints, a second-order recursion relation fixing the subleading coefficients $g^{(k)}_{ab}$, $k<d$, as unique local functions of the data encoded in the boundary metric (see \textit{e.g.} \cite{Fischetti:2012rd,deHaro:2000vlm,Papadimitriou:2010as}). In this recursion process, a particular behavior is observed for the step aimed at constraining $g^{(d)}_{ab}$ for each dimension $d$, because this coefficient does not intervene in the algebraic constraint at this order. Although this algebraic relation is tautologic for odd $d$, it brings an additional and unwelcome constraint on lower order coefficients in even $d$. This motivates the introduction of a logarithmic term in \eqref{preliminary FG} for even $d$ only in order to maintain the consistency of the ansatz after imposing the equations of motion. The new field $\tilde g_{ab}^{[d]}$ that is introduced is also a local function of the boundary metric. The fact that $g^{(d)}_{ab}$ is absent from the algebraic constraint means that it provides a new set of free data. The latter does not share the same nature as $g^{(0)}_{ab}$ because it appears further in the expansion in $\rho$ and, therefore, does care about the bulk dynamics. In other words, $g^{(d)}_{ab}$ is subjected to Hamiltonian and momentum constraints on the leaves of constant $\rho$ and sees thus its trace and covariant divergence fixed by the equations of motion. Beyond this subtlety, the recursion process resumes and gives the higher order coefficients $g^{(k)}_{ab}$ for $k>d$ as local functions of $g^{(0)}_{ab}$ and $g^{(d)}_{ab}$ and the expansion \eqref{preliminary FG} is completely determined on-shell by specifying $g_{ab}^{(0)}$ and ${g}_{ab}^{(d)}$. For future purposes, we define the \textit{holographic stress-energy tensor} as
\begin{equation}
\boxed{
T_{ab}^{[d]} = \frac{d}{16 \pi G} \frac{\eta}{\ell} \left( g_{ab}^{(d)} + X^{[d]}_{ab}[g^{(0)}] \right),
\label{holographic stress-energy tensor}
}
\end{equation}
where $X_{ab}^{[d]} [g^{(0)}]$ is a function which vanishes for $d$ odd \cite{deHaro:2000vlm, Balasubramanian:1999re} and is aimed at ensuring that \eqref{holographic stress-energy tensor} is divergence free (see \eqref{X2} and \eqref{X4} for expressions in $d=2$ and $d=4$).

Finally, let us see how the (A)dS global vacua are written in the SFG gauge \eqref{FG gauge}. The interesting point is that they are conformally flat (the Weyl tensor vanishes identically). Consequently, it can be shown that $g^{(d)}_{ab} = 0$, which is related to the fact that $\mathscr I$ has no extrinsic curvature when embedded into the unphysical spacetime $\bar{\mathscr M}$. Within these hypotheses, the recursion process terminates with a finite number of terms and $\gamma_{ab} = \rho^{-2}g^{(0)}_{ab} + g^{(2)}_{ab} + \rho^2 g^{(4)}_{ab}$ is the exact expansion in $\rho$ for the induced metric. We have
\begin{equation}
g^{(0)}_{ab}\D x^a \D x^b = -\frac{\eta}{\ell^2}\D t^2 + \mathring q_{AB}\D x^A \D x^B
\end{equation}
in a set of boundary coordinates $(t,x^A)$ foliating $\mathscr I$ by constant $t$ $(d-1)$ unit-round spheres covered by the angles $x^A$. The subleading pieces are determined in terms of the boundary metric as \eqref{g2} and $g^{(4)}_{ab} = \frac{1}{4}g^{(2)}_{ac}g_{(0)}^{cd}g^{(2)}_{bd}$.

\subsection{Solution space}
\label{sec:Solution space}

Let us make this discussion more precise by entering into the explicit resolution of the Einstein equations. We closely follow \cite{Henningson:1998gx,deHaro:2000vlm} up to some different choices of conventions and notations. In the SFG gauge $g_{\rho\rho} = \eta \frac{\ell^2}{\rho^2}$, $g_{\rho a} = 0$, the coefficients of the Levi-Civita connection can be expressed in terms of the induced metric $\gamma_{ab}(\rho,x^c)$ on leaves of constant $\rho$ as
\begin{equation}
\begin{split}
&\Gamma^\rho_{\rho\rho} = -\frac{1}{\rho}, \quad \Gamma^\rho_{\rho a} = 0,\quad \Gamma^a_{\rho\rho} = 0, \\
&\Gamma^a_{b\rho} = \frac{1}{2}\gamma^{ac}\partial_\rho \gamma_{bc}, \quad \Gamma^\rho_{ab} = -\frac{1}{2}\eta \frac{\rho^2}{\ell^2}\partial_\rho \gamma_{ab}, \quad \Gamma^a_{bc} = \Gamma^a_{bc}[\gamma].
\end{split}
\end{equation}
For practical computations, it is useful to notice that $\Gamma^{a}_{\rho a} = -\eta \frac{\ell^2}{\rho^2}\gamma^{ab}\Gamma^\rho_{ab}$. Denoting by $\mathcal D_a$ the Levi-Civita connection associated with $\gamma_{ab}(\rho,x^c)$, the components of the Ricci tensor read as
\begin{equation}
\begin{split}
R_{ab} &= R_{ab}[\gamma] + (\partial_\rho+\Gamma^\rho_{\rho\rho})\Gamma^\rho_{ab} + \Gamma^\rho_{ab} \Gamma^c_{c\rho} - 2 \Gamma^{\rho}_{c(a}\Gamma^c_{b)\rho} \\
&= R_{ab}[\gamma] - \frac{1}{2}\eta \frac{\rho^2}{\ell^2} \left[ \Big(\partial_\rho  + \frac{1}{\rho}\Big)\partial_\rho \gamma_{ab} - \partial_\rho \gamma_{c(a}\gamma^{cd}\partial_\rho \gamma_{b)d} + \frac{1}{2}(\gamma^{cd}\partial_\rho \gamma_{cd})\partial_\rho \gamma_{ab} \right], \\
R_{\rho a} &= -2 \mathcal D_{[a} \Gamma^b_{b]\rho} = \frac{1}{2}\left[ \mathcal D^b (\partial_\rho \gamma_{ab}) - \mathcal D_{a} (\gamma^{bc}\partial_\rho \gamma_{bc}) \right], \\
R_{\rho\rho} &= -(\partial_\rho - \Gamma^\rho_{\rho\rho})\Gamma^a_{a\rho} - \Gamma^{a}_{b\rho}\Gamma^b_{a\rho} = -\frac{1}{2}\left( \partial_\rho + \frac{1}{\rho} \right) (\gamma^{ab}\partial_\rho \gamma_{ab}) - \frac{1}{4}\gamma^{ac}\gamma^{bd}\partial_\rho \gamma_{ab}\partial_\rho \gamma_{cd}  .
\end{split}
\end{equation}
The Ricci curvature is
\begin{equation}
R = \gamma^{ab} R_{ab} [\gamma] - \eta\frac{\rho^2}{\ell^2} \left[ \gamma^{ab}\left(\partial_\rho  + \frac{1}{\rho}\right)\partial_\rho\gamma_{ab} - \frac{3}{4}\gamma^{ac}\gamma^{bd}\partial_\rho\gamma_{ab}\partial_\rho\gamma_{cd} + \frac{1}{4}(\gamma^{ab}\partial_\rho\gamma_{ab})^2\right].
\end{equation}

\subsubsection{Organization of the Einstein equations}
Due to the particular form of the bulk metric, vacuum Einstein's equations read as follows: \begin{align}
R_{\rho\rho} &= -\frac{d}{\rho^2}  , \label{Einstein rhorho} \\
R_{\rho a} &= 0 , \label{Einstein rhoa} \\
R_{ab} &= -\eta\frac{d}{\ell^2}\gamma_{ab} \label{Einstein ab}
\end{align}
when the Ricci curvature is set to its on-shell value
\begin{equation}
R = -\eta \frac{d(d+1)}{\ell^2}  . \label{Einstein supp}
\end{equation}
The four equations \eqref{Einstein rhorho}-\eqref{Einstein supp} can be solved order by order in $\rho$ if $\gamma_{ab}(\rho,x^c)$ is expressed as \eqref{preliminary FG}. We can already eliminate $g_{ab}^{(1)}$ since the leading order $\mathcal O(1/\rho)$ of \eqref{Einstein ab} imposes that $(d-1)g^{(1)}_{ab} = 0$, hence this field is zero for any interesting case $d\geq 2$. The ansatz \eqref{preliminary FG} is designed to ensure consistency of the Einstein equations at each order in $\rho$, including the logarithmic branch about which we will give some comments below. Hence the radial dependence does not have to be determined and is spoiled out directly by the power series. As a consequence, in the following, all fields under consideration are local functions of the boundary coordinates $\{ x^a \}$ and indices are lowered and raised by the boundary metric $g^{(0)}_{ab}$ and its inverse $g_{(0)}^{ab}$, \textit{e.g.} $g_{(2)}^{cd}= g^{ca}_{(0)} g^{db}_{(0)}  g^{(2)}_{ab}$.
 
The purely radial constraint \eqref{Einstein rhorho} at order $\mathcal O(\rho^{k-2})$ fixes the trace of $g_{ab}^{(k)}$ with respect to $g^{(0)}_{ab}$ for $k\geq 3$. The $\mathcal O(\rho^0)$ of the equation is tautologic, hence the first non-trivial trace, $g_{(0)}^{ab}g^{(2)}_{ab}$, is not actually fixed by \eqref{Einstein rhorho} and requires a particular treatment. In fact, using $R[\gamma] = \rho^2 R^{(0)}+\mathcal O(\rho^3)$, that trace can be extracted easily from \eqref{Einstein supp} at leading order as
\begin{equation}
g^{ab}_{(0)} g^{(2)}_{ab} = -\frac{\eta\,\ell^2 }{2(d-1)}R^{(0)}. \label{Tr g2}
\end{equation}
For any dimension $d$, the leading logarithmic term in \eqref{Einstein rhorho} imposes that $\tilde g^{[d]}_{ab}$ is trace-free. 

Once the traces of the various coefficients have been fixed, the contribution of the momentum constraint \eqref{Einstein rhoa} at order $\mathcal O(\rho^{k-1})$  fixes the covariant divergence of $g^{(k)}_{ab}$ with respect to $g^{(0)}_{ab}$ for $k\geq 2$. In particular, $D^b g^{(k)}_{ab} = 0$ is identically zero when $k$ is odd. For any dimension $d$, the leading logarithmic term in \eqref{Einstein rhoa} imposes that $\tilde g^{[d]}_{ab}$ is divergence-free. 

Once the traces and divergences of the various coefficients have been fixed by \eqref{Einstein rhorho} and \eqref{Einstein rhoa}, the transverse components \eqref{Einstein ab} are solved algebraically to get the explicit expressions of the aforementioned coefficients in terms of lower order coefficients. Starting from $g^{(1)}_{ab}=0$, one can show inductively that $(d-k)g^{(k)}_{ab} = 0$ for $k$ odd at order $\mathcal O(\rho^{k-2})$, hence only even powers of $\rho$ appear in \eqref{preliminary FG} up to $\mathcal O(\rho^{2[\frac{d-1}{2}]})$ where $[x]$ indicates the integer part of $x$. For $k$ even, one gets generically $(d-k)g^{(k)}_{ab} = \mathcal C^{(k)}[g^{(0)},...,g^{(k-1)}]$ for $k < d$, ${\mathcal C'}^{(d)}[g^{(0)},...,g^{(d-1)},\tilde g^{[d]}]=0$ for $k=d$ and $(d-k)g^{(k)}_{ab} = {\mathcal C''}^{(k)}[g^{(0)},...,g^{(d)},\tilde g^{[d]},...,g^{(k-1)}]$ for $k>d$. In any dimension $d$, we see that the expression $g_{ab}^{(d)}$ is always left unfixed by the equations of motion, only its trace and divergence can be computed from \eqref{Einstein rhorho} and \eqref{Einstein rhoa} respectively. We also remark that the logarithmic piece $\rho^{(d-2)}\ln\rho^2 \tilde g^{[d]}_{ab}$ is necessary to ensure the consistency of Einstein's equations when $d$ is even. Indeed, if we do not introduce the field $\tilde g^{[d]}_{ab}$, the equation $\tilde{\mathcal C}^{(d)}[g^{(0)},...,g^{(d-1)}]=0$ brings in general an additional constraint on the coefficients $g^{(0)}_{ab},...,g^{(d-1)}_{ab}$ which are supposed to have been fixed at lower order in $\rho$. Rather than that, ${\mathcal C'}^{(d)}[g^{(0)},...,g^{(d-1)},\tilde g^{[d]}]=0$ gives in these cases the explicit solution for $\tilde g^{[d]}_{ab}$ in terms of lower order coefficients. To summarize,
\begin{equation}
\tilde g^{[2k+1]}_{ab} = 0 , \quad \tilde g^{[2k]}_{ab} \neq 0.
\end{equation}
A notable exception occurs for $d=2$ as we will see below in more details, simply because ${\mathcal C'}^{(2)}[g^{(0)}]=0$ on-shell without the help of any additional field.

\subsubsection{Trace of metric coefficients} 
Apart of the trace of $g^{(2)}_{ab}$ set as \eqref{Tr g2}, we extract the various traces from the equation \eqref{Einstein rhorho}, starting at $\mathcal O(\rho)$ order:
\begin{align}
g^{ab}_{(0)} g^{(3)}_{ab} &= 0  , \label{Tr g3} \\
g^{ab}_{(0)} g^{(4)}_{ab} &= \frac{1}{4} g^{ab}_{(2)} g^{(2)}_{ab}   , \label{Tr g4} \\
g^{ab}_{(0)} g^{(5)}_{ab} &= 0   , \label{Tr g5} \\
g^{ab}_{(0)} g^{(6)}_{ab} &= \frac{2}{3} g_{(2)}^{ab}g^{(4)}_{ab} - \frac{1}{6}(g_{(0)}^{ab}g^{(2)}_{ab})^3  ,\  \dots \label{Tr g6}
\end{align}
These equations help to compute the trace of the holographic stress-tensor as
\begin{equation}
\mathcal{T}^{[2k+1]} \equiv  g_{(0)}^{ab} T_{ab}^{[2k+1]} = 0, \, \forall \, k \in \mathbb N_0. \label{trace even}
\end{equation} Furthermore, for $d=2$, we have 
\begin{equation}
\mathcal{T}^{[2]} \equiv g_{(0)}^{ab} T_{ab}^{[2]} = \frac{c}{24 \pi} R^{(0)} = \frac{\ell}{\sqrt{|g^{(0)}|}} L_{EH}[g^{(0)}]  \label{trace 3d}
\end{equation} where $c = \frac{3 \ell}{2G}$ is the Brown-Henneaux central charge \cite{Brown:1986nw} and $L_{EH}$ is the Einstein-Hilbert Lagrangian density. For $d=4$, we have
\begin{equation}
\begin{split}
\mathcal{T}^{[4]} \equiv g_{(0)}^{ab} T_{ab}^{[4]} =& - \frac{1}{16 \pi G}\frac{\eta}{\ell}\left[(g^{cd}_{(0)} g_{cd}^{(2)})^2 - g_{(2)}^{cd} g^{(2)}_{cd}\right] = \frac{\eta\,\ell^3}{64 \pi G} \left( R_{ab}^{(0)} R^{ab}_{(0)} - \frac{R_{(0)}^2}{3} \right) \\
=& \frac{\eta\,\ell^3}{4\sqrt{|g^{(0)}|}} \left( L_{QCG(1)}[g^{(0)}] - \frac{1}{3} L_{QCG(2)}[g^{(0)}] \right) 
\end{split} \label{trace 4d}
\end{equation}  where we used \eqref{Tr g4} and \eqref{g2} explicitly. We see that \eqref{trace 3d} and \eqref{trace 4d} reproduce the conformal anomaly in three and five dimensions \cite{Henningson:1998gx}. Interestingly, $\mathcal{T}^{[4]}$ involves the Lagrangian densities $L_{QCG(1)}[g^{(0)}]$ and $L_{QCG(2)}[g^{(0)}]$ of the quadratic curvature gravity on the boundary \cite{Salvio:2018crh,Stelle:1977ry}. The expressions of $L_{QCG(1)}$ and $L_{QCG(2)}$ are included in appendix \ref{New massive gravity}.

\subsubsection{Divergence of metric coefficients} 
Developing the momentum constraint \eqref{Einstein rhoa} leads to the generic form $D^{b} g^{(k)}_{ab} = -D^{b} X^{[k]}_{ab} + V_a^{[k]}$ where $D_a$ is the Levi-Civita connection with respect to $g^{(0)}_{ab}$. This equation is crucial because it allows to check explicitly that the boundary stress-tensor $T_{ab}^{[d]}$ is indeed divergence-free. For any odd $k$, we simply have $X_{ab}^{[k]} = 0$ and $V_a^{[k]} = 0$. For even $k$, we can show that
\begin{align}
X^{[2]}_{ab} &= -(g^{cd}_{(0)} g_{cd}^{(2)})g_{ab}^{(0)}, \quad  V_a^{[2]} = 0, \label{X2} \\
X^{[4]}_{ab} &= -\frac{1}{8}g_{ab}^{(0)} \left[ (g_{(0)}^{cd} g^{(2)}_{cd})^2 - g_{(2)}^{cd} g^{(2)}_{cd}  \right] - \frac{1}{2} g_{(2)}^{ac} g_{(0)}^{cd} g^{(2)}_{bd} + \frac{1}{4} (g_{(0)}^{cd} g^{(2)}_{cd}) g^{(2)}_{ab} ,  \quad V_a^{[4]} = 0,  \ \dots \label{X4}
\end{align}
These results show that the combination $g_{ab}^{(d)}+X^{[d]}_{ab}$ is sufficient to get the conservation of the holographic stress-tensor for dimensions $d\leq 4$, 
\begin{equation}
\boxed{ D^a T_{ab}^{[d]} = 0. }
\label{divergence free}
\end{equation}
One can notice the presence of a non-vanishing $V_a$ covector for $d=6$ and generically beyond. It makes the construction of candidates for $T_{ab}^{[d]}$ from the equations of motion slightly more involved. Moreover, one can also check that $X_{ab}^{[6]}$ contains an anti-symmetric part, which was not the case for lower dimensions. However one can perform a field redefinition from $g_{ab}^{(6)}$ in order to get a covariantly conserved symmetrical tensor \cite{deHaro:2000vlm}. 

\subsubsection{Explicit expressions of metric coefficients} 
Now that we have determined the trace and the divergence of the metric coefficients we are interested in, the last equation \eqref{Einstein ab} allows to derive their explicit form in terms of lower order coefficients. At leading order $\mathcal O(\rho^0)$, we find
\begin{equation}
g_{ab}^{(2)} = -\frac{\eta \ell^2}{d-2} \left( R_{ab}^{(0)} - \frac{R^{(0)}}{2(d-1)}g_{ab}^{(0)}\right), \ \forall d \geq 3. \label{g2}
\end{equation}
For $d=2$, the $\mathcal O(\rho^0)$ contribution of \eqref{Einstein ab} does not constrain $g_{ab}^{(2)}$ but fixes the logarithmic term $\tilde g_{ab}^{[2]}$ in the expansion and states nothing but $\tilde g^{[2]}_{ab} = \frac{1}{2} \eta \ell^2 (R_{ab}^{(0)} - \frac{1}{2}R^{(0)}g_{ab}^{(0)})$. This leads to
\begin{equation}
\tilde g_{ab}^{[2]} = 0,
\end{equation}
since the Einstein tensor vanishes identically in two dimensions. At next order $\mathcal O(\rho)$, we find
\begin{equation}
(d-3)g^{(3)}_{ab} = 0,
\end{equation}
which means that $g^{(3)}_{ab}$ is identically zero except for $d=3$ where it is unconstrained. Going further at order $\mathcal O(\rho^2)$, we find for $d\neq 4$
\begin{equation}
(d-4)g^{(4)}_{ab} + g_{ac}^{(2)} g_{(0)}^{cd} g_{bd}^{(2)} - \frac{1}{4}g_{(2)}^{cd}g^{(2)}_{cd}g_{ab}^{(0)} + \frac{1}{2}\eta\ell^2 R^{(2)}_{ab}[g^{(0)}] = 0 \label{g4 temp}
\end{equation}
where we used the notation $R_{ab}[\gamma] = R^{(0)}_{ab}[g^{(0)}] + \rho^2 R_{ab}^{(2)}[g^{(0)}] + \mathcal O(\rho^3)$. In order to extract some information from this equation, we need to compute the first subleading piece of the Ricci tensor. First of all, we can remark that ${\Gamma^a}_{bc}[\gamma] = {\Gamma^a}_{bc}[g^{(0)}] + \rho^2 \Gamma^a_{(2)bc}[g^{(0)}] + \mathcal O(\rho^3)$, where
\begin{equation}
\begin{split}
\Gamma^a_{(2)bc}[g^{(0)}] &= \frac{1}{2}g_{(0)}^{ad} (\partial_b g_{dc}^{(2)} + \partial_c g_{bd}^{(2)} - \partial_d g^{(2)}_{bc} ) - \frac{1}{2}g_{(2)}^{ad} (\partial_b g_{dc}^{(0)} + \partial_c g_{bd}^{(0)} - \partial_d g^{(0)}_{bc} ) \\
&= \frac{1}{2}g_{(0)}^{ad} (D_b g_{cd}^{(2)} + D_c g_{bc}^{(2)} - D_d g^{(2)}_{bc})
\end{split}
\end{equation}
is a covariant $(1,2)$-tensor with respect to the boundary geometry and only depends on $g^{(0)}_{ab}$ and its curvature because of \eqref{g2}. As a consequence,
\begin{align}
&R_{ab}^{(2)}[g^{(0)}] = D_c \Gamma^c_{(2)ab}[g^{(0)}] - D_b \Gamma^c_{(2)ac}[g^{(0)}] \nonumber \\
&\quad = D^c D_{(a} g^{(2)}_{b)c} - \frac{1}{2}D^2 g_{ab}^{(2)} - \frac{1}{2}D_aD_b (g^{cd}_{(0)}g^{(2)}_{cd}) \\
&\quad = -\frac{\eta\,\ell^2}{d-2}\left[R_{ca}^{(0)}R^{c}_{(0)b} - R_{acbd}^{(0)}R^{cd}_{(0)} + \frac{(d-2)}{4(d-1)}D_aD_bR^{(0)} - \frac{1}{2} D^c D_c R_{ab}^{(0)} + \frac{D^c D_c R^{(0)}}{4(d-1)} g^{(0)}_{ab} \right].\nonumber 
\end{align}
We can deduce that
\begin{equation}
R[\gamma] = \gamma^{ab}R_{ab}[\gamma] = \rho^2 R^{(0)} + \rho^4 \frac{\eta\,\ell^2}{(d-2)}\left[ R^{ab}_{(0)}R_{ab}^{(0)} - \frac{1}{2(d-1)}(R^{(0)})^2\right] + \mathcal O(\rho^5).
\end{equation}
Finally computing
\begin{equation}
\begin{split}
g_{(2)a}^{c} g_{bd}^{(2)} &= \frac{\ell^4}{(d-2)^2}\left[ R_{(0)a}^{c} R_{bd}^{(0)} - \frac{R^{(0)}}{d-1}R_{ab}^{(0)} + \frac{(R^{(0)})^2}{4(d-1)^2}g^{(0)}_{ab} \right], \\
g_{(2)}^{cd}g^{(2)}_{cd} &= \frac{\ell^4}{(d-2)^2}\left[ R^{cd}_{(0)} R_{cd}^{(0)} + \frac{(4-3d)}{4(d-1)^2}(R^{(0)})^2\right],
\end{split} \label{g2square}
\end{equation}
we can solve \eqref{g4 temp} for $g^{(4)}_{ab}$ if $d\neq 4$ to get
\begin{equation}
\begin{aligned}
g^{(4)}_{ab} = \frac{\ell^4}{(d-4)}  & \left[ \frac{1}{8(d-1)}D_a D_b R^{(0)} - \frac{1}{4(d-2)}D_c D^c R_{ab}^{(0)} + \frac{D_c D^c R^{(0)}}{8(d-1)(d-2)}g_{ab}^{(0)} \right.\\
& \quad - \frac{1}{2(d-2)}R_{acbd}^{(0)}R^{cd}_{(0)} + \frac{(d-4)}{2(d-2)^2}R_a^{(0)c} R^{(0)}_{cb} + \frac{R^{(0)}}{(d-1)(d-2)^2}R_{ab}^{(0)} \\
& \quad  \left. +\, \frac{1}{4(d-2)^2}R^{cd}_{(0)} R_{cd}^{(0)} g^{(0)}_{ab} - \frac{3d}{16(d-1)^2(d-2)^2}(R^{(0)})^2 g^{(0)}_{ab} \right] . 
\end{aligned} \label{g4}
\end{equation}
When $d=4$, \eqref{g4 temp} fixes the logarithmic term $\tilde g^{[4]}_{ab}$ in the expansion. After some algebra we obtain
\begin{equation}
\begin{split}
\tilde g^{[4]}_{ab} &= \frac{1}{2} g_{ac}^{(2)}g^{cd}_{(0)}g^{(2)}_{bd} - \frac{1}{8} g_{(2)}^{cd} g^{(2)}_{cd} g^{(0)}_{ab} + \frac{\eta\ell^2}{4}R^{(2)}_{ab}  \\
&= \frac{1}{2} g_{ac}^{(2)}g^{cd}_{(0)}g^{(2)}_{bd} - \frac{1}{8} g_{(2)}^{cd} g^{(2)}_{cd} g^{(0)}_{ab} + \frac{\eta\ell^2}{8}\left[ 2 D^c D_{(a} g^{(2)}_{b)c} - D^2 g_{ab}^{(2)} - D_aD_b (g^{cd}_{(0)}g^{(2)}_{cd}) \right],
\end{split}
\end{equation}
or explicitly in terms of $g^{(0)}_{ab}$
\begin{equation}
\begin{aligned}
\tilde g^{[4]}_{ab} &= \frac{\ell^4}{8}R_{acbd}^{(0)}R^{cd}_{(0)} - \frac{\ell^4}{48}D_aD_b R^{(0)} + \frac{\ell^4}{16}D_c D^c R_{ab}^{(0)} - \frac{\ell^4}{24}R^{(0)}R^{(0)}_{ab}  \\
&\quad\ + \frac{\ell^4}{96}\left[ (R^{(0)})^2 - D_c D^c R^{(0)} - 3 R^{cd}_{(0)} R^{(0)}_{cd}\right]g^{(0)}_{ab}. 
\end{aligned}
 \label{tilde 4}
\end{equation}
The algorithm can be pursued as follows. We can check that $(d-5)g^{(5)}_{ab}$ is linear in $g^{(3)}_{ab}$, so $g^{(5)}_{ab}=0$ except for $d=5$ as expected. We have $(d-6)g^{(6)}_{ab} = \mathcal C^{(6)}[g^{(0)}]$ for $d<6$, $(d-6)g^{(6)}_{ab} = \mathcal C^{(6)}[g^{(0)},\tilde g^{(6)}_{ab}]$ for $d>6$ and the same equation fixes $\tilde g^{[6]}_{ab}$ in terms of $g^{(0)}_{ab}$ when $g^{(2)}_{ab}$ and $g^{(4)}_{ab}$ hit their on-shell values. We will not present the results in details here because we have already derived all that we need. The interested reader can refer to \cite{deHaro:2000vlm} for the explicit expressions in $d=5,6$.

\subsection{Residual gauge diffeomorphisms}

The second step in the path towards the formulation of the Al(A)dS$_{d+1}$ phase space consists in determining the residual symmetries among the diffeomorphisms on $\mathscr M$ that survive after the gauge fixing \eqref{FG gauge}. Such diffeomorphisms are generated at the infinitesimal level by vector fields $\xi = \xi^\rho \partial_\rho + \xi^a \partial_a$  tangent to $\mathscr M$ such that the transformed metric $g_{\mu\nu}\to g_{\mu\nu}+\mathcal L_\xi g_{\mu\nu}$, expressed in the same coordinates $\{\rho,x^a\}$, still belongs to the SFG solution space, \textit{i.e.} reads as \eqref{FG gauge} for another dynamical field $\gamma'_{ab}$. We have the following $d+1$ conditions 
\begin{equation}
\mathcal{L}_\xi g_{\rho \rho} = 0, \quad \mathcal{L}_\xi g_{\rho a} = 0
\end{equation}
which are first order differential constraints on the gauge parameter $\xi$. The first condition leads to the equation 
\begin{equation}
\partial_\rho \xi^\rho = \frac{1}{\rho} \xi^\rho, \label{eq:CstVector1}
\end{equation}
which can be solved for $\xi^\rho$ as 
\begin{equation}
\xi^\rho = \sigma (x^a) \rho, \label{AKV 1}
\end{equation}
where $\sigma (x^a)$ is a possibly field-dependent arbitrary function of the boundary coordinates that parametrizes the \textit{Weyl rescalings} on the boundary. The second condition leads to the equation 
\begin{equation}
\rho^2 \gamma_{ab} \partial_\rho \xi^b + \eta\ell^2\partial_a \xi^\rho =0, \label{eq:CstVector2}
\end{equation}
which can be solved for $\xi^a$ as 
\begin{equation}
\xi^a = \bar \xi ^a (x^b) -\eta\ell^2\partial_b \sigma \int_0^\rho \frac{\D\rho'}{\rho'} \gamma^{ab}(\rho', x^c),
\label{AKV 2}
\end{equation} where $\bar \xi ^a (x^b)$ is a possibly field-dependent arbitrary \textit{diffeomorphism} on the spacetime boundary. The subset of residual gauge diffeomorphisms parametrized by $\sigma$ are referred as the \textit{Penrose-Brown-Henneaux (PBH) transformations} \cite{Brown:1986nw,Penrose:1986ca,Imbimbo:1999bj}. General diffeomorphisms $\chi_1,\chi_2$ on $\mathscr M$ are known to form a Lie algebra for the Lie bracket defined as
\begin{equation}
[\chi_1,\chi_2] = (\chi_1^\mu \partial_\mu \chi_2^\nu - \chi_2^\mu \partial_\mu \chi_1^\nu)\partial_\nu.
\end{equation}
The subclass of residual gauge diffeomorphisms $\xi$ satisfying \eqref{AKV 1} and \eqref{AKV 2} also form a Lie algebra but under the modified Lie bracket \eqref{modified bracket general} that takes into account the field-dependence of the vectors fields\cite{Barnich:2007bf,Barnich:2010eb,Schwimmer:2008yh},
\begin{equation}
[\xi_1, \xi_2 ]_\star = [\xi_1, \xi_2 ] - \delta_{\xi_1} \xi_2 + \delta_{\xi_2} \xi_1 . 
\label{modified bracket}
\end{equation}
The modification has to be interpreted as follows. In gravity, the gauge parameter $\xi$ in a whole is rarely field-independent precisely because the gauge-preserving conditions involves the metric field explictly. Here a joint solution of \eqref{AKV 1} and \eqref{AKV 2} is clearly depending on the dynamical field $\gamma_{ab}$, hence the gauge parameters $\xi$ have to be considered as functions of the coordinates as well as on the solution space. We can thus give sense to the variation $\delta_{\xi_1}\xi_2$ for any couple of residual gauge diffeomorphisms $\xi_1,\xi_2$: it represents the result of the action of $\xi_1$ on the dynamical field implicitly present in $\xi_2$. This action ensures the preservation of the gauge when performing $\xi_1$ and $\xi_2$ successively. When computing the commutator $[\xi_1,\xi_2]$, one includes this action unintentionally but then the usual Lie bracket does not reflect the true commutation relation between the vectors themselves. This explains why the bracket \eqref{modified bracket} subtracts these contributions in order to correctly close the algebra of vectors: this contextualizes the discussion of section \ref{sec:Asymptotic symmetry algebra}.

When \eqref{AKV 1} and \eqref{AKV 2} are solved, the residual diffeomorphisms are uniquely defined if a set of gauge parameters $(\sigma,\bar \xi^a)$ is given. We denote this as $\xi = \xi(\sigma,\bar \xi^a)$. How is the algebra \eqref{modified bracket} encoded at the level of the parameters $\sigma$ and $\bar \xi^a$? Let $\xi_1 \equiv \xi (\sigma_1, \bar\xi_1)$ and $\xi_2 \equiv \xi (\sigma_2, \bar\xi_2)$ be two residual gauge diffeomorphisms of the SFG gauge. Both satisfy the condition \eqref{eq:CstVector2}, hence the computation of $[\xi_1,\xi_2]^\rho_\star$ is straightforward and gives
\begin{equation}
\frac{1}{\rho}[\xi_1,\xi_2]^\rho_\star = \left( \xi^a_1 \partial_a\sigma_2 - \xi^a_2 \partial_a\sigma_1\right) - \delta_{\xi_1}\sigma_2 + \delta_{\xi_2}\sigma_{1}.
\end{equation}
Taking a derivative with respect to $\rho$ and using again \eqref{eq:CstVector2}, we get
\begin{equation}
\partial_\rho \left( \frac{1}{\rho}[\xi_1,\xi_2]^\rho_\star \right) = \partial_\rho \xi^a_1 \partial_a \sigma_2 -\partial_\rho \xi_2^a\partial_a\sigma_{1} = 0,
\end{equation}
which shows that $[\xi_1,\xi_2]^\rho_\star = \rho \hat \sigma(x^a)$ for
\begin{equation}
\hat \sigma = \frac{1}{\rho}[\xi_1,\xi_2]^\rho_\star \Big|_{\rho=0} =  \bar\xi_1^a\partial_a \sigma_2 - \bar\xi_2 \partial_a \sigma_1 - \delta_{\xi_1} \sigma_2 + \delta_{\xi_2}\sigma_1. \label{eq:braAdS}
\end{equation}
Let us now consider the transverse components. By evaluating the commutator at leading order, we derive that
\begin{equation}
\hat{\bar \xi}^a = \lim_{\rho\to 0} [\xi_1,\xi_2]^a_\star = [\bar\xi_1,\bar\xi_2]^a - \delta_{\xi_1} \bar\xi_2^a + \delta_{\xi_2} \bar\xi_1^a.
\end{equation}
Recalling that $\delta_\xi \gamma^{ab} = \mathcal L_\xi \gamma^{ab} = \rho\sigma_\xi \partial_\rho \gamma^{ab} + \xi^c \partial_c \gamma^{ab} - 2 \gamma^{c(a}\partial_c \xi^{b)}$ and explicitly using \eqref{eq:CstVector2} to express $\partial_\rho \xi^a_1$ and $\partial_\rho \xi^b_2$ in terms of $\sigma_1$ and $\sigma_2$, respectively, a direct computation yields
\begin{equation}
\partial_\rho \left( [\xi_1,\xi_2]^a_\star \right) = -\eta\ell^2\frac{1}{\rho}\gamma^{ab} \partial_b \hat \sigma.
\end{equation}
The reasoning above transposes a similar proof given in \cite{Barnich:2010eb} for BMS symmetries. 

As a result, the residual gauge diffeomorphisms $\xi_1$ and $\xi_2$ of the SFG gauge satisfy 
\begin{equation}
\boxed{
\begin{gathered}
{}[\xi(\sigma_1,\bar \xi_{1}^a),\xi(\sigma_2,\bar \xi_{2}^a)]_\star =  \xi (\hat{\sigma},\hat{\bar \xi}^a) , \quad
 \text{where} \ 
	\left\lbrace
	\begin{gathered}
		\,\,\hat \sigma = \bar{\xi}_{1}^a\partial_a \sigma_2  - \delta_{\xi_1} \sigma_2 - (1 \leftrightarrow 2 ), \\
        \,\,\hat{\bar \xi}^a =  \bar \xi^b_{1} \partial_b \bar \xi^a_{2} - \delta_{\xi_1} \bar \xi_{2}^a - (1 \leftrightarrow 2).
	\end{gathered}
	\right.
\end{gathered}
}
\label{eq:VectorAlgebra}
\end{equation} 
This extends the analysis of \cite{Schwimmer:2008yh} where the bracket \eqref{eq:braAdS} was applied for the subclass of PBH transformations. In this derivation, we do not assume that the residual gauge diffeomorphism parameters $\sigma$ and $\bar \xi^a$ are field-independent, although the majority of previous analyses in the field made this assumption. This is due to the fact that usual boundary conditions in gravity fix enough boundary degrees of freedom of the gravitational field in order to forbid the residual gauge constraints to be field-dependent, hence the solution for the residual gauge parameters is also field-independent. However, we will see in section \ref{Application to more restrictive boundary conditions} below that it is crucial to remain the most general possible in treating the actual dependence of the parameters if we want to consider generalized sets of boundary conditions that are less restrictive than the usual ones and allow for some leaks at infinity. Therefore, the terms $\delta_{\xi}\sigma$ and $\delta_\xi\bar\xi$ have to be kept in \eqref{eq:VectorAlgebra}. Finally, if one assumes that the parameters are field-independent at this stage ($\delta_{\xi} \sigma =0$ and $\delta_{\xi} \bar \xi^a =0$), then the commutation relations \eqref{eq:VectorAlgebra} reduce to those of the semi-direct sum $\text{Diff}(\mathscr{I}) \loplus \mathbb{R}$ where Diff($\mathscr{I}$) denotes the diffeomorphisms on the boundary $\mathscr{I}$, parametrized by $\bar{\xi}^a$ and $\mathbb{R}$ denotes the abelian Weyl rescalings on the boundary, parametrized by $\sigma$. 

\subsection{Variations of the solution space}
\label{Variations of the solution space}
We conclude our presentation of the kinematics by specifying the action of the residual gauge diffeomorphisms $\xi(\sigma,\bar\xi)$ on the solution space described in section \ref{sec:Solution space}. Since any Al(A)dS solution in the SFG gauge is characterized by the couple of free data $(g^{(0)}_{ab},T_{ab}^{[d]})$, we just have to understand how these tensors transform under the action of $\xi$. Up to some conventions, the derivation is a rephrasing of the one provided in \cite{Imbimbo:1999bj,deHaro:2000vlm}.

Under the full set of residual gauge diffeomorphisms $\xi = \xi^\rho\partial_\rho + \xi^a \partial_a$, the metric varies as
\begin{equation}
\delta_\xi \gamma_{ab} (\rho,x^c) = \mathcal L_\xi \gamma_{ab} (\rho,x^c) = \left( \sigma\rho\partial_\rho + \mathcal L_{\xi^c} \right) \gamma_{ab}(\rho,x^c). \label{delta gamma ab}
\end{equation} 
The condition \eqref{AKV 1} has been incorporated directly in the first term. Owing to \eqref{AKV 2}, we can obtain the transformation of the boundary metric by extracting the leading order $\mathcal O(1/\rho^2)$ of \eqref{delta gamma ab}. We get \cite{Imbimbo:1999bj,Papadimitriou:2010as}
\begin{equation}
\boxed{
\delta_\xi g^{(0)}_{ab} = \mathcal{L}_{\bar\xi} g^{(0)}_{ab} - 2 \sigma g_{ab}^{(0)} . } \label{eq:action solution space}
\end{equation} 
This justifies \textit{a posteriori} the codification adopted for the parameters $(\sigma,\bar\xi)$. The first term in \eqref{eq:action solution space} is the expected action of a boundary diffeomorphism $\bar\xi$ on the induced metric at the boundary $\mathscr I$. The second term is the infinitesimal implementation of a Weyl rescaling on $\mathscr I$. Indeed, if $\xi = \xi(\sigma,\bar\xi=0)$, \eqref{AKV 1} can be readily exponentiated and generates the transformation $g^{(0)}_{ab} \to e^{-2\sigma}g^{(0)}_{ab}$ on $g^{(0)}_{ab}$, by looking at the leading $\mathcal O(1/\rho^2)$ of \eqref{FG gauge}. 

The transformation of the holographic stress-energy needs to go further into the power series in $\rho$. Writing the solution \eqref{AKV 2} as $\xi^a(\rho,x^b) = \bar\xi^a(x^b) + \sum_{k=1}^{+\infty} \xi^a_{(k)}(\rho,x^b) \rho^k$, we have
\begin{equation}
\xi^a_{(1)} = 0, \quad \xi^a_{(2)} = -\frac{1}{2}\eta\ell^2 g^{ab}_{(0)}\partial_b \sigma, \quad \xi^a_{(3)} = 0, \quad \xi^a_{(4)} = \frac{1}{4}\eta\ell^2 g^{ab}_{(2)}\partial_b \sigma.
\end{equation}
Evaluating \eqref{delta gamma ab} order by order in $\rho$ yields the following tower of variations:
\begin{align}
\delta_\xi g^{(2)}_{ab} &= \mathcal L_{\bar\xi} g^{(2)}_{ab} +2  D_{(a} \xi^{(2)}_{b)}, \label{delta g2} \\
\delta_\xi g^{(3)}_{ab} &= \mathcal L_{\bar\xi} g^{(3)}_{ab} + \sigma g^{(3)}_{ab}, \label{delta g3} \\
\delta_\xi g^{(4)}_{ab} &= \mathcal L_{\bar\xi} g^{(4)}_{ab} + 2 \sigma (g^{(4)}_{ab}+\tilde g^{[4]}_{ab})+ 2 D_{(a} \xi_{b)}^{(4)} + \xi_{(2)}^c D_c g^{(2)}_{ab}  + 2 g^{(2)}_{c(a} D_{b)}\xi^c_{(2)}, \label{delta g4} \dots
\end{align} 
As a consistency check, we can verify by some straightforward algebra that \eqref{delta g2}-\eqref{delta g4} are compatible with their explicit solutions presented in Section \ref{sec:Solution space} when \eqref{eq:action solution space} holds. For each dimension $d$, one takes benefit from these variations to compute case by case the transformation of the holographic stress-energy tensor. It yields in general \cite{deHaro:2000vlm,Papadimitriou:2010as}
\begin{equation}
\boxed{
\delta_{\xi} T_{ab}^{[d]} = \mathcal{L}_{\bar\xi}  T_{ab}^{[d]} + (d-2) \sigma  T_{ab}^{[d]}  + A_{ab}^{[d]}[\sigma] }
\label{eq:action solution space 2}
\end{equation} 
where $A_{ab}^{[d]}[\sigma]$ denotes the inhomogeneous part of the transformation due to Weyl rescalings. We have $A_{ab}^{[2k+1]}[\sigma] = 0$ for $k \in\mathbb N_0$. For $d = 2$, we find
\begin{equation}
A_{ab}^{[2]}[\sigma] = - \frac{\ell}{8 \pi G} (D_a D_b\sigma - g^{(0)}_{ab} D^c D_c \sigma ), 
\end{equation} while for $d=4$, we obtain
\begin{equation}
\begin{split}
A_{ab}^{[4]}[\sigma] = \frac{\eta}{4\pi G\ell} &\left[ 2 \sigma \tilde{g}^{[4]}_{ab} +  \frac{\ell^4}{4} D^c \sigma \Big( D_c R_{ab}^{(0)} - D_{(a} R_{b)c}^{(0)}- \frac{1}{6} D_c R^{(0)} g_{ab}^{(0)} \Big) \right. \\ 
&\phantom{\Big[} + \frac{\ell^4}{24} D_{(a} \sigma D_{b)} R^{(0)} + \frac{\ell^4}{12} R^{(0)} \Big( D_a D_b \sigma - g_{ab}^{(0)} D^c D_c \sigma  \Big) \\
&\phantom{\Big[} \left. + \frac{\ell^4}{8} \Big(  R_{ab}^{(0)} D^c D_c \sigma - 2 R_{c(a}^{(0)} D^c D_{b)} \sigma + g^{(0)}_{ab} R_{cd}^{(0)} D^c D^d \sigma  \Big) \right] ,
\end{split}
\end{equation} where $\tilde{g}^{[4]}_{ab}$ is given explicitly in equation \eqref{tilde 4}. From general considerations \cite{Henneaux:1992ig,Barnich:2010xq,Barnich:2018gdh,Barnich:2007bf,Compere:2018aar,Ruzziconi:2019pzd} briefly reviewed in section \ref{sec:Representation on the solution space}, the action on the solution space prescribed by \eqref{eq:action solution space} and \eqref{eq:action solution space 2} forms a representation of the symmetry algebra \eqref{eq:VectorAlgebra}, namely
\begin{equation}
[\delta_{\xi_1} ,  \delta_{\xi_2}] (g^{(0)}_{ab}, T^{[d]}_{ab})  = - \delta_{[\xi_1, \xi_2]_\star} (g^{(0)}_{ab}, T^{[d]}_{ab})  
\end{equation} where $[\delta_{\xi_1} ,  \delta_{\xi_2}] = \delta_{\xi_1}  \delta_{\xi_2} - \delta_{\xi_2}  \delta_{\xi_1}$. This is a particular realization of \eqref{representation on solution space general}.

\section{Holographic renormalization}
\label{sec:Holographic renormalization}
In this section, we review the formulation of the variational principle for Al(A)dS$_{d+1}$ gravity in the SFG gauge. In our framework and conventions, we describe the holographic renormalization procedure that yields a finite on-shell action \cite{deHaro:2000vlm,Bianchi:2001kw,Skenderis:2000in}, while providing many additional details in order to be transparent and complete concerning the various assumptions and derivations. We then report the prescribed counter-terms at the level of the presymplectic structure to obtain finite expressions on the phase space \cite{Compere:2008us,Papadimitriou:2005ii}.

\subsection{Variational principle}
\label{sec:Variational principle AdSd}

The variational principle for General Relativity without matter in Al(A)dS$_{d+1}$ spacetimes is generically given by \cite{deHaro:2000vlm,Bianchi:2001kw,Skenderis:2000in}
\begin{equation}
S_{ren} = \int_{\mathscr{M}}  \bm L_{EH} + \int_{\mathscr{I}} \bm  L_{GHY}+  \int_{\mathscr{I}}   \bm L_{ct} + \int_{\mathscr{I}} \bm L_\circ .
\label{renormalized action}
\end{equation} The first term in \eqref{renormalized action} is the Einstein-Hilbert action whose Lagrangian $(d+1)$-form is
\begin{equation}
\bm L_{EH}[g] = \frac{1}{16 \pi G} \sqrt{-g} \left( R[g] + \eta \frac{d(d-1)}{\ell^2} \right) \D^{d+1}x . \label{LEH}
\end{equation} The second term is the Gibbons-Hawking-York boundary term \cite{York:1972sj,Gibbons:1976ue}
\begin{equation}
\bm L_{GHY}[\gamma] = \frac{1}{8\pi G} \eta \sqrt{|\gamma|} K \D^d x  \label{LGHY}
\end{equation} 
that allows to have a well-defined variational principle (\textit{i.e.} $\delta S_{ren} = 0$ on-shell) when all induced fields on $\mathscr I$ are fixed \cite{Papadimitriou:2005ii}. It involves the second fundamental form $K_{ab} = \frac{1}{2} \mathcal{L}_N \gamma_{ab} = \nabla_{(a} N_{b)}$ of $\mathscr I$ and its trace $K = \gamma^{ab} K_{ab}$, the extrinsic curvature. These objects are built up from the outward normal vector $N = N^\mu \partial_\mu = - \sqrt{|g^{\rho\rho}|} \partial_\rho$, $N^\mu N_\mu = \eta$, which is the natural background structure induced by the SFG foliation \eqref{FG gauge}. In summary, these two first pieces in \eqref{renormalized action} represent the standard variational principle for Einstein's gravity considering manifolds with boundaries and are sufficient when the boundary structure is frozen. As long as the conformally compact manifold $\mathscr M$ is equipped with a metric $g$ diverging at infinity, we expect that the Einstein-Hilbert contribution has similar poles around infinity which are, in that case, perfectly canceled by the Gibbons-Hawking-York counterterm: this is what one can observe by developing extensively the computations. 

When the boundary metric is allowed to fluctuate (which is somehow natural observing that the SFG expansion does not impose any constraint on the boundary field $g^{(0)}_{ab}$), the Gibbons-Hawking-York term cannot maintain the action stationary on solutions -- which is generically not a problem, see sections \ref{sec:Leaky boundary conditions theo} and \ref{sec:Boundary conditions and the Cauchy problem} -- but also fails to render the whole variational principle finite on-shell. The third piece in \eqref{renormalized action} is thus precisely meant to renormalize the on-shell action when allowing for arbitrary variations of the boundary structure. Since the latter is now dynamical, nothing prevents us from adjusting the finite piece of the variational principle in accordance with the actual value of the boundary fields: this is the role of the fourth and last term $\bm L_\circ=\mathcal O(\rho^0)$ included in the action \eqref{renormalized action}. We require that $\bm L_\circ$ is made up with covariant objects with respect to the boundary manifold $(\mathscr I,g^{(0)}_{ab})$ and invariant under boundary Weyl rescalings. Its precise form depends on the motivations and the particular boundary conditions under consideration \cite{Compere:2008us,Compere:2020lrt}. 

\subsubsection{Regulated variational principle}
\label{sec:Regulated variational principle}
Let us give some details about the procedure and derive the expressions of the various contributions in \eqref{renormalized action} in our framework. Following the method introduced in \cite{deHaro:2000vlm}, we can control the divergences of the on-shell action by introducing an infrared cut-off $\epsilon>0$ (called the \textit{regulator}) and perform the integration towards the boundary up to $\rho=\epsilon$. This defines the regularized action
\begin{equation}
S_{reg}^\epsilon = \int_{\rho \ge \epsilon} \bm L_{EH} + \int_{\rho = \epsilon}  \bm L_{GHY} ,
\end{equation} 
When the equations of motion hold, we have $R = -\eta \frac{d(d+1)}{\ell^2}$ and $\sqrt{-g} = \frac{\ell}{\rho}\sqrt{|\gamma|}$. The second fundamental form is evaluated to $K = \gamma^{ab}\nabla_a N_b = -\gamma^{ab} \Gamma^\rho_{ab}N_\rho = \frac{1}{2}g^{\rho\rho}\gamma^{ab}\partial_\rho\gamma_{ab}N_\rho$, allowing us to derive the expression of $S_{reg}^\epsilon$ in terms of the boundary volume form only:
\begin{equation}
S_{reg}^\epsilon = \frac{\eta}{16\pi G\ell} \int \D^d x \left[ \int_{\epsilon}^\infty d\rho \left( -\frac{2d}{\rho} \sqrt{|\gamma|} \right) - \Big[ 2\rho\partial_\rho \sqrt{|\gamma|} \Big]\Big|_{\rho=\epsilon} \right]. \label{Sreg}
\end{equation}
Note that the upper bound $\infty$ of the integral should be intended as some cut-off in the bulk within the validity range of the SFG coordinates and such contributions will be ignored, as usual, in the formulation of the variational principle. The evaluation of the radial divergences amounts to plug the polyhomogeneous expansion of $\sqrt{|\gamma|} = \rho^{-d}\sqrt{|g^{(0)}|} \sqrt{\det(\delta^c_b+h^c_b)}$, $h^c_b = \mathcal O(\rho^2)$ into \eqref{Sreg}. We get $S_{reg}^{\epsilon} = S_{reg}^{\epsilon,div}[g^{(0)}] + \mathcal O(\varepsilon^0)$ with
\begin{equation}
S_{reg}^{\epsilon,div}[g^{(0)}] = \frac{\eta}{16\pi G\ell} \int \D^d x \frac{\sqrt{|g^{(0)}|} }{\epsilon^d} \left( a_{(0)} + \epsilon^2 a_{(2)}+ \dots + \epsilon^{d-2} a_{(d-2)} - \ln \epsilon^2\,  \tilde a_{[d]} \right). \label{Radial div}
\end{equation}
The coefficients $a_{(k)}$, $k\neq d$, are local covariant expressions involving the boundary metric $g^{(0)}_{ab}$ and its curvature tensor $R_{abcd}^{(0)}$. We have
\begin{equation}
\begin{split}
a_{(0)} &= 2 (d-1), \quad a_{(2)} = \frac{(d-4)(d-1)}{d-2} g_{(0)}^{ab} g^{(2)}_{ab} \quad (d > 2), \\
a_{(4)} &= \frac{16-9d+d^2}{4(d-4)}\left[ (g_{(0)}^{ab} g^{(2)}_{ab})^2 - g_{(2)}^{ab} g^{(2)}_{ab} \right]\quad  (d>4), \ \dots
\end{split}
\end{equation} 
The logarithmic divergence is due to the integration in the first term of \eqref{Sreg} when considering the finite part of $\sqrt{|\gamma|}$ times $1/\rho$. The associated coefficients read as
\begin{equation}
\tilde{a}_{[2]} = - g^{ab}_{(0)} g^{(2)}_{ab} , \quad \tilde{a}_{[4]} = -\frac{1}{2}\left[ (g_{(0)}^{ab} g^{(2)}_{ab})^2 - g_{(2)}^{ab} g^{(2)}_{ab} \right] .
\end{equation}
Since this part is shown to be precisely the conformal anomaly in $d$ dimensions \cite{Henningson:1998gx} (up to some numerical factor), 
\begin{equation}
\mathcal T^{[2]} = \frac{\eta}{8\pi G\ell}\tilde{a}_{[2]} , \quad \mathcal T^{[4]} = \frac{\eta}{8\pi G\ell}\tilde{a}_{[4]} .
\end{equation}
we can understand why the coefficient $\tilde a_{[d]}$ appears thus only for even $d$. Notice that one should make good use of \eqref{Tr g4} to get the final form of $a_{(4)}$ and $\tilde a_{[4]}$. As a consequence, the regulated variational principle must be supplied by a counterterm action such that
\begin{equation}
S_{ren}^\epsilon \equiv  S_{reg}^\epsilon + S_{ct}^\epsilon   , \quad \  S_{ct}^\epsilon= \int_{\rho=\epsilon} \D^dx \,  \bm L_{ct}[\gamma;\epsilon]  
\end{equation}
is finite and the limit $\epsilon\to 0$ can be safely taken. This yields the renormalized action $S_{ren} = \lim_{\epsilon\to 0}S_{ren}^\epsilon = \mathcal O(\epsilon^0)$ \cite{deHaro:2000vlm}. Although $\bm L_{ct}[\gamma;\epsilon]$ can be inferred directly from \eqref{Radial div}, the resulting expression will not reveal the fact that the counterterm Lagrangian is in fact a covariant object with respect to the induced metric $\gamma_{ab}(\epsilon,x^c)$. But one can guess the form of $\bm L_{ct}$ in terms of the volume form $\sqrt{|\gamma|}$ and a power series in the Ricci tensor $R_{ab}[\gamma]$. We give here the first pieces of the development \cite{deHaro:2000vlm,Bianchi:2001kw}
\begin{equation}
\begin{split}
\bm L_{ct}[\gamma;\epsilon] = \frac{1}{16\pi G}\frac{\eta}{\ell} \Big[ & -2(d-1)\sqrt{|\gamma|} - \frac{\eta \ell^2}{(d-2)}R[\gamma]\sqrt{|\gamma|} \\
& - \frac{\ell^4}{(d-4)(d-2)^2}\sqrt{|\gamma|}\left(R^{ab}[\gamma]R_{ab}[\gamma] - \frac{d}{4(d-1)}R[\gamma]^2\right) \\
& + \sqrt{|\gamma|}\, \tilde a_{[d]} \ln \epsilon^2 + \mathcal O(R[\gamma]^3) \Big]  (\D^d x)  .
\end{split} \label{Lct in terms of gamma}
\end{equation}
One can check by a straighforward computation, involving \eqref{Tr g2}, \eqref{g2square} and 
\begin{equation}
(g_{(0)}^{ab} g^{(2)}_{ab})^2 - g_{(2)}^{ab} g^{(2)}_{ab} = -\frac{\ell^4}{(d-2)^2}\left[R^{ab}_{(0)}R^{(0)}_{ab} - \frac{d}{4(d-1)}(R^{(0)})^2\right]  ,
\end{equation}
that \eqref{Lct in terms of gamma} encompasses the boundary-covariant counterterms needed to substract the radial divergences in \eqref{Radial div} up to $d=6$, while the neglected $\mathcal O(R[\gamma]^3)$ higher-curvature terms are intented for renormalizing in higher dimensions. Obviously, the number of terms needed for each $d$ depends on $d$ and one does not have to worry about the fact that the second and third terms are singular for $d=2$ and $d=4$ respectively, because they are not supposed to be introduced for these particular dimensions. For $d=2k$, the renormalization requires the participation of the first $k$ counterterms only, together with the logarithmic divergence. For $d=2k+1$, $\tilde a_{[d]}$ is set to zero by the Einstein equations and only the first $k+1$ counterterms participate in the renormalization procedure.

\subsubsection{Holographic stress-tensor}
The procedure described in the previous paragraphs has provided us with a finite on-shell action in the limit $\epsilon\to 0$ that we denote by
\begin{equation}
\bar S_{ren} \equiv \lim_{\epsilon\to 0} (S^\epsilon_{reg}+S^\epsilon_{ct}).
\end{equation}
We delay a bit the discussion about the incorporation of the model-dependent finite piece $\bm L_\circ$ to the renormalized action principle and the bar in the notation $\bar S_{ren}$ indicates that we are always free to set up the finite piece. Now we can define the holographic stress-tensor as \cite{Balasubramanian:1999re,deHaro:2000vlm}
\begin{equation}
T_{ab}^{[d]} \equiv -\frac{2}{\sqrt{|g^{(0)}|}} \frac{\partial \bar S_{ren}}{\partial g^{ab}_{(0)}} \label{Tab general}
\end{equation}
where $\bar S_{ren}$ is evaluated on-shell before taking the functional derivative with respect to $g^{(0)}_{ab}$ and we ignore for the moment the contributions brought by the finite term $S_{\circ}^\epsilon$. It was proven in \cite{deHaro:2000vlm} that, up to some tuning in the finite piece of the counterterms leading to the renormalized action principle, the definitions of the holographic stress-energy tensor \eqref{holographic stress-energy tensor} and \eqref{Tab general} perfectly coincide. Practical computations at the level of the variational principle are often made clearer and more efficient if we recall that the on-shell regulated action principle \eqref{Sreg} as well as the counterterm Lagrangian \eqref{Lct in terms of gamma} are local functions of the induced metric $\gamma_{ab}(\epsilon,x^c)$ on the leaves $\rho=\epsilon$. Hence, an alternative definition of $T_{ab}^{[d]}$ is
\begin{equation}
T_{ab}^{[d]} = \lim_{\epsilon \to 0} \left( \frac{1}{\rho^{d-2}}\frac{-2}{\sqrt{|\gamma|}} \frac{\partial \bar S^\epsilon_{ren}}{\partial \gamma^{ab}}\Big|_{\rho=\epsilon} \right) \equiv \lim_{\epsilon \to 0} \left( \frac{1}{\rho^{d-2}} \texttt T_{ab}^{[d]}[\gamma]\Big|_{\rho=\epsilon}\right)
\end{equation}
where $\texttt T_{ab}^{[d]}[\gamma]$ denotes the stress-energy tensor of the theory living on the $\rho=\epsilon$ hypersurface described by $S_{ren}^\epsilon$ and computed with respect to the dynamical field $\gamma_{ab}(\epsilon,x^c)$. It is divided into four pieces as $\texttt T_{ab}^{[d]}[\gamma] = \texttt T_{ab}^{reg,[d]}[\gamma] + \texttt T_{ab}^{ct,[d]}[\gamma]  - \ln \epsilon^2 \, \texttt T_{ab}^{log,[d]}[\gamma],$ where
\begin{equation}
\texttt T_{ab}^{reg,[d]}[\gamma] = -\frac{2}{\sqrt{|\gamma|}} \frac{\partial S^\epsilon_{reg}}{\partial \gamma^{ab}}\Big|_{\rho=\epsilon} = \frac{\eta}{8\pi G} ( K_{ab} - K\,\gamma_{ab} ) \Big|_{\rho=\epsilon}
\end{equation}
is the contribution of the regulated Einstein-Hilbert action,
\begin{equation}
\begin{split}
\texttt T_{ab}^{ct,[d]}[\gamma] &=  -\frac{2}{\sqrt{|\gamma|}} \frac{\partial S^\epsilon_{ct}}{\partial \gamma^{ab}}\Big|_{\rho=\epsilon} \\
&= \frac{1}{8\pi G}\frac{\eta}{\ell} \left[ -(d-1)\gamma_{ab} + \frac{\eta\,\ell^2}{(d-2)}\left(R_{ab}[\gamma] - \frac{1}{2}R[\gamma]\gamma_{ab}\right) + \mathcal O(R[\gamma]^2) \right] 
\end{split}
\end{equation}
is the piece coming from the counterterm action and finally,
\begin{equation}
\texttt T_{ab}^{log,[d]}[\gamma] = -\frac{2}{\sqrt{|\gamma|}} \frac{\delta (\sqrt{|\gamma|}\,\tilde a_{[d]})}{\delta \gamma^{ab}}\Big|_{\rho=\epsilon} 
\end{equation}
is the stress-tensor associated with the action whose integral kernel is the conformal anomaly (up to a numerical factor). The terms we are interested in go to the $\mathcal O(\epsilon^2)$ order. We can develop order by order, check at each step that all divergences cancel out and finally pick the leading order. For $d=2$, we get immediately
\begin{equation}
\texttt T^{[2]}_{ab}[\gamma] = \frac{\eta}{8\pi G \ell} \left[ g^{(2)}_{ab} - (g^{cd}_{(0)} g^{(2)}_{cd})g^{(0)}_{ab} \right] + \mathcal O(\epsilon^2)
\end{equation}
which verifies \eqref{holographic stress-energy tensor} with \eqref{X2}. For $d=3$, the only surviving term is
\begin{equation}
\texttt T^{[3]}_{ab}[\gamma] = \epsilon\frac{3\eta}{16\pi G \ell} g_{ab}^{(3)} + \mathcal O(\epsilon^2),
\end{equation}
This result can be lifted easily to any odd dimension $d$ since there is not $X^{[d]}_{ab}$ field for odd $d$ and the trace of $T_{ab}^{[d]}$ is zero on-shell as well. For the treatment of the $d=4$ case, we need to reverse-engineer the relations between $g^{(2)}_{ab}$, $\tilde g^{[4]}_{ab}$ and $R^{(2)}_{ab}$. We can show that
\begin{equation}
\begin{split}
R_{ab}[\gamma] &= R^{(0)}_{ab} + \epsilon^2 \frac{4\eta}{\ell^2} \left[ \tilde g_{ab}^{[4]} - \frac{1}{2} g_a^{(2)c}g_{bc}^{(2)} + \frac{1}{8} (g^{cd}_{(2)}g^{(2)}_{cd})g^{(0)}_{ab} \right] + \mathcal O(\epsilon^3)  , \\
R[\gamma] &= \epsilon^2 R^{(0)} + \epsilon^4 \frac{2\eta}{\ell^2} \left[ g^{cd}_{(2)} g^{(2)}_{cd} + \frac{1}{2}(g^{cd}_{(0)}g^{(2)}_{cd})^2 \right] + \mathcal O(\epsilon^5)   .
\end{split}
\end{equation}
This helps in deriving
\begin{equation}
\texttt T^{[4]}_{ab}[\gamma] = \epsilon^2\frac{\eta}{4\pi G\ell} \left( g^{(4)}_{ab} + X^{[4]}_{ab} + \frac{3}{2}\tilde g_{ab}^{[4]} \right)+\mathcal O(\epsilon^4) \label{T4_with_gtilde}
\end{equation}
where $X^{[4]}_{ab}$ is given by \eqref{X4}. The last term in \eqref{T4_with_gtilde} can be removed by means of the freedom to add a finite boundary term $\bm L_{ct}^{(fin)} = \mathcal O(\epsilon^0)$ to $\bm L^\epsilon_{ct}$. The key point for this procedure is the conformal anomaly: since it can be shown \cite{deHaro:2000vlm} that
\begin{equation}
\tilde g_{ab}^{[d]} = \frac{1}{\sqrt{|g^{(0)}|}} \frac{2}{d} \frac{\delta}{\delta g_{(0)}^{ab}} \Big( \sqrt{|g^{(0)}|}\, \tilde a_{[d]} \Big)  ,
\label{gtilde in terms of action tilde a}
\end{equation}
we see that it is sufficient to incorporate a finite counterterm of the form
\begin{equation}
S_{ct}^{\epsilon\,(fin)} = \int_{\rho=\epsilon} \bm L_{ct}^{(fin)} \equiv \frac{\eta\, \kappa_{[d]}}{16\pi G\ell} \int_{\rho=\epsilon} \D^d x\sqrt{|g^{(0)}|}\, \tilde a_{[d]} \label{Sa}
\end{equation}
in the on-shell regulated variational principle in order to cure the discrepancy between \eqref{holographic stress-energy tensor} for $d=4$ and \eqref{T4_with_gtilde}. Here $\kappa_{[d]}\in\mathbb R$ that has to be fixed for each $d$ separately. Notice that this can also be seen as coming from a contribution of $\bm L_\circ$, but we do not want to assimilate this universal subtraction in the definition of the stress tensor \cite{deHaro:2000vlm} and further finite adjustments related to particular configurations or boundary conditions \cite{Compere:2020lrt}, in order to avoid confusion about what we are doing. A straightforward chain of algebraic manipulations leads to the choice
 \begin{equation}
\bm L_{ct}^{(fin)} [g^{(0)}]  = \kappa_{[d]} \sqrt{|g^{(0)}|} \tilde{a}_{[d]} \D^d x, \quad \kappa_{[2k+1]} = 0, \, \kappa_{[2]} = 0, \, \kappa_{[4]} = \frac{3}{2},\dots
\end{equation}
As a side remark, the equation \eqref{gtilde in terms of action tilde a} can be explicitly checked for the range of dimensions we are interested in. For $d=2$, $\tilde a_{[2]}$ is a topological invariant (\textit{i.e.} the Euler characteristic) and has thus no variation on the solution space, giving $\tilde g^{[2]}_{ab} = 0$. For $d=3$, $\tilde a_{[3]}=0$ so this leads to $\tilde g^{[3]}_{ab}$ immediately. For $d=4$, \eqref{gtilde in terms of action tilde a} is shown to match with \eqref{tilde 4} using the equations of motion of quadratic gravity summarized in the appendix \ref{New massive gravity}. 

Before entering more deeply in the construction of the phase space, let us recall that the finite piece $\bm L_{\circ}$ remains to be fixed in the action principle. This fixation is model-dependent and left at our best convenience regarding the boundary conditions we are dealing with. After setting this boundary Lagrangian to a particular expression, the total renormalized action principle reads as \cite{deHaro:2000vlm,Bianchi:2001kw,Skenderis:2000in}
\begin{equation}
\boxed{
S_{ren} = \lim_{\epsilon\to 0}(S_{reg}^\epsilon + S_{ct}^\epsilon + S^\epsilon_\circ) = \bar S_{ren} + \lim_{\epsilon\to 0} S_\circ^\epsilon. }
\label{ren action principle}
\end{equation}
The boundary stress-tensor also contain the finite piece $T^\circ_{ab}$ evaluated as
\begin{equation}
T^\circ_{ab} \equiv -\frac{2}{\sqrt{|g^{(0)}|}} \frac{\partial S^{\epsilon}_\circ}{\partial g^{ab}_{(0)}} \Rightarrow \boxed{ T_{ab}^{[d],(tot)} \equiv -\frac{2}{\sqrt{|g^{(0)}|}} \frac{\partial S_{ren}}{\partial g^{ab}_{(0)}} = T_{ab}^{[d]} + T_{ab}^\circ.}
\label{Tab tot}
\end{equation}
Since we assumed that $S^{\epsilon}_\circ$ is diffeomorphism and Weyl invariant on the boundary, $T^\circ_{ab}$ is constrained as \cite{Compere:2020lrt}
\begin{equation}
g_{ab}^{(0)}T^{ab}_\circ = 0,\qquad D_a T^{ab}_\circ = 0. \label{conditions on Tab circ}
\end{equation}
We will give a particular example of such a stress-tensor in section \ref{sec:Flat limit of the action and corner terms} when discussing the flat limit of the Al(A)dS$_{4}$ phase space.

\subsection{Renormalized presymplectic structure}

Following the Comp\`ere-Marolf prescription \cite{Compere:2008us} (see also \cite{Papadimitriou:2005ii}), we use the holographically renormalized variational principle \eqref{renormalized action} to remove the divergences of the presymplectic potential and fix the ambiguities of the covariant phase space formalism introduced in section \ref{sec:Some residual ambiguities}. This is the last step needed for the construction of the Al(A)dS$_{d+1}$ phase space. 

\subsubsection{Dynamical fields and background structures}
A subtlety that has to be mentioned and that will be showing its importance later in the text is that the renormalization procedure summarized in section \ref{sec:Regulated variational principle} is formulated in a particular gauge of coordinates where radial cut-offs are everywhere. The derivation and the form of the various counterterms are intrinsic to leaves of constant $\rho=\epsilon$, implying that these terms are not built to be covariant with the respect of the bulk geometry. In other words, fixing a $\rho=\epsilon$ hypersurface to write down all quantities needed for renormalization and then take the limit to $\mathscr I$ by pushing $\epsilon$ to $0$ is responsible for an explicit breaking of general covariance. Physically speaking, by choosing the SFG gauge to evaluate the poles of the action at infinity, we isolate a class of observers living on leaves at constant $\rho$ and, therefore, their formulation of the various divergences and the counterterms aimed at curing them are not bulk covariant. However, the differential objects formulated by these observers can be extended into the bulk thanks to a lift along the holographic coordinate $\rho$. 

The foliation driven by $\rho$, emanating from the gauge fixing, has to be considered as a background structure when formulating the variational principle: it is immutable by design since the variations on the solution space do not touch it and the breaking of covariance occurring when we define regulating hypersurfaces at $\rho=\epsilon$ is naturally formulated thanks to the existence of the foliation. But this is not the end of the story, because beyond the gauge fixing, we have to impose boundary conditions that will bring additional background structures, this time intrinsic to the boundary $\mathscr I$, such as a foliation on a time direction in $\mathscr I$ or a fixed volume on codimensions 2 sections of $\mathscr I$. Hence we would like to take care of the various boundary structures introduced by this fixation of boundary conditions that can enter into the definition of $\bm L_\circ$. In this picture, the renormalized presymplectic structure is not only a function of the bulk metric $g$ but also depends upon all of these additional structures. We denote the collection of fields and background structures as $\phi \equiv \{ g_{\mu\nu},\gamma_{ab},g^{(0)}_{ab},\dots \}$. The notation here is far from being innocent: for example, making explicitly the distinction between the bulk metric $g_{\mu\nu}$ and the induced one $\gamma_{ab}$ implies the existence of the gauge foliation on $\rho$. The dots refer to the possible additional boundary structures whose nature and precise definitions are not meant to be developed at this stage but will be discussed in chapter \ref{chapter:LambdaBMS}.

\subsubsection{Renormalization of the presymplectic potential}
\label{sec:Renormlization of the presymplectic potential}
Taking an arbitrary variation on all Lagrangians appearing in the renormalized action \eqref{renormalized action} formally defines an associated set of equations of motion multiplying the variation of the fields and the total derivative of a presymplectic potential. More explicitly, we have
\begin{equation}
\begin{split}
&\delta \bm L_{EH} = \frac{\delta \bm L_{EH}}{\delta g^{\mu\nu}}\delta g^{\mu\nu} + \D\bm\Theta_{EH}   , \quad \delta \bm L_{GHY} = \frac{\delta \bm L_{GHY}}{\delta \gamma^{ab}}\delta \gamma^{ab} + \D\bm\Theta_{GHY}, \\
&\delta \bm L_{ct} = \frac{\delta \bm L_{ct}}{\delta \gamma^{ab}}\delta \gamma^{ab} + \D\bm\Theta_{ct}    , \quad \delta \bm L_{\circ} = \frac{\delta \bm L_{\circ}}{\delta g^{ab}_{(0)}}\delta g^{ab}_{(0)} + \D\bm\Theta_{\circ}   .
\end{split} \label{all potentials}
\end{equation}
The codimension 1 form $\bm\Theta_{EH}[g;\delta g]$ is the Einstein-Hilbert presymplectic potential expressed as \eqref{EH potential} in terms of the bulk metric $g$. Since the Einstein-Hilbert action is divergent when approaching $\mathscr I$, $\bm\Theta_{EH}$ has to be renormalized as well. Since $\delta\sqrt{|\gamma|} \propto \gamma_{ab}\delta \gamma^{ab}$ we see that $\bm\Theta_{GHY}$ can be taken as to be identically zero. All quantities are geometrical objects with respect to the bulk geometry. Now let us establish a method telling us how to incorporate \eqref{all potentials} into the presymplectic potential to renormalize it. 

For that purpose, let us model the situation as follows. We have a bulk Lagrangian $\bm L_B$ at our disposal (say, the Einstein-Hilbert Lagrangian) and we supply it with a collection of boundary counter-terms $\bm L_b$ that are necessary and sufficient to render the action finite on-shell. In that model, the renormalized Lagrangian has the form $\bm L_{ren} = \bm L_B +\D \bm L_b$. Denoting as $(\chi^i)$ the collection of boundary fields and background structures entering in the formulation of $\bm L_b$, we have
\begin{equation}
\begin{split}
\delta\bm L_{ren} = \delta \bm L_B+\D \delta \bm L_b &= \frac{\delta \bm L_B}{\delta g^{\mu\nu}}\delta g^{\mu\nu} + \D\bm\Theta_B + \D \left( \frac{\delta \bm L_b}{\delta \chi^i}\delta \chi^i + \D\bm\Theta_b \right) \\
&= \frac{\delta \bm L_B}{\delta g^{\mu\nu}}\delta g^{\mu\nu}  + \D \left(\bm\Theta_B + \frac{\delta \bm L_b}{\delta \chi^i}\delta \chi^i \right)
\end{split}
\end{equation}
Taking into account
\begin{equation}
\delta \bm L_{ren} = \frac{\delta \bm L_{ren}}{\delta g^{\mu\nu}} + \D \bm\Theta_{ren}\ ,\quad \frac{\delta \bm L_{ren}}{\delta g^{\mu\nu}} = \frac{\delta \bm L_{B}}{\delta g^{\mu\nu}},
\end{equation}
we can identify the canonical renormalized presymplectic potential as
\begin{equation}
\bm\Theta_{ren} \equiv \bm \Theta_B + \frac{\delta \bm L_b}{\delta \chi^i}\delta \chi^i = \bm\Theta_B - \delta \bm L_b + \D \bm\Theta_b. \label{Theta ren full general}
\end{equation}
Obviously, this identification of $\bm\Theta_{ren}$ is ambiguous because it comes from the datum of an exact $n$-form. Here we just gave the ``canonical'' choice that allows for the renormalization of $\bm\Theta_B$ using the boundary equations of motion only. Since we do not want to set the boundary fields $\chi^i$ on-shell but only the bulk field $g$, this is the most clever choice to make because the contributions of these equations of motion are automatically substracted from the final variational principle which, therefore, keeps the equations of motion for $g$ intact without constraining the boundary fields $\chi^i$. Adapting from this modelization of \eqref{renormalized action}, we can write down our prescription for the renormalized presymplectic potential as the following master equation \cite{Compere:2008us}
\begin{equation}
\boxed{
\bm \Theta_{ren} [\phi ; \delta \phi] \equiv \bm \Theta_{EH} - \delta \bm L_{GHY} - \delta \bm L_{ct} - \delta \bm L_\circ + \D \bm \Theta_{ct} + \D \bm \Theta_{\circ}. }
\label{renormalized presympelctic}
\end{equation}
In the framework of the covariant phase space methods \cite{Ashtekar:1990gc,Crnkovic:1986be,Crnkovic:1986ex,Lee:1990nz,Iyer:1994ys}, this renormalization procedure involves the two types of ambiguities arising in the formalism. Indeed, as we reviewed in section \ref{sec:Some residual ambiguities}, adding a boundary term $\bm A \equiv \bm{L}_{GHY} + \bm{L}_{ct} + \bm{L}_\circ$ to the Einstein-Hilbert Lagrangian modifies the presymplectic potential as $\bm{\Theta}_{EH} \to \bm{\Theta}_{EH} - \delta \bm{A}$. This has no consequence at the level of the presymplectic current and so does not modify the surface charges; it is however necessary to incorporate these $\delta$-exact terms to obtain a finite presymplectic flux through $\mathscr I$. Furthermore, the presymplectic potential is defined up to an exact $d$-form, $\bm{\Theta}_{EH} \to \bm{\Theta}_{EH} + \D \bm{Y}$, which is taken as $\bm Y \equiv \bm \Theta_{ct} + \bm \Theta_{\circ} $, in such a way that the remaining finite piece is linear in $\delta g^{(0)}_{ab}$ and only involves the holographic stress-tensor, thanks to \eqref{Tab general}. This requirement allows the action to be stationary on-shell when restricting to the standard case of boundary conditions with fixed boundary metric \cite{Papadimitriou:2005ii}. This choice for the two ambiguities of the Iyer-Wald formalism \cite{Iyer:1994ys} reproduces precisely \eqref{renormalized presympelctic}. The resulting pull-back on $\mathscr I$ of the renormalized presymplectic potential reads as \cite{Compere:2008us,Compere:2020lrt,Fiorucci:2020xto}
\begin{equation}
\boxed{
\bm \Theta_{ren}[\phi;\delta \phi]\Big|_{\mathscr I} = - \frac{1}{2} \sqrt{|g^{(0)}|}  T^{ab}_{[d],(tot)} \delta g^{(0)}_{ab} (\D^{d}x)  .
} \label{Theta ren pullback}
\end{equation} 
Using \eqref{renormalized presympelctic}, a straightforward computation shows that \eqref{Theta ren pullback} encodes the variation of the action when evaluated on a solution, namely
\begin{equation}
\delta S_{ren} = -\int_{\mathscr{I}} \bm \Theta_{ren} [\phi ; \delta \phi]\Big|_{\mathscr{I}}  
\label{var principle theta}
\end{equation} where we have considered only the conformal boundary $\mathscr{I}$, see \eqref{symplectic flux general}. The minus sign is due to the fact that, when formulating the variational principle, we integrate on $\rho$ from the boundary $\mathscr I$ located at $\rho=0$ to the bulk $\rho >0$, which gives the negative orientation to the Stokes formula. The resulting integral is identified with the $\rho$ component of $\bm\Theta_{ren}$ since the outward normal to the regulating surface is collinear to $\partial_\rho$. The renormalized presymplectic current is now defined as \eqref{omega grassmann even} that is
\begin{equation}
\bm \omega_{ren} [ \phi ; \delta_1 \phi, \delta_2 \phi] = \delta_1 \bm \Theta_{ren} [\phi ; \delta_2 \phi] - \delta_2 \bm \Theta_{ren} [\phi ; \delta_1 \phi]  . \label{omega ren in terms of theta ren}
\end{equation}
It is finite by design and its pullback to $\mathscr I$ is evaluated to 
\begin{equation}
\boxed{
\bm \omega_{ren} [\phi ; \delta_1 \phi, \delta_2 \phi] \Big|_{\mathscr I} = - \frac{1}{2} \delta_1 \left( \sqrt{|g^{(0)}|}  T^{ab}_{[d],(tot)} \right) \delta_2 g^{(0)}_{ab} \, (\D^{d}x) - (1 \leftrightarrow 2) . 
}
\label{renormalized presymplectic current}
\end{equation} 
As explained in section \ref{sec:Conservation of the charges theo}, this formula encodes the flux of charges through the spacetime boundary $\mathscr{I}$. This is a manifestation of the fundamental theorem \eqref{FUNDAMENTAL THEOREM} of the covariant phase space formalism
\begin{equation}
\D \bm k_{\xi,ren}[\phi; \delta \phi] = \bm \omega_{ren} [ \phi;\delta_\xi \phi, \delta \phi]. \label{fundamental formula}
\end{equation}
The actual value of the presymplectic current is in fact controlled by the particular boundary conditions imposed at $\mathscr I$ that we discuss now.

\subsubsection{Boundary conditions and the Cauchy problem}
\label{sec:Boundary conditions and the Cauchy problem}
In this section, we address the problem of fixing meaningful boundary conditions for the gravitational field in Al(A)dS$_{d+1}$ spacetimes. At first glance, AdS and dS asymptotics are different and call naturally for different choices of boundary conditions (see sections \ref{sec:dSintro}, \ref{sec:holo} and \ref{sec:Asymptotic properties of (A)dS spacetimes}). However, it is possible to reunite them in a picture that is also in line with the asymptotically flat case (previously discussed through chapter \ref{chapter:Charges}), by a convenient fixation of a boundary structure. 

\begin{figure}[!ht]
\centering
\begin{tikzpicture}[scale=0.8]
\draw (0,8)node[]{\hypertarget{fig:AdS}{}};
\draw (8,3)node[]{\hypertarget{fig:dS}{}};
\draw (0,-8) node[below]{\footnotesize (a) \ AlAdS \textit{case}.};
\draw (8,-8) node[below]{\footnotesize (b) \ AldS \textit{case}.};


\draw[white] (-4,-9) -- (-4,8) -- (12,8) -- (12,-9) -- cycle;
 
\def\dx{.2};
\def\dy{.4};
\coordinate (A) at (-2,-7+\dy);
\coordinate (B) at (2,-7+\dy);
\coordinate (C) at (2,-\dy);
\coordinate (D) at (-2,-\dy);

\coordinate (E) at (-2,\dy);
\coordinate (F) at (2,\dy);
\coordinate (G) at (2,7-\dy);
\coordinate (H) at (-2,7-\dy);

\draw (B) arc[x radius=2, y radius=0.5, start angle=0, end angle=-180];
\draw [dashed] (B) arc[x radius=2, y radius=0.5, start angle=0, end angle=180];
\draw (A) -- (D);
\draw (B) -- (C);
\draw (C) arc[x radius=2, y radius=0.5, start angle=0, end angle=-360];

\draw (E) -- (H);
\draw (F) -- (G);
\fill [white] (F) arc[x radius=2, y radius=0.5, start angle=0, end angle=360];
\draw (F) arc[x radius=2, y radius=0.5, start angle=0, end angle=-180];
\draw [dashed] (F) arc[x radius=2, y radius=0.5, start angle=0, end angle=180];
\draw (G) arc[x radius=2, y radius=0.5, start angle=0, end angle=360];

\draw [dashed] (D) -- (E);
\draw [dashed] (C) -- (F);
\draw [dashed] (A) -- ($(A)-(0,2*\dy)$);
\draw [dashed] (B) -- ($(B)-(0,2*\dy)$);
\draw [dashed] (G) -- ($(G)+(0,2*\dy)$);
\draw [dashed] (H) -- ($(H)+(0,2*\dy)$);

\def\decal{0.1};
\draw [decorate,decoration={brace,amplitude=10pt}] ($(E)-(\decal,0)$) -- ($(H)-(\decal,0)$) node[midway,xshift=-25,draw=black,align=center,rotate=90,text width=2.3cm,fill=black!5] {\textbf{Leaky b.c.}}; 
\draw [decorate,decoration={brace,amplitude=10pt}] ($(A)-(\decal,0)$) -- ($(D)-(\decal,0)$) node[midway,xshift=-25,draw=black,align=center,rotate=90,text width=3.5cm,fill=black!5] {\textbf{Conservative b.c.}}; 

\path [-{Latex[width=2mm]},thick,draw=red] (0.5,-5.5) -- (2,-4) -- (0,-2);
\path [-{Latex[width=2mm]},thick,draw=red] (0.25,-5.25) -- (2,-3.5) -- (0.25,-1.75);
\path [-{Latex[width=2mm]},thick,draw=red] (0,-5) -- (2,-3) -- (0.5,-1.5);
\path [blue,thick] (-2,-4.5) edge[bend left=30] (0,-4.5) -- (0,-4.5) edge[bend right=30] (2,-4.5);
\draw (-1,-4) node[above,blue]{$\Sigma$};

\path [-{Latex[width=2mm]},thick,draw=red] (0.5,1.5) -- (2,3) -- (3.25,4.25);
\path [-{Latex[width=2mm]},thick,draw=red] (0.25,1.75) -- (2,3.5) -- (3,4.5);
\path [-{Latex[width=2mm]},thick,draw=red] (0,2) -- (2,4) -- (2.75,4.75);
\path [blue,thick] (-2,2.5) edge[bend left=30] (0,2.5) -- (0,2.5) edge[bend right=30] (2,2.5);
\draw (-1,3) node[above,blue]{$\Sigma$};

\coordinate (C1) at ($(F)!0.9!(G)$);
\coordinate (C2) at ($(F)!0.5!(G)$);
\coordinate (C3) at ($(F)!0.1!(G)$);
\coordinate (D1) at ($(C1)+(3*\decal,0)$);
\coordinate (D2) at ($(C2)+(2.5+3*\decal,0)$);
\coordinate (D3) at ($(C3)+(3*\decal,0)$);

\draw[black!50,very thick] (D1) -- node[black,above,midway]{$\qquad\mathscr I^+$} (D2)node[right,black]{$i^0$} -- node[black,below,midway]{$\qquad\mathscr I^-$} (D3);

\draw ($(D2)!0.5!(D3)$) node[rotate=45,above,outer sep=2pt]{\textbf{Reservoir}};

\draw ($(A)!0.5!(B)$) node[]{$i^-_{\text{AdS}}$};
\draw ($(G)!0.5!(H)$) node[]{$i^+_{\text{AdS}}$};
\draw ($(B)!0.8!(C)$) node[right]{$\mathscr I_{\text{AdS}}$};



\coordinate (P) at (5,-7);
\coordinate (Q) at (11,-7);
\coordinate (R) at (11,-1);
\coordinate (S) at (5,-1);
\draw (P) -- (Q) -- (R) -- (S) -- cycle;
\draw ($(P)!0.8!(Q)$) node[below]{$\mathscr I^-_{\text{dS}}$};
\draw ($(R)!0.8!(S)$) node[above]{$\mathscr I^+_{\text{dS}}$};
\draw ($(P)!0.5!(S)$) node[above,rotate=90]{North pole};
\draw ($(Q)!0.5!(R)$) node[above,rotate=-90]{South pole};
\path [blue,thick] (5,-3) edge[bend left=30] (8,-3) -- (8,-3) edge[bend right=30] (11,-3);
\draw (9.5,-4) node[blue]{$\Sigma$};
\draw[black!50,dashed] (P) -- (R);

\path [-{Latex[width=2mm]},thick,draw=red] (6,-3.5) -- (10,0.5);
\path [-{Latex[width=2mm]},thick,draw=red] (6.25,-3.75) -- (10.25,0.25);
\path [-{Latex[width=2mm]},thick,draw=red] (6.5,-4) -- (10.5,0);

\draw (8,1.5) node[draw=black,align=center,text width=2.3cm,fill=black!5] {\textbf{Leaky b.c.}};
\draw [->] (8,1) -- (8,0);

\end{tikzpicture}
\caption{Conservative \emph{vs.} leaky boundary conditions in Al(A)dS spacetimes.}
\label{fig:Conservative vs leaky}
\end{figure}
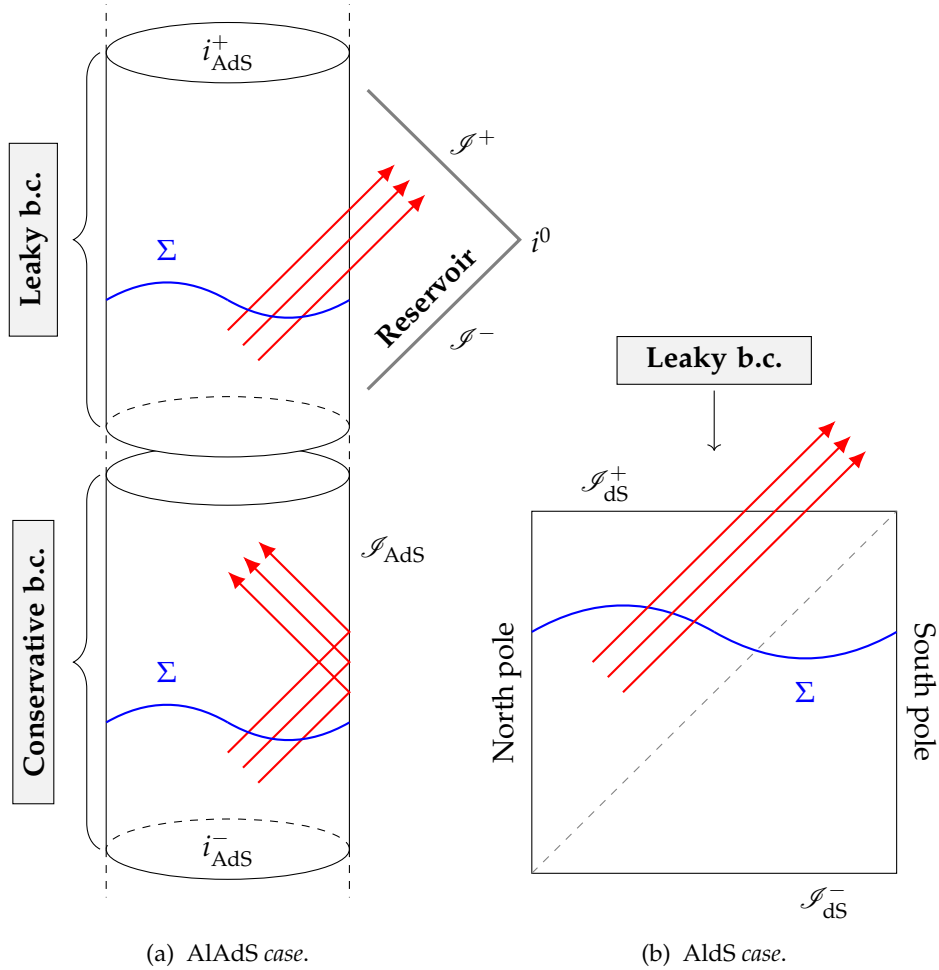

\paragraph{Conservative boundary conditions for AdS} For AdS asymptotics, the fact that null rays hit the conformal boundary $\mathscr I_{\text{AdS}}$ after a finite amount of time is fundamentally due to the fact that the far future of light is also a timelike surface (vertical in the Penrose-Carter diagram, see figure \ref{fig:Conservative vs leaky}\hyperlink{fig:dS}{.(a)}). On $\mathscr I_{\text{AdS}}$, we can choose a coordinate system $(x^a) = (t,x^A)$ where $t$ is a timelike coordinate and $x^A$ are $d-1$ compact coordinates (angles). Giving some initial conditions on a Cauchy surface $\Sigma$ at $t=t_0$ is only sufficient to prescribe the evolution up to the time $t=t_1>t_0$ when the first null ray emitted at $t=t_0$ hits $\mathscr I_{\text{AdS}}$. For any $t>t_1$, the initial conditions on $t=t_0$ are not able to predict the motion of the null rays: will they be reflected back to the bulk of $\mathscr M$ or be transmitted in the environment? This is precisely determined by the boundary conditions. The AlAdS$_{d+1}$ manifolds are said to be non-globally hyperbolic, because the solutions to hyperbolic evolution equations are not completely determined by initial data on a Cauchy slice $\Sigma$ in the past, but also need additional boundary conditions at infinity. In particular, the Cauchy problem in $\mathscr M$ is well-posed if and only if one sets reflexive boundary conditions on $\mathscr I_{\text{AdS}}$ \cite{Ishibashi:2004wx}. Natural boundary conditions for AlAdS$_{d+1}$ spacetimes are thus \textit{conservative}, in the terminology introduced in section \ref{sec:Conservative boundary conditions theo}, for which
\begin{equation}
\bm \omega_{ren} [\phi ; \delta_1 \phi, \delta_2 \phi] \Big|_{\mathscr I_{\text{AdS}}} = 0. \label{Vanishing of the symp form}
\end{equation}
So $\delta S_{ren} = 0$ with a potential adjustment of the finite counter-term. By virtue of \eqref{fundamental formula}, this implies that there is no flux leaking through $\mathscr I$ (the gravitational waves are bouncing on the spacetime boundary) and the charges are conserved. Two prototypic examples of such conservative boundary conditions are:
	\begin{itemize}[label=$\rhd$]
	\item The \textit{Dirichlet boundary conditions} \cite{Brown:1986nw,Hawking:1983mx,Ashtekar:1984zz,Henneaux:1985tv,Henneaux:1985ey,Ashtekar:1999jx} for which the boundary metric is frozen, \textit{i.e.} $\delta g^{(0)}_{ab}=0$. In that case, the presymplectic flux in \eqref{var principle theta} vanishes identically and the action $S_{ren}$ is immediately stationary on-shell \cite{Papadimitriou:2005ii}. Broadly discussed in the literature, they allow for holographic discussions of quantum gravity in AlAdS$_{d+1}$ spacetimes. The asymptotic symmetries are reduced to the group of conformal isometries of the fixed boundary geometry, which is finite-dimensional for any $d>2$ \cite{Henneaux:1985ey,Henneaux:1985tv,Ashtekar:1984zz} and consists of two copies of the De Witt algebra for $d=2$ \cite{Brown:1986nw}. We will particularize our formalism to this important subclass of boundary conditions in section \ref{Dirichlet}.
	\item The \textit{Neumann boundary conditions} \cite{Compere:2008us} which permit arbitrary variations of $g^{(0)}_{ab}$ while keeping the holographic stress-tensor $T_{ab}^{[d]}$ fixed. Again thanks to \eqref{Theta ren pullback} and \eqref{var principle theta}, we see that $\delta S_{ren} = 0$ is immediate for odd $d$, but requires the adjunction of a counter-term proportional to the trace $\mathcal{T}^{[d]} = g^{ab}_{(0)} T_{ab}^{[d]}$ of the stress-tensor for even $d$. These conditions lead to a theory of gravity with trivial infinite-dimensional asymptotic group, as reviewed in section \ref{sec:Neumann}. 
	\end{itemize}
Let us ponder a bit the definition of the Dirichlet boundary conditions. In AlAdS$_{d+1}$ spacetimes, the boundary metric $g^{(0)}_{ab}$ is always defined up to the choice of a conformal frame, which is the relic of the definition of $g^{(0)}_{ab}$ through the conformal compactification process, see section \ref{sec:Conformally compact manifolds}. Therefore, a natural boundary condition to impose in at conformal infinity is to fix the conformal class of the boundary metric rather than a particular representative \cite{Papadimitriou:2005ii}. More precisely, this amounts to impose that the boundary metric is fixed, up to a conformal factor, 
\begin{equation}
\delta g^{(0)}_{ab} = \lambda^2(x^c) g^{(0)}_{ab} ,
\label{conformal class fixed}
\end{equation}
where $\lambda(x^c)$ is a smooth scalar field on $\mathscr I_{\text{AdS}}$. Taking \eqref{conformal class fixed} into account, the general result of the variation of the on-shell action encoded in \eqref{Theta ren pullback} and \eqref{var principle theta} reduces to
\begin{equation}
\delta S_{ren} = \frac{1}{2} \int_{\mathscr I_{\text{AdS}}} \sqrt{- g^{(0)}} \lambda^2 \mathcal{T}^{[d]},
\end{equation} which reproduces the \textit{integrated Weyl anomaly} \cite{Henningson:1998gx,Papadimitriou:2005ii}. As a consequence of \eqref{trace even}, the action is stationary on solutions when $d$ is odd. However, this is generically not true when $d$ is even and one has to pick up a specific representative so that $\delta S_{ren} = 0$ on-shell. This is the point of view adopted here and inspired by the seminal works\cite{Brown:1986nw ,  Henneaux:1985tv , Henneaux:1985ey,Ashtekar:1984zz}, where the leading order of the bulk metric is taken to be a specific boundary metric. 

\paragraph{Leaky boundary conditions for dS} The situation is completely different in AldS$_{d+1}$ spacetime, see Figure \ref{fig:Conservative vs leaky}\hyperlink{fig:dS}{.(b)}. The Cauchy problem is well-defined if we give initial conditions on the Cauchy slice $\Sigma$ in the past and let the equations of motion act on them. The future of light $\mathscr I^+_{\text{dS}}$ is the future conformal boundary that is now spacelike. It represents another Cauchy surface on which it would be unwelcome to impose restrictions analogous to the AdS case. Indeed, conservative boundary conditions at $\mathscr I^+_{\text{dS}}$ will eliminate late-time radiation \cite{Ashtekar:2015lla} and definitely violate causality by constraining the Cauchy problem back in time \cite{Anninos:2011jp}. So conservative boundary conditions are meaningless when $\Lambda>0$ because, just like in the flat case, they will kill the dynamical degrees of freedom. For instance, the analysis of gravitational radiation linearized around the global dS vacuum shows that freezing the boundary metric at $\mathscr I^+_{\text{dS}}$ removes half of the radiative degrees of freedom \cite{Ashtekar:2015lla}. We are thus naturally led to allow some non-vanishing flux going through the spacetime boundary
\begin{equation}
\bm \omega_{ren} [\phi ; \delta_1 \phi, \delta_2 \phi] \Big|_{\mathscr I^+_{\text{dS}}} \neq 0 
\end{equation}
to include all of the radiative solutions in the phase space and \textit{leaky boundary conditions} become an essential ingredient!

Observing \eqref{Theta ren pullback}, the radiative modes generate some flux through the future conformal boundary by sourcing the fluctuations of the boundary metric $g_{ab}^{(0)}$. A general variation of $g_{ab}^{(0)}$ allows for boundary diffeomorphisms as well as Weyl rescalings, see \eqref{eq:action solution space}. But some components of the boundary metric are in fact kinematical and can be gauge-fixed such that the remaining fluctuations are only dynamical and \eqref{Theta ren pullback} encodes the physical symplectic flux. As in the flat case, a rescaling of the holographic coordinate driven by \eqref{AKV 1} can fix the boundary volume $\sqrt{g^{(0)}}$ to correspond to that of global dS$_{d+1}$ boundary. This is a practical choice that allows to directly incorporate the global vacuum dS$_{d+1}$ solution in the orbit of the asymptotic group without performing any further Weyl rescaling. Considering \eqref{eq:action solution space}, the residual diffeomorphisms preserving this first gauge fixing are all Diff($S^d$) symmetries \cite{Anninos:2011jp,Anninos:2010zf}. Next, on a $d$ dimensional manifold, one can adjust $d$ diffeomorphism parameters to fix $d$ components of the metric, only leaving $\frac{1}{2}(d+1)(d-2)$ dynamical boundary degrees of freedom in $g_{ab}^{(0)}$, since the gauge freedom has been completely used. They are many ways to perform this boundary gauge fixing depending on the particular motivations for building the solution space. The subclass of boundary gauge fixings consisting in the fixation of a radial foliation of the hypersphere $S^d$ by $(d-1)$ spheres is very natural and will be explored in full details in the dedicated chapter \ref{chapter:LambdaBMS}. By construction, it does not eliminate any radiative solution in the phase space and is merely a way to distinguish the dynamical (physical) pieces among the boundary degrees of freedom, following the terminology introduced in section \ref{sec:Physical content of the presymplectic flux}. Finally, we will show that a non-trivial infinite-dimensional asymptotic group survives to the boundary gauge fixing, enhancing the asymptotic structure for any $d>2$ (for $d=2$, there is no radiative degree of freedom, hence the boundary gauge fixing leads to the usual Dirichlet boundary conditions).

\paragraph{Leaky boundary conditions for AdS} Our intimate willing in this thesis is to treat AdS and dS asymptotics on the same footing despite the conceptual differences on which we shed some light in section \ref{sec:Asymptotic properties of (A)dS spacetimes}. We are thus interested in leaky boundary conditions for AlAdS$_{d+1}$ as well. Their status is similar as in the previous cases and allow for some flux of gravitational radiation through $\mathscr I_{\text{AdS}}$. But due to the different nature of the conformal boundary, they are not essential, as in AldS$_{d+1}$ spacetimes, to have radiation. Indeed, conservative boundary conditions in AlAdS$_{d+1}$ spacetimes do not eliminate all of the radiative solutions of Einstein's equation. Gravitational waves are allowed to evolve in the bulk but are meant to be reflected at infinity, which physically means that the repulsive gravitational potential present in AdS due to $\Lambda<0$ is strong enough to constrain the radiation to remain inside $\mathscr M$. This contrasts with the $\Lambda\geq 0$ case where radiation must be free to flow through null infinity. This being said, in general, one can conceive radiative solutions for which the gravitational waves modify the leading order of the metric around infinity, violating the Dirichlet boundary conditions. This is the case, for instance, for Robinson-Trautman waves \cite{Robinson:1960zzb} with negative cosmological constant \cite{Bakas:2014kfa,Stephani:2003tm,Bicak:1999ha,Bicak:1999hb}. Hence fixing a set of conservative boundary conditions is not free of consequences and rules out solutions that we could qualify of \textit{strong radiative solutions}, by opposition of weak perturbations that are repelled at infinity.

There are also current developments in the literature that suggest to investigate beyond the strong Dirichlet boundary conditions and consider leaky boundary conditions in AdS.  A first example occurs in the recent analysis of the black hole information paradox aimed at deriving the Page curve from quantum gravity path integral arguments \cite{Almheiri:2019yqk , Almheiri:2019qdq , Almheiri:2020cfm}. In this context, it has been useful to allow some radiation to escape the spacetime boundary so that the black hole can evaporate in AdS. This was implemented in practice by gluing an asymptotically flat region to the AdS boundary (acting as a reservoir for the outgoing radiation) and coupling the dual theory to a thermal bath (see Figure \ref{fig:Conservative vs leaky}\hyperlink{fig:dS}{.(a)}). Another example appears when considering brane worlds interacting with ambient higher-dimensional spacetimes \cite{Randall:1999ee, Randall:1999vf}. This picture naturally yields holographic dualities with fluctuating boundary metric and induced quantum gravity on the boundary \cite{Compere:2008us}. These are concrete motivations to treat both signs of the cosmological constant at once in the chapter \ref{chapter:LambdaBMS} devoted to discussing leaky boundary conditions within a suitable boundary gauge fixing, as announced in the previous paragraph. 

However, it is too early to restrict ourselves to certain classes of boundary conditions right now. In the next section, we compute the canonical surface charges for any residual diffeomorphism of the SFG gauge. This can be achieved without stipulating any stronger assumption on the boundary since the presymplectic flux \eqref{Theta ren pullback} has already been renormalized and the associated variational principle \eqref{var principle theta} is a well-defined, action principle, although non-stationary in general.

\section{Charge algebra in asymptotically locally (A)dS spacetimes}
\label{sec:Charge algebra in asymptotically locally}

This section aims at deriving the infinitesimal charges of Al(A)dS$_{d+1}$ spacetimes on the strength of the fundamental relation \eqref{fundamental formula} and the expression of the most general finite presymplectic current \eqref{renormalized presymplectic current} in the Starobinsky/Fefferman Graham gauge. The charges associated with boundary diffeomorphisms are generically non-vanishing, while those associated with Weyl rescalings are non-vanishing only in odd spacetime dimensions. We also derive the charge algebra using the Barnich-Troessaert bracket \eqref{BT definition} and show that a field-dependent $2$-cocycle appears in odd spacetime dimensions.

\subsection{Canonical surface charges}
\label{sec:Canonical surface charges}
Let us recall that in the covariant phase space formalism presented in section \ref{sec:Covariant phase space formalism}, the infinitesimal charges associated with residual gauge diffeomorphisms $\xi$ are obtained by integrating the codimension $2$ forms $\bm k_{\xi ,ren}[g; \delta g ]$, constrained as \eqref{fundamental formula}, on a codimension $2$ section $S_\infty \equiv \lbrace \rho,t=\text{constant}\rbrace$ of $\mathscr{I}$:
\begin{equation}
\ndelta H_\xi [g] = \oint_{S_\infty} \bm k_{\xi,ren}[g; \delta g] =  \oint_{S_\infty} (\D^{d-1} x) k^{\rho t}_{\xi,ren}[g; \delta g]. 
\label{integration of krhoa}
\end{equation} Here $t \equiv \ell x^1$ denotes the first coordinate among the transverse SFG coordinates (given in units of length), timelike if $\eta = 1$, spacelike if $\eta =-1$, and driving the (time) evolution along the cylinder in the AdS case and the sphere radius in the dS case. 

Starting from the SFG element \eqref{FG gauge} with the suitable fall-offs \eqref{preliminary FG}, the codimension 2 form \eqref{IyerWald explicit in GR} is divergent, since the Einstein-Hilbert action suffers from radial divergences. From the considerations of section \ref{sec:Some residual ambiguities}, one can deduce a canonical procedure to bring the renormalization \eqref{renormalized presympelctic} at the level of the codimension 2 form $\bm k_{\xi,ren}[\phi;\delta\phi]$ \cite{Iyer:1994ys} (see also \textit{e.g.} (5.32) of \cite{Compere:2018ylh} or (2.3.8.4) of \cite{Ruzziconi:2020cjt}) but the concrete manipulations are evolved and cautious. Another route, more direct, has been chosen: we start from the fundamental theorem \eqref{fundamental formula} and compute the charges simply by seeking for the boundary term in the contracted presymplectic current. Because of the definition \eqref{integration of krhoa}, we are mainly interested in the radial components of the codimension 2 form $\bm k_{\xi,ren}[\phi; \delta \phi]$ that satisfy
\begin{equation}
\partial_a k_{\xi,ren}^{\rho a}[\phi; \delta \phi] = \omega_{ren}^\rho [ \phi;\delta_\xi \phi, \delta \phi]. \label{fundamental formula 2}
\end{equation} 
This defines $k_{\xi,ren}^{\rho a}[\phi; \delta \phi]$ up to total derivative terms, \textit{i.e.} 
\begin{equation}
k_{\xi,ren}^{\rho a}[\phi; \delta \phi] \to k_{\xi,ren}^{\rho a}[\phi; \delta \phi] + \partial_b M_\xi^{[\rho ab]}[\phi;\delta \phi],
\end{equation}
where $M_\xi^{[\rho ab]}[\phi; \delta \phi]$ are the components of a codimension $3$ form. This ambiguity does not play any role when the integration \eqref{integration of krhoa} on $S_\infty$ is performed. Note also, that for the purpose of our present analysis, we assume that the finite term $\bm L_\circ$ in \eqref{renormalized action} is absent, hence $T_{ab}^{[d],(tot)} \equiv T_{ab}^{[d]}$. We postpone the discussion on the insertion of such finite contributions modifying the holographic stress-tensor to section \ref{sec:Flat limit of the action and corner terms}.

We start by computing the right-hand side of \eqref{fundamental formula 2} thanks to \eqref{renormalized presymplectic current}. Taking \eqref{eq:action solution space} and \eqref{eq:action solution space 2} into account, the variations $\delta_\xi \sqrt{|g^{(0)}|}$ and $\delta_\xi T^{ab}_{[d]}$ are given by
\begin{align}
\delta_\xi \sqrt{|g^{(0)}|} &= \frac{1}{2}\sqrt{|g^{(0)}|} \, g^{ab}_{(0)}\delta_\xi g_{ab}^{(0)} = \sqrt{|g^{(0)}|} ( D_a \bar\xi^a - d\, \sigma ), \label{eq:deltaG0} \\
\delta_\xi T^{ab}_{[d]} &= \mathcal L_{\bar\xi} T^{ab}_{[d]} + (d+2) \sigma T^{ab}_{[d]} + A^{ab}_{[d]}[\sigma]. \label{eq:deltaTab}
\end{align}
Recalling that $D_a T^{ab}_{[d]} =0$ on-shell and writing $g^{ab}_{(0)} T_{ab}^{[d]} = \mathcal{T}^{[d]}$, we get
\begin{align}
-\delta \Theta^\rho_{{ren}}[\phi; \delta_\xi \phi] =& \, \delta\left(\sqrt{|g^{(0)}|} T^{ab}_{[d]}\right)D_a \bar\xi_b +  \sqrt{|g^{(0)}|} T^{ab}_{[d]} \delta \left(D_a \bar\xi_b\right) - \delta \left( \sqrt{|g^{(0)}|} \mathcal{T}^{[d]} \sigma  \right) + \mathcal O(\rho),\nonumber \\
\delta_\xi \Theta_{{ren}}^\rho [\phi ; \delta \phi] =& - \frac{1}{2} \sqrt{|g^{(0)}|} \left( D_c \bar\xi^c \, T^{ab}_{[d]} + \mathcal L_{\bar\xi} T^{ab}_{[d]}\right)\delta g_{ab}^{(0)} -  \sqrt{|g^{(0)}|} T^{ab}_{[d]}\delta (D_a\bar\xi_b) \\
& - \frac{1}{2} \sqrt{|g^{(0)}|} A^{ab}_{[d]}[\sigma] \delta g_{ab} + \sqrt{|g^{(0)}|} \mathcal{T}^{[d]} \delta \sigma + \mathcal O(\rho).\nonumber 
\end{align} Putting all these contributions together in \eqref{omega ren in terms of theta ren}, we obtain
\begin{equation}
\boxed{
\begin{aligned}
\omega^\rho_{ren} [\phi; \delta_\xi \phi, \delta \phi] &=  \delta\left(\sqrt{|g^{(0)}|} T^{ab}_{[d]}\right)D_a \bar\xi_b  - \frac{1}{2} \sqrt{|g^{(0)}|} \left( D_c \bar\xi^c\, T^{ab}_{[d]} + \mathcal L_{\bar\xi} T^{ab}_{[d]}\right)\delta g_{ab}^{(0)} \\
&\phantom{=}\, - \delta \left( \sqrt{|g^{(0)}|} \mathcal{T}^{[d]}  \right) \sigma  - \frac{1}{2} \sqrt{|g^{(0)}|} A^{ab}_{[d]}[\sigma] \delta g_{ab}^{(0)} + \mathcal{O}(\rho).
\end{aligned}
} \label{explicit omega rho ren}
\end{equation} 
The presymplectic current can be fragmented in two groups of two terms whose origin and meaning are different. The first line in \eqref{explicit omega rho ren} is universal and both terms are present, with the same formal expression, for any dimension $d$. They only involves the boundary diffeomorphisms $\bar\xi$ and pure variations of boundary tensorial fields under $\bar\xi$. For Dirichlet boundary conditions, only the first term is present and is also integrable because the allowed boundary diffeomorphisms are in finite number and consist of the exact isometries of the (A)dS global vacuum. Hence there is no field-dependence in $\bar\xi$. Up to a partial integration, the codimension 2 form is very simple in this case and reduces to 
\begin{equation}
k_{\xi,ren}^{\rho a}[\phi; \delta \phi] = \sqrt{|g^{(0)}|} {g^{ac}_{(0)} T^{[d]}_{bc}}\bar\xi^b.
\end{equation}
By opposition, the second line in \eqref{explicit omega rho ren} is not universal since the quantities $\mathcal T^{[d]}$ and $A_{ab}^{[d]}[\sigma]$ are manifestly dimension-dependent. It is not a surprise since it involves the Weyl parameter $\sigma$: so it goes under the name of ``Weyl part'' of the presymplectic current. In particular, it is identically zero for odd $d$ where the holographic stress tensor has no trace and transforms homogenenously under Weyl rescalings. The trace term can be integrated on the phase space if $\delta\sigma=0$ (\textit{i.e.} $\sigma$ is field-independent) and gives then the contribution of the conformal anomaly to the presymplectic flux, while the last term is non-integrable in general even if $\delta\sigma=0$ but cancels immediately if one freeze the boundary structure.  

After a lengthy computation, we are able to extract a boundary term $W^{[d]\rho a}_\sigma [\phi;\delta\phi]$ from the Weyl part of \eqref{explicit omega rho ren}, \textit{i.e.}
\begin{equation}
\partial_a W^{[d]\rho a}_\sigma [\phi;\delta\phi] =  - \delta \left( \sqrt{|g^{(0)}|} \mathcal{T}^{[d]}  \right) \sigma  - \frac{1}{2} \sqrt{|g^{(0)}|} A^{ab}_{[d]}[\sigma] \delta g_{ab}. \label{flux W rho a}
\end{equation}
This equation is tautologic for $d=2k+1$, $k\in\mathbb N_0$, hence $W^{[2k+1]\rho a}_\sigma [\phi;\delta\phi] = 0$. However, $W^{[2k]\rho a}_\sigma$ is not generically zero. Since there is no algorithm giving us the expressions of $\mathcal T^{[d]}$ and $A_{ab}^{[d]}[\sigma]$ in terms of $d$ apart from computing them case by case as we did above, a generic expression for $W^{[d]\rho a}_\sigma$ cannot be found. We have just computed this boundary term in the relevant cases for most applications. For $d= 2$, we have
\begin{equation}
W_\sigma^{[2]t} [\phi; \delta \phi] = -\frac{\ell}{16 \pi G} D_b \sigma \left[ \sqrt{|g^{(0)}|} \delta g^{t b}_{(0)} + 2 \delta \sqrt{|g^{(0)}|} g^{tb}_{(0)} \right] - \ell \sigma  \Theta^t_{EH} [g^{(0)}; \delta g^{(0)}] .
\end{equation} For $d= 4$, we find 
\begin{align}
W^{[4] t}_\sigma[\phi;\delta \phi] =& \frac{\eta \, \ell^3}{16 \pi G} \left[ \frac{1}{6} \sqrt{|g^{(0)}|} R^{(0)} D_b \sigma \delta g^{tb}_{(0)} + \frac{1}{3} R^{(0)} D^t \sigma \delta \sqrt{|g^{(0)}|}   \right. \nonumber \\
& \left. \quad\quad\quad - \frac{1}{2} R^{tc}_{(0)} D_c \sigma \delta \sqrt{|g^{(0)}|} + \frac{1}{4} \sqrt{|g^{(0)}|} R_{cb}^{(0)} D^t \sigma \delta g^{bc}_{(0)} - \frac{1}{2} \sqrt{|g^{(0)}|} R^{(0)t}_c D_b \sigma \delta g^{bc}_{(0)} \right] \nonumber \\
&- \eta \frac{\ell^3}{4} \sigma \left[  \Theta^t_{QCG(1)}[g^{(0)}; \delta g^{(0)}]  - \frac{1}{3} \Theta^t_{QCG(2)}[g^{(0)}; \delta g^{(0)}]  \right] .
\end{align} In these expressions, the presymplectic potentials of the boundary Einstein-Hilbert theory $\Theta_{EH}^a$ and the quadratic curvature gravity $\Theta^a_{QCG(1)}$ and $\Theta^a_{QCG(2)}$, appear naturally (see appendix \ref{New massive gravity}). The relevant components of the codimension 2 form $\bm k_{ren,\xi}[\phi;\delta\phi]$ are
\begin{equation}
k_{ren, \xi}^{\rho a}[\phi;\delta\phi] = K^{\rho a}_{\bar \xi} [\phi;\delta\phi]+ W^{(d)\rho a}_\sigma [\phi;\delta\phi] 
\label{krhoa}
\end{equation}
where $\partial_a K^{\rho a}_{\bar \xi} [\phi;\delta\phi]$ equates the universal part of the presymplectic current given by the first line of \eqref{explicit omega rho ren}. A non-trivial computation shows that
\begin{equation}
K^{\rho a}_{\bar \xi}[\phi;\delta\phi] =  \delta \left( \sqrt{|g^{(0)}|} g_{(0)}^{ac} T^{[d]}_{bc} \right) \bar\xi^b - \frac{1}{2} \sqrt{|g^{(0)}|} \, \bar\xi^a \, T^{bc}_{[d]} \delta g_{bc}^{(0)}. \label{universal Krhoa}
\end{equation} 
Rather than presenting this quite long and not so illuminating calculation, we prefer to verify relatively easily that the flux of $K^{\rho a}_{\bar \xi}[\phi;\delta\phi]$ is consistent with \eqref{fundamental formula 2}. Taking one derivative while recalling that $T_{ab}^{[d]}$ is conserved on-shell yields
\begin{equation}
\begin{split}
\partial_a K^{\rho a}_{\bar \xi}[g ; \delta g] &=  \delta \left(\sqrt{|g^{(0)}|}T^{ab}_{[d]}\right)D_a  \bar\xi_b + \sqrt{|g^{(0)}|} T^{ab}_{[d]} \delta g_{bc}^{(0)} D_a \bar\xi^c \\
&\quad - \frac{1}{2} \sqrt{|g^{(0)}|} D_a \bar\xi^a T^{bc}_{[d]} \delta g_{bc}^{(0)} - \frac{1}{2} \sqrt{|g^{(0)}|} \bar\xi^a D_a  T^{bc}_{[d]}\delta g_{bc}^{(0)}.
\end{split}
\end{equation} 
Using now \eqref{flux W rho a} and $\mathcal L_{\bar\xi} (T^{ab}_{[d]}) = \bar\xi^c D_c T^{ab}_{[d]} - 2 T_{[d]}^{c(a}D_c \bar\xi^{b)}$, we check that
\begin{equation}
\begin{split}
&\partial_a k^{\rho a}_{ren ,\xi} [g; \delta g] - \omega^\rho_{ren} [g; \delta_\xi g , \delta g]\\
&\qquad\qquad = \frac{1}{2}\sqrt{|g^{(0)}|} \left(\mathcal L_{\bar\xi} T^{ab}_{[d]}\delta g_{ab}^{(0)} + 2 T^{ab}_{[d]}\delta g_{bc}^{(0)}D_a  \bar\xi^c - \bar\xi^c D_c T^{ab}_{[d]} \delta g_{ab}^{(0)}\right) = 0,
\end{split}
\end{equation}
and we are finished with the verification of the fundamental flux formula. Integrating \eqref{krhoa} on a codimension $2$ section of $\mathscr{I}$ ($t = \text{constant}$), as in \eqref{integration of krhoa}, gives the explicit expression of the infinitesimal charges in Al(A)dS$_{d+1}$ spacetimes. We end up with
\begin{equation}
\boxed{
\ndelta H_\xi [\phi] = \oint_{S_\infty} (\D^{d-1} x)\left[ \delta \left( \sqrt{|g^{(0)}|} {g^{tc}_{(0)} T^{[d]}_{bc}} \right) \bar\xi^b - \frac{1}{2} \sqrt{|g^{(0)}|} \bar\xi^t T^{bc}_{[d]} \delta g_{bc}^{(0)} + W^{[d]t}_\sigma [g; \delta g] \right].  
}
\label{surface charge expression}
\end{equation} 
The first piece is the most famous one: it gives, as excepted, the integrable Komar term when the boundary diffeomorphisms $\bar\xi$ are field-independent. The second term is the Iyer-Wald contribution \cite{Iyer:1994ys,Wald:1999wa}, \textit{i.e.} the pull-back of $-i_\xi\bm\Theta_{ren}$ to the boundary \cite{Papadimitriou:2005ii,Compere:2008us} and gives a first non-integrable piece directly linked to the symplectic flux through $\mathscr I$. The third piece is the most unusual and represents the charges conjugated to the Weyl symmetry on the boundary, driven by the gauge parameter $\sigma$. While the first two terms are universal by design, \textit{i.e.} independent on the dimension of the spacetime, the Weyl charges only appear for even $d$ and have a different expression in each dimensionality. Let us now make some general comments. 

\begin{itemize}[label=$\rhd$]
\item The infinitesimal charges \eqref{surface charge expression} are associated with the most generic Al(A)dS$_{d+1}$ spacetimes written in the SFG gauge \eqref{FG gauge}. They are finite even if the boundary metric is varied, as a direct consequence of the holographic renormalization of the symplectic structure controlled by \eqref{renormalized presymplectic current}. This result generalizes previous considerations \cite{Compere:2013bya} by allowing field-dependence of the parameters \cite{Barnich:2007bf,Adami:2020ugu}, both signs of the cosmological constant ($\eta = \pm 1$) and non-vanishing Weyl parameters $\sigma$. It includes all previous analyses with more restrictive boundary conditions \cite{Brown:1986nw,Henneaux:1985tv,Henneaux:1985ey,Ashtekar:1999jx,Troessaert:2013fma,Compere:2013bya,Papadimitriou:2005ii,Perez:2016vqo,Aros:1999kt,Fischetti:2012rd,Alessio:2020ioh,Ashtekar:1984zz}. 

\item Furthermore, the infinitesimal charges \eqref{surface charge expression} are generically non-integrable. As discussed in section \ref{sec:Integrability of the charges theo}, obtaining finite charges requires both the prescription to select a preferred integrable part \cite{Wald:1999wa,Compere:2018ylh,Chandrasekaran:2020wwn} and the integration on a path in the solution space \cite{Barnich:2007bf}. The reason for the non-integrability here is two-fold: it is due to the non-vanishing symplectic flux through the boundary (this contribution is encoded in the Iyer-Wald term) as well as the (possible) field-dependence of the gauge parameters. It is believed that when the symplectic flux is turned off one can choose a slicing of the phase space (including a field-dependent redefinition of the gauge parameters) in order to render the charges integrable. We decide to accept this non-integrability as a manifestation of the non-equilibrium dynamics induced by leaks at infinity, without forgetting the ``improper'' removable part which can be absorbed in stationary (conservative) configurations.

\item In addition, the infinitesimal charges \eqref{surface charge expression} are generically not conserved (see equations \eqref{renormalized presymplectic current} and \eqref{fundamental formula}). The source of the non-conservation can be related to radiating degrees of freedom but also to Weyl anomalies. In particular, for $d=2$, the breaking in the conservation law was interpreted in \cite{Alessio:2020ioh} as an anomalous Ward-Takahashi identity for the Weyl symmetry in the dual theory.

\item Finally, an important observation is that the charges associated with the Weyl parameter $\sigma (x)$ vanish for odd $d$, but are generically non-vanishing for even $d$. Let us provide an interpretation of this phenomenon. The key of the argument consists in recalling the definition of the asymptotic symmetry group as the quotient between residual gauge diffeomorphisms and trivial gauge diffeomorphisms. Here, a residual gauge diffeomorphism $\xi$ is trivial if \eqref{surface charge expression} vanishes, \textit{i.e.} $\ndelta H_\xi [\phi] = 0$. The action of the asymptotic symmetries modifies the state of the system, while the trivial gauge diffeomorphisms do not affect it and are pure redundancies of the theory. Therefore, in odd $d$, since the Weyl charges vanish, we are free to perform Weyl rescalings without affecting the physics. This corresponds to the freedom to choose the finite part of the conformal factor in the conformal compactification process. On the contrary, for even $d$, Weyl charges are non vanishing. Henceforth, Weyl rescalings are not pure redundancies of the theory and different conformal factors in the conformal compactification process lead to physically inequivalent situations. Of course, this statement is natural from the holographic perspective because of the presence of Weyl anomalies in even $d$ \cite{Henningson:1998gx,Papadimitriou:2005ii}. Indeed, in this case, we are not free to perform Weyl rescalings on the induced boundary metric because of the Weyl anomaly, which is consistent with the bulk result. 
\end{itemize}

\subsection{Charge algebra}

\subsubsection{Computation of the current algebra}

The infinitesimal charge expression \eqref{surface charge expression} being non-integrable, one cannot use the standard results of the representation theorem to derive the charge algebra but instead the Barnich-Troessaert bracket, as explained in full details in section \ref{sec:Non-integrable case: the Barnich-Troessaert bracket}. The algorithm to define the algebra starts from the choice of a split between integrable and non-integrable parts in \eqref{surface charge expression}. We write
\begin{equation}
\ndelta H_\xi [\phi]  = \delta  H_\xi [\phi] + \Xi_\xi [\phi; \delta \phi] 
\label{total charge}
\end{equation} where
\begin{equation}
\begin{split}
H_\xi [\phi] &= \oint_{S_\infty}  (\D^{d-1} x) \left[  \sqrt{|g^{(0)}|} g^{tc}_{(0)} T^{[d]}_{bc} \bar{\xi}^b \right], \\
\Xi_\xi  [\phi; \delta \phi] &= \oint_{S_\infty}  (\D^{d-1} x) \left[ - \frac{1}{2} \sqrt{|g^{(0)}|} \bar{\xi}^t T^{bc}_{[d]} \delta g_{bc}^{(0)} + W^{[d]t}_\sigma [\phi; \delta \phi] \right] - H_{\delta \xi} [\phi] .
\end{split}
\label{split of the charge}
\end{equation} We recall that this split is ambiguous since one can always shift the integrable part and the non-integrable part by some function $\Delta H_\xi[\phi]$ on the phase space as
\begin{equation}
H_\xi [\phi] \to H_\xi [\phi] + \Delta H_\xi [\phi] , \quad \Xi_\xi  [\phi; \delta \phi] \to \Xi_\xi  [\phi; \delta \phi] - \delta (\Delta H_\xi [\phi]) ,
\label{shift}
\end{equation} without affecting the total charge \eqref{total charge} (see the discussion around \eqref{shift theo}). The charge algebra under the Barnich-Troessaert bracket
\begin{equation}
\{ H_{\xi_1} [\phi] , H_{\xi_2} [\phi] \}_\star = \delta_{\xi_2} H_{\xi_1}[\phi] + \Xi_{\xi_2} [\phi; \delta_{\xi_1} \phi] ,
\label{BT bracket}
\end{equation}
still holds after the redistribution $\Delta H_\xi[\phi]$, up to a modification of the 2-cocycle, see \eqref{cocycle shift}. Hence the split has to be motivated and we will justify our choice below. 

Now let us compute the charge algebra by evaluating the right-hand side of \eqref{BT bracket} with \eqref{total charge}-\eqref{split of the charge} and the variations of the solution space given in section \ref{Variations of the solution space}. We work at the level of the currents instead of the charges, which allows us to keep track of the boundary terms in the computation. Rewriting \eqref{total charge} and \eqref{split of the charge} in terms of the boundary currents, we have
\begin{equation}
\ndelta J^a_\xi [\phi] = \delta J^a_{\xi} [\phi] + \Xi^a_\xi [\phi;\delta \phi]  
\end{equation} where
\begin{equation}
\begin{split}
J^a_\xi [\phi] &=  \sqrt{|g^{(0)}|} g^{ac}_{(0)} T^{[d]}_{bc} \bar{\xi}^b , \\
\Xi_\xi^a  [\phi; \delta \phi] &=  - \frac{1}{2} \sqrt{|g^{(0)}|} \bar{\xi}^a T^{bc}_{[d]} \delta g_{bc}^{(0)} + W^{[d]a}_\sigma[\phi; \delta \phi]  - J^a_{\delta \xi} [\phi] .
\end{split}
\label{currents expressions}
\end{equation}
Integrating these currents on a $(d-1)$-sphere at infinity $S_\infty$ gives the infinitesimal charge expressions \eqref{total charge} and \eqref{split of the charge}:
\begin{equation}
\begin{split}
H_\xi [\phi] &= \oint_{S_\infty} (\D^{d-1} x)~ J^t_\xi [\phi], \\
 \Xi_\xi  [\phi; \delta \phi] &= \oint_{S_\infty} (\D^{d-1} x)~ \Xi_\xi^t [\phi; \delta \phi] , \\ 
W^{[d]}_\sigma[\phi; \delta \phi] &= \oint_{S_\infty} (\D^{d-1} x)~ W^{[d]t}_\sigma [\phi; \delta \phi].
\end{split}
\end{equation} 
At the level of the currents, a straightforward although cautious computation shows that
\begin{equation}
\begin{split}
&\delta_{\xi_2} J^a_{\xi_1}[\phi] - \frac{1}{2} \sqrt{|g^{(0)}|} \bar{\xi}^a_2 T^{bc}_{[d]} \delta_{\xi_1} g_{bc}^{(0)}  - J^a_{\delta_{\xi_1} \xi_2} [\phi] \\
&\quad = J^a_{[\xi_1, \xi_2]_\star}[\phi]+ \sqrt{|g^{(0)}|} \left( \sigma_1 \bar{\xi}^a_2 \mathcal{T}^{[d]} + g^{ac}_{(0)} A_{bc}^{[d]} \bar{\xi}^b_1 \right) + \partial_b L_{\xi_1, \xi_2}^{[ab]}[\phi] 
\end{split}
\end{equation} where the bracket of vector fields $[\xi_1, \xi_2]_\star$ is given by \eqref{eq:VectorAlgebra} and 
\begin{equation}
L^{[ab]}_{\xi_1,\xi_2}[\phi] = 2 \sqrt{|g^{(0)}|} T_{cd}^{[d]}g^{d[b}_{(0)} \bar{\xi}^{a]}_2 \bar{\xi}^c_1
\end{equation}
is a total derivative term that will not contribute when integrating on the $(d-1)$-sphere. From this expression, we already notice that, for odd $d$, the algebra with the modified Lie bracket closes without any extension by a 2-cocycle. Indeed, in that case, the holographic stress-tensor transforms homogeneously, its trace is zero and the Weyl part of the charge vanishes: 
\begin{equation}
A_{bc}^{[2k+1]}[\sigma] = 0,\ \mathcal T^{[2k+1]} = 0 \text{ and } W^{[2k+1]a}_\sigma[g; \delta g] = 0 \text{ for any }k \in\mathbb N_0.
\end{equation}
Let us now include the $W^{[d]a}_\sigma[\phi; \delta \phi]$ term appearing in $\Xi_\xi^a  [\phi; \delta \phi]$ (see equation \eqref{currents expressions}). After a second lengthy computation, we find
\begin{equation}
\delta_{\xi_2} J^a_{\xi_1}[\phi] + \Xi^a_{\xi_2} [\phi; \delta_{\xi_1} \phi] = J^a_{[\xi_1, \xi_2]_\star}[\phi] + K_{\xi_1, \xi_2}^{[d]a}[\phi] + \partial_b L^{[ab]}_{\xi_1, \xi_2} [\phi] + \partial_b M^{[ab]}_{\xi_1, \xi_2} [\phi].
\label{current algebra AdSd}
\end{equation}
As already announced, for $d= 2k+1$ dimensions ($k \in\mathbb N_0$), we have $M^{[ab]}_{\xi_1,\xi_2}[\phi] = 0$ and $K_{\xi_1, \xi_2}^{[2k+1]a}[\phi]=0$. Now, for $d= 2$, the total derivative term takes the form
\begin{equation}
M^{[ab]}_{\xi_1, \xi_2} [\phi] = \frac{\ell}{8\pi G} \sqrt{|g^{(0)}|} \left( 2 \bar{\xi}^{[a}_1 D^{b]} \sigma_2 + D^{[a} \bar{\xi}^{b]}_1 \sigma_2 \right)
\end{equation} while the field-dependent 2-cocycle is given by 
\begin{equation}
\begin{split}
K_{\xi_1, \xi_2}^{[2]a}[\phi] &= {}_{(1)} K_{\xi_1, \xi_2}^{[2]a} [g]  + {}_{(2)} K_{\xi_1, \xi_2}^{[2]a} [g] ,\\
{}_{(1)} K_{\xi_1, \xi_2}^{[2]a} [\phi] &= \frac{\ell }{8 \pi G} \sqrt{|g^{(0)}|} \left( \sigma_1 D^a \sigma_2 - \sigma_2 D^a \sigma_1  \right) , \\
{}_{(2)} K_{\xi_1, \xi_2}^{[2]a} [\phi] &= \frac{\ell }{16 \pi G} \sqrt{|g^{(0)}|}~ R_{(0)} \left( \sigma_1 \bar{\xi}^a_2 - \sigma_2 \bar{\xi}^a_1 \right).
\end{split} \label{2 cocycle d=3}
\end{equation}
For $d=4$, the total derivative term takes the form 
\begin{equation}
\begin{split}
M^{[ab]}_{\xi_1, \xi_2} [\phi] = \frac{\eta\,\ell^3}{16\pi G}&\sqrt{|g^{(0)}|}  \left[ \frac{1}{6} D^{[a}(\sigma_2 R^{(0)}) \bar\xi_1^{b]} - \frac{1}{3} \sigma_2 R^{(0)} D^{[a}\bar\xi^{b]}_1 + \sigma_2 R^{c[a}_{(0)} D_c \bar\xi_1^{b]} \right. \\
&\,\ \left. - \sigma_2 D^{[a} {R^{b]}_{(0)c}} \bar\xi^c_1 + \frac{1}{2} R^{(0)} D^{[a} \sigma_2 \bar\xi_1^{b]} + R^{c[a}_{(0)} ( D^{b]}\sigma_2 \bar\xi^1_c -D_c\sigma_2\bar\xi^{b]}_1 ) \right]
\end{split}
\end{equation}
while the field-dependent 2-cocycle is now given by  
\begin{equation}
\begin{split}
K_{\xi_1, \xi_2}^{[4]a}[\phi] &= {}_{(1)} K_{\xi_1, \xi_2}^{[4]a} [g]  + {}_{(2)} K_{\xi_1, \xi_2}^{[4]a} [g] ,\\
{}_{(1)} K_{\xi_1, \xi_2}^{[4]a} [\phi] &= \frac{\eta\,\ell^3}{16\pi G}\sqrt{|g^{(0)}|}  \left( R^{ab}_{(0)}-\frac{1}{2}R^{(0)}g^{ab}_{(0)}\right) \left(\sigma_1 D_b \sigma_2 - \sigma_2 D_b \sigma_1\right) , \\
{}_{(2)} K_{\xi_1, \xi_2}^{[4]a} [\phi] &= \frac{\eta\,\ell^3}{64\pi G}\sqrt{|g^{(0)}|} \left( R^{bc}_{(0)}R_{bc}^{(0)} - \frac{1}{3} R^2_{(0)}\right) \left( \sigma_1 \bar{\xi}^a_2 - \sigma_2 \bar{\xi}^a_1 \right).
\end{split} \label{2 cocycle d=4}
\end{equation}

The subscript $(1)$ labels the part in $K_{\xi_1, \xi_2}^{[d]a}[\phi]$ which is pure Weyl, \textit{i.e.} mixes $\sigma_1,\sigma_2$, and the subscript $(2)$ designates the part involving the boundary diffeomorphisms $\bar \xi^a$. Up to our knowledge, there is no algorithm giving a general expression of $K_{\xi_1, \xi_2}^{[d]a}[\phi]$ and the total derivative term $M^{[ab]}_{\xi_1,\xi_2}[\phi]$ in any $d$ because the transformation law of the holographic stress-tensor cannot be given for any $d$ without computing explicitly the variation of the solution space case by case. For the sake of conciseness and because our interest here is mainly focused on the $d=2,3,4$ cases, we will not present the expressions for $K_{\xi_1, \xi_2}^{[d]a}[\phi]$ and $M^{[ab]}_{\xi_1,\xi_2}[\phi]$ beyond $d=4$. 

\subsubsection{Properties of the charge algebra}

After integrating \eqref{current algebra AdSd} on $S_\infty$ and throwing away the total derivative terms, we end up with the expected result
\begin{equation}
\boxed{\{ H_{\xi_1} [\phi] , H_{\xi_2} [\phi] \}_\star = H_{[\xi_1, \xi_2]_\star}[\phi] + K^{[d]}_{\xi_1, \xi_2} [\phi].   }
\label{charge algebra}
\end{equation}
This gives the surface charge algebra in generic Al(A)dS$_{d+1}$ spacetimes which represents the vector algebra \eqref{eq:VectorAlgebra} by means of the Barnich-Troessaert bracket up to a field-dependent $2$-cocycle $K^{[d]}_{\xi_1, \xi_2} [\phi]$. The latter is obtained by integrating $K_{\xi_1, \xi_2}^{[d]a}[\phi]$ on $S_\infty$ and thus only involves its $t$ component. It vanishes for odd $d$, \textit{i.e.} $K^{[2k+1]}_{\xi_1, \xi_2} [\phi]=0$ ($k\in\mathbb N_0$). For $d=2$, we have explicitly
\begin{equation}
\begin{split}
K_{\xi_1, \xi_2}^{[2]}[\phi] = \frac{\ell }{16 \pi G} \oint_{S_\infty} (\D^{d-1} x) \sqrt{|g^{(0)}|} \Big[ 2\left(\sigma_1 D^t \sigma_2 - \sigma_2 D^t \sigma_1 \right)+R^{(0)} \left(\sigma_1 \bar\xi_2^t - \sigma_2 \bar\xi^t_1 \right) \Big]. 
\end{split} 
\label{cocycle 3d}
\end{equation} In section \ref{Dirichlet}, we show that \eqref{cocycle 3d} reproduces the Brown-Henneaux central extension in three dimensions \cite{Brown:1986nw}, indicating the presence of a holographic Weyl anomaly \cite{Henningson:1998gx,deHaro:2000vlm,Papadimitriou:2005ii}. For $d = 4$, we obtain 
\begin{equation}
\begin{split}
K_{\xi_1, \xi_2}^{[4]}[\phi] = \frac{\eta\,\ell^3}{16\pi G} \oint_{S_\infty} (\D^{d-1} x) \sqrt{|g^{(0)}|}  &\left[ \left( R^{tb}_{(0)}-\frac{1}{2}R^{(0)}g^{tb}_{(0)}\right) \left(\sigma_1 D_b \sigma_2 - \sigma_2 D_b \sigma_1\right) ,\right. \\
&\quad\quad+ \frac{1}{4} \left. \left( R^{bc}_{(0)}R_{bc}^{(0)} - \frac{1}{3} R^2_{(0)}\right) \left( \sigma_1 \bar{\xi}^t_2 - \sigma_2 \bar{\xi}^t_1 \right)\right].
\end{split} \label{cocycle 5d}
\end{equation} As a consequence of the general results presented in section \ref{sec:Field-dependent 2-cocycle}, the field-dependent $2$-cocycle is antisymmetric, $K_{\xi_1, \xi_2}^{[d]}[\phi] = - K_{\xi_2, \xi_1}^{[d]}[\phi]$ and satisfies the cocycle condition \eqref{cocycle condition theory}, \textit{i.e.}
\begin{equation}
K_{[\xi_1, \xi_2]_\star, \xi_3}^{[d]}[g] + \delta_{\xi_3} K_{\xi_1, \xi_2}^{[d]}[g] + \text{cyclic(1,2,3)} = 0. 
\label{2 cocycle condition}
\end{equation}
The explicit form of the field-dependent $2$-coycle relies on the choice of split between integrable and non-integrable parts \eqref{split of the charge}. Indeed, under a shift \eqref{shift}, the field-dependent $2$-cocycle transforms as \eqref{cocycle shift} without affecting the structure of the algebra \eqref{charge algebra}. The split \eqref{split of the charge} leads to a centerless algebra of charges in the odd $d$ cases, among them one finds the physical case $d=3$. This is one good motivation to consider it among the huge class of possible splits.

To conclude this derivation, we should like to present an explicit proof of \eqref{2 cocycle condition} for $d=2$ which mostly provides us with a cross-check of our computations. The proof is similar for $d=4$ but less transparent because of the growing analytical complexity. We work again at the level of the current algebra, hence we check that \eqref{2 cocycle condition} holds up to total derivatives on the sphere $S_\infty$. Let us first consider the part (1) of \eqref{2 cocycle d=3}. We have
\begin{equation}
\begin{split}
&{}_{(1)} K_{[\xi_1, \xi_2]_\star, \xi_3}^{[2]a}[\phi] + \delta_{\xi_3} \left( {}_{(1)} K_{\xi_1, \xi_2}^{[2]a}[\phi]\right) + \text{cyclic(1,2,3)} \\
&\quad = \partial_b \left( 2~ \bar\xi^{[b}_3 {}_{(1)} K^{a]}_{\xi_1, \xi_2} [\phi]  \right) + \frac{\ell }{8 \pi G} \sqrt{|g^{(0)}|} \bar\xi^a_3 \left(\sigma_1 D^c D_c \sigma_2 - \sigma_2 D^c D_c \sigma_1 \right) + \text{cyclic(1,2,3)}.
\end{split}
\label{cocycle condition 3d1}
\end{equation} Now, the part (2) yields
\begin{equation}
\begin{split}
&{}_{(2)} K_{[\xi_1, \xi_2]_\star, \xi_3}^{[2]a} [\phi] + \delta_{\xi_3} \left( {}_{(2)} K_{\xi_1, \xi_2}^{[2]a} [\phi]\right) + \text{cyclic(1,2,3)} \\
&\quad = \partial_b \left( \bar\xi^{[b}_3 {}_{(2)} K^{a]}_{\xi_1, \xi_2}[\phi]  \right) - \frac{\ell }{8 \pi G} \sqrt{|g^{(0)}|}\bar\xi^a_3 \left(\sigma_1 D^c D_c \sigma_2 - \sigma_2 D^c D_c \sigma_1 \right) + \text{cyclic(1,2,3)}.
\end{split}
\label{cocycle condition 3d2}
\end{equation}
We observe some symmetry between both expressions, since the total derivative term which appears naturally involves $\bar\xi_3$ antisymmetrized with ${}_{(i)} K_{\xi_1, \xi_2}[\phi]$ in both cases, $i=1,2$. Putting \eqref{cocycle condition 3d1} and \eqref{cocycle condition 3d2} together, we finally obtain
\begin{equation}
\begin{split}
&K_{[\xi_1, \xi_2]_\star, \xi_3}^{[2]a}[\phi] + \delta_{\xi_3} K_{\xi_1, \xi_2}^{[2]a}[\phi] + \text{cyclic(1,2,3)} \\
&\quad = \partial_b \left( 2 \bar\xi^{[b}_3 {}_{(1)} K^{a]}_{\xi_1, \xi_2} [\phi]  + \bar\xi^{[b}_3 {}_{(2)} K^{a]}_{\xi_1, \xi_2} [\phi]  \right) + \text{cyclic(1,2,3)} 
\end{split}
\end{equation} where the right-hand side is a total derivative term that will disappear after integration on the $(d-1)$-sphere. This concludes the demonstration.

\section{Application to more restrictive boundary conditions}
\label{Application to more restrictive boundary conditions}
The three previous sections have developed a general formalism to treat asymptotically locally (A)dS gravity in a well-adapted coordinate system which keeps trace of the conformal structure naturally associated with conformally compact manifolds. We will take the full benefit from the generality of the framework presented so far, but before entering into the core of our quest and discuss the most natural leaky boundary conditions one can define in this formalism, we would like to apply our general results to specific cases of boundary conditions that have been considered in previous analyses. This presentation does not pretend to be exhaustive but instead is designed to show that the results of section \ref{sec:Charge algebra in asymptotically locally} reduce consistently to some well-known results of the literature by focusing on conservative boundary conditions. More specifically, we consider the largely celebrated Dirichlet \cite{Brown:1986nw,Hawking:1983mx,Ashtekar:1984zz,Henneaux:1985tv,Henneaux:1985ey,Ashtekar:1999jx,Papadimitriou:2005ii} boundary conditions in asymptotically AdS$_{d+1}$ spacetimes, since the leaky boundary conditions in which we will be interested in the next chapter can be seen as a direct generalization of them with a slight although crucial relaxed hypothesis. We end by giving some comments on the Neumann boundary conditions \cite{Compere:2008us} that also lead to a well-defined variational principle up to supplying the on-shell action by a finite counterterm.

\subsection{Dirichlet boundary conditions}
\label{Dirichlet}
Dirichlet boundary conditions in asymptotically AdS$_{d+1}$ spacetimes consist in freezing the boundary metric $g^{(0)}_{ab}$ on the phase space, \textit{i.e.} $\delta g^{(0)}_{ab}=0$, see section \ref{sec:Boundary conditions and the Cauchy problem} for more details. On the general phase space, the datum of a fixed $g^{(0)}_{ab}$ defines an orbit under the asymptotic group we still have to determine. If we want this orbit to encompass the global vacuum AdS$_{d+1}$ spacetime, we have to impose that $g^{(0)}_{ab}$ is the metric of the $d$-dimensional cylinder $\mathbb R\times S^{d-1}$. Setting $x^a = (t/\ell,x^A)$, where $t$ is the (dimension-full) coordinate along the invariant direction of the cylinder and $x^A$, $A=2,\dots,d$, are the usual angles on the $S^{d-1}$ sphere, the fixation of Dirichlet boundary conditions yields
\begin{equation}
g^{(0)}_{ab} \D x^a \D x^b = -\frac{1}{\ell^2} \D t^2 + \mathring{q}_{AB} \D x^A \D x^B 
\label{Dirichlet boundary conditions}
\end{equation}
where $\mathring{q}_{AB}$ is the unit-round $(d-1)$-sphere metric. For $d=2$, the metric $\mathring{q}_{AB}$ has only one component that we take $\mathring{q}_{\phi \phi} = 1$. From a geometrical point of view, the boundary condition \eqref{Dirichlet boundary conditions} requires to give a foliation $\bm T = \ell\partial_t$ on the boundary as well as a fixed transverse metric induced on constant $t$ spheres $S^{d-1}$. 

\begin{figure}[!ht]
\centering
\begin{tikzpicture}[scale=0.8]
\draw[opacity=0] (-2.5,-4.5) -- (3.5,-4.5) -- (3.5,3) -- (-2.5,3) -- cycle;
\coordinate (A) at (0,-4);
\coordinate (B) at (3,-3);
\coordinate (C) at (3,1);
\coordinate (D) at (0,0);
\coordinate (M) at ($(C)!0.5!(A)$);
\def\dx{-0.3};
\def\dy{0.3};
\foreach \k in {5,4,...,1}
{
\def\fillopa{0.5/\k};
\def\opa{0.5/\k}
\draw[black,opacity=\opa,fill=black!50,fill opacity=\fillopa] ($(A)+(\k*\dx,\k*\dy)$) -- ($(B)+(\k*\dx,\k*\dy)$) -- ($(C)+(\k*\dx,\k*\dy)$) -- ($(D)+(\k*\dx,\k*\dy)$) -- cycle;
}
\draw[black,fill=black!50,fill opacity=0.7] (A) -- (B) -- (C) -- (D) -- cycle;
\draw[very thick,black,-latex] (M) -- ($(M)-3*(\dx,\dy)$) node[below]{$\bm N$};
\draw[] (D)node[anchor=north west]{$\mathscr I$};
\coordinate (R) at ($(A)-(-\dx,\dy)$);
\draw[-latex] (R) -- ($(R)+5*(\dx,\dy)$) node[anchor=south east]{$\rho$};
\draw[very thick,orange,-latex] (M) -- ($(M)+(0,1)$) node[above]{$\bm T$};
\coordinate (dockL) at ($(D)!0.5!(A)$);
\coordinate (dockR) at ($(B)!0.5!(C)$);
\path [blue,thick] (dockL) edge[bend right=0] (M) -- (M) edge[bend left=0] (dockR);
\draw[blue] ($(dockL)!0.5!(M)-(0,0.5)$) node[]{$\mathring q_{AB}$};

\end{tikzpicture}
\caption{Universal boundary structure \\for Dirichlet boundary conditions.}
\label{fig:Dirichlet}
\end{figure}
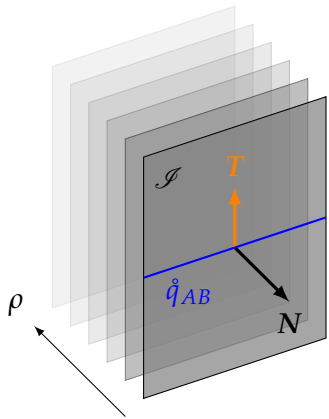

\subsubsection{Asymptotic symmetry algebra}
\label{Asymptotic symmetry algebra Dirichlet}

The residual gauge diffeomorphisms \eqref{AKV 1} and \eqref{AKV 2} preserving the boundary conditions \eqref{Dirichlet boundary conditions} are constrained through $\delta_\xi g^{(0)}_{ab} = 0$. Using \eqref{eq:action solution space}, this yields
\begin{equation}
\mathcal{L}_{\bar{\xi}} g^{(0)}_{ab} = 2 \sigma g^{(0)}_{ab} , \quad \sigma = \frac{1}{d} D_c \bar{\xi}^c , 
\label{CKV eq}
\end{equation} meaning that the boundary diffeomorphisms $\bar{\xi}^a$ are conformal Killing vectors of $g^{(0)}_{ab}$. A remarkable subgroup is constituted by boundary Killing vectors that satisfy also $D_c \bar \xi^c = 0$, leading to the exact symmetries of the $d$-dimensional cylinder. For $d> 2$, the equation \eqref{CKV eq} can be rewritten equivalently as
\begin{equation}
\begin{split}
&\partial_t \bar{\xi}^t = \frac{1}{(d-1)} D_A \bar{\xi}^A , \qquad \partial_t \bar{\xi}^A = \frac{1}{\ell^2} \mathring{q}^{AB} D_B \bar{\xi}^t , \qquad \sigma = \frac{1}{(d-1)} D_A \bar{\xi}^A, \\
&D_A \bar{\xi}_B + D_B \bar{\xi}_A = \frac{2}{(d-1)} D_C \bar{\xi}^C \mathring q_{AB}
\end{split}
\label{CKV eq decomposed}
\end{equation} where the last equation is the conformal Killing equation on the unit $(d-1)$-sphere metric. As discussed in \cite{Henneaux:1985tv,Henneaux:1985ey,Barnich:2013sxa}, the asymptotic symmetry algebra formed by the residual gauge diffeomorphisms \eqref{AKV 1} and \eqref{AKV 2} satisfying \eqref{CKV eq decomposed} is the conformal algebra in $d$ dimensions, namely $SO(d, 2)$. Assuming the field-independence of the parameters $\bar{\xi}^a$, which is consistent with the constraints \eqref{CKV eq decomposed}, the algebra \eqref{eq:VectorAlgebra} reduces to
\begin{equation}
[\xi(\bar \xi_{1}^a),\xi(\bar{\xi}_{2}^a)]_\star =  \xi (\hat{\bar \xi}^a) , \qquad \hat{\bar{\xi}}^a = \bar{\xi}_{1}^b \partial_b  \bar{\xi}_{2}^a  - \bar{\xi}_{2}^b \partial_b  \bar{\xi}_{1}^a . \label{Dirichlet algebra xi}
\end{equation}
For $d=2$, the equation \eqref{CKV eq} infers
\begin{equation}
\partial_t \bar{\xi}^t = \partial_\phi \bar{\xi}^\phi , \qquad\partial_t \bar{\xi}^\phi = \frac{1}{\ell^2} \partial_\phi \bar{\xi}^t , \qquad \sigma =  \partial_\phi \bar{\xi}^\phi.
\label{Dirichlet 2d}
\end{equation} Performing the coordinate transformation $x^\pm = \frac{t}{\ell} \pm \phi$ and expressing the parameters $\bar{\xi}^t$, $\bar{\xi}^\phi$ as 
\begin{equation}
\bar{\xi}^t = \frac{\ell}{2} (Y^+ + Y^- ), \qquad \bar{\xi}^\phi = \frac{1}{2} (Y^+ - Y^-),
\label{redefinition generators}
\end{equation} the constraints \eqref{Dirichlet 2d} imply $Y^\pm = Y^\pm (x^\pm)$ \cite{Barnich:2012aw} and \eqref{Dirichlet algebra xi} becomes
\begin{equation}
[\xi(Y^\pm),\xi(Y^\pm)]_\star =  \xi (\hat{Y}^\pm) , \qquad \hat{Y}^\pm = Y_1^\pm \partial_\pm Y_2^\pm - Y_2^\pm \partial_\pm Y_1^\pm.
\label{vector algebra 2d}
\end{equation} Finally, expanding the parameters in modes as $Y^\pm = \sum_{m \in \mathbb{Z}} Y_m^\pm  l_m^\pm$, with $l_m^\pm = e^{i m x^\pm}$, the commutation relations \eqref{vector algebra 2d} yield
\begin{equation}
i [ l_m^\pm , l_n^\pm ] = (m-n) l_{m+n}^\pm  , \qquad [ l_m^\pm , l_n^\mp ] = 0  ,
\end{equation} which corresponds to the double copy of the Witt algebra, nambley Diff$(S^1)$ $\oplus$ Diff$(S^1)$ \cite{Brown:1986nw}.

\subsubsection{Charge algebra}
\label{charge algebraaa diri}

Inserting \eqref{Dirichlet boundary conditions} into \eqref{surface charge expression}, we deduce that the infinitesimal charges associated with Dirichlet boundary conditions are integrable, \textit{i.e.} $\ndelta H_\xi [\phi] = \delta H_\xi [\phi]$ with
\begin{equation}
H_\xi [\phi] =  \frac{1}{\ell}\oint_{S_\infty}  (\D^{d-1} x) \sqrt{\mathring{q}}\left( {T^t}_b \bar{\xi}^b \right) 
\label{Bare Dirichlet charges}
\end{equation} 
and $\mathring{q} = \det (\mathring{q}_{AB})$. This corresponds to the Noether charge of a conformal field theory obtained by contracting the stress-energy tensor with a conformal Killing vector. Here $\phi = \{g,\bm N,\bm T,\mathring q_{AB}\}$ where $\bm N$ stands for the SFG foliation, $\bm T$ is the boundary foliation and $\mathring q_{AB}$ the choice for the transverse boundary metric. Now integrating on a path in the solution space and requiring that the charges vanish for global AdS$_{d+1}$, we obtain
\begin{equation}
\tilde H_\xi [g] = \frac{1}{\ell}\oint_{S_\infty}  (\D^{d-1} x) \sqrt{\mathring{q}}\left( {T^t}_b \bar{\xi}^b \right) - N_\xi , \qquad N_\xi \equiv H_\xi [g]\Big|_{\text{AdS}}.
\label{Dirichlet charges}
\end{equation}
Here $N_\xi$ denotes \eqref{Bare Dirichlet charges} evaluated for global AdS$_{d+1}$. As a consequence of \eqref{fundamental formula} and \eqref{Vanishing of the symp form}, the charges \eqref{Dirichlet charges} are conserved in time.  

Now, we bring the boundary conditions \eqref{Dirichlet boundary conditions} at the level of the charge algebra \eqref{charge algebra}. Since the charges \eqref{Dirichlet charges} are integrable, the standard results of the representation theorem \cite{Barnich:2001jy, Barnich:2007bf,Brown:1986ed,Koga:2001vq} are recovered. Indeed, the Barnich-Troessaert bracket \eqref{BT bracket} reduces to the standard Poisson bracket for integrable charges ($\Xi_{\xi_2}[g; \delta_{\xi_1} g ] =0$), namely
\begin{equation}
\{ H_{\xi_1} [g], H_{\xi_2} [g] \} = \delta_{\xi_2} H_{\xi_1} [g]  .
\end{equation} Henceforth, the charge algebra \eqref{charge algebra} yields
\begin{equation}
\{ \tilde H_{\xi_1} [g], \tilde H_{\xi_2} [g] \} = \tilde H_{[\xi_1, \xi_2]_\star}[g] + \tilde K_{\xi_1, \xi_2}^{[d]}, \qquad \tilde K_{\xi_1, \xi_2}^{[d]} \equiv K_{\xi_1, \xi_2}^{[d]} + N_{[\xi_1,\xi_2]_\star}
\label{charge algebra 2d}
\end{equation} where $\tilde K_{\xi_1, \xi_2}^{[d]}$ vanishes for odd $d$, \textit{i.e.} $\tilde K_{\xi_1, \xi_2}^{[2k+1]} = 0$ ($k \in \mathbb{N}_0$). In odd spacetime dimensions (even $d$), taking \eqref{Dirichlet boundary conditions} into account, the $2$-cocycle $\tilde K_{\xi_1, \xi_2}^{[d]}$ is field-independent and becomes a central extension that satisfies the standard $2$-cocycle condition 
\begin{equation}
\tilde K_{[\xi_1, \xi_2]_\star, \xi_3}^{[d]} + \text{cyclic(1,2,3)} = 0 
\end{equation} as a direct consequence of \eqref{2 cocycle condition} and \eqref{Dirichlet boundary conditions}. 

In particular, for $d=2$, the $2$-cocycle reduces to the Brown-Henneaux central extension \cite{Brown:1986nw}. Indeed, inserting \eqref{Dirichlet boundary conditions} and \eqref{Dirichlet 2d} into \eqref{cocycle 3d} and adding the contribution of the global AdS$_3$ background as in \eqref{charge algebra 2d}, we readily obtain 
\begin{equation}
\tilde K^{[2]}_{\xi_1, \xi_2} = -\frac{1}{8\pi G} \int_0^{2\pi} \D \phi \Big[\partial_\phi \bar \xi_1^\phi \partial_\phi^2 \bar \xi^t_2 - \frac{1}{2}\bar\xi^t_1 \partial_\phi \bar \xi_2^\phi -\frac{1}{2} \bar \xi^\phi_1 \partial_\phi \bar \xi^t_2 - (1\leftrightarrow 2) \Big].
\label{intermediate step}
\end{equation} 
Integrating by parts and throwing away the total derivatives on $\phi$, the central extension \eqref{intermediate step} can be expressed in terms of the parameters $Y^+$ and $Y^-$ defined in \eqref{redefinition generators} as 
\begin{equation}
\tilde K^{[2]}_{\xi_1, \xi_2} = \frac{\ell}{16\pi G} \int_0^{2\pi} \D \phi \,  \Big[ Y^+_1 (\partial_+^3 Y^+_2 + \partial_+ Y^+_2) + Y^-_1 (\partial^3_- Y^-_2 + \partial_- Y^-_2) \Big] .
\end{equation} Finally, writing $\texttt L_m^\pm = \tilde H_{\xi(l^\pm_m)}[g]$ in \eqref{charge algebra 2d}, we recover the double copy of the Virasoro algebra
\begin{equation}
i \{\texttt  L_m^\pm ,\texttt  L_n^\pm \} = (m-n) \texttt L_{m+n}^\pm - \frac{c^\pm}{12} m (m^2-1) \delta^0_{m+n} , \qquad \{ \texttt L_m^\pm , \texttt L_n^\mp \} = 0 
\end{equation} where 
\begin{equation}
\boxed{
c^\pm = \frac{3\ell}{2G},
}
\end{equation}
which corresponds to the results of \cite{Brown:1986nw}.

For $d=4$, the $2$-cocycle $\tilde K_{\xi_1,\xi_2}^{[4]}$ vanishes \cite{Henneaux:1985ey}. In fact, inserting \eqref{Dirichlet boundary conditions} and \eqref{CKV eq decomposed} into \eqref{cocycle 5d} and adding the contribution of the global AdS$_5$ background as in \eqref{charge algebra 2d}, one readily finds
\begin{equation}
\tilde K^{[4]}_{\xi_1, \xi_2} = \frac{\ell^2}{48 \pi G} \oint_{S_\infty} (\text{d}^{d-1}x) \, \sqrt{\mathring q} \left[D_A \bar\xi^A_1 D_B D^B \bar\xi_2^t - (1\leftrightarrow 2)\right] - \frac{3\ell^2}{64 \pi G}  \oint_{S_\infty} (\text{d}^{d-1}x) \, \sqrt{\mathring q}\, \hat{\bar{\xi}}^t .
\end{equation} Integrating by parts and using $D_B D^B (D_A \bar\xi^A_1) = -3 D_A \bar\xi^A_1$, which is a consequence of the conformal Killing equation \eqref{CKV eq decomposed}, we get
\begin{equation}
\begin{split}
\tilde K^{[4]}_{\xi_1, \xi_2} &= -\frac{\ell^2}{16 \pi G} \oint_{S_\infty} (\text{d}^{d-1}x) \, \sqrt{\mathring q} \left[D_A \bar\xi^A_1 \bar\xi_2^t - (1\leftrightarrow 2)\right] - \frac{3\ell^2}{64 \pi G}  \oint_{S_\infty} (\text{d}^{d-1}x) \, \sqrt{\mathring q}\, \hat{\bar{\xi}}^t = 0. \label{cocycle 4 zero}
\end{split} 
\end{equation} To obtain the second equality, we have integrated by parts and used the $t$-component of the commutation relations \eqref{Dirichlet algebra xi}, namely $\hat{\bar{\xi}}^t = \bar{\xi}_{1}^A D_A \bar{\xi}_{2}^t  + \frac{1}{3} \bar{\xi}_{1}^t D_A \bar{\xi}_{2}^A   - (1 \leftrightarrow 2 )$. From \eqref{cocycle 4 zero}, we see that the $2$-cocycle \eqref{cocycle 5d} with Dirichlet boundary conditions satisfies $K^{[4]}_{\xi_1, \xi_2} = - N_{[\xi_1, \xi_2]_\star}$. This means that this $2$-cocycle is a coboundary that is reabsorbed by adjusting the zero of the charges as in \eqref{Dirichlet charges} (see \textit{e.g.} \cite{Barnich:2007bf,Barnich:2001jy,Compere:2018aar}). As discussed in \cite{Henneaux:1985ey}, one can actually show that the $2$-cocycle $\tilde K^{[d]}_{\xi_1, \xi_2}$ appearing in \eqref{charge algebra 2d} vanishes for any $d> 2$.

\subsection{Neumann boundary conditions}
\label{sec:Neumann}
Neumann boundary conditions are the complementary branch of boundary conditions that ensure that the presymplectic current vanishes everywhere on the conformal boundary \eqref{Vanishing of the symp form} in AlAdS$_{d+1}$ spacetimes. As recalled in section \ref{sec:Boundary conditions and the Cauchy problem}, they require that $\delta T_{ab}^{[d]} = 0$ while keeping the boundary metric $g^{(0)}_{ab}$ free. Geometrically, these boundary conditions do not involve the intrinsic geometry of $\mathscr I_{\text{AdS}}$ but needs a bulk information about the embedding of $\mathscr I_{\text{AdS}}$ as an hypersurface of $\mathscr M$. Forbidding the stress tensor to vary is equivalent to give a frozen value to the extrinsic curvature $K$ of $\mathscr I_{\text{AdS}}$ into $\mathscr M$. Following \cite{Compere:2008us}, we impose the stronger condition 
\begin{equation}
T_{ab}^{[d]} = 0 ,
\label{Neumann}
\end{equation}
which translates the natural fixation $K=0$. The condition \eqref{Neumann} allows to derive clearer constraints on the residual gauge diffeomorphisms. We repeat briefly the discussion of \cite{Compere:2008us} to apply our general framework.

\subsubsection{Residual diffeomorphisms}

In the odd $d$ case, the holographic stress-energy tensor \eqref{holographic stress-energy tensor} transforms homogeneously under the residual gauge diffeomorphisms \eqref{AKV 1} and \eqref{AKV 2} (see equation \eqref{eq:action solution space 2}). Hence, the boundary condition \eqref{Neumann} does not imply any constraint on the parameters $\sigma$ and $\bar{\xi}^a$. However, in the even $d$ case, the transformation of the holographic stress-energy tensor involves inhomogeneous terms in $A_{ab}^{[d]}[\sigma]$. Henceforth, one has to impose $\sigma = 0$ for \eqref{Neumann} to be satisfied in even $d$. 

\subsubsection{Charge algebra}

In odd $d$, inserting the condition \eqref{Neumann} into \eqref{surface charge expression} readily yields $\ndelta H_\xi [g] = 0$. Therefore, the residual gauge diffeomorphisms \eqref{AKV 1} and \eqref{AKV 2} are trivially represented. Similarly, in even $d$, inserting the condition \eqref{Neumann} into \eqref{surface charge expression} and taking into account that $\sigma = 0$, we obtain that the charges are zero. The asymptotic symmetry group associated with Neumann boundary conditions is therefore trivial. From the point of view of the dual theory, the boundary diffeomorphisms are pure gauge transformations, which indicates the presence of quantum gravity on the boundary. The latter is Weyl invariant for odd $d$.\hfill{\color{black!40}$\blacksquare$}

%
%

\chapter{\texorpdfstring{$\Lambda$}{Lambda}-BMS and the flat limit}
\label{chapter:LambdaBMS}

In the previous chapter, we introduced a phase space of Al(A)dS$_{d+1}$ spacetimes without imposing any boundary condition more restrictive than the minimal fall-offs required for the conformal compactification. We showed how to take advantage of the ambiguities in the covariant phase space formalism to bring the holographic renormalization at the level of the presymplectic structure and deduced finite surface charges evaluated on the phase space. Since the boundary metric is allowed to fluctuate, there is a non-vanishing presymplectic current through the conformal boundary responsible for the non-conservation of these charges. Although the charges are non-integrable, which is an expected feature when dealing with non-equilibrium physics like the radiative spacetimes we are considering, we showed that the Barnich-Troessaert prescription for the charge bracket leads to a charge algebra representing the asymptotic vector algebra up to a field-dependent 2-cocycle. The latter is non-trivial in general: it contains the information about the central charge in $3d$ gravity. We finally applied this general formalism to conservative boundary conditions that are widely studied in the literature on AlAdS spacetimes. These boundary conditions are interesting for the usual conception of holography, where they appear as the ineluctable conditions for the dual quantum theory to be unitary. 

At the fundamental level, conservative boundary conditions restore the global hyperbolicity of the AlAdS spacetimes by transforming the conformal boundary in a sort of mirror that reflects any null wave back to the center. This peculiar feature is also believed to keep the holographic building in place, because the bouncing waves causally link the boundary piece of information with its bulk counterpart. However, conservative boundary conditions are completely meaningless for AldS spacetimes, as there is no room for reflexion at the future conformal boundary, which would violate causality. As a consequence, if one kills the outgoing presymplectic flux with a conservation requirement like in AdS, one imposes a ``future constraint'' on null rays that can be integrated back in time and constrains the data on the initial Cauchy slice. Such boundary conditions definitively destroy (part of) the dynamics of null matter and gravitational waves in the bulk of spacetime. 

In this thesis, our purpose is to treat AdS and dS asymptotics on the same footing and to discuss leaky boundary conditions that allow for some flux through the conformal boundary. Allowing for leaks at infinity seems completely unavoidable for dS, but largely less canonical for AdS. The scope of our work does not include discussions of the implications of these boundary conditions in holography, which we hope to be numerous and interesting. We limit our study to demonstrating that they are natural from the gravitational point of view. The goal of this chapter is to discuss a new set of boundary conditions for the Al(A)dS$_{d+1}$ phase space that merely amount to a boundary gauge fixing in order to leave the Cauchy problem in the AldS case completely free of constraints. 

The setup, presented in section \ref{sec:Leaky boundary conditions and LBMSd}, slightly generalizes the Dirichlet boundary structure to allow for some flux at infinity and we justify that the new conditions are suitable for both AdS and dS configurations. The associated asymptotic symmetry algebra is infinite-dimensional and presents the pleasant feature of reducing to the Generalized BMS algebra in the flat limit. Since the proposed boundary gauge fixing represents the most natural extrapolation of asymptotically locally flat boundary conditions with fluctuating metric on the celestial sphere, our leaky boundary conditions lead to the exact analog of the famous asymptotic algebra for flat asymptotics, earning the name ``$\Lambda$-BMS.'' After discussing the physical content of these boundary conditions in section \ref{sec:Aspects of the LBMS4 phase space}, we dedicate section \ref{sec:Flat limit of the LBMS4 phase space} to explicitly matching the phases spaces and their associated asymptotic symmetries for both $\Lambda\neq 0$ and $\Lambda=0$ in $d=3$ dimensions. For this purpose, we must work in a coordinate system that exists for any value of $\Lambda$, in which we can define an algorithm for a safe flat limit process. The appropriate system turns out to be the Bondi gauge, for which we have to obtain the solution space before matching the fundamental dynamical quantities with the Starobinsky/Fefferman-Graham boundary data thanks to a diffeomorphism between both gauges. We use this dictionary to give sense to the flat limit at the level of the phase space and to recover the asymptotically locally flat phase space discussed in chapter \ref{chapter:Charges}.

The developments of this chapter are taken from \cite{Fiorucci:2020xto} (for section \ref{sec:Leaky boundary conditions and LBMSd}) and \cite{Compere:2019bua,Compere:2020lrt} (for sections \ref{sec:Aspects of the LBMS4 phase space} and \ref{sec:Flat limit of the LBMS4 phase space}). 

\section{Leaky boundary conditions and \texorpdfstring{$\Lambda$}{Lambda}-BMS\texorpdfstring{$_{d+1}$}{(d+1)}}
\label{sec:Leaky boundary conditions and LBMSd}
In this section, we start from the definition of a geometric structure leading to the leaky boundary conditions we are advocating for. We proceed by generalizing the boundary structure related to Dirichlet boundary conditions and show that the components of the boundary metric can be partially gauge fixed without any cost on the Cauchy problem. Next, we establish the constraints on the gauge parameters imposed by the boundary gauge fixing, study the structure of the associated asymptotic symmetry algebra and discuss its flat limit $\ell\to+\infty$ giving the Generalized BMS$_{d+1}$ symmetry algebra. 

\subsection{Boundary conditions and dynamical degrees of freedom}
\label{Boundary conditions and asymptotic symmetry algebra LBMS}

In the SFG gauge, the boundary metric $g^{(0)}_{ab}$ is left completely free by Einstein's equations and there is no preferred choice of coordinates on the boundary geometry $(\mathscr I,g_{ab}^{(0)})$. But from an intrinsic point of view, nothing prevents us from selecting a coordinate system on the boundary by using part of the diffeomorphism freedom encoded in the codimension 1 field $\bar \xi^a$ appearing into the definition of the asymptotic vector \eqref{AKV 2}. This possibly reduces the class of allowing gauge parameters but not the set of solutions as we can show by a simple counting of desired constraints \textit{versus} the parameters at our disposal. In section \ref{Dirichlet}, we defined the boundary background structure needed to impose Dirichlet boundary conditions. It consists of a foliation $\bm T$ and a fixed boundary metric on the codimension 2 sections of $\mathscr I$. The choice of $\bm T$ is physically motivated in the AdS case since it gives the direction of time evolution along the boundary cylinder which also helps to define the Hamiltonian in the dual quantum theory. In the dS case, $\bm T$ gives the natural foliation of the asymptotic $S^d$ by $S^{d-1}$ spheres of constant volume. For these reasons, we do not want to renounce to this foliation, taking into account that requiring its presence merely amounts to defining Gaussian normal coordinates on $\mathscr I$. This is achieved by employing the gauge freedom at the boundary to eliminate $d$ kinematical degrees of freedom thanks to a gauge fixing diffeomorphism $\bar \xi^a_{(GF)}$ defined intrinsically on $\mathscr{I}$. 

But we can do even better: thanks to \eqref{AKV 1} and \eqref{AKV 2}, this boundary diffeomorphism is lifted to the bulk in order to preserve the SFG gauge. The whole transformation on the spacetime also involves a Weyl rescaling of the boundary metric generated by some $\sigma_{(GF)}$, as it can be seen from \eqref{eq:action solution space}. We can use this bulk information to gauge-fix one further quantity in the boundary metric \cite{Compere:2019bua}, namely the volume of the codimension 2 spherical sections of $\mathscr I$, in perfect analogy with the asymptotically locally flat boundary conditions when $\Lambda=0$ \cite{Compere:2018ylh,Flanagan:2019vbl,Campiglia:2015yka}. The boundary gauge fixing that we impose is thus
\begin{equation}
g_{tt}^{(0)} = -\frac{\eta}{\ell^2} , \qquad g^{(0)}_{tA} = 0 , \qquad \sqrt{|g^{(0)}|} = \frac{1}{\ell} \sqrt{\mathring{q}} 
\label{BC Lambda BMS}
\end{equation} 
where $\mathring{q}$ is a fixed volume of a codimension 2 surface taken to be the determinant of the unit round $(d-1)$-sphere metric $\mathring{q}_{AB}$ in order to include the global (A)dS$_{d+1}$ spacetime in the phase space. 

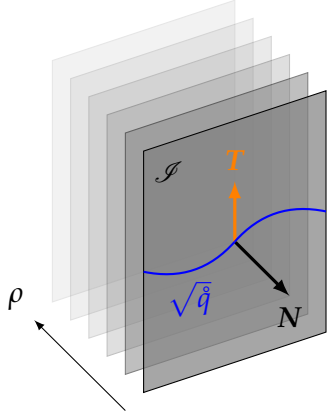
\begin{figure}[!ht]
\centering
\begin{tikzpicture}[scale=0.8]
\draw[opacity=0] (-2.5,-4.5) -- (3.5,-4.5) -- (3.5,3) -- (-2.5,3) -- cycle;
\coordinate (A) at (0,-4);
\coordinate (B) at (3,-3);
\coordinate (C) at (3,1);
\coordinate (D) at (0,0);
\coordinate (M) at ($(C)!0.5!(A)$);
\def\dx{-0.3};
\def\dy{0.3};
\foreach \k in {5,4,...,1}
{
\def\fillopa{0.5/\k};
\def\opa{0.5/\k}
\draw[black,opacity=\opa,fill=black!50,fill opacity=\fillopa] ($(A)+(\k*\dx,\k*\dy)$) -- ($(B)+(\k*\dx,\k*\dy)$) -- ($(C)+(\k*\dx,\k*\dy)$) -- ($(D)+(\k*\dx,\k*\dy)$) -- cycle;
}
\draw[black,fill=black!50,fill opacity=0.7] (A) -- (B) -- (C) -- (D) -- cycle;
\draw[very thick,black,-latex] (M) -- ($(M)-3*(\dx,\dy)$) node[below]{$\bm N$};
\draw[] (D)node[anchor=north west]{$\mathscr I$};
\coordinate (R) at ($(A)-(-\dx,\dy)$);
\draw[-latex] (R) -- ($(R)+5*(\dx,\dy)$) node[anchor=south east]{$\rho$};
\draw[very thick,orange,-latex] (M) -- ($(M)+(0,1)$) node[above]{$\bm T$};
\coordinate (dockL) at ($(D)!0.5!(A)$);
\coordinate (dockR) at ($(B)!0.5!(C)$);
\path [blue,thick] (dockL) edge[bend right=30] (M) -- (M) edge[bend left=30] (dockR);
\draw[blue] ($(dockL)!0.5!(M)-(0,0.7)$) node[]{$\sqrt{\mathring q}$};

\end{tikzpicture}
\caption{Universal boundary structure \\for $\Lambda$-BMS$_{d+1}$ boundary conditions.}
\label{fig:LBMS}
\end{figure}

From a geometrical point of view, the boundary conditions \eqref{BC Lambda BMS} are induced by the fixation of the foliation $\bm T$ on the boundary and a volume form $\sqrt{\mathring q}(\D^{d-1}x)$ on the leaves orthogonal to $\bm T$ -- see figure \ref{fig:LBMS}. For $d=2$, these leaves are $S^1$ circles on which we take $\mathring{q} = 1$. The requirement \eqref{BC Lambda BMS} is thus equivalent to the Dirichlet boundary conditions in $3d$ gravity. For any $d>2$, the fluctuating transverse components $g^{(0)}_{AB}(t,x^C)$ of the boundary metric contain the remaining $\frac{1}{2}(d+1)(d-2)$ degrees of freedom in $g_{ab}^{(0)}$. This is the most natural relaxation of the Dirichlet boundary conditions we presented for AlAdS$_{d+1}$ spacetimes, borrowing their foliation structure but in a weaker sense that allows for some fluctuations of the transverse metric components while keeping part of the boundary structure fixed. This is done in order to single out the dynamical modes among all variations of $g_{ab}^{(0)}$. 

All in all, the boundary conditions \eqref{BC Lambda BMS} are well-suitable for our original purpose because they enjoy the following properties:
\begin{enumerate}
\item As we will show in the next section \ref{Lambda BMS charge algebra section}, they are \textit{leaky} in the sense previously given in sections \ref{sec:Leaky boundary conditions theo} and \ref{sec:Boundary conditions and the Cauchy problem}. Indeed, the boundary degrees of freedom will act as sources for the presymplectic flux and will thus be responsible for non-integrability and non-conservation of the gravitational surface charges. 
\item They are also \textit{relevant for both signs of the cosmological constant}. Indeed, the boundary gauge fixing \eqref{BC Lambda BMS} is built to be always reachable using the freedom we have on the residual gauge diffeomorphisms \eqref{AKV 1} and \eqref{AKV 2}, namely $d$ parameters that can be fixed in the SFG expansion. This means that any solution written down in this coordinate system can be transformed by diffeomorphism in order to satisfy \eqref{BC Lambda BMS}. Therefore, these conditions do not constrain the Cauchy problem in asymptotically dS$_{d+1}$ spacetimes, in the sense that the flux allowed through the future conformal boundary is left completely arbitrary and will only depend on the field configuration set on a Cauchy surface at early times.
\end{enumerate}
To complete the list of good news, the asymptotic symmetry algebra associated with them is infinite-dimensional and has a strong connection with the symmetry structure appearing at null and spatial infinity of asymptotically locally flat spacetimes. This is what we will discover in the next section!

\subsection{Residual gauge symmetries: the \texorpdfstring{$\Lambda$}{Lambda}-BMS\texorpdfstring{$_{d+1}$}{(d+1)} algebroid}
\label{sec:Lambda BMS d}
Requiring the boundary conditions \eqref{BC Lambda BMS} to be preserved under the residual gauge diffeomorphisms generated by \eqref{AKV 1} and \eqref{AKV 2} yields the following conditions on the parameters:
\begin{equation}
\partial_t \bar{\xi}^t = \frac{1}{(d-1)} D_A \bar{\xi}^A , \qquad \partial_t \bar{\xi}^A = \frac{\eta}{\ell^2} g^{AB}_{(0)} D_B \bar{\xi}^t , \qquad \sigma = \frac{1}{(d-1)} D_A \bar{\xi}^A .
\label{equation Lambda BMS}
\end{equation} Since these equations explicitly involve the transverse metric $g^{AB}_{(0)}$, which is a dynamical field of the theory, the parameters $\bar{\xi}^t$ and $\bar{\xi}^A$ are \textit{field-dependent}. Therefore, the two first partial differential equations cannot be readily integrated for any metric $g_{AB}^{(0)}$, but in practice, any solution of \eqref{equation Lambda BMS} should involve $d-1$ ``integration constants'' which are distributed in two codimension 2 fields (one scalar and one vector on $S^{d-1}$) depending only on the angles. Due to the coupling between the constraint equations, even in simplest cases (the vacuum orbit for example), we cannot give a closed form for these codimension 2 fields but only a development in terms of $(d-1)$ spherical harmonics whose coefficients are time-dependent. We differ the presentation of these explicit solutions to section \ref{sec:Explicit solutions and flat limit} because the local information encoded in \eqref{equation Lambda BMS} is sufficient to discuss the properties of the asymptotic symmetry algebra.

Precisely taking these constraints into account, the residual gauge diffeomorphisms satisfy the following commutation relations with the modified Lie bracket \eqref{modified bracket}:
\begin{equation}
[{\xi}(\bar{\xi}_{1}^t,\bar{\xi}_{1}^A ),{\xi}(\bar{\xi}_{2}^t ,\bar{\xi}_{2}^A )]_\star =  \xi (\hat{\bar{\xi}}_{1}^t,\hat{\bar{\xi}}_{1}^A) \label{eq:VectorAlgebra1Lambda}
\end{equation}
where
\begin{equation}
\begin{split}
\hat{\bar{\xi}}^t &= \bar{\xi}_{1}^A D_A \bar{\xi}_{2}^t  + \frac{1}{(d-1)} \bar{\xi}_{1}^t D_A \bar{\xi}_{2}^A   - \delta_{\xi_1} \bar{\xi}_{2}^t - (1 \leftrightarrow 2 ), \\
\hat{\bar{\xi}}^A &=  \bar{\xi}^B_{1} D_B \bar{\xi}^A_{2} + \frac{\eta}{\ell^2} \bar{\xi}^t_{1} g^{AB}_{(0)} D_B \bar{\xi}_{2}^t   - \delta_{\xi_1} \bar{\xi}_{2}^A - (1 \leftrightarrow 2).
\end{split} \label{eq:VectorAlgebraLambda}
\end{equation} This is a corollary of \eqref{eq:VectorAlgebra}. These commutation relations are field-dependent for generic $d$ and depend therefore on the position in the solution space. This algebra of asymptotic symmetries constitutes rather a \textit{Lie algebroid} \cite{Crainic,Barnich:2010xq,Barnich:2017ubf} that we call $\Lambda$-BMS$_{d+1}$. For the definition given in section \ref{sec:Representation on the solution space}, the base manifold is the set of metrics $g^{(0)}_{ab}$ on $\mathscr I$ verifying \eqref{BC Lambda BMS}, the field-dependent Lie algebra is the Lie algebra of asymptotic symmetries \eqref{eq:VectorAlgebraLambda} and the anchor is the map $\bar\xi\mapsto\delta_{\bar\xi}g^{(0)}_{ab}$. In contrast to the standard BMS case, we cannot get rid of the precise reference to a particular point of the solution space to discuss separately the $\Lambda$-BMS$_{d+1}$ asymptotic algebroid. In flat spacetime, it is natural to consider leaky boundary conditions while keeping the gauge parameters field-independent as seen earlier, but this is mainly due to the null character of the conformal boundary. Indeed, the explicit field-dependence in the right-hand side of the second equation of \eqref{equation Lambda BMS} is precisely coming from the metric coefficient $g_{tt}^{(0)} = -\eta/\ell^2$. The latter would be zero on a hypersurface with $\bm T$ null, which is the case at flat limit $\ell\to+\infty$. Let us mention some properties of the $\Lambda$-BMS$_{d+1}$ algebroid:
\begin{itemize}[label=$\rhd$]
\item At each point of the solution space (\textit{i.e.} for a given field $g_{AB}^{(0)}$), $\Lambda$-BMS$_{d+1}$ forms an infinite-dimensional algebra. Indeed, it always contains the infinite-dimensional algebra of area-preserving diffeomorphisms on the $(d-1)$-sphere as a subalgebra, because $\bar{\xi}^t = 0$, $\partial_t \bar{\xi}^A = 0$, $D_A \bar{\xi}^A = 0$ is a trivial set of solutions of \eqref{equation Lambda BMS}. The last equality is solved by vector fields $\bar\xi^A = \varepsilon^{AB}\partial_B \Psi$, $\partial_t \Psi = 0$ for some arbitrary $\Psi(x^A)$ on the $S^{d-1}$ sphere.

\item In the flat limit $\ell \to +\infty$, $\Lambda$-BMS$_{d+1}$ reduces to the asymptotic symmetry algebra of asymptotically (locally) flat spacetimes, namely the (Generalized) BMS in $d+1$ dimensions, written BMS$_{d+1}$. Indeed, taking $\ell \to +\infty$ in \eqref{equation Lambda BMS}, we find
\begin{equation}
\partial_t\bar\xi^t = \frac{1}{(d-1)}D_A\bar\xi^A, \qquad \partial_t\bar\xi^A = 0,
\end{equation}
which are the constraints on the Generalized BMS$_{d+1}$ generators (\textit{i.e.} the higher dimensional uplift of \eqref{generalized BMS 4 differential eq}). These differential constraints can be explicitly solved in terms of codimension 2 fields as
\begin{equation}
\bar{\xi}^t = T + \frac{t}{2} D_A Y^A, \quad \bar{\xi}^A = Y^A 
\label{solution BMS}
\end{equation} where $T=T(x^A)$ is the supertranslation parameter and $Y^A = Y^A (x^B)$ is the super-Lorentz parameter. At the level of the commutation relations, the flat limit gives $\delta_{\xi} \bar{\xi}^t = 0$, $\delta_\xi \bar{\xi}^A = 0$ and
\begin{equation}
\begin{split}
\hat{\bar{\xi}}^t &= \bar{\xi}_{1}^A D_A \bar{\xi}_{2}^t  + \frac{1}{(d-1)} \bar{\xi}_{1}^t D_A \bar{\xi}_{2}^A   - (1 \leftrightarrow 2 ), \\
\hat{\bar{\xi}}^A &=  \bar{\xi}^B_{1} D_B \bar{\xi}^A_{2}  - (1 \leftrightarrow 2), 
\end{split} \label{eq:VectorAlgebraBMS4}
\end{equation} or, taking \eqref{solution BMS} into account,
\begin{equation}
\begin{split}
\hat T &= Y_1^A D_A T_2  + \frac{1}{(d-1)} T_1 D_A Y^A_2   - (1 \leftrightarrow 2 ), \\
\hat Y^A &=  Y^B_1 D_B Y^A_2  - (1 \leftrightarrow 2).
\end{split}
\end{equation} These commutation relations precisely correspond to those of the Generalized BMS$_{d+1}$ algebra given by Supertranslations $\loplus$ Diff$(S^{d-1})$. Note that the flat limit taken here is merely a contraction of Lie algebroids by sending the parameter $\ell$ to infinity. This is only the first hint of the link between $\Lambda$-BMS and BMS through a flat limit process. The latter is ill-defined in the SFG gauge and we need a bit more work to discuss all aspects of it. 

\item For $d= 2$, the codimension 2 boundary metric $g_{AB}^{(0)}$ with fixed determinant $\mathring{q} = 1$ has only one component $g^{(0)}_{\phi\phi} = 1$ and the boundary conditions \eqref{BC Lambda BMS} reduce to the Dirichlet boundary conditions \eqref{Dirichlet boundary conditions}. Henceforth, the $\Lambda$-BMS$_3$ algebroid is an algebra and corresponds to the infinite-dimensional conformal algebra in two dimensions Diff$(S^1)$ $\oplus$ Diff$(S^1)$ discussed in section \ref{Asymptotic symmetry algebra Dirichlet}. For $d = 3$, the structure of $\Lambda$-BMS$_4$ can be investigated in more details. In particular, some explicit solutions for the generators \eqref{equation Lambda BMS} will be obtained in section \ref{sec:Explicit solutions and flat limit} using the Helmholtz decomposition of $\bar{\xi}^A$ into a curl-free and a divergence-free part, available on the $2$-sphere. 

\item Owing to \eqref{Theta ren pullback}, we will see in the next section \ref{Lambda BMS charge algebra section} that the flux arising through the spacetime boundary is present when coupling the holographic stress-energy tensor to the fluctuating transverse boundary metric $g_{AB}^{(0)}$. As already mentioned, this field plays the role of source in the same way that the asymptotic shear $C_{AB}$ in asymptotically flat spacetimes. In the Generalized BMS phase space, the metric $q_{AB}$ on the celestial sphere is also allowed to vary and should also be considered as a source. Whatever their nature, these sources yield non-equilibrium physics and interactions between the radiating gravitational system and the environment, inducing non-integrability and non-conservation of the surface charges, see sections \ref{sec:Conservation of the charges theo}--\ref{sec:Physical content of the presymplectic flux}. Turning off the sources eliminates the leaks: the system is considered as isolated and the charges are integrable and conserved. In the Al(A)dS$_{d+1}$ phase space, starting from the boundary gauge fixing \eqref{BC Lambda BMS}, the requirement that the flux is canceled translates into $\delta g_{AB}^{(0)}=0$. In this case, we recover the Dirichlet boundary conditions \eqref{Dirichlet boundary conditions} and  the infinite-dimensional Lie algebroid $\Lambda$-BMS$_{d+1}$ ($d>2$) reduces to the finite-dimensional symmetry algebras $SO(d+1,1)$ for $\eta = -1$ and $SO(d,2)$ for $\eta = +1$ of the (A)dS$_{d+1}$ vacuum \cite{Barnich:2013sxa} (see also appendix \ref{app:ExactVectors}).
\begin{equation}
\begin{aligned}
\begin{tikzpicture}
\coordinate (A) at (0,0);
\coordinate (B) at (3,0);
\draw[] (A) node[draw=black,outer sep=5,left]{$\Lambda\text{-BMS}_{d+1}$};
\draw[-latex] (0,0) -- (B);
\draw[] (B) node[draw=black,outer sep=5,right]{$\begin{array}{c}
SO(d+1,1) \text{ if } \Lambda>0 \\
SO(d,2) \text{ if } \Lambda<0
\end{array}$};
\draw ($(A)!0.5!(B)$) node[above]{\scriptsize freezing sources};
\end{tikzpicture}
\label{symmetry breaking AdS}
\end{aligned}
\end{equation}
In the asymptotically locally flat phase space, the condition of freezing the sources, $\delta C_{AB}$ $=$ $0$ ($\delta q_{AB} = 0$), yields a similar symmetry breaking 
\begin{equation}
\begin{aligned}
\begin{tikzpicture}[baseline=(current  bounding  box.center)]
\coordinate (A) at (0,0);
\coordinate (B) at (3,0);
\draw (A) node[draw=black,outer sep=5,left]{\normalsize $\text{(Generalized) BMS}_{d+1}$};
\draw[-latex] (0,0) -- (B);
\draw (B) node[draw=black,outer sep=5,right]{\normalsize $SO(d,1) \loplus \mathbb{R}^4$};
\draw ($(A)!0.5!(B)$) node[above]{\scriptsize freezing sources};
\end{tikzpicture}
\label{Symmetry breaking flat space}
\end{aligned}
\end{equation}
where $SO(d,1) \loplus \mathbb{R}^4$ is the Poincaré algebra. In other words, the infinite-dimensional (Generalized) $\text{BMS}_{d+1}$ algebra reduces to the finite-dimensional symmetry group of the Minkowski vacuum when turning off the sources. Recalling that $\Lambda$-BMS$_{d+1}$ gives Generalized BMS$_{d+1}$ if $\Lambda\to 0$, it is not hard to be convinced that the symmetry breaking \eqref{Symmetry breaking flat space} is the flat limit of the symmetry breaking \eqref{symmetry breaking AdS}. These two diagrams commute, namely
\begin{equation}
\begin{aligned}
\begin{tikzpicture}[baseline=(current  bounding  box.center)]
\node[draw=black,outer sep=5] at (-5,0) (A)  {\normalsize $\Lambda\text{-BMS}_{d+1}$};
\node[draw=black,outer sep=5] at (0,1) (B) {
$\begin{array}{c}
SO(d+1,1) \text{ if } \Lambda>0 \\
SO(d,2) \text{ if } \Lambda<0
\end{array}$
};
\node[draw=black,outer sep=5] at (0,-1) (C) {\normalsize $\text{(Generalized) BMS}_{d+1}$};
\node[draw=black,outer sep=5] at (5,0) (D) {\normalsize $SO(d,1) \loplus \mathbb{R}^4$.};

\draw[-latex] (A) |- (B);
\draw[-latex] (A) |- (C);
\draw[latex-] (D) |- (B);
\draw[latex-] (D) |- (C);

\draw[] let \p1 = (A), \p2 = (B) in (\x1,\y2) node[above right]{\scriptsize freezing sources};

\draw[] let \p1 = (A), \p2 = (C) in (\x1,\y2) node[below right]{\scriptsize flat limit $\Lambda\to 0$};

\draw[] let \p1 = (D), \p2 = (B) in (\x1,\y2) node[above left]{\scriptsize flat limit $\Lambda\to 0$};

\draw[] let \p1 = (D), \p2 = (C) in (\x1,\y2) node[below left]{\scriptsize freezing sources};

\end{tikzpicture}
\end{aligned}
\label{fig:diagram}
\end{equation}
This gives another argument for considering relaxed boundary conditions as \eqref{BC Lambda BMS}: in order to recover an infinite dimensional algebra in the flat limit, one needs to have the sources turned on to follow the lower leg of the diagram \eqref{fig:diagram}. In other words, the flat limit of the asymptotically (A)dS phase space with Dirichlet boundary conditions \eqref{Dirichlet boundary conditions} only gives the subsector of stationary solutions with asymptotically flat boundary conditions in the flat limit.
\end{itemize}

We conclude here our discussion about the mathematical properties of the $\Lambda$-BMS$_{d+1}$ algebroid and get deeper into the physics it underlies in the next section.

\subsection{\texorpdfstring{$\Lambda$}{Lambda}-BMS\texorpdfstring{$_{d+1}$}{(d+1)} charge algebra}
\label{Lambda BMS charge algebra section}

Taking the boundary conditions \eqref{BC Lambda BMS} into account, the renormalized presymplectic potential \eqref{Theta ren pullback} reduces to
\begin{equation}
\bm\Theta_{ren}^{\Lambda\text{-BMS}}[\phi;\delta \phi]\Big|_{\mathscr I} = - \frac{\sqrt{\mathring{q}}}{2 \ell} T^{AB}_{TF} \delta g_{AB}^{(0)} \, (\D^d x) \label{presymplectic flux Lambda BMS d}
\end{equation} where $T^{AB}_{TF} = T^{AB} - \frac{1}{d-1} g^{AB}_{(0)} T^C_C$ is the trace-free part of the $(d-1)$-dimensional tensor $T^{AB} \stackrel{\text{not}}{=} T^{AB}_{[d],(tot)}$. Here the fields $\phi = \{g,\bm N,\bm T,\sqrt{\mathring q}\}$ are the bulk metric tensor $g$, the SFG foliation $\bm N$ and the boundary background structure $(\bm T,\sqrt{\mathring q})$ leading to the $\Lambda$-BMS$_{d+1}$ boundary conditions. The associated presymplectic current is given by
\begin{equation}
\bm \omega_{ren}^{\Lambda\text{-BMS}}[\phi; \delta_1 \phi, \delta_2 \phi]\Big|_{\mathscr I} = - \frac{\sqrt{\mathring{q}}}{2 \ell} \delta_1 T^{AB}_{TF} \delta_2 g_{AB}^{(0)} \,(\D^d x) - (1 \leftrightarrow 2)  .
\label{non vanishing flux}
\end{equation} This implies that the boundary conditions \eqref{BC Lambda BMS} are well leaky for $d>2$, \textit{i.e.} there is some flux going through the spacetime boundary. In the asymptotically dS$_{d+1}$ case, an arbitrary radiation crossing $\mathscr{I}^+_{\text{dS}}$ is expected since, as explained in section \ref{Boundary conditions and asymptotic symmetry algebra LBMS}, the boundary conditions \eqref{BC Lambda BMS} do not restrict the Cauchy problem. In the asymptotically AdS$_{d+1}$ case, the presence of a non-vanishing flux \eqref{non vanishing flux} through $\mathscr{I}_{\text{AdS}}$ yields a non-globally hyperbolic spacetime \cite{Ishibashi:2004wx}. In the holographic perspective, this translates into the fact that the dual theory couples to an external system \cite{Giddings:2020usy,Jana:2020vyx,Compere:2008us,Troessaert:2015nia,Almheiri:2014cka,Maldacena:2016upp}.

Similarly, the $\Lambda$-BMS$_{d+1}$ charges are obtained by inserting the boundary conditions \eqref{BC Lambda BMS} into the general charge expressions \eqref{surface charge expression}. We have explicitly
\begin{equation}
\ndelta H^{\Lambda\text{-BMS}} [\phi] = \delta H_\xi^{\Lambda\text{-BMS}} [\phi] + \Xi^{\Lambda\text{-BMS}}_\xi [\phi;\delta\phi]  \label{ndelta H Lambda BMS}
\end{equation} where
\begin{equation}
\begin{split}
H_\xi^{\Lambda\text{-BMS}} [\phi] &= - \eta \ell \oint_{S_\infty} (\D^{d-1} x) \sqrt{\mathring{q}} \left[  {T}_{tt} \bar{\xi}^t + T_{tB} \bar{\xi}^B \right], \\
\Xi_\xi^{\Lambda\text{-BMS}}  [\phi; \delta \phi] &= \oint_{S_\infty} (\D^{d-1} x) \left[ - \frac{1}{2\ell} \sqrt{\mathring{q}}\, \bar{\xi}^t \, T^{BC}_{TF} \delta g_{BC}^{(0)} +W^{[d]}[\phi; \delta \phi] \right] - H^{\Lambda\text{-BMS}}_{\delta \xi} [\phi] .
\end{split}
\label{split of the charge Lambda BMS}
\end{equation} The third first terms are universal and present for any dimension $d$. Only the relic of the Weyl term $W^{[d]}_{\Lambda\text{-BMS}}[\phi; \delta \phi]$, obtained by inserting \eqref{BC Lambda BMS} into $W^{[d]t}_\sigma [\phi; \delta \phi]$, depends on the dimension. In particular, this contribution vanishes when $d$ is odd. For $d=2$, the conditions \eqref{BC Lambda BMS} become the Dirichlet boundary conditions and $W^{[2]}_{\Lambda\text{-BMS}} = 0$. Finally, for $d=4$, taking into account that \eqref{BC Lambda BMS} holds, one obtains 
\begin{equation}
\begin{split}
R^{(0)}_{tt} &= -\partial_t l - l^{AB}l_{AB}\ , \quad R^{(0)}_{tA} = D^B l_{AB} - \partial_A l  , \\
R^{(0)}_{AB} &= R_{AB}[q] + \eta\,\ell^2 \left( (\partial_t+l)l_{AB} - 2 {l_A}^C l_{BC} \right) , \\
R^{(0)} &= R[q] + \eta\,\ell^2 \left( (\partial_t+l)l + l^{AB}l_{AB} \right) 
\end{split}
\end{equation}
for the usual notations $q_{AB}\equiv g_{AB}^{(0)}$, $l_A^B \equiv \frac{1}{2} q^{AC}\partial_t q_{BC}$ and $l = l^A_A = 0$. We have
\begin{align}
W^{[4]}_{\Lambda\text{-BMS}}[g^{(0)};\delta g^{(0)}] &= \int_{S_\infty} (\D^{d-1}x) \,\sqrt{\mathring q}\, \Big( 4 \bar\xi^t M_{\sigma}[g^{(0)};\delta g^{(0)}] + 2 \bar\xi^A \partial_A N_\sigma[g^{(0)};\delta g^{(0)}] \Big)  , \nonumber \\
M_{\sigma}[g^{(0)};\delta g^{(0)}] &= \frac{1}{768\pi G} D_A D^A \Big[ R_{BC}[q]\delta q^{BC} + \eta\ell^2 (\partial_t l_{BC}-2 {l_A}^C l_{BC})\delta q^{BC} \Big]  , \nonumber \\
N_{\sigma}[g^{(0)};\delta g^{(0)}] &= \frac{\eta\ell^3}{16\pi G} \Big[ \frac{1}{12} \eta\ell D_B(D^D l_{CD}\delta q^{BC}) + \frac{\ell^4}{12} R^{BC}[q]\delta l_{BC} + \frac{\eta\ell^6}{24}(\partial_t l^{BC}+2l^{BD} {l_D}^C)\delta l_{BC} \nonumber \\
&\phantom{= \frac{\eta\ell^3}{16\pi G} \Big[} - \frac{\ell^4}{12} D_C D^D l_{BD}\delta q^{BC} + \frac{\ell^4}{24}\partial_t R_{BC}[q]\delta q^{BC} \nonumber \\
&\phantom{= \frac{\eta\ell^3}{16\pi G} \Big[} + \frac{\ell^4}{24} (\partial_t^2 l_{BC} - 2 {l_B}^D \partial_t l_{CD} + 4 {l_B}^D {l_D}^E l_{EC} )\delta q^{BC} \nonumber \\
&\phantom{= \frac{\eta\ell^3}{16\pi G} \Big[} + \frac{\ell^4}{36} (R[q] + \eta\ell^2)l_{DE}l^{DE})l_{BC}\delta q^{BC}\Big]  \label{relations d4}
\end{align} 
after performing several integrations by parts on the boundary $3$-sphere and using \eqref{equation Lambda BMS} explicitly.

As a corollary of \eqref{charge algebra}, one obtains the $\Lambda$-BMS$_{d+1}$ charge algebra:
\begin{equation}
\boxed{\{ H_{\xi_1}^{\Lambda\text{-BMS}} [\phi] , H^{\Lambda\text{-BMS}}_{\xi_2} [\phi] \}_\star = H^{\Lambda\text{-BMS}}_{[\xi_1, \xi_2]_\star}[\phi] + K^{[d],\Lambda\text{-BMS}}_{\xi_1, \xi_2} [\phi].   }
\label{charge algebra Lambda BMS d}
\end{equation}
For $d=2k+1$ ($k \in\mathbb N_0$), the algebra closes without central extensions
\begin{equation}
K^{[2k+1],\Lambda\text{-BMS}}_{\xi_1, \xi_2} [\phi] \equiv 0.   
\label{cocycle Lambda BMS odd d}
\end{equation}
For even $d$, the 2-cocycle present in \eqref{charge algebra Lambda BMS d} is generalically non-zero. The case $d=2$ has already been discussed in section \ref{charge algebraaa diri}. Finally, for $d=4$, using \eqref{relations d4}, we obtain the following expression for the field-dependent $2$-cocycle:
\begin{align}
K^{[4],{\Lambda\text{-BMS}}}_{\xi_1,\xi_2}[\phi] &= \frac{\eta\,\ell^2}{16\pi G} \int (\D^{d-1}x)\,\sqrt{\mathring q} \, \times \ldots \nonumber \\
\ldots \times &\left\lbrace \frac{1}{18}(R[q]-\eta\ell^2 l_{AB}l^{AB})D_C \bar\xi^C_1 D_D D^D \bar\xi^t_2 - \frac{\eta\ell^2}{9}D_B l^{AB} D_C \bar\xi_1^C D_A D_D \bar\xi_2^D \right. \nonumber \\
&\quad \Big[ - \frac{\eta\ell^2}{12} D_B l^{AB} D^C l_{BC} + \frac{1}{12} R^{AB}[q]R_{AB}[q]  + \frac{\eta\ell^2}{6}R^{AB}[q](\partial_t l_{AB}-2 {l_A}^C l_{BC})  \nonumber \\
&\quad + \frac{\ell^4}{12}\partial_t l_{AB}\partial_t l^{AB} + \frac{\eta\ell^2}{12}\partial_t l_{AB} l^{AC}{l_C}^B - \frac{\eta\ell^2}{6}l^{AC}{l_C}^B\partial_t l_{AB}  + {l_A}^C {l_C}^B {l_D}^A {l_B}^D \nonumber \\
&\quad \left. - \frac{1}{36}R[q]^2 - \frac{\eta\ell^2}{18} l^{AB}l_{AB} + \frac{\ell^4}{18}(l_{AB}l^{AB})^2 \Big] D_E \bar\xi^E_1 \bar\xi^t_2 - (1\leftrightarrow 2) \right\rbrace .
\end{align} 
This concludes our exploration of the properties of the $\Lambda$-BMS algebroid in $d+1$ dimensions. From now on, we will only consider the case $d=3$ and pursue the analysis in that context.

\section{Aspects of the \texorpdfstring{$\Lambda$}{Lambda}-BMS\texorpdfstring{$_4$}{4} phase space}
\label{sec:Aspects of the LBMS4 phase space}
We investigate here several interesting aspects of the Al(A)dS$_4$ phase space in Einstein's gravity admitting the $\Lambda$-BMS$_{d+1}$ algebroid as asymptotic symmetries. To make more transparent the natural analogies between this phase space and the asymptotically locally flat phase space episodically discussed through chapter \ref{chapter:Charges}, we work from the beginning in the outgoing Bondi coordinates in which we will construct a flat limit algorithm in section \ref{sec:Flat limit of the LBMS4 phase space}. The full solution space of Einstein's gravity with cosmological constant is derived in this gauge and the parameters on the phase space as well as the residual gauge symmetries are matched with their counterparts intervening in the SFG gauge through a dictionary between both asymptotic choices of coordinates. Next, the analog of Bondi news and Bondi mass are identified, further properties of the $\Lambda$-BMS$_4$ algebroid are explored in the Bondi gauge and some particular solutions for the $\Lambda$-BMS$_4$ generators are obtained by integrating the constraint equations around the unit-round sphere. We end by discussing the conservative subsector of the $\Lambda$-BMS$_4$ phase space and discover unexpected new stationary solutions of gravity with locally AdS boundary conditions.

\subsection{Solution space in Bondi gauge}

In the following we derive the general solution space of Einstein gravity coupled to a cosmological constant $\Lambda$ of either sign in the Bondi gauge $(u,r,x^A)$. Doing so, we generalize the work of \cite{Poole:2018koa} which only focused on axisymmetric configurations, also completing the work of \cite{He:2015wfa}. The Newman-Penrose version of our derivation can be found in \cite{Mao:2019ahc}. 

We recall that the line element is given by \eqref{Bondi line element} where $\beta$, $U^A$, $g_{AB}$ and $V$ are arbitrary functions of the coordinates and the $2$-dimensional metric $g_{AB}$ satisfies the determinant condition
\begin{equation}
\partial_r \left(\frac{\det (g_{AB})}{r^4} \right) = 0. \label{eq:DetCond}
\end{equation} 
As justified in section \ref{sec:Gauge fixing conditions}, any metric can be brought in this gauge irrespectively of the value of the cosmological constant. For example, global (A)dS$_4$ is obtained by choosing $\beta = 0$, $U^A = 0$, $V/r = (\Lambda r^2/3)-1$, $g_{AB} = r^2 \mathring q_{AB}$, where $\mathring q_{AB}$ is the unit round-sphere metric. Global Minkowski spacetime is simply obtained be setting $\Lambda=0$ into the Bondi parameters of global (A)dS$_4$, namely $V/r = -1$. Note crucially that the Bondi gauge fixing \eqref{Bondi gauge conditions} does not make any reference to the value of $\Lambda$, in contrast with the SFG gauge fixing \eqref{FG gauge} which is ill-defined in the flat limit $\ell\to+\infty$ ($\Lambda\to 0$). This means that one can treat the three solution spaces for $\Lambda<0$, $\Lambda=0$ and $\Lambda>0$ on the same footing in the Bondi gauge and directly observe the relations between the dynamical fields in the three configurations. The residual gauge diffeomorphisms preserving Bondi's choice of coordinates have been obtained in section \ref{sec:Residual gauge transformations} without any assumption on $\Lambda$ and are still given by \eqref{eq:xir}.

Now we are ready to solve Einstein's equations $G_{\mu\nu} + \Lambda g_{\mu\nu} = 0$ for pure gravity with the line element \eqref{Bondi line element} in order to determine the functions $\beta,U^A,g_{AB}$. We follow the solving scheme and the notations of \cite{Barnich:2010eb,Tamburino:1966zz,Compere:2019bua}. In particular, we use the Christoffel symbols that have been derived in the first reference. We just need to give some minimal fall-off conditions on the codimension 2 metric $g_{AB}$ as a starting point for the solving algorithm.

\subsubsection{Minimal fall-off requirements}
We are interested in conformally compact solutions of the Einstein equations, so we impose the fall-off condition $g_{AB} = \mathcal{O}(r^2)$ allowing for conformal compactification. In addition of that, we assume an analytic expansion for $g_{AB}$, namely 
\begin{equation}
g_{AB} = r^2 \, q_{AB}  + r\, C_{AB} + D_{AB} + \frac{1}{r} \, E_{AB} + \frac{1}{r^2} \, F_{AB} + \mathcal{O}(r^{-3})\label{eq:gABFallOff}
\end{equation} 
where each term involves a symmetric tensor whose components are functions of $(u,x^C)$. For $\Lambda \neq 0$, the Fefferman-Graham theorem  \cite{Starobinsky:1982mr,Fefferman:1985aa,Skenderis:2002wp,2007arXiv0710.0919F,Papadimitriou:2010as}, together with the map between SFG and Bondi gauges that will be presented in section \ref{sec:FGg}, ensures that the expansion \eqref{eq:gABFallOff} leads to the most general solution to the vacuum Einstein equations. Nevertheless, recall that for $\Lambda = 0$, the analytic expansion \eqref{polynomial gAB} is an hypothesis since additional logarithmic branches might occur \cite{Winicour1985,Chrusciel:1993hx,ValienteKroon:1998vn}. Hence, in the flat limit, we will just reproduce the analytical subsector of asymptotically flat gravity but we will show that it is sufficient for our analysis. Crucially, the fall-off conditions \eqref{eq:gABFallOff} do not impose any constraint on the generators of residual diffeomorphisms in the Bondi gauge \eqref{eq:xir}. As in section \ref{sec:Solution space in Bondi gauge}, $\mathcal D_A$ represents the Levi-Civita connection for $g_{AB}$ and upper case Latin indices are lowered and raised by the $2$-dimensional metric $q_{AB}$ and its inverse when the $r$-dependence has been made completely explicit. The gauge condition \eqref{eq:DetCond} imposes successively that $\det(r^{-4}g_{AB}) = \det (q_{AB})$, $C_{AB}$ is tracefree and the traces of the subleading pieces $D_{AB},E_{AB},F_{AB},\dots$ are fixed as \eqref{traces of D E F}.

\subsubsection{Organization of Einstein's equations}
\label{sec:Bondi Organization of Einstein's equations}

We organize the equations of motion as follows. First, we solve the equations that do not involve the cosmological constant. The radial constraint $G_{rr} = R_{rr} = 0$ fixes the $r$-dependence of $\beta$, while the cross-term constraint $G_{rA} = R_{rA} = 0$ fixes the $r$-dependence of $U^A$. Next, we treat the equations that do depend upon $\Lambda$. The equation $G_{ur} + \Lambda g_{ur} = 0$ determines the $r$-dependence of $V/r$ in terms of the previous variables. Noticing that $R = g^{\mu\nu} R_{\mu\nu} = 2 g^{ur} R_{ur} + g^{rr} R_{rr} + 2 g^{rA} R_{rA} + g^{AB} R_{AB}$ and taking into account that $R_{rr} = 0 = R_{rA}$, one gets $G_{ur} + \Lambda g_{ur} = R_{ur} - \frac{1}{2} g_{ur} R + \Lambda g_{ur} = -\frac{1}{2} g_{ur} (g^{AB} R_{AB}-2\Lambda) = 0$ so that we can solve equivalently $g^{AB} R_{AB} = 2\Lambda$. Thereafter, we concentrate on the pure angular equation, $G_{AB} + \Lambda g_{AB} = 0$, which can be splitted into a tracefree part 
\begin{equation}
G_{AB} - \frac{1}{2} g_{AB} \, g^{CD}G_{CD} = 0 \label{eq:GABTF}
\end{equation}
and a pure-trace part
\begin{equation}
g^{CD} G_{CD} + 2\Lambda = 0. \label{eq:GABTFull}
\end{equation}
We consider now the Bianchi identities $\nabla_\mu G^{\mu\nu} = 0$ that can be rewritten as
\begin{equation}
2 \sqrt{-g} \nabla_\mu G^\mu_\nu = 2 \partial_\mu (\sqrt{-g}G^\mu_\nu) - \sqrt{-g} G^{\mu\lambda} \partial_\nu g_{\mu\lambda} = 0. \label{rewrite Bianchi}
\end{equation}
Since $\partial_\nu g_{\mu\lambda} = - g_{\mu\alpha}g_{\lambda\beta}\partial_\nu g^{\alpha\beta}$, we have
\begin{equation}
2 \partial_\mu (\sqrt{-g}G^\mu_\nu) + \sqrt{-g} G_{\mu\lambda} \partial_\nu g^{\mu\lambda} = 0.
\end{equation}
Taking $\nu = r$ and noticing that $G_{r\alpha} +\Lambda g_{r\alpha} = 0$ have already been solved, one gets
\begin{equation}
G_{AB} \partial_r g^{AB} = \frac{4\Lambda}{r}. \label{eq:int1}
\end{equation}
Recalling that \eqref{eq:GABTF} holds and that the determinant condition implies $g^{AB}\partial_r g_{AB} = 4/r$, we see that  \eqref{eq:int1} is equivalent to \eqref{eq:GABTFull}. As a consequence, the constraint $G_{AB} + \Lambda g_{AB} = 0$ is completely obeyed if \eqref{eq:GABTF} is solved. Indeed, once the tracefree part \eqref{eq:GABTF} has been set to zero, the tracefull part \eqref{eq:GABTFull} is automatically constrained by the Bianchi identity. Another way to see this is as follows. Imposing that $G_{r\alpha} +\Lambda g_{r\alpha} = 0$ holds, \eqref{eq:GABTF} is equivalent to
\begin{equation}
(M^{TF})^A_{\,\,B} \equiv M^A_{\,\,B} - \frac{1}{2} \delta^A_B M^C_{\,\,C} = 0, \quad M^A_{\,\,B} \equiv g^{AC}R_{CB}, \label{eq:MABTF}
\end{equation}
since the trace of $M^A_{\,\,B}$ has already been set to $2\Lambda$ while fixing the radial dependence of $V/r$. 

At this stage, Einstein's equations $(r,r)$, $(r,A)$, $(r,u)$ and $(A,B)$ have been solved. It remains to solve the $(u,u)$ and $(u,A)$ components. Doing so, we derive the evolution equations for the Bondi mass and angular momentum aspects. Expressing the $A$ component of the contracted Bianchi identities \eqref{rewrite Bianchi} yields
\begin{equation}
\partial_r \Big[ r^2 \Big( G_{uA} + \Lambda g_{uA} \Big) \Big] = \partial_r \Big[ r^2 \Big( R_{uA} - \Lambda g_{uA} \Big) \Big] = 0. \\
\end{equation}
The equation of motion $r^2 (G_{uA} + \Lambda g_{uA}) = 0$ can therefore be solved for a single value of $r$. We can isolate the only non-trivial equation to be the $1/r^2$ part of $R_{uA} - \Lambda g_{uA} = 0$. This determines the evolution of the Bondi angular momentum aspect, denoted by $N^{(\Lambda)}_A(u,x^B)$. Assuming that $G_{uA} + \Lambda g_{uA} = 0$ is solved, the last Bianchi identity \eqref{rewrite Bianchi} for $\nu = u$ becomes
\begin{equation}
\partial_r \Big[ r^2 \Big( G_{uu} + \Lambda g_{uu} \Big) \Big] = \partial_r \Big[ r^2 \Big( R_{uu} - \Lambda g_{uu} \Big) \Big] = 0,
\end{equation}
and the reasoning is similar. We solve the $r$-independent part of $r^2 (R_{uu} - \Lambda g_{uu})$, which uncovers the equation governing the time evolution of the Bondi mass aspect $M^{(\Lambda)}(u,x^A)$. Repeating this reasoning while assuming $\Lambda=0$ from the beginning leads to the solution space summarily described in section \eqref{sec:Solution space in Bondi gauge}. We thus adapted the solving procedure of \cite{Tamburino:1966zz,Barnich:2010eb} to $\Lambda\neq 0$ configurations.

\subsubsection{Solution to the algebraic equations}

We define several auxiliary fields as in \cite{Barnich:2010eb}. Starting from \eqref{eq:gABFallOff}, we can build $k_{AB} = \frac{1}{2} \partial_r g_{AB}$, $l_{AB} = \frac{1}{2} \partial_u g_{AB}$ and $n_A = \frac{1}{2} e^{-2\beta}g_{AB}\partial_r U^B$. The determinant condition \eqref{eq:DetCond} allows us to split the tensors $k_{AB}$ and $l_{AB}$ in leading trace-full parts and subleading trace-free parts as
\begin{equation}
\begin{split}
k^A_B &\equiv g^{AC} k_{BC} = \frac{1}{r} \delta^A_B + \frac{1}{r^2} K^A_B, \qquad K^A_A = 0, \\
l^A_B &\equiv g^{AC} l_{BC} = \frac{1}{2} q^{AC}\partial_u q_{BC} + \frac{1}{r} L^A_B, \qquad L^A_A = 0.
\end{split}
\end{equation}
Note that 
\begin{eqnarray}
l = l^A_A = \frac{1}{2} q^{AB}\partial_u q_{AB} = \partial_u \ln \sqrt{q}.
\end{eqnarray} 
Let us start by solving $R_{rr} = 0$ which leads to
\begin{equation}
\partial_r \beta = -\frac{1}{2r} + \frac{r}{4} k^A_B k^B_A = \frac{1}{4r^3} K^A_B K^B_A.
\end{equation}
Expanding $K^A_B$ in powers of $1/r$, we get
\begin{align}
\beta(u,r,x^A) &= \beta_0 (u,x^A) + \frac{1}{r^2} \Big[ -\frac{1}{32} C^{AB} C_{AB} \Big] + \frac{1}{r^3} \Big[ -\frac{1}{12} C^{AB} \mathcal{D}_{AB} \Big] \label{eq:EOM_beta} \\
&\qquad + \frac{1}{r^4}\Big[ - \frac{3}{32} C^{AB}\mathcal{E}_{AB} - \frac{1}{16} \mathcal{D}^{AB}\mathcal{D}_{AB} + \frac{1}{128} (C^{AB}C_{AB})^2 \Big] + \mathcal{O}(r^{-5}). \nonumber
\end{align}
Up to the integration ``constant'' $\beta_0 (u,x^A)$, the condition \eqref{eq:gABFallOff} uniquely determines  $\beta$. In particular, the $1/r$ order is always zero on-shell. This equation also holds for $\Lambda =0$ but standard asymptotically flat conditions set $\beta_0 = 0$. We shall keep it arbitrary here. 

Next, we develop $R_{rA} = 0$, which gives
\begin{equation}
\partial_r (r^2 n_A) = r^2 \Big( \partial_r - \frac{2}{r} \Big) \partial_A \beta - \mathcal{D}_B K^B_A.
\end{equation}
We now expand the transverse covariant derivative $\mathcal{D}_A$:
\begin{equation}
\Gamma^B_{AC}[g_{AB}] = \Gamma^B_{AC}[q_{AB}] + \frac{1}{r} \Big[ \frac{1}{2} (D_A C^B_C + D_C C^B_A - D^B C_{AC}) \Big] + \mathcal{O}(r^{-2}),
\end{equation}
in terms of the transverse covariant derivative $D_A$ defined with respect to the leading transverse metric $q_{AB}$. This implies in particular that
\begin{equation}
\mathcal{D}_B K^B_A = -\frac{1}{2} D^B C_{AB} + \frac{1}{r} \Big[ -D^B \mathcal{D}_{AB} + \frac{1}{8} \partial_A (C_{BC}C^{BC}) \Big] + \mathcal{O}(r^{-2}).
\end{equation}
Using explicitly \eqref{eq:EOM_beta}, we find
\begin{equation}
n_A = -\partial_A \beta_0 + \frac{1}{r}\Big[ \frac{1}{2}D^B C_{AB} \Big] + \frac{1}{r^2} \Big[ \ln r \,  D^B \mathcal{D}_{AB} + N_A \Big] + o(r^{-2})
\end{equation}
where $ N_A$ is a second integration ``constant'' (\textit{i.e.} $\partial_r  N_A = 0$), which corresponds to the Bondi angular momentum aspect in the asymptotically flat case. After inverting the definition of $n_A$, integrating one time further on $r$ and raising the index $A$, we end up with
\begin{equation}
\begin{split}
U^A = \,\, & U^A_0(u,x^B) +\overset{(1)}{U^A}(u,x^B) \frac{1}{r} + \overset{(2)}{U^A}(u,x^B) \frac{1}{r^2} \\
&+ \overset{(3)}{U^A}(u,x^B) \frac{1}{r^3} + \overset{(\text{L}3)}{U^A}(u,x^B) \frac{\ln r}{r^3} + o(r^{-3})
\end{split} \label{eq:EOM_UA}
\end{equation}
with
\begin{eqnarray}
\overset{(1)}{U^A}(u,x^B)\hspace{-6pt} &=&\hspace{-6pt} 2 e^{2\beta_0} \partial^A \beta_0 ,\nonumber \\
\overset{(2)}{U^A}(u,x^B)\hspace{-6pt} &=&\hspace{-6pt} - e^{2\beta_0} \Big[ C^{AB} \partial_B \beta_0 + \frac{1}{2} D_B C^{AB} \Big], \nonumber\\
\overset{(3)}{U^A}(u,x^B)\hspace{-6pt} &=& \hspace{-6pt}- \frac{2}{3} e^{2\beta_0} \Big[ N^A - \frac{1}{2} C^{AB} D^C C_{BC} +   (\partial_B \beta_0 - \frac{1}{3} D_B) \mathcal{D}^{AB} - \frac{3}{16} C_{CD}C^{CD} \partial^A \beta_0  \Big], \nonumber\\
\overset{(\text{L}3)}{U^A}(u,x^B) \hspace{-6pt}&=&\hspace{-6pt} -\frac{2}{3}e^{2\beta_0}D_B \mathcal{D}^{AB}, \label{eq:EOM_UA2}
\end{eqnarray}
where $U^A_0(u,x^B)$  is a new integration ``constant''. Again, this equation also holds if $\Lambda$ is absent, but standard asymptotic flatness sets this additional parameter to zero. As in the flat case analysis, the presence of $\mathcal{D}_{AB}$ is responsible of logarithmic terms in the expansion of $U^A$. We will shortly derive that $\mathcal{D}_{AB}$ vanishes on-shell if $\Lambda \neq 0$. The hypothesis \eqref{eq:gABFallOff} is thus saved!

Given that
\begin{equation}
\begin{split}
M^A_{\,\,B} &= e^{-2\beta} \Big[ (\partial_r + \frac{2}{r}) (l^A_B + k^A_B \frac{V}{r} + \frac{1}{2} \mathcal{D}_B U^A + \frac{1}{2} \mathcal{D}^A U_B) \\
&\qquad\qquad + k^A_C \mathcal{D}_B U^C - k^C_B \mathcal{D}_C U^A + (\partial_u + l)k^A_B + \mathcal{D}_C (U^C k^A_B) \Big] \\
&\qquad + R^A_B[g_{CD}] - 2(\mathcal{D}_B \partial^A \beta + \partial^A \beta \partial_B \beta + n^A n_B),
\end{split}
\end{equation}
we extract the $r$-dependence of $V/r$ thanks to $M^A_{\,\,A} = 2\Lambda$, which reads as
\begin{equation}
\begin{split}
\partial_r V = & - 2 r (l + D_A U^A) + \\ &e^{2\beta} r^2 \Big[ D_A D^A \beta + (n^A - \partial^A \beta) (n_A - \partial_A \beta) - D_A n^A - \frac{1}{2} R[g_{AB}] + \Lambda \Big] .
\end{split}
\end{equation}
Taking into account \eqref{eq:gABFallOff}, \eqref{eq:EOM_beta} and \eqref{eq:EOM_UA}, we get after integration on $r$
\begin{align}
\frac{V}{r} = &\frac{\Lambda}{3} e^{2\beta_0} r^2 - r (l + D_A U^A_0) \label{eq:EOMVr} \\
&- e^{2\beta_0} \Big[ \frac{1}{2}\Big( R[q] + \frac{\Lambda}{8}C_{AB} C^{AB} \Big) + 2 D_A \partial^A \beta_0 + 4 \partial_A \beta_0 \partial^A \beta_0 \Big] + \frac{2  M}{r} + \mathcal{O}(r^{-1}) \nonumber 
\end{align}
where $ M(u,x^A)$ is an integration ``constant'' which, in flat asymptotics, is recognized as the Bondi mass aspect. 

Afterwards, we solve \eqref{eq:MABTF} order by order, which provides us the constraints imposed on each independent order of $g_{AB}$. The leading $\mathcal{O}(r^{-1})$ order of that equation yields
\begin{equation}
\frac{\Lambda}{3} C_{AB} = e^{-2\beta_0} \Big[ (\partial_u - l) q_{AB} + 2 D_{(A} U^0_{B)} - D^C U^0_C q_{AB} \Big].
\label{eq:CAB}
\end{equation}
This result shows that there is a discrete bifurcation between the asymptotically flat case and the case $\Lambda \neq 0$. Indeed, when $\Lambda = 0$, the left-hand-side vanishes, which leads to a constraint on the time-dependence of $q_{AB}$, see \eqref{EOM qAB time evolution}. As a consequence, the field $q_{AB}$ is constrained while $C_{AB}$ is completely free and interpreted as the shear. For (A)dS$_4$ asymptotics, $C_{AB}$ is entirely determined by $q_{AB}$ and $U^A_0$, while the boundary metric $q_{AB}$ is left completely undetermined by the equations of motion. This is consistent with previous analyses \cite{Ashtekar:2014zfa,He:2015wfa,Saw:2017amv,He:2018ikd,Poole:2018koa}.

Going to $\mathcal{O}(r^{-2})$, we get
\begin{equation}
\frac{\Lambda}{3} \mathcal{D}_{AB} = 0,\label{eq:DAB}
\end{equation}
which removes the logarithmic term in \eqref{eq:EOM_UA} for $\Lambda \neq 0$, but not for $\Lambda = 0$. The condition at next $\mathcal{O}(r^{-3})$ order
\begin{equation}
\partial_u \mathcal{D}_{AB} + U_0^C D_C \mathcal{D}_{AB} + 2 \mathcal{D}_{C(A} D_{B)}U_0^C = 0, \label{eq:partial u DAB}
\end{equation}
is thus trivial for $\Lambda \neq 0$, but reduces to $\partial_u \mathcal{D}_{AB} = 0$ in the flat limit, consistently with previous results \cite{Barnich:2010eb,Tamburino:1966zz}. 

Using an iterative argument as in \cite{Poole:2018koa}, we now make the following observation. If we decompose $g_{AB} = r^2 \sum_{n\geq 0} g_{AB}^{(n)} r^{-n}$, we see that the iterative solution of \eqref{eq:MABTF} is organized as $\Lambda g_{AB}^{(n)} = \partial_u g_{AB}^{(n-1)} + (...)$ at order $\mathcal{O}(r^{-n})$, $n\in\mathbb{N}_0$. Accordingly, the form of $\mathcal{E}_{AB}$ should have been fixed by the equation \eqref{eq:partial u DAB} found at $\mathcal{O}(r^{-3})$, but it is not the case, since both contributions of $\mathcal{E}_{AB}$ cancel between $G_{AB}$ and $\Lambda g_{AB}$. Moreover, the equation $\Lambda g_{AB}^{(4)} = \partial_u g_{AB}^{(3)} + (...)$ at next order turns out to be a constraint for $g_{AB}^{(4)} \sim \mathcal{F}_{AB}$, determined with other subleading data such as $C_{AB}$ or $\partial_u g_{AB}^{(3)} \sim \partial_u \mathcal{E}_{AB}$. It shows that $\mathcal{E}_{AB}$ is a set of two free data on the boundary, built up from two arbitrary functions of $(u,x^A)$. It shows moreover that there are no additional data to be uncovered for $\Lambda \neq 0$. 

As a conclusion, Einstein's equations $(r,r)$, $(r,A)$, $(r,u)$ and $(A,B)$ can be solved iteratively in the asymptotic expansion for $\Lambda \neq 0$. We identified 11 independent functions $\{ \beta_0 (u,x^A)$, $U^A_0 (u,x^B)$, $q_{AB} (u,x^C)$, $ M (u,x^C)$, $ N_A(u,x^C)$, $\mathcal{E}_{AB} (u,x^C)\}$ that determine the asymptotic solution. We will see below that the remaining equations are equivalent to evolution equations for $M(u,x^A)$ and $ N_A(u,x^B)$. Hence the solution space is
\begin{equation}
\mathcal S_\Lambda = \left\{ g_{\mu\nu}\left[\beta_0,U^A_0,q_{AB},M,N_A,\mathcal D_{AB},\mathcal E_{AB}\right] \ \Big| \ G_{\mu\nu}[g]+\Lambda g_{\mu\nu} = 0 \right\}. \label{SLambda}
\end{equation}
This contrasts with the asymptotically flat case $\Lambda = 0$ where an infinite series of functions appear in the radial expansion of $g_{AB}$, see \eqref{S0}. We will give more details about this discrepancy while discussing the flat limit.

\subsubsection{\texorpdfstring{$\Lambda$}{Lambda}-BMS\texorpdfstring{$_4$}{4} boundary gauge fixing}
At this point, we solved all of the algebraic relations among the Einstein equations. As we said, it remains to give the dynamical equations for the Bondi mass and angular momentum aspects. In full generality, these constraint equations look like a complete mess due to the unbecoming presence of the boundary fields $\beta_0$ and $U^A_0$. In that case, why do we not simplify our analysis by imposing a $\Lambda$-BMS type of boundary gauge fixing in order to get the expressions tractable and readable? As we stressed earlier, there is a more fundamental reason for which this reduction is appealing: the non-trivial charges will obey flux-balance laws at conformal infinity and the natural conjugated diffeomorphism parameters are codimension 2 functions on the angles, which appear as integration constants of the constraint equations \eqref{equation Lambda BMS} recast in the Bondi gauge as \eqref{eq0a}--\eqref{eq2a}. 

Let us see how the boundary gauge fixing acts on Bondi variables. The pullback of the most general Bondi metric satisfying \eqref{eq:gABFallOff} to the conformal boundary gives
\begin{equation}
\left. \D s^2 \right|_{\mathscr{I}} = \Big[ \frac{\Lambda}{3}e^{4\beta_0} + U_0^A U^0_A \Big] \D u^2 - 2 U_A^0 \D u \D x^A + q_{AB} \D x^A \D x^B. \label{eq:ds2I}
\end{equation}
Defining a set of Gaussian normal coordinates foliated along $u$ and fixing the transverse volume thanks to a Weyl rescaling on the boundary as in \eqref{BC Lambda BMS} yield
\begin{equation}
\beta_0 = 0,\quad U^A_0 = 0,\quad \sqrt{q} = \sqrt{\bar q}\label{bndgauge}
\end{equation}
where $\sqrt{\bar q}$ is an arbitrary measure on the sphere. Generically, it may depend on time, but it satisfies $\delta\sqrt{\bar q}$ for any linearized variation. The gauge \eqref{bndgauge} is a temporal boundary gauge for $\Lambda < 0$, a radial boundary gauge for $\Lambda > 0$ and a null boundary gauge for $\Lambda = 0$ with $g_{ur} = -1+ \mathcal{O}(r^{-1})$ in \eqref{Bondi line element}. But is this always reachable for any value of $\Lambda$ in the Bondi gauge? Computing the Lie derivative on the Bondi metric on-shell and retaining only the leading $\mathcal{O}(r^2)$ terms, we get the transformation laws of the boundary fields $q_{AB}$, $\beta_0$ and $U^A_0$ under the set of residual gauge transformations \eqref{eq:xir}:
\begin{align}
\delta_\xi q_{AB} &= f(\partial_u - l)q_{AB} + (\mathcal{L}_Y - D_C Y^C +2\omega)q_{AB} \nonumber \\
&\quad - 2 (U_{(A}^0 \partial_{B)}f - \frac{1}{2} q_{AB} U^C_0 \partial_C f),\label{eq56} \\
\delta_\xi \beta_0 &= (f \partial_u + \mathcal{L}_Y) \beta_0 + \frac{1}{2}\Big[ \partial_u - \frac{1}{2}l + \frac{3}{2} U^A_0 \partial_A \Big] f - \frac{1}{4} (D_A Y^A  - 2 \omega),\label{eq57} \\
\delta_\xi U^A_0 &= (f\partial_u + \mathcal{L}_Y) U_0^A - \Big[ \partial_u Y^A - \frac{1}{\ell^2} e^{4\beta_0} q^{AB} \partial_B f \Big] + U_0^A (\partial_u f + U_0^B \partial_B f).\label{eq58}
\end{align}
The first equation implies that $q^{AB}\delta_\xi q_{AB} = 4\omega$. We can therefore adjust the Weyl generator $\omega$ in order to reach the gauge $ \sqrt{q} = \sqrt{\bar q}$. This is similar to the boundary gauge fixing \eqref{BC Lambda BMS} in the SFG gauge where one has to use the bulk information of the boundary diffeomorphism to fix the area of the transverse spaces. Although the parameters are completely redistributed in the Bondi gauge, the establishment of a dictionary in section \ref{sec:FGg} will show that both boundary gauge fixings are identical. The form of the infinitesimal transformations \eqref{eq57}-\eqref{eq58} involves $\p_u f$ and $\p_u Y^A$. This ensures that a finite gauge transformation labeled by $f,Y^A$ can be found by integration over $u$ in order to reach $\beta_0=0$, $U^A_0=0$, at least in a local patch. As a result, the conditions \eqref{bndgauge} can always be reached by gauge fixing, at least locally. The vanishing of the inhomogeneous contributions in the transformation laws \eqref{eq57}-\eqref{eq58} constrains the parameters $f,Y^A$ and reduces the set of allowed vectors as it has been observed in the SFG gauge. When $\Lambda\neq 0$, the residual diffeomorphisms constitute the $\Lambda$-BMS$_4$ algebroid as we will review below. When $\Lambda=0$, these residual diffeomorphisms define the Generalized BMS$_4$ algebra. This is corollary of the results presented in section \ref{sec:Lambda BMS d}, up to the subtlety  that here the flat limit is also well-defined at the level of the solution space and not at the level of the symmetries only. A more precise discussion on the properties of the $\Lambda$-BMS$_4$ algebroid as well as the explicit flat limit of the symmetry parameters is postponed to section \ref{sec: Lambda BMS 4 in Bondi}.

\subsubsection{Constraint equations as Bondi evolution equations}

Assuming the gauge fixing conditions \eqref{bndgauge}, we are now ready to present the evolution equations that follow from the remaining Einstein equations. As justified before, the $\mathcal{O}(r^0)$ part of $r^2 (R_{uA} - \Lambda g_{uA}) = 0$ fixes the temporal evolution of $N_A$. From the Christoffel symbols found in \cite{Barnich:2010eb}, we can develop the first term as 
\begin{align}
R_{uA} = &-(\partial_u - l) \partial_A \beta - \partial_A l - (\partial_u + l) n_A  \\
&+ n_B \mathcal{D}^B U_A - \partial_B \beta \mathcal{D}_A U^B + 2 U^B (\partial_A \beta \partial_B \beta + n_A n_B) \nonumber \\
&+ \mathcal{D}_B \Big[ l^B_A + \frac{1}{2} ( \mathcal{D}^B U_A - \mathcal{D}_A U^B ) + U^B (\partial_A \beta - n_A) \Big] + 2 n_B l^B_A \nonumber \\
&- \frac{1}{2}(\partial_r + 2\partial_r \beta + \frac{2}{r})  \partial_A \frac{V}{r} - \frac{V}{r} (\partial_r + \frac{2}{r})n_A + k_A^B (\partial_B \frac{V}{r} + 2 \frac{V}{r} n_B) \nonumber \\
&- e^{-2\beta} (\partial_r + \frac{2}{r}) \Big[ U^B(l_{AB} + \frac{V}{r} k_{AB} + \mathcal{D}_{(A}U_{B)}) \Big] \nonumber \\
&- e^{-2\beta} U^B \Big[ (\partial_u + l) k_{AB} - 4 l^C_{(A} k_{B)C} - 2 k^C_A k_{BC} \frac{V}{r} + \mathcal{D}_C (k_{AB}U^C) - 2 k_{C(A}\mathcal{D}^C U_{B)} \Big]. \nonumber
\end{align}
Let us emphasize that the $r$-dependence of the fields is not yet explicit in this expression: the upper case Latin indices are lowered and raised by the full metric $g_{AB}$ and its inverse. Expanding all fields in power series of $1/r$ in $R_{uA}$ and $\Lambda g_{uA}$, 	and selecting the $1/r^2$ terms yields
\begin{equation}
(\partial_u + l)  N_A^{(\Lambda)} - \partial_A  M^{(\Lambda)} - \frac{\Lambda}{2} D^B  J_{AB} = 0. \label{eq:EvolutionNA}
\end{equation}
Here, we defined with hindsight the Bondi mass and angular momentum aspects for $\Lambda \neq 0$ as
\begin{align}
M^{(\Lambda)} &=  M + \frac{1}{16} (\partial_u + l)(C_{CD}C^{CD}), \label{eq:hatM} \\
N^{(\Lambda)}_A &=  N_A - \frac{3}{2\Lambda} D^B (N_{AB} - \frac{1}{2} l C_{AB}) - \frac{3}{4} \partial_A (\frac{1}{\Lambda} R[q] - \frac{3}{8}  C_{CD}C^{CD}), \label{eq:hatNA}
\end{align}
and the traceless symmetric tensor $J_{AB}$ ($q^{AB} J_{AB} = 0$) as
\begin{align}
J_{AB} = &-\mathcal{E}_{AB} -\frac{3}{\Lambda^2} \Big[ \partial_u (N_{AB} - \frac{1}{2} lC_{AB})  -\frac{\Lambda}{2} q_{AB} C^{CD}(N_{CD} - \frac{1}{2} l C_{CD}) \Big] \nonumber \\
&\quad +\frac{3}{\Lambda^2} (D_A D_B l - \frac{1}{2} q_{AB} D_C D^C l) \nonumber \\
&\quad -\frac{1}{\Lambda} (D_{(A}D^C C_{B)C} - \frac{1}{2} q_{AB} D^C D^D C_{CD}) \nonumber \\
&\quad +C_{AB} \Big[ \frac{5}{16} C_{CD}C^{CD} + \frac{1}{2\Lambda}R[q]\Big] . \label{eq:hatJAB}
\end{align}
We used the notation $N_{AB} \equiv \partial_u C_{AB}$. This tensor is symmetric and obeys $q^{AB}N_{AB} = \frac{\Lambda}{3} C^{AB}C_{AB}$. When $\Lambda = 0$, $N_{AB}$ is thus traceless and represents the Bondi news tensor. We will justify the definitions \eqref{eq:hatM}-\eqref{eq:hatNA} in section \ref{sec:FGg}. Notice also that $\partial_u q_{AB}$ has been eliminated using \eqref{eq:CAB}. The transformations of these fields under the residual gauge symmetries $\xi$ preserving the Bondi gauge \eqref{Bondi line element} and the boundary gauge \eqref{bndgauge} are given by
\begin{align}
\delta_\xi M^{(\Lambda)} &= [f\partial_u + \mathcal{L}_Y + \frac{3}{2}(D_A Y^A + f l - 2 \omega)]M^{(\Lambda)} - \frac{\Lambda}{3} N_A^{(\Lambda)} \partial^A f, \label{eq:VarM} \\
\delta_\xi N_A^{(\Lambda)} &= [f\partial_u + \mathcal{L}_Y + D_B Y^B + f l - 2 \omega] N_A^{(\Lambda)} + 3 M^{(\Lambda)} \partial_A f + \frac{\Lambda}{2} J_{AB} \partial^B f, \label{eq:VarNA} \\
\delta_\xi J_{AB} &= [f\partial_u + \mathcal{L}_Y + \frac{1}{2}(D_C Y^C + f l - 2 \omega)] J_{AB} \nonumber \\
&\quad\, - \frac{4}{3} (N_{(A}^{(\Lambda)}\partial_{B)}f - \frac{1}{2} N_C^{(\Lambda)} \partial^C f q_{AB}). \label{eq:VarJAB}
\end{align}
Let us now derive the temporal evolution of $ M$, encoded in the $r$-independent part of $r^2 (R_{uu} - \Lambda g_{uu})=0$. The first term is worked out to be 
\begin{equation}
\begin{split}
R_{uu} = \hspace{0.2cm} &(\partial_u + 2 \partial_u \beta + l) \Gamma^u_{uu} + (\partial_r + 2\partial_r \beta + \frac{2}{r}) \Gamma^r_{uu} + (\mathcal{D}_A + 2\partial_A \beta) \Gamma^A_{uu}  \\
&- 2\partial_u^2 \beta - \partial_u l - (\Gamma^u_{uu})^2 - 2 \Gamma^u_{uA}\Gamma^A_{uu} - (\Gamma^r_{ur})^2 -2\Gamma^r_{uA}\Gamma^A_{ur} - \Gamma^A_{uB}\Gamma^B_{uA}
\end{split}
\end{equation}
where all Christoffel symbols can be found in page 26 of \cite{Barnich:2010eb}. We finally get
\begin{equation}
(\partial_u + \frac{3}{2}l) M^{(\Lambda)} + \frac{\Lambda}{6} D^A  N^{(\Lambda)}_A + \frac{\Lambda^2}{24} C_{AB} J^{AB} = 0. \label{eq:EvolutionM}
\end{equation}

As a conclusion, in the Bondi gauge \eqref{Bondi line element}, assuming the fall-off condition \eqref{eq:gABFallOff} and the boundary gauge fixing \eqref{bndgauge}, the general solution to Einstein's equations with $\Lambda \neq 0$ is entirely determined by the following 7 free functions of $(u,x^A)$: $q_{AB}$ with fixed area $\sqrt{\bar q}$, $ M$, $N_A$ and $J_{AB}$ tracefree, $M$ and $N_A$ being constrained by the evolution equations \eqref{eq:EvolutionM}-\eqref{eq:EvolutionNA}, \textit{i.e.}
\begin{equation}
\bar{\mathcal S}_\Lambda = \left\{ g_{\mu\nu}\left[q_{AB},M,N_A,J_{AB}\right] \ \Big| \ G_{\mu\nu}[g] + \Lambda g_{\mu\nu} = 0 \, ; \bm T,\sqrt{q}=\sqrt{\bar q}\right\}, \label{SLambdabar}
\end{equation}
where we indicate clearly the boundary background structure associated with the solution space, namely the foliation $\bm T$ and the codimension 2 volume $\sqrt{\bar q}$. Remark that in general, such a solution space does not contain the (A)dS$_4$ global vacuum, since the boundary volume is not necessarily the unit-round sphere volume $\sqrt{\mathring q}$. In the following, as in the flat case, we will mainly be interested in the solution spaces
\begin{equation}
\mathring{\mathcal S}_\Lambda = \left\{ g_{\mu\nu}\left[q_{AB},M,N_A,J_{AB}\right] \ \Big| \ G_{\mu\nu}[g] + \Lambda g_{\mu\nu} = 0 \, ; \bm T,\sqrt{q}=\sqrt{\mathring q}\right\}, \label{SLambdaring}
\end{equation}
which all contain (A)dS$_4$ as particular vacuum solution for any value of $\Lambda$. For the purpose to match these solution spaces to with our previous analysis in the SFG gauge, we have to study how to relate the Bondi parameters of $\bar{\mathcal S}_\Lambda$ to the SFG variables $(g_{ab}^{(0)},T^{ab}_{[3]})$.

\subsection{Dictionary between Starobinsky/Fefferman-Graham and Bondi gauges}
\label{sec:FGg}

In this section, we find the explicit change of coordinates that maps a general vacuum solution of asymptotically locally (A)dS$_4$ gravity in the Bondi gauge to the SFG gauge. This procedure will lead to the explicit dictionary linking the free functions \eqref{SLambda} defined in the Bondi gauge to the holographic functions defined in SFG gauge, namely the boundary metric $g_{ab}^{(0)}$ and the boundary stress-tensor 
\begin{equation}
T^{[3]}_{ab} = \eta\frac{\sqrt{3|\Lambda|}}{16\pi G} g^{(3)}_{ab} \label{eq:T3ab}
\end{equation}
which particularizes \eqref{holographic stress-energy tensor} for $d=3$.

We follow and further develop the procedure introduced in \cite{Poole:2018koa}. The intermediate steps are presented for the AdS case ($\Lambda<0$) for the sake of conciseness, but we restore $\Lambda$ with an indeterminate sign in the final expressions, providing thus the change of coordinate for both dS and AdS cases. We first note that one can map the AdS$_4$ vacuum metric in Bondi coordinates
\begin{equation}
\D s^2_{\text{AdS}_4} = -\left(\frac{r^2}{\ell^2}+1\right)\D u^2 - 2\D u\D r +r^2\mathring q_{AB}\D x^A \D x^B \label{AdSBondi}
\end{equation}
to the global static patch \eqref{AdS static} $(t_\star,r_\star,x^A_\star)$ by using $u = t_\star - r_\star$ where the tortoise coordinate is $r_\star \equiv \ell [\arctan \left( \frac{r}{\ell} \right)  -\frac{\pi}{2}]$. This maps the conformal infinity $\mathscr I$ situated at $r=\infty$ to $r_\star=0$, which is very convenient since this also matches immediately the location of $\mathscr I$ at the roots of the holographic coordinate $\rho$ in the SFG coordinate system. Therefore the change of coordinates from $(t_\star,r_\star,x^A_\star)$ to SFG gauge $(t,\rho,x^A)$ can be performed perturbatively in series of $\rho$ around $\rho = 0$, identified with $r_\star = 0$. At leading order, the time $t_\star$ of the global patch is equivalent to the SFG time $t$, which justifies a bit more to use the global coordinates as an intermediate coordinate system trading the retarded time $u$ for $t$. The general algorithm is thus the following:
\begin{enumerate}
\item Starting from any asymptotically locally AdS$_4$ solution formulated in the Bondi gauge $(u,r,x^A)$, we perform the preliminary change to the coordinate system $(t_\star,r_\star,x^A_\star)$,
\begin{equation}
\begin{split}
u &\to t_\star-r_\star, \quad x^A \to x^A_\star, \\
r &\to \ell \tan \Big[\frac{r_\star}{\ell}+\frac{\pi}{2}\Big] = -\frac{\ell^2}{r_\star} + \frac{r_\star}{3} + \frac{r_\star^3}{45\ell^2} + \mathcal{O}(r_\star^{5})\\
&\phantom{\to \ell \tan \Big[\frac{r_\star}{\ell}+\frac{\pi}{2}\Big] }= \frac{3}{\Lambda}\frac{1}{r_\star} + \frac{r_\star}{3} - \frac{\Lambda}{135} r_\star^3+ \mathcal{O}(r_\star^{5})
\end{split} \label{eq:Bondi to tortoise}
\end{equation}
where the last equality uses $\Lambda = -3/\ell^2$. This restores $\Lambda$ instead of $\ell$ and one can check that this perturbative change of coordinate $r\to r_\star$ transforms both dS and AdS global vacua from the Bondi gauge to the line elements \eqref{dS static} and \eqref{AdS static} respectively. Hence the expression where $\Lambda$ replaces $\sim \ell^{-2}$ is valid for both signs of the cosmological constant. The exact change for $\Lambda>0$ is $r\to \ell \coth(r_\star/\ell)$ whose asymptotic series for $r_\star\to 0$ is precisely \eqref{eq:Bondi to tortoise} when $\ell$ is traded for $\Lambda$.
\item In a second time, we reach perturbatively the SFG gauge up to order $N \in\mathbb N$,
\begin{equation}
g_{\rho\rho} = -\frac{3}{\Lambda}\frac{1}{\rho^2} \Big( 1 + \mathcal{O}(\rho^{N+1}) \Big), \quad g_{\rho t} = \frac{1}{\rho^2}\mathcal{O}(\rho^{N+1}), \quad g_{\rho A} = \frac{1}{\rho^2}\mathcal{O}(\rho^{N+1}), \label{eq:FGgaugecond}
\end{equation} thanks to a second change of coordinates,
\begin{equation}
\begin{split}
r_\star &\to \sum_{n=1}^{N+1} R_n (t,x^A)\rho^n, \\
t_\star &\to t + \sum_{n=1}^{N+1} T_n (t,x^A)\rho^n, \\
x^A_\star &\to x^A + \sum_{n=1}^{N+1} X^A_n (t,x^B)\rho^n.
\end{split} \label{tortoise to FG}
\end{equation}
\end{enumerate}
In order to obtain all of the free functions in $\gamma_{ab}$ for $d=3$, we need to proceed up to order $N=3$. For each $n$, each gauge condition \eqref{eq:FGgaugecond} can be solved separately and determines algebraically $R_n$, $T_n$ and $X_n^A$ respectively. Only the function $R_1(t,x^A)$ remains unconstrained by these conditions, since it represents a Weyl transformation on the boundary metric that is allowed within the SFG gauge. This Weyl transformation is constrained however by the choice of the luminosity distance $r$ in Bondi coordinates which ensures that $g_{AB}^{(0)}=q_{AB}$. The explicit change of coordinates \eqref{tortoise to FG} can be found in appendix \ref{app:chgt}, here we just plan to present the final result of the diffeomorphism on the free functions of the solution space. 

The boundary metric in the SFG gauge is related to the functions in the Bondi gauge through
\begin{equation}
g_{tt}^{(0)} = \frac{\Lambda}{3} e^{4 \beta_0} + U_0^C U_{C}^0 , \qquad g_{tA}^{(0)} = - U_A^0, \qquad g_{AB}^{(0)} =   q_{AB},
\label{g0 in term of Bondi}
\end{equation} 
in accordance with \eqref{eq:ds2I}, where all quantities on the right-hand sides are now evaluated as functions of $(t,x^A)$. The parameters $\lbrace \sigma,\bar\xi^t, \bar\xi^A \rbrace$ of the residual gauge diffeomorphisms in the SFG gauge \eqref{AKV 1} and \eqref{AKV 2} can be related to those of the Bondi gauge appearing in \eqref{eq:xir} through
\begin{equation}
\begin{split}
\bar\xi^t &= f ,\\
\bar\xi^A &= Y^A, \\
\sigma &= \frac{1}{2} (D_A Y^A + f l - U_0^A \partial_A f - 2\omega),
\end{split}
\label{translation parameters}
\end{equation} 
where all parameters on the right-hand sides are also evaluated as functions of $(t,x^A)$. Considering \eqref{g0 in term of Bondi}, the boundary gauge fixings \eqref{BC Lambda BMS} (particularized here for $d=3$) and \eqref{bndgauge} are equivalent: for $\Lambda < 0$ (resp. $\Lambda > 0$), there are exactly the temporal (resp. radial) gauge for the boundary metric, with a fixed area form on the 2-dimensional transverse spaces. 

Pursuing the change of coordinates to SFG gauge up to $N=3$, it can be shown that the holographic stress tensor $T_{ab}^{[3]} \stackrel{\text{not}}{=} T_{ab}$ is given, in terms of Bondi variables, by
\begin{equation}
T_{ab} = \frac{\sqrt{3 |\Lambda|}}{16\pi G} \left[
\begin{array}{cc}
-\frac{4}{3} M^{(\Lambda)} & -\frac{2}{3}  N^{(\Lambda)}_B \\ 
-\frac{2}{3}  N^{(\Lambda)}_A &  J_{AB} + \frac{2}{\Lambda} M^{(\Lambda)} q_{AB}
\end{array} 
\right] \label{eq:RefiningTab}
\end{equation}
where $M^{(\Lambda)} (t,x^A)$ and $N^{(\Lambda)}_A (t,x^B)$ are the boundary fields defined as \eqref{eq:hatM}-\eqref{eq:hatNA} and $J_{AB}$ is precisely the tensor \eqref{eq:hatJAB}, all evaluated as functions of $t$ instead of $u$. In order to complete the picture, let us see how the conservation equation $D_a T^{ab} = 0$ as well as the trace condition $g^{ab}_{(0)}T_{ab}=0$ are translated in the Bondi gauge. Owing to the $\Lambda$-BMS boundary gauge fixing \eqref{BC Lambda BMS} for $d=3$, the latter reads as 
\begin{equation}
q^{AB}T_{AB} = -\frac{3}{\Lambda} T_{tt}\Leftrightarrow q^{AB}J_{AB} = 0.
\end{equation}
We define $T^{TF}_{AB}$ as the tracefree part of $T_{AB}$, \textit{i.e.} $T_{AB} = T^{TF}_{AB} -\frac{3}{2\Lambda} T_{tt} q_{AB}$. The conservation equation yields
\begin{equation}
\begin{split}
(\partial_t + \frac{3}{2} l) T_{tt} + \frac{\Lambda}{3} D^A T_{tA} - \frac{\Lambda}{6} \partial_t q_{AB} T^{AB}_{TF} &= 0, \\
(\partial_t + l) T_{tA} - \frac{1}{2} \partial_A T_{tt} + \frac{\Lambda}{3} D^B T^{TF}_{AB} &= 0,
\end{split}
\label{EOM FG}
\end{equation}
which can be seen as time-evolution constraints for $T_{tt}$ and $T_{tA}$ respectively. The parallel is pretty clear: the conservation equations \eqref{EOM FG} are in fact equivalent to \eqref{eq:EvolutionM} and \eqref{eq:EvolutionNA}. The proof amounts to using the dictionary \eqref{eq:RefiningTab} and solve $\partial_t q_{AB}$ in terms of $C_{AB}$ using \eqref{eq:CAB}. Morever, we checked that the transformation laws \eqref{eq:VarM}-\eqref{eq:VarJAB} are equivalent to the variation
\begin{equation}
\delta_\xi T_{ab} = \mathcal L_{\bar\xi} T_{ab} + \sigma\,T_{ab}, \label{eq:TransformationTab}
\end{equation}
which is \eqref{eq:action solution space 2} particularized for $d=3$. We therefore identified the Bondi mass aspect $M^{(\Lambda)}$ and the Bondi angular momentum aspect $N^{(\Lambda)}_A$ as the components $T_{tt}$ and $T_{tA}$ of the holographic stress-tensor, up to a normalization constant.

\subsection{Bondi news and Bondi mass in (A)dS\texorpdfstring{$_4$}{4}}
\label{sec:Bondi news in AdS}

Now that the matching of the solution space parameters has been executed, let us understand what happens at the level of the presymplectic structure coming from $\bm \Theta_{ren}^{\Lambda\text{-BMS}_4}[\phi;\delta\phi]$. The latter has been computed in the SFG gauge: we thus need to invert (at least at leading order) the change of coordinates constructed in section \ref{sec:FGg} to translate this expression into the Bondi gauge. Assuming the $\Lambda$-BMS$_4$ gauge fixing, from \eqref{tortoise to FG} and the expressions $R_1(t,x^A) = -\frac{3}{\Lambda}$, $T_1(t,x^A) = X_1^A(t,x^B) = 0$ given in appendix \ref{app:chgt}, we have
\begin{equation}
\rho = -\frac{\Lambda}{3}r_\star + \mathcal O(r_\star^3), \qquad t = t_\star + \mathcal O(r_\star ), \qquad x^A = x^A_\star + \mathcal O(r_\star ).
\end{equation}
Provided that the field-dependence involved in \eqref{tortoise to FG} appears only at subleading orders in $\rho$ (or equivalently in $r_\star$), we can observe that the variational operator $\delta$ is not modified at leading order. Coupling both pieces of information, we see that the leading contributions in $\bm \Theta_{ren}^{\Lambda\text{-BMS}_4}[\phi;\delta\phi]$ are not affected by the change of coordinates, hence $\Theta^{r_\star}_{\Lambda\text{-BMS}_4} = \Theta^{\rho}_{\Lambda\text{-BMS}_4} + \mathcal O(r_\star)$, $\Theta^{a}_{\Lambda\text{-BMS}_4} = (\Theta^{t_\star}_{\Lambda\text{-BMS}_4},\Theta^{A_\star}_{\Lambda\text{-BMS}_4}) + \mathcal O(r_\star)$. We compose this transformation with the inverse of \eqref{eq:Bondi to tortoise} to reach the Bondi gauge $(u,r,x^A)$. We have
\begin{equation}
r_\star = \frac{3}{\Lambda}\frac{1}{r}+\mathcal O(r^{-3}),\quad t_\star = u +\mathcal O(r^{-1}), \quad x^A_\star = x^A,
\end{equation}
which implies finally that
\begin{equation}
\rho = -\frac{1}{r} + \mathcal O(r^{-3}),\quad t = u + \mathcal O(r^{-1}), \quad x^A = x^A + \mathcal O(r^{-1}).
\end{equation}
As a result, $\Theta^{r}_{\Lambda\text{-BMS}_4} = \Theta^{\rho}_{\Lambda\text{-BMS}_4} + \mathcal O(\rho)$. Particularizing \eqref{Theta ren pullback} for $d=3$ and using \eqref{bndgauge} as well as the parametrization \eqref{eq:RefiningTab}, we get the analog of \eqref{presymplectic flux Lambda BMS d} in Bondi variables, that is
\begin{equation}
\boxed{
\Theta_{\Lambda\text{-BMS}_4}^r [\phi;\delta\phi] = \frac{\Lambda}{2}\frac{\sqrt{\mathring q}}{16\pi G} J^{AB}\delta q_{AB} + \mathcal O(r^{-1}). \label{eq:potential LBMS4}
}
\end{equation}
Since all dynamical quantities can be computed from this presymplectic potential, it shows the equivalence between both gauges at the level of the dynamics. 

The action principle \eqref{var principle theta} becomes
\begin{equation}
\delta S_{ren}^{\Lambda\text{-BMS}_4}[\phi] = \frac{\Lambda}{2}\frac{1}{16\pi G} \int_{\mathscr I} (\D^3x) \sqrt{\mathring q}  \, J^{AB}\delta q_{AB}.\label{action lambda bms 4}
\end{equation}
The sources for the leaks are identified to be $q_{AB}$ and $J^{AB}$. The presymplectic form associated with \eqref{eq:potential LBMS4} is
\begin{equation}
\boxed{
\omega_{\Lambda\text{-BMS}_4}^r [\phi;\delta_1\phi,\delta_2\phi] = \frac{\Lambda}{2}\frac{\sqrt{\mathring q}}{16\pi G} \delta_1 J^{AB} \wedge \delta_2 q_{AB} + \mathcal O(r^{-1}). \label{eq:omega LBMS4}
}
\end{equation}
Again thanks to the fundamental theorem \eqref{FUNDAMENTAL THEOREM}, or \eqref{fundamental formula}, \eqref{eq:omega LBMS4} shows that the information on gravitational energy fluxes passing through $\mathscr{I}$ is entirely contained in the couple $(J^{AB},q_{AB})$. We will therefore call this couple of functions the \textit{news} for asymptotically locally (A)dS$_4$ spacetimes. The expression \eqref{eq:omega LBMS4} is perfectly reminiscent to the flat presymplectic form \eqref{omega flat complete} which is proportional to $\delta N^{AB}\wedge \delta C_{AB}$ when evaluated for $\delta q_{AB} = 0$. This is in fact the direct analogue in asymptotically locally (A)dS$_4$ spacetimes of the asymptotically flat expression of the presymplectic flux. The role of  Bondi news and shear $(N^{AB},C_{AB})$ is now played by $(J^{AB},q_{AB})$. Note that while $N_{AB} = \partial_u C_{AB}$, there is no such relationship between $J^{AB}$ and $q_{AB}$ in asymptotically locally dS$_4$ spacetimes since otherwise the Cauchy data would be constrained. In asymptotically locally AdS$_4$ spacetimes, a well-defined Cauchy problem will require a condition at conformal infinity and traditionally for conservative (reflexive) boundary conditions, we would like to cancel the flux induced by \eqref{eq:omega LBMS4}. These conservative subsectors of the $\Lambda$-BMS$_4$ phase space will be treated in section \ref{sec:Conservative subsectors and stationary solutions}.

Let us for the moment delay the determination of the surface charges induced by \eqref{eq:omega LBMS4} (see section \ref{sec:Lambda BMS 4 charges and algebra}) but discuss qualitatively the physical content of this symplectic form. For that purpose, we recall that for standard asymptotically flat spacetimes, the Bondi mass is given by $\mathcal M = \oint_{S_\infty} (\D^2 x) \, \sqrt{\mathring q} \, M$ and obeys the Bondi mass loss formula \eqref{Bondi mass loss formula}, implying that $\mathcal M(u)$ is a decreasing function of $u$. The integrations are performed on a transverse 2-surface $S_\infty$ at $\mathscr I^+$. The definition of Bondi mass for $\Lambda \neq 0$ is instead $\mathcal M^{(\Lambda)} = \oint_{S_\infty} (\D^2 x) \, \sqrt{\mathring q} \, M^{(\Lambda)}$ and it evolves as
\begin{equation}
\frac{\D}{\D u}  \mathcal M^{(\Lambda)} = - \frac{\Lambda^2}{24} \oint_{S_\infty} (\D^2 x) \, \sqrt{\mathring q} \, J^{AB}C_{AB}  = - \frac{\Lambda}{8} \oint_{S_\infty} (\D^2 x) \, \sqrt{\mathring q} \, J^{AB}\p_u q_{AB} , \label{eq:Mass loss AdS}
\end{equation}
after using \eqref{eq:CAB} and \eqref{bndgauge}. The major subtlety here is that the \textit{right-hand side of \eqref{eq:Mass loss AdS} is not manifestly non-positive!} It remains an important problem to determinate whether the Bondi mass $\mathcal M^{(\Lambda)}$ decreases with $u$, what is currently under investigation. 

\subsection{\texorpdfstring{$\Lambda$}{Lambda}-BMS\texorpdfstring{$_4$}{4} algebroid in Bondi gauge}
\label{sec: Lambda BMS 4 in Bondi}

In this section, we use the dictionary provided in section \ref{sec:FGg} to repeat the analysis of the set of residual gauge transformations that preserve the boundary gauge fixing \eqref{bndgauge} in the Bondi gauge and form the $\Lambda$-BMS$_4$ algebroid, as a corollary of the various definitions given in section \ref{sec:Lambda BMS d}. We give some explicit solutions for the $\Lambda$-BMS$_4$ generators around the unit-round sphere metric, where the constraint equations can be solved analytically in terms of infinite expansions in spherical harmonics. We also comment on the direct link between the codimension 2 functions appearing in these generators and the BMS$_4$ supertranslations and super-Lorentz parameters in the flat limit. Thereafter, we give some additional comments on the $\Lambda$-BMS$_4$ charge algebra, particularizing the results of the above section \ref{Lambda BMS charge algebra section}, and the associated flux-balance laws at conformal infinity.

\subsubsection{Bondi gauge parameters and constraints}
\label{sec:Bondi gauge parameters and constraints}

As announced, we study here the consequences of the boundary gauge fixing \eqref{bndgauge} on the residual gauge transformations intrinsically in the Bondi gauge. We start from \eqref{eq:xir} and impose \eqref{bndgauge}, which is \eqref{BC Lambda BMS} specified for $d=3$ and interpreted in the Bondi gauge thanks to the diffeomorphism constructed in the previous section \ref{sec:FGg}. Doing so, we should like to make a slight generalization, considering any fixed boundary volume form instead of $\sqrt{\mathring q}$ of the unit-round sphere. Using \eqref{eq56}-\eqref{eq58}, the condition $\delta_\xi \sqrt{q} = 0$ leads to
\begin{equation}
\omega = 0.\label{eq0a}
\end{equation} The condition $\delta_\xi \beta_0 = 0$ leads to 
\begin{equation}
\Big(\partial_u - \frac{1}{2}l\Big)f = \frac{1}{2} D_A Y^A,\label{eq1a}
\end{equation} while $\delta_\xi U_0^A = 0$ gives
\begin{equation}
\partial_u Y^A = -\frac{\Lambda}{3} q^{AB}\partial_B f.\label{eq2a}
\end{equation} 
The solution to \eqref{eq1a}-\eqref{eq2a} admits three integration ``constants'' $T(x^A),V^A(x^B)$, though the aforementioned constraints cannot be explicitly integrated for an arbitrary transverse metric $q_{AB}$ in terms of these functions. In the SFG notation, the equations \eqref{eq0a}-\eqref{eq2a} are equivalent to 
\begin{align}
\sigma &= \frac{1}{2}(D_A^{(0)} \xi^A_0 +f l) , \label{sig}\\
\Big(\partial_t - \frac{1}{2} l\Big) \xi^t_0 &= \frac{1}{2} D_A^{(0)}  \xi^A_0, \\ \partial_t \xi^A_0 &= -\frac{\Lambda}{3}g^{AB}_{(0)}\partial_B \xi^t_{(0)}.
\end{align}
Up to the incorporation of the $l$ contributions appearing only when $\partial_u \sqrt{\bar q}\neq 0$, we find exactly the same differential structure as in \eqref{equation Lambda BMS}. The generators $f,Y^A$ span the $\Lambda$-BMS$_4$ algebroid written in the Bondi gauge. They satisfy the commutation relations
\begin{equation}
[\bar{\xi}_1 , \bar{\xi}_2]_\star = \hat{\bar{\xi}},
\end{equation} where $\hat{\bar{\xi}} = \hat{f} \partial_u + \hat{Y}^A \partial_A$ with
\begin{align}
\hat{f} &= Y_1^A \partial_A f_2 + \frac{1}{2} f_1 D_A Y_2^A - (1 \leftrightarrow 2) \label{bms like algebra 1}, \\
\hat{Y}^A &= Y^B_1 \partial_B Y_2^A - \frac{\Lambda}{3} f_1 q^{AB} \partial_B f_2 - (1 \leftrightarrow 2).
\label{bms like algebra 2}
\end{align} 
as a corollary of \eqref{eq:VectorAlgebra1Lambda}-\eqref{eq:VectorAlgebraLambda}. In the $4d$ case, the analogy with the standard (Generalized) BMS$_4$ algebra, reviewed in section \ref{sec:Application to asymptotically locally flat spacetimes at null infinity}, is very stringent and we may call the vectors generated by $T(x^A)$ and $V^A(x^B) $ the $\Lambda$-\textit{supertranslation} and the $\Lambda$-\textit{super-Lorentz} generators, respectively. In the asymptotically flat limit $\Lambda = 0$ and for time-independent transverse metric (\textit{i.e.} $\sqrt{\bar q} = \sqrt{\mathring q}$), the functions $Y^A$, $f$ reduce to $Y^A = V^A(x^B)$, $f = T(x^A) + \frac{u}{2} D_A V^A$ and the structure constants reduce to those of the Generalized BMS$_4$ algebra. For $\Lambda \neq 0$, the fragmentation between ``supertranslations'' and ``super-Lorentz'' transformations is not clear and this terminology has to be intended more at the level of the pedagogical analogy than at a more fundamental level. In particular, the $\Lambda$-supertranslations do not commute and the structure constants depend explicitly on $q_{AB}$. As before, when the transverse metric $q_{AB}$ is equal to the unit-round sphere metric $\mathring{q}_{AB}$ (\textit{i.e.} for Dirichlet boundary conditions), the $\Lambda-$BMS$_4$ algebra contains the $SO(3,2)$ algebra for $\Lambda <0$ and the $SO(1,4)$ algebra for $\Lambda > 0$, see \cite{Barnich:2013sxa}. For the interested reader, we review the explicit expressions of the $SO(3,2)$ Killing vectors of AdS$_4$ in Appendix \ref{app:ExactVectors}.

\subsubsection{Explicit solutions and flat limit}
\label{sec:Explicit solutions and flat limit}
We mentioned in the previous paragraph that the joint solution of \eqref{eq1a} and \eqref{eq2a} in terms of codimension 2 fields $T(x^A)$ and $V^A(x^B)$ for arbitrary background fields $q_{AB},l$ is yet unknown to us. In what follows, we show how to get a flavor of the analytical structure of $f$ and $Y^A$ in terms of codimension 2 functions by attempting to integrate this differential system around the unit-round sphere $q_{AB} = \mathring q_{AB}$, $l=0$. We believe that a argument by covariance could allow to obtain the solution for the generic case $q_{AB}\neq \mathring q_{AB}$, $l\neq 0$ from the results we present here, but we do not plan to discuss it in details. Instead, we close the section by presenting a beautiful construction of the flat limit straightforwardly taken on the explicit solutions. Note that our derivation is here performed in the Bondi gauge. The exact analogous procedure in the SFG gauge can be found in the original article \cite{Compere:2020lrt}.

We set $\ell^2 = -3\eta/\Lambda$ as usual, $\sqrt{q}=\sqrt{\mathring q}$ and $l=0$, but we conserve $q_{AB}$ general until we need to particularize to the unit-round sphere metric. The first goal is to obtain scalar differential equations for the parameters. Taking a second time derivative on \eqref{eq1a} and using \eqref{eq2a}, we obtain that
\begin{equation}
\partial_u^2 f - \frac{\eta}{2\ell^2}D_A D^A f=0, \label{eq:WaveF}
\end{equation}
\textit{i.e.} $f$ is governed by a wave equation whose group velocity is commensurable to $\Lambda$. We further separate $Y^A$ into a scalar field $\Phi$ and a pseudo-scalar field $\Psi$ thanks to the Helmholtz decomposition $Y^A = q^{AB} \partial_B \Phi (u,x^C) + \epsilon^{AB}\partial_B \Psi(u,x^C)$, where $\epsilon^{AB} = \sqrt{\mathring q}\varepsilon^{AB} = \mathring \epsilon^{AB}$ is the Levi-Civita tensor on the unit-round sphere. Since the evolution equation for $Y^A$ is a vector constraint on the sphere, we can solve it for its curl and its divergence separately. This yields respectively
\begin{align}
& D_A D^A (\partial_u \Psi) = \varepsilon_{AB} D^A W^B, \label{eq:WavePsi} \\
& D_A D^A \left( \partial_u \Phi -\frac{\eta}{\ell^2} f\right) = -D_A W^A, \label{eq:WavePhi}
\end{align}
where $W^A \equiv \partial_u q^{AB} \partial_B \Phi$ (recall that, at this stage, we have not yet imposed $q_{AB} = \mathring{q}_{AB}$). It is worth noticing that the wave equation \eqref{eq:WaveF} does not fully determine $f$, since this is a second order equation with respect to $u$, while the original constraint on $f$ \eqref{eq1a} is a first order equation. The information we have lost by applying the second derivative is encoded in the remaining condition
\begin{equation}
\partial_u f = \frac{1}{2}D_A D^A \Phi
\label{eq:SupplCstf}
\end{equation}
that we need to take into account. Hence the system is rephrased as  \eqref{eq:WaveF}--\eqref{eq:SupplCstf}.

The equations for $\partial_u f$, $\partial_u\Phi$ and $\partial_u\Psi$ determine the fields $f$, $\Phi$ and $\Psi$ up to three arbitrary function of the angles $(\hat f,\hat \Phi,\hat \Psi)$. Let us show that these functions are constrained. We obtain $D_A D^A \hat f = 0$ directly from \eqref{eq:WaveF} and $D_A D^A \hat \Phi = 0$ from \eqref{eq:SupplCstf}. Recalling that only constants are harmonic functions on compact manifolds, we can set $\hat f = f_0$ ($f_0\in\mathbb R$) and $\hat \Phi = 0$, since a constant value of $\Phi$ does not appear in the vector $Y^A$. Hence we see that the solution for $\partial_u f$ and $\partial_u \Phi$ fully determines $f$ (up to a residual constant) and $\Phi$. However, we also observe that nothing constrains $\hat \Psi$, showing that it will remain an arbitrary function on the angles in $\Psi$. Finally the wave equation involving $f$ gives rise to two arbitrary functions of the angles as integration constants, which will be brought to $\Phi$ thanks to \eqref{eq:WavePhi}. Hence the number of arbitrary functions of $x^A$ is shown to be three, as expected! 

We now set $q_{AB} \equiv \mathring q_{AB}$ and $W^A$ vanishes identically. The solution of \eqref{eq:WavePsi} is $\partial_u\Psi = c$ for some real constant $c$. This constant can be removed as an ambiguity in defining the Helmholtz fields, since it is responsible for a linear term $cu$ in $\Psi$ which will never contribute to the actual vector $Y^A$. The solution for $\Psi$ is thus simply $\Psi=\Psi(x^A)$. We can directly solve \eqref{eq:WaveF} by using Fourier transform methods. We obtain
\begin{equation}
f(u,x^A) = \int_0^\infty \text d\omega \, \left[ f_E (x^A) \cos \left(\frac{\omega \,u}{\ell}\right) + f_O(x^A) \sin \left(\frac{\omega \,u}{\ell}\right)\right]
\end{equation}
if $\Lambda <0$ and 
\begin{equation}
f(u,x^A) = \int_0^\infty \text d\omega \, \left[ f_E (x^A) \cosh \left(\frac{\omega \,u}{\ell}\right) + f_O(x^A) \sinh \left(\frac{\omega \,u}{\ell}\right)\right]
\end{equation}
if $\Lambda >0$. In both cases, the Fourier coefficients are constrained as follows:
\begin{equation}
\mathring D_A \mathring D^A f_{E} = -2\omega^2 f_{E},\quad \mathring D_A \mathring D^A f_{O} = -2\omega^2 f_{O}.
\end{equation}
These equations select discrete values for $\omega$ which satisfy $\omega_{\texttt{\textit{l}}}^2 = \frac{1}{2} \texttt{\textit{l}}(\texttt{\textit{l}}+1)$ with $\texttt{\textit{l}}\in\mathbb N$ and the solution for $f$ is given in terms of the real spherical harmonics $Y_{\texttt{\textit{lm}}}(x^A)$ ($\texttt{\textit{m}}\in\mathbb N$, $|\texttt{\textit{m}}|\leq \texttt{\textit{l}}$). Indeed,
\begin{equation}
f(u,x^A) = \left\lbrace
\begin{aligned}
\, & \sum_{\texttt{\textit{l,m}}} \left[ a_{\texttt{\textit{lm}}} \cos \left(\frac{\omega_\texttt{\textit{l}} \,u}{\ell}\right) + b_{\texttt{\textit{lm}}} \sin \left(\frac{\omega_\texttt{\textit{l}} \,u}{\ell}\right)\right] Y_{\texttt{\textit{lm}}} (x^A) \text{ if } \Lambda < 0,\\
\, & \sum_{\texttt{\textit{l,m}}} \left[ a_{\texttt{\textit{lm}}} \cosh \left(\frac{\omega_\texttt{\textit{l}} \,u}{\ell}\right) + b_{\texttt{\textit{lm}}} \sinh \left(\frac{\omega_\texttt{\textit{l}} \,u}{\ell}\right)\right] Y_{\texttt{\textit{lm}}} (x^A) \text{ if } \Lambda > 0,
\end{aligned}
\right. \label{eq:GeneralSolutionf}
\end{equation}
where $\lbrace a_{\texttt{\textit{lm}}} \rbrace$ and $\lbrace b_{\texttt{\textit{lm}}} \rbrace$ are two sets of real constants. From \eqref{eq:WavePhi}, we can deduce that
\begin{equation}
\partial_u\Phi - \frac{\eta}{\ell^2}f = 0 \label{eq:PartialPhi}
\end{equation}
up to some real constant that can again be removed as an ambiguity in defining $\Phi$. Taking one more derivative with respect to $u$ and recalling that \eqref{eq:SupplCstf} holds, we obtain that $\Phi$ satisfies the same wave equation as $f$,
\begin{equation}
\partial_u^2 \Phi - \frac{\eta}{2\ell^2}\mathring D_A \mathring D^A \Phi=0. \label{eq:BoxPhi}
\end{equation}
The general solution has already been derived (see \eqref{eq:GeneralSolutionf}) and reads as
\begin{equation}
\boxed{
\Phi(u,x^A) = \left\lbrace
\begin{aligned}
\, & \sum_{\texttt{\textit{l,m}}} \left[ A_{\texttt{\textit{lm}}} \cos \left(\frac{\omega_\texttt{\textit{l}} \,u}{\ell}\right) + \frac{1}{\ell} \, B_{\texttt{\textit{lm}}} \sin \left(\frac{\omega_\texttt{\textit{l}}  \,u}{\ell}\right)\right] Y_{\texttt{\textit{lm}}} (x^A)  \text{ if } \Lambda < 0,\\
\, & \sum_{\texttt{\textit{l,m}}} \left[ A_{\texttt{\textit{lm}}} \cosh \left(\frac{\omega_\texttt{\textit{l}}  \,u}{\ell}\right) + \frac{1}{\ell}\, B_{\texttt{\textit{lm}}} \sinh \left(\frac{\omega_\texttt{\textit{l}}  \,u}{\ell}\right)\right] Y_{\texttt{\textit{lm}}} (x^A)  \text{ if } \Lambda > 0,
\end{aligned}
\right.
}
\label{eq:PhiLBMS4}
\end{equation}
involving two new sets of real constants $\lbrace A_{\texttt{\textit{lm}}}\rbrace$ and $\lbrace B_{\texttt{\textit{lm}}}\rbrace$. The $1/\ell$ factor in front of the $B$'s is for now purely conventional, but will ensure later that the coefficients $A_{lm}$ and $B_{lm}$ do not depend on $\ell$ in the flat limit process $\ell\to \infty$. These new sets of constants are not independent of $\lbrace a_{lm}\rbrace$ and $\lbrace b_{lm}\rbrace$, since we must require that the remaining constraints \eqref{eq:SupplCstf} and \eqref{eq:PartialPhi} hold. The first one has been implemented in the derivation of the wave equation \eqref{eq:BoxPhi}, so we just have to impose the second one. This yields
\begin{equation}
a_{\texttt{\textit{lm}}} = \eta\,B_{\texttt{\textit{lm}}}\,\omega_\texttt{\textit{l}}, \quad b_{\texttt{\textit{lm}}} = -\ell \,A_{\texttt{\textit{lm}}}\,\omega_\texttt{\textit{l}}.
\end{equation}
Hence, the gauge parameter $f$ reads as 
\begin{equation}
\boxed{
f(u,x^A) = \left\lbrace
\begin{aligned}
\,  & \sum_{\texttt{\textit{l,m}}} \left[ B_{\texttt{\textit{lm}}} \cos \left(\frac{\omega_\texttt{\textit{l}} \,u}{\ell}\right) - \ell \, A_{\texttt{\textit{lm}}} \sin \left(\frac{\omega_\texttt{\textit{l}} \,u}{\ell}\right)\right] \omega_\texttt{\textit{l}} \, Y_{\texttt{\textit{lm}}} (x^A)  \text{ if } \Lambda < 0,\\
\,  & -\sum_{\texttt{\textit{l,m}}} \left[ B_{\texttt{\textit{lm}}} \cosh \left(\frac{\omega_\texttt{\textit{l}} \,u}{\ell}\right) + \ell \, A_{\texttt{\textit{lm}}} \sinh \left(\frac{\omega_\texttt{\textit{l}} \,u}{\ell}\right)\right] \omega_\texttt{\textit{l}} \, Y_{\texttt{\textit{lm}}} (x^A)  \text{ if } \Lambda > 0,
\end{aligned}
\right.
}
\label{eq:fLBMS4}
\end{equation}
The exact isometries of global (A)dS$_4$, reviewed in appendix \ref{app:ExactVectors}, are recovered if we consider the lowest modes $\texttt{\textit{l}}=0$ and $\texttt{\textit{l}}=1$. Indeed, if we further require that $\delta_{\xi} q_{AB} = 0$, it comes
\begin{equation}
\mathring D_A \xi^{(0)}_B+\mathring D_B \xi^{(0)}_A-\mathring D_C \xi_{(0)}^C \mathring q_{AB}=0 \Leftrightarrow \left\lbrace 
\begin{aligned}
(\mathring D_A \mathring D_B \Phi)^{TF} &= 0, \\
\varepsilon_{C(A}\mathring D_{B)}\partial^C \Psi &= 0.
\end{aligned}
\right.
\end{equation}
In stereographic coordinates $(z,\bar z)$, if we introduce the auxiliary fields $\phi \equiv (1+z\bar z)\Phi $ and $\psi \equiv (1+z\bar z)\Psi $, these equations become simply $\partial_z^2 \phi = 0 = \partial_{\bar z}^2 \phi$ and $\partial_z^2 \psi = 0 = \partial_{\bar z}^2 \psi$. Hence $\phi$ and $\psi$ are at most linear in $z$ and $\bar z$, the only non-linear piece that can appear being the squared modulus $z\bar z$. In conclusion, the solution to the conformal Killing equation developed for the Helmholtz fields $\Phi$ and $\Psi$ only involve the lowest (real) spherical harmonics with $\texttt{\textit{l}}=0,1$.

Finally, let us show that the flat limit obtained by taking directly $\ell\to\infty$ in \eqref{eq:PhiLBMS4} and \eqref{eq:fLBMS4} reproduces the expression of the Generalized BMS$_4$ generators, which are solutions of \eqref{generalized BMS 4 differential eq}. The flat limit of \eqref{eq:PhiLBMS4} is immediate and gives
\begin{equation}
\Phi(u,x^A) = \sum_{\texttt{\textit{l,m}}} A_{\texttt{\textit{lm}}} Y_{\texttt{\textit{lm}}}(x^A) = \Phi(x^A)
\end{equation}
as expected. The two Helmholtz fields $\Phi$ and $\Psi$ are arbitrary and define a general super-Lorentz transformation $Y^A$ around $q_{AB}=\mathring q_{AB}$. The same limit of \eqref{eq:fLBMS4} leads to
\begin{equation}
\begin{split}
f(u,x^A) &= \sum_{\texttt{\textit{l,m}}} \left[ \eta B_{\texttt{\textit{lm}}} - \omega_\texttt{\textit{l}} \, u\, A_{\texttt{\textit{lm}}} \right] \omega_\texttt{\textit{l}} \, Y_{\texttt{\textit{lm}}}(x^A) \\
&= \sum_{\texttt{\textit{l,m}}} \left[ \tilde B_{\texttt{\textit{lm}}} \, Y_{\texttt{\textit{lm}}}(x^A) - \omega_\texttt{\textit{l}}^2 \, u\, A_{\texttt{\textit{lm}}} \, Y_{\texttt{\textit{lm}}}(x^A) \right], \, \left(\tilde B_{\texttt{\textit{lm}}} \equiv \eta \, \omega_\texttt{\textit{l}} B_{\texttt{\textit{lm}}}\right) \\
&= \sum_{\texttt{\textit{l,m}}} \tilde B_{\texttt{\textit{lm}}} \, Y_{\texttt{\textit{lm}}}(x^A) + \frac{u}{2}\sum_{\texttt{\textit{l,m}}} A_{\texttt{\textit{lm}}} \mathring D_B \mathring D^B Y_{\texttt{\textit{lm}}}(x^A) \\
&= T(x^A) + \frac{u}{2}\mathring D^B \mathring D_B \Phi(x^A) \\
&= T(x^A) + \frac{u}{2} \mathring D_B Y^B(x^A),
\end{split}
\end{equation}
where $T(x^A)$ is an arbitrary scalar field on the celestial sphere.

\subsubsection{\texorpdfstring{$\Lambda$}{Lambda}-BMS\texorpdfstring{$_4$}{4} charges and algebra}
\label{sec:Lambda BMS 4 charges and algebra}
Before proceeding further, we want to particularize the results of section \ref{Lambda BMS charge algebra section} in this paragraph. We are back to consider the charges \eqref{ndelta H Lambda BMS} for the split \eqref{split of the charge Lambda BMS}. For $d=3$, the Weyl non-integrable charge vanishes. Using $\D^2 \Omega = \sqrt{\mathring q}(\D^{2}x)$ and the dictionary elaborated in section \ref{sec:FGg} to express the charges with Bondi quantities, we find
\begin{align}
\ndelta H_\xi^{\Lambda\text{-BMS}_4}[\phi;\delta\phi] &= \delta H_\xi^{\Lambda\text{-BMS}_4}[\phi] + \Xi_\xi^{\Lambda\text{-BMS}_4}[\phi;\delta\phi] \label{split lambda bms 4}
\end{align}
with
\begin{align}
H_\xi^{\Lambda\text{-BMS}_4}[\phi] &= \frac{\eta}{16\pi G} \oint_{S_\infty} \D^2 \Omega \ \left[ 4\, f\, M^{(\Lambda)} + 2\, Y^A\, N_A^{(\Lambda)} \right], \label{I for L BMS 4} \\
\Xi_\xi^{\Lambda\text{-BMS}_4}[\phi;\delta\phi] &= \frac{\Lambda}{32\pi G} \oint_{S_\infty} \D^2 \Omega \  \left(f J_{AB}\delta q^{AB}\right)  - H_{\delta\xi}^{\Lambda\text{-BMS}_4}[\phi] \label{NI for L BMS 4}
\end{align}
and all the involved quantities are intended as functions of $(t,x^A)$ in the SFG gauge. This is the last incursion of this coordinate system in our analysis but we prefer keep it while discussing the charges for the sake of clarity. Written in that way, the $\Lambda$-BMS$_4$ charges are reminiscent of the BMS$_4$ charges for asymptotically Minkowskian spacetimes \cite{Wald:1999wa,Barnich:2010eb}, see \eqref{FNCharges}-\eqref{FNCharges non int} with fixed boundary metric $\delta q^{AB}=0$. In particular, the non-integrable part involves the symplectic pairing $(J_{AB},q^{AB})$ as defined in section \ref{sec:Bondi news in AdS}. As a corollary of \eqref{charge algebra Lambda BMS d} and \eqref{cocycle Lambda BMS odd d} for $d=3$, the algebra of charges \eqref{split lambda bms 4} represent the $\Lambda$-BMS$_4$ algebroid without any central extension to the Barnich-Troessaert bracket:
\begin{equation}
\boxed{\{ H_{\xi_1}^{\Lambda\text{-BMS}_4} [\phi] , H^{\Lambda\text{-BMS}_4}_{\xi_2} [\phi] \}_\star = H^{\Lambda\text{-BMS}_4}_{[\xi_1, \xi_2]_\star}[\phi] .  }
\label{charge algebra Lambda BMS 4}
\end{equation}
It is interesting to observe that there is no 2-cocycle at the level of the charge algebra for the immediate extension of the Generalized BMS$_4$ symmetries for $\Lambda\neq 0$ although the flat limit of \eqref{charge algebra Lambda BMS 4} is supposed to exhibit the 2-cocycle previously calculated as \eqref{cocycle flat}. Any shortcut reasoning of this kind will be pondered by the explicit flat limit of the $\Lambda$-BMS$_4$ phase space described in section \ref{sec:Flat limit of the action and corner terms}, because this process is not well-defined for dynamical quantities such as the presymplectic structure or the charges. Both of them contain poles in $\Lambda^{-1}$ that will require further renormalization of the action principle before sending $\Lambda$ to zero. 

As a crucial remark, let us mention that \eqref{charge algebra Lambda BMS 4} controls the non-conservation of the $\Lambda$-BMS$_4$ charges. The proof of this statement has already been elaborated for the Generalized BMS$_4$ charges in section \ref{sec:The Generalized BMS$_4$ charge algebra}, and reads as follows. The total evolution of the Hamiltonian associated with $\xi$, considered as the integrable part selected in \eqref{split lambda bms 4}, is governed by the derivative along the foliation $\bm T$ on $\mathscr I$, given by $\frac{\D}{\D u}\equiv \frac{\D}{\D t} + \mathcal O(r^{-1})$. By definition, we have
\begin{equation}
\frac{\D}{\D u} H_\xi^{\Lambda\text{-BMS}_4}[\phi] = \frac{\partial}{\partial u}H_\xi^{\Lambda\text{-BMS}_4}[\phi] + \delta_{\partial_u}H_\xi^{\Lambda\text{-BMS}_4}[\phi] - H_{\delta_{\partial_u}\xi}^{\Lambda\text{-BMS}_4}[\phi]
\end{equation}
which constitutes a slight generalization of \eqref{flux from algebra 1} while dealing with field-dependent diffeomorphism parameters. The third term is indeed zero for asymptotically flat configurations since $\delta_\xi f = 0 = \delta_\xi Y^A$. We have
\begin{equation}
\frac{\partial}{\partial u}H_\xi^{\Lambda\text{-BMS}_4}[\phi] - H_{\delta_{\partial_u}\xi}^{\Lambda\text{-BMS}_4}[\phi] = H_{\partial_u \xi - \delta_{\partial_u}\xi}^{\Lambda\text{-BMS}_4}[\phi] = -H_{[\xi,\partial_u]_\star}^{\Lambda\text{-BMS}_4}[\phi]
\end{equation}
because the charges depend linearly upon the gauge parameter. The conclusion of the proof is similar to what we did before. Using the algebra \eqref{charge algebra Lambda BMS 4}, we obtain easily
\begin{equation}
\frac{\D}{\D u}H_\xi^{\Lambda\text{-BMS}_4}[\phi] = -\Xi_{\partial_u}^{\Lambda\text{-BMS}_4}[\phi;\delta_\xi\phi] = -\oint_{S_\infty} \iota_{\partial_u} \bm\Theta_{ren}^{\Lambda\text{-BMS}_4}[\phi;\delta_\xi\phi], \label{eq:flux balance for LBMS4}
\end{equation}
recognizing the renormalized presymplectic flux \eqref{action lambda bms 4} in the non-integrable part \eqref{NI for L BMS 4}. This flux formula is a particular case of Eq. (4.9) of \cite{Anninos:2010zf} for $\Lambda>0$ and is the analogue of the flux formula \eqref{flux from algebra final} in asymptotically locally flat spacetimes \cite{Wald:1999wa,Ashtekar:1981bq}.

The computation of the charges \eqref{split lambda bms 4} allows us to conclude by giving a striking argument in favor of the $\Lambda$-BMS$_4$ algebroid as candidate for the role of natural asymptotic symmetries of Al(A)dS$_4$ gravity. While experiencing some leaks through the conformal boundary, the fundamental theorem \eqref{FUNDAMENTAL THEOREM} indicates that each charge is associated with a flux-balance law controlling its non-conservation. By a counting of boundary degrees of freedom, we see that there are three ordinary differential equations in time at the boundary, which are given by \eqref{EOM FG} as a consequence of the conservation of the holographic stress tensor on $\mathscr I$. Written in Bondi coordinates, these equations are \eqref{eq:EvolutionM}-\eqref{eq:EvolutionNA} and give thus three flux-balance laws for the Bondi mass as well as the Bondi angular momentum for $\Lambda\neq 0$. It can be shown that they control the time-derivative of the asymptotic charges \eqref{I for L BMS 4} in accordance with \eqref{eq:flux balance for LBMS4}. We conclude that the $\Lambda$-BMS$_4$ algebroid, whose elements are parametrized by three functions $f,Y^A$ on $\mathscr I$, is the non-trivial sub-algebra of Diff($\mathscr I$) which encapsulates the three towers of non-trivial charges that obey the right number of flux-balance laws (or Ward identities at quantum level, see section \ref{sec:Scattering}). Assorted with the argument according to which \eqref{bndgauge} is a boundary gauge fixing which does not reduce the solution space, we have in fact proven that the $\Lambda$-BMS$_4$ phase space is \textit{complete} in the sense that it arises from the most stringent gauge fixing that does not rule out any solution while admitting as many asymptotic generators as necessary to reproduce the three flux-balance laws naturally present in the solution space. In other words, this is the canonical subset of Diff($\mathscr I$) associated with the physical flux given locally by \eqref{eq:EvolutionM} and \eqref{eq:EvolutionNA}. 

Two last remarks should be made at this stage:
\begin{itemize}[label=$\rhd$]
\item Changes of foliation $\bm T$ span an equivalence class of boundary gauge fixings and $\Lambda$-BMS$_4$ algebroids: this amounts to saying that we can select Gaussian normal coordinates at will on $\mathscr I$ when $\Lambda\neq 0$. However, in flat space, the null direction on $\mathscr I^+$ is given unambiguously. As we will observe in section \ref{sec:Flat limit of the LBMS4 phase space}, only the consistency with the flat limit process breaks this equivalence relation by selecting a preferred direction on $\mathscr I$ that becomes the null direction when $\Lambda\to 0$.
\item Also in flat case, the Generalized BMS$_4$ symmetries are complete only at second order in the Bondi expansion, in the sense that no other evolution equation than \eqref{duM} and \eqref{EOM1} arises from Einstein’s equations up to that order. But there is an infinite tower of subleading flux-balance laws beyond that order, related to $\mathcal E_{AB}$, $\mathcal F_{AB}$ \textit{etc.} \cite{Barnich:2010eb}, see section \ref{sec:Solution space in Bondi gauge}. This is in contrast with what happens here, where \eqref{eq:EvolutionM} and \eqref{eq:EvolutionNA} are the only flux-balance laws appearing in the Bondi expansion at large radii. 
\end{itemize}


\subsection{Conservative subsectors and stationary solutions}
\label{sec:Conservative subsectors and stationary solutions}
In this short section, we investigate a bit some conservative subsectors of the leaky boundary conditions leading to the $\Lambda$-BMS$_4$ phase space. Irrelevant for dS$_4$ asymptotics, they ensure the stationarity of the variational problem as well as the well-definiteness of the Cauchy problem in the case of asymptotically locally AdS$_4$ spacetimes. They need a supplementary boundary condition to be implemented at the spatial boundary $\mathscr I_{\text{AdS}}$ of AdS$_4$ \cite{Ishibashi:2004wx} and amounts to requiring that the symplectic flux \eqref{action lambda bms 4} at spatial infinity is identically zero. Along with the famous Dirichlet boundary conditions \cite{Brown:1986nw,Hawking:1983mx,Ashtekar:1984zz,Henneaux:1985tv,Henneaux:1985ey,Ashtekar:1999jx,Papadimitriou:2005ii} that are the natural conservative subcase of the $\Lambda$-BMS$_4$ boundary conditions fixing completely the boundary structure, we derive new boundary conditions for asymptotically locally AdS$_4$ spacetimes that admit the Schwarzschild-AdS black hole and new stationary rotating solutions distinct from the Kerr-AdS black hole \cite{Carter:1968ks} (see also \cite{Gibbons:2004js,Gibbons:2004uw} for the higher dimensional generalization). The asymptotic symmetry group is shown to be a subgroup of the  $\Lambda$-BMS$_4$ group consisting of time translations and area-preserving diffeomorphisms. 

\subsubsection{Mixed Neumann conservative boundary conditions}

A vanishing symplectic flux in \eqref{action lambda bms 4} is observed when $\delta q_{AB} = 0$ for $J^{AB}$ completely free. This subsector of the $\Lambda$-BMS$_4$ phase space consists of solutions with Dirichlet boundary conditions defined in section \ref{Dirichlet}. The fixation of the boundary metric reduces the asymptotic symmetry group to the group of exact symmetries of AdS$_4$, namely $SO(3,2)$, see section \ref{sec: Lambda BMS 4 in Bondi}. Neumann boundary conditions defined in section \ref{sec:Neumann} freeze the components of the holographic stress-tensor while leaving the whole boundary metric free to fluctuate. The resulting asymptotic symmetry group is empty: all residual gauge transformations have vanishing charges. This is due to the fact that there are kinematical degrees of freedom in the fluctuations of the boundary metric that should be fixed by some boundary gauge fixing, allowing the symplectic conjugate parts of the stress tensor to be non-zero while enforcing a vanishing symplectic flux. This is precisely what the $\Lambda$-BMS$_4$ boundary gauge fixing \eqref{bndgauge} does: the equation \eqref{eq:potential LBMS4} shows that we can set partial Neumann boundary conditions 
\begin{equation}
J^{AB} = 0  \label{BCads}
\end{equation}
to kill the symplectic flux without constraining the momentum components $T^{tt}$, $T^{tA}$ of the holographic stress-tensor, which contain the mass and angular momentum aspects, see \eqref{eq:RefiningTab}. These are conservative boundary conditions compatible with a fluctuating boundary structure. More concretely, \eqref{BCads} coupled with \eqref{bndgauge} provides new mixed Dirichlet-Neumann boundary conditions, where the latter condition is a Dirichlet boundary condition on part of the boundary metric, but which the good property of being always reachable by gauge fixing. The additional fixation \eqref{BCads} restricts the phase space of solutions as well as the asymptotic symmetries. Let us find out what remains of the $\Lambda$-BMS$_4$ algebroid when \eqref{BCads} holds.

\subsubsection{Asymptotic symmetry algebra}

We derive the asymptotic symmetries preserving the boundary conditions and compute the associated charge algebra. We show that the subgroup of asymptotic symmetries preserving \eqref{BCads} is the direct product $\mathbb{R} \times \mathcal{A}$ where $\mathbb{R}$ are the time translations on the boundary (along the foliation vector $\bm T$) and $\mathcal{A}$ is the group of 2-dimensional area-preserving diffeomorphisms on the transverse 2-spheres. As expected for a set of conservative boundary conditions, the associated surface charges are finite, integrable, conserved and generically non-vanishing on the phase space.  

Everything starts from the transformation law of $J_{AB}$. Using \eqref{eq:RefiningTab}, $\delta_\xi J_{AB}$ can be extracted in the SFG gauge from the variation \eqref{eq:TransformationTab}, then reinterpreted in the Bondi gauge with the dictionary \eqref{translation parameters}. The result is
\begin{equation}
\delta_\xi {J}_{AB} = (f\partial_u + \mathcal{L}_{Y} + \frac{1}{2}D_C Y^C) {J}_{AB} - \frac{4}{3} \Big[  N_{(A} \partial_{B)}f - \frac{1}{2} N_{C} \partial^C f q_{AB} \Big]. \label{delta JAB}
\end{equation} 
Imposing $\delta_\xi J_{AB} = 0$ leads to the following constraint on the residual gauge diffeomorphisms:
\begin{equation}
\partial_A f =0.
\label{angle independance}
\end{equation}
This is the minimal field-independent condition to cancel the inhomogeneous part of the transformation \eqref{delta JAB}. Therefore, the asymptotic symmetry generators satisfy the relations
\begin{equation}
\partial_u f = \frac{1}{2} D_A Y^A , \qquad \partial_u Y^A = 0.
\end{equation}
as a corollary of \eqref{eq1a}-\eqref{eq2a}. The second equation implies $Y^A = Y^A(x^B)$, while the first gives
\begin{equation}
f = T + \frac{u}{2} D_A Y^A,
\end{equation} where $T$ is a constant by virtue of \eqref{angle independance} and $D_A Y^A \equiv c$ where $c$ is also a constant. The structure of $f$ and $Y^A$ is reminiscent of the Generalized BMS$_4$ algebra in the flat case. We use again the Helmholtz decomposition for the vector $Y^A$ into a divergence-free and a curl-free part, $Y^A = \epsilon^{AB} \p_B \Psi + q^{AB} \p_B \Phi$ where $\Psi$ and $\Phi$ are functions of $x^C$. We obtain thus that $\Psi$ is unconstrained while $D_AD^A\Phi = c$. On the sphere, this last equation admits a solution if and only if $c=0$ which is given by $\Phi = 0$. Therefore, the asymptotic symmetry generators are given by 
\begin{equation}
f = T, \qquad Y^A =  \epsilon^{AB} \p_B \Psi(x^C),
\end{equation} where $T$ is a constant and $\Psi(x^C)$ is arbitrary and defines an area-preserving diffeomorphism on the 2-sphere. Indeed, we have $\Lie_Y \sqrt{q} = \partial_A (\sqrt{q}Y^A) = \sqrt{q}D_AY^A = 0$. Writing the boundary diffeomorphism as $\bar{\xi} = \bar \xi^a \p_a =  T\partial_t + \epsilon^{AB} \p_B \Psi \partial_A$, we have $[\bar{\xi}_1,  \bar{\xi}_2 ]_\star = [\bar{\xi}_1,  \bar{\xi}_2 ] = \hat{\bar{\xi}}$ where
\begin{equation}
\hat{T} = 0, \qquad \hat{\Psi} = \epsilon^{AB}\p_A \Psi_2 \p_B \Psi_1. 
\label{algebra are preserving}
\end{equation} Hence, after imposing the boundary condition \eqref{BCads}, the $\Lambda$-BMS$_4$ algebroid reduces to the $\mathbb{R} \oplus \mathcal{A}$ algebra where $\mathbb{R}$ denotes the abelian time translations and $\mathcal{A}$ is the algebra of 2$d$ area-preserving diffeomorphisms. The latter symmetries are an infinite-dimensional extension of the $SO(3)$ rotations. They are in addition field-independent, since \eqref{angle independance} has decoupled the constraint \eqref{eq2a} in which the boundary metric $q_{AB}$ appeared naturally.

Let us now obtain the associated charges. Again, the computation is more practical in SFG gauge, but we can translate the results into the Bondi gauge afterwards. We particularize the charge formula \eqref{ndelta H Lambda BMS} with \eqref{split of the charge Lambda BMS} to $d=3$. The Weyl part $W^{[3]}_\sigma$ is absent and the first non-integrable term vanishes because $T_{AB}^{TF} \propto J_{AB} = 0$. Hence the charges are integrable and read as
\begin{equation}
Q_{\xi(T,\Psi)}[g] = \oint_{S_\infty } \D^2 \Omega \, \sqrt{\bar q} \, [ T^t_{\; \, \, t}\, T + T^t_{\; \, \, A} \epsilon^{AB} \p_B \Psi ] .
\label{charge expression}
\end{equation}
Translating this result in the Bondi gauge thanks to \eqref{eq:RefiningTab}, we get
\begin{equation}
Q_{\xi(T,\Psi)}[g] = \frac{\ell}{16\pi G}\oint_{S_\infty } \D^2 \Omega \, \sqrt{\bar q} \, \left[4\, T\, M^{(\Lambda)} +  2\epsilon^{AB} N_A^{(\Lambda)}\p_B \Psi \right] .\label{charge expression Bondi}
\end{equation}
From this expression, we see that the charges associated with the symmetry $\mathbb{R} \oplus \mathcal{A}$ are generically non-vanishing. Taking $T = 1$ and $\Psi = 0$ gives the total energy. The first harmonic modes of $\Psi$ give three components of the total angular momentum, while the higher modes give an infinite tower of charges. Using the commutation relations \eqref{algebra are preserving} and the equations of motion \eqref{EOM FG}, a simple computation shows that the charges \eqref{charge expression} satisfy the algebra
\begin{equation}
-\delta_{\xi (T_1, \Psi_1)} Q_{\xi (T_2, \Psi_2)} = Q_{\xi (\hat T,\hat \Psi)} .
\end{equation} The charges $Q_{\xi(T,\Psi)}[g]$ form a representation of $\mathbb{R} \oplus \mathcal{A}$ without central extension.


\subsubsection{Stationary solutions}

Let us study the stationary sector of the phase space associated with the boundary condition \eqref{BCads}. For these solutions, we have successively $\partial_t q_{AB}=0$ or equivalently $C_{AB} = 0$ and $\partial_t M^{(\Lambda)}=0$, $\partial_t N_A^{(\Lambda)}$. Using \eqref{eq:EvolutionNA} and \eqref{eq:EvolutionM} we get
\begin{equation}
D_A N^A_{(\Lambda)} = 0 \Leftrightarrow N^A_{(\Lambda)} = \epsilon^{AB} D_B \alpha (x^C), \quad \partial_A M^{(\Lambda)} = 0
\label{soll}
\end{equation} for $\alpha (x^C)$ an arbitrary function of the angles. As a result, even for a stationary class of solutions, we see that the charges \eqref{charge expression Bondi} associated with the area-preserving diffeomorphisms are generically non-vanishing. It would be interesting to study the regularity of the general solutions \eqref{soll} in the bulk of spacetime, but answering this question is definitely beyond the scope of our analysis. 

We now discuss some prototypical examples of stationary solutions: the family of Kerr-AdS$_4$ metric \cite{Carter:1968ks}, describing celestial bodies in stationary rotation with negative cosmological constant. Clearly the static Schwarzschild-AdS$_4$ solution is included in the phase space with $J_{AB}=0$. Indeed, it can be set in the SFG gauge, which allows to identify $q_{AB} = \mathring q_{AB}$ the unit metric on the sphere, as well as ${T^t}_t = \frac{M}{4 \pi G}$, $T_{tA} = 0$ and $T_{AB} = 0$, which finally implies $J_{AB} = 0$. This is a good point. Now let us introduce rotation of the central body. The boundary metric and holographic stress-tensor of Kerr-AdS$_4$ are given in the conformally flat frame by \cite{Awad:1999xx,Bhattacharyya:2007vs,Bhattacharyya:2008ji}
\begin{align}
g^{(0)}_{ab}\D x^a \D x^b &= -\ell^{-2}\D t^2 + \D \theta^2 +  \sin^2\theta \D \phi^2,\label{flat} \\
T^{ab} &= T_{\text{Kerr}}^{ab}  \equiv - \frac{m \gamma^3\ell }{8 \pi} (3 u^a u^b + g_{(0)}^{ab}),
\end{align}
where $\Xi = 1-a^2 \ell^{-2}$ and 
\begin{equation}
u^a \p_a = \gamma \ell (\p_t +\frac{a}{\ell^2} \p_\phi),\qquad \gamma^{-1} \equiv \sqrt{1-\frac{a^2}{\ell^2} \sin^2\theta}. 
\end{equation} 
The total mass and angular momentum are $\mathcal M = \int \sqrt{\bar q} \, T^t_{\,\, t} = \frac{m}{\Xi^2}$,  $\mathcal J = Ma = -\int \sqrt{\bar q} \, T^t_{\,\,\phi} = \frac{ma}{\Xi^2}$. We observe that $J_{AB} \neq 0$. Therefore, the Kerr-AdS$_4$ solution is not included in the phase space with $J_{AB}=0$ (although it is contained in the phase space with Dirichlet boundary conditions). However, it is possible to obtain a stationary axisymmetric solution with $J_{AB} = 0$ as follows. The most general diagonal traceless and divergence free stationary $T^{ab}$ is given by 
\begin{equation}
\begin{split}
&T^{tt}_{\text{corr}} = \ell^2 [2 T^{\theta\theta}(\theta) + \tan \theta ~{T^{\theta \theta}}^\prime  (\theta)], \quad T^{\theta \theta}_{\text{corr}}  = T^{\theta \theta}(\theta), \\
&T^{\phi\phi}_{\text{corr}}  = \frac{1}{\sin^2 \theta} [T^{\theta\theta}(\theta) + \tan \theta ~{T^{\theta\theta}}^\prime(\theta) ]
\end{split}
\end{equation} and the other components are set to zero. We consider the sum of $T_{\text{Kerr}}+T_{\text{corr}}$. We solve for $T^{\theta\theta}(\theta)$ to set $J^{AB} = 0$. The regular solution at $\mathscr I$ is unique and given by
\begin{equation}
T^{tt}= -\frac{m \ell^3}{4\pi},\qquad T^{t\phi} = -\frac{3a m \ell \gamma^5}{8 \pi }, \qquad T^{AB} = -\frac{m \ell}{8 \pi} q^{AB}. 
\end{equation}
The total mass and angular momentum of the new solution are $\mathcal M  = \int \sqrt{-g_{(0)}} \, T^t_{\,\, t} = m$,  $\mathcal J= -\int \sqrt{-g_{(0)}} \, T^t_{\,\,\phi} = \frac{ma}{\Xi^2}$. Again, to know whether this solution is regular in the bulk of spacetime remains an open question. This ends our discussion of some interesting aspects of the $\Lambda$-BMS$_4$ phase space.

\section{Flat limit of the \texorpdfstring{$\Lambda$}{Lambda}-BMS\texorpdfstring{$_4$}{4} phase space}
\label{sec:Flat limit of the LBMS4 phase space}
The correspondence between both formulations of the $\Lambda$-BMS$_4$ phase space formulated in the SFG and the Bondi gauges, as well as the straightforward link between $\Lambda$-BMS$_4$ and Generalized BMS$_4$ algebras at flat limit $\Lambda\to 0$, give us confidence to build a rigorous and meaningful flat limit process to map the leaky phase spaces in the presence of $\Lambda\neq 0$ discussed thus far onto the asymptotically flat phase space discussed in chapter \ref{chapter:Charges} when sending $\Lambda$ to zero. One of the major advantages is to exploit the better control on renormalization that we have for (A)dS asymptotics to asymptotically flat spacetimes. For instance, this would provide a firm base for the Iyer-Wald counter-term \eqref{Yur}-\eqref{YrA} postulated when attempting to define a finite action principle for the Generalized BMS$_4$ phase space. We proceed step by step: we first comment on the various subtleties of the flat limit at a conceptual level and how we treat them; next, we prove that the solution space admits a well-defined flat; finally, we analyze the same limit at the level of the presymplectic structure, which needs a further regularization by a corner-term subtraction.

\subsection{The philosophy of the flat limit}

\begin{figure}[!htb]
\centering
\resizebox{\textwidth}{!}{
\subfloat[][Asymptotically locally \emph{dS}$_4$ case.]{
	\resizebox{0.5\textwidth}{!}{
	\begin{tikzpicture}
	\draw[white] (-5,-4) -- (-5,5) -- (5,5) -- (5,-4);
	\draw[very thick] (-4,4) -- (4,4);
	\draw[] (0,4) node[above]{$\mathscr{I}^+_{\text{dS}}$};
	\fill[black!10] (-4,3) -- (-1,0) -- (-4,-3);
	\draw[blue,thick] (-4,3) node[right]{\phantom{..}$\mathscr{I}^+$} -- (-1,0);
	\node[black!50,right,text width=5cm,rotate=45] at (-2.5,-3) {Asymptotically flat\\ ($r \ll \Lambda^{-1}$)};
	\draw[blue,thick] (-1,0) -- (-4,-3) node[right]{\phantom{..}$\mathscr{I}^-$};
	\node[circle,fill=black,inner sep=0pt,minimum size=5pt] at (-1,0) {};
	\node[right] at (-1,0) {$i^0$};
	\foreach \k in {-3.75,-3.50,...,-2.25} 
	{
	\coordinate (pk) at (\k,-\k-2);
	\coordinate (qk) at (6+2*\k,4);
	\draw[red,thick] (pk) -- (qk);
	\draw[red,thick,-Latex] (pk) -- ++(1.0,1.0);
	}
	\end{tikzpicture}
	}
}\subfloat[][Asymptotically locally \emph{AdS}$_4$ case.]{
	\resizebox{0.5\textwidth}{!}{
	\begin{tikzpicture}
	\draw[white] (-3,-4.5) -- (-3,4.5) -- (7,4.5) -- (7,-4.5);
	\draw[very thick] (4,-4) -- (4,4);
	\draw[] (4,0) node[right]{$\mathscr{I}_{\text{AdS}}$};
	\fill[black!10] (0,2) -- (3,-1) -- (0,-4);
	\draw[blue,thick] (0,2) node[right]{\phantom{..}$\mathscr{I}^+$} -- (3,-1);
	\draw[blue,thick] (3,-1) -- (0,-4) node[right]{\phantom{..}$\mathscr{I}^-$};
	\node[circle,fill=black,inner sep=0pt,minimum size=5pt] at (3,-1) {};
	\node[right] at (3,-1) {$i^0$};
	\node[black!50,right,text width=5cm,rotate=45] at (1.4,-4) {Asymptotically flat \\ ($r \ll |\Lambda|^{-1}$)};
	\foreach \k in {-3.5,-3.25,...,-2.25} 
	{
	\coordinate (pk) at (\k+4,-\k-3);
	\coordinate (qk) at (4,-3-2*\k);
	\draw[red,thick] (pk) -- (qk);
	\draw[red,thick,-Latex] (pk) -- ++(1.0,1.0);
	}
	\end{tikzpicture}
	}
}
}
\caption{Energy flow in the bulk and flat limit of (A)dS asymptotics.}
\label{Fig1}
\end{figure}
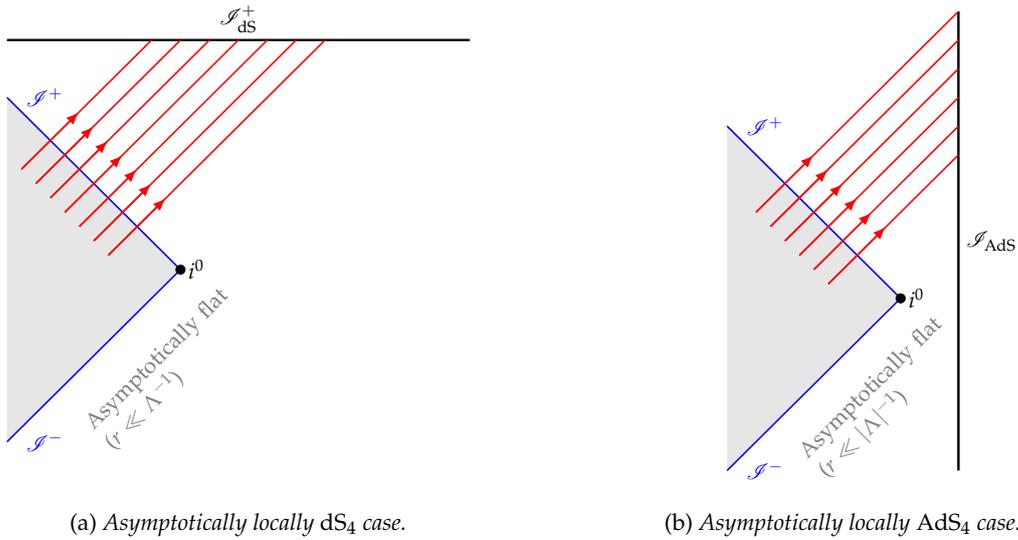

The fact that the Bondi coordinates \eqref{Bondi line element} can be defined for any value of the cosmological constant is not the only reason for using them to define the flat limit procedure. The choice is motivated by the intuition according to which null-type coordinates are more adapted than others to take the flat limit and the naive argument is as follows. Let us consider an AlAdS$_4$ spacetime. It is well-known that the cosmological constant is manifesting itself at cosmological scales, \textit{i.e.} for phenomena occurring at relative distance $r\sim \ell$, where $\ell\sim \sqrt{1/\Lambda}$ is the AdS radius defined previously. On astrophysical scales $r\ll \ell$, the gravitation field can be described in good approximation with $\Lambda\approx 0$. We can thus imagine a little causal diamond representing an asymptotically flat spacetime sitting far deep in the bulk of the AlAdS$_4$ spacetime under consideration, as represented in Figure \ref{Fig1}.

We further assume that the Bondi coordinate system continues to be valid far enough from the conformal boundary $\mathscr I_{\text{AdS}}$ in order to cover this little central diamond. This pictorial description shows that the outgoing null rays naturally map the ``points'' of $\mathscr I^+$ of the asymptotically flat spacetime in the deep bulk to the ``points'' of the conformal infinity $\mathscr I_{\text{AdS}}$. Following the flow of null radiation shows that in for any compact submanifolds of $\mathscr I^+$ and $\mathscr I_{\text{AdS}}$, this relation is bijective if we assume that all of the sources of gravitational radiation are beyond $\mathscr I^+$ in the bulk. This is precisely why there is a link between the dynamical quantities, such as the Bondi mass and angular momentum aspects, defined for the $\Lambda$-BMS$_4$ phase space as codimension 1 functions living on $\mathscr I_{\text{AdS}}$ and the homologuous functions in the context of the Generalized BMS$_4$ phase space. Considering the asymptotically flat limit as a zoom on the central diamond also allows to understand why the symmetries have a well-defined flat limit, but the dynamical quantities have not. Indeed, symmetries are not only defined on the boundary but extends to the bulk, see \eqref{eq:xir}. Assuming that the power series in $1/r$ still holds for values of $r\ll \ell$, the action of a symmetry at the vicinity of $\mathscr I_{\text{AdS}}$ is transmitted along the outgoing null rays (that are immutable because they define the Bondi coordinate $u$) to $\mathscr I^+$. However, the dynamical quantities are not expected to have a well-defined flat limit since energy flows along the null lines from $\mathscr I^+$ to $\mathscr I_{\text{AdS}}$. This observation also applies for the symplectic structure, the surface charges and their algebra: their bare expressions are all singular in the flat because of divergences in the limit $\Lambda\to 0$. The construction of a full flat limit process must therefore be assorted with some further renormalization of the presymplectic structure, this time not aimed at curing radial divergences, but parametric poles in $\ell \sim 1/\Lambda$ in the limit $\Lambda\to 0$ or $\ell\to+\infty$. Note that all of these features can be discussed for AldS$_4$ spacetimes with perfectly equivalent conclusions.

We are done with the naive picture that can merely give a flavor of what we can expect from the flat limit. The major technical obstruction to consider the flat limit as a zoom in the far bulk is the necessity to understand the link between the data in the Bondi gauge defined around the conformal infinity $\mathscr I$ of the host spacetime and the data in the radically different Bondi gauge describing the gravitational field around the null hypersurface $\mathscr I^+$ delimiting the central diamond at finite distance. This would demand to resum the Bondi expansion in the far bulk of the Al(A)dS$_4$ spacetime and again perform a matching between gauges, which is something we do not want to do and which is probably impossible in general. What we can do is to consider the $\Lambda$-BMS$_4$ solution spaces, $\bar{\mathcal S}_\Lambda$, defined in \eqref{SLambdabar}, as a one-parameter family of solution spaces whose $\bar{\mathcal S}_0$ defined in \eqref{S0} would be an adherence point in the limit $\Lambda\to 0$, at the price, of course, of performing some field redefinitions in the parameters of $\bar{\mathcal S}_\Lambda$ to ensure that the limit is not singular. The ``future of light'' $\mathscr I$ for solutions in $\bar{\mathcal S}_\Lambda$ is mapped to the ``future of light'' $\mathscr I^+$ for solutions in $\bar{\mathcal S}_0$ while keeping the same boundary structure. For instance, the norm of the boundary foliation vector $\bm T$ is proportional to $\Lambda$: $\bm T$ is timelike for AdS asymptotics, spacelike for dS asymptotics and the gap between both cases is filled by the regular limit $\Lambda=0$ where $\bm T$ is null, as expected. The $u$ coordinate evolving on the flow of $\bm T$ changes its nature, from the intrinsic point of view of the boundary, but remains a null coordinate with respect to the bulk (by definition of the Bondi gauge) and the limit is smooth. The other coordinates $r$ and $x^A$ are not touched by the flat limit, since the background structure related to them (a radial null foliation in the bulk and a parametrization of the codimension 2 spaces transverse to $\bm T$) is invariant under modifications of $\Lambda$. This is also the point of view of \cite{Barnich:2012aw}, where a similar process was successfully performed in 3$d$ gravity. In summary, with the parametric limit $\Lambda\to 0$ we perform below, we just map boundaries on boundaries, which is a complementary picture of the naive ``zoom'' setup, but offers a better analytical control. Let us explain how to implement this process first at the level of the solution space.

\subsection{Flat limit of the solution space}
\label{sec:Flat limit of the solution space}

For any non-vanishing value of the cosmological constant, the mere hypothesis of conformal compactification with analyticity \eqref{eq:gABFallOff} leads to the most general solution space in the Bondi gauge by virtue of the Fefferman-Graham theorem \cite{Fefferman:1985aa}. As announced, the analyticity requirement breaks down when in general $\Lambda=0$ but was maintained in our presentation of the Generalized BMS$_4$ solution space in section \ref{sec:Solution space in Bondi gauge}. The flat limit of any Al(A)dS$_4$ solution space can thus lead only to the subsector of asymptotically (locally) flat gravity for which the logarithmic branch has been neglected. Since we have been only concerned in the analytic branch in the flat case, this is not a restriction for our analysis. 

The cornerstone of the flat limit process is the constraint equation \eqref{eq:CAB} which reads as
\begin{equation}
(\partial_u-l) q_{AB} = \frac{\Lambda}{3}C_{AB} \label{switch flat limit}
\end{equation}
after the boundary gauge fixing \eqref{bndgauge} in $\bar{\mathcal S}_\Lambda$. Like all of the analytical constraints brought by Einstein's equations \eqref{eq:MABTF}, it acts as a switch leading to radically different conclusions depending on whether $\Lambda$ is vanishing or not. 
\begin{itemize}[label=$\rhd$]
\item When $\Lambda\neq 0$, the equation \eqref{switch flat limit} determines algebraically $C_{AB}$ in terms of the time evolution of the boundary metric, which is completely unconstrained.
\item When $\Lambda = 0$, the meaning of \eqref{switch flat limit} drastically changes. It imposes a strong constraint on the boundary metric, prescribing its time evolution as \eqref{EOM qAB time evolution}, while keeping the shear $C_{AB}$ completely free.
\end{itemize}
We see that imposing this equation of motion does not commute with the flat limit $\Lambda\to 0$, simply because, depending upon the way we solve it (\textit{i.e.} for $\partial_u q_{AB}$ or $C_{AB}$), the limit $\Lambda\to 0$ can be either well-defined or ill-defined. Our prescription for the flat limit process is to conserve the well-behaved field $C_{AB}$ explicit and trade, thanks to \eqref{switch flat limit}, any time derivative of the boundary metric for the expression $\Lambda C_{AB}$ which is vanishing in the flat limit. This allows us to be systematic and track down any contribution in $\Lambda$ coupled to well-behaved fields like $C_{AB}$. Doing so, we \textit{assume} that the latter does not depend upon $\Lambda$, in order to end up with expressions in which any dependency in $\Lambda$ is clear, and take the limit $\Lambda\to 0$ analytically afterwards, as we could do for numerical functions $f(\Lambda)$. This also amounts to extracting explicitly the trace of $N_{AB} = \partial_u C_{AB}$, which contains an hidden contribution in $\Lambda$. Defining
\begin{equation}
N_{AB}^{TF} \equiv N_{AB} - \frac{\Lambda}{6}q_{AB} C_{CD}C^{CD}, \quad q^{AB}N_{AB}^{TF} = 0,
\end{equation}
we get a tensor whose limit for $\Lambda\to 0$ is smooth: it is assimilated to the Bondi news tensor in the flat case.

A second switch, less crucial although nevertheless important, involves $\mathcal D_{AB}$. The relevant equations of motion are \eqref{eq:DAB}, \eqref{eq:partial u DAB} and $D_B \mathcal D^{AB} = 0$ by virtue of the Fefferman-Graham theorem \cite{Starobinsky:1982mr,Fefferman:1985aa,2007arXiv0710.0919F} which forbids the appearance of a logarithmic term in \eqref{eq:EOM_UA}. In the boundary gauge fixing \eqref{bndgauge}, the second equation states nothing but $\partial_u \mathcal D_{AB}=0$ which has a trivial flat limit. The divergence-free condition also admits a direct flat limit and is in accordance with \eqref{eq:DivDAB=0}. However, the condition $\Lambda \mathcal D_{AB}=0$ indicates that $\mathcal D_{AB}$ itself has a poor behavior in the flat limit, since it can be non-zero only if $\Lambda=0$. Again, imposing that the product $\Lambda \mathcal{D}_{AB}$ is zero does not commute with the flat limit process $\Lambda\to 0$, because if we impose it, for instance, before sending $\Lambda$ to zero, we would get only the subsector of the flat theory with $\mathcal D_{AB} = 0$. The trick is to write all the equations with an apparent time-independent divergence-free $\mathcal D_{AB}$ for any value of $\Lambda$ and impose the additional constraint $\Lambda \mathcal D_{AB}=0$ only if $\Lambda$ is fixed. However, the flat limit $\Lambda\to 0$ has to be understood as the limit of the expressions with the apparent $\mathcal D_{AB}$ while maintaining the constraint $\Lambda\mathcal D_{AB}=0$ without solving it, in such a way that $\mathcal D_{AB}$ is present in the flat limit and unconstrained by the equations of motion. Denoting again $g_{AB} = r^2\sum_{n\geq 0}g^{(n)}_{AB}r^{-n}$, the switch conditions involving $\mathcal E_{AB},\mathcal F_{AB},\dots$ formally read as $\Lambda g^{(n)}_{AB} = \partial_u g_{AB}^{(n-1)}+(\dots)$ at subleading orders. In the flat limit, they degenerate when into the evolution equations $\partial_u g_{AB}^{(n-1)}+(\dots) = 0$. Hence the symmetric traceless tensor fields $\mathcal E_{AB},\mathcal F_{AB},\dots$ are no longer algebraically constrained by the fields of lower order: only their time derivative is kept constrained, in accordance with the discussion below the equation \eqref{EOM qAB time evolution}. We obtain a infinite-dimensional countable tower of fields determined by a function of the angles at some retarded time $u = u_0$ and whose value of $\mathscr I^+$ is obtained by the integration of the associated evolution equations in $u$ \cite{Barnich:2010eb,Tamburino:1966zz,10.2307/2415610}.

With this knowledge, the flat limit of the other algebraic equations is immediate. Indeed, imposing the boundary gauge fixing \eqref{bndgauge} in \eqref{eq:EOM_beta} rules out the leading $\beta_0$ field and nothing depends upon $\Lambda$ except the constraint \eqref{eq:DAB} that we do not impose before taking the limit in $\Lambda$. Hence the flattened version of \eqref{eq:EOM_beta} is \eqref{eq:EOM_beta flat} as it should. Working in the same way on \eqref{eq:EOM_UA} gives straightforwardly \eqref{eq:EOM_UA flat} in the flat limit without the logarithmic term since $D_B \mathcal D^{AB}=0$ has been maintained during the process. Finally \eqref{eq:EOMVr} is sent on \eqref{eq:EOM_Vr flat} since $\beta_0$ and $U_0^A$ disappear when the conditions \eqref{bndgauge} hold and the linear pieces in $\Lambda$ at $\mathcal O(r^2)$ and $\mathcal O(r^0)$ orders are consistently brought to zero.  The asymptotically flat limit of the evolution equations \eqref{eq:EvolutionNA} and \eqref{eq:EvolutionM} requires a bit more care, provided that the quantities \eqref{eq:hatM}, \eqref{eq:hatNA} and \eqref{eq:hatJAB} defined intrinsically for $\Lambda\neq 0$ configurations are not meant to be well-defined in the flat limit. We start by collecting the terms in $1/\Lambda$ hidden in \eqref{eq:EvolutionNA}:
\begin{equation}
\begin{split}
-\frac{3}{2\Lambda} \Big[ &(\partial_u + l) D^B (\partial_u C_{AB} - \frac{1}{2} l C_{AB}) - D^B \partial_u (\partial_u C_{AB} - \frac{1}{2} l C_{AB}) \\
&\quad + \frac{1}{2} (\partial_u + l) \partial_A R[q] + D^B (D_A D_B l - \frac{1}{2} D_C D^C l q_{AB}) \Big]. \label{eq:ProblematicTerms}
\end{split}
\end{equation}
We recall that there are two subtle steps in the massaging of the evolution equation needed before taking the limit $\Lambda\to 0$. First, we have to develop the remaining derivatives in $u$ acting on covariant derivatives and solve the switch constraint \eqref{switch flat limit} for $\partial_u q_{AB}$ to highlight all the factors in $\Lambda$. Second, we need to extract the trace of $N_{AB}$, which also contains a residual contribution $\sim \Lambda$. We end up with
\begin{equation}
\begin{split}
\eqref{eq:ProblematicTerms} = \,\, &\frac{1}{2} D_C (N_{AB}^{TF} C^{BC}) + \frac{1}{4} N_{BC}^{TF} D_A C^{BC} - \frac{1}{4} D_A D_B D_C C^{BC} \\
&+ \frac{1}{8} C^B_C C^C_B \partial_A l - \frac{3}{16} l \partial_A (C^B_C C^C_B).
\end{split}
\label{eq:Probl2}
\end{equation}
The following identities turn out to be useful for the computation:
\begin{equation}
\begin{split}
(\partial_u + 2l) H^{AB} &= q^{AC}\partial_u H_{CD}q^{BD} + \frac{2\Lambda}{3} C^{C(A}H^{B)}_C,   \\
(\partial_u+l) (D^B H_{AB}) &= D^B \partial_u H_{AB} - \frac{1}{2} q^{CD}H_{CD}\partial_A l   \\
& \quad\, - \frac{\Lambda}{3}\Big[ D_B (C^{BC}  H_{AC}) + \frac{1}{2} H^{BC}D_A C_{BC}\Big], \\
(\partial_u+l) C^{AB}C_{AB} &= 2 N^{AB}C_{AB} - l C_{AB}C^{AB}, \\
(\partial_u + l) \partial_A R[q] &= - (D^B D_B + \frac{1}{2} R[q]) \partial_A l + \frac{\Lambda}{3} D_A D_B D_C C^{BC}
\end{split}
\end{equation}
for an arbitrary symmetric rank 2 transverse tensor $H_{AB}(u,x^C)$. Using \eqref{identite glenn}, the first term of \eqref{eq:Probl2} can be rewritten as
\begin{equation}
\frac{1}{2} D_C (N_{AB}^{TF} C^{BC}) = \frac{1}{4} D_B (N_{AC}^{TF} C^{BC}-C_{AC} N^{BC}_{TF}) + \frac{1}{4} \partial_A (C_{BD}N^{BD}_{TF}).
\end{equation}
We can now present \eqref{eq:EvolutionNA} in a way that makes terms in $\Lambda$ explicit:
\begin{align}
(\partial_u + l)  N_A &- \partial_A  M - \frac{1}{4} C_{AB} \partial^B R[q] - \frac{1}{16} \partial_A (N_{BC}^{TF} C^{BC}) \label{eq:EvolutionNA_Explicit} \\
&- \frac{1}{32} l \partial_A (C_{BC}C^{BC}) +\frac{1}{4} N_{BC}^{TF} D_A C^{BC} + \frac{1}{4} D_B (C^{BC} N_{AC}^{TF} - N^{BC}_{TF} C_{AC}) \nonumber \\
&+\frac{1}{4} D_B (D^B D^C C_{AC} - D_A D_C C^{BC}) + \frac{\Lambda}{2} D^B(\mathcal{E}_{AB} - \frac{7}{96} C^C_D C^D_C C_{AB}) = 0.\nonumber 
\end{align}
As a result, the asymptotically flat limit can be safely taken and \eqref{eq:EvolutionNA_Explicit} reduces to
\begin{align}
(\partial_u + l)  N_A &- \partial_A  M - \frac{1}{4} C_{AB} \partial^B R[q] - \frac{1}{16} \partial_A (N_{BC}C^{BC}) \nonumber \\
&- \frac{1}{32} l \partial_A (C_{BC}C^{BC}) +\frac{1}{4} N_{BC} D_A C^{BC} + \frac{1}{4} D_B (C^{BC} N_{AC} - N^{BC} C_{AC}) \nonumber \\
&+\frac{1}{4} D_B (D^B D^C C_{AC} - D_A D_C C^{BC}) = 0, \label{duNA flat limit}
\end{align}
where we identified $N_{AB} = N_{AB}^{TF}$ when $\Lambda = 0$. This equation fully agrees with \eqref{duNA flat}. Finally, the asymptotically flat limit of the second evolution equation \eqref{eq:EvolutionM} is straightforward and gives
\begin{equation}
\begin{split}
(\partial_u + \frac{3}{2}l )  M +\frac{1}{8} N_{AB} N^{AB} - \frac{1}{8} l N_{AB} C^{AB} + \frac{1}{32} l^2 C_{AB} C^{AB} - \frac{1}{8} D_A D^A R[q] \\
- \frac{1}{4} D_A D_B N^{AB} + \frac{1}{4} C^{AB} D_A D_B l + \frac{1}{4} \partial_{(A} l D_{B)} C^{AB} + \frac{1}{8} l D_A D_B C^{AB} = 0,
\end{split}\label{duM flat limit}
\end{equation}
in perfect agreement with \eqref{duM flat}. Note that \eqref{duNA flat limit} and \eqref{duM flat limit} are also in agreement with (4.49) and (4.50) of \cite{Barnich:2010eb} respectively, after a change of conventions: the Bondi news tensor is defined in \cite{Barnich:2010eb} as $N_{AB}^{\text{\cite{Barnich:2010eb}}} = \partial_u C_{AB} - l C_{AB}$ while we always define $N_{AB} = \partial_u C_{AB}$. 

Summarizing the results, we find that
\begin{equation}
\lim_{\Lambda\to 0} \bar{\mathcal S}_\Lambda = \bar{\mathcal S}_0 \label{limits of solution spaces bar}
\end{equation}
where the solution spaces are defined in \eqref{SLambdabar} and \eqref{S0} respectively and the limit $\Lambda\to 0$ is defined formally with the prescriptions enunciated at the beginning of the current section. Fixing the boundary volume form to include the global vacua (A)dS$_4$ and Minkowski in the solution spaces -- which amounts to selecting the $\Lambda$-BMS$_4$ orbit for which $\sqrt{q}=\sqrt{\mathring q}$, we obtain
\begin{equation}
\boxed{
\lim_{\Lambda\to 0}\mathring{\mathcal S}_\Lambda = \mathring{\mathcal S}_0.
}
\label{limits of solution spaces ring}
\end{equation}
Although not surprising since the Bondi gauge is defined for any value of $\Lambda$ and the derivation of the respective solution spaces when $\Lambda=0$ and $\Lambda\neq 0$ is similar, it is yet remarkable to observe that a straightforward flat limit process (up to some purely analytical subtleties) maps the evolution equations of Bondi aspects in Al(A)dS$_4$ spacetimes to the standard time evolution constraints of the Bondi aspects defined for flat asymptotics. Provided that the evolution of mass and angular momentum for $\Lambda\neq 0$ is governed by the (covariant) conservation of a boundary stress-tensor, this fact can help as a building block for the understanding of a putative 4$d$-3$d$ formulation of \textit{flat holography}, where $M$ and $N_A$ would evolve as momenta on the codimension 1 null (or Carrollian) manifold at the boundary (see \textit{e.g.} \cite{Duval:2014uva,Duval:2014uoa,Hopfmuller:2018fni,Korovin:2017xqu,Ciambelli:2019lap,Ciambelli:2018ojf}). Note also that the flat limit process is intimately related to the choice of a preferred boundary foliation $\bm T$. When $\Lambda\neq 0$, as noted previously, the fixation of $\bm T$ is merely a choice of Gaussian normal coordinates on $\mathscr I$ which can always be made without ruling out any solution in the phase space. However, when taking the limit $\Lambda\to 0$ while keeping $\bm T$ untouched, this vector determines the (degenerate) null direction along which the flat $\mathscr I^+$ will be aligned. As a result, considering a particular foliation $\bm T$ gives one particular flat limit for which the hypersurfaces at constant retarded time have been preserved and $\bm T$ has becomed null on the boundary.

\subsection{Flat limit of the action and corner terms}
\label{sec:Flat limit of the action and corner terms}

Finally we turn our interest on the flat limit of the symplectic structure on the $\Lambda$-BMS$_4$ phase space. The most rigorous way to proceed is to reconsider the on-shell renormalized variational principle \eqref{action lambda bms 4}, isolate the explicit dependency in $\Lambda$ and see if the flat limit $\Lambda\to 0$ is readily possible or not. If a whole section has been devoted to discussing this process, one can already expect that it is neither direct nor even well-defined \textit{a priori}. Indeed, we prove here that a further renormalization of the on-shell action principle, although finite in $r$, is mandatory in order to obtain a meaningful asymptotically flat limit. 

\subsubsection{Corner terms and prescription for renormalization}

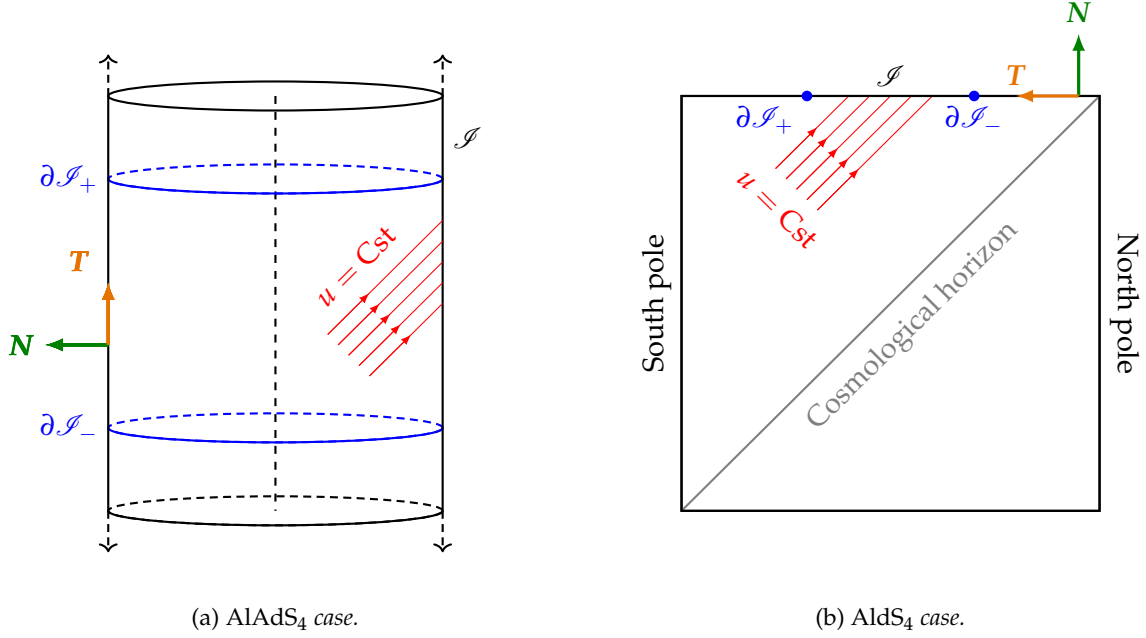
\begin{figure}[!ht]
\centering
\subfloat[\emph{AlAdS$_4$} case.]{
\begin{tikzpicture}[scale=0.55]
	\draw[white] (-7,-7) -- (-7,7) -- (7,7) -- (7,-7) -- cycle;
	\draw[thick, dashed] (0,5) -- (0,-5);
	\draw[thick, blue, densely dashed] (0,-3) ellipse (4 and 0.35);
	\draw[thick,blue] (4,3) arc (0:-180:4 and 0.35);
	\draw[thick, blue, densely dashed] (0,3) ellipse (4 and 0.35);
	\draw[thick,blue] (4,-3) arc (0:-180:4 and 0.35);
	\draw[blue] (-4, 3) node[left]{$\partial\mathscr{I}_+$};
	\draw[blue] (-4,-3) node[left]{$\partial\mathscr{I}_-$};
	\draw[thick] (4,-5) -- (4,5);
	\draw[thick] (-4,-5) -- (-4,5);
	\draw[thick, densely dashed,->] (4,5) -- (4,6);
	\draw[thick, densely dashed,->] (-4,5) -- (-4,6);
	\draw[thick, densely dashed,->] (4,-5) -- (4,-6);
	\draw[thick, densely dashed,->] (-4,-5) -- (-4,-6);
	\draw[thick] (0,5) ellipse (4 and 0.35);
	\draw[thick, densely dashed] (0,-5) ellipse (4 and 0.35);
	\draw[thick] (4,-5) arc (0:-180:4 and 0.35);
	\draw[] (4,4) node[right]{$\mathscr{I}$};
	\node[red,rotate=45] at (1.9,0.9) {$u = \text{Cst}$};
	\foreach \k in {-4.25,-4.0,...,-3.25} 
	{
	\coordinate (pk) at (\k+5.5,-\k-5);
	\coordinate (qk) at (4,-6.5-2*\k);
	\draw[red] (pk) -- (qk);
	\draw[red,-latex] (pk) -- ++(1.0,1.0);
	\draw[very thick,green!50!black,-latex] (-4,-1) -- (-5.5,-1)node[left]{$\bm N$};
	\draw[very thick,orange!90!black,-latex] (-4,-1) -- (-4,0.5)node[above]{$\bm T\qquad$};
	}
	\end{tikzpicture}
}
\hfill
\subfloat[\emph{AldS$_4$} case.]{
\begin{tikzpicture}[scale=0.55]
	\draw[white] (-7,-7) -- (-7,7) -- (7,7) -- (7,-7) -- cycle;
	\coordinate (tl) at (-5, 5);
	\coordinate (tr) at ( 5, 5);
	\coordinate (bl) at (-5,-5);
	\coordinate (br) at ( 5,-5);
	\draw[thick] (tl) -- (tr) -- (br) -- (bl) -- cycle;
	\draw[] ($(tl)!0.5!(bl)$) node[above,rotate=90]{South pole};
	\draw[] ($(tr)!0.5!(br)$) node[above,rotate=-90]{North pole};
	\draw[black!50, thick] (tr) -- (bl);
	\draw[black!50] ($(tr)!0.5!(bl)$) node[below,rotate=45,outer sep=3pt]{Cosmological horizon};
	\node[red,rotate=-45] at (-2.7,2.3) {$u = \text{Cst}$};
	\foreach \k in {-4.25,-4.0,...,-3.25} 
	{
	\coordinate (pkp) at (\k+1.5,-\k-1);
	\coordinate (qkp) at (7.5+2*\k,5);
	\coordinate (arp) at ($(pkp)+(1.0,1.0)$);
	\draw[red] (pkp) -- (qkp);
	\draw[red, -latex] (pkp) -- (arp);
	}
	\draw[] ($(tl)!0.5!(tr)$) node[above]{$\mathscr{I}$};
	\draw[thick] (tl) -- (tr) -- (br) -- (bl) -- cycle;
	\draw[] ($(tl)!0.3!(tr)$) node[circle, fill=blue, minimum size=4pt,inner sep=0pt]{};
	\draw[blue] ($(tl)!0.2!(tr)$) node[below]{$\partial\mathscr I_+$};
	\draw[] ($(tl)!0.7!(tr)$) node[circle, fill=blue, minimum size=4pt,inner sep=0pt]{};
	\draw[blue] ($(tl)!0.7!(tr)$) node[below]{$\partial\mathscr I_-$};
	\draw[very thick,green!50!black,-latex] (4.5,5) -- (4.5,6.5)node[above]{$\bm N$};
	\draw[very thick,orange!90!black,-latex] (4.5,5) -- (3,5)node[above]{$\bm T$};
\end{tikzpicture}
}
\caption{Geometry of Al(A)dS$_4$ boundaries with background structure.}
\label{fig:GeometryBndry}
\end{figure}

We reformulate the $\Lambda$-BMS$_4$ variational principle \eqref{action lambda bms 4} in a way similar to the asymptotically flat variational principle discussed in section \ref{sec:Flat variational principle}, except that the radial regulation in $r$ has already been taken into account by holographic renormalization, see section \ref{sec:Regulated variational principle}. Recall that the construction of the $\Lambda$-BMS$_4$ phase space assumes the existence of a foliation $\bm T$ on the conformal boundary $\mathscr I$ with topological spheres as orthogonal surfaces of measure $\sqrt{\mathring q}$. Considering a portion of $\mathscr I$, the initial and final values of the foliation parametrized by $u$ define the ``corner'' boundaries denoted as $\p\!\mathscr I_+$ (at $u = u^+$) and $\p\!\mathscr I_-$ (at $u = u^-$), see Figure \ref{fig:GeometryBndry}. 

Ignoring for the moment the boundary terms at fixed $u = u^\pm$, the variation of the renormalized action \eqref{renormalized action} is given on-shell by
\begin{equation}
\delta S_{ren}^{\Lambda\text{-BMS}_4}[\phi] = \int_{\mathscr I} \boldsymbol \Theta_{ren}^{\Lambda\text{-BMS}_4}[\phi;\delta \phi]= \frac{\Lambda}{32 \pi G}\int_{\mathscr I} (\D^3x) \sqrt{\mathring q}  \, J_{(tot)}^{AB}\delta q_{AB}, 
\end{equation}
assuming the boundary gauge fixing \eqref{bndgauge} and incorporating with hindsight the contribution of some $\bm L_{\circ}$ to the stress-tensor, written as in \eqref{Tab tot} in the SFG gauge but easily transposed in the Bondi gauge with the algorithm described in section \ref{sec:FGg}. We have $J_{AB}^{(tot)} = J_{AB} + J_{AB}^\circ$ where $J_{AB}$ is still defined in \eqref{eq:RefiningTab} and $J_{AB}^\circ = T_{AB}^{\circ} - \frac{1}{2}q_{AB}q^{CD}T_{CD}^\circ$. 

As a boundary Lagrangian for the gravitational field, we assume that
\begin{equation}
\bm L_\circ = L_\circ [q_{AB},\bm T,\mathring q] \,(\D^2 x)
\end{equation}
is only built from the boundary background structure $\bm T$, $\sqrt{\mathring q}$ and the metric on the transverse space $q_{AB}$. Using these new assumptions and the substitution $ \sqrt{\mathring{q}} (\D^3 x) = \D u \, \D^2 \Omega$ we have, equivalently, 
\begin{equation}
\delta S_{ren}^{\Lambda\text{-BMS}_4}[\phi] = \int_{\mathscr I} \bm \Theta^{\Lambda\text{-BMS}_4}_{ren}[\phi;\delta \phi]=  \int_{\mathscr I} \D u \, \D^2 \Omega  \, \left(- \frac{\Lambda}{32 \pi G} J_{AB} +\frac{1}{\sqrt{\mathring{q}}}\frac{\delta L_\circ}{\delta q^{AB}} \right) \delta q^{AB}. \label{delta S with delta L circ flat limit}
\end{equation}
The flat limit $\Lambda \to 0$ is obtained by massaging the first term of \eqref{delta S with delta L circ flat limit} in a way analogous to the procedure already described in section \ref{sec:Flat limit of the solution space}. Thanks to \eqref{eq:hatJAB}, we develop $J_{AB}$ in terms of $C_{AB}$, $q_{AB}$, $N_{AB}^{TF}$, $M$ and $N_{A}$ where $N_{AB}^{TF}$ is the trace-free part of the Bondi news tensor $N_{AB}$. In particular, all derivatives of the boundary metric $\p_u q_{AB}$ need to be expressed in terms of $C_{AB}$ using \eqref{switch flat limit}. The computation requires a bit of dexterity while dealing with variations of the fields $q_{AB}$, $C_{AB}$, $N_{AB}^{TF}$\dots It will not be reproduced here but the reader interested in verifying the algebraic development would find a list of useful relations for that task in appendix \ref{Useful relations}. Using the form notation $\bm L_\circ = \D u \frac{\D^2 \Omega}{\sqrt{\mathring{q}}} L_\circ$, we obtain
\begin{align}
\boldsymbol \Theta_{ren}^{\Lambda\text{-BMS}_4}[\phi;\delta \phi] \Big|_{\mathscr I} &= \frac{\D u\, \D^2\Omega }{16\pi G} \left[ \frac{3}{2\Lambda}\partial_u N_{AB}^{TF}- D_{(A}\mathcal U_{B)} - \frac{1}{4} R[q] C_{AB} \right] \delta q^{AB} \nonumber \\
& \quad + \frac{\delta \bm L_\circ}{\delta q^{AB}} \delta q^{AB}+ \mathcal O(\Lambda), \label{intermediate step flat limit}
\end{align} 
where $\mathcal O(\Lambda)$ denotes terms that vanish in the $\Lambda \to 0$ limit. We observe that there is a pole $\sim \Lambda^{-1}$ and the flat limit is not well-defined. However, if we write \eqref{intermediate step flat limit} in the equivalent form
\begin{align}
\boldsymbol \Theta_{ren}^{\Lambda\text{-BMS}_4}[\phi;\delta \phi] \Big|_{\mathscr I}  &=\frac{\D u\, \D^2\Omega }{16\pi G}  \left[ \frac{3}{2\Lambda}\partial_u (N_{AB}^{TF}\delta q^{AB}) + \frac{1}{2}\Big( N^{AB}_{TF} + \frac{1}{2} R[q]q^{AB}\Big) \delta C_{AB} - D_{(A}\mathcal U_{B)}\delta q^{AB}\right] \nonumber \\ \label{intermediate step flat limit2}
& \quad + \frac{\delta \bm L_\circ}{\delta q^{AB}} \delta q^{AB} + \mathcal O(\Lambda ),
\end{align} 
it becomes clear that the divergence in the variation of the action is a corner term defined at the boundaries $\p\! \mathscr I_+$, $\p \!\mathscr I_-$ of the conformal boundary $\mathscr I$. In order to make these corner terms explicit, we consider the total derivative boundary Lagrangian
\begin{equation}
L_\circ= \p_u L_C[q_{AB}, \bm T, \mathring{q}] , 
\label{boundary lagrangian total derivative}
\end{equation} where the corner Lagrangian is
\begin{equation}
L_C[q_{AB},\bm T, \mathring{q}] \equiv \frac{\sqrt{\mathring{q}}}{64 \pi G} C^{AB} C_{AB} =  \frac{\sqrt{\mathring{q}}\ell^4}{64 \pi G} \p_u q_{AB} q^{AC}q^{BD}\p_u q_{CD}. \label{cornerL}
\end{equation}
We claim that the Lagrangian $L_\circ$ is of the form considered in section \ref{sec:Variational principle AdSd}. It is covariant on $\mathscr I$ since we can rewrite it as 
\begin{equation}
\boxed{
L_\circ =  \frac{\sqrt{\mathring{q}}\ell}{64 \pi G} \mathcal L_{\bm T} \left[ \mathcal L_{\bm T} g^{(0)}_{ab} P^{ac} P^{bd}\mathcal L_{\bm T}  g^{(0)}_{cd} \right]. 
} \label{eq:tensor Lcirc}
\end{equation}
Here $P_{ab}=g_{ab}^{(0)}+ \eta T_a T_b$ is the projector onto surfaces orthogonal to $\bm T$ and \eqref{eq:tensor Lcirc} is indeed a tensorial expression on the conformal boundary. Since $\bm L_\circ = \D \bm L_C$ where $\bm L_C = \frac{\D^2 \Omega}{\sqrt{\mathring{q}}} L_C \equiv \D^2x L_C$, its stress-energy tensor \eqref{Tab tot} vanishes, $T_\circ^{ab} = 0$, which implies that the last term of \eqref{intermediate step flat limit2} vanishes. In particular, it obeys trivially our hypotheses \eqref{conditions on Tab circ}. However, this peculiarity indicates that boundary Lagrangians like $\bm L_\circ = \D \bm L_C$ have no contributions in \eqref{renormalized presympelctic}, hence this prescription fails to be helpful for corner Lagrangians. 

Let us see how to overcome this difficulty. In the action \eqref{renormalized action}, the term associated with the boundary Lagrangian \eqref{boundary lagrangian total derivative} can be written as
\begin{equation}
\int_{\mathscr{I}}  \D^2x \, \D u \, L_\circ = \int_{\p\! \mathscr I_+}  \D^2 x \, L_C - \int_{\p\! \mathscr I_-}  \D^2 x \,  L_C  . 
\end{equation} It is manifestly the difference of two corner terms at $\p\! \mathscr I_+$ and $\p \!\mathscr I_-$. The variation of the Lagrangian $L_\circ$ gives
\begin{align}
\delta L_\circ &=  \p_u \left( \frac{\delta L_C}{\delta q_{AB}}\delta q_{AB} \right) + \p_u \Theta^C_\circ, \label{delta L circ flat limit} \\
\frac{\delta L_C}{\delta q_{AB}} &= - \frac{3 \sqrt{\mathring q} }{32\pi G \Lambda}N_{TF}^{AB},\qquad \Theta^C_\circ =  \frac{3  \sqrt{\mathring q} }{32\pi G \Lambda} \p_u \left( C^{AB}\delta q_{AB} \right). \label{divL}
\end{align}
To obtain this result, we used \eqref{switch flat limit} and $\delta \sqrt{\mathring{q}}=0$ (see also the useful relations of Appendix \ref{Useful relations}). With these definitions, we can now rewrite the divergent term in \eqref{intermediate step flat limit2} as 
\begin{equation}
\frac{3\sqrt{\mathring q}}{32\pi G \Lambda} \partial_u (N_{AB}^{TF}\delta q^{AB}) = \p_u \left( \frac{\delta L_C}{\delta q_{AB}}\delta q_{AB} \right)  = \delta L_\circ - \p_u \Theta^C_\circ. \label{divergent terms at flat limit}
\end{equation}
Note that despite the notation, $\Theta^C_\circ$ appearing in \eqref{divL} is \textit{not} the presymplectic potential associated with $L_\circ$, the latter being not defined since the sphere has no boundary. This is a pure corner term living on each section of $\mathscr I$, in particular on $\p\!\mathscr I_-$ and $\p\!\mathscr I_+$. From the datum of $\Theta^C_\circ$, we build the \emph{corner presymplectic potential} and \emph{corner presymplectic form} as
\begin{equation}
\bm{\Theta}^C_\circ = \D^2 x\,  \Theta^C_\circ, \qquad \boldsymbol \omega_\circ^C(\delta_1 q_{AB}, \delta_2 q_{AB}; q_{AB}) = \delta_1 \boldsymbol \Theta^C_\circ(\delta_2 q_{AB}; q_{AB}) - (1 \leftrightarrow 2).   \label{eq:DefCornerPS}
\end{equation}
We emphasize that they are defined on each sphere of the boundary $\mathscr I$, not only at $\partial \!\mathscr I_\pm$. The presymplectic potential $\bm{\Theta}_\circ$ of $L_\circ$ defined in \eqref{all potentials} is given by the sum $ \frac{\delta \bm L_C}{\delta q_{AB}}\delta q_{AB} + \bm{\Theta}^C_\circ$. We observe that $-\delta\bm L_\circ+\D \bm\Theta_\circ = 0$ hence there is no contribution brought to the prescription \eqref{renormalized presympelctic} for renormalizing the presymplectic potential. This is consistent with the discussion below \eqref{eq:tensor Lcirc}. Inspired by the presence of the corner terms in \eqref{divergent terms at flat limit}, we supply this prescription with
\begin{equation}\boxed{
\boldsymbol  \Theta_{ren,tot}^{\Lambda\text{-BMS}_4}\Big|_{\mathscr I} =\bm \Theta_{ren}^{\Lambda\text{-BMS}_4} \Big|_{\mathscr I} -\delta  \D \bm L_C + \D \boldsymbol \Theta^C_\circ . \label{thetarentot}}
\end{equation}
As already discussed in section \ref{sec:Renormlization of the presymplectic potential}, such a prescription fixes the two standard ambiguities that arise in the covariant phase space formalism, here for codimension 2 objects on sections of $\mathscr I$. Taking a variation on \eqref{thetarentot}, we obtain the associated presymplectic form as
\begin{equation}
\bm \omega_{ren,tot}^{\Lambda\text{-BMS}_4}\Big|_{\mathscr I} = \bm \omega_{ren}^{\Lambda\text{-BMS}_4} \Big|_{\mathscr I} + \D \boldsymbol \omega^C_\circ , \label{omegatot}
\end{equation}
where $\bm\omega_{ren}^{\Lambda\text{-BMS}_4}$ is given by \eqref{omega ren in terms of theta ren} with the $\Lambda$-BMS$_4$ gauge fixing and its pull-back to $\mathscr I$ is the leading order of \eqref{eq:omega LBMS4}. Our construction, in particular \eqref{divergent terms at flat limit}, ensures that the inclusion of the corner terms according to \eqref{thetarentot} in the renormalization process of the presymplectic potential are sufficient to remove the pole in $\Lambda^{-1}$ in \eqref{intermediate step flat limit2}: as a result, the asymptotically flat limit $\Lambda\to 0$ can now be taken safely in \eqref{omegatot}, which provides the most important quantity in the definition of the covariant phase space.

\subsubsection{Corner contributions to the charge algebra}
Before taking the flat limit, let us study how the incorporation of these corner terms impacts the $\Lambda$-BMS$_4$ charge algebra \eqref{charge algebra Lambda BMS 4}. The codimension $2$ form $\bm k_{\xi,(tot)}^{\Lambda\text{-BMS}_4}[\phi;\delta\phi]$ including the contribution of the corner presymplectic form verifies
\begin{equation}
\D \bm k_{\xi,(tot)}^{\Lambda\text{-BMS}_4}[\phi;\delta\phi]\Big|_\mathscr{I} = \bm\omega_{ren,tot}^{\Lambda\text{-BMS}_4}[\phi;\delta_\xi\phi,\delta\phi]\Big|_\mathscr{I},
\end{equation}
from which we deduce
\begin{equation}
\ndelta H_{\xi,\text{tot}}^{\Lambda\text{-BMS}_4}[\phi] = \delta H_{\xi}^{\Lambda\text{-BMS}_4}[\phi] + \Xi_{\xi}^{\Lambda\text{-BMS}_4} [\phi;\delta \phi ] + \int_{S^2_\infty} \D^2\Omega \,\, \omega_\circ^C [\delta_\xi q_{AB},\delta q_{AB}].
\end{equation}
because of \eqref{omegatot}. We dropped the dependence on $q_{AB}$ in $\omega_\circ^C$ because it involves only its variation $\delta q_{AB}$, see \eqref{divL} and  \eqref{eq:DefCornerPS}. Since the corner presymplectic structure term is not integrable, we keep our definition \eqref{I for L BMS 4} of the Hamiltonian and add this term to the non-integrable term \eqref{NI for L BMS 4}. Again, the computation of the charge algebra under the Barnich-Troessaert bracket gives
\begin{equation}
\begin{split}
&\left\lbrace H_{\xi_1}^{\Lambda\text{-BMS}_4}[\phi],H_{\xi_2}^{\Lambda\text{-BMS}_4}[\phi] \right\rbrace_{\star,(tot)} \\
&\qquad = \delta_{\xi_2} H_{\xi_1}^{\Lambda\text{-BMS}_4}[\phi] + \Xi_{\xi_2}^{\Lambda\text{-BMS}_4} [\delta_{\xi_1} \phi ; \phi] + \int_{S^2_\infty} \D^2\Omega \, \omega_\circ^C [\delta_{\xi_2} q_{AB},\delta_{\xi_1} q_{AB}] \\
&\qquad \equiv H^{\Lambda\text{-BMS}_4}_{[\xi_1,\xi_2]_{\star}}[\phi] + K^{\Lambda\text{-BMS}_4,(tot)}_{\xi_1,\xi_2}[q_{AB}]
 \label{lambda bms corner}
\end{split}
\end{equation}
as a corollary of \eqref{charge algebra Lambda BMS 4}. The additional term
\begin{equation}
K^{\Lambda\text{-BMS}_4,(tot)}_{\xi_1,\xi_2}[q_{AB}] = -\int_{S^2_\infty} \D^2\Omega \, \omega_\circ^C [\delta_{\xi_1} q_{AB},\delta_{\xi_2} q_{AB}]
\label{K lambda BMS 4 corner}
\end{equation}
is obviously antisymmetric under the exchange $\xi_1\leftrightarrow\xi_2$ and satisfies the cocycle condition \eqref{cocycle condition theory}. Hence, the corner terms naturally give rise to a field-dependent 2-cocycle in the right-hand side of the $\Lambda$-BMS$_4$ charge algebra.


\subsubsection{Symplectic structure at flat limit and discussion}

Finally, let us consider the flat limit of the symplectic structure. Taking $\Lambda\to 0$, the equations \eqref{thetarentot} and \eqref{omegatot} give respectively 
\begin{equation}
\lim_{\Lambda\to 0}\boldsymbol \Theta_{ren,tot}^{\Lambda\text{-BMS}_4}[\phi;\delta\phi] \Big|_{\mathscr I} = \frac{\D u\, \D^2\Omega }{16\pi G}  \left[  \frac{1}{2}\Big( N^{AB}_{TF} + \frac{1}{2} R[q]q^{AB}\Big) \delta C_{AB} - D_{(A}\mathcal U_{B)}\delta q^{AB}\right]   
\end{equation}
and more importantly
\begin{equation}
\boxed{
\begin{aligned}
&\lim_{\Lambda\to 0}\bm \omega_{ren,tot}^{\Lambda\text{-BMS}_4}[\phi;\delta_1 \phi,\delta_2 \phi]\Big|_{\mathscr I}   \\
&\qquad = \frac{\D u \, \D^2\Omega}{16\pi G} \left[ \frac{1}{2}\delta_1 \left(N^{AB} + \frac{1}{2}R[q]q^{AB}\right) \wedge \delta_2 C_{AB} - \delta_1 \left( D_{(A}\mathcal U_{B)} \right) \wedge \delta_2 q^{AB}\right].
\end{aligned}
}
\label{presymplectic current flat}
\end{equation} 
The presymplectic current \eqref{presymplectic current flat} is the Generalized BMS$_4$ presymplectic structure after pullback on the future null infinity $\mathscr I = \mathscr{I}^+$ in asymptotically locally flat spacetimes. This expression exactly matches with the presymplectic structure obtained by regularization methods for asymptotically locally flat spacetimes as an example of covariant phase space methods in section \ref{sec:Asymptotically locally flat radiative phase spaces}. More precisely, \eqref{presymplectic current flat} and \eqref{omega flat complete} are stricly identical, which is a highly non-trivial agreement! Recall that the regularization procedure used in sections \ref{sec:Flat variational principle} and \ref{sec:Symplectic structure and charges FLAT} \cite{Compere:2018ylh} was not explicitly covariant \cite{Flanagan:2019vbl}: the corner counterterms \eqref{Yur}-\eqref{YrA} were indeed given in a coordinate dependent way ($r$ is clearly apparent) and we did not give a proof that they were part of a covariant Iyer-Wald counterterm $\bm Y$ as we introduced to renormalize the presymplectic potential. But since the procedure used here is explicitly covariant (in terms of the boundary structure for $\Lambda\neq 0$) and the flat limit process is consistent either at the level of the solution space, the symmetries or the phase space, this therefore justifies \textit{a posteriori} the counterterm prescription for subtracting a radial divergence used in \cite{Compere:2018ylh} and reviewed in chapter \ref{chapter:Charges}. We find curious that the radially diverging term encountered in the flat case, namely \eqref{Yur}, is structurally similar to the $\Lambda \to 0$ diverging term found here \eqref{divL}. Here, using the prescription \eqref{thetarentot}, we are able to track the origin of this term coming from a corner Lagrangian \eqref{cornerL}. The latter is a kinetic action for the boundary metric of Al(A)dS$_4$ spacetimes. Let us conclude by making a few comments.
\begin{itemize}[label=$\rhd$]
\item Even though the flat limit of the action only requires to consider the corner action $\int_{\p \mathscr I_+} L_C$ $-$ $\int_{\p \mathscr I_-} L_C$ at $u=u^\pm$, the total presymplectic potential \eqref{thetarentot} and current \eqref{omegatot} are defined from the corner action at any $u$ along $\mathscr I$, which also shifts the surface charges at any $u$. In particular, this brings a non-zero field-dependent 2-cocycle \eqref{K lambda BMS 4 corner} in the surface charge algebra. The link between this modified $\Lambda$-BMS$_4$ charge algebra \eqref{lambda bms corner} that takes the presence of corner terms into account and the Generalized BMS$_4$ charge algebra discussed in section \ref{sec:The Generalized BMS$_4$ charge algebra} is not direct. Indeed, due to the subleading field-dependence of the diffeomorphism between SFG and Bondi gauges, the asymptotic Killing vectors do not transform as simple vectors (see \textit{e.g.} Eq. (70) of \cite{Compere:2016hzt}). This implies that the surface charge codimension 2 form transforms non-trivially, which leads to a shift of the objects appearing in the charge algebra that is hard to track. For example, the 2-cocycle in \eqref{lambda bms corner} does not admit a well-defined flat limit and is therefore not directly related to the 2-cocycle \eqref{cocycle flat} obtained in the flat case. It is simpler to take the flat limit at the level of the symplectic structure that determines all dynamical quantities rather than at the level of the charge algebra. Since the flat limit of the renormalized $\Lambda$-BMS$_4$ presymplectic structure is the Generalized BMS$_4$ presymplectic structure, this proves indirectly that the $\Lambda$-BMS$_4$ charges (and charge algebra) are flattened into the Generalized BMS$_4$ charges (and charge algebra) even though we do not present the intricate computation making the link explicit. But since all dynamical quantities are derived from the presymplectic structure, understanding the flat limit at the level of it is necessary and sufficient to consider the flat limit of the phase space.

\item The next observation has already been mentioned but will be rephrased here for summarizing perhaps the most important result of this section. The Compère-Marolf prescription \eqref{renormalized presympelctic} defined in \cite{Compere:2008us} fixes the usual Iyer-Wald ambiguities in the definition of the presymplectic potential \cite{Lee:1990nz, Wald:1993nt, Iyer:1994ys} using as an input the boundary counterterms defined at $\mathscr I$ (see also \cite{Papadimitriou:2005ii}). Such a prescription fails to give a renormalized symplectic structure in the flat limit because there subsists a pole in $\Lambda^{-1}$ in the presymplectic potential which can only be subtracted by a corner counterterm living on codimension 2 sections of $\mathscr I$. We argued that the existence of a corner Lagrangian defined for each sphere on $\mathscr I$ naturally leads to the additional prescription \eqref{thetarentot} which gives a well-defined symplectic structure in the flat limit. 

\item Finally, we already observed \textit{e.g.} in section \ref{sec:Bondi gauge parameters and constraints} that we need the $\Lambda$-BMS$_4$ algebroid to recover the (Generalized) BMS$_4$ group at flat limit. These symmetries are related to leaky boundary conditions putting some non-vanishing symplectic flux through the conformal boundary $\mathscr I$, while conservative boundary conditions with non-trivial asymptotic group such as Dirichlet boundary conditions $\delta q_{AB} = 0$ only give a finite-dimensional asymptotic group in the flat limit, the exact Poincaré group (see \eqref{fig:diagram}). Here we have seen that the leaks through $\mathscr I$ allowed by the $\Lambda$-BMS$_4$ boundary conditions are essential to recover the presymplectic form \eqref{presymplectic current flat} which must be non-zero even when considering strict Minkowskian asymptotics with $\delta q_{AB}=0$. The particular structure of the conformal boundary of (A)dS$_4$ implies that one cannot extend the finite symmetry groups $SO(3,2)$ or $SO(4,1)$ when $\Lambda\neq 0$ with some avatar of ``supertranslations'' without gaining ``superrotations'' for free, which is the take-away message from the solution of \eqref{eq1a}-\eqref{eq2a}. As a result, the flat limit of the $\Lambda$-BMS$_4$ phase space does naturally include the super-Lorentz transformations introduced in section \ref{sec:Asymptotic symmetries and the BMS4 group} in a natural way beside the traditional supertranslations. This is another striking evidence that ``superrotations'' are as legitimate asymptotic symmetries than the supertranslations in the flat case and can be interestingly included in the asymptotically flat phase space!

\end{itemize}

This last observation puts an end to our fruitful journey at the boundary of (A)dS spacetimes in Einstein's gravity. Besides the results intrinsically interesting for radiative solutions with cosmological constant, we have acquired a fundamental and highly non-trivial theoretical confirmation of our heuristic renormalization in the asymptotically locally flat phase space about which it remains a lot to be said. The next chapter is thus an invitation to go back to the flat land to pursue the analysis started in the chapter \ref{chapter:Charges} as a motivating example for the covariant phase space formalism.\hfill{\color{black!40}$\blacksquare$}

%
%

\chapter{Super-Lorentz transformations and Gravitational Memories}
\label{chapter:Flat}

After our captivating peregrinations in Al(A)dS spacetimes concluded by a controlled landing in the form of a flat limit process, Chapter \ref{chapter:Flat} recovers the well-known asymptotically flat landscapes about which several points require clarification. Section \ref{sec:Vacua} derives a closed-form expression of the orbit of Minkowski spacetime under Generalized BMS$_4$ transformations, \textit{i.e.} arbitrary Diff$(S^2)$ super-Lorentz transformations and smooth supertranslations. The resulting metrics are labelled by the superboost, superrotation and supertranslation fields: they determine the vacuum structure of asymptotically locally flat spacetimes. Section \ref{sec:Gravitational memory effects} then discusses the associated gravitational memory effects and details physical processes that can lead to transitions among the above-mentioned vacua. Impulsive vacuum transitions driven by overleading cosmic events are related to the refraction memory effect and the usual displacement memory effect, which also receives a detailed review in this chapter. Finally, in section \ref{sec:Generalized BMS4 finite charges}, we conclude the discussion about the radiative phase spaces admitting the whole Generalized BMS$_4$ group as the asymptotic symmetry group that we initiated in section \ref{sec:Asymptotically locally flat radiative phase spaces} by providing a set of meaningful physical prescriptions to isolate finite Hamiltonian canonically conjugated to these symmetries. We show that our final surface charge expressions are consistent with the leading and subleading soft graviton theorems and we contrast the leading infrared triangle structure to a new mixed overleading/subleading square structure.  

This chapter mixes the parts of \cite{Compere:2018ylh} that were not presented in chapter \ref{chapter:Charges} with refined results given in \cite{Compere:2020lrt} concerning the Generalized BMS$_4$ charge algebra as well as the suitable choice for the finite Hamiltonians.

\section{Vacua of the gravitational field}
\label{sec:Vacua}

Let $g_{\mu\nu}$ be a state of the gravitational field that belongs to the solution space \eqref{S0ring} in the Bondi gauge. It is said to be a \textit{vacuum state} if it is diffeomorphic to the Minkowski spacetime
\begin{equation}
\D s^2 = -\D u^2 + 2\D u\D r + r^2 \mathring q_{AB}\D x^A\D x^B \label{Mink4 in bondi}
\end{equation}
that, written in retarded coordinates, is trivially part of \eqref{S0ring}. This metric is Riemann-flat $R_{\mu\nu\alpha\beta} = 0$ by construction and the Weyl tensor is also vanishing, $W_{\mu\nu\alpha\beta} = 0$, which is expected from a solution without gravitation. The statement of vanishing Weyl is a tensorial proposition that is untouched by the action of the allowed residual gauge diffeomorphisms preserving \eqref{S0ring}, which constitutes the orbit of the Generalized BMS$_4$ group for which \eqref{sqrt q fixed in Bondi} is obeyed with $\sqrt{\bar q}=\sqrt{\mathring q}$. This is the only orbit that contains \eqref{Mink4 in bondi} as particular solution for the diffeomorphism generators $f=0$ and $Y^A=0$. The whole Generalized BMS$_4$ orbit of vacua is obtained in practice by exponentiating a general BMS$_4$ transformation (generated by vectors \eqref{eq:xir} parametrized by $f$, $Y^A$, $\omega=0$, $\sqrt{q}=\sqrt{\mathring q}$) acting on the Minkowski spacetime \eqref{Mink4 in bondi}. The subset of this orbit for which only supertranslations are turned on contains the non-equivalent vacua of asymptotically flat spacetimes, already covered in section \ref{sec:Asymptotic symmetries and the BMS4 group}: they are labeled by a single fundamental field labeling inequivalent vacua: the supertranslation field $C(x^A)$ defined in \eqref{CAB parametrized}-\eqref{C field variation}. Here we derive the orbit of Minkowski under supertranslations as well as super-Lorentz transformations that we exponentiate following the method developed in \cite{Compere:2016jwb}. This leads to the appearance of new boundary fields beside of $C$ that are coined as the super-Lorentz fields. 

\subsection{Generation of the vacua}
\label{Generation of the vacua}

The exponentiation leading to finite Generalized BMS$_4$ transformations simplifies considerably if we start from the \textit{complex plane coordinates} $(u_C,r_C,z_C,\bar z_C)$ in which the flat metric reads as 
\begin{equation}
\D s^2 = -2\text{d}u_C \text{d}r_C + 2 r_C^2 \D z_C \D\bar{z}_C.  \label{MinkowskiCC}
\end{equation}
One obtains this line element from \eqref{Mink4 in bondi} by the following change of coordinates \cite{Compere:2016jwb}
\begin{equation}
\begin{split}
r_C &= \frac{\sqrt{2}r}{1+z\bar z}+\frac{1}{\sqrt{2}}u \, , \quad u_C = \frac{1+z\bar z}{\sqrt{2}}u - \frac{z\bar z}{2 r_C}u^2\, , \quad z_C = z - \frac{zu}{\sqrt{2}r_C}u \, , \quad \bar z_C = z_C^*,
\end{split}
\end{equation}
where $(z,\bar z)$ are the stereographic coordinates on the celestial sphere. We define the background structures
\begin{equation}
\gamma_{\underline{A}\underline{B}} = \left[ \begin{array}{cc}
0 & 1 \\ 
1 & 0
\end{array} \right], \quad \epsilon_{\underline{A}\underline{B}} = \left[ \begin{array}{cc}
0 & 1 \\ 
-1 & 0
\end{array} \right]
\end{equation}
with inverse $\gamma^{\underline{A}\underline{B}} = \gamma_{\underline{A}\underline{B}}$, $\epsilon^{\underline{A}\underline{B}} = \epsilon_{\underline{A}\underline{B}}$. In the following, all underlined indices are related to the complex plane metric $\gamma_{\underline{A}\underline{B}}$ on the codimension 2 surfaces of constant $u_C,r_C$. The goal is to introduce a diffeomorphism to Bondi gauge $(u_C,r_C,z_C,\bar{z}_C)\to (u,r,z,\bar{z})$ that exponentiates $\text{Diff}(S^2)$ super-Lorentz transformations and supertranslations. Requiring that $(u,r,z,\bar z)$ are Bondi coordinates leads to the conditions \eqref{Bondi gauge conditions}. The algebraic conditions $g_{rr} = 0$, $g_{rA} = 0$ are solved if
\begin{align}
r_C &= r_C (r,u,z,\bar z), \label{diff bondi 1} \\
u_C &= W(u,z,\bar{z}) - \frac{1}{r_C} \gamma_{\underline{A}\underline{B}} H^{\underline{A}} (u,z,\bar z) H^{\underline{B}} (u,z,\bar z), \label{diff bondi 2} \\
z_C^{\underline{A}} &= G^{\underline{A}} (z,\bar z) - \frac{1}{r_C} H^{\underline{A}} (u,z,\bar z), \quad
H^{\underline{A}} (u,z,\bar z) = - \frac{1}{D_G} \epsilon^{\underline{A}\underline{B}} \gamma_{\underline{B}\underline{C}}\epsilon^{AB} \partial_A W \partial_B G^{\underline{C}} \label{diff bondi 3}
\end{align}
where $D_G = \det (\partial_A G^{\underline{B}}) = \frac{1}{2!}\epsilon_{\underline{A}\underline{B}}\epsilon^{AB}\partial_A G^{\underline{A}} \partial_B G^{\underline{B}}$. The determinant condition $\partial_r(r^{-4}\det g_{AB})=0$ fixes the functional dependence of $r_C$ as \cite{Compere:2016jwb}
\begin{align}
r_C (r,u,z,\bar z) &= R_0 (u,z,\bar z) + \sqrt{\frac{r^2}{(\partial_u W)^2} + R_1 (u,z,\bar z)}.\label{rc}
\end{align}
The function $r_C$ admits a well-defined series in inverse powers of $r$:
\begin{equation}
r_C(r,u,z^c) = \frac{r}{(\partial_u W)^2} + R_0 (u,z,\bar z) + \frac{1}{2}(\partial_u W)^2 R_1 (u,z,\bar z)\frac{1}{r} + \mathcal O(r^{-3}). \label{rc power}
\end{equation}
The expressions of the unknown codimension 1 functions $R_0$ and $R_1$ are respectively obtained by requiring that the determinant condition is obeyed up to second and third order in $1/r$. The information is propagated at subleading orders with the power expansion \eqref{rc power} of the functional dependence in $r$. 

Acting with the change of coordinates \eqref{diff bondi 1}-\eqref{diff bondi 3} on \eqref{MinkowskiCC} gives the Minkowski metric in the Bondi gauge up to a finite Generalized BMS$_4$ transformation. Expanding the resulting $g_{AB}$ in powers of $r$ as in \eqref{polynomial gAB}, we can read the boundary metric $q_{AB}$ in the $\mathcal O(r^2)$ contribution. The boundary conditions \eqref{preliminary BC Bondi} and \eqref{sqrt q round sphere} leading to the Generalized BMS$_4$ solution space \eqref{S0ring} require that $\partial_u q_{AB} = 0$ and $g_{uu}$ is finite in $r$ by virtue of \eqref{pu qAB = 0} and \eqref{Vr flat with sphere det}. This yields $\partial^2_u W = 0$, so $W$ is at most linear in $u$. Moreover, regularity of the finite diffeomorphism \eqref{rc} implies that $\partial_u W$ is nowhere vanishing. Therefore,
\begin{equation}
W(u,z^c) =  \exp \left[\frac{1}{2}\Phi (z,\bar z)\right] (u + C(z,\bar z)).
\label{Wform}
\end{equation}
and the boundary metric reads as
\begin{equation}
q_{AB} =q_{AB}^{vac} \equiv e^{-\Phi} \partial_A G^{\underline{A}} \partial_B G^{\underline{B}} \gamma_{\underline{A}\underline{B}}.
\label{qABform}
\end{equation}
It is indeed the result of a large diffeomorphism and a Weyl transformation. The latter is obviously constrained since $\det q_{AB} = \mathring q$ or $\Phi = \ln D_G-\ln\sqrt{\mathring q}$ using \eqref{qABform}. The asymptotic shear $C_{AB}$ appearing at $\mathcal O(r)$ in $g_{AB}$ is found to be the trace-free part ($TF$) of the following tensor
\begin{equation}
C_{AB} = C_{AB}^{vac} \equiv \left[ \frac{2}{(\partial_u W)^2} \partial_u \left( D_A W D_B W \right) - \frac{2}{\partial_u W} D_A D_B W \right]^{TF}.
\label{CABGeneral}
\end{equation}
Introducing \eqref{Wform}, it comes 
\begin{equation}
C^{vac}_{AB}[\Phi,C] =  (u +C) N^{vac}_{AB} + C^{(0)}_{AB}, \quad \left\lbrace \begin{array}{ccl}
N^{vac}_{AB} & = & \left[\frac{1}{2}D_A \Phi D_B \Phi - D_A D_B \Phi  \right]^{TF} \label{PhiL} \\ 
C^{(0)}_{AB} & = &   - 2 D_A D_B C + q_{AB} D^2 C.
\end{array} \right.
\end{equation}
We find that all explicit reference on $\gamma_{\underline{A}\underline{B}}$ or $G^{\underline{A}}$ disappeared. The tensor $C_{AB}^{(0)}$ is the standard asymptotic shear for the BMS$_4$ vacua. Moreover, the news tensor of the vacua $N^{vac}_{AB}$ is only built up with $\Phi$. It can be checked that the boundary Ricci scalar is given in terms of $\Phi$ as 
\begin{equation}
R[q] = D^2 \Phi,
\label{Liouville}
\end{equation}
which implies 
\begin{equation}
D_A N_{vac}^{AB}=-\frac{1}{2}D^B R[q]. \label{Li2}
\end{equation} 
We can therefore add a trace to $N^{vac}_{AB}$ to form the conserved stress-tensor
\begin{equation}
T_{AB}[\Phi]= \frac{1}{2}D_A \Phi D_B \Phi - D_A D_B \Phi + \frac{1}{2}q_{AB} \left( 2 D^2 \Phi  -\frac{1}{2} D^C \Phi D_C \Phi \right). \label{TAB Liouville}
\end{equation}
Its trace is equal to $D^2 \Phi$. The tensor $T_{AB}$ is precisely the stress-tensor of Euclidean Liouville theory \cite{DHoker:1982wmk,Jackiw:1982hg,Seiberg:1990eb,Teschner:2001rv,Deser:1983nh}
\begin{equation}
L[\Phi ; q_{AB}] = \sqrt{q}\left( \frac{1}{2} D^A \Phi D_A \Phi +\alpha e^\Phi + R[q] \Phi  \right)  \label{var principle Liouville}
\end{equation}
for the field $\Phi$, where the parameter $\alpha$ is zero in order to satisfy \eqref{Liouville}. Note that in order to derive the stress-tensor \eqref{TAB Liouville} from the Lagrangian \eqref{var principle Liouville}, one needs to consider the Liouville field $\Phi$ off-shell but fix the metric $q_{AB}$ as a background field. The equation of motion for $\Phi$ derived from the variational principle \eqref{var principle Liouville} on a fixed background field $q_{AB}$ is \eqref{Liouville}, coined as the \textit{Liouville equation}. Under a super-Lorentz transformation 
\begin{equation}
\delta_Y (D^2 \Phi -  R[q]) = (\mathcal L_Y + D_A Y^A)  (D^2 \Phi - R[q]).\label{actionRL}
\end{equation} 
Therefore, imposing the Liouville equation is consistent with the action of super-Lorentz transformations and we can demand that it holds for any metric $q_{AB}$ obtained by super-Lorentz transformation from the sphere metric $\mathring q_{AB}$. Note also that aside from its divergence, see \eqref{Li2}, the curl of $N_{AB}^{vac}$ is also fixed. Indeed, one can show that 
\begin{equation}
D_{[A}N^{vac}_{B]C} = -\frac{1}{2}D_{[A}R[q] \, q_{B]C} \label{curl NABvac}
\end{equation}
is equivalent to \eqref{Liouville}, see \textit{e.g.} \cite{Campiglia:2020qvc}. In \cite{1977asst.conf....1G}, Geroch showed that there exists a unique curl-free tensor $\rho_{AB}$ such that $q^{AB}\rho_{AB}=R[q]$. The condition \eqref{curl NABvac} implies that $\rho_{AB} = \frac{1}{2}R[q]q_{AB} + N_{AB}^{vac}$, hence $N_{AB}^{vac}$ is uniquely defined when the super-Lorentz frame $q_{AB}$ is fixed. In particular, when $q_{AB}=\mathring q_{AB}$, $N_{AB}^{vac}$ is identically vanishing. A second thing that has to be mentioned about $N_{AB}^{vac}$ and particularly about $\Phi$, is that the latter can always be modified by a conformal isometry $Y^A$ of $q_{AB}$ without changing the vacuum news tensor. Indeed, using $\mathcal L_Y q_{AB} = D_C Y^C q_{AB}$, it can be shown that $\Phi$ and $\Phi+D_A Y^A$ lead to the same $N_{AB}^{vac}$ because the differential operator in \eqref{PhiL} is blind to such a shift in $\Phi$. This ambiguity is irrelevant once the reference metric $\gamma_{\underline{A}\underline{B}}$ is fixed in \eqref{qABform}. Moreover, $\Phi$ never appears alone in the following but only through the differential operator defining $N_{AB}^{vac}[\Phi]$. This expresses the fact that the Minkowski vacuum \eqref{Mink4 in bondi} is preserved by the Lorentz transformations and $N_{AB}^{vac}$ is only turned up by pure superboost transformations.

Using the boundary metric \eqref{qABform} and shear \eqref{CABGeneral}, one can work out the covariant expressions for $R_0$ and $R_1$ in \eqref{rc}. They are given by
\begin{equation}
R_0 = \frac{1}{2} e^{-\Phi}D^AD_A W \qquad \text{and} \qquad R_1 = \frac{1}{8} e^{-\Phi} C_{AB} C^{AB}.
\end{equation}
Finally, after some algebra, one can write the full metric as
\begin{equation}
\boxed{
\D s^2_{vac} = - \frac{R[q]}{2}\text{d}u^2 - 2 \text{d}\rho \text{d}u + (\rho^2 q_{AB} + \rho C^{vac}_{AB} + \frac{1}{8}C_{vac}^2 q_{AB})\text{d}x^A \text{d}x^B + D^B C^{vac}_{AB} \text{d}x^A \text{d}u
}\label{metf}
\end{equation}
where $\rho = \sqrt{r^2+ \frac{1}{8}C_{vac}^2}$, $C_{vac}^2\equiv C_{AB}^{vac}C^{AB}_{vac}$, is a derived quantity in terms of the Bondi radius $r$ which is nothing but the radial Newman-Unti coordinate (see section \ref{sec:Gauge fixing conditions} for a definition). We observe that the metric of the vacua is thus more natural in the Newman-Unti gauge $(u,\rho,z^A)$ where $g_{\rho \mu} = -\delta_\mu^u$. 

Let us also comment on the meromorphic extension of the Lorentz group. When super-Lorentz transformations reduce to local conformal Killing vectors on $S^2$ \textit{i.e.} $G^z = G(z)$ and $G^{\bar z} \equiv \bar{G}(\bar z) $, the boundary metric after a diffeomorphism is the unit round metric on the sphere 
\begin{eqnarray}
\mathring{q}_{AB} \text{d}z^A \text{d}z^B = 2 \mathring{q} \text{d}z \text{d}\bar{z},\qquad \mathring{q} = \frac{2}{(1+z\bar z)^2}
\end{eqnarray} 
(and $R[q] = \mathring R = 2$) \emph{except at the singular points of }$G(z)$. The Liouville field reduces to the sum of a meromorphic and an anti-meromorphic part minus the unit sphere factor 
\begin{equation}
\Phi = \phi(z)+\bar \phi(\bar z)- \ln \sqrt{\mathring{q}}.\label{Phim}
\end{equation}
The metric \eqref{metf} then exactly reproduces the expression of \cite{Compere:2016jwb} with the substitution $T^{\text{\cite{Compere:2016jwb}}}_{AB} = \frac{1}{2} N^{vac}_{AB}$. We have therefore found the generalization of the metric of the vacua for arbitrary $\text{Diff}(S^2)$ super-Lorentz transformations together with arbitrary supertranslations.

\subsection{The superboost, superrotation and supertranslation fields}
\label{sec:The superboost, superrotation and supertranslation fields}

A general vacuum metric is parametrized by a boundary metric $q^{vac}_{AB}$, the field $C$ that we call the \emph{supertranslation field} and $\Phi$ that we will call either the \emph{Liouville field} or the \emph{superboost field}. Under a BMS transformation, the bulk metric transforms into itself, with the following transformation law of its boundary fields,
\begin{align}
\delta_{T,Y}q^{vac}_{AB} &= D_A Y_B + D_B Y_A - q^{vac}_{AB} D_C Y^C, \\
\delta_{T,Y}\Phi &= Y^A \p_A \Phi + D_A Y^A,\\
\delta_{T,Y}C &= T + Y^A \p_A C - \frac{1}{2}C D_A Y^A. \label{deltaC}
\end{align}
Only the divergence of a general super-Lorentz transformation sources the Liouville field. Since rotations are divergence-free but boosts are not, we call $\Phi$ the \emph{superboost field}. In general, one can decompose a vector on the 2-sphere as a divergence and a curl part. For a generic superrotation, there should be a field that is sourced by the curl of $Y^A$. We call this field the \emph{superrotation field} $\widetilde\Phi$ and we postulate its transformation law
\begin{eqnarray}
\delta_{T,Y}\widetilde\Phi &=& Y^A \p_A \widetilde\Phi + \eps^{AB} D_A Y_B.
\end{eqnarray}
Where is that field in \eqref{metf}? In fact, the boundary metric $q^{vac}_{AB}$ is not a fundamental field. It depends upon the Liouville field $\Phi$ and the background metric $\gamma_{\underline A\underline B}$. Since it transforms under superrotations, the metric \eqref{qABform} should also depend upon the superrotation field $\widetilde\Phi$. The explicit form $q^{vac}_{AB}[\gamma_{\underline A\underline B},\Phi,\widetilde\Phi]$ is not known to us and will not be needed in the following. We will call the set of boundary fields $(\Phi,\widetilde\Phi)$ the super-Lorentz fields.

Under a Generalized BMS$_4$ transformation, the news of the vacua $N^{vac}_{AB}$ and the tensor $C^{(0)}_{AB}$ transform inhomogenously as 
\begin{align}
\label{dRN}
\delta_{T,Y} N^{vac}_{AB} &= \mathcal L_Y N^{vac}_{AB} -D_A D_B D_C Y^C + \frac{1}{2}q_{AB}D_CD^C D_D Y^D,\\\delta_{T,Y}C^{(0)}_{AB}  &=  \mathcal L_Y C^{(0)}_{AB} - \frac{1}{2}D_C Y^C C^{(0)}_{AB}    -2 D_A D_B T + q_{AB}D_CD^C T.\label{baret}
\end{align}
From \eqref{metf}, one can read off the explicit expressions of the Bondi mass and angular momentum aspects of the vacua
\begin{equation}
\begin{split}
M &= -\frac{1}{8} N^{vac}_{AB} C_{vac}^{AB} ,\\
N_{A} &= -\frac{3}{32} D_A (C^{vac}_{BC}C_{vac}^{BC}) - \frac{1}{4} C^{vac}_{AB} D_C C_{vac}^{BC}. 
\end{split}\label{valvac}
\end{equation}
The Bondi mass is time-dependent and its spectrum is not bounded from below because $\p_u M = -\frac{1}{8}N_{AB}^{vac}N^{AB}_{vac}$ as observed in \cite{Compere:2016jwb}. Yet, the Weyl tensor is identically zero and the standard Newtonian potential vanishes. This indicates that the physical mass should be identically zero, or alternatively, that the Bondi mass aspect should be refined in some sense, provided that $M$ is also diverging in $u$ at spatial infinity $u\to -\infty$. For asymptotically Minkowskian spacetimes, the integral of $M$ on the asymptotic sphere $u \to -\infty$ is identified with the total (ADM) mass of the spacetime computed from spatial infinity. Keeping $M$ as the faithful mass aspect would yield an divergent value for the total mass of spacetime even without bulk sources, which is not suitable. This discussion is overlapping the question of the finiteness of the charges as well as the behavior in $u$ of the various Bondi fields around the corners $u\to -\infty$ (spatial infinity) and $u\to +\infty$ (future timelike infinity) that we will address in section \ref{sec:Initial and late data in general super-Lorentz frames}.

\section{Gravitational memory effects}
\label{sec:Gravitational memory effects}

Like any gauge theory, Einstein's gravity possesses an infinite-dimensional class of degenerate though unequivalent vacua \cite{Strominger:2013jfa}. They are represented by the action of finite residual gauge diffeomorphisms on the global Minkowski vacuum, in any kind of gauge fixing deleting the kinematical redundancies in order to single out the only dynamic degrees of freedom in the gauge field (here, the metric tensor). In this point of view, the vacua derived in the previous section are simply stationary configurations without zero modes such as mass, linear or angular momenta. For localized radiative phenomena which occur only in a bounded range of (retarded) time $u_i < u < u_f$, one can expect that the gravitational field, coming from some vacuum configuration for $u<u_i$ relaxes back to a vacuum state for $u>u_f$ when the source for radiation has been turned off (modulo some zero modes like mass or angular momentum). For generic non-stationary phases, there is no reason for the early and late vacua to be equivalent: the radiative process is responsible for a \textit{vacuum transition}. The existence of non-equivalent vacua related by diffeomorphism indicates that the final vacuum is not arbitrary and should be related by a residual gauge transformation to the original vacuum. Incidentally, this gives one supplementary motivation to study infinite-dimensional extensions of the natural asymptotic group which would be, in the case of interest, the Poincaré group. The observation of the vacuum transition between ``early'' and ``late'' times, \textit{i.e.} $u<u_i$ and $u>u_f$, as a permanent relic of the radiative phenomenon occured in the time interval $u_i < u < u_f$, is possible thanks to (at least conceptually) simple experimental devices that we describe below. This is called a \textit{gravitational memory effect} \cite{Christodoulou:1991cr,Thorne:1992sdb,Zeldovich:1974gvh,Blanchet:1987wq,Blanchet:1992br}. The concern of the present section is to derive the memory effects associated with Generalized and Extended BMS$_4$ symmetries and identify their possible sources in the evolution equations governing the transitions.

\subsection{Axiomatic definition and relation to gauge symmetries}
\label{sec:Axiomatic definition and relation to gauge symmetries}
Before considering concrete examples, we want to give an axiomatic definition of a gravitational memory effect that would include all of the particular cases discussed later in this section and also other known effects in the literature. This definition is taken from the inspirational essay \cite{Compere:2019odm}.

Let us consider a class of asymptotically locally flat spacetimes for which $\mathscr I^+$ exists, assorted with a class of observables $O(\theta,\phi)$ defined on $\mathscr I^+$ and required to be diffeomorphism-invariant. $O(\theta,\phi)$ arises from the integration on the whole boundary $\mathscr I^+$ of some functional of gauge-invariant fields living in the asymptotic region (a canonical example is the Weyl tensor for pure gravity). The boundaries of $\mathscr I^+$ are $\mathscr I^+_+$ in the future (future timelike infinity) and $\mathscr I^+_-$ (spatial infinity). One says that $O$ is a \textit{memory observable} if and only if it can be written as the difference between a local field $\varphi_+$ defined at $\mathscr I^+_+$ and a local field $\varphi_-$ at $\mathscr I^+_-$, \textit{i.e.} $O = \varphi_+ -\varphi_-$, in some coordinate system. The fields $\varphi_+$ and $\varphi_-$ might not be defined locally in any gauge but there must exists a particular gauge where this is the case. In the (possibly unique) gauge $(x^\mu)$ where these fields are local functions of the boundary coordinates, they are coined as the \textit{memory fields}. 

As mentioned in the introduction of the section, we are interested in dynamical processes that occur in a finite time lapse. Hence, we restrict our analysis to the class of spacetime manifolds relaxing back to some stationary configurations at ``early time'' (\textit{i.e.} around $\mathscr I^+_-$) and at ``late time'' (\textit{i.e.} around $\mathscr I^+_+$). These stationary configurations are vacua possibly endowed with zero modes like constant mass or angular momentum. At early and late times, the diffeomorphism-invariant quantities are expected to be zero in the absence of radiation, but the memory fields $\varphi_\pm$ can generically be non-zero. The gauge-invariant fields (keep the example of the Weyl tensor in mind) decay to zero in the stationary zones around $\mathscr I^+_-$ and $\mathscr I^+_+$. Therefore, there must exist a residual gauge transformation $\xi$ to shift the memory fields back to zero for early and late times. Let us give an illustrative example: let us assume for instance that the spacetime asymptotes to Minkowski at late times. Hence there exists at least one coordinate system around $\mathscr I^+_-$ in which the memory field $\varphi_+$, absent by design in Minkowski, is identically vanishing. So the stationary line element at late time is diffeomorphic to Minkowski, and the corresponding diffeomorphism verifies $\delta_\xi \varphi_+ = 0$. The same reasoning can be applied in the presence of stationary zero modes such as constant mass or angular momentum (more relevant at early time to allow \textit{i.e.} for total (ADM) charges). 

Recalling now the gauge invariance of $O$, \textit{i.e.} $\delta_\xi O = 0$, we have $\delta_\xi \varphi_+ = \delta_\xi \varphi_-$ and $\xi$ shifts the memory fields identically. This is not a surprise since residual gauge parameters surviving simultaneously the gauge fixing and the imposition of some boundary conditions are codimension 2 fields that depend only upon the angles and no longer on time. For example, the solution space $\mathring{\mathcal S}_0$ of Einstein's gravity in the Bondi gauge with asymptotically locally boundary conditions admits the Generalized BMS$_4$ transformations as residual diffeomorphisms parametrized by $T(\theta,\phi)$ and $Y^A(\theta,\phi)$. A Generalied BMS$_4$ transformation thus acts globally on $\mathscr I^+$, regardless of the local value of the retarded time $u$. More generally, once the gauge has been fixed, we have the following result: 
\begin{center}
	\begin{minipage}[c]{0.75\textwidth}
	\vspace{5pt}
		\begin{center}
			\textit{The set of memory fields is in one-to-one correspondence with the codimension 2 parameters of the set of residual gauge diffeomorphisms}.
		\end{center}
	\vspace{5pt}
	\end{minipage}
\end{center}
 In the case under interest, we expect to observe a memory effect associated with supertranslations, encoded in the supertranslation field $C$ and a memory effect associated with super-Lorentz transformations, encoded in the super-Lorentz fields. We shall start by illustrating the axiomatic definition we have just reviewed for the displacement memory effect associated with the supertranslation symmetry, before focusing on the superboost field for which we have derived the explicit dependency in the gravitational field.

\subsection{Linear displacement memory effect}
\label{sec:Linear displacement memory effect}
As a warm-up, let us briefly review the first historical prediction of a gravitational memory effect that is the linear displacement memory effect \cite{Zeldovich:1974gvh,Thorne:1992sdb,Christodoulou:1991cr,Blanchet:1987wq,Blanchet:1992br} related to the supertranslation symmetry. In this subsection, we focus on the subspace $\mathring{\mathcal S}_0^{\text{Mink}}$ of asymptotically Minkowskian solutions and the assorted global BMS$_4$ symmetry group.

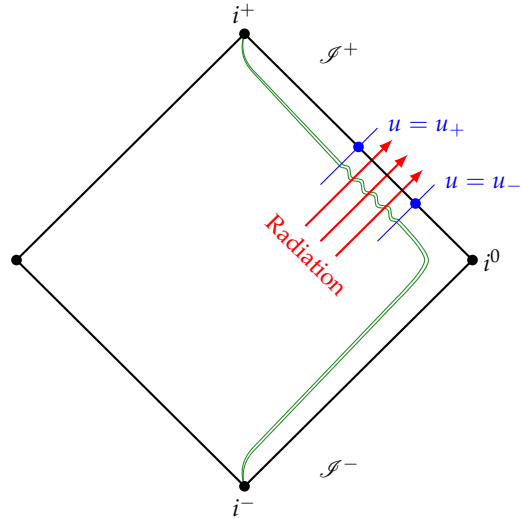
\begin{figure}[!ht]
    \begin{center}
        \begin{tikzpicture}[scale=1.0]
				\draw[ForestGreen] (0,-3) [bend left=30] to (0.1,-2.5) -- (2.0,-0.5) -- plot [smooth,tension=0.5] coordinates { (2.0,-0.5) (2.37,0) (2.0,0.5)} decorate [decoration={snake, segment length=7.2pt, amplitude=1pt}] {(2.0,0.5) -- (1.25,1.25)} (1.25,1.25) -- (0.1,2.5) [bend left=30] to (0,3);
				\draw[ForestGreen] (0,-3) [bend left=30] to (0.15,-2.5) -- (2.05,-0.5) -- plot [smooth,tension=0.5] coordinates { (2.05,-0.5) (2.42,0) (2.03,0.53)} decorate [decoration={snake, segment length=7.2pt, amplitude=1pt}] {(2.03,0.53) -- (1.28,1.28)} (1.28,1.28) -- (0.13,2.5) [bend left=30] to (0,3);
				\draw[red,thick,-latex] (1.2,0.05) -- (2.35,1.20);
				\draw[red,thick,-latex] (0.8,0.45) -- (1.95,1.60);
				\draw[red,thick,-latex] (1.0,0.25) -- (2.15,1.40);
				\node[below,red,rotate=-45] at (1.0,0.25) {\footnotesize Radiation};
        		\draw[thick] (0, 3) -- (3, 0) node[right] {\footnotesize $i^0$} -- (0,-3) -- (-3,0) -- cycle;
	            \node[] (1) at (1.25,  2.75) {\footnotesize $\cI^+$};
    	        \node[] (2) at (1.25, -2.75) {\footnotesize $\cI^-$};
				\fill[black] ( 3, 0) circle [radius=2pt];
				\fill[black] ( 0, 3) circle [radius=2pt] node[above] {\footnotesize $i^+$};
				\fill[black] ( 0,-3) circle [radius=2pt] node[below] {\footnotesize $i^-$};
				\fill[black] (-3, 0) circle [radius=2pt]; 
				\fill[blue] (2.25,0.75) circle [radius=2pt];
				\fill[blue] (1.50,1.50) circle [radius=2pt];
				\draw[blue] (1.75,0.25) -- (2.50,1.00) node[right] {\footnotesize $u=u_-$};
				\draw[blue] (1.00,1.00) -- (1.75,1.75) node[right] {\footnotesize $u=u_+$};
        \end{tikzpicture}
    \end{center}
    \caption{Model of detector for gravitational memories.}
    \label{fig:Detector}
\end{figure}

Let us consider a couple of inertial observers (that we will refer to as the ``detector'') travelling near future null infinity $\cI^+$. The detector is localized in a region with no gravitational radiation, or more generally no null signal, at both late and early (retarded) times. Let us declare that the radiation is turned on at $u = u_-$ and stops at $u=u_+$. For any value of retarded time, exclusive of the interval $[u_-,u_+]$, the Bondi news tensor and the matter stress-tensor are identically zero by hypothesis: we are in the vacuum. The metric in the regions $u<u_-$ and $u>u_+$ is \eqref{metf} (written in the more convenient Newmann-Unti coordinates), particularized for $q_{AB}=\mathring q_{AB}$ and therefore $C_{AB} = C_{AB}^{(0)}$ according to \eqref{PhiL}. Both vacua are labeled by the supertranslation field $C_-$ for $u \leq u_-$ and $C_+$ for $u \geq u_+$ (see also section \ref{sec:Asymptotic symmetries and the BMS4 group}).

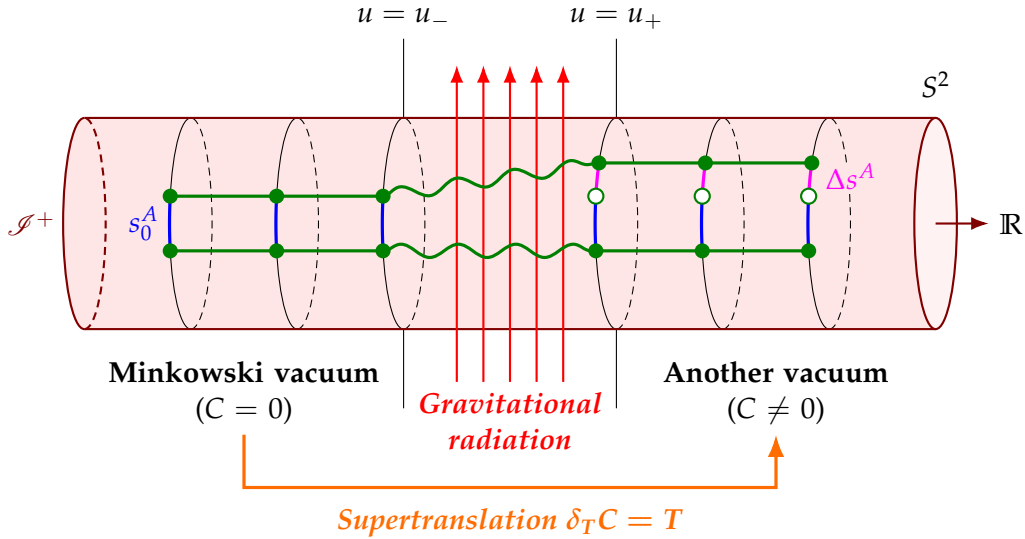
\begin{figure}[h!]
	\centering
    \begin{tikzpicture}[scale=0.7]
    \draw[opacity=0] (-11,-6.5) -- (11,-6.5) -- (11,5) -- (-11,5) -- cycle;
    \fill[red!10] (-8,-2) -- (8,-2) -- (8,2) -- (-8,2) -- cycle;
    \fill[red!10] (-8,0) ellipse (0.4 and 2);
    \draw[thick,red!50!black] (-8,-2) arc (-90:90:-0.4 and 2);
    \draw[thick,red!50!black,densely dashed] (-8,-2) arc (-90:90:0.4 and 2);
    \draw[] (-6,-2) arc (-90:90:-0.4 and 2);
    \draw[densely dashed] (-6,-2) arc (-90:90:0.4 and 2);
    \draw[] (-4,-2) arc (-90:90:-0.4 and 2);
    \draw[densely dashed] (-4,-2) arc (-90:90:0.4 and 2);
    \draw[] (-2,-2) arc (-90:90:-0.4 and 2);
    \draw[densely dashed] (-2,-2) arc (-90:90:0.4 and 2);
    \draw[] (2,-2) arc (-90:90:-0.4 and 2);
    \draw[densely dashed] (2,-2) arc (-90:90:0.4 and 2);
    \draw[] (4,-2) arc (-90:90:-0.4 and 2);
    \draw[densely dashed] (4,-2) arc (-90:90:0.4 and 2);
    \draw[] (6,-2) arc (-90:90:-0.4 and 2);
    \draw[densely dashed] (6,-2) arc (-90:90:0.4 and 2);
    \draw[thick,red!50!black,fill=red!5] (8,0) ellipse (0.4 and 2);
    \draw[thick,red!50!black,fill=red!5,-Latex] (8,0) -- (9,0)node[right]{{\color{black}$\mathbb R$}};
    \draw[] (8,2.2)node[above]{{\color{black}$S^2$}};
    \draw[] (-2,-3.5) -- (-2,-2);
    \draw[] (-2, 3.5)node[above]{$u = u_-$} -- (-2, 2);
	\draw[] ( 2,-3.5) -- ( 2,-2);
    \draw[] ( 2, 3.5)node[above]{$u = u_+$} -- ( 2, 2);
	\draw[thick,red!50!black] (-8,-2) -- (8,-2) (-8,2) -- (8,2);
    \draw[red,thick] plot ({-6+0.4*cos(-10)}, {2*sin(-10)});
    \tikzset{spot/.style={fill=green!50!black,circle,inner sep=2pt}};
    \tikzset{oldspot/.style={fill=white,draw=green!50!black,thick,circle,inner sep=2pt}};
    \def\ang{15};
    \def\dang{35};
    \coordinate (LB1) at ($({-6-0.4*cos(-\ang)}, {2*sin(-\ang)})$);
    \coordinate (LB2) at ($({-4-0.4*cos(-\ang)}, {2*sin(-\ang)})$);
    \coordinate (LB3) at ($({-2-0.4*cos(-\ang)}, {2*sin(-\ang)})$);
    \coordinate (LA1) at ($({-6-0.4*cos( \ang)}, {2*sin( \ang)})$);
    \coordinate (LA2) at ($({-4-0.4*cos( \ang)}, {2*sin( \ang)})$);
    \coordinate (LA3) at ($({-2-0.4*cos( \ang)}, {2*sin( \ang)})$);
    \coordinate (RB1) at ($({ 2-0.4*cos(-\ang)}, {2*sin(-\ang)})$);
    \coordinate (RB2) at ($({ 4-0.4*cos(-\ang)}, {2*sin(-\ang)})$);
    \coordinate (RB3) at ($({ 6-0.4*cos(-\ang)}, {2*sin(-\ang)})$);
    \coordinate (RA1) at ($({ 2-0.4*cos(\dang)}, {2*sin(\dang)})$);
    \coordinate (RA2) at ($({ 4-0.4*cos(\dang)}, {2*sin(\dang)})$);
    \coordinate (RA3) at ($({ 6-0.4*cos(\dang)}, {2*sin(\dang)})$);
    \coordinate (SA1) at ($({ 2-0.4*cos( \ang)}, {2*sin( \ang)})$);
    \coordinate (SA2) at ($({ 4-0.4*cos( \ang)}, {2*sin( \ang)})$);
    \coordinate (SA3) at ($({ 6-0.4*cos( \ang)}, {2*sin( \ang)})$);
    \foreach \k in {-6,-4,-2,2,4,6}{
    \draw[blue,very thick,domain=-\ang:\ang] plot ({\k-0.4*cos(\x)}, {2*sin(\x)});
    }
    \foreach \k in {2,4,6}{
    \draw[magenta,very thick,domain=\ang:\dang] plot ({\k-0.4*cos(\x)}, {2*sin(\x)});
    }
    \draw[very thick,green!50!black] (LB1) -- (LB2) -- (LB3) (LA1) -- (LA2) -- (LA3);
    \draw[very thick,green!50!black] (RB1) -- (RB2) -- (RB3) (RA1) -- (RA2) -- (RA3);
    \foreach \n in {LB1,LB2,LB3,LA1,LA2,LA3,RB1,RB2,RB3,RA1,RA2,RA3}{
    \node[spot] at (\n){};
    }
    \foreach \n in {SA1,SA2,SA3}{
    \node[oldspot] at (\n){};
    }
    \node[blue,left] at (-6.4,0){$s^A_0$};
    \node[magenta,left,fill=red!10,inner sep=0pt] at (6.9,0.9){$\Delta s^A$};
    \foreach \j in {-1.0,-0.5,0,0.5,1.0}{
    \draw[red,thick,-Latex] (\j,-3) -- (\j,3);
    }
    \node[red,below,align=center,text width=3cm] at (0,-3){\textit{\textbf{Gravitational radiation}}};
    \tikzset{snake it/.style={decorate, decoration=snake}};
    \draw[very thick,green!50!black,snake it,segment length=21.5] (LB3) -- (RB1);
    \draw[very thick,green!50!black,snake it,segment length=21.5] (LA3) -- (RA1);
    \node[text width=4.75cm,align=center,above] at (-5,-4){\textbf{Minkowski vacuum}\\ (\textit{C} $=$ 0)};
    \node[text width=4.75cm,align=center,above] at ( 5,-4){\textbf{Another vacuum}\\ (\textit{C} $\neq$ 0)};
    \draw[very thick,-Latex,orange!85!red] (-5,-4) |- (0,-5) -| ( 5,-4);
    \node[orange!85!red,below] at (0,-5.2){\textbf{\textit{Supertranslation}} $\bm{\delta_T C = T}$};
    \node[red!50!black] at (-9,0){$\cI^+$};
    \end{tikzpicture}
    \caption{Displacement memory effect.}
    \label{fig:Memory}
\end{figure}

The detector, which moves on a timelike trajectory in the far region depicted in green on Figure \ref{fig:Detector} \cite{Strominger:2017zoo}, experiences null radiation only during the time interval $\Delta u = u_+ - u_-$. The two inertial observers forming the detector follow a timelike geodesics in the vicinity of $\cI^+$, characterized by a $4$-velocity $v = v^\mu\partial_\mu$. Since their trajectories are located near $\cI^+$, we can admit that $v^\mu \partial_\mu = \partial_u + \mathcal O(\rho^{-2})$ up to subleading corrections necessary for $v$ to verify $v^\mu v_\mu = -1$ asymptotically. The separation between the two geodesic trajectories is given by the deviation vector $s = s^\mu\partial_\mu$ which is solution of the \textit{equation of geodesic deviation} :
\begin{equation}
\nabla_v \nabla_v s^{\mu} = R^\mu_{\phantom{\mu} \alpha\beta\gamma} v^\alpha v^\beta s^\gamma \label{geodesic deviation}
\end{equation}
where $\nabla_v = v^\mu \nabla_\mu$ is the directional derivative along $v^\mu$. We suppose that both detectors move on the same celestial sphere, so $s^\rho = 0$, with an initial angular separation $s^A = s^A_0$ for $u<u_-$. We get
\begin{equation}
\rho^2 \mathring q_{AB} \partial_u^2 s^B = R_{uAuB} s^B \Leftrightarrow \mathring q_{AB} \partial_u^2 s^B = \frac{1}{2\rho} \partial_u^2 C_{AB} s^B + \mathcal O(\rho^{-2})
\end{equation}
after using $R_{uAuB}  = -\frac{\rho}{2} \partial_u^2 C_{AB} + \mathcal{O}(\rho^0)$. One can check the latter relation using the metric in the Bondi gauge then translating the result in the Newman-Unti gauge, which is easy because $\rho = r + \mathcal O(r^{-1})$ when $\beta = \mathcal O(r^{-2})$. We introduce some angular perturbation $s^A = s^A_0 + \rho^{-1} \Delta s^A + \mathcal O(\rho^{-2})$ in the deviation vector for any $u\geq u_+$. Using $\Delta u \ll \rho$, we can integrate on $u$ to get \cite{Strominger:2014pwa}
\begin{equation}
 \mathring q_{AB} \Delta s^B = \frac{1}{2} \Delta C_{AB}^{(0)} s_0^B + \mathcal{O}(\rho^{-1}) . \label{kinematic shift}
\end{equation}
Therefore, if the phenomena occurring during the non-stationary phase $[u_-,u_+]$ are responsible for a shift in the asymptotic shear, the angular separation of the inertial observers will be irreversibly shifted: this is the \textit{displacement memory effect}! The spacetime is able to keep track of the radiative phase in terms of a \textit{DC} effect. Flashing a light between these two observers will measure the shift $\Delta s^A$, which is therefore detectable. This was first observed by Zeldovich and Polnarev in 1974 \cite{Zeldovich:1974gvh} and nicely reviewed in the Bondi gauge by Strominger and Zhiboedov \cite{Strominger:2014pwa}. The figure \ref{fig:Memory} provides a schematic picture of the process at null infinity.

Making contact with our discussion in section \ref{sec:Vacua}, the displacement memory effect is equivalent to a change of vacuum between the two stationary phases $u<u_-$ and $u>u_+$, for which the supertranslation field $C_-$ has been shifted into $C_+ = C_- + \Delta C$. By virtue of \eqref{deltaC}, we see that there exists a smooth supertranslation parametrized by $T(x^A)$ such that $\Delta C = T$. The kinematical equation \eqref{kinematic shift} informs us how to ``measure'' the action of a supertranslation. The presence of the infinite-dimensional extension of the Poincaré group as BMS$_4$ supertranslations is therefore mandatory if we allow radiative configurations in the solution space \cite{Ashtekar:1981hw,Ashtekar:2014zsa}. In the absence of pure supertranslations, the only shifts allowed in the field $C$ are annihilated by the operator $(D_AD_B)^{TF}$ because the four translation parameters $T$ precisely verify $(D_AD_B T)^{TF} =0$ by definition. Therefore $\Delta C_{AB}^{(0)}$ would be zero for any vacuum transition and the shift \eqref{kinematic shift} would never be observed. But this statement is equivalent to assume that there is no radiative mode in the phase space.

Let us now study the causes of the displacement \eqref{kinematic shift}. Any physical process that can change the tensor $C_{AB}$ leads to a displacement memory effect. To analyze the possible sources, let us consider the time evolution of the Bondi mass aspect introducing some matter stress-tensor $T_{\mu\nu}^M$ in the bulk of spacetime. The equation \eqref{duM} is modified as 
\begin{equation}
\partial_u M = -\frac{1}{8}N_{AB}N^{AB} + \frac{1}{4}D_AD_B N^{AB} - 4\pi G\lim_{r\to\infty}(r^2 T^M_{uu})
\end{equation}
if we are on the unit-round sphere metric (see \textit{e.g.} \cite{Strominger:2013jfa,Flanagan:2015pxa}). The demonstration follows the reasoning presented in section \eqref{sec:Solution space in Bondi gauge} and requires to extract the $r$-independent part of $r^2 G_{uu} = 8\pi G T_{uu}^M$. Integrating between $u_-$ and $u_+$, this equation simply yields \cite{Strominger:2014pwa}
\begin{equation}
-\frac{1}{4} (D^2 +2) D^2 \Delta C = \Delta M + \int_{u_-}^{u_+}  \D u \: \left[ \frac{1}{8} N_{AB} N^{AB} + 4\pi G \lim_{r\rightarrow\infty} (r^2 T_{uu}^M) \right]
\label{eq:SourcingC}
\end{equation}
for $D^2 \equiv D^A D_A$ and using \eqref{PhiL} for both values of $u = u_-$, $u = u_+$.
The shift of supertranslation field $\Delta C$ obeys a quartic elliptic equation which is sourced by 3 qualitatively distinct terms, hence the displacement memory detector will trigger for each of the following causes:  
\begin{enumerate}
\item \textit{If the Bondi mass aspect varies between $u_-$ and $u_+$.} This is sometimes called \textit{ordinary memory} \cite{Zeldovich:1974gvh,1987Natur.327..123B}. For example, a single massive body containing a string that suddently separate into two parts due to a trigger will modify the Bondi mass aspect $M$ because the mass will suddenly possess a strong dipolar component. What Einstein gravity tells us is that a signal is sent at null infinity with that information and the memory effect follows.
\item \textit{If some null matter represented by $T_{uu}^M = \mathcal O(r^{-2})$ reaches $\cI^+$ between $u_-$ and $u_+$.} This is sometimes called the \textit{null memory effect}. For example, electromagnetic radiation causes the displacement memory effect. 
\item And finally, \textit{if gravitational waves, whose flux is encoded in the news tensor $N_{AB}$, pass through $\cI^+$ between $u_-$ and $u_+$.} This is called the \textit{Christodoulou effect} \cite{Christodoulou:1991cr,Thorne:1992sdb}, which appeared also earlier in perturbative form in post-Newtonian formalism \cite{Blanchet:1987wq,Blanchet:1992br}. 
\end{enumerate}
In the literature, one also find the epithet ``non-linear'' to qualify the Christodoulou effect since it is produced by the gravitational field itself, by opposition with both other effects which are qualified as ``linear'' because the sources are material.

Let us finish by two concluding observations. First of all, the description of a gravitational memory effect always comes with two equations, \eqref{kinematic shift} and \eqref{eq:SourcingC} in the present case. 
\begin{itemize}[label=$\rhd$]
\item The first equation is \textit{kinematical} and tells us how the radiative phase affects the detectors. It allows to determine what is the memory field $\varphi$ that will be permanently shifted (\textit{i.e.} $\varphi_-\to\varphi_+$) after the end of the non-stationary phase at $u>u_+$. It will thus be the depositary of the ``memory'' of what happened in the time interval $[u_-,u_+]$. This role is played here by the supertranslation field $C$. The inhomogeneous part $\delta^I_\xi \varphi$ of the transformation law of the memory field gives the asymptotic symmetry $\xi$ relating both vacua for $u<u_-$ and $u>u_+$. Indeed, $\delta^I_\xi C = T$ from \eqref{deltaC} and the final vacuum is the supertranslated vacua with $T = \Delta C$. Matching with the notations of section \ref{sec:Axiomatic definition and relation to gauge symmetries}, $\varphi\equiv C$ is well a local field in Bondi (or Newman-Unti) gauge and $O(\theta,\phi)\equiv \Delta C$. The latter is gauge-invariant: performing a global supertranslation $\xi$ of amplitude $T_{glob}$ on the whole solution will modify $\delta_\xi\varphi_+ = T_{glob} = \delta_\xi\varphi_-$ hence $\delta_\xi \Delta C = 0$. Note crucially that $T_{glob}$ is an actual transformation on the solution space while $T$ is the mapping between early and late stationary configurations: $T_{glob}$ represents any supertranslation that can be performed on the whole solution obtained by gluing continuously the three regions $u<u_i$, $u_i < u < u_f$ and $u>u_f$, while $T$ is determined as the diffeomorphism needed to map the late region $u<u_i$ onto the late region $u>u_f$.
\item The second equation is \textit{dynamical} and is used to compute the exact shift $\Delta C$ of the memory field from the data of the sources (distribution of mass, radiation,\dots). This information has to come from the equations of motion for the gravitational field, eventually coupled with a matter stress-tensor. The evolution of the memory field is dictated by a relevant evolution equation implying quantities of the same nature: hence the evolution of the scalar field $C$ is governed by the constraint equation on the Bondi mass aspect. 
\end{itemize}
Note finally that the displacement memory effect has not yet been observed at the time of writing. Nevertheless, we are hopeful that the experimental confirmation will be obtained soon from gravitational astronomy, by means of gravitational wave detectors \cite{Lasky:2016knh} or pulsar timing arrays \cite{2010CQGra..27h4013H,2011ASSP...21..229H}.

\subsection{Superboost transitions}
\label{sec:Superboost}
Let us now focus on gravitational memory effects related to super-Lorentz transformations. We start by describing the constraints on the bulk data producing generic superboost transitions that change the boundary metric and the vacuum Bondi news tensor at leading order. We also explain why we are mainly interested in the impulsive limit of such exotic events and characterize their imprint on asymptotic detectors of free-falling observers. We conclude that superboost vacuum transitions allow to deduce the \textit{velocity kick/refraction memory effects} \cite{Podolsky:2002sa,Podolsky:2010xh,Podolsky:2016mqg} from a symmetry principle in the asymptotic region. These effects latter are qualitatively distinct from the \textit{spin memory effect} \cite{Pasterski:2015tva} which has no local associated memory field in the Bondi gauge (see also \cite{Compere:2018aar} for a review). It is in fact related to subleading diffeomorphism transitions in (retarded) harmonic gauge \cite{Himwich:2019qmj}. They are also distinct from the \textit{center-of-mass memory effect} \cite{Nichols:2018qac}, also lacking of a local memory field in Bondi coordinates and related to changes in the center-of-mass part of the angular momentum.

\subsubsection{Penrose's gravitational shockwaves}
\label{sec:Shockwaves}
In the previous section, we reviewed the effect of permanent displacement of asymptotic free-falling observers due to a radiative process leading to a transition between two supertranslated vacua in $\mathring{\mathcal S}_0^{\text{Mink}}$. Analogously, we define here a \textit{superboost transition} as a transition between two vacua associated with two different superboost fields. The configuration is as follows: 
\begin{itemize}[label=$\rhd$]
\item We consider a spacetime in $\mathring{\mathcal S}_0$ for which we assume that the radiative phase has compact support in the time interval $[u_-,u_+]$.
\item For $u<u_-$, the spacetime is Minkowski in some coordinate system where $\Phi = \Phi_-(x^A)$, $q_{AB} = q_{AB}^{vac}[\mathring{q},\Phi_-,\widetilde\Phi]$, $N_{AB}^{vac} = N_{AB}^{vac}[\Phi_-]$ as initial conditions.
\item For $u>u_+$, the spacetime is also Minkowski in another coordinate system for $\Phi = \Phi_+(x^A)$, $q_{AB} = q_{AB}^{vac}[\mathring{q},\Phi_+,\widetilde\Phi]$, $N_{AB}^{vac} = N_{AB}^{vac}[\Phi_+]$ as late data.
\end{itemize}
During the non-stationary phase $u_-<u<u_+$, the metric field must vary in time at leading order. This information, combined with the equation of motion \eqref{pu qAB = 0}, implies that a superboost transition cannot be observed as a vacuum solution of Einstein's equation but needs the coupling with some matter with exotic fall-offs in its stress-energy tensor in a way to modify the leading structure at $\mathscr I^+$. A simple power counting in \eqref{EOM qAB time evolution} gives that $T_{AB} = \mathcal O(r)$ for $u_-<u<u_+$ is required to observe a superboost transition. Note also that \eqref{EOM qAB time evolution} implies that the time-dependency of the metric can only be turned on in the vacuum by a transition in some conformal factor in front of $q_{AB}$ for which the boundary condition \eqref{sqrt q fixed in Bondi} cannot be obeyed, but it is definitely not a superboost transition despite some similarities. We plan to study these generalized transitions (probably related to Robinson-Trautman waves \cite{Robinson:1960zzb} and the Weyl symmetry \cite{Freidel:2021cbc,Barnich:2010eb}) in future works. Instead of entering into these considerations, we will rather discuss here superboost transitions driven by topological defects for which the vacuum Einstein equation, in particular \eqref{pu qAB = 0}, is obeyed locally except at some singular points. We will see that we can yet extract a lot of information from these processes that are in fact describing gravitational shockwaves in the bulk.

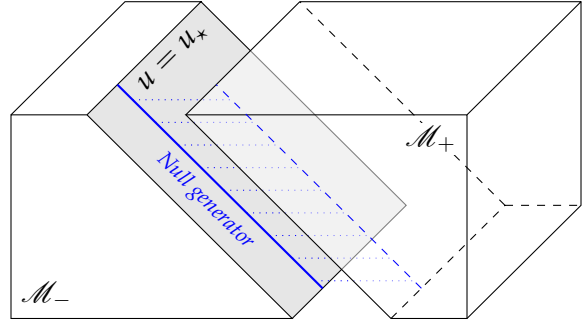
\begin{figure}[!ht]
\centering
\begin{tikzpicture}[scale=1]
\draw[opacity=0] (0,0) -- (0,4.5);
\def\hx{3.7};
\def\tx{1};
\def\decal{1.5};
\def\u{0.4};
\def\shift{1.3};
\def\step{0.5};
\coordinate (A) at (0,0);
\coordinate (B) at (\hx,0);
\coordinate (C) at (\tx,\hx-\tx);
\coordinate (D) at (0,\hx-\tx);
\coordinate (E) at ($(B)+(\shift,0)$);
\coordinate (F) at ($(E)+(\tx,0)$);
\coordinate (G) at ($(F)+(0,\hx-\tx)$);
\coordinate (H) at ($(C)+(\shift,0)$);
\fill[black!10] (B) -- ($(B)+(\decal,\decal)$) -- ($(C)+(\decal,\decal)$) -- (C) -- cycle;
\draw[] (A) -- (B) -- (C) -- (D) -- cycle;
\draw[] (C) -- ($(C)+(\decal,\decal)$) -- ($(D)+(\decal,\decal)$) -- (D) ($(C)+(\decal,\decal)$) -- ($(B)+(\decal,\decal)$) -- (B);
\draw[thick,blue] ($(B)+(\u,\u)$) -- ($(C)+(\u,\u)$);
\foreach \k in {0.1,0.4,...,2.7}{
\draw[blue,dotted] ($(B)+(\u,\u)+(-\k,\k)$) -- ($(E)+(\u,\u)+(-\k,\k)$);
}
\fill[white,opacity=0.5] (E) -- ($(E)+(\decal,\decal)$) -- ($(H)+(\decal,\decal)$) -- (H) -- cycle;
\draw[] (E) -- (F) -- (G) -- (H) -- cycle;
\draw[] (F) -- ($(F)+(\decal,\decal)$) -- ($(G)+(\decal,\decal)$) -- ($(H)+(\decal,\decal)$) (G) -- ($(G)+(\decal,\decal)$) (H) -- ($(H)+(\decal,\decal)$);
\draw[dashed] (E) -- ($(E)+(\decal,\decal)$) ($(E)+(\decal,\decal)$) -- ($(F)+(\decal,\decal)$) ($(E)+(\decal,\decal)$) -- ($(H)+(\decal,\decal)$);
\draw[dashed,blue] ($(E)+(\u,\u)$) -- ($(H)+(\u,\u)$);
\coordinate (A1) at ($(C)+(\u,\u)$);
\coordinate (A2) at ($(C)+(\decal,\decal)$);
\coordinate (A3) at ($(B)+(\u,\u)$);
\draw[] ($(A1)!0.5!(A2)$)node[rotate=45,below]{$u=u_\star$};
\draw[blue] ($(A1)!0.5!(A3)$)node[rotate=-45,below]{\scriptsize \textit{Null generator}};
\draw[] (A)node[anchor=south west]{$\mathscr M_-$};
\draw[] (G)node[anchor=north east,fill=white,inner sep=1pt,outer sep=3pt]{$\mathscr M_+$};
\end{tikzpicture}
\caption{Penrose's ``scissor and paste'' construction.}
\label{fig:scissor}
\end{figure}

Let us briefly review the way Penrose introduced the notion of impulsive gravitational wave in \cite{Penrose:1972aa} thanks to the suggestively called ``scissor and paste'' construction, that we translate here in our formalism. The impulsive gravitational wave, or \textit{shockwave}, is modelized by a local vacuum solution of Einstein's field equation obtained by cutting the spacetime along a null hypersurface located at $u=u_\star$ and gluing two different patches $\mathscr M_- = (u_-,r_-,z_-,\bar z_-)$ for $u<u_\star$ and $\mathscr M_+ = (u_+,r_+,z_+,\bar z_+)$ for $u>u_\star$ of Minkowski spacetime $\mathscr M$. The construction is schematized in Figure \ref{fig:scissor} \cite{Penrose:1972aa}.

Assuming that, for $u<u_\star$, the spacetime is the global Minkowski vacuum in Bondi coordinates where $x^A_- = (z_-,\bar z_-)$ are the stereographic coordinates on the sphere, we can impose the continuity conditions $r_-^2 q^{vac}_{AB}[\mathring q,\Phi_-,\widetilde\Phi]\D x^A_- \D x^B_- = r_+^2 q^{vac}_{AB}[\mathring q,\Phi_+,\widetilde\Phi]\D x^A_+ \D x^B_+$ almost everywhere on the celestial sphere. This is solved locally by the holomorphic transformation $z_+ = G(z_-)$ compensated by a suitable Weyl rescaling $r_- \to r_+$ in order to keep the sphere area preserved across $u=u_\star$. By hypothesis $\Phi_- = -\ln \sqrt{\mathring q}$, $q^{vac}_{AB}\D x^A_- \D x^B_- = \mathring q_{AB}\D x^A_- \D x^B_- = 2r_- \sqrt{\mathring q} \D z_- \D \bar z_-$ and $N_{AB}^{vac}[\Phi_-] = 0$. Taking the convenient choice of origin for the retarded time such that $u_\star = 0$, the line element resulting from the scissor and paste construction can be written as \cite{Nutku:1992aa}
\begin{equation}
\text ds^2 = -\text{d}u^2 - 2 \text{d}\rho \text{d}u + \left[ \rho^2 q_{AB} + u \rho \Theta(u) N^{vac}_{AB} + \frac{u^2}{8}\Theta(u) N^{vac}_{CD}N_{vac}^{CD} q_{AB} \right] \text{d}x^A \text{d}x^B .\label{imp}
\end{equation}
in the Newmann-Unti gauge. Here $N^{vac}_{AB}  = \left[\frac{1}{2}D_A \Phi_+ D_B \Phi_+ - D_A D_B \Phi_+  \right]^{TF}$ and $\Theta(u)$ is the Heaviside distribution. The vacuum news tensor coincides with \eqref{PhiL} after substituting $\Phi = -\ln \sqrt{\mathring q} + \phi_+(z)+\bar\phi_+(\bar z) = \Phi_+$ as in \eqref{Phim}. The presence of a gravitational shockwave located on the null hypersurface $u=0$ is responsible for a superboost transition between two vacua which differ by a \textit{meromorphic} superboost transformation. By construction, \eqref{imp} is globally Riemann-flat for $u< 0$ and locally for $u>0$ except at the singularities of the meromorphic function $\phi_+(z)$, located at the poles $z=0,\infty$ of the celestial sphere. At these points, one should not forget that $q_{AB}\neq \mathring q_{AB}$ is singular and $g_{uu} = -1-\frac{1}{2}D^2 \phi_+$ but otherwise $q_{AB}=\mathring q_{AB}$ and $g_{uu}=  -1$. One can thus prove that \eqref{imp} is Ricci-flat for any $u$ including the location of the shockwave, also except at the singular points. At $u=0$ the Riemann-flatness, is only broken by the shockwave which contributes to the Weyl curvature as $W_{uAuB} = r\delta (u) N_{AB}^{vac}[\phi_+]$. 

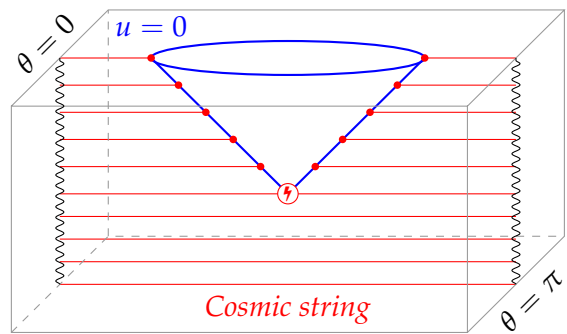
\begin{figure}[!ht]
\centering
\begin{tikzpicture}[scale=1.5]
\draw[opacity=0] (0,0) -- (0,3.3);
\coordinate (A) at (0,0);
\coordinate (B) at (4,0);
\coordinate (C) at (4,2);
\coordinate (D) at (0,2);
\def\psx{0.85};
\def\psy{0.85};
\coordinate (Ap) at ($(A)+(\psx,\psy)$);
\coordinate (Bp) at ($(B)+(\psx,\psy)$);
\coordinate (Cp) at ($(C)+(\psx,\psy)$);
\coordinate (Dp) at ($(D)+(\psx,\psy)$);
\coordinate (MG) at ($(D)!0.5!(Dp)$);
\coordinate (MD) at ($(C)!0.5!(Cp)$);
\coordinate (MC) at ($(MG)!0.5!(MD)$);
\coordinate (CG) at ($(A)!0.5!(Ap)$);
\coordinate (CD) at ($(B)!0.5!(Bp)$);
\def\el{1.2};
\draw[blue,thick] (MC) ellipse (1.2 and 0.15);
\draw[blue,thick] ($(MC)-(0,\el)$) -- ($(MC)-(\el,0)$) ($(MC)-(0,\el)$) -- ($(MC)+(\el,0)$);
\foreach \n in {0.2,0.4,...,1.0}{
\draw[red] ($(MG)-(0,\el-\n*\el)$)--($(MC)-(\n*\el,\el-\n*\el)$)node[circle,fill,inner sep=1pt]{} ($(MD)-(0,\el-\n*\el)$)--($(MC)-(-\n*\el,\el-\n*\el)$)node[circle,fill,inner sep=1pt]{};
}
\foreach \n in {0,0.2,0.4,0.6,0.8}{
\draw[red] ($(CG)+(0,\n)$) -- ($(CD)+(0,\n)$);
}
\draw[black!40] (A)--(B)--(C)--(D)--cycle (C)--(Cp)--(Dp)--(D) (B)--(Bp)--(Cp);
\draw[dashed,black!40] (A)--(Ap)--(Bp) (Ap)--(Dp);
\draw[red] ($(MC)-(0,\el)$)node[draw,fill=white,circle,inner sep=0.5pt]{\tiny\text{\faBolt}};
\node[blue,above] at ($(MC)-(\el,-0.1)$) {$u=0$};
\draw[decorate, decoration={snake,segment length=4pt,amplitude=1.5pt}] (CG) -- (MG);
\draw[decorate, decoration={snake,segment length=4pt,amplitude=1.5pt}] (CD) -- (MD);
\node[rotate=45,above] at (MG) {$\theta=0$};
\node[rotate=45,below] at (CD) {$\theta=\pi$};
\node[red,below] at ($(CG)!0.5!(CD)$) {\textit{Cosmic string}};
\end{tikzpicture}
\caption{Gravitational shockwave induced by cosmic string snapping.}
\label{fig:Cosmic string}
\end{figure}

To summarize, the singular impulsive limit describing transitions induced by shockwaves requires to consider singular diffeomorphisms transitions that are meromorphic superboost transitions. The metric $q_{AB}$ is the unit sphere metric globally for $u<0$ and locally for $u > 0$, but it contains singularities at isolated points for $u > 0$. These singularities have been understood as cosmic string decays \cite{Nutku:1992aa,Griffiths:2002hj,Griffiths:2002gm,Strominger:2016wns} that provide a concrete physical process to be associated with superboost transitions. The metric \eqref{imp} for $u>0$ is thought to represent a cosmic string supported at the end points $z=0,\infty$ if the superboost relating the two vacua is driven by a transformation of the form $G(z) = z^{1+\varepsilon}$ \cite{Nutku:1992aa,Strominger:2016wns}. This is a topological defect in the Minkowski spacetime which induces a deficit angle linear in $\varepsilon$ along the polar axis $z=0,\infty$ (or $\theta =0,\pi$ for the standard spherical colatitude) of the null hypersurface $u=0$. Because the string pierces null infinity, the boundary metric is singular at the anchor points, what is in line with the analysis above. The time-reversion $u\to -u$ of \eqref{imp} is thus seen as the snapping of such a cosmic string at $u=0$, the two free ends starting from $r=u=0$ in Minkowski space and travel along the singular points $z=0,\infty$ at the speed of light towards null infinity, see Figure \ref{fig:Cosmic string} \cite{Nutku:1992aa}. When the string snaps, a gravitational shockwave is emitted and travels along the null front $u=0$. The free extremities of the string support particles or, better, black holes in the full non-linear theory. An impulsive superboost is thus a toy-model to analyze the radiative consequences of a black hole pair creation in the bulk. Some explicit computations and results in that direction can be found in \cite{Strominger:2016wns}.

\subsubsection{General impulsive gravitational wave transitions}

In general, both the supertranslation field $C$ and the superboost field $\Phi$ can change with \textit{hard} (finite energy) processes involving null radiation reaching $\mathscr I^+$. This null radiation can have its origin in matter or in gravity itself. Such processes, occurring during a finite amount of time $[u_-,u_+]$, induce vacuum transitions among initial $(C_-,\Phi_-)$ ($u<u_-$) and final $(C_+,\Phi_+)$ ($u<u_+$) boundary fields. The difference between these fields can be expressed in terms of components of the matter stress-tensor and metric potentials reaching $\mathscr I^+$. The simplest possible transition between vacua are shockwaves (say, at $u=0$) that carry a matter stress-tensor and a Weyl curvature proportional to a $\delta(u)$ function, as in the original Penrose construction \cite{Penrose:1972aa}. A distinct vacuum lies on each side of the shockwave and the transition between the boundary fields is dictated by the matter stress-tensor as well as the radiative components of the Weyl tensor. This picture allows us to include both types of transitions previously discussed in sections \ref{sec:Linear displacement memory effect} and \ref{sec:Shockwaves}. For $u<0$ and $u>0$, the line element describes a vacuum situation which reads as \eqref{metf} in Newmann-Unti coordinates. 

A general shockwave takes the form 
\begin{equation}
\text ds^2 = - \frac{R[q]}{2}\text{d}u^2 - 2 \text{d}\rho \text{d}u + (\rho^2 q_{AB} + \rho C_{AB} + \frac{1}{8}C_{CD}C^{CD} q_{AB})\text{d}x^A \text{d}x^B + D^B C_{AB} \text{d}x^A \text{d}u\label{metf2}
\end{equation}
where 
\begin{align}
q_{AB}&=\Theta(-u) q_{AB}^{vac}[\Phi_-] + \Theta(u) q_{AB}^{vac}[\Phi_+],\\
C_{AB}&= \Theta(-u) C_{AB}^{vac}[\Phi_-, C_-] + \Theta(u) C_{AB}^{vac}[\Phi_+,C_+]
\end{align}
where $q_{AB}^{vac}[\Phi]$ and $C_{AB}^{vac}[\Phi,C]$ are given in \eqref{qABform} and \eqref{PhiL}. The metric \eqref{imp} is recovered for $\Phi_- = -\ln\sqrt{\mathring{q}}$, $\Phi_+ = \phi_+(z)+\bar\phi_+(\bar z) - \ln\sqrt{\mathring{q}}$ as in \eqref{Phim} and $C_+ = C_- = 0$. Our goal now is to study the memory effects associated with such a vacuum transition. Among them, there is obviously the shift of the supertranslation field previously analyzed in section \ref{sec:Linear displacement memory effect}.

\subsubsection{Evolution of the Bondi mass aspect and the center-of-mass}
In the absence of superboost transitions and assuming the standard boundary condition $q_{AB}=\mathring q_{AB}$, the integral between initial $u_-$ and final retarded times $u_+$ of the evolution equation \eqref{duM} for the Bondi mass aspect can be re-expressed as the differential equation \eqref{eq:SourcingC} determining the difference between the supertranslation field $\Delta C = C_+ - C_-$ between initial and final retarded times. This is one of the two master equations governing the vacuum transition. The four lowest spherical harmonics $\texttt{\textit{l}}=0,1$ are zero modes of the differential operator appearing on the left-hand side of \eqref{eq:SourcingC}. Recall that translations precisely shift the supertranslation field as \eqref{deltaC} or more particularly \eqref{C field variation}. The 4 lowest harmonics of $C$ can thus be interpretated as the \textit{center-of-mass} of the asymptotically flat system. This center-of-mass is not constrained by the conservation law \eqref{eq:SourcingC} as observed previously.

A new feature arises in the presence of a superboost transition. The four zero modes of the supertranslation field $C$ are now determined by the evolution equation. This can be seen in the context of impulsive transitions \eqref{metf2}. For simplicity, we take $C_- = 0$ and $\Phi_- = -\ln\sqrt{\mathring q}$ ($q_{AB}[\Phi_-] = \mathring q_{AB}$). Given that the Bondi mass aspect and the Bondi news of the vacua are non-zero \eqref{valvac}, we first define the ``normalized'' Bondi mass aspect and Bondi news as
\begin{eqnarray}
\bar M &=& M + \frac{1}{8} C_{AB} N^{AB}_{vac}[\Phi_+],\\
\hat N_{AB} &=& N_{AB} - \Theta(u) N_{AB}^{vac}[\Phi_+],
\end{eqnarray}
which are zero for the vacua \eqref{metf}. We will give a precise meaning for these shifted quantities in sections \ref{sec:Initial and late data in general super-Lorentz frames} and \ref{sec:Finite Hamiltonians}, where we will show that $\bar M$ is the momentum canonically conjugated to the supertranslation parameter $T$ in the finite Generalized BMS$_4$ Hamiltonian at $\mathscr I^+$ while $\hat N_{AB}$ is the physical notion of Bondi news which is not improperly sourced by super-Lorentz transformations. After integration over $u$ of \eqref{duM}, we take the corollary \eqref{Li2} of the Liouville equation \eqref{Liouville} into account to obtain 
\begin{equation}
-\frac{1}{4} D^2 (D^2 + R[q])  C_+ +\frac{1}{4}N^{AB}_{vac}[\Phi_+] D_A D_B C_+  + \frac{1}{8} C_+ D_A D^A R[q]  = \Delta \bar M +  \int_{u_-}^{u_+} \text{d}u \: T_{uu} \label{eq4}
\end{equation}
where 
\begin{equation}
T_{uu}= \frac{1}{8}\hat N^{AB} \hat N_{AB} + 4\pi G \lim_{r\rightarrow\infty} (r^2 T_{uu}^M),
\end{equation}
$\Delta \bar M$ act as sources for $C_+$ and all quantities are evaluated on the final metric $q_{AB}[\Phi_+]$. We have that $\Delta \bar M=0$ for transitions between vacua but we included it (1) for making the comparison with \eqref{eq:SourcingC} more manifest and (2) for leaving the possiblity to easily adapt the result in the presence of zero modes like some ADM mass when $u<0$. Here is the main observatio: the lowest $\texttt{\textit{l}}=0,1$ spherical harmonics of the supertranslation field $C$ are not zero modes of the quartic differential operator on the left-hand side of \eqref{eq4} for any inhomogeously curved boundary metric, \textit{i.e.} in an arbitrary super-Lorentz frame where neither $q_{AB}$ nor $N_{AB}^{vac}$ are trivial. Therefore, the center-of-mass is also determined by the equation of motion for the Bondi mass aspect. We are now about to discuss the second aspect of the memory effect, which is the predictions of the geodesic deviation equation.

\subsubsection{Refraction/Velocity kick memory}

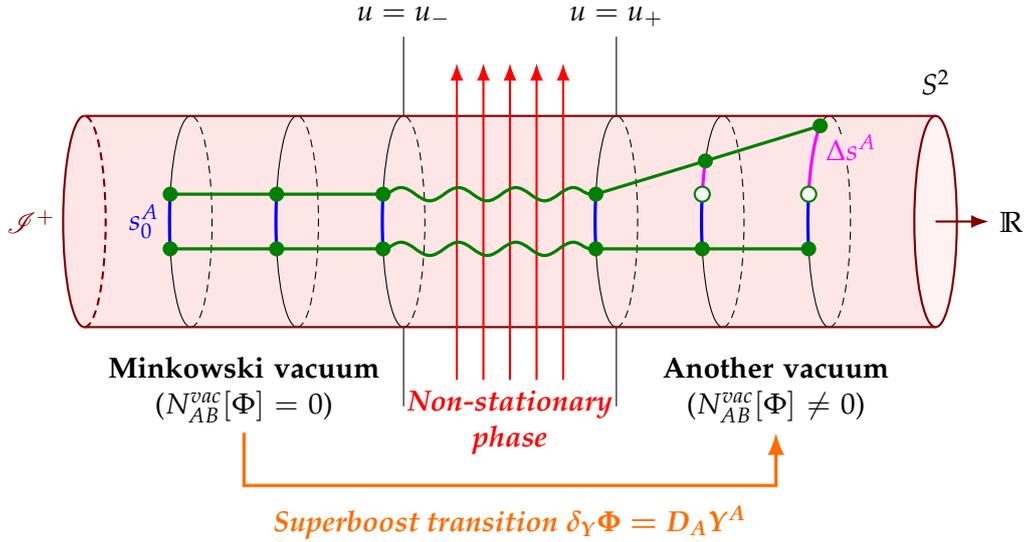
\begin{figure}[h!]
\centering
    \begin{tikzpicture}[scale=0.7]
    \draw[opacity=0] (-11,-6.5) -- (11,-6.5) -- (11,5) -- (-11,5) -- cycle;
    \fill[red!10] (-8,-2) -- (8,-2) -- (8,2) -- (-8,2) -- cycle;
    \fill[red!10] (-8,0) ellipse (0.4 and 2);
    \draw[thick,red!50!black] (-8,-2) arc (-90:90:-0.4 and 2);
    \draw[thick,red!50!black,densely dashed] (-8,-2) arc (-90:90:0.4 and 2);
    \draw[] (-6,-2) arc (-90:90:-0.4 and 2);
    \draw[densely dashed] (-6,-2) arc (-90:90:0.4 and 2);
    \draw[] (-4,-2) arc (-90:90:-0.4 and 2);
    \draw[densely dashed] (-4,-2) arc (-90:90:0.4 and 2);
    \draw[] (-2,-2) arc (-90:90:-0.4 and 2);
    \draw[densely dashed] (-2,-2) arc (-90:90:0.4 and 2);
    \draw[] (2,-2) arc (-90:90:-0.4 and 2);
    \draw[densely dashed] (2,-2) arc (-90:90:0.4 and 2);
    \draw[] (4,-2) arc (-90:90:-0.4 and 2);
    \draw[densely dashed] (4,-2) arc (-90:90:0.4 and 2);
    \draw[] (6,-2) arc (-90:90:-0.4 and 2);
    \draw[densely dashed] (6,-2) arc (-90:90:0.4 and 2);
    \draw[thick,red!50!black,fill=red!5] (8,0) ellipse (0.4 and 2);
    \draw[thick,red!50!black,fill=red!5,-Latex] (8,0) -- (9,0)node[right]{{\color{black}$\mathbb R$}};
    \draw[] (8,2.2)node[above]{{\color{black}$S^2$}};
    \draw[] (-2,-3.5) -- (-2,-2);
    \draw[] (-2, 3.5)node[above]{$u = u_-$} -- (-2, 2);
	\draw[] ( 2,-3.5) -- ( 2,-2);
    \draw[] ( 2, 3.5)node[above]{$u = u_+$} -- ( 2, 2);
	\draw[thick,red!50!black] (-8,-2) -- (8,-2) (-8,2) -- (8,2);
    \draw[red,thick] plot ({-6+0.4*cos(-10)}, {2*sin(-10)});
    \tikzset{spot/.style={fill=green!50!black,circle,inner sep=2pt}};
    \tikzset{oldspot/.style={fill=white,draw=green!50!black,thick,circle,inner sep=2pt}};
    \def\ang{15};
    \def\dang{35};
    \def\ddang{65};
    \coordinate (LB1) at ($({-6-0.4*cos(-\ang)}, {2*sin(-\ang)})$);
    \coordinate (LB2) at ($({-4-0.4*cos(-\ang)}, {2*sin(-\ang)})$);
    \coordinate (LB3) at ($({-2-0.4*cos(-\ang)}, {2*sin(-\ang)})$);
    \coordinate (LA1) at ($({-6-0.4*cos( \ang)}, {2*sin( \ang)})$);
    \coordinate (LA2) at ($({-4-0.4*cos( \ang)}, {2*sin( \ang)})$);
    \coordinate (LA3) at ($({-2-0.4*cos( \ang)}, {2*sin( \ang)})$);
    \coordinate (RB1) at ($({ 2-0.4*cos(-\ang)}, {2*sin(-\ang)})$);
    \coordinate (RB2) at ($({ 4-0.4*cos(-\ang)}, {2*sin(-\ang)})$);
    \coordinate (RB3) at ($({ 6-0.4*cos(-\ang)}, {2*sin(-\ang)})$);
    \coordinate (RA1) at ($({ 2-0.4*cos( \ang)}, {2*sin( \ang)})$);
    \coordinate (RA2) at ($({ 4-0.4*cos(\dang)}, {2*sin(\dang)})$);
    \coordinate (RA3) at ($({ 6-0.4*cos(\ddang)}, {2*sin(\ddang)})$);
    \coordinate (SA1) at ($({ 2-0.4*cos( \ang)}, {2*sin( \ang)})$);
    \coordinate (SA2) at ($({ 4-0.4*cos( \ang)}, {2*sin( \ang)})$);
    \coordinate (SA3) at ($({ 6-0.4*cos( \ang)}, {2*sin( \ang)})$);
    \foreach \k in {-6,-4,-2,2,4,6}{
    \draw[blue,very thick,domain=-\ang:\ang] plot ({\k-0.4*cos(\x)}, {2*sin(\x)});
    }
    \draw[magenta,very thick,domain=\ang:\dang] plot ({4-0.4*cos(\x)}, {2*sin(\x)});
    \draw[magenta,very thick,domain=\ang:\ddang] plot ({6-0.4*cos(\x)}, {2*sin(\x)});
    \draw[very thick,green!50!black] (LB1) -- (LB2) -- (LB3) (LA1) -- (LA2) -- (LA3);
    \draw[very thick,green!50!black] (RB1) -- (RB2) -- (RB3) (RA1) -- (RA2) -- (RA3);
    \foreach \n in {LB1,LB2,LB3,LA1,LA2,LA3,RB1,RB2,RB3,RA1,RA2,RA3}{
    \node[spot] at (\n){};
    }
    \foreach \n in {SA2,SA3}{
    \node[oldspot] at (\n){};
    }
    \node[blue,left] at (-6.4,0){$s^A_0$};
    \node[magenta,left,fill=red!10,inner sep=0pt] at (6.9,1.4){$\Delta s^A$};
    \foreach \j in {-1.0,-0.5,0,0.5,1.0}{
    \draw[red,thick,-Latex] (\j,-3) -- (\j,3);
    }
    \node[red,below,align=center,text width=3cm] at (0,-3){\textit{\textbf{Non-stationary phase}}};
    \tikzset{snake it/.style={decorate, decoration=snake}};
    \draw[very thick,green!50!black,snake it,segment length=21.5] (LB3) -- (RB1);
    \draw[very thick,green!50!black,snake it,segment length=21.5] (LA3) -- (RA1);
    \node[text width=4.75cm,align=center,above] at (-5,-4){\textbf{Minkowski vacuum}\\ ($N_{AB}^{vac}[\Phi]$ $=$ 0)};
    \node[text width=4.75cm,align=center,above] at ( 5,-4){\textbf{Another vacuum}\\ ($N_{AB}^{vac}[\Phi]$ $\neq$ 0)};
    \draw[very thick,-Latex,orange!85!red] (-5,-4) |- (0,-5) -| ( 5,-4);
    \node[orange!85!red,below] at (0,-5.2){\textbf{\textit{Superboost transition}} $\bm{\delta_Y \Phi = D_A Y^A}$};
    \node[red!50!black] at (-9,0){$\cI^+$};
    \end{tikzpicture}
\caption{Velocity kick memory effect.}
\label{fig:VelocityKick}
\end{figure}

Let us consider a Bondi detector formed by two free-falling observers near $\cI^+$ that evolve at some finite large radius $\rho$ in the impulsive gravitational wave spacetime \eqref{imp}. We continue to work in the Newmann-Unti coordinate system and we just study pure superboost transitions for the moment, \textit{i.e.} transitions with $\Delta C = C_+ = C_- = 0$. We also choose to work with observers away from the singular points on the celestial sphere at any retarded time $u$, so that we can ignore the singularities due to the meromorphic superboosts. By definition, the observers have a 4-velocity like $v^\mu \p_\mu = \p_u + \mathcal{O}(\rho^{-1})$ as before. The deviation vector $s^\mu$ between two neighboring geodesics obeys again \eqref{geodesic deviation}. As in section \ref{sec:Linear displacement memory effect}, we assume that $s^\rho=0$ and $s^A = s^A_0$ at early times. Since we are away from the singularities, the boundary metric can be assumed to stick at $\mathring q_{AB}$ and we have $R_{uA uB}= -\frac{\rho}{2} \p^2_u C_{AB} + \mathcal{O}(\rho^0)$ as before. Here $C_{AB} = u \Theta(u) N_{AB}^{vac}$ and therefore 
\begin{equation}
q_{AB} \p_u^2 s^B = \frac{1}{2\rho} \delta(u) N_{AB}^{vac}s^B + \mathcal{O}(\rho^{-2}). \label{geodesic for velocity kick} 
\end{equation}
When dealing with impulsive transitions, any primitive or derivative in $u$ involving $\Theta(u)$ or $\delta(u)$ is intended in the sense of distributions. For instance, $\partial_u \Theta(u)\sim \delta(u)$ and $\partial_u (u\Theta(u)) \sim u\delta(u) + \Theta(u) \sim \Theta(u)$ when integrated against any numerical test function in $u$. We deduce from \eqref{geodesic for velocity kick} that the angular deviation is given by $s^A =s^A_{0} + \rho^{-1} \Delta s^A (u,x^A)+\mathcal{O}(\rho^{-2})$ and after two integrations in $u$, $\delta(u)\to\Theta(u)\to u\Theta(u)$ so 
\begin{equation}
\Delta s^A = \frac{u}{2} \Theta(u) q^{AB } N_{BC}^{vac}s^C_{0} . \label{devG}
\end{equation}
Before the shockwave, there is no relative angular velocity between observers. After the shockwave, this relative angular velocity is turned on at order $\propto \rho^{-1}$, since the right-hand side of \eqref{devG} is linear in $u$. We observe a \textit{velocity kick} between two such neighboring geodesics due to the shockwave \cite{Podolsky:2002sa,Podolsky:2010xh,Podolsky:2016mqg}, see figure \ref{fig:VelocityKick} for a schematic picture. This is a qualitatively distinct effect from the displacement memory effect \cite{Zeldovich:1974gvh,Blanchet:1987wq,Christodoulou:1991cr,Blanchet:1992br,0264-9381-9-6-018} reviewed in section \ref{sec:Linear displacement memory effect}, the spin effect \cite{Pasterski:2015tva} and the center-of-mass effect \cite{Nichols:2018qac}.

Let us analyze if the same is true for incoming light rays traveling in the asymptotic region. One can consider a congruence of null geodesics which admits a constant leading angular velocity $\Omega^A \p_A$, with total 4-impulsion 
\begin{equation}
v^\mu \p_\mu = \left(\sqrt{\Omega^A q_{AB} \Omega^B} + \mathcal{O}(\rho^{-1})\right)\p_u + \mathcal{O}(\rho^{-1})\p_\rho + \frac{1}{\rho}\left( \Omega^A + \mathcal{O}(\rho^{-1})\right)\p_A . 
\end{equation}
We consider again a deviation vector of the form $s^A =s^A_{0} + \frac{1}{\rho}\Delta s^A (u,x^A)+\mathcal{O}(\rho^{-2})$. A simple computation convinces that the deviation vector obeys again \eqref{devG}. Null geodesics are thus refracted by the shockwave. This is the \textit{refraction memory effect} usually described in the bulk of spacetime \cite{Podolsky:2002sa,Podolsky:2010xh,Podolsky:2016mqg}. We identified here the class of null geodesics which displays the refraction memory effect close to future null infinity.   

Let us now shortly comment on the case where the change of boundary metric is not localized at individual points. The main point is that timelike geodesics will now admit non-trivial deviation vector already at leading order $\propto\, \rho^0$, $s^A = s^A_{lead} +\mathcal{O}(\rho^{-1})$, with  
\begin{equation}
\frac{1}{2} q_{AB} \p_u^2 s^B_{lead} + \frac{1}{2} \p_u^2 (q_{AB} s^B_{lead}) = - \frac{1}{2} \p_u^2 q_{AB} s_{lead}^B  .
\end{equation}
A velocity kick will therefore already occur at order $\rho^0$.

\subsubsection{A new non-linear displacement memory}
\label{sec:memory}

Until now, we have considered impulsive vacuum transitions \eqref{imp} without any shift in the supertranslation field. We explain now that there is a non-linear displacement memory induced by a superboost transition, when it is accompanied by a supertranslation transition. This case was not considered in \cite{Podolsky:2002sa,Podolsky:2010xh,Podolsky:2016mqg} where all supertranslation transitions were absent. In order to describe the effect, we can consider either timelike or null geodesics. For definiteness, we consider a congruence of timelike geodesics that evolve at some finite large radius $\rho$ in the general impulsive gravitational wave spacetime \eqref{metf2}. For simplicity, we assume that the metric is Minkowski in the past and we consider again the simplified case of a meromorphic superboost transition, assuming that the detector singled out in the timelike congruence is away from the singularities on the celestial sphere. In other words, we assume $\Phi_- = -\ln\sqrt{\mathring q}$ ($q^{vac}_{AB}[\Phi_-]=\mathring q_{AB}$), $C_- = 0$, $\Phi_+ = \phi(z)+\bar \phi(\bar z)-\ln\sqrt{\mathring q}$ and $C_+=C_+(x^A)$ is arbitrary. The 4-velocity is now $v^\mu \p_\mu = \sqrt{\frac{2}{R[q]}}\p_u +\mathcal{O}(\rho^{-1})$ and $R_{uA uB}= -\frac{\rho}{2} \p^2_u C_{AB} + \mathcal{O}(\rho^0)$ is not modified. Following the same procedure, we provide the power expansion $s^A =s^A_{0}+ \frac{1}{\rho}\Delta s^A (u,x^A)+\mathcal{O}(\rho^{-2})$ and away from the singular points on the sphere, we get
\begin{equation}
\begin{split}
\Delta s^A &= \frac{1}{2} q^{AB}C_{BC} s^C_{0}  \\
&= \frac{1}{2} \Theta(u)  \mathring q^{AB}C_{BC}^{vac} s^C_{0} \\
&= \frac{1}{2} \mathring q^{AB} \left(u \Theta(u) N_{BC}^{vac} +  \Theta(u) C^{(0)}_{BC}  + \Theta(u) C N_{BC}^{vac}  \right) s^C_{0}.
\end{split}
\end{equation}
The first term $\propto\, u \Theta(u)$ leads to the velocity kick memory effect. The second term $\propto\, \Theta(u) C^{(0)}_{BC} $ leads to the displacement memory effect due to a change of supertranslation field $C$ between the final and initial states \cite{Strominger:2014pwa}.  The third and last term $\propto\, \Theta(u) C N^{vac}_{BC} $ is a new type of non-linear displacement memory effect due to change of both the superboost field $\Phi$ and the supertranslation field $C$. The four lowest spherical harmonics $\texttt{\textit{l}}=0,1$ of $C$, interpretated as the center-of-mass, do not contribute to the standard displacement memory effect because they are zero modes of the differential operator $C_{AB}^{(0)}$ in \eqref{PhiL}. Here, they do contribute to the non-linear displacement memory effect. The modification of the supertranslation field, in particular of the center-of-mass, is determined by \eqref{eq4}, as discussed earlier. Note crucially that these observations are only relevant in the presence of superboost transitions driven by cosmic events that are able to modify the leading structure of the Bondi metric at null infinity. It is therefore not expected to be relevant for localized events such as compact binary mergers, the latter being perfectly understood in the reduced phase space built up from $\mathring{\mathcal S}_0^{\text{Mink}}$. We finally close this discussion of memory effects associated with generalized vacuum transitions in the presence of both supertranslation and superboost symmetries and are on track to address the last point of this thesis.

\section{Generalized BMS\texorpdfstring{$_4$}{4} finite charges}
\label{sec:Generalized BMS4 finite charges}
In section \ref{sec:Vacua}, we derived a closed-form expression of the orbit of Minkowski spacetime under arbitrary super-Lorentz transformations. Thanks to this knowledge, we were able to describe the gravitational memory effects related to these leading symmetries in the previous section, \ref{sec:Gravitational memory effects}. In this last section, we describe the construction of finite integrable charges for the Generalized BMS$_4$ asymptotic symmetries of flat spacetime. We also compute the associated fluxes of charges and match them with quantum results at the linear level of the theory.

\subsection{Initial and late data in general super-Lorentz frames}
\label{sec:Initial and late data in general super-Lorentz frames}
For the purpose of this analysis, it is primordial to study the behavior of the field at $\mathscr I^+_-$ and $\mathscr I^+_+$, which are respectively the past and future boundaries of $\mathscr I^+$ for $u\to -\infty$ and $u\to +\infty$. The asymptotic sphere $\mathscr I^+_-$ is bordering spatial infinity while its future counterpart $\mathscr I^+_+$ touches future timelike infinity. Prescribing some fall-off conditions in $u$ and asymptotic values when $|u|\to+\infty$ for the various fields parametrizing the phase space amounts to defining a set of \textit{initial} and \textit{late} data for any solution in the phase space which consists of an additional physically motivated input of boundary conditions. Here we focus on physically relevant solutions that start from a stationary configuration without radiation in the past and revert back to stationarity in the future. Note that, \textit{a priori}, this assumption rules out black hole formation.

Let us describe the consequences of assuming early and late stationarity in the restricted context of the asymptotically Minkowskian solution space $\mathring{\mathcal S}_0^{\text{Mink}}$ \eqref{S0Mink}. There is no allowed pure super-Lorentz transformation, hence the class of spacetimes under consideration is assumed to start from a stationary configuration in the far past $u\to -\infty$, solely labeled by a supertranslation field $C_-(x^A)$ and some zeros modes (ADM charges) and decay to another stationary configuration in the far future $u\to +\infty$, labeled by another supertranslation field $C_+(x^A)$ and another bunch of zero modes (frequently assumed to be zero). We have $C_{AB} \to C^{(0)}_{AB}[C_\pm]$ for $u\to\pm\infty$, given by \eqref{PhiL}. In particular, $N_{AB} \to 0$ in this limit and $C_{AB}$ is purely electric \cite{Strominger:2013jfa}, \textit{i.e.}
\begin{equation}
\left. (D_B D^C C_{AC} - D_A D^C C_{BC}) \right|_{\cI^+_\pm} = 0.
\label{Electricity condition MINK}
\end{equation}
This is a differential statement equivalent to say that $C_{AB}$ is \textit{pure-gauge} since it is only built from the supertranslation field $C_-$ or $C_+$. Since we have stationary configurations as temporal boundary conditions, we can equivalently say that the radiative phase is assumed to have compact support on $\mathscr I^+$, because the Weyl tensor is required to fall off to zero at large retarded times. Such radiative solutions have been defined in a rigorous way by Christodoulou and Klainerman \cite{Christodoulou:1993uv}. They showed that there exists a class of Cauchy data which decay sufficiently fast at spatial infinity such that the Cauchy problem leads to a smooth geodesically complete solution. In fact, they prove that the non-linear stability of Minkowski spacetime (\textit{i.e.} the hypothesis to be in a -- possibly supertranslated -- vacuum for $|u| \to+\infty$) requires that the Bondi news falls off as 
\begin{equation}
N_{AB} = \mathcal{O} \left( |u|^{-(1+\varepsilon)} \right) \label{CK conditions}
\end{equation}
when $ u \rightarrow \pm \infty$, for any $\varepsilon>0$ arbitrarily fixed, while the Bondi functions $M$ and $N_A$ remain finite in the two limits and we can endow the asymptotic vacua by these zero modes. Now, even if black holes form in the spacetime, we do not expect that these quantities will behave differently since they do not emit any radiation at early or late retarded times. We can thus safely assume that all physical asymptotically flat spacetimes worth of interest in $\mathring{\mathcal S}_0^{\text{Mink}}$ obey these additional boundary conditions. Using \eqref{duM} and \eqref{EOM1} restricted on this solution space, we can check explicitly that $M$ and $N_A$ become constant while approaching spatial infinity, forming the total ADM mass and angular momentum of the spacetime after integration on $\mathscr I^+_-$. The constraint \eqref{Electricity condition MINK}, obeyed by the boundary tensor $C_{AB}^{(0)}[C_-]$, has to be used explicitly in \eqref{EOM1} to show that $\partial_u N_A\to 0$ at spatial infinity. 

Let us now see how these considerations can be extrapolated (or not) in the Generalized BMS$_4$ solution space $\mathring{\mathcal S}_0$ \eqref{S0ring}. Again, we demand that any spacetime under consideration decays to stationary configurations when $|u|\to+\infty$. By consistency, the spacetime in these stationary regions is one of the Generalized BMS$_4$ vacua possibly endowed with zeros modes. Hence, the boundary conditions are described by a triplet of boundary fields $(C_-,\Phi_-,\widetilde\Phi_-)$ in the past and by another triplet of boundary fields $(C_+,\Phi_+,\widetilde\Phi_+)$ in the future, according to the results of section \ref{sec:The superboost, superrotation and supertranslation fields}. We write 
\begin{align}
q_{AB}^\pm  \equiv \lim_{u \rightarrow \pm\infty } q_{AB} &= q^{vac}_{AB}[ \gamma_{\underline{AB}}, \sqrt{\mathring{q}} ,\Phi_\pm ,\widetilde\Phi_\pm] +  o(u^{0}), \label{BCC3}\\ 
\lim_{u \rightarrow \pm\infty } C_{AB} &= C^{vac}_{AB}[q_{AB}^\pm, \Phi_\pm ,C_\pm ] +  o(u^{0}),\label{BCC2}
\end{align}
where $q_{AB}^{vac}$ and $C^{vac}_{AB}$ are defined in \eqref{qABform} and \eqref{PhiL} respectively. It follows that
\begin{equation}
\lim_{u \rightarrow \pm\infty } N_{AB} = N^{vac}_{AB}[ q^\pm_{AB}, \Phi_\pm ] +  o(u^{-1}) \label{BCC}
\end{equation}
where $N^{vac}_{AB}$ is defined in \eqref{PhiL}. When $u$ evolves, the initial fields $(C_-,\Phi_-,\Psi_-)$ can change with hard processes involving null radiation reaching $\mathscr I^+$. Even though this null radiation can originate from matter or from gravity itself, here we want to consider only gravity. This implies in particular that \eqref{pu qAB = 0} holds, hence 
\begin{equation}
q_{AB}\equiv q_{AB}^- = q_{AB}^+\, ,\quad \Phi \equiv \Phi_+ = \Phi_-\, , \quad \widetilde\Phi \equiv \widetilde\Phi_+ = \widetilde\Phi_-. \label{prevent superboost transitions}
\end{equation}
These conditions forbid transitions between the initial and final superboost and superrotation fields. The class of spacetimes that we are considering is therefore more general than those considered in \cite{Campiglia:2015yka,Christodoulou:1993uv} but lacks to include transitions by gravitational radiation in the boundary metric, since the latter can only be driven by a dynamical Weyl rescaling $q_{AB}\to e^{2\varphi(u,x^C)} q_{AB}$ \cite{Barnich:2010eb,Barnich:2019vzx}, as indicated by \eqref{EOM qAB time evolution}. Let us mention at this point that the Weyl rescaling can always be gauge fixed by a Weyl transformation allowed within the Bondi gauge to get \eqref{sqrt q round sphere}. This repackages the arbitrary time dependence in $\varphi(u,x^A)$ into the news tensor whose fall-offs in $u$ can thus be arbitrary. Here, since we assume the precise asymptotic behavior \eqref{BCC} for $N_{AB}$, we rule out solutions with arbitrary time-dependence in the Weyl factor. For instance, general Robinson-Trautman solutions \cite{Robinson:1960zzb,Stephani:2003tm}, whose only degree of freedom is precisely such a Weyl factor with an arbitrary dependency in $u$, are not present in the phase space. Recently, a proposal for more general phase spaces including Weyl-sourced Bondi news tensors has been given in \cite{Freidel:2021yqe,Freidel:2021cbc} but without allowing the Weyl factor to be time-dependent. It should be interesting to extend further these analyses, then gauge-fix the Weyl factor as \eqref{sqrt q round sphere} for the purpose to obtain a suitable generalization of \eqref{BCC} in that context.  

In an arbitrary superboost frame, \eqref{BCC} reads as $N_{AB} \to N^{vac}_{AB}[q_{AB},\Phi] + o(u^{-1})$ which, of course, does not obey the Christodoulou-Klainermann fall off condition \eqref{CK conditions}. This is fundamentally due to the fact that the Bondi news tensor transforms inhomogenenously under superboost, as stated by \eqref{dNAB}. Indeed, the term $(D_A D_B D_C Y^C)^{TF}$, shown to be zero for any global boost (annihilated by three derivatives because of the conformal Killing equation \eqref{Killing eq for rotations}), is not vanishing when extending the symmetries to superboosts. Starting from the Minkowski global spacetime (for which the Weyl tensor is identically zero) where $N_{AB} = 0$, any superboost generates a non-vanishing news tensor $N_{AB}^{vac}$ in the orbit of Minkowski while keeping a vanishing Weyl tensor. Hence, $N_{AB}$ does not formalize, strictly speaking, the notion we have of a ``physical'' Bondi news tensor that should be zero in the absence of gravitational radiation. We define the shifted tensor
\begin{equation}
\hat N_{AB} \equiv N_{AB} - N_{AB}^{vac}[\Phi] \label{shift NAB}
\end{equation}
and we call it the \textit{physical news tensor} for the following reasons. First of all, from \eqref{BCC} and \eqref{prevent superboost transitions}, we observe that $\hat N_{AB}$ obeys the fall-off requirement \eqref{CK conditions} that is suitable to describe well-behaved gravitational radiation around spatial infinity. Moreover, owing to \eqref{dNAB} and \eqref{dRN}, we observe that 
\begin{equation}
\delta_{T,Y}\hat N_{AB} = (T\partial_u + \mathcal L_Y + \frac{u}{2}D_C Y^C)\hat N_{AB}, \label{delta hat NAB}
\end{equation}
hence $\hat N_{AB}$ transforms homogeneously. Any vacuum configuration, with a vanishing Weyl tensor, will verify, in particular, the suitable condition $\hat N_{AB} = 0$. This refines the notion of stationarity in a covariant manner under the full Generalized BMS$_4$ group. Inspired by the shift \eqref{shift NAB} we also define
\begin{equation}
\hat C_{AB} \equiv C_{AB} - u N_{AB}^{vac}[\Phi] \label{shift CAB}
\end{equation}
whose transformation law is
\begin{equation}
\delta_{T,Y}\hat C_{AB} = [(T+\frac{u}{2}D_CY^C)\partial_u + \mathcal L_Y - \frac{1}{2}D_CY^C]\hat C_{AB} - 2(D_AD_BT)^{TF} + T N_{AB}^{vac}, \label{hatCAB transfo}
\end{equation}
and $\hat N_{AB} = \partial_u \hat C_{AB}$ is at most finite in $u$. In the vacua, by virtue of \eqref{PhiL}, we have
\begin{equation}
\lim_{u\to\pm\infty} \hat C_{AB} = C_\pm N_{AB}^{vac}[\Phi] + C_{AB}^{(0)}[C_\pm] = C_\pm N_{AB}^{vac}[\Phi] - (2 D_A D_B C_\pm)^{TF}
\end{equation}
which the generalization of a pure-supertranslation shear tensor (see \eqref{CAB parametrized} and \eqref{C field variation}) in an arbitrary superboosted frame. It is also purely electric in the sense that
\begin{equation}
\left. \left[ (D_B D_C - \frac{1}{2}N_{BC}^{vac})\hat C {{}_A}^C - (D_A D_C - \frac{1}{2}N_{AC}^{vac}) \hat C {{}_B}^C \right] \right|_{\mathscr I^+_\pm} = 0. \label{Electricity general}
\end{equation}
This generalizes \eqref{Electricity condition MINK} for any non-trivial superboost field $\Phi$ and can be proved thanks to \eqref{Li2} and \eqref{curl NABvac}. Using \eqref{Li2}, one can prove that $N_{AB}^{vac}$ trivially satisfies \eqref{Electricity general}, hence we can trade $\hat C_{AB}$ for $C_{AB}$ in this condition. Hence, to say that $C_{AB}$ is pure-gauge or purely electric as \eqref{Electricity general} are equivalent statements. The incorporation of ADM zero modes at spatial infinity does not change any of these conclusions, so we can demand that the spacetime decays to vacuum configurations for $|u|\to+\infty$ modulo these zero modes which however produce non-trivial contributions in the Weyl tensor (the Newtonian potential for instance). This ends our discussion on the behavior of the fields at $\mathscr I^+_\pm$. Let us explain now how to use these observations to build finite and well-behaved charges for the Generalized BMS$_4$ symmetries.

\subsection{Finite Hamiltonians from the Generalized BMS\texorpdfstring{$_4$}{4} charges}
\label{sec:Finite Hamiltonians}
As we explained in section \ref{sec:Symplectic structure and charges FLAT}, the Generalized BMS$_4$ infinitesimal surface charges \eqref{decomposition charge} are not integrable because the radiative flat spacetimes emit some flux through future null infinity $\mathscr I^+$. In that context, the definition of an Hamiltonian (\textit{i.e.} an integrable part) associated with any asymptotic symmetry $\xi$ is ambiguous because there is no preferred split between integrable and non-integrable parts in the infinitesimal surface charge in the absence of further boundary conditions allowing to fix this ambiguity. In other words, the covariant phase space methods only fix the infinitesimal charge variation, not the finite charge variation due to the lack of integrability: some prescription is therefore required to define the finite charge related to the asymptotic symmetries.

To that end, we introduce four requirements that are physically motivated and provide a set of minimal conditions to be obeyed by the suitable finite Hamiltonian $\bar H_\xi^{\text{GBMS}_4}[\phi]$. We denote by $\phi$ the whole collection of metric parts as well as foliations and boundary data we need, including the vacuum fields $\Phi,C_\pm$ previously defined. Here are the requirements:
\begin{enumerate}[leftmargin=1.5cm,label=\Roman*., ref=\Roman*]
\item $\bar H_\xi^{\text{GBMS}_4}[\phi]$ is finite in the non-compact Bondi coordinates $r$ and $u$. \label{item:I} 
\item The flux of $\bar H_\xi^{\text{GBMS}_4}[\phi]$ vanishes identically for stationary configurations $\hat N_{AB} = 0$. \label{item:II} 
\item $\bar H_\xi^{\text{GBMS}_4}[\phi]$ is zero in any vacuum configuration for which the Weyl tensor vanishes. \label{item:III} 
\item The algebra of charges $\bar H_\xi^{\text{GBMS}_4}[\phi]$ closes under the Poisson bracket \eqref{Poisson bracket jet space} at $\cI^+_\pm$. \label{item:IV} 
\end{enumerate}
The requirement \ref{item:I} is very natural since we want to manipulate well-defined Hamiltonians on each point of $\mathscr I^+ = \{r\to +\infty\}$ and consider the limits to spatial and (future) timelike infinities. Thanks to the regularization procedure developed in section \ref{sec:Asymptotically locally flat radiative phase spaces} and confirmed by the robust construction explained in section \ref{sec:Flat limit of the action and corner terms}, the finiteness in $r$ is already implemented for any Hamiltonian deducible from \eqref{decomposition charge}, including the minimal choice $H_\xi^{\text{GBMS}_4}[\phi]$ given by \eqref{FNCharges}. But using \eqref{BCC3}--\eqref{BCC}, one can show that the latter is not finite in $u$, hence it has to be denied as a good choice for the desired finite Hamiltonian. Recalling \eqref{split ambiguity N}, we define
\begin{equation}
\bar H_\xi^{\text{GBMS}_4}[\phi] \equiv H_\xi^{\text{GBMS}_4}[\phi] + \Delta H_\xi^{\text{GBMS}_4}[\phi] = \mathcal O(u^0). \label{shift flat hamiltonian}
\end{equation}

Among the prospective finite Hamiltonians, Wald and Zoupas \cite{Wald:1999wa} proposed to adjust the definition of $\bar H_\xi^{\text{GBMS}_4}[\phi]$ such that the local flux $\partial_u \bar H_\xi^{\text{GBMS}_4}[\phi]=0$ for stationary configurations. In the standard BMS$_4$ phase space, a stationary configuration verifies $N_{AB} = 0$. The latter is not a statement that is invariant under super-Lorentz transformations, see \eqref{dNAB} or the discussion above \eqref{shift NAB}. Instead we can impose that the physical news tensor $\hat N_{AB}$ vanishes in any super-Lorentz frame since it transforms homogeneously under all asymptotic symmetries, according to \eqref{delta hat NAB}. We are led to devise a generalized notion of stationarity demanding the vanishing of $\hat N_{AB}$ instead of $N_{AB}$. Hence the requirement \ref{item:II} is the extension of Wald-Zoupas' prescription to arbitrary super-Lorentz frames \cite{Compere:2018ylh}. 

Particular stationary configurations are the vacuum configurations which are locally flat (\textit{i.e.} the Weyl tensor vanishes). The vacua of gravitational fields are not expect to carry any mass or angular momentum that cannot be created by diffeomorphism acting on Minkowski spacetime. From this point of view, the suggestion \ref{item:III} is the reasonable requirement that the Riemann-flat vacuum is the reference that fixes the ``offset'' of the Hamiltonian $\bar H_\xi^{\text{GBMS}_4}[\phi]$. The curvature being a tensor field, a good choice of charges will give zero for this vacuum written in any Generalized BMS$_4$ frame. 

Finally recalling the theorem \eqref{Rep theo general} and its modification under some shift of the Hamiltonian as \eqref{cocycle shift}, we deduce that \eqref{shift flat hamiltonian} satisfies the following algebra under the standard Poisson bracket \cite{Compere:2020lrt}
\begin{equation}
\left\lbrace \bar H_{\xi_1}^{\text{GBMS}_4}[\phi] , \bar H_{\xi_2}^{\text{GBMS}_4}[\phi] \right\rbrace = \bar H_{[\xi_1,\xi_2]_\star}^{\text{GBMS}_4}[\phi] + R_{\xi_1,\xi_2}[\phi],
\end{equation}
where the residue
\begin{equation}
R_{\xi_1,\xi_2}[\phi] = K^{\text{GBMS}_4}_{\xi_1,\xi_2}[\phi] - \Delta H_{[\xi_1,\xi_2]_\star}^{\text{GBMS}_4}[\phi] + \delta_{\xi_2}\Delta H_{\xi_1}^{\text{GBMS}_4}[\phi] - \Xi_{\xi_2}^{\text{GBMS}_4}[\phi;\delta_{\xi_1}\phi] \label{residue flat}
\end{equation}
is defined from \eqref{cocycle flat} and \eqref{FNCharges non int}. Written in that way, it is not manifestly antisymmetric. At early and late times, the absence of radiation and physical flux suggests that the charge algebra should close under the standard Poisson bracket which does not involve any non-integrable contribution, since the latter would be intimately related to some non-equilibrium physics at the boundary, nonexistent by hypothesis at the corners $\mathscr I^+_\pm$. Hence demanding that \ref{item:IV} is respected is equivalent to affirm that there exists a shift \eqref{shift flat hamiltonian} such that $R_{\xi_1,\xi_2}[\phi] \to 0$ when $|u|\to+\infty$ for any couple of Generalized BMS$_4$ generators $\xi_1,\xi_2$. Considering \eqref{residue flat}, this brings an involved constraint on the shift $\Delta H_\xi^{\text{GBMS}_4}[\phi]$. Unlike the three constraints \ref{item:I}--\ref{item:III} that allow for a considerable ambiguity on the finite Hamiltonian (\textit{i.e.} any finite term vanishing in the vacuum and whose time derivative is zero when $\hat N_{AB}$ can be added without ruining any constraint), the last requirement \ref{item:IV} sharply restricts the set of admissible terms to be added to the Hamiltonian. We do not attempt here to find the general solution of \eqref{residue flat} in terms of $\Delta H_{\xi}^{\text{GBMS}_4}[\phi]$ which is a highly non-trivial task. But we can even restrict the possible ambiguity by demanding that the shift \eqref{shift flat hamiltonian} is a local function of $C_{AB}$, the initial data $q_{AB}$, $N_{AB}^{vac}$ and the gauge parameters $f,Y^A$ only. The dependency in the gauge parameters is assumed to be linear while the dependency in $C_{AB}$ or $N_{AB}^{vac}$ is at most quadratic.

With these clarifications in mind, we claim that the finite Hamiltonians satisfying the four prescriptions \ref{item:I}--\ref{item:IV} are given by
\begin{equation}
\boxed{
\bar H_\xi^{\text{GBMS}_4}[\phi] = \oint_{S_\infty} \frac{\D^2\Omega}{16\pi G} \left[ 4\, T\, \bar M + 2\,Y^A\,\bar N_A \right] \label{hamiltonian flat final}
}
\end{equation}
for
\begin{align}
\bar M &= M + \frac{1}{8}C_{AB}N^{AB}_{vac}, \label{bar M} \\
\bar N_A &= N_A - u\partial_A \bar M + \frac{1}{4}C_{AB}D_C C^{BC}+\frac{3}{32}\partial_A (C_{BC}C^{BC}) \nonumber \\
&\quad\, + \frac{u}{4}  D^B(D_B D_C - \frac{1}{2}N_{BC}^{vac}){C_A}^C - \frac{u}{4} D^B(D_A D_C - \frac{1}{2}N_{AC}^{vac}) {C_B}^C . \label{bar NA}
\end{align}
This is a new prescription for the Generalized BMS$_4$ Hamiltonians that reunites the expressions of \cite{Compere:2018ylh} and \cite{Compere:2020lrt}, the first paper focusing on both requirements \ref{item:I} and \ref{item:II} and the second one further discussing both requirements \ref{item:III} and \ref{item:IV} in the last section. We managed to reduce the amount of parentheses in \eqref{bar NA} but the $\mathcal O(u)$ should be read as $D^B S_{[AB]}$ for $S_{AB} = (D_AD_C-\frac{1}{2}N_{AC}^{vac}){C_B}^C$. It is worth noticing for the discussions below that the differential operator defining $S_{[AB]}$ first appeared in the generalized electricity condition \eqref{Electricity general}.

We can immediately check by a straightforward computation that the charges \eqref{hamiltonian flat final} are finite in $u$ at the corners $\cI^+_\pm$. Indeed, we can identify the leading terms in \eqref{duM} and \eqref{EOM1} when $|u|\to+\infty$, providing the divergences in $\mathcal O(u)$ in $M$ and $N_A$ after integration in $u$. This information can be used to prove that \eqref{bar M} and \eqref{bar NA} go to constants at early and late times, because $\partial_u\bar M\to 0$ and $\partial_u\bar N_A\to 0$. We do not give the detailed demonstration here. 

The explicit shift that must be introduced in \eqref{shift flat hamiltonian} is
\begin{equation}
\boxed{
\begin{aligned}
\Delta H_\xi^{\text{GBMS}_4}[\phi] = &\oint_{S_\infty} \frac{\D^2\Omega}{16\pi G}  \left[\frac{1}{2}f C_{AB}N^{AB}_{vac} + \frac{1}{2}Y^A C_{AB}D_C C^{BC} + \frac{1}{8} Y^A \partial_A (C^{BC}C_{BC}) \right. \\
&+ \frac{u}{2} \left. Y^A D^B(D_B D_C - \frac{1}{2}N_{BC}^{vac}){C_A}^C -  \frac{u}{2} \, Y^A D^B(D_A D_C - \frac{1}{2}N_{AC}^{vac}) {C_B}^C \right].  \label{shift flat final}
\end{aligned}
}
\end{equation}
The first term aims at subtracting the vacuum contribution to the Bondi mass aspect \eqref{valvac}. The second and third terms operate similarly on the vacuum Bondi angular momentum aspect, which demonstrates that \eqref{hamiltonian flat final} is identically zero if the Weyl curvature is zero and thus obeys the prescription \ref{item:III}. The last terms, linear in $C_{AB}$, are vanishing in the vacua for which $\hat C_{AB}$ is purely electric, see \eqref{Electricity general} and do not contribute in the limit $|u|\to+\infty$ for our choice of boundary conditions in $u$. Let us stress again that $C_{AB}$ can be traded for $\hat C_{AB}$ without cost thanks to \eqref{Li2}. These terms were not included in the charge expression given in \cite{Compere:2020lrt} but do not change any conclusion of this paper since they vanish identically in the vacuum as well as at $\mathscr I^+_\pm$. Nevertheless, they are necessary to obey the prescription \ref{item:II} in general at any retarded time $u$, as shown in \cite{Compere:2018ylh}. Indeed, nothing prevents $\hat C_{AB}$ from getting a non-trivial magnetic part in the non-radiative phases for $|u|<+\infty$ when $\hat N_{AB}=0$ and the latter can contribute to the flux, which should contradict our hypothesis \ref{item:II}. We claim that the shift \eqref{shift flat final} is the unique local function of $q_{AB}$, $C_{AB}$ and $N_{AB}^{vac}$ only such that \eqref{hamiltonian flat final} fulfills the requirements \ref{item:I}--\ref{item:IV}. 

In a Bondi frame where $q_{AB}=\mathring q_{AB}$ and $N_{AB}^{vac} = 0$, we have
\begin{align}
\bar M &= M, \\
\bar N_A &= N_A - u\partial_A M + \frac{1}{4}C_{AB}D_C C^{BC}+\frac{3}{32}\partial_A (C_{BC}C^{BC}) \nonumber \\
&\quad\, + \frac{u}{4}  D^BD_B D_C {C_A}^C - \frac{u}{4} D^BD_A D_C {C_B}^C . \label{second line}
\end{align}
The ``soft'' $\mathcal O(u)$ terms still appear in \eqref{second line} and only depend on the magnetic part of the shear. Performing integrations by parts on the sphere, we observe that these terms contribute as $\sim uC_{zz}D_z^3Y^z+\text{c.c.}$ in the charges when working in stereographic coordinates and therefore vanish identically for global Lorentz symmetries. They are shown to be mandatory in the discussion of the subleading soft theorem \cite{Distler:2018rwu,Compere:2018ylh}, see section \ref{sec:Leading and subleading soft theorems}. But if we restrict only to Lorentz generators $Y$, we have
\begin{equation}
\boxed{
\mathscr L_Y \equiv H_{\xi(0,Y)}^{\text{GBMS}_4}[\phi] = \oint_{S_\infty} \frac{\D^2\Omega}{16\pi G} Y^A \Big[ N_A - u\partial_A M + \frac{1}{4}C_{AB}D_C C^{BC}+\frac{3}{32}\partial_A (C_{BC}C^{BC}) \Big].}
\end{equation}
Our proposal for the angular momentum thus coincides with \cite{Strominger:2014pwa,Pasterski:2015tva}, see the matching of conventions regarding the Bondi angular momentum aspect below \eqref{eq:DivDAB=0}.

\subsection{Flux algebra and realization at spatial infinity}
\label{sec:Flux algebra and realization at spatial infinity}
Inserting the shift \eqref{shift flat final} into \eqref{residue flat} while evaluating \eqref{FNCharges non int} and \eqref{cocycle flat}, we can show after a quite long computation that $R_{\xi_1,\xi_2}[\phi] = 0$ at the corners $\mathscr I^+_-$ and $\mathscr I^+_+$. This is precisely what we wanted by imposing the requirement \ref{item:IV} and the Generalized BMS$_4$ surface charge algebra closes under the standard Poisson bracket at $\mathscr I^+_-$ and $\mathscr I^+_+$ \textit{without} any central extension, \textit{i.e.}
\begin{equation}
\boxed{
\lbrace \bar H_{\xi_1}^{\text{GBMS}_4}[\phi],\bar H_{\xi_2}^{\text{GBMS}_4} [\phi] \rbrace \Big\vert_{\mathscr I^+_\pm} = \bar H_{[\xi_1,\xi_2]_\star}^{\text{GBMS}_4}[\phi] \Big\vert_{\mathscr I^+_\pm}.
} 
\label{alg5}
\end{equation}
This is a highly non-trivial result. For vacuum configurations, this is direct because of the hypothesis \ref{item:III} since all Generalized BMS$_4$ charges $\bar H_{\xi}^{\text{GBMS}_4}[\phi]$ are zero. But the result obtained here is stronger and means that this algebra is valid also for non-vacuum configurations with Poincar\'e charges such as mass and angular momentum, in the presence of displacement memory and in arbitrary Lorentz and super-Lorentz frames. It also applies at the corners of the past null boundary $\mathscr I^-$ under similar boundary conditions. In fact, since there is no flux at spatial infinity, all Generalized BMS$_4$ Hamiltonians $\bar H_{\xi}^{\text{GBMS}_4}[\phi]$ are \textit{conserved} at spatial infinity and the Generalized BMS$_4$ charge algebra \eqref{alg5} is therefore \textit{realized} also at spatial infinity, including the super-Lorentz transformations. Since the Generalized BMS$_4$ Hamiltonians are all generically non-vanishing from their explicit expressions (5.1) and (5.9) of \cite{Compere:2019gft}, we have proven the following theorem:
\resuwt{
The asymptotic symmetry algebra of asymptotically locally flat spacetimes in $\mathring{\mathcal S}_0$ with non-radiative boundary conditions at early and late times is the Generalized BMS$_4$ charge algebra \eqref{alg5} without central extension.
}
This means that the Generalized BMS$_4$ asymptotic symmetry group acts on the whole boundary of flat spacetime, including spatial infinity. This extends the result derived in \cite{Troessaert:2017jcm,Henneaux:2018cst} in the Hamiltonian formalism and includes super-Lorentz asymptotic symmetries as well. 

Now let us consider the local fluxes of the charges. They are given at any $u$ by
\begin{equation}
\boxed{
\begin{aligned}
\partial_u &H^{\text{GBMS}_4}_{\xi(T,Y)}[\phi] \\
&= -\frac{1}{32\pi G}\oint_S \left[ f\left( N^{AB} \hat N_{AB} - 2 D^A D^B \hat N_{AB} \right) - 2 Y^A C_{AB}D_C \hat N^{BC} - Y^A D_B (C_{AC}\hat N^{BC})    \right. \\
&\phantom{-\frac{1}{32\pi G}Y^A \Big[}\quad \left. - u Y^A D_B (D^B D^C - \frac{1}{2}N^{BC}_{vac})\hat N_{AC}+ u Y^A D_B (D_A D_C - \frac{1}{2}N_{AC}^{vac})\hat N^{BC} \right]
\end{aligned}
}\label{flux STR}
\end{equation}
where the parameter $f = T+\frac{u}{2}D_AY^A$ has been retained against $T$ for the sake of conciseness. To obtain these expressions, several manipulations are necessary. They involve the equations of motion \eqref{duM} and \eqref{EOM1}, the boundary conditions \eqref{BCC3}--\eqref{BCC}, the redefinitions \eqref{shift NAB} and \eqref{shift CAB} as well as the identities \eqref{identite glenn} and \eqref{identite FN}. Thanks to the ``Christodoulou-Klainermann fall-off condition'' $\hat N_{AB} = o(u^{-1})$, it is easy to show that the local fluxes \eqref{flux STR} are vanishing in the limit $|u|\to +\infty$, in accordance with the requirements \ref{item:I}--\ref{item:III}. Hence the Hamiltonians \eqref{hamiltonian flat final} go to constants at the corners, are finite in $u$ and their fluxes go to zero when the spacetime reverse back to stationary configurations at early and late times. We also observe that these fluxes vanish identically on any interval $\mathscr U\subset \cI^+$ if and only if $\hat N_{AB} = 0$ for all $u\in \mathscr U$. They are thus zero for stationary configurations and obey the requirement \ref{item:III}! Note crucially that the two last terms in \eqref{shift flat hamiltonian} are necessary to have the corresponding differential operator $D_AD_B-\frac{1}{2}N_{AB}^{vac}$ acting on $\hat N_{AB}$ in \eqref{flux STR} instead of the whole $C_{AB}$ which may contains a magnetic part contributing to the flux in generic radiative situations. 

We define the \textit{total integrated fluxes} at $\mathscr I^+$ as
\begin{equation}
\bar F_{\xi}[\phi] = \bar H^{\text{GBMS}_4}_{\xi}[\phi]  \Big|_{\mathscr I_+^+} - \bar H^{\text{GBMS}_4}_{\xi}[\phi]  \Big\vert_{\mathscr I_-^+} = \int_{-\infty}^{+\infty} \D u\,  \p_u \bar H^{\text{GBMS}_4}_{\xi}[\phi]. \label{eq:Fluxdiff}
\end{equation}
We denote the supermomenta fluxes and super-Lorentz fluxes as $\bar{P}_{T}[\phi] \equiv \bar F_{\xi(T,0)}[\phi]$ and $\bar{J}_{Y}[\phi] \equiv \bar F_{\xi(0,Y)}[\phi]$, respectively. An immediate consequence of \eqref{alg5} is the following algebra of Generalized BMS$_4$ fluxes, 
\begin{equation}
\boxed{
\lbrace \bar{P}_{T_1},\bar{P}_{T_2} \rbrace = 0, \quad \lbrace \bar{J}_{Y_1},\bar{P}_{T_2}\rbrace = \bar{P}_{Y_1(T_2)}, \quad \lbrace \bar{J}_{Y_1},\bar{J}_{Y_2} \rbrace = \bar{J}_{[Y_1,Y_2]} 
}
\end{equation} 
where $Y_1(T_2)\equiv (Y_1^A\partial_A - \frac{1}{2} D_A Y^A_1)T_2$. Therefore, the algebra of Generalized BMS$_4$ fluxes represents the Generalized BMS$_4$ algebra of asymptotic symmetries \eqref{eq:BMSCommu} without central extension. This confirms the construction of \cite{Campiglia:2020qvc}, which can now be deduced from covariant phase space methods where renormalization is provided by the flat limit of the $\Lambda \neq 0$ holographic renormalization scheme and our new treatment of corner terms, presented in section \ref{sec:Flat limit of the action and corner terms}.

\subsection{Infrared structures and soft theorems}

In this section, we give important comments on the various relationships existing in the low energy sector of the asymptotically flat theory of gravity, between physical observable predictions (gravitational memory effects), mathematical features of the radiative phase spaces (the Generalized BMS$_4$ symmetries) and infrared factorization properties of the scattering quantum amplitudes (the soft theorems) \cite{Strominger:2017zoo,Strominger:2014pwa}. We review the junction conditions that must be imposed around spatial infinity in order to be provided with a well-defined gravitational scattering problem \cite{Strominger:2013jfa}, implying asymptotic conservation laws for between past and future null infinities for the finite Hamiltonian derived in section \ref{sec:Finite Hamiltonians}. Under these additional boundary constraints, we show that the flux of charges computed in section \ref{sec:Flux algebra and realization at spatial infinity} reproduces the flux-balance laws needed for the matching with the leading \cite{Weinberg:1965nx,He:2014laa} and subleading \cite{Cachazo:2014fwa,Campiglia:2014yka} soft theorems. Finally, we review Strominger's leading infrared triangle of relationships \cite{Strominger:2017zoo}, involving the supertranslation symmetry and contrasts it with an hybrid overleading/subleading incomplete square of relationships, this time involving the super-Lorentz symmetries.

\subsubsection{Gravitational scattering problem and junction conditions}
\label{sec:Scattering}


The null motion in asymptotically flat spacetimes at null infinity is governed by hyperbolic equations of motion for which initial conditions are given at past null infinity $\cI^-$ and the evolution of null rays propagates this information to future null infinity $\cI^+$. Both limiting hypersurfaces come with the same structure, namely a null normal $\bm T$ and a boundary volume $\sqrt{\mathring q}$. These structures are preserved by two copies of the Generalized BMS$_4$ group, one living on $\mathscr I^+$ and denoted as G-BMS$_4^+$ and an homologous one, denoted as G-BMS$_4^-$, living on $\mathscr I^-$ where the whole analysis performed so far can be repeated up to considering advanced Bondi coordinates $(v,r,x^A)$ instead of retarded Bondi coordinates $(u,r,x^A)$. The evolution of null matter as well as gravitational waves from data imposed around $\mathscr I^-$, crossing the bulk of spacetime and reaching $\mathscr I^+$ defines the (\textit{null}) \textit{scattering problem} in asymptotically flat spacetimes. It gives the fundamental element to build the S-matrix for gravity, at least in the low-energy (infrared) regime, for which the ``in'' states are incoming at $\mathscr I^-$ and the ``out'' states are outgoing at $\mathscr I^+$. As argued by Strominger in \cite{Strominger:2013jfa}, a prescription should be given to relate the ``in'' and the ``out'' states in order to ensure the well-definiteness of the scattering problem. In particular, without such matching conditions, the two copies of Generalized BMS$_4$ are independent, which is not suitable: indeed, the action of a G-BMS$_4^-$ transformation on the initial data should be reflected at some point in the late data arriving at $\cI^+$. In other words, the two asymptotic groups cannot be independent symmetries acting on both the initial and final states. Moreover, there is one and only one copy of the Generalized BMS$_4$ group acting at spatial infinity, which also suggests that a diagonal subgroup of G-BMS$_4^-$ $\times$ G-BMS$_4^+$ has to be selected and manifests its presence at spatial infinity \cite{Compere:2017knf,Troessaert:2017jcm,Compere:2020lrt,Henneaux:2018cst,Henneaux:2019yax}.

The remaining question is: which diagonal subgroup is selected? Or in other words, how to identify the Generalized BMS$_4^+$ generators at $\mathscr I^+$ with the ones at $\mathscr I^-$? The crucial clue is that propagating null fields in the Minkowski vacuum obey universal \textit{antipodal matching conditions} at spatial infinity. For each bulk field $\phi^i$, one can define its limit at $\mathscr I^+$ and then take $u \rightarrow -\infty$ which defines the field at $\mathscr I^+_-$. The field at $\mathscr I^-_+$ is obtained in a similar way up to the standard trading of $u$ for $v$. An antipodal matching condition would mean that 
\begin{equation}
\phi^i(\theta,\varphi)\Big|_{\mathscr I^+_-} = \phi^i(\pi - \theta,\varphi + \pi )\Big|_{\mathscr I^-_+} 
\end{equation}
where $(\theta,\varphi)$ are the colatitude and azimuth angles on the sphere at $|u|\to+\infty$. For instance, it turns out that the electromagnetic Li\'enard-Wiechert field describing the retarded electromagnetic field of a uniformly moving source obeys these antipodal matching conditions \cite{Kapec:2015ena,Campiglia:2015qka}. The metric of the boosted Kerr black hole \cite{Soares:2018yym} is also expected to obey these conditions. Moreover, these conditions are $CPT$ invariant and Lorentz invariant. Since the seminal work \cite{Strominger:2013jfa}, it is admitted that the metric field in asymptotically flat spacetimes obeys the antipodal matching conditions. Yet, it has not been derived whether or not the antipodal map is generally valid for any subleading field in the asymptotic expansion close to null infinity and for other types of causal structures, such as spacetimes containing a black hole formed from stellar collapse. For the leading order fields, consistent boundary conditions, which admit antipodal matching boundary conditions, were found both in $3d$ \cite{Compere:2017knf} and $4d$ Einstein gravity \cite{Troessaert:2017jcm,Henneaux:2018cst,Henneaux:2019yax}. 

From the perspective of perturbative quantum gravity, the antipodal matching conditions imply the BMS invariance of the gravitational S-matrix in the sense that acting with the Generalized BMS$_4$ fluxes $\bar P_T$, $\bar J_Y$ on the ``in'' states or ``out'' states are two commuting operations: 
\begin{equation}
\bar P_T^+ \text{S}  = \text{S} \bar P_T^-, \qquad \bar J_Y^+ \text{S}  = \text{S} \bar J_Y^-, \label{Smatrix commu}
\end{equation}
where the superscript $+$ (resp. $-$) denotes the evaluation of the fluxes through $\cI^+$ (resp. $\cI^-$). The study for global BMS$_4$ has been developed in \cite{Strominger:2013jfa,He:2014laa} and extended to Diff($S^2$) super-Lorentz symmetries in \cite{Campiglia:2014yka,Campiglia:2015yka}. The invariance of the gravitational S-matrix under the asymptotic symmetry groups leads to Ward identities for supertranslations and Lorentz transformations or their extension. The interesting -- although not so suprising -- thing is that the supertranslation Ward identities are identical, after a change of notation, to the \textit{soft graviton theorem} \cite{He:2014laa}, derived by Weinberg in 1965 \cite{Weinberg:1965nx} for excitations around the Minkowski vacuum. At subleading order, it was shown in \cite{Campiglia:2014yka} that the Diff$(S^2)$ superrotation Ward identities are identical, after rewriting, to the newly found Cachazo-Strominger \textit{subleading soft graviton theorem} \cite{Cachazo:2014fwa}. The antipodal matching conditions are therefore compatible with the soft theorems, which validates their range of applicability around Minkowski spacetime. Indeed, sandwiching \eqref{Smatrix commu} between some ``in'' and ``out'' quantum states at $\cI^-$ and $\cI^+$ in order to compute the scattering amplitude $\mathcal M_n (p_1,\dots,p_n)$ for a $n$-particle process, one gets
\begin{equation}
\mathcal M_{n+1}(p_0,p_1,\dots,p_n) = \left[\frac{1}{p_0} S^{(0)}+S^{(1)}+ \mathcal O(p_0)\right]\mathcal M_n (p_1,\dots,p_n)  \label{soft thm}
\end{equation}
in the limit $p_0\to 0$. This states that in any scattering process involving $n+1$ particules among which one finds a massless particle with very weak momentum $p_0$ (a ``soft particle''), the contribution of the latter factorizes out in the scattering amplitude in the limit $p_0 \to 0$. The soft factors $S^{(0)}$ and $S^{(1)}$ are universal and only depend on the nature of the soft particle. The presence of $S^{(0)}$ is due to the supertranslation symmetry while the second factor $S^{(1)}$ is tightly related to super-Lorentz symmetries. Proposals to continue the Taylor series in $p_0$ in the soft limit have also been discussed \textit{e.g.} in \cite{DiVecchia:2016amo,Zlotnikov:2014sva} although the underlying tower of symmetry principles remains to be found.

Assuming that the antipodal matching conditions hold in generality directly leads to conservation laws. Indeed, physical quantities depend upon the fields and if all relevant fields (including the symmetry parameters) are antipodally matched, the finite Hamiltonian charges $\bar H_\xi^{\text{GBMS}_4}$ at $\mathscr I^+_-$ and $\mathscr I^-_+$ (either supertranslations or super-Lorentz) are related by the antipodal map symbolically denoted by $\text{AntiPodMap}(\circ)$, 
\begin{equation}
\bar H_\xi^{\text{GBMS}_4}\Big|_{\mathscr I^+_-} = \text{AntiPodMap}(\bar H_\xi^{\text{GBMS}_4})\Big|_{\mathscr I^-_+} 
\end{equation}
Using the flux-balance laws of these charges, we deduce by integration along $u$ and $v$ the conservation laws 
\begin{equation}
\begin{gathered}
\int_{\cI^+} \D u\: \partial_u \bar H_\xi^{\text{GBMS}_4} + \bar H_\xi^{\text{GBMS}_4}\Big|_{\mathscr I^+_+} = \text{AntiPodMap} \Big( \int_{\mathscr I^-} \D v\: \partial_v \bar H_\xi^{\text{GBMS}_4} + \bar H_\xi^{\text{GBMS}_4}\Big|_{\mathscr I^-_-} \Big)\\
\Longleftrightarrow\\
\bar F_\xi^{\text{GBMS}_4} \Big|_{\cI^+} + \bar H_\xi^{\text{GBMS}_4}\Big|_{\mathscr I^+_+} = \text{AntiPodMap} \Big( \bar F_\xi^{\text{GBMS}_4} \Big|_{\cI^-} + \bar H_\xi^{\text{GBMS}_4}\Big|_{\mathscr I^-_-} \Big)
\end{gathered} \label{eq:antipod}
\end{equation}
including the eventual contribution of matter originating at past timelike infinity $\mathscr I^-_-$ and terminating at future timelike infinity $\mathscr I^+_+$. These supertranslation and super-Lorentz conservation laws are the \textit{conservation of energy} and \textit{angular momentum at each angle} on the celestial sphere \cite{Strominger:2013jfa}.

\subsubsection{Leading and subleading soft theorems}
\label{sec:Leading and subleading soft theorems}
Let us show how to match our expressions for the fluxes with the expressions of the literature used in the Ward identities that are equivalent to the leading \cite{Weinberg:1965nx} and subleading \cite{Cachazo:2014fwa} soft graviton theorems. The final flux \eqref{flux STR} can be decomposed in \textit{soft} (\textit{i.e.} linear in $\hat C_{AB}$, $\hat N_{AB}$) and \textit{hard} parts (\textit{i.e.} quadratic in $\hat C_{AB}$, $\hat N_{AB}$). Let us first consider the flux associated with supertranslations $\xi(T,0)$. We have
\begin{equation}
\bar F_{\xi(T,0)}^{\text{GBMS}_4}[\phi] = \int_{\mathscr I^+} \text{d}u \, \p_u \bar H^{\text{GBMS}_4}_{\xi(T,0)}[\phi] = F_S[T] + F_H[T],
\label{eq:FinalFlux}
\end{equation}
that is, after integrations by parts on the sphere,
\begin{equation}
F_S[T] = \frac{1}{16\pi G}\int_{\mathscr I^+} \text{d}u \, \text{d}^2\Omega\; \hat N^{AB} D_A D_B T = -\frac{1}{32\pi G}\int_{\mathscr I^+} \text{d}u \, \text{d}^2\Omega\; \hat N^{AB}\delta_T^I \hat C_{AB} \label{SoftT}
\end{equation}
and
\begin{equation}
F_H[T] = -\frac{1}{32\pi G}\int_{\mathscr I^+} \text{d}u \, \text{d}^2\Omega\; T\, \hat  N^{AB} N_{AB} = -\frac{1}{32\pi G}\int_{\mathscr I^+} \text{d}u \, \text{d}^2\Omega\; \hat N^{AB}\delta_T^H \hat C_{AB}. \label{HardT}
\end{equation}
Here $\delta_T^H \hat C_{AB}$ and $\delta_T^I \hat C_{AB}$ represents the homogeneous and inhomogeneous parts of the transformation under $\xi=\xi(T,0)$ of $\hat C_{AB}$ defined as \eqref{shift CAB}, which can be straightforwardly deduced from \eqref{hatCAB transfo}. In the standard case where $N_{AB}^{vac} = 0$, the flux of supermomentum \eqref{SoftT}-\eqref{HardT} reproduces (2.11) of \cite{He:2014laa} up to a conventional overall sign, which itself agrees with previous results \cite{Ashtekar:1978zz}. After imposing the antipodal matching condition on the renormalized Bondi mass aspect $\bar M$ \eqref{bar M} at spatial infinity, one can equate the flux on $\mathscr I^+$ with the antipodally related flux on $\mathscr I^-$. The result of \cite{He:2014laa} is precisely that the quantum version of this identity is the Ward identity of the leading soft graviton theorem. We have now obtained a generalization in the presence of a superboost background field.

The treatment of super-Lorentz fluxes requires a bit more care. Starting again from \eqref{flux STR}, one can isolate the terms associated with vectors $\xi(0,Y)$. After a non-trivial computation involving the identities \eqref{identite glenn} and \eqref{identite FN}, we can decompose the flux as
\begin{equation}
\bar F_{\xi(0,Y)}^{\text{GBMS}_4}[\phi] = \int_{\mathscr I^+} \text{d}u \, \p_u \bar H^{\text{GBMS}_4}_{\xi(0,Y)}[\phi] = F_S[Y] + F_H[Y] + Q_{\mathscr I^+_\pm}[Y].
\end{equation}
The hard term is completely analogous to \eqref{HardT} and reads as
\begin{equation}
\begin{split}
F_H[Y] &= -\frac{1}{32\pi G} \int_{\mathscr I^+} \text{d}u \, \text{d}^2\Omega\; \hat N^{AB}\left(\mathcal L_Y \hat C_{AB} + \frac{u}{2}D_CY^C\hat N_{AB} - \frac{1}{2}D_CY^C\hat C_{AB}\right)\\
&= -\frac{1}{32\pi G} \int_{\mathscr I^+} \text{d}u \, \text{d}^2\Omega\; \hat N^{AB}\delta^H_Y\hat C_{AB} 
\end{split} \label{HardY}
\end{equation}
where $\delta^H_Y$ is the homogeneous part of the transformation of $\hat C_{AB}$. The expression matches (up to the overall conventional sign) with equation (40) of \cite{Campiglia:2015yka} (see also (4.2) of \cite{Campiglia:2020qvc}). The soft part simplifies to
\begin{equation}
F_S[Y] = \frac{1}{16\pi G} \int_{\mathscr I^+} \D u\D^2\Omega \, u \, \hat N^{AB}\Sigma_{AB}
\end{equation}
for a symmetric traceless tensor $\Sigma_{AB}$ given by
\begin{equation}
\begin{split}
\Sigma_{AB} &= \left[D_AD_BD_CY^C-\Lie_Y N_{AB}^{vac} + \frac{1}{2}R[q]D_{(A}Y_{B)} - \frac{1}{2}D_{(A}(D^CD_C+\frac{1}{2}R[q])Y_{B)}\right]^{TF}\\
&= \left[ -\delta_Y N_{AB}^{vac} + \frac{1}{4} R[q]\delta_Y q_{AB} - \frac{1}{2}D_{(A} D^C \delta_Y q_{B)C} \right]^{TF}
\end{split}
 \label{SigmaAB}
\end{equation}
using \eqref{dRN} and \eqref{dqAB} in the form $D^C \delta_Y q_{AC} = D_C D^C Y_A + \frac{1}{2}R[q] Y_A$. The tensor \eqref{SigmaAB} is recognized as the generalization of (47) of \cite{Campiglia:2015yka} in the presence of a non-trivial boundary curvature and $N_{AB}^{vac}\neq 0$ (see also (4.14) of \cite{Campiglia:2020qvc}). Since $N_{AB}^{vac}$ and $q_{AB}$ are invariant under Lorentz transformations, $\Sigma_{AB} = 0$ for any boost or rotation. The last piece is the boundary term 
\begin{equation}
Q_{\mathscr I^+_\pm}[Y] = \frac{1}{32\pi G}\int_{\mathscr I^+_\pm} \D^2\Omega \left( Y^A \hat C_{AB} D_C\hat C^{BC} + \frac{1}{4}Y^A\partial_A (\hat C_{BC}\hat C^{BC}) \right)
\end{equation} 
where the notation $\int_{\mathscr I^+_\pm}$ stands for the difference of the sphere integrals between $\mathscr I^+_+$ and $\mathscr I^+_-$ (also proposed in \cite{Campiglia:2020qvc} in a different form). This is precisely the flux contribution of the shift needed to subtract the non-trivial super-Lorentz charges of the vacua and vanishes when $Y^A$ generates a Lorentz symmetry \cite{Compere:2016jwb,Compere:2018ylh}. Hence the flux of angular momentum is only \eqref{HardY} in agreement with \cite{Ashtekar:1981bq}. Furthermore, as a boundary term in the flux, it cancels out in the Ward identity and does not affect the soft theorems \cite{Campiglia:2020qvc}. 

Now that we identified our expressions with the ones of \cite{Campiglia:2015yka,Campiglia:2020qvc}, we can use their results. After imposing the antipodal matching condition on $\bar N_A$ at spatial infinity, one can equate the flux of super-Lorentz charge on $\mathscr I^+$ with the antipodally related flux on $\mathscr I^-$ as originally proposed in \cite{Hawking:2016sgy} (but where the expression for $\bar N_A$ should be modified to \eqref{bar NA}). The result of \cite{Campiglia:2015yka} is precisely that this identity is the Ward identity of the subleading soft graviton theorem \cite{Cachazo:2014fwa}. The same result holds here in an arbitrary super-Lorentz frame.

Finally, considering only the background Minkowski spacetime ($q_{AB} =\mathring {q}_{AB}$ and $N_{AB}^{vac} = 0$), one can check that in stereographic coordinates one has $\Sigma_{zz} =  \partial_z^3 Y^z =  D_z^3 Y^z$. The soft piece of the super-Lorentz flux then reads as 
\begin{equation}
F_S [Y] = \frac{1}{16\pi G}\int_{\mathscr I^+} \text{d}u \, \text{d}^2 z \, \sqrt{\mathring q} \, (u \hat N^{zz} D_{{z}}^3 Y^{{z}} + u \hat N^{\bar{z}\bar{z}} D_{\bar{z}}^3 Y^{\bar{z}} )
\end{equation}
where we keep $Y^A \partial_A = Y^z (z,\bar{z}) \partial_z + Y^{\bar{z}} (z,\bar{z}) \partial_{\bar{z}}$ arbitrary. In the case of meromorphic super-Lorentz tranformations, this reproduces equation (5.3.17) of \cite{Strominger:2017zoo} (up to the conventional global sign).  This concludes our checks with the literature concerning the soft theorems. It shows that the Ward identities of supertranslations and super-Lorentz transformations are equivalent to the leading and subleading soft graviton theorems following the arguments of \cite{He:2014laa,Campiglia:2014yka}. 

\subsubsection{Infrared relationships}
We close this chapter by summarizing the results discussed so far while giving some diagrammatic relationships between them. 

Since the works of Strominger and collaborators in the previous decade, we know that the BMS$_4$ supertranslation symmetry, the leading soft graviton theorem and the displacement memory effect form three corners of a triangle describing the leading infrared structure of asymptotically flat spacetimes at null infinity \cite{Strominger:2017zoo}. The three edges of the triangle can be described in the language of vacuum transitions, Ward identities and Fourier transforms, see Figure \ref{fig:leading}, in the following sense:
\begin{itemize}[label=$\rhd$]
\item A supertranslation $T$ induces a vacuum transition mapping the Minkowski vacuum $\eta_{\mu\nu}$ on a vacuum metric $g_{\mu\nu}[C]$ now labeled by a time-independent field whose non-trivial value is generated by \eqref{deltaC}. Although both vacuum carry zero charges, it is possible to probe the vacuum transition by the mean of a Bondi detector described in section \ref{sec:Linear displacement memory effect} built up from two geodesic observers far away from the sources of radiation near $\cI^+$. The permanent shift they will experience after the vacuum transition is the linear displacement memory effect that we reviewed in section \ref{sec:Linear displacement memory effect} and is precisely controlled by the (angle-dependent) amplitude of the supertranslation involved in the process. That gives the left arrow of the triangle.
\item Any symmetry is associated with a finite Hamiltonian as defined in section \ref{sec:Finite Hamiltonians} which obeys conservation laws for the scattering problem around spatial infinity assuming some matching conditions between past and future null infinities. We gave the motivations for the choice of antipodal matching conditions \cite{Strominger:2013jfa} in section \ref{sec:Scattering}. Under such conditions, the conservation laws can be sandwiched by ``in'' and ``out'' quantum states defined respectively at past and future null infinity and lead to a Ward identity for the supertranslation symmetry at quantum level \cite{Strominger:2014pwa,He:2014laa}. The Ward identity implies the factorization in quantum scattering amplitudes of the low-energy (soft) contributions, which is nothing but Weinberg's soft graviton theorem \cite{Weinberg:1965nx}. That gives the right arrow of the triangle.
\item Finally, as shown in \cite{Strominger:2014pwa}, if we refine the soft factor as $p_0^{-1}S^{(0)} \equiv \pi^{\mu\nu}S_{\mu\nu}$ where $\pi_{\mu\nu}$ is the polarization tensor of the soft graviton, the symmetric transverse traceless tensor $S_{\mu\nu}$ can be recast as the leading stationary phase approximation of the Fourier transform between time and momentum variables $u$ and $p_0$ of the metric shift occasioned by the displacement memory effect. The latter shall be expressed as the difference in the linear perturbation metric across the vacuum transition, in transverse traceless de Donder gauge \cite{1987Natur.327..123B}. In short we can say that the Weinberg's soft graviton theorem \textit{is} the Fourier transform of the kinematic equation governing the memory effect, written as \eqref{kinematic shift} in the Bondi gauge, which closes the triangle by the bottom arrow.
\end{itemize}
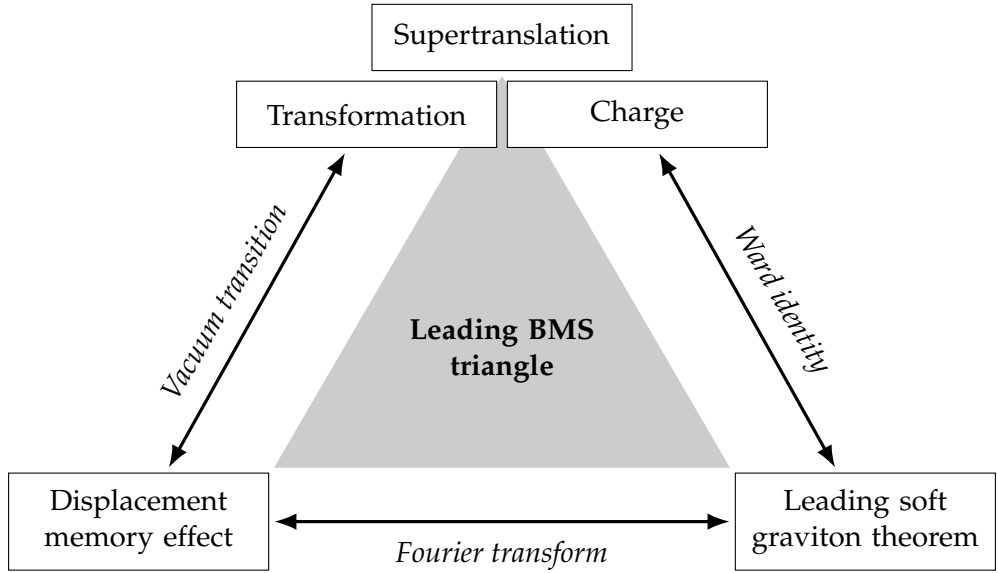
\begin{figure}[!ht]
\centering
\vspace{5pt}
\begin{tikzpicture}[boxx/.style={draw,fill=white,align=center,text width=3cm,inner sep=6pt,outer sep=2pt,minimum height=25pt}]
\def\lg{6};
\coordinate (L) at (0,0);
\coordinate (R) at (\lg,0);
\coordinate (T) at (60:\lg);
\fill[black!20] (L) -- (R) -- (T) -- cycle;
\draw (L) node[boxx,anchor=north east] (memory) {Displacement memory effect};
\draw (R) node[boxx,anchor=north west] (soft) {Leading soft graviton theorem};
\draw (T) node[boxx,anchor=south] (ST) {Supertranslation};
\draw (T) node[boxx,anchor=north east] (transf) {Transformation};
\draw (T) node[boxx,anchor=north west] (charge) {Charge};
\draw[very thick,Latex-Latex] (memory) -- (transf);
\draw[very thick,Latex-Latex] (memory) -- (soft);
\draw[very thick,Latex-Latex] (soft) -- (charge);
\draw[] ($(memory)!0.5!(transf)$) node[rotate=60,above,minimum height=25pt]{\textit{Vacuum transition}}; 
\draw[] ($(charge)!0.5!(soft)$) node[rotate=-60,above,minimum height=25pt]{\textit{Ward identity}}; 
\draw[] ($(memory)!0.5!(soft)$) node[below,minimum height=25pt]{\textit{Fourier transform}}; 
\coordinate (C) at ($(L)!0.5!(R)$);
\draw[] ($(C)!0.3!(T)$) node[text width=3cm,align=center]{\textbf{Leading BMS triangle}};
\end{tikzpicture}
\caption{Leading infrared triangular relationship.}
\label{fig:leading}
\end{figure}
The exceptional feature of the leading infrared triangle is that the diagram closes and can be looped in any direction. This is due to the fact that the supertranslation transformation as well as the associated charge describe objects at the same order in $1/r$. Indeed, the supertranslation symmetry does not modify the metric field at leading order on $\cI^+$, \textit{i.e.} $\delta_T q_{AB} = 0$. It impacts only the subleading field $C_{AB}$ and the more suppressed orders in $1/r$ in the power series of $g_{AB}$. On the other hand, the supertranslation charge is related to the subleading structure of the gravitational field, \textit{i.e.} the Bondi mass aspect, which appears also as a $1/r$ field in the metric tensor. Therefore, there is no bifurcation at the vertices of the diagram. 

The situation is radically different in the case of super-Lorentz symmetry. Indeed, for this complementary part of the Generalized BMS group, one needs to distinguish the symmetry -- which is ``overleading'' in the sense that it modifies the leading order metric at future null infinity -- and the charges that still describe the subleading structure of gravitational fields, \textit{i.e.} the angular momentum aspect. If we attempt to build a triangle as in Figure \ref{fig:leading} with the super-Lorentz symmetry sitting at the apex, this will be the locus of a bifurcation. The branch induced to the left from the ``transformation'' node will be a vacuum transition responsible for a shift in $q_{AB}$ at overleading order with respect to the shift induced by supertranslations, while the other branch running to the right from the ``charge'' node will lead to a subleading Ward identity which is definitely not in the same footing that the left branch. Hence we expect to get an open diagram which cannot be a cycle anymore. Let us find out what are the edges of this novel diagram.

\begin{figure}[!ht]
\vspace{5pt}
\centering
\begin{tikzpicture}[boxx/.style={draw,fill=white,align=center,text width=3cm,inner sep=6pt,outer sep=2pt,minimum height=25pt}]
\def\lg{4};
\coordinate (L) at (-\lg,0);
\coordinate (R) at (\lg,0);
\coordinate (T) at (0,\lg);
\coordinate (B) at (0,-\lg);
\fill[black!20] (L) -- (B) -- (R) -- (T) -- cycle;
\draw (L) node[boxx,left] (memory) {Refraction memory effect};
\draw (R) node[boxx,right] (soft) {Subleading soft graviton theorem};
\draw (T) node[boxx,anchor=south] (SL) {Super-Lorentz};
\draw (T) node[boxx,anchor=north east] (transf) {Transformation};
\draw (T) node[boxx,anchor=north west] (charge) {Charge};
\draw (B) node[boxx,anchor=north] (spin) {Spin \\ memory effect};
\draw[very thick,Latex-Latex] (memory.north) -- (transf.south west);
\draw[very thick,Latex-Latex] (soft.north) -- (charge.south east);
\draw[very thick,Latex-Latex] (spin.east) -- (soft.south);
\draw[] ($(memory.north)!0.5!(transf.south west)$) node[rotate=45,above,minimum height=35pt,text width=2.5cm,align=center]{\textit{Superboost transition}}; 
\draw[] ($(charge.south east)!0.5!(soft.north)$) node[rotate=-45,above,minimum height=35pt,text width=1.5cm,align=center]{\textit{Ward identity}}; 
\draw[] ($(spin.east)!0.5!(soft.south)$) node[rotate=45,below,minimum height=25pt,text width=3cm,align=center]{\textit{Fourier transform}}; 
\draw[] ($(L)!0.5!(R)$) node[text width=4.5cm,align=center]{\textbf{Overleading/Subleading BMS square}};
\end{tikzpicture}
\caption{Overleading/Subleading infrared relationship.}
\label{fig:subleading}
\end{figure}
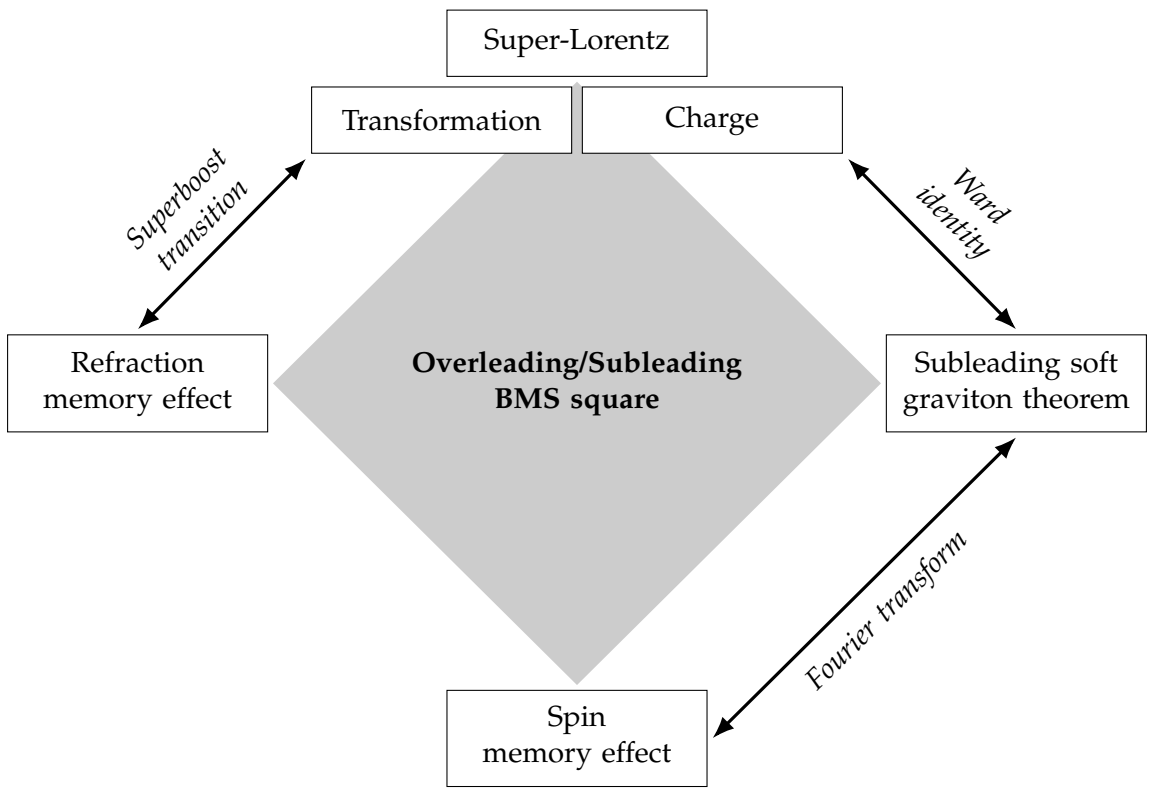
 
Two edges have been previously drawn relating super-Lorentz charges to the subleading soft graviton theorem by Ward identities \cite{Kapec:2014opa,Campiglia:2014yka,Campiglia:2015yka} and relating the subleading soft graviton theorem to the \textit{spin memory effect} by a Fourier transformation \cite{Pasterski:2015tva} in a perfect analogy with the leading set of relatioships. The spin memory effect is not described by a local memory field in the Bondi gauge, which is the reason of its absence in our description of gravitational memories. The spin memory effect is related to the time integral of the magnetic part of $C_{AB}$ during radiative phases, see \cite{Compere:2018aar} for more explanations. It was also shown that a local memory field leading to the spin memory effect can be found in retarded harmonic gauge among the radiative components of the Weyl tensor \cite{Himwich:2019qmj}. This completes the right chain of arrows in the Figure \ref{fig:subleading} connecting nodes which involve consistent objects at the same order in $1/r$ (namely $N_A$ and the magnetic part of $C_{AB}$). In section \ref{sec:Superboost}, we clarified how the superboost transitions lead to the refraction or velocity kick memory effect at null infinity. The latter are different from the spin memory effect and are encoded with local functions in the Bondi gauge. This gives the left arrow in the Figure \ref{fig:subleading} involving now exclusively overleading quantities with respect to supertranslations. This definitively suggests to describe the infrared structure related to super-Lorentz transformations by an hybrid overleading/subleading structure which looks like an (incomplete) square instead of a triangle.

It could be expected that a similar overleading/subleading square structure will also appear in the description of other gauge and gravity theories when the symmetry transformations and the associated charges do not probe and involve the fields at the same order in $1/r$. This is the last comment that we shall make to close this dissertation on infrared structures of asymptotically flat spacetimes.\hfill{\color{black!40}$\blacksquare$}

%
%

\chapter{Conclusion and perspectives}
\label{chapter:Conclusion}

In this final chapter, we summarize the main points from the various results discussed throughout the thesis, and then offer some ideas for further examination that we propose to follow in the future.

\section{Summary of the main results} 
We begin with a summary of the ideas and results defended in this manuscript, providing an overview of what has been discussed throughout the whole dissertation, which could be used as a memory refresh after reading.

\paragraph{Gravity with leaky boundary conditions} We began with a theoretical presentation of covariant phase space methods in gravity, mixing a review of well-known results and some brand new improvements.
\begin{itemize}[label=$\rhd$]
\item We introduced the dichotomic terminology to distinguish between conservative boundary conditions leading to stationary variational principles and leaky boundary conditions designed to model non-stationary situations for which some gauged degrees of freedom are fluxing through the boundary under consideration. The second class is the mandatory mathematical framework to formulate radiative solutions in General Relativity which lies at the core of our discussion and interests.
\item The permeability of the boundary yields non-equilibrium physics giving rise to non-conserved and non-integrable charges. The algebraic properties of non-integrable charges are encapsulated by the Barnich-Troessaert bracket, for which the representation theorem has been extended and proven using few assumptions, namely the existence of a phase space provided with a differential structure. 
\end{itemize}
This long technical introduction was required to understand the physical results obtained further in two (not so) distinct contexts, the Al(A)dS and the asymptotically flat spacetimes.
  
\paragraph{Asymptotically locally (A)dS spacetimes} First, we focused on the class of radiative Al(A)dS spacetimes in $d+1$ dimensions.
\begin{itemize}[label=$\rhd$]
\item In the absence of further boundary conditions beyond the minimal fall-offs allowing for conformal compactification, Al(A)dS$_{d+1}$ spacetimes admit a permeable boundary similar to the null boundary of asymptotically flat spacetimes. Arbitrary boundary diffeomorphisms are associated with finite surface Hamiltonian charges that obey a flux-balance law while Weyl rescalings admit a non-vanishing Hamiltonian charge only for even $d$. For these particular cases, this peculiar feature is seen as the manifestation of the Weyl anomaly in the dual conformal field theory. 
\item We proved that the Hamiltonians for the class of boundary diffeomorphisms represent the diffeomorphism algebra under the Barnich-Troessaert bracket without non-trivial central extensions for odd $d$. In even $d$ the non-vanishing contribution of the Weyl symmetry to the charges brings a non-trivial field-dependent 2-cocycle at the level of the algebra, reducing to the Brown-Henneaux central charge when $d=2$ with Dirichlet boundary conditions.
\item In order to distinguish the radiative boundary degrees of freedom among the components of the boundary metric and holographic stress-tensor, we proposed a boundary gauge fixing which can be reached locally without constraining the initial value problem. It consists in fixing a boundary foliation and measure, which reduces the boundary diffeomorphism algebra to the $\Lambda$-BMS$_{d+1}$ algebroid. 
\item The boundary gauge fixing is a weaker condition than Dirichlet boundary conditions in bulk dimensions higher than three, since for $d=2$, they reduce to the Dirichlet boundary conditions and the asymptotic symmetry group is therefore nothing else than the standard two copies of the De Witt algebra. Instead, in $d+1>3$ dimensions, the codimension two boundary metric $q_{AB}$ orthogonal to the foliation still admits $d-1$ arbitrary functions (since its determinant is fixed). These free functions determine the structure constants of the $\Lambda$-BMS$_{d+1}$ algebroid.
\item The flat limit of the $\Lambda$-BMS$_{d+1}$ algebroid gives the Generalized BMS algebra in $d+1$ dimensional asymptotically flat spacetimes, consisting of an abelian ideal of smooth supertranslations in semi-direct product with super-Lorentz transformations, both being field-independent.
\end{itemize}
We then particularized this analysis to the physical case of four spacetime dimensions.
\begin{itemize}[label=$\rhd$]
\item We obtained the most general solution space in the Bondi gauge with a non-vanishing cosmological constant, written within the convenient $\Lambda$-BMS boundary gauge fixing. Thanks to a diffeomorphism between the SFG and Bondi gauges, we matched the radiative data naturally defined in the Bondi gauge with the holographic fields provided by the SFG expansion as well as the asymptotic symmetry parameters. 
\item In particular, we found that the role of the couple $(N_{AB},C^{AB})$ formed by the asymptotic shear and the Bondi news tensors when $\Lambda=0$ is played, in the presence of $\Lambda\neq 0$, by the symplectic pairing $(T_{AB}^{TF},g^{AB}_{(0)})$ in SFG coordinates or $(J_{AB},q^{AB})$ in Bondi coordinates. We have deduced flux-balance laws for the fluxes of $\Lambda$-BMS$_4$ charges through conformal infinity driven by these radiative data and checked their consistency with the conservation equation of the holographic stress-tensor.
\item In the Bondi gauge, we performed the consistency check to map the Al(A)dS$_4$ solution space on the asymptotically locally flat solution space invariant under Generalized BMS$_4$ transformations through a flat limit process. In order to promote the latter at the level of the action, the symplectic structure and the Hamiltonian charges, we found that corner terms must supply the action principle in addition to the standard holographic counterterms. More precisely, we introduced a codimension two kinetic Lagrangian for $q_{AB}$ and prescribed the boundary terms to be added to the symplectic structure, which completes the standard prescription \eqref{Theta ren full general} to bring effects of boundary-covariant renormalization schemes, such as holographic renormalization in the SFG gauge, at the level of the presymplectic structure. We expect that our procedure for treating corner boundaries in the presence of fluxes could also be useful in the context of finite boundaries (see \textit{e.g.} \cite{Freidel:2015gpa,Hopfmuller:2016scf,Donnelly:2016auv,Geiller:2017xad,Chandrasekaran:2018aop,Harlow:2019yfa,Horowitz:2019dym}).
\end{itemize}

\paragraph{Asymptotically locally flat spacetimes} The second axis of research was devoted to addressing various issues and questions induced by the extension of the BMS$_4$ group into the Generalized BMS$_4$ group including infinite-dimensional super-Lorentz transformations. 
\begin{itemize}[label=$\rhd$]
\item Because the super-Lorentz transformations modify the boundary structure at null infinity, the well-definiteness of the Generalized BMS$_4$ surface charge algebra requires a renormalization procedure \cite{Compere:2018ylh}, which cannot be derived from a covariant prescription in terms of the bulk metric alone \cite{Flanagan:2019vbl}. We circumvented this difficulty by obtaining the Generalized BMS$_4$ surface charge algebra as a contraction of the $\Lambda$-BMS$_4$ charge algebroid induced by a flat limit $\Lambda\to 0$ at the level of the phase space. The renormalization of the surface charges follows from holographic renormalization of the embedding Al(A)dS$_4$ spacetimes combined with a new prescription for treating corner terms in the presence of asymptotic fluxes. 
\item Such a renormalization is based on the introduction of a background structure that consists of an asymptotic bulk foliation by codimension one hypersurfaces, a boundary foliation by codimension two hypersurfaces and a boundary measure. It therefore provides a geometrical and covariant framework for the asymptotically flat renormalization procedure introduced in components in \cite{Compere:2018ylh}, which gives another motivation to consider the $\Lambda$-BMS$_4$ extension of the standard finite-dimensional asymptotic groups for $\Lambda\neq 0$.
\item We derived the asymptotic symmetry algebra at the corners $\mathscr I^+_+,\mathscr I^+_-,\mathscr I^-_+,\mathscr I^-_-$ of future and past null infinity of asymptotically flat spacetimes that are non-radiative at early and late times. We showed that the Generalized BMS$_4$ algebra including super-Lorentz transformations is realized there as asymptotic symmetry algebra without non-trivial central extension. Moreover, we gave the prescription for fixing the ambiguity in the definition of the BMS$_4$ surface charges such that these charges identically vanish for vacuum configurations and such that the central extension explicitly vanishes. Conservation of the Hamiltonian charges at spatial infinity implies that the asymptotic symmetry group is realized at spatial infinity as well. As a corollary of the representation theorem, the fluxes at null infinity obey the BMS$_4$ algebra without central extension.
\item Finally, we computed the class of vacuum solutions invariant under the Generalized BMS$_4$ group. It is parametrized by three fields on the sphere, transforming independently under superrotations, superboosts and supertranslations. We then analyzed the kinematics and the dynamics of gravitational memory effects observed for transitions among these vacua and provided physical processes inducing these transitions. Like supertranslations that give a symmetry principle underlying the usual displacement memory effect, superboost transitions yield the refraction and velocity kick effects on asymptotic Bondi detectors.
\end{itemize}

\section{Some future perspectives}

We conclude our journey by examining some perspectives and research opportunities left ahead of us by our research program. 

\paragraph{$\Lambda$-BMS phase space in arbitrary dimensions} In this thesis, we demonstrated how to gain control on radiative data as well as the associated complete set of non-trivial symmetries thanks to a diffeomorphism between the SFG coordinates in which Al(A)dS gravity has been built and understood for decades and the Bondi gauge well-suited to tackle radiative problems (see section \ref{sec:FGg}). Moreover, the flat limit, impossible in the SFG gauge, works almost directly in Bondi coordinates because the latter are defined irrespectively of the cosmological constant (see sections \ref{sec:Flat limit of the LBMS4 phase space} and \ref{sec:Flat limit of the action and corner terms}). One might ask if the procedure is exportable to higher dimensional cases. 

In the presence of a cosmological constant, the structure lifts easily and includes the $\Lambda$-BMS algebroid as well as the same pattern of radiative fields $(T_{AB}^{TF},g^{AB}_{(0)})$. The asymptotically flat phase space, however, does not call for an infinite-dimensional extension of Poincaré like BMS to include interesting radiative solutions \cite{Hollands:2016oma,Hollands:2004ac,Hollands:2003ie,Tanabe:2009va,Tanabe:2010rm,Tanabe:2011es,Tanabe:2012fg,Garfinkle:2017fre,Barnich:2011ct}, in contrast with the $d=3$ case, where at least the presence of proper supertranslations is required to allow for an arbitrary $C_{AB}$. The appearance of radiative data at subleading orders in the Bondi gauge could be explained by the fact that the holographic stress-tensor also appears at deeper orders in the SFG expansion for $d>3$, modulo the generalization of the diffeomorphism to the Bondi gauge and the flat limit process. However, the reduction of the asymptotic symmetry group is less clear. Other crucial questions should be addressed, such as the fact that the Weyl charges are non-vanishing in the flat case  \cite{Barnich:2019vzx,Freidel:2021cbc,Freidel:2021yqe} although our general analysis in Al(A)dS$_4$ without boundary conditions \cite{Compere:2020lrt} tells us that they are zero before taking the flat limit. In higher odd bulk dimensions, it would also be interesting to study the impact of the Weyl anomalies in (A)dS in the flat phase space obtained by a suitable flat limit process and to compute the Weyl charges and charge algebra in that context.

\paragraph{Content of the $\Lambda$-BMS phase space and Bondi mass loss} Another perspective is the analysis of particular solutions on which one could perform a stress-test of our formalism and results, in order to effectively observe the practical utility of our leaky boundary conditions. In particular, if the boundary gauge fixing presented in section \ref{sec:Lambda BMS d} does not rule any radiative solution, it is rare to find such solutions already written in these adequate coordinates. One could think, for instance, of Robinson-Trautman solutions \cite{Robinson:1960zzb} in AdS \cite{Bakas:2014kfa,Bicak:1999ha,Bicak:1999hb}, for which two approaches are possible. The first approach is to keep in force the Bondi gauge without assuming the boundary gauge fixing, in particular allowing the boundary volume form to be variable and arbitrarily time-dependent. This approach would require a further extension of the phase space, in the spirit of \cite{Barnich:2010eb,Freidel:2021cbc,Freidel:2021yqe}, to include Weyl symmetries on transverse spheres as part of the asymptotic group in (A)dS. The second one takes the more direct route to perform a diffeomorphism within the Bondi gauge to put the Robinson-Trautman waves in $\Lambda$-BMS boundary coordinates. A sketchy analysis confirms that maintaining the boundary foliation while fixing the boundary measure is possible at the price to transfer the time dependence in the remaining degrees of freedom of $q_{AB}$. This would be an interesting arena to investigate, for example, the question of the positivity of the mass loss in (A)dS and the implications for non-equilibrium holography.

\paragraph{Infrared structure of gravity with $\Lambda\neq 0$} In the context of asymptotically flat spacetimes, we have seen that memory effects observed by inertial detectors traveling near null infinity are related to vacuum transitions driven by non-stationary phenomena. Consistency of the phase space under the action of Generalized BMS$_4$ implies that the kinematics of these memory effects is encoded into the asymptotic symmetry group (\textit{i.e.} proper supertranslations or super-Lorentz transformations) relating both vacua before and after the non-stationary phase. In the presence of a cosmological constant, the same kind of memory effects are expected, eventually enhanced by correcting terms in $\Lambda$ (see \textit{e.g.} \cite{Bieri:2015jwa , Chu:2015yua, Tolish:2016ggo , Chu:2016qxp , Chu:2016ngc , Hamada:2017gdg,Bieri:2017vni, Chu:2019ssw}). It would be interesting to see how these effects are related to the $\Lambda$-BMS symmetries derived in section \ref{sec:Lambda BMS d}. Although we treated AdS and dS asymptotics on the same footing because of the analytical similarities between both cases, any physical discussion, such as the derivation of a memory effect, will be radically different and should exhibit all the complexity of physics in dS. On one hand, the description of the memory effect as perceived by a pair of inertial timelike observers near the conformal boundary of AdS should not pose any conceptual issue and should be very reminiscent of its homologous effect in the flat limit. On the other hand, in dS, physical free-falling observers travel towards the timelike boundary and the setup is radically different. Thus, it will also be interesting to consider memory effects in higher dimensions, as has been done in the flat context \cite{Pate:2017fgt}. 
\newpage
To complete a tentative infrared triangle, it would be worth investigating how suitable past boundary conditions could yield a well-defined scattering problem in (A)dS as is the case for $\Lambda=0$ thanks to the remarkable work of Strominger and collaborators in the field. The Ward identities that would stand for the quantum version of the obtained conservation laws for $\Lambda$-BMS charges could provide symmetry principles underlying the consistency conditions for correlation functions \cite{Maldacena:2002vr,Creminelli:2004yq,Cheung:2007sv,Hinterbichler:2013dpa,Horn:2014rta,Hui:2018cag,Mirbabayi:2016xvc,Kehagias:2016zry,Berezhiani:2013ewa}, which are the closest analog for $\Lambda\neq 0$ of Weinberg's factorization properties derived for $\Lambda=0$.

\paragraph{Looking towards flat holography} The holographic principle experienced its most brilliant realization through the AdS/CFT correspondence. Even at the classical level, we observe that all of the ingredients are present to ensure its good functioning, namely the conformal compactness of the bulk spacetime bounded by a codimension one spacelike surface on which a clear notion of chronology can be defined everywhere. Otherwise, in the flat case, an analogous holographic correspondence is much less known and mastered, mainly because of the null nature of the conformal boundary. Even if the conformal compactification process can be performed, there is no more SFG expansion allowing to parametrize the gravitational phase space uniquely by codimension one boundary data in a transparent fashion. This thesis reviewed the usual method to approach null infinity, which consists in following the directions of null cones in the bulk that precisely define the Bondi gauge null foliation. In (A)dS, we provided all technicities to translate the boundary data commonly used in holography to Bondi radiative fields, at least in four dimensions. We also proposed a well-defined flat limit process within the Bondi gauge at the level of the phase space, which is enough to obtain all dynamical information about the bulk gravity that we need to build a more robust notion of flat holography. We have shown that the renormalization of the asymptotically flat spacetime can be understood as a relic of the holographic renormalization procedure in (A)dS coupled to a careful management of the corner terms before pushing the cosmological constant to zero. We believe that all of the technical elements to build a consistent notion flat holography on the shoulders of the famous AdS/CFT correspondence have been derived through to the $\Lambda$-BMS program. 

We plan to pursue this research direction in the future. Crucial progress toward this endeavor has been made recently, namely the study of the coadjoint representation of BMS$_4$ \cite{Barnich:2021dta} under which the non-radiative solution space in the Bondi gauge transforms. In three-dimensional gravity genuinely exempt of radiation, the coadjoint representation of BMS$_3$ is necessary and sufficient to encapsulate the whole solution space \cite{Barnich:2015uva}, and the theory has been shown to be perfectly described by a geometric action \cite{Barnich:2017jgw} for a theory living on the boundary. The inclusion of radiation in four-dimensional gravity would amount to couple the dual geometric action defined on the coadjoint orbits of BMS$_4$ spanning the stationary subsector of the theory to external sources encoded in the asymptotic shear and Bondi news tensors. This would lead to a holographic description of asymptotically flat gravity at null infinity. 

\paragraph{Gauge theories with external sources} Beyond its concrete contributions to a better understanding of the asymptotic symmetries of General Relativity, this thesis was an opportunity to gather knowledge about the leaky phase spaces that can be constructed for this theory and to emphasize their importance in the theoretical discussion of the phenomena involving gravitational radiation. Although this is a very old research topic first initiated by Bondi, Metzner, Sachs and van der Burg almost 60 years ago, several technical results are still missing or incompletely understood in the literature. For instance, gauge theories with non-conserved and non-integrable charges need more than ever renewed attention from a purely mathematical point of view. The algebraic properties of the surface charges as postulated by Barnich and Troessaert for the very restricted case of the BMS$_4$ algebra \cite{Barnich:2011mi} were re-analyzed very recently considering radiative gravity as a Hamiltonian open system (see \textit{e.g.} \cite{Wieland:2017zkf,Wieland:2021eth,Troessaert:2015nia}) with more care given to the treatment of sources of non-conservation and non-integrability. New prescriptions for the charge bracket in the presence of anomalies due to background structures have also been studied \cite{Freidel:2021cbc,Freidel:2021yqe}. In the future, we would like to extend the analysis to any gauge theory coupled with external sources, with a framework that clearly separates the dynamical degrees of freedom that contribute to the physical flux and are responsible for non-integrability as well as non-conservation of the charges among all gauged degrees of freedom. These investigations could help in proving the conjecture of \cite{Adami:2020ugu}, discussed in section \ref{sec:Physical content of the presymplectic flux} for gravity. Moreover, considering the low number of hypotheses and the generality of the developments presented in section \ref{sec:Non-integrable case: the Barnich-Troessaert bracket}, we would also like to address the relevance of the Barnich-Troessaert bracket for theorizing the dynamics of open systems in classical Hamiltonian and Lagrangian mechanics, where dissipative phenomena and exchanges with the environment in the grand-canonical ensemble are often incorporated by hand in the equations of motion without completely controlling the modified phase space features and symmetries disturbed by the presence of the external sources. This ambitious research endeavor has already been started \cite{ToAppear}, encouraged by the current stimulating interest around these fundamental problematics.

\vfill
\paragraph{A few last words} This concludes our exploration of leaky covariant phase spaces in General Relativity as well as four years of intense personal and professional enrichment. The exercise of completing a Ph.D. thesis is not a series of solitary computations, but a living adventure during which one acquires a more solid technical mastery while sharing ideas and projects with wonderful people. I hope that all doctoral students will have as fruitful and pleasant an experience as I had. Although what lies ahead contains that bundle of unexpectedness and hope that makes it so appealing, it should never be forgotten that it is built on the beautiful experiences and fond memories of the past. In writing these last lines I now take the step to forever put my studies behind me, with a firm desire to do justice to what I have been taught and to continue to contribute honorably to a quest definitely promised a bright future!\hfill{\color{black!40}$\blacksquare$}

%
%

\appendix
\chapter*{Appendices}
\markboth{Appendices}{Appendices}
\addcontentsline{toc}{chapter}{Appendices}
\renewcommand{\thesection}{\Alph{section}}

\section{Killing vectors of (A)dS\texorpdfstring{$_4$}{4} in Bondi gauge}
\label{app:ExactVectors}

In this appendix, we review the expression of the Killing vectors for the global vacua of 4$d$ gravity (Minkowksi, (A)dS$_4$) in the Bondi gauge. Once the results have been presented for AdS$_4$ (with radius $\ell$), one can readily extrapolate to the dS$_4$ case thanks to the analytical continuation $\ell\to i\ell$. The isometries of flat space are finally obtained by taking the flat limit $\ell\to+\infty$.

For the purpose to compute the symmetries of AdS$_4$, it is convenient to see this spacetime isometrically immersed into $\mathbb{R}^{(2,3)}$ as the hypersurface
\begin{equation}
\lbrace X^\mu \in \mathbb{R}^{(2,3)} | -X_0^2 - X_{\underline{0}}^2 + X_1^2 +X_2^2+X_3^2 = -\ell^2 \rbrace , \label{hypersurface defining AdS4}
\end{equation}
see section \ref{sec:Asymptotic properties of (A)dS spacetimes} for a review. In this point of view, the exact symmetry group of AdS$_4$ is the homogeneous part of the isometry group of $\mathbb{R}^{(2,3)}$, $ISO(2,3)$, which is $SO(2,3)$. The generators of $SO(2,3)$ algebra 
\begin{equation}
\mathcal J_{ab} = \mathcal J_{[ab]} = X_b \partial_a - X_a \partial_b,
\end{equation}
$a,b\in\{0,\underline 0,1,2,3\}$, directly lead to the Killing vectors of AdS$_4$ after pullback on \eqref{hypersurface defining AdS4}.

In retarded Bondi coordinates $(u,r,x^A)$, the AdS$_4$ line element takes the form \eqref{AdSBondi} where we take $x^A=(z,\bar z)$ as the stereographic coordinates on the 2-sphere with unit-round metric $\mathring q_{AB}$. The Minkowski metric in retarded coordinates \eqref{Mink4 in bondi} is recovered in the flat limit $\ell \rightarrow \infty$. In these coordinates, the Killing vectors are given by $\mathcal J = \mathcal J^\mu \partial_\mu$ with
\begin{equation}
\mathcal J_{ab}^u = f, \qquad \mathcal J_{ab}^r = -\frac{r}{2} \mathring D_A \mathcal J_{ab}^A,\qquad \mathcal J_{ab}^A = Y^A - \frac{1}{r} \partial^A f 
\end{equation}
where $\mathring D_A$ is the covariant derivative on the sphere and the functions $f, Y^A$ obey \cite{Barnich:2013sxa} 
\begin{equation}
\partial_u f = \frac{1}{2} \mathring D_A  Y^A \, , \qquad
\partial_u Y^A = \frac{1}{\ell^2} \partial^A f \, , \qquad 2\mathring  D_{(A} Y_{B)} - \mathring q_{AB} \mathring D_C Y^C = 0. \label{SO32 constraints}
\end{equation}
The last equation fragments the set of Killing vectors of AdS$_4$ in two categories, namely the 4 vectors associated with exact isometries of the boundary cylinder (with $\mathring D_A Y^A=0$):
\begin{enumerate} 
\item Rotations:
\begin{equation}
\begin{split}
\mathcal J_{12} &= [0,0,-i z,-i \bar{z}] = \partial_\phi, \\
\mathcal J_{13} &= \Big[0,0,\frac{1}{2} (-1-z^2),\frac{1}{2} (-1-\bar{z}^2)\Big], \\
\mathcal J_{23} &= \Big[0,0,\frac{1}{2} i (-1 + z^2),\frac{1}{2} i (-1 + \bar{z}^2)\Big];
\end{split}
\end{equation}
\item Time translation: 
\begin{equation}
\mathcal J_{0\underline{0}} = \partial_u;
\end{equation}
\end{enumerate}
and the 6 vectors associated with the conformal isometries of the boundary cylinder, for which $\mathring D_A Y^A\neq 0$:
\begin{enumerate}
\setcounter{enumi}{2}
\item Boosts on the first timelike coordinate: with the shorthand notation $\mu(u/\ell,r)$ $\equiv$ $ r$ $ \cos (u/\ell)$ $ +$ $ \ell \sin (u/\ell)$, we have
\begin{equation}
\begin{split}
\mathcal J_{01} &= \Big[ \frac{(z+\bar{z}) \ell \sin (\frac{u}{\ell})}{1+z\bar{z}}, -\frac{(z+\bar z)\mu(\frac{u}{\ell},r)}{(1+z\bar{z})},\frac{(-1+z^2) \mu(\frac{u}{\ell},r)}{2 r},-\frac{(-1+\bar{z}^2) \mu(\frac{u}{\ell},r)}{2 r}  \Big], \\
\mathcal J_{02} &= \Big[ -\frac{(z-\bar{z}) \ell \sin (\frac{u}{\ell})}{1+z\bar{z}}, \frac{i(z-\bar z)\mu(\frac{u}{\ell},r)}{(1+z\bar{z})},-\frac{i(1+z^2) \mu(\frac{u}{\ell},r)}{2 r},\frac{i(1+\bar{z}^2) \mu(\frac{u}{\ell},r)}{2 r }  \Big], \\
\mathcal J_{03} &= \Big[ \frac{(-1+z\bar{z}) \ell \sin (\frac{u}{\ell})}{1+z\bar{z}}, \frac{(-1+z\bar z)\mu(\frac{u}{\ell},r)}{(1+z\bar{z})},-\frac{z \mu(\frac{u}{\ell},r)}{r},-\frac{\bar z \mu(\frac{u}{\ell},r)}{r}  \Big] ;\\
\end{split}
\end{equation}

\item Boosts on the second timelike coordinate: by virtue of the $SO(2,3)$ algebra,
\begin{equation}
\mathcal{J}_{\underline 0 i} = -[\mathcal{J}_{0\underline 0},\mathcal{J}_{0i}] = - \partial_u \mathcal{J}_{0i}.
\end{equation}

\end{enumerate}
The Lorentz algebra can be explicitly checked for the rotations $V_i = {\epsilon_i}^{jk} \mathcal J_{jk}$ together with the boosts along $0$ time direction, $K_i = \mathcal  J_{0i}$, or the boosts along $\underline 0$ time direction, $K_i=\ell \mathcal J_{\underline{0}i}$:
\begin{equation}
[ V_i,V_j] = \epsilon_{ijk} V_k \, , \quad [ K_i,K_j] = -\epsilon_{ijk} V_k \, , \quad [V_i,K_j] = \epsilon_{ijk} K_k.
\end{equation}
The time translation and the three rotations have a trivial flat limit. When $\ell\to+\infty$ the boosts $\mathcal J_{0i} $ become the Lorentz boosts 
\begin{equation}
\begin{split}
\mathcal J_{01} &\rightarrow \Big[  \frac{u(z+\bar z)}{1+z\bar z}, -\frac{(r+u)(z+\bar{z})}{1+z\bar{z}} , \frac{(r+u)(-1+z^2)}{2 r} , \frac{(r+u)(-1+\bar{z}^2)}{2 r}  \Big], \\
\mathcal J_{02} &\rightarrow \Big[  -i\frac{u(z-\bar z)}{1+z\bar z}, i\frac{(r+u)(z-\bar{z})}{1+z\bar{z}} , -i\frac{(r+u)(1+z^2)}{2 r} , i\frac{(r+u)(1+\bar{z}^2)}{2 r}  \Big], \\
\mathcal J_{03} &\rightarrow \Big[  \frac{u(-1+z\bar z)}{1+z\bar z}, -\frac{(r+u)(-1+z\bar{z})}{1+z\bar{z}} , -\frac{(r+u)z}{r} , -\frac{(r+u)\bar{z}}{r}  \Big],
\end{split}
\end{equation}
while the boosts $\mathcal J_{\underline 0i} $ become the spatial translations
\begin{equation}
\begin{split}
\mathcal J_{\underline 01} &\rightarrow \Big[ \frac{z+\bar{z}}{1+z\bar{z}} , - \frac{z+\bar{z}}{1+z\bar{z}} , \frac{-1+z^2}{2r} , \frac{-1+\bar{z}^2}{2r} \Big], \\
\mathcal J_{\underline 02} &\rightarrow \Big[ -i \frac{z-\bar{z}}{1+z\bar{z}} , i \frac{z-\bar{z}}{1+z\bar{z}} , -i \frac{1+z^2}{2r}, i \frac{1+\bar{z}^2}{2r} \Big], \\
\mathcal J_{\underline 03} &\rightarrow \Big[ - \frac{1-z\bar{z}}{1+z\bar{z}} , \frac{1-z\bar{z}}{1+z\bar{z}} , -\frac{z}{r}, -\frac{\bar{z}}{r} \Big].
\end{split}
\end{equation}
One can check explicitly that they verify the constraints \eqref{SO32 constraints} in the limit $\ell \to +\infty$.

\section{Map from Bondi to Starobinsky/Fefferman-Graham gauge}
\label{app:chgt}

This appendix contains the detailed material complementing the section \ref{sec:FGg}. It is devoted to discuss the change of coordinates that maps a general vacuum asymptotically locally (A)dS$_4$ spacetime ($\Lambda \neq 0$) in the Bondi gauge to the Starobinsky/Fefferman-Graham gauge. In particular, we give here the explicit forms of the various coefficients appearing in the perturbative change of coordinate \eqref{tortoise to FG} in terms of the metric fields. 

We use the following shorthand notations for subleading fields in the Bondi gauge:
\begin{equation}
\begin{split}
\frac{V}{r} &= \frac{\Lambda}{3}e^{2\beta_0} r^2 + r \ V_{(1)} (t,x^A) + V_{(0)} (t,x^A) + \frac{2M}{r} + \mathcal{O}(r^{-2}), \\
U^A &= U_{0}^A(t,x^B) + \frac{1}{r}U_{(1)}^A(t,x^B) + \frac{1}{r^2} U_{(2)}^A(t,x^B) + \frac{1}{r^3} U_{(3)}^A(t,x^B) + \mathcal{O}(r^{-4}), \\
\beta &= \beta_{0}(t,x^A) + \frac{1}{r^2}\beta_{(2)}(t,x^A)+ \mathcal{O}(r^{-4}).
\end{split}
\end{equation}
whose explicit on-shell values can be read off in \eqref{eq:EOMVr} and \eqref{eq:EOM_UA2}. That allows to write the equations in a more compact way. All fields are now evaluated on $(t,x^A)$ since the boundary time coordinate can be defined as $t$ as well as $u$. We also define  some recurrent structures appearing in the diffeomorphism as differential operators on boundary scalar fields $\varphi(t,x^A)$, $\chi(t,x^A)$:
\begin{equation}
\begin{split}
P[\varphi] &= \frac{1}{2} e^{-4\beta_0} (\partial_t \varphi + U_0^A \partial_A \varphi), \\
Q[\varphi;\chi] &= P[\varphi] - 2  P[\chi] \varphi, \\
B_A[\varphi] &= \frac{1}{2} e^{-2\beta_0} (\partial_A - 2\partial_A \beta_0) \varphi.
\end{split}
\end{equation}
Below, $P^n[\varphi]$ denotes $n$ applications of $P$ on $\varphi$, for example $P^2[\varphi] \equiv P[P[\varphi]]$. Now we can write down the perturbative change of coordinate to Starobinsky/Fefferman-Graham gauge:

\newpage
\begin{flalign*}
R_1 (t,x^A) &= -\frac{3}{\Lambda}, && \\ 
R_2 (t,x^A) &= \frac{9}{2\Lambda^2} e^{-2\beta_0} V_{(1)}, && \\ 
R_3 (t,x^A) &= \frac{3}{2\Lambda} \beta_{(2)} - \frac{3}{\Lambda^2} \Big( 1 + \frac{3}{4} e^{-2\beta_0} V_{(0)} \Big) +\frac{27}{2\Lambda^3}  \Big( Q[V_{(1)};\beta_0] - \frac{3}{8} e^{-4\beta_0} V_{(1)}^2 \Big), && \\ 
R_4 (t,x^A) &= \frac{3}{\Lambda^2} e^{-2\beta_0} \Big( M + 2 e^{4\beta_0} P[\beta_{(2)}] - \frac{5}{2} V_{(1)} \beta_{(2)}  \Big) && \\ 
	&\hspace{-40pt}\quad -\frac{9}{\Lambda^3} \Big\lbrace Q[V_{(0)};\beta_0] + \frac{1}{4} e^{-4\beta_0} \Big[ U^A_{(1)} \partial_A V_{(1)} - 2 V_{(1)} U^A_{(1)} \partial_A\beta_0 - 3 V_{(1)} (2 e^{2\beta_0} + V_{(0)}) \Big] \Big\rbrace && \\ 
	&\hspace{-40pt}\quad + \frac{27}{\Lambda^4} e^{2\beta_0} \Big[ P^2[V_{(1)}] - 2 V_{(1)} \Big( P^2[\beta_0]+ \frac{1}{2}e^{-4\beta_0} Q[V_{(1)};\beta_0] - \frac{3}{32} e^{-8\beta_0} V_{(1)}^2 \Big) - 2 P[\beta_0]P[V_{(1)}] \Big], &&
\end{flalign*}
\begin{flalign*}
T_1 (t,x^A) &= (1 - e^{-2\beta_0}) R_1(t,x^A),&& \\ 
T_2 (t,x^A) &=  (1 - e^{-2\beta_0}) R_2 (t,x^A) -\frac{18}{\Lambda^2} \Big( P[\beta_0] - \frac{1}{4}e^{-4\beta_0} V_{(1)} \Big),&& \\ 
T_3 (t,x^A) &=  (1- e^{-2\beta_0})R_3(t, x^A) -\frac{3}{\Lambda^2} e^{-2\beta_0} (1 + e^{-2\beta_0} V_{(0)} - 2 \partial^A \beta_0 \partial_A \beta_0) && \\ 
	&\quad + \frac{9}{\Lambda^3} e^{-2\beta_0} \Big( Q[V_{(1)};		\beta_0] - 4 e^{4\beta_0} P^2[\beta_0]- \frac{1}{2}		e^{-4\beta_0} V_{(1)}^2 \Big), && \\ 
T_4 (t,x^A) &= (1- e^{-2\beta_0})R_4(t, x^A) && \\ 
& \quad+ \frac{9}{2\Lambda^2} \Big[ e^{-4\beta_0} \Big( M - \beta_{(2)} V_{(1)} - \frac{1}{3} U_{(2)}^A \partial_A \beta_0 \Big) -\frac{1}{2} (P[\beta_{(2)}] - 8 \beta_{(2)} P[\beta_0]) \Big] && \\ 
	&\quad - \frac{27}{\Lambda^3} \Big\lbrace \frac{1}{8} e^{-2\beta_0} \Big( 3 Q[V_{(0)};\beta_0] - \frac{8}{3} P[\beta_0]V_{(0)} - 2 e^{-4\beta_0} V_{(1)}V_{(0)} \Big) && \\ 
		&\qquad\qquad + \frac{1}{3} e^{-2\beta_0} \Big( P[U_{(1)}^A]\partial_A\beta_0 + \frac{3}{2} U_{(1)}^A \partial_A P[\beta_0] \Big) && \\ 
		&\qquad\qquad - \frac{1}{12}e^{-4\beta_0} \Big[ U_{(1)}^A B_A [V_{(1)}] + 6 V_{(1)} - 2 (V_{(1)}\partial_A \beta_0 + 2 \partial_B \beta_0 \partial_A U_0^B)\partial^A \beta_0 \Big] \Big\rbrace && \\ 
	&\quad + \frac{81}{\Lambda^4} \Big\lbrace -\frac{2}{3} e^{4\beta_0} \Big( P^3[\beta_0] + 2 P[\beta_0]P^2[\beta_0] \Big) + \frac{1}{4} \Big(P^2[V_{(1)}] - 2 V_{(1)} P^2[\beta_0]\Big) && \\ 
	&\qquad\qquad +\frac{1}{6} P[\beta_0] \Big[\Big(\frac{13}{4}e^{-4\beta_0} V_{(1)} - 8 P[\beta_0] \Big) V_{(1)} + P[V_{(1)}] \Big] && \\ 
	&\qquad\qquad - \frac{1}{16} e^{-4\beta_0} V_{(1)} \Big(5P[V_{(1)}]- e^{-4\beta_0} V_{(1)}^2\Big) \Big\rbrace, 
\end{flalign*}
\begin{flalign*}
X_1^A (t,x^B) &= (T_1 - R_1) U_0^A,&& \\ 
X_2^A (t,x^B) &= (T_2 - R_2) U_0^A - \frac{3}{2\Lambda} e^{-2\beta_0} U_{(1)}^A + \frac{9}{\Lambda^2} P[U_0^A],&& \\ 
X_3^A (t,x^B) &= (T_3 - R_3) U_0^A + \frac{1}{\Lambda} e^{-2\beta_0} U_{(2)}^A && \\ 
	&\quad - \frac{6}{\Lambda^2} \Big[ Q[U_{(1)}^A;\beta_0] + \frac{1}{2} B^A[V_{(1)}] + \frac{1}{4} e^{-4\beta_0} (U_{(1)}^B \partial_B U^A_0 - V_{(1)}U_{(1)}^A) \Big] && \\ 
		&\quad + \frac{18}{\Lambda^3} e^{2\beta_0} Q[P[U_0^A];\beta_0], && \\ 
X_4^A (t,x^B) &= (T_4 - R_4) U_0^A - \frac{3}{4\Lambda} e^{-2\beta_0} \Big[ U_{(3)}^A + \frac{1}{2}e^{2\beta_0} (\partial^A \beta_{(2)} - 8 \beta_{(2)}\partial^A\beta_0) \Big] && \\ 
& + \frac{9}{2\Lambda^2} \Big[ Q[U_{(2)}^A;\beta_0] - \frac{1}{2} e^{-4\beta_0} \Big( V_{(1)} U_{(2)}^A - \frac{1}{3} U_{(2)}^B \partial_B U_{(0)}^A \Big) - 2 \beta_{(2)} P[U_0^A] && \\ 
		&\qquad\qquad + \frac{1}{4} B^A[V_{(0)}] + \frac{1}{2} C^{AC} B_C[V_{(1)}] + \frac{1}{2} e^{-2\beta_0} U_{(1)}^C B_C[ U^A_{(1)} ] \Big] && \\ 
		& -\frac{27}{\Lambda^3} \Big\lbrace  e^{2\beta_0} \Big( P[Q[U_{(1)}^A;\beta_0]] + P[B^A[V_{(1)}]] - \frac{1}{2} q^{AC} P[B_C[V_{(1)}]] \Big) && \\ 
		&\qquad\qquad - \frac{1}{2}e^{-2\beta_0} \Big[ V_{(1)} P[U_{(1)}^A] - \frac{2}{3}\Big( P[U_{(1)}^C] + \frac{1}{2} B^C[V_{(1)}] - 5 P[\beta_0]U_{(1)}^C \Big)\partial_C U_0^A && \\ 
		&\qquad\qquad + \frac{1}{2} P[V_{(1)}]U_{(1)}^A - \frac{2}{3}(V_{(0)} - 8 e^{2\beta_0} \partial^B \beta_0 \partial_B \beta_0)P[U_0^A] - U_{(1)}^B P[\partial_B U_0^A] && \\ 
		&\qquad\qquad + \frac{1}{2} (\partial^A U_0^C)B_C[V_{(1)}] \Big] + 3 P[\beta_0] V_{(1)}\partial^A \beta_0 && \\ 
		&\qquad\qquad - e^{-4\beta_0} \Big[ \frac{3}{32} \Big(\partial^A (V_{(1)}^2) - \frac{20}{3} \partial^A \beta_0 V_{(1)}^2 \Big) + \frac{1}{6} (V_{(1)}  \partial^B \beta_0 - \partial^C\beta_0 \partial_C U_0^B) \partial_B U_0^A \Big] \Big\rbrace && \\ 
		&\hspace{-3pt} +\frac{81}{\Lambda^4} \Big\lbrace \frac{1}{3} e^{4\beta_0} P^3[U_0^A] + \Big[ \frac{1}{4}e^{-4\beta_0} V_{(1)}^2 - \frac{1}{3} Q[V_{(1)};\beta_0] - \frac{4}{3} e^{4\beta_0} (P^2[\beta_0] + P[\beta_0]^2)  \Big] P[U_0^A] \Big\rbrace. \nonumber
\end{flalign*}
The function $R_1$ is not constrained by the gauge conditions \eqref{eq:FGgaugecond} but is simply fixed by requiring that the Weyl rescalings of the transverse metric on the sphere agree between both gauges, \textit{i.e.} $g^{(0)}_{AB} = q_{AB}$. Several consistency checks can be performed at each stage of the computation. The boundary metric in the SFG gauge must be equivalent to the pulled-back metric on the hypersurface $\lbrace r\to \infty \rbrace$ in the Bondi gauge, up to the usual replacement $u\to t$. This is the case because of \eqref{g0 in term of Bondi}. At subleading orders, $g^{(1)}_{ab}$ and $g^{(2)}_{ab}$ must be algebraically determined by $g_{ab}^{(0)}$ and its first and second derivatives, what turns out to be the case. The constraint \eqref{eq:CAB} enforces $g^{(1)}_{ab} = 0$ while the annulation of $\mathcal{D}_{AB}(t,x^C)$ \eqref{eq:DAB} results in
\begin{equation}
g^{(2)}_{ab} = \frac{3}{\Lambda} \Big[ R^{(0)}_{ab} - \frac{1}{4} R_{(0)} g^{(0)}_{ab} \Big].
\end{equation}
This agrees with the $d=3$ version of the equation of motion \eqref{g2}. We will not give the full general form of $g^{(3)}_{ab}$, but it can be proven that this tensor is traceless with respect to $g^{(0)}_{ab}$ and that the equations of motion in the Bondi gauge are necessary and sufficient to show its conservation $D^{(0)}_a g_{(3)}^{ab} = 0$, as we argued in the main text.

After partial boundary gauge fixing $\beta_0 = 0$, $U_0^A = 0$ (Gaussian normal coordinates on the boundary), but leaving $\sqrt{q}$ arbitrary, the expressions of each coefficient in the diffeomorphism simplifies drastically:
\begin{flalign*}
R_1 (t,x^A) &= -\frac{3}{\Lambda},&& \\ 
R_2 (t,x^A) &= \frac{9}{2\Lambda^2} V_{(1)}, && \\
R_3 (t,x^A) &= \frac{3}{2\Lambda} \beta_{(2)} - \frac{3}{\Lambda^2} \Big( 1 + \frac{3}{4} V_{(0)} \Big) +\frac{27}{2\Lambda^3}  \Big( \frac{1}{2} \partial_t V_{(1)} - \frac{3}{8} V_{(1)}^2 \Big),&& \\
R_4 (t,x^A) &= \frac{3}{\Lambda^2} \Big( M + \partial_t \beta_{(2)} - \frac{5}{2} V_{(1)} \beta_{(2)} \Big) -\frac{9}{\Lambda^3} \Big[ \frac{1}{2} \partial_t V_{(0)} - \frac{3}{4} V_{(1)} (2 + V_{(0)}) \Big] && \\
	&\quad + \frac{27}{\Lambda^4} \Big( \frac{1}{4}\partial_t^2[V_{(1)}] - \frac{1}{4}\partial_t V_{(1)}^2 + \frac{6}{32} V_{(1)}^3\Big). && 
\end{flalign*}
\begin{flalign*}
T_1 (t,x^A) &= 0, && \\
T_2 (t,x^A) &=  \frac{9}{2\Lambda^2} V_{(1)},&& \\
T_3 (t,x^A) &=  -\frac{3}{\Lambda^2} (1 + V_{(0)}) + \frac{9}{2\Lambda^3} \Big( \partial_t V_{(1)} - V_{(1)}^2 \Big), && \\
T_4 (t,x^A) &= \frac{9}{2\Lambda^2} \Big[ M - \frac{1}{4} (\partial_t \beta_{(2)} + 4 V_{(1)}\beta_{(2)}) \Big] + \frac{9}{\Lambda^3} \Big[ - \frac{9}{16} \partial_t V_{(0)} + \frac{3}{2} V_{(1)} (1+\frac{1}{2} V_{(0)}) \Big] && \\
	&\quad + \frac{81}{\Lambda^4} \Big(  \frac{1}{16} \partial_t^2 V_{(1)}  - \frac{5}{64} \partial_t V_{(1)}^2 + \frac{1}{16}V_{(1)}^3 \Big). && 
\end{flalign*}
\begin{flalign*}
X_1^A (t,x^B) &= X_2^A (t,x^B) = 0, && \\
X_3^A (t,x^B) &= \frac{1}{\Lambda} U_{(2)}^A - \frac{3}{2\Lambda^2} \partial^A V_{(1)}, && \\
X_4^A (t,x^B) &= - \frac{3}{4\Lambda} \Big( U_{(3)}^A + \frac{1}{2}\partial^A \beta_{(2)} \Big) + \frac{9}{2\Lambda^2} \Big( \frac{1}{2}\partial_t U_{(2)}^A - \frac{1}{2}  V_{(1)} U_{(2)}^A + \frac{1}{8} \partial^A V_{(0)} \Big) && \\
		&\quad -\frac{27}{16\Lambda^3}  \ q^{AB} \Big( \partial_t \partial_B V_{(1)} + \frac{1}{2} V_{(1)} \partial_B V_{(1)} \Big).
\end{flalign*}
In particular, $t = u + \mathcal O(\rho)$ and both time coordinates can be identified on the boundary. 

\paragraph{Note.} The expressions of the diffeomorphism and the metric elements $g_{ab}^{(0)}$, $g_{ab}^{(2)}$ and $g_{ab}^{(3)}$ (the latter within the additional boundary gauge fixing) can be found in the \texttt{Mathematica} file \texttt{BMS\_TO\_FG.nb} attached to \cite{Compere:2019bua} on the \textit{arXiv} publication. The package \texttt{RGtensors} \cite{RGtensorslink} is needed to run this notebook.

\section{Useful relations for computations in the Bondi gauge}
\label{Useful relations}

In this appendix, we give some useful relations in the Bondi gauge that are widely used in sections \ref{sec:Asymptotically locally flat radiative phase spaces}, \ref{sec:The Generalized BMS$_4$ charge algebra} and \ref{sec:Flat limit of the action and corner terms}. The metric on the celestial sphere is written as $q_{AB}$. From $\delta q^{AB} = -q^{AC}q^{BD} \delta q_{CD}$, we have
\begin{equation}
T_{AB} \delta q^{AB} = - T^{AB}\delta q_{AB}. \label{prop1}
\end{equation} for any symmetric tensor $T_{AB}$. For any $C_{AB}$ symmetric traceless tensor (\textit{i.e.} $C_{AB}=C_{(AB)}$, $q^{AB}C_{AB}=0$), it follows that 
\begin{equation}
q_{AB} \delta C^{AB} = C_{AB} \delta q^{AB}.
\end{equation} Requiring further that $\delta\sqrt{q}=0$ one finds
\begin{align}
\delta (C_{AB} C^{AB}) &= 2 C_{AB}\delta C^{AB}, \\
\delta (C_{AB} C^{AB}) &= q_{AB} \delta (C^{AC} C_C^B ), \\
M_{AB} \delta C^{AB} &= M^{AB} \delta C_{AB}, \label{prop5}
\end{align}
where $M_{AB}$ is also an arbitrary symmetric traceless tensor. Considering two variations $\delta,\delta'$ one can prove the following identity, again with $\delta\sqrt{q}=0$ :
\begin{equation}
T_{AB} \delta\delta' q^{AB} = - T^{AB} \delta \delta' q_{AB} - T \delta q_{AB} \delta' q^{AB}, 
\end{equation} where $T \equiv q^{AB} T_{AB}$.
It follows from (\ref{prop1}) that
\begin{equation}
\delta T_{AB} \delta' q^{AB} = - \delta T^{AB} \delta' q_{AB} + T \delta q_{AB} \delta' q^{AB}. \label{prop9}
\end{equation}
For a traceless tensor such as $M_{AB}$, we have
\begin{equation}
\delta M_{AB} \delta' q^{AB} = - \delta M^{AB} \delta' q_{AB}.
\end{equation}
For the metric $q_{AB}$ itself, $T=2$ and
\begin{equation}
\delta q_{AB} \delta' q^{AB} = \delta q^{AB} \delta' q_{AB}.
\end{equation}
Finally from (\ref{prop5}) we get
\begin{equation}
\delta M_{AB} \delta' C^{AB} = \delta M^{AB}\delta' C_{AB}. \label{eq:DeltaSymCC}
\end{equation}
While manipulating rank 2 symmetric traceless tensors $C_{AC}$, $M_{AB}$ on the sphere, the following identities turn out to be very useful:
\begin{align}
&\frac{1}{2}(C_{AB}M^{BC}+M_{AB} C^{BC}) = \frac{1}{2}{\delta_A}^B (C_{CD}M^{CD}), \label{identite glenn} \\
&D_A C_{BC}M^{BC} = D_B C_{AC}M^{BC} + M_{AB}D_C C^{BC}. \label{identite FN}
\end{align}

\section{Technical material from quadratic curvature gravity}
\label{New massive gravity}

In this appendix, we review some material about the quadratic curvature gravity theory \cite{Stelle:1977ry , Salvio:2018crh} (also coined as new massive gravity theory \cite{Alkac:2012bz}). Indeed, the patterns of this theory appear naturally when considering the Weyl charges in the $d = 4$ case.

The quadratic gravity Lagrangian in $d$ dimensions contains generally three pieces
\begin{equation}
\bm L_{QCG}[g] = \bm L_{EH}[g]  + \beta_1 \, \bm L_{QCG(1)}[g]  + \beta_2 \, \bm L_{QCG(2)}[g] 
\end{equation} where $g$ denotes an arbitrary metric tensor in $d$ dimensions, $\beta_1,\beta_2$ are real constants and
\begin{equation}
\begin{split}
\bm L_{EH}[g]  &=\frac{\sqrt{|g|}}{16 \pi G} \left( R + \eta \frac{d (d-1)}{\ell^2} \right) ~\D^dx, \\
\bm L_{QCG(1)}[g]  &=\frac{\sqrt{|g|}}{16 \pi G} R_{ab} R^{ab} ~\D^dx, \qquad
\bm L_{QCG(2)}[g]  = \frac{\sqrt{|g|}}{16 \pi G} R^2  ~\D^dx.
\end{split}
\end{equation} Taking the variation of these Lagrangians, we have
\begin{equation}
\begin{split}
\delta \bm L_{EH}[g]  &= \frac{\delta \bm L_{EH}[g]}{\delta g^{ab}} \delta g^{ab} + \D \bm \Theta_{EH}[g;\delta g] ,\\
\delta \bm L_{QCG(1)}[g]  &= \frac{\delta \bm L_{QCG(1)}[g]}{\delta g^{ab}} \delta g^{ab} + \D \bm \Theta_{QCG(1)}[g;\delta g], \\
\delta \bm L_{QCG(2)}[g]  &= \frac{\delta \bm L_{QCG(2)}[g]}{\delta g^{ab}} \delta g^{ab} + \D \bm \Theta_{QCG(2)}[g;\delta g]  
\end{split}
\end{equation} where
\begin{align}
 \frac{\delta \bm L_{EH}[g] }{\delta g^{ab}} &= \frac{\sqrt{|g|}}{16 \pi G}\left[ R_{ab} - \frac{1}{2} g_{ab} - \eta \frac{d(d-1)}{2 \ell^2} g_{ab} \right] ~\D^d x ,\nonumber \\
\frac{\delta \bm L_{QCG(1)}[g] }{\delta g^{ab}} &= \frac{\sqrt{|g|}}{16 \pi G} \left[ 2 R_{acbd} R^{cd} - D_a D_b R + D^c D_c R_{ab} + \frac{1}{2} g_{ab} \left( D^c D_c R - R_{cd} R^{cd} \right) \right] ~\D^dx, \nonumber \\
\frac{\delta \bm L_{QCG(2)}[g] }{\delta g^{ab}} &=\frac{\sqrt{|g|}}{16 \pi G} \left[ 2 R R_{ab} - 2 D_a D_b R + g_{ab} \left( 2 D^c D_c R - \frac{1}{2} R^2 \right) \right] ~\D^dx 
\end{align} and $\bm\Theta_{EH}[g;\delta g] = \Theta_{EH}^a[g;\delta g]~ (\D^{d-1} x)_a$ (idem for $\bm\Theta_{QCG(1)}[g;\delta g],\bm\Theta_{QCG(2)}[g;\delta g]$) with
\begin{align}
\Theta^a_{EH}[g;\delta g] &= \frac{\sqrt{|g|}}{16 \pi G} \left[ D_b (\delta g)^{ab} - D^a (\delta g)^c_c \right] ,\nonumber  \\
\Theta^a_{QCG(1)}[g;\delta g] &= \frac{\sqrt{|g|}}{16 \pi G}  \left[ 2 R^{bc} \delta \Gamma^a_{bc} - 2 R^{ab} \delta \Gamma^c_{bc} + D^a R \delta \ln \sqrt{|g|} + 2 D_c {R^a}_b \delta g^{bc} - D^a R_{bc} \delta g^{bc} \right] , \nonumber \\
\Theta^a_{QCG(2)}[g;\delta g] &=2R \bm \Theta_{EH}[g;\delta g] +\frac{\sqrt{|g|}}{16 \pi G} \left[ 4 D^a R \delta \ln \sqrt{|g|} + 2 D_b R \delta g^{ab} \right] . \label{NMG potentials}
\end{align}
These formulae for the following identifications: $d = 4$, $g_{ab} = g^{(0)}_{ab}$, $\beta_1 = -\frac{\ell^3}{4}$, $\beta_2 = -\frac{1}{3}\beta_1 = \frac{\ell^3}{12}$ appear in the sections \ref{sec:Solution space} and \ref{sec:Canonical surface charges}.\hfill{\color{black!40}$\blacksquare$}

%
%


\providecommand{\href}[2]{#2}\begingroup\raggedright\endgroup

%
%

\newpage
\thispagestyle{empty}
$ $
\newpage
\thispagestyle{empty}
\newgeometry{top=2cm, bottom=3cm, left=2.5cm, right=2.5cm}

\begin{tikzpicture}[overlay,remember picture]
	\draw [xshift=4mm,line width=1.5pt,rounded corners=15pt]
        ($ (current page.north west) + (.5cm,-.5cm) $)
        rectangle
        ($ (current page.south east) + (-.5cm,.5cm) $);
    \draw [line width=1pt,rounded corners=15pt,black!75]
        ($ (current page.north west) + (.75cm,-.75cm) $)
        rectangle
        ($ (current page.south east) + (-.75cm,.75cm) $);
    \draw [line width=0.5pt,rounded corners=15pt,black!50]
        ($ (current page.north west) + (1cm,-1cm) $)
        rectangle
        ($ (current page.south east) + (-1cm,1cm) $);
\end{tikzpicture}

\begin{center}

{\Large \textbf{\centering --- Adrien Fiorucci ---}}\\[25pt]
\begin{center}
	{\bfseries\huge Leaky covariant phase spaces}\\[5pt]
	{\bfseries\Large\itshape Theory and application to $\bm\Lambda$-BMS symmetry}
	\end{center}
\vspace{15pt}

Ph.D. thesis. June 2021.\\
\textit{Supervisors:} Prof. G. Comp\`ere \&  Prof. G. Barnich.\\[10pt]
\begin{minipage}[c]{0.65\textwidth}
\begin{center}
\textit{Universit\'e Libre de Bruxelles and International Solvay Institutes
CP 231, B-1050 Brussels, Belgium}
\end{center}
\end{minipage}
$ $\\[30pt]
\begin{minipage}[c]{0.85\textwidth}
\begin{center}
{\small
The present thesis aims at providing a unified description of radiative phase spaces in General Relativity for any value of the cosmological constant using covariant phase space methods. We start by considering generic asymptotically locally (A)dS spacetimes with leaky boundary conditions in the Starobinsky/Fefferman-Graham gauge. The boundary structure is allowed to fluctuate and plays the role of source yielding some flux of gravitational radiation at the boundary. The holographic renormalization procedure is employed to obtain finite surface charges for the whole class of boundary diffeomorphisms and Weyl rescalings. We then propose a boundary gauge fixing isolating the radiative boundary degrees of freedom without constraining the Cauchy problem in asymptotically dS spacetimes. The residual gauge transformations form the infinite-dimensional $\Lambda$-BMS algebroid, which reduces to the Generalized BMS algebra of smooth supertranslations and super-Lorentz transformations in the flat limit. The analysis is repeated in the Bondi gauge in which we identify the analogues of the Bondi news, mass and angular momentum aspects in the presence of a cosmological constant. We give a prescription to perform the flat limit of the phase space and demonstrate how to use this connection to renormalize the corresponding phase space of asymptotically locally flat spacetimes at null infinity including smooth super-Lorentz transformations. In that context, we discuss the memory effects associated with super-Lorentz vacuum transitions and finally provide a new definition of the BMS charges whose fluxes are compatible with soft theorems.} \\[20pt]
\textit{\textbf{Work supported by the F.R.S.-FNRS (Belgium).}}
\end{center}
\end{minipage}
\vfill
\begin{minipage}[c]{0.75\textwidth}
	\begin{minipage}[c]{0.25\textwidth}
		\centering
		\includegraphics[width=0.8\textwidth]{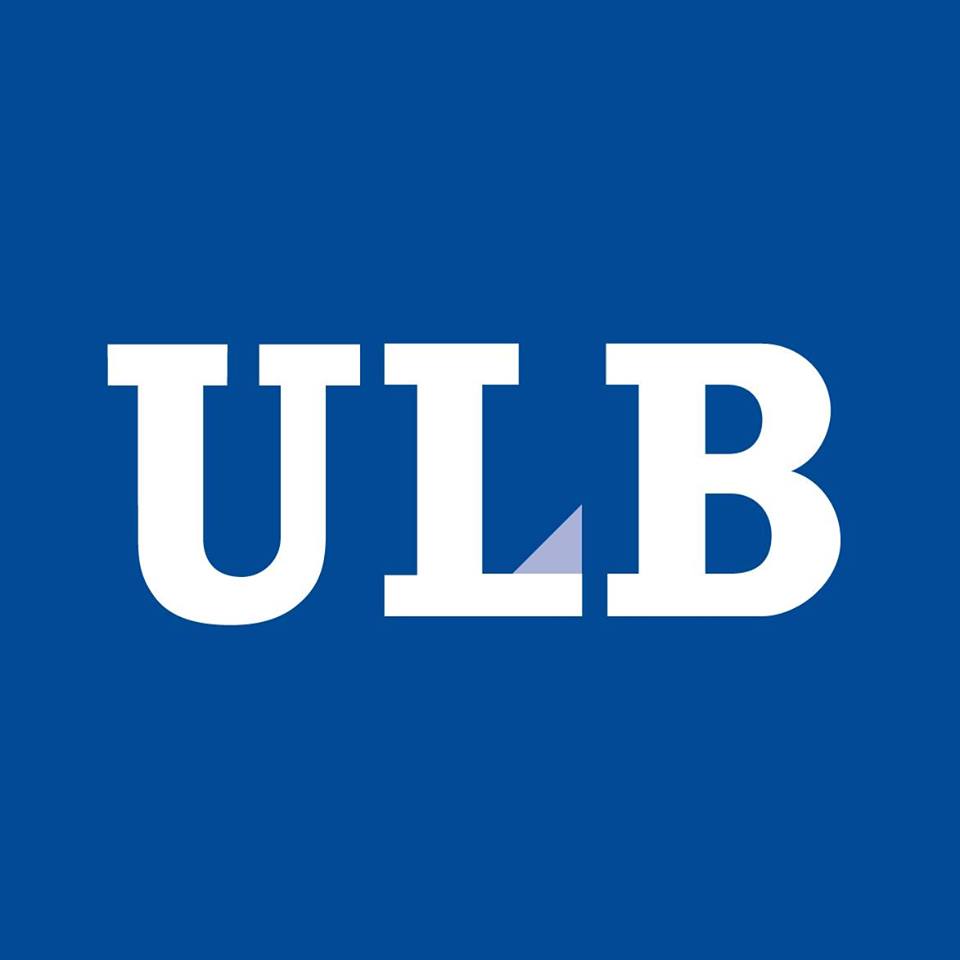}
	\end{minipage}
	\hfill
	\begin{minipage}[c]{0.25\textwidth}
		\centering
		\includegraphics[width=0.8\textwidth]{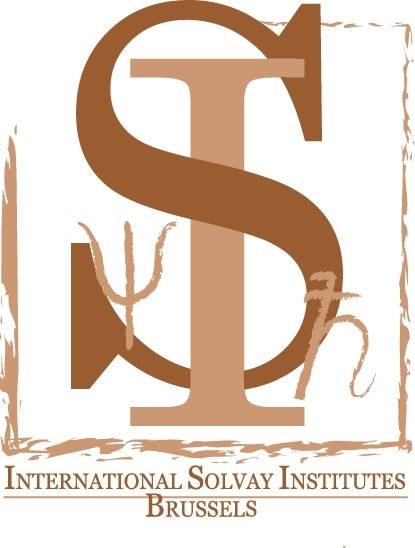}
	\end{minipage}
	\hfill
	\begin{minipage}[c]{0.25\textwidth}
		\centering
		\includegraphics[width=0.85\textwidth]{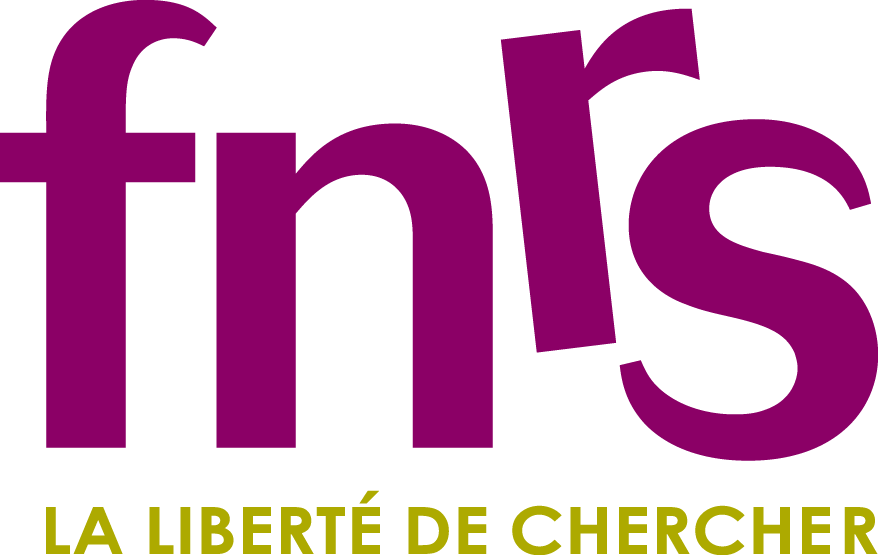}
	\end{minipage}
\end{minipage}

\end{center}

%
%

\end{document}